\documentclass[useAMS,usenatbib]{mnras}
\usepackage{times}
\usepackage{textgreek}
\usepackage{graphicx, latexsym, amssymb, amscd, psfrag, cancel}
\usepackage{epsfig}
\extrafloats{100}
\usepackage{float}
\usepackage{placeins}
\usepackage{color}
\usepackage{amsmath}
\usepackage[upgreek, LGRgreek]{mathastext}
\usepackage{ulem}
\usepackage{longtable}

\usepackage[switch]{lineno}

\title[Diffuse emission in MeerKAT-GCLS clusters]{The MeerKAT Galaxy Cluster Legacy Survey - II. Catalogue of the diffuse radio emission in MeerKAT-GCLS clusters}

\author[Konstantinos Kolokythas et al.]{Konstantinos Kolokythas,$^{1,2}$\thanks{E-mail: kkolok.astro@gmail.com (KK)} Tiziana Venturi,$^{3,1}$ Kenda Knowles,$^{1,2}$ Marcus Br\"{u}ggen,$^{4}$ \newauthor{Francesco de Gasperin,$^{3}$ S. Precious Sikhosana,$^{5,6}$ Tracy E. Clarke,$^{7}$ Oleg Smirnov,$^{1,2,3}$} \newauthor{S. Ilani Loubser,$^{8}$ and Kavilian Moodley$^{5,6}$} \\ \\ 
$^{1}$Centre for Radio Astronomy Techniques and Technologies, Department of Physics and Electronics, Rhodes University,\\ P.O. Box 94, Makhanda 6140, South Africa\\
$^{2}$South African Radio Astronomy Observatory, 2 Fir Street, Observatory
7925, South Africa\\
$^{3}$ INAF -- Istituto di Radioastronomia, via Gobetti 101, I–40129 Bologna, Italy\\
$^{4}$University of Hamburg, Hamburger Sternwarte, Gojenbergsweg 112, 21029 Hamburg, Germany\\
$^{5}$Astrophysics Research Centre, University of KwaZulu-Natal, Durban 3696, South Africa\\
$^{6}$School of Mathematics, Statistics and Computer Science, University of KwaZulu-Natal, Westville Campus, Durban 4041, South Africa\\
$^{7}$U.S. Naval Research Laboratory, 4555 Overlook Avenue SW, Washington, DC 20375, US\\
$^{8}$Centre for Space Research, North-West University, Potchefstroom 2520, South Africa}

\newcommand{\kms}{~km~s$^{-1}$}

\begin{document}

\date{Accepted 2025 September 4. Received 2025 July 27; in original form 2025 May 11}

\pagerange{\pageref{firstpage}--\pageref{lastpage}} \pubyear{2025}
\maketitle

\label{firstpage}
 \begin{abstract}
We present a follow-up study focused on cluster-scale diffuse radio emissions in 115 galaxy clusters in the Southern sky, using full-resolution ($7.8''$) and tapered low-resolution ($15''$) images from the MeerKAT Galaxy Cluster Legacy Survey (MGCLS). In this MGCLS follow-up paper, we update and quantitatively characterise the presence of diffuse radio emission and provide detailed catalogue information on its radio properties at 1.28~GHz.  As the MGCLS sample is heterogeneous and was initially used as a test bed to reveal the scientific potential of MeerKAT, the reported numbers are subject to this special case. More than half ($\sim$54\%; 62/115) of the observed MGCLS clusters present diffuse cluster radio emission, with the total number of diffuse radio sources or candidates detected being 103. Including candidates, we find that radio relics are the most frequently detected diffuse sources in MGCLS at 53\% (55/103), followed by halos at 32\% (33/103) and mini-halos at 10\% (11/103), with only 3\% (3/103) being Phoenixes. The sizes of the diffuse radio structures and candidates range from $\sim$~55~kpc to over 2~Mpc, with P$_{1.28GHz}$ radio power ranging from $\sim$~10$^{22}$ W~Hz$^{-1}$ to greater than 10$^{25}$ W~Hz$^{-1}$. In-band radio spectral index estimates reveal revived radio plasma emissions that exhibit steep radio spectral indices down to $\alpha_{908}^{1656}\sim-3.5$. Mini-halos and their respective candidates are detected mainly in low-mass (M$_{500}$ $\leq$ 5$\times10^{14}$ M$_{\odot}$) and low-power (P$_{1.4\,GHz}$ $\leq$ 10$^{23}$ W~Hz$^{-1}$) systems. We suggest the presence of a statistically significant correlation between the 1.4~GHz radio power and the cluster mass for mini-halo (and candidate) systems. 

 \end{abstract}
 
 \begin{keywords}
   galaxies: clusters: general ---  galaxies: active --- galaxies: clusters: individual ---  radio continuum: galaxies
 \end{keywords}

 
 \section{Introduction} 
 
 The advances in radio interferometers over the past decade are considerably broadening the parameter space available for discoveries in radio astronomy. Over a frequency range from $\sim$~200~MHz to 8~GHz and above, broadband receivers are available on the Very Large Array (VLA; \citealt{Perley11}), upgraded Giant Metrewave Radio Telescope (uGMRT; \citealt{Guptaetal17}), MeerKAT \citep{Jonas2016,Camilo2018} and the Australian SKA Pathfinder (ASKAP; \citealt{Johnsonetal07,Hotanetal21}) allowing new and upgraded radio telescopes a substantial jump in sensitivity. 

The study of diffuse, extended radio emission in galaxy clusters is taking full advantage of these telescope capabilities. The number of antennae and the configuration of ASKAP and MeerKAT provide excellent sensitivity to angular scales ranging from a few arcsecs to arcmin at frequencies close to $\sim$~1~GHz. ASKAP's ongoing deep Evolutionary Map of the Universe continuum survey (EMU; \citealt{Norrisetal11,Norrisetal21}) covers the Southern Sky at 943~MHz, aiming, amongst others, to explore an uncharted region of observational parameter space, revealing new or helping characterising diffuse radio emission in galaxy clusters (e.g., \citealt{Duchesneetal2021,Duchesneetal2021b,Bruggenetal21,Riseleyetal22b,Loietal23}). MeerKAT's Galaxy Cluster Legacy Survey (MGCLS; \citealt{Knowlesetal22}, hereafter K22) has provided data products of selected galaxy clusters with unprecedented sensitivity and resolution, allowing in-depth studies of cluster diffuse radio emission (e.g., \citealt{Sikhosanaetal23,Riseleyetal23,Trehaevenetal23}). At this frequency range, the uGMRT plays a key role in the detection of cluster radio sources (or revealing new features) and the multi-frequency radio studies of galaxy clusters, as it provides the required bandwidth to investigate their wideband spectral properties (e.g., \citealt{DiGennaroetal21,Kaleetal22,Schellenberger22}).

At frequencies below 200~MHz, observations of the LOw Frequency ARray (LOFAR; \citealt{vanHaarlemetal13}) LBA (Low Band ANtenna) and HBA (High Band Antenna) have also unveiled galaxy clusters with unprecedented sensitivity and angular resolution. The recent first data release from the LOFAR LBA Sky Survey (LoLSS; \citealt{deGasperin23}) at 54~MHz can be used to derive unique information on the low-frequency spectral properties of many thousands of radio sources (e.g., \citealt{Pasini24}) whereas at 144~MHz the LOFAR Two-metre Sky Survey (LoTSS-DR2; \citealt{Shimwelletal22}) can be used for surveys identifying new diffuse cluster sources (e.g., \citealt{Hoangetal22} or constructing samples of clusters hosting diffuse radio sources \citep{Botteonetal22} and detailed statistical analyses (e.g., \citealt{Cassanoetal23,Cucitietal23}). In the Southern Hemisphere, the extensive bandwidth of the Murchison Widefield Array (MWA; \citealt{Tingayetal23} \citealt{Waythetal18}) is ideal for investigating spectral properties of steep-spectrum cluster radio sources (e.g., \citealt{Giacintuccietal20,Duchesneetal22}).

More specifically, on the detection of diffuse radio sources, radio halos and relics are found in an increasing number of galaxy clusters (see \citealt{Botteonetal22} and the most recent observational review by \citealt{2019SSRv..215...16V}). These are Mpc-scale, steep synchrotron spectrum ($\alpha \leq -1$, for S$\propto \nu^{\alpha}$) sources located at the centre or in the outskirts of the host cluster. Mini-halos are smaller in size (typically below 500~kpc) and are characterized by steep spectra, too \citep{2017ApJ...841...71G}.  A close connection between cluster dynamics and the presence (or lack) of radio halos and relics is an established result \citep{2013ApJ...777..141C, Kale2015EGRHS,2021A&A...647A..50C,2021A&A...647A..51C}, with radio halos and relics typically found in merging clusters (e.g., the Toothbrush, \citealt{2012A&A...546A.124V}; ZwCl008, \citealt{2019ApJ...873...64D}; MACS J0717.5+3745, \citealt{2021A&A...646A..56R}), and mini-halos in relaxed systems \citep{Govonietal09,Giacintuccietal14,Giacintuccietal19}. 
 
 The spatial coincidence of halos with the X--ray emission of the cluster, the location of relics at X--ray discontinuities and the confinement of mini-halos by cold fronts in the inter-cluster medium (ICM) clearly show a tight link between the thermal and non-thermal components of galaxy clusters \citep[see][]{2014IJMPD..2330007B}. Radio halos and relics are thought to originate from re-acceleration processes induced by turbulence and shocks, respectively during the formation of large-scale structures \citep{2014IJMPD..2330007B}; mini-halos, too, can be explained in the re-acceleration scenario, although the energy supplied may come from core sloshing at the centre of relaxed systems \citep{Giacintuccietal14,2017ApJ...841...71G,2014IJMPD..2330007B}.

 LOFAR and uGMRT observations have shown that the connection between cluster dynamics and the type of diffuse radio source may be more complex, as extended emission beyond the cold fronts and on a scale broader than that of mini-halos has been found in a few cases \citep{Savinietal18,Biavaetal21,Hoangetal25}.
 Recent observations with LOFAR have enabled the discovery of significantly extended, faint `megahalos' that extend out to $R_{500}$, as well as the discovery of inter-cluster bridges of emission spanning Mpc scales, connecting galaxy clusters. \citet{2022Natur.609..911C} observed four clusters whose radio halos are embedded in a more extended diffuse radio emission, filling a volume 30 times larger than typical radio halos. In addition, LOFAR has detected a radio emission bridge between Abell~399 and Abell~401 \citep{Govonietal19}, and a radio bridge connecting two galaxy subclusters in Abell~1758 \citep{2020MNRAS.499L..11B}, while MeerKAT made the first GHz detection of a bridge connecting Abell~3562 and the galaxy SC1329--313 in the Shapley Concentration \citep{Venturietal22}. 
 
 
 The advent of MeerKAT offers high sensitivity, dynamic range, and resolution, dramatically improving the image quality and fidelity of diffuse cluster radio emission in complex fields at frequencies around GHz. The MGCLS observations presented by K22 was carried out in 2018 on a heterogeneous sample of 115 galaxy clusters. Apart from making various legacy products available to the astronomical community and providing information on data processing, K22 also offered radio source classification and initial science findings in cluster diffuse emission, radio galaxy physics, star-forming systems, and neutral hydrogen (H\textsc{i}) mapping capabilities. In this follow-up paper to K22, we focus on cluster-scale diffuse radio emissions in the MGCLS. By examining MGCLS full-resolution and tapered images for all sources, our purpose is to provide a detailed catalogue with the measured observational properties of the detected cluster-scale diffuse radio emission sources (including flux densities) in the MGCLS and reveal a) the scientific potential of MeerKAT observations in our understanding of the origin of these sources, b) MeerKAT's ability to enable new detections, c) the challenges in characterising and classifying diffuse radio sources, and d) the potential necessity in the near future to introduce new classes of diffuse radio emission that have not been seen before.
 
 
 
 The paper is organized as follows: Section~\ref{Sec2} summarizes the selection criteria, and the sample properties; Section~\ref{Sec3} provides an overviewiew of the data and details off the image analysis;  Section~\ref{Sec4} describes the diffuse radio emission classification; Statistics on the sample, discoveries and a general discussion are presented in Section~\ref{sec5}. The summary and conclusions are given in Section~\ref{sec6}. Literature reviews and descriptions of each galaxy cluster that presents diffuse structures are provided in the Appendix~\ref{AppB}. The images for the sample galaxy clusters with diffuse radio emission are presented in the Appendix~\ref{AppC}.  We assume a $\Lambda$CDM
cosmology of H$_o$ = 70 Mpc$^{-1}$, $\Omega_m$ = 0.30, $\Omega_\Lambda$ = 0.70 for all MGCLS clusters. We adopt the convention of S $\propto\nu^\alpha$ where S is the flux density, $\nu$ is the observing frequency, and $\alpha$ is the spectral index. All uncertainties are quoted at the 1$\sigma$ level of significance.

\section{The Sample}
\label{Sec2}
The MGCLS is a sample of 115 galaxy clusters with a mean redshift of $\sim 0.14$ (only four clusters have \textit{z} $> 0.4$) that span the Southern sky between declination $-$80$^{\circ}$ and $\sim$~0$^{\circ}$. K22 provides a detailed description of the sample, observations and data; we only give a summary. The targeted clusters were observed between June 2018 and June 2019 and were scheduled as `fillers' during
observing schedule gaps. For observations, the complete MeerKAT radio telescope array (minimum of 59 antennas per observation) with the L-band receiver (900–1670 MHz) in 4K mode (4096 channels) was used, with an integration time of 8 seconds. The duration of the observations was between 8 and 12 hours, repeating a cycle between the target cluster and the various calibrator sources, with the integrated time on source spanning from 5.5 to 9.5 hours.

For the MGCLS sample, no specific selection criteria were applied, in mass, redshift, or luminosity, making MGCLS a heterogeneous sample consisting of clusters that were drawn from i) a group of “radio-selected” clusters and ii) a group of “X--ray-selected” ones. The radio subsample consists of 40 clusters, most of which exhibit known diffuse cluster radio emission from several earlier studies \citep[e.g.,][]{1999NewA....4..141G,Feretti2012,Lindner2014,Kale2015EGRHS,Shakouri2016ARDES,2017ApJ...841...71G,2017MNRAS.472..940K,2017MNRAS.470.3465B,Golovich2019} but also includes systems without any prior known diffuse emission that exhibit radio loud AGN emission or candidate structures.  The “radio-selected” clusters span redshifts between 0.018 and 0.870, with a median redshift of \textit{z}$\sim0.22$. Due to this selection, the radio sub-sample contains only high-mass clusters ($M_{500} \geq 6\times10^{14}~M_{\odot}$) and is a priori strongly biased towards clusters that present diffuse emission structures (radio halos and relics). 

On the other hand, the remaining clusters (75 clusters or 65\% of the MGCLS sample) consist of the X--ray sub-sample that was selected from the MCXC catalogue \citep{2011A&A...534A.109P}, in an effort to balance the MGCLS sample cluster selection and avoid any prior biases towards or against cluster radio properties including clusters that were not radio selected. The X--ray subsample includes systems that exhibit luminosities L$_X$ between $\sim10^{43} - 3\times10^{45} erg~s^{-1}$ spanning redshifts between 0.011 and 0.370 with a median of \textit{z}$\sim0.13$. We note that X--ray selection is very different from the optical selection. In the future, eROSITA could provide new X--ray samples \citep{Bulbuletal22,Merlonietal24,Bulbuletal24}. For more information on sample selection, see K22.

The MGCLS sample, therefore, was drawn to test MeerKAT's performance at the specifications of the L-band frequency. Focusing on known clusters that are ideal for diffuse emission studies and those in which the diffuse emission is potential allows one to evaluate the telescope's image product quality, sensitivity, and ability to detect new radio sources.

\section{Data and image analysis}
\label{Sec3}

All MGCLS datasets were calibrated and imaged with the basic procedure described in \citet{DEEP2} using the Obit package 2 \citep{OBIT}. The images were corrected for the primary beam at each frequency, as described in \citet{DEEP2}, both at the full-resolution of the image ($\sim$ 7.5 $-$ 8$''$) and at a convolved 15$''$ resolution image to aid in the recovery of low-surface-brightness features. As in K22, diffuse radio sources in our sample were classified by visually inspecting MGCLS 1.28~GHz images at different resolutions without subtracting point radio sources. The excellent short baseline spacing of MeerKAT (29~m is the minimum baseline length; \citealt{Camilo2018}) is key for detecting diffuse radio structures and allows the full recovery of up to 10$'$ extended emission in angular scales. For more details on the MGCLS data and their analysis, see \S 3 and 4 of K22.

For the flux density calculation of the diffuse radio sources, we used a similar strategy as in K22 for the highlighted systems, estimating the total flux density at the 3$\sigma$ level of significance of a radio structure using the 15$''$ resolution image and then extracting the contribution of any central AGN or background point radio sources within this region using the full-resolution image. Physical projected sizes were calculated based on the 15$''$ resolution image (see Table~\ref{tab:diffuse}), assuming that every diffuse structure has a relevant 2-D geometrical shape, with the largest physical projected sizes estimated in the form of a minor $\times$ major axis of a fitted 2D Gaussian. The largest physical linear size (LLS) was calculated as the distance between the two points with maximum separation, which are considered part of the diffuse radio emission, following \citet{Jonesetal23}. This line of maximum separation is then used to calculate the minor axis, by drawing a line perpendicular to the LLS and calculating the maximum distance (or width in the case of elongated structures) between the pixels that are part of the radio emission. The error in the LLS corresponds to one beam width. We note that this can be considered a lower limit for the LLS due to the available resolution used from the products and similarly for the width measurements of the radio relics since the downstream extent may be too faint to detect. We used the same approach for all other size measurements presented in this paper, apart from the arc-shaped (bent) structure cases in relics, for which an approximate linear size estimation was not feasible. In these cases, we estimated the largest physical size by following the path between the two points of the diffuse radio emission with maximum separation (connecting two lines) to better constrain the extent of these radio sources. Despite our efforts, we note that these are rough estimates. We adopted a typical 6\% uncertainty (for more details see MGCLS~I; K22) in the flux density scale of our MeerKAT images that includes systematic and calibration uncertainties. 
We have provided estimates of the integrated primary-beam-corrected $\alpha_{908}^{1656}$ spectral index for the diffuse radio emission in the MGCLS sample, which was possible to extract by inspecting the in-band spectral index maps. The description of the in-band spectral indices for the relevant systems in which a spectral index estimation was possible is mentioned in the literature review text of the Appendix~\ref{AppB}. We note here that these are only tentative estimates of the spectral indices for each radio source, as the in-band spectral index estimation requires splitting into subbands that reduce the image sensitivity by nearly a factor of 4, causing parts of the faint diffuse radio sources to fall below detection. The coordinates for the radio halo and mini-halo radio sources are considered coincident with the NED\footnote{NASA/IPAC Extragalactic Database (NED, http://ned.ipac.caltech.edu)} position of their clusters (see Table~\ref{tab:diffuse}, columns 2-3); however, for the radio sources (radio relics or Phoenixes) whose location does not coincide with the NED position of their cluster (e.g., they are located at their periphery), their position is determined based on the local epicentre of each radio source based on their measured minor $\times$ major axis.

\section{Classification of diffuse radio emission}
\label{Sec4}

One aspect of observing galaxy clusters is the search for diffuse, cluster-scale radio synchrotron emission. The detection of such structures plays a key role as it can reveal information regarding the evolution and the formation history of a cluster (for an observational detailed review, see, e.g., \citealt{2019SSRv..215...16V}, and for a theoretical one, see \citealt{2014IJMPD..2330007B}). 

Our aim in this work is to present the detected diffuse synchrotron sources from the MGCLS sample in detail, along with relevant information on their host clusters. Classifying diffuse radio emission is challenging using only radio data because of the presence of contaminating discrete radio sources (either AGN or star-forming galaxies) and certain calibration artefacts. The MGCLS radio data products were also inspected while overlaid with optical and X--ray data, where available, to better understand the nature of a diffuse source. 

However, the distinction between radio-emitting plasma of AGN origin and a cluster diffuse radio source is not always straightforward, and the classification remained uncertain in several clusters. MGCLS observations are very sensitive to ICM relativistic electrons, revealing faint halos, but MGCLS images also show that AGN contribute a significant fraction of ionized radio plasma. This means that radio emission from bent and/or tailed AGN can spatially coincide with radio emission from relics and halos, making the observational separation between a first-hand accelerated AGN plasma and an in-situ re-accelerated ICM plasma very challenging. This highlights the critical role of the initial (seed as often called) electrons in forming cluster-scale diffuse sources, a key aspect in (re)acceleration models \citep{2019SSRv..215...16V,Botteonetal22}. 

Spatially resolved radio spectral index mapping and comparison with the thermal X--ray properties of the ICM play a significant role in correctly classifying and understanding the origin of the extended diffuse radio emission. A great variety of low surface brightness and steep radio spectra ($\alpha\leq-1.0$) diffuse radio morphologies has been detected to date in clusters; however, historically, the most widely studied and dominant classes into which almost every detected diffuse cluster radio emission falls, are: a) \textit{Radio Halos}, b) \textit{Radio mini-halos}, and c) \textit{Radio relics}.  In more detail:

a) \textit{Radio Halos (RH)} are diffuse sources that are typically correlated with the distribution of the X--ray-emitting ICM, extending on cluster scales. The main mechanism that gives rise to these structures that span scales greater than 500~kpc (up to a few Mpc) is turbulence-driven particle re-acceleration as a result of massive cluster mergers, with radio halos exhibiting observed correlations between the radio source's power and its host cluster mass, its thermal X--ray--emitting ICM properties, and cluster morphological parameters, with the occurrence of the latter increasing with lower frequencies \citep{2013ApJ...777..141C,Kale2015EGRHS,2019SSRv..215...16V,2021A&A...647A..50C,2021A&A...647A..51C,Botteonetal22}.

b) \textit{Radio Mini-Halos (MH)} are structures smaller than radio halos, with projected sizes ranging from a few tens to a few hundreds of kpc, situated in the central area of dynamically relaxed, cool-core clusters \citep{2017ApJ...841...71G}, most of the time confined within cold fronts observed at the cluster core. A radio active brightest cluster galaxy (BCG) is always present in mini-halo clusters that are contributing, at minimum, part of the seed electrons necessary to generate the observed diffuse emission (e.g., \citealt{2020MNRAS.499.2934R}). As is the case in radio halos, particle re-acceleration induced by gas sloshing is most likely the driving mechanism for the production of radio mini-halos (e.g., \citealt{2011ApJ...743...16Z,2013ApJ...762...78Z}).


c) \textit{Radio relics} are a class in which elongated kpc-to Mpc-scale structures are usually observed at the outskirts of merging galaxy clusters. One of the main observed properties of radio relics is that they present a high degree of polarisation (e.g., \citealt{2010Sci...330..347V}), which denotes that their origin is closely related to the presence of merger-induced shocks in the ICM. For this reason, numerous clusters exhibit relics in opposite directions in the periphery of a cluster, also known as double radio relics (e.g., Abell 3667; \citealt{1997MNRAS.290..577R}) either with the detection of
a radio halo in-between them or not (see \citealt{2012MNRAS.426...40B,Lindner2014}). This class also includes a sub-class of revived fossil emission from radio AGN in the region of the cluster known as radio Phoenixes \citep{2019SSRv..215...16V}.

In addition to these three well-known categories, radio sources in this study for which the diffuse emission is not consistent with any of the current categories of radio halos and relics, also based on available ancillary data, are classified as \textit{Unclassified/Unknown diffuse emission (U)}. This classification mainly collects Phoenixes or (revived) fossil plasma diffuse radio sources \citep{Kempneretal04}, i.e., an extended radio source that is not a radio galaxy or AGN and falls outside of the established radio halo and relic classifications without obvious optical identification. 

Lastly, we assign the category of candidate sources \textit{candidate radio halo, mini-halo, or candidate relic (cMHalo or cR)} that present a marginal detection or an uncertain feature that is, however, in agreement with the main properties of these types of source also based on the location of the radio emission on the optical image and the available X--ray data. This class can include systems that cannot be adequately classified due to poor data quality, e.g., those affected by calibration artefacts or systems, or where the nature of their diffuse emission is uncertain due to severe contamination from superimposed compact sources in the cluster.

Although alternative classification approaches have been used to provide reproducible classifications, such as the decision tree classification by \citet{Botteonetal22}, our classification scheme is less strict and similar to the visual inspection classification approach by \citet{Hoangetal22} applied to non-Planck Sunyaev-Zeldovich (SZ) cluster catalogue (PSZ2) clusters in the LoTSS-DR2 data and the \citet{Duchesneetal24} pilot search of diffuse sources in 71 PSZ2 clusters from archival ASKAP observations. When comparing classification methods using the same decision tree, both approaches yield similar results for classifying radio halos and relics \citep{Hoangetal22}. However, for the classification of the remaining unknown diffuse sources and AGN-related emission, the two approaches may be less consistent.

\begin{figure*}
\centering
\includegraphics[width=0.98\textwidth]{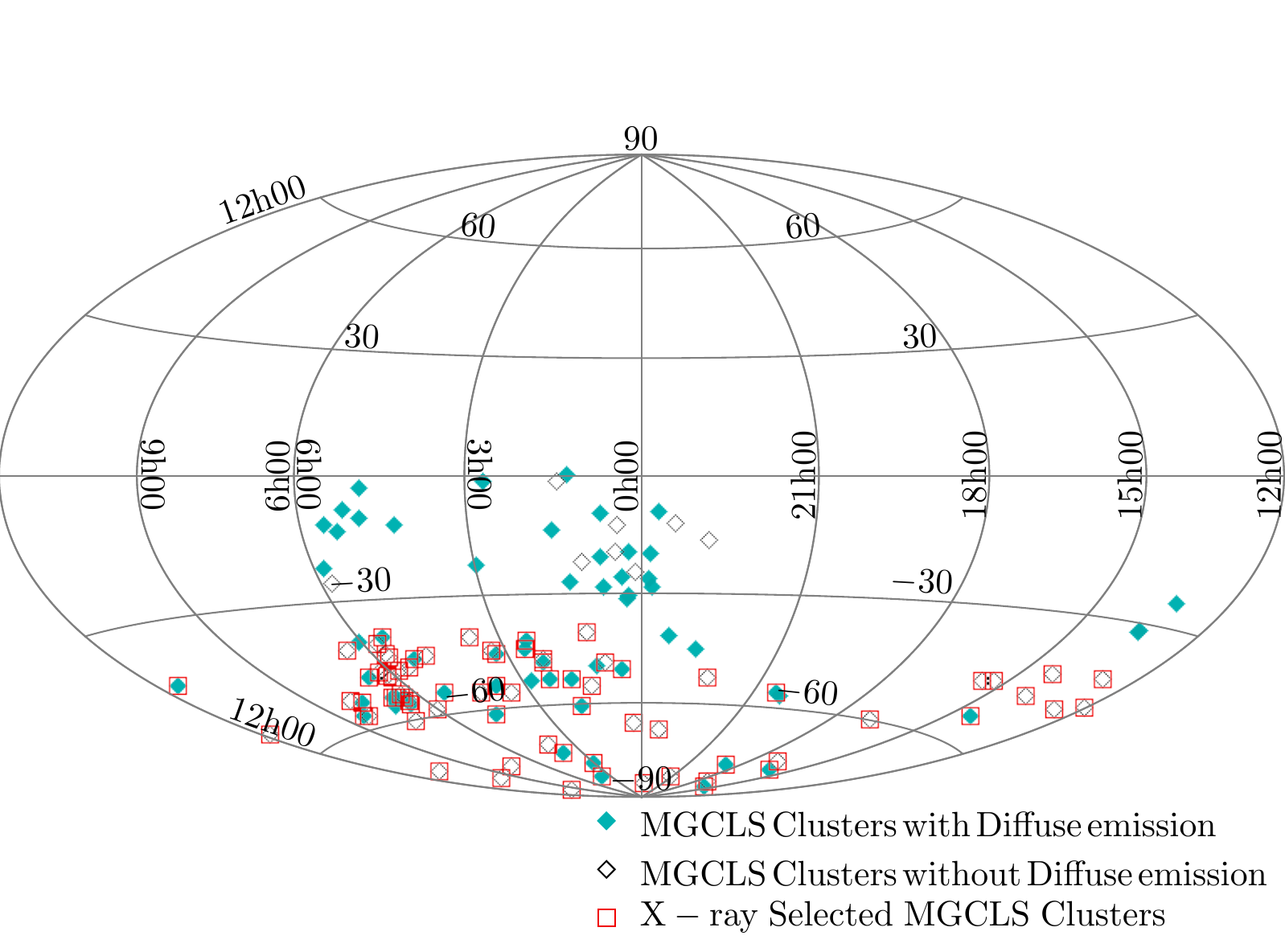}
\caption{Position of the 115 MGCLS clusters. The clusters where some kind of diffuse radio emission is detected are shown in cyan rhombi, whereas open rhombi are used to show the ones without any diffuse radio emission detected. Open red squares show the X--ray selected MGCLS clusters.}
\label{Skydiffuse}
\end{figure*}

\section{Results and Discussion} 
\label{sec5}
This work focuses on the MGCLS clusters where diffuse radio emission has been reported in K22 (see their Table~4), providing a detailed description and complete analysis of their properties. The features reported here are the sizes and flux densities visible in the 15$^{\prime\prime}$ resolution images. Separating the diffuse emission by filtering or source subtraction, followed by convolution, might lead to larger values in some cases. The properties of the 62 MGCLS clusters with diffuse emission, along with relevant information on the radio diffuse structures and image quality data, are shown in Table~\ref{tab:diffuse}. 
We underline that in this study, we have expanded and specified the characterisation for 8 out of 9 ambiguous/unknown diffuse radio sources mentioned in K22. To the best of our knowledge, the classification of the nature of 8/9 systems that were characterised as unknown or uncertain in K22 is updated, along with updates in a few more systems. Characterising and updating the classification of radio sources in these systems has increased the total number of MGCLS diffuse radio structures from 99 in K22 to 103.  Detailed information on 56 individual MGCLS systems is presented in the Appendix~\ref{AppB}, excluding the six systems (Abell~85, Abell~3667, RXC~J520.7-1328, J0351.1-8212, J0352.4-7401, and J0631.3-5610) which have been presented and their cases extensively highlighted in K22. Lastly, radio images at different resolutions for the 56 MGCLS systems are overlaid on optical images of the Digital Sky Survey (DSS) and are shown in the Appendix~\ref{AppC}.

In this section, we present and discuss the number and detection fraction statistics of the diffuse radio emission in the MGCLS sample, examining in detail the detection fractions of individual radio structures with properties such as radio power ($P_{1.28GHz}$), redshift, LLS and cluster mass. We determine the $M_{500} - P_{1.28GHz}$ scaling relation for the detected radio halos with estimated SZ masses from Planck PSZ2 survey \citep{Planckcollab16} and also examine the existence of a similar correlation for the mini-halos.

\subsection{Detection fractions and distribution of diffuse radio sources}

The MGCLS sample comprises 115 clusters that form a non-statistically (heterogeneously) selected sample. However, as mentioned earlier, to balance the cluster selection and bypass any biases towards or against known cluster radio properties, the MGCLS sample consists of two sub-samples: a radio-selected class of clusters and an X--ray-selected one. We focus on the MGCLS clusters where diffuse radio emission has been detected. We describe the total sample results and each sub-sample's detection fractions.

More than half ($\sim$~54\%; 62/115) of the observed MGCLS clusters in this Legacy Survey (K22) present some diffuse cluster radio emission (see Fig.~\ref{Skydiffuse}). In this work, we find the total number of detected diffuse radio structures or candidate radio structures being 103. This denotes that several MGCLS clusters host more than one radio structure or candidate radio structures, with 58\% (60/103) of these detected radio structures reported as new in K22. Our 103 diffuse cluster emission detections can be summarised as follows: three mini-halos (all listed as new in K22) and eight mini-halo candidates (all new), 26 halo detections (eight new), and seven halo candidates (five new), 31 relics (15 new), and 24 relic candidates (18 new), one radio Phoenix and two Phoenix candidates (all new), and one new diffuse source with ambiguous or unknown classification.

Our findings reveal that (including the candidates) only $\sim$~10\% (11/115) of the observed MGCLS clusters present a radio mini-halo. In contrast, the detection rate for radio halos in MGCLS clusters is much higher at 29\% (33/115). The detection rate for clusters that exhibit only a radio halo without the presence of a relic is 16\% (18/115). The detection rate of MGCLS clusters that present at least one radio relic is 29\% (33/115) as some clusters present more than one relic, whereas the detection rate for the MGCLS clusters with radio halo that present double relics is $\sim$~7\% (8/115; Abell~521, Abell~2744, El Gordo, MACS~J0417.5-1154, RXC~J1314.4-2515, J0232.2-4420, J0352.4-7401, J0516.6-5430). Examining the 103 detected radio sources in all clusters revealed that radio relics are the most commonly detected diffuse structures in MGCLS with an occurrence rate of 53\% (55/103), followed by halos at 32\% (33/103) and mini-halos at 11\% (11/103). Only 3\% (3/103) are found to be Phoenixes, with only 1\% of the detected radio structures listed as ambiguous/unknown.

\begin{figure}
\centering
\includegraphics[width=0.49\textwidth]{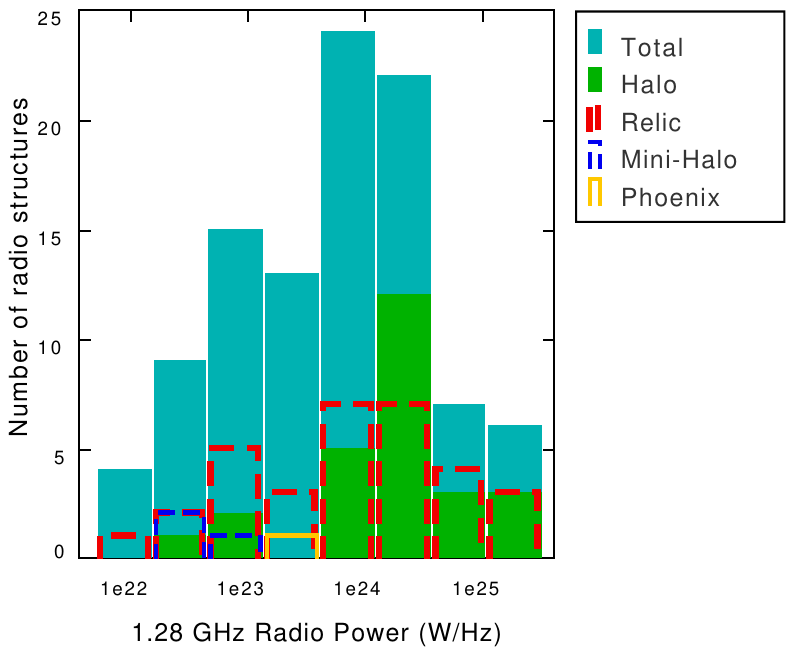}
\caption{Radio power distribution of the detected diffuse radio emission structures in the MGCLS sample at 1.28~GHz. Total refers to the number of detected radio structures and candidates (cyan columns). The radio halos are shown in overlapping green columns, the radio relics in red dashed line, the radio mini-halos in blue dashed line, whereas the detected radio Phoenix with a yellow solid line.}
\label{HistP1280structures}
\end{figure}

\begin{figure}
\centering
\includegraphics[width=0.49\textwidth]{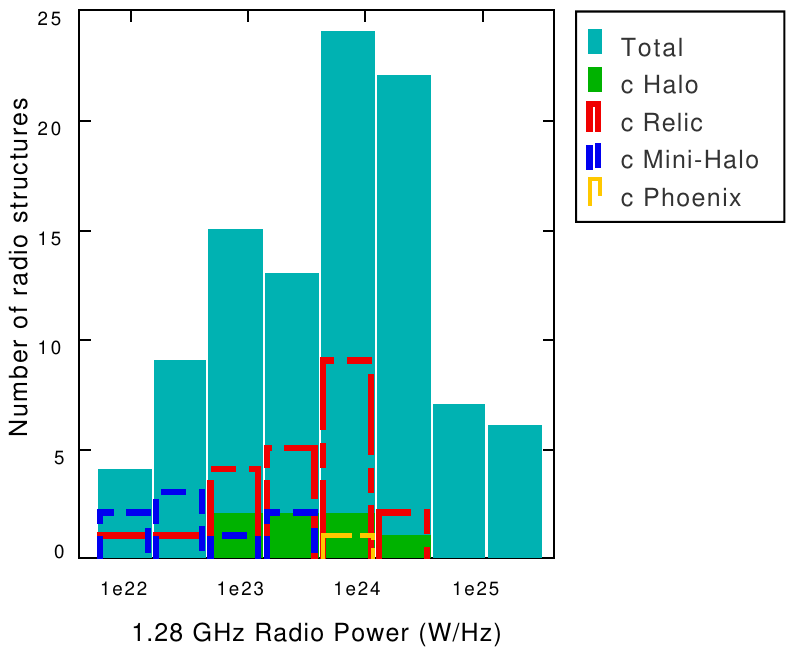}
\caption{Radio power distribution of the detected diffuse radio emission candidate structures in the MGCLS sample at 1.28~GHz. Total refers to the number of detected radio structures and candidates (cyan columns). Colour description is the same as in Figure~\ref{HistP1280structures} but for candidates.}
\label{HistP1280candidates}
\end{figure}

\begin{figure}
\centering
\includegraphics[width=0.49\textwidth]{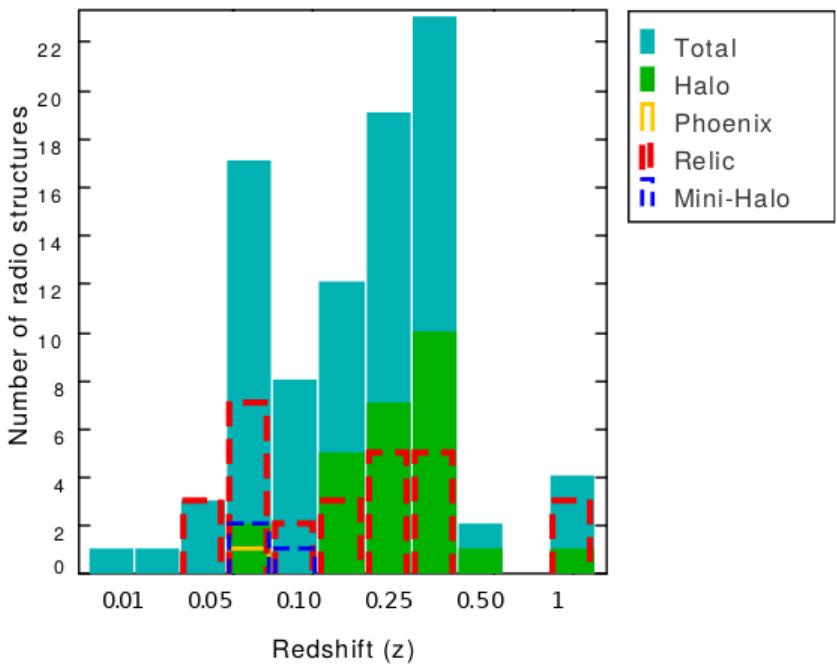}
\caption{Redshift distribution of the detected diffuse radio emission structures in the MGCLS sample at 1.28~GHz. Similarly to Fig.~\ref{HistP1280structures}, Total refers to the number of detected radio structures and candidates (cyan columns).  Colour description is the same as in Figure~\ref{HistP1280structures}.}
\label{Histzstructures}
\end{figure}

\begin{figure}
\centering
\includegraphics[width=0.49\textwidth]{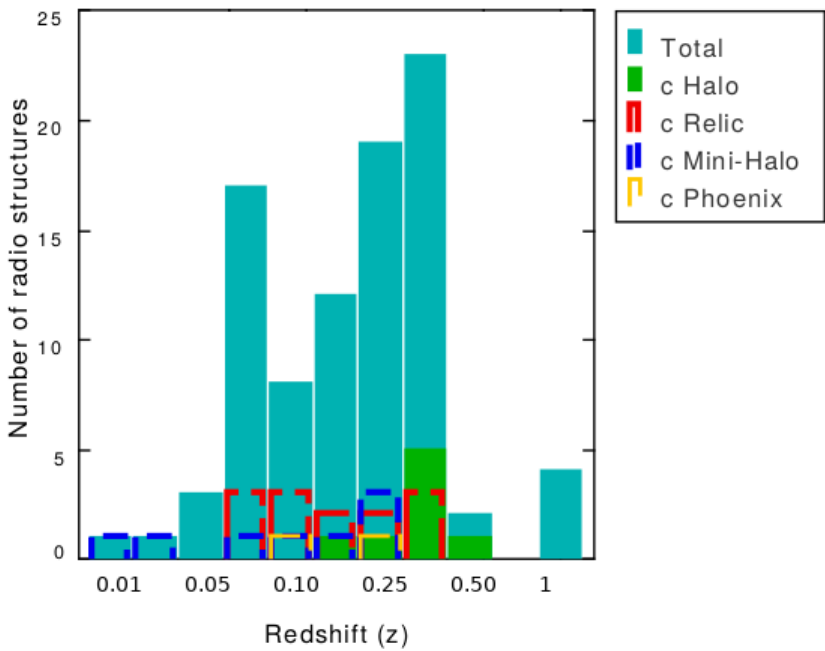}
\caption{Redshift distribution of the detected diffuse radio emission structures in the MGCLS sample at 1.28~GHz. Similarly to Fig.~\ref{HistP1280structures}, Total refers to the number of detected radio structures and candidates (cyan columns). Colour description is the same as in Figure~\ref{HistP1280structures} but for candidates.}
\label{Histzcandidates}
\end{figure}

The detected MGCLS diffuse radio structures and candidates vary significantly in size, flux density, spectral index, and radio power, indicating the diversity in their properties. Figure~\ref{HistP1280structures} presents the 1.28~GHz radio power distribution of the MGCLS detected structures, whereas Figure~\ref{HistP1280candidates} presents the equivalent histogram for the candidate radio structures. Figures~\ref{Histzstructures} and \ref{Histzcandidates} present the redshift distribution of the structures and candidates detected by MGCLS, respectively.

\begin{figure}
\centering
\includegraphics[width=0.49\textwidth]{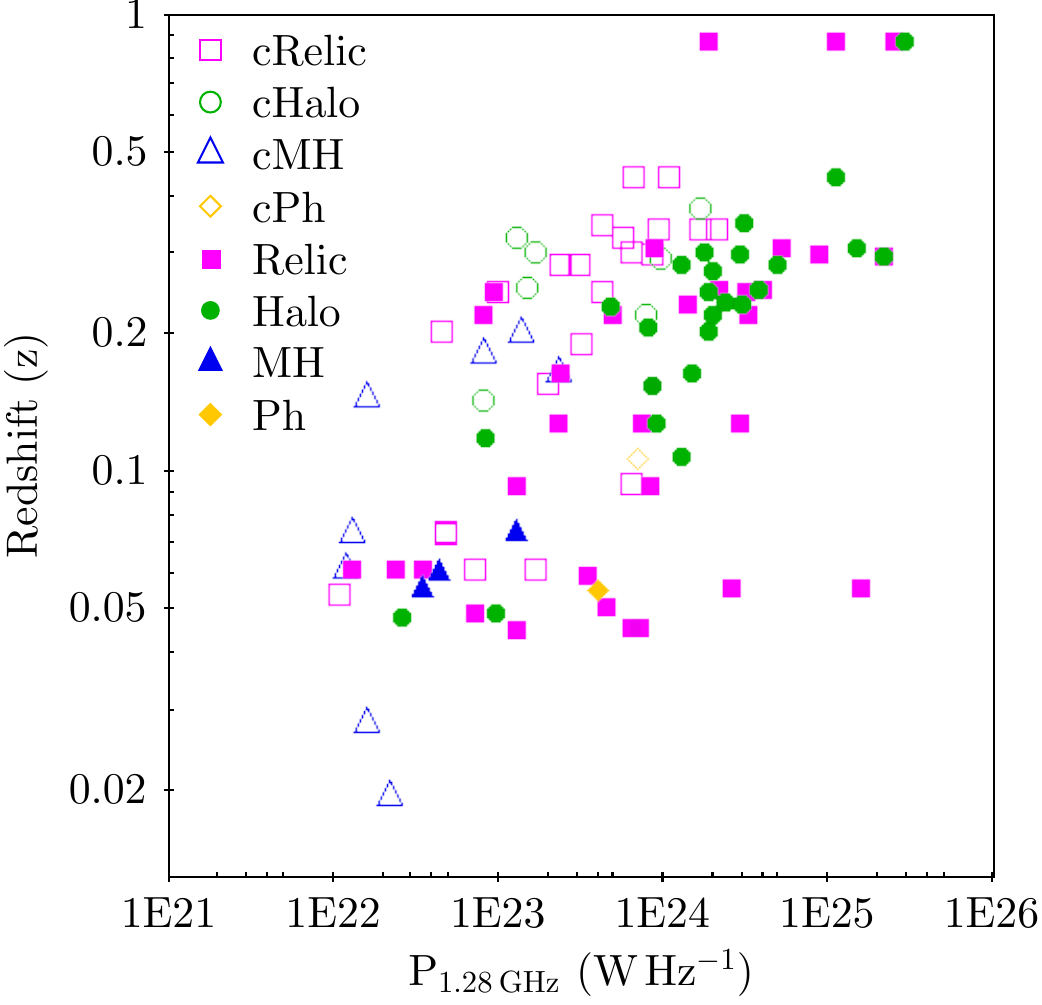}
\caption{1.28~GHz radio power vs redshift $\textit{z}$ of the 103 diffuse emission structures detected in MGCLS clusters. The different symbols shown indicate the corresponding radio morphologies with candidate structures shown in equivalent open symbols.}
\label{P1280vsRedshift}
\end{figure}

We find that the projected sizes in the MGCLS sample range from as small as $\sim$~55~kpc (J1840.6$-$7709; candidate mini-Halo) to as big as $\sim$~2.3~Mpc (Abell~3667; relic). The estimated flux densities at 1.28~GHz range between as faint as $\sim$~0.4 mJy (MACS~J0257.6$-$2209; candidate radio halo) and as strong as $\sim$~400 mJy (A3667; relic), with the radio power (P$_{1.28GHz}$) of the detected structures spanning three orders of magnitude (from $\sim$~10$^{22}$ W~Hz$^{-1}$ to greater than 10$^{25}$ W~Hz$^{-1}$). Examining the in-band radio spectral index distribution, we find structures that have regions with relatively flat spectral index ($\alpha_{908}^{1656}\sim-0.50$) with radio relics exhibiting steep spectra indices such as $-$3.5. From Figure~\ref{P1280vsRedshift}, we see that diffuse radio structures are detected all over the redshift extent of the MGCLS sample, with the majority of the detected radio halos and the respective candidates found mainly between redshift $\sim$ 0.2 $-$ 0.4. Naturally, this result is constrained by the redshift span of the sample selection. However, since the MGCLS sample is heterogeneous and is not based on the redshift of the clusters, the radio halo occurrence in this range is in line with the observed abundance of interacting/merging systems in intermediate redshift clusters (e.g., \citealt{1994ApJ...430..107D,Cassanoetal06}) where galaxy clusters are growing by the infall of less massive groups with much lower velocity dispersions, thus enabling strong, slow encounters \citep{2004cgpc.symp..277M}.
 


\begin{figure}
\centering
\includegraphics[width=0.49\textwidth]{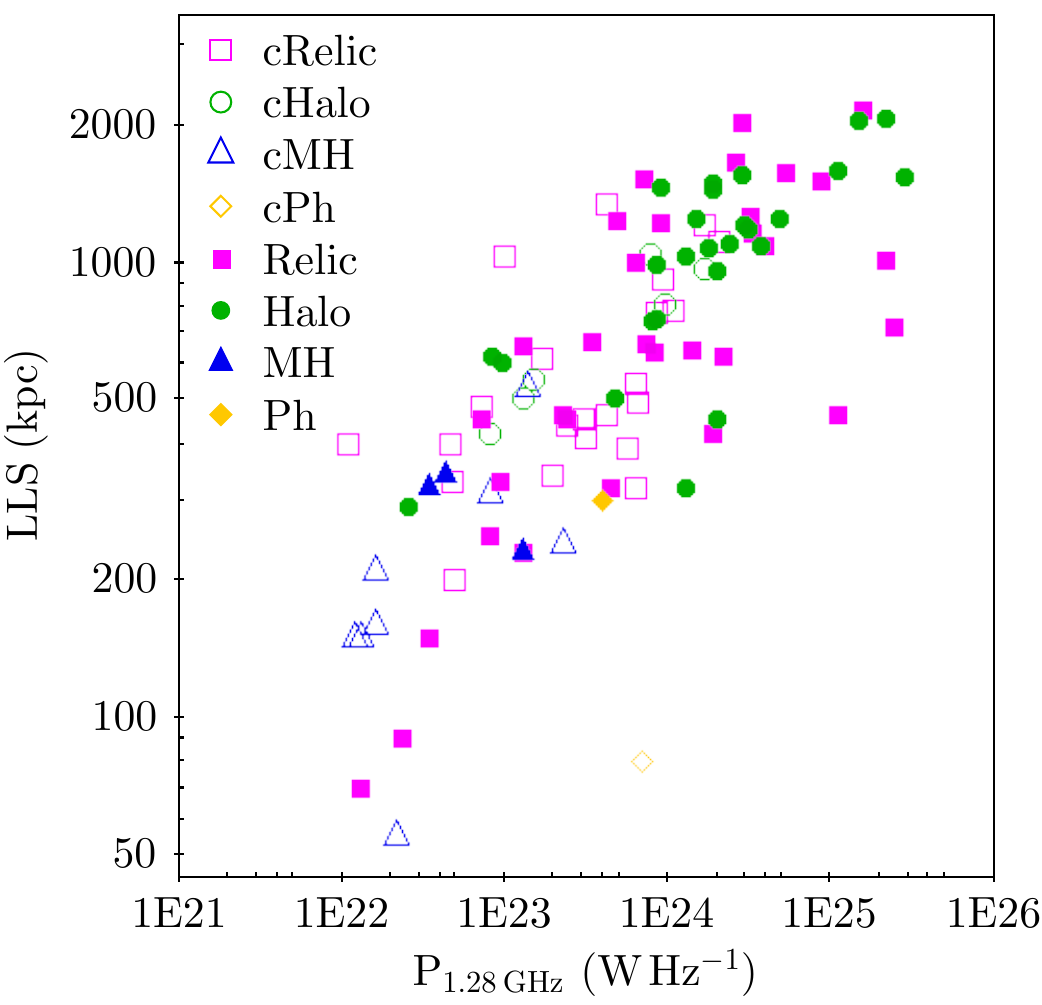}
\caption{1.28~GHz radio power vs Largest Linear Size (LLS) of the 103 diffuse emission structures detected in MGCLS clusters. The different symbols shown indicate the corresponding radio morphologies with candidate structures shown in equivalent open symbols.}
\label{P1280vsLLS}
\end{figure}

We also examined the relationship between the 1.28~GHz power of the different diffuse MGCLS radio structures against their LLS. Figure~\ref{P1280vsLLS} shows the relation for all diffuse radio sources resolved morphologically, with their LLS calculated along the longest extent of the detected diffuse radio emission, assuming a geometry that represents each structure based on a projected minor $\times$ major axis within 3$\sigma$. We find that the MGCLS radio halo and mini-halo diffuse radio sources are in agreement with the linear correlation between size and radio power \citep{Cassanoetal07,Murgiaetal09} for cluster environments that stands for almost four orders of magnitude at 1.28~GHz.

 We find that 33/40 clusters in the radio-selected sample present diffuse radio emission, whereas 29/70 clusters in the X--ray-selected sample present diffuse radio emission. Focusing only on the new radio detections in the MGCLS-selected subsamples, we find that 31\% of these new structures are discovered in the radio-selected sample (19/61), whereas the majority (69\%; 42/61) of the newly detected radio structures are seen in the X--ray--selected sample. This suggests that existing X--ray--selected cluster samples in the Southern sky are more likely to reveal new radio structures complementing earlier radio surveys/studies, as most X--ray--selected clusters in the sample are located at declinations below $-$40$^{\circ}$, which was the limit for existing radio facilities in the Southern sky before the advent of MeerKAT. In other words, the X--ray subsample was largely unexplored before the advent of MeerKAT. However, this is also possible due to MeerKAT's unique ability as an instrument to detect new radio structures in a `blind' search. MeerKAT's dense core configuration at relatively high resolution is optimized towards this direction and, in combination with its unprecedented high sensitivity, makes it an ideal tool for performing searches for the detection of diffuse radio emission, opening the path to discoveries. However, due to the much improved surface brightness sensitivity of the MGCLS at 1.28~GHz, we also note that a significant number of new faint radio structures have been confirmed/found even in the radio-selected sample containing known radio detections.

\subsection{Updates on MGCLS radio diffuse catalogue}
\label{AppA}

Based on a detailed investigation of the radio images at full (7.8$^{\prime\prime}$) and low (15$^{\prime\prime}$) resolution in this catalogue, along with the aid of optical and X--ray data (where available), we suggest updates on the possible nature of the observed diffuse radio sources in ten MGCLS clusters, compared to the ones reported in Table~1 of K22. To our knowledge, the definition of the nature of diffuse radio structures in the 8/9 systems that were characterized as unknown or uncertain in K22 is updated. In this section, we list the morphology classification updates of these systems in this catalogue; for more details on each system, see the Appendix~\ref{AppB}. Characterizing these eight systems has led to the inclusion of six more diffuse radio structures. In addition to these nine systems, for one system, RXC~J0510.7-0801, we do not report the existence of a candidate radio relic, as it is a possible Wide-Angle Tail (WAT) radio source. Hence, the total number of MGCLS diffuse radio structures from 99 in K22 is 103. 

In more detail, the 10 MGCLS systems with updates on the classification of their diffuse radio sources are: i) Abell~S1121 (fig.~\ref{fig:AS1121}), for which we now report the presence of a southeast candidate radio relic; ii) MACS J0257.6--2209 (fig.~\ref{fig:MACSJ0257}), for which we now report the presence of a possible candidate mini-halo; iii) RXC~J0510.7-0801 (fig.~\ref{fig:RXCJ0510}), in which we do not report the existence of a candidate radio relic as we exclude it as a possible WAT radio source; iv) RXC~J0520.7--1328, which is highlighted in K22, we report the presence of an eastern candidate relic and a southern candidate relic; v) RXC~J2351.0--1954 (fig.~\ref{fig:RXCJ2351}), for which we do not report the existence of a candidate radio halo as no diffuse radio emission detection is visible in this system based on the MGCLS products; vi) J0351.1--8212, which is also highlighted in K22, we report the presence of a candidate mini-halo, a southeast relic, a western relic and a southern relic; vii) J0637.3--4828 (fig.,~\ref{fig:J0637.3}), for which we now report the presence of a radio halo; viii) J0745.1--5404 (fig.~\ref{fig:J0745.1}), for which we now report the presence of a southern candidate radio relic; ix) J0820.9--5704 (fig.~\ref{fig:J0820.9}), for which we now report the presence of a southeast candidate radio relic and a southwest candidate radio relic; and, x) J1423.7--5412 (fig.~\ref{fig:J1423.7}), for which we now report the presence of a possible candidate radio halo and a west candidate radio relic.

\subsection{Challenges in the classification of diffuse cluster radio emission}
Recovering diffuse radio emission in clusters is a challenging process with different methods being used either by `clearing' the radio images (e.g., compact radio source subtraction and minimising the presence of radio artefacts) or boosting the presence of any existing diffuse radio emission, i.e., low-resolution imaging or angular-scale filtering (a multi-resolution filtering method; \citealt{2002PASP..114..427R}). The MGCLS images used in this work are the direct output of the basic procedure described in \citet{DEEP2} using the Obit package 2 \citep{OBIT}. The MGCLS survey provided radio image products at the full-resolution of the image ($\sim$ 7.5 $-$ 8$''$) but also at a convolved 15$''$ to optimize the recovery of low surface brightness features aiding in the recovery of any diffuse radio emission and also allow a reliable flux density estimate based on the MGCLS products. We suggest that both image products form an approximate representation of the existing classes of diffuse radio emission in these systems.



 In addition to recovering diffuse radio emission, its classification is equally challenging.  Although we have attempted with the MGCLS products at hand to carefully classify the radio sources based on their morphology, available X--ray data, and their relative location from the MGCLS images, we end up with several sources (41) that remain as candidates and one source (in J0431.4-6126; Fig.~\ref{fig:J0431.4}) marked as unknown for a few reasons. The first reason is the limitation that comes from the radio data itself, the sensitivity of which in some cases is affected by the presence at the cluster core of a bright, compact radio source, creating residuals in the analysis due to imperfect calibration, which leaves the classification of faint radio sources unclear, such as the cases of candidate radio halos, for example, in MACS J0257.6-2209 (fig.~\ref{fig:MACSJ0257}) and J1423.7–5412 (fig.~\ref{fig:J1423.7}). Another reason is the dense blending and emission contamination of compact radio sources in confined space areas that also affects the revealing and classification of faint diffuse sources, which is the case for several candidate radio halos (e.g., Abell~22; Fig.~\ref{fig:A22}, Abell~370; Fig.~\ref{fig:A370}, Abell~2645; Fig.~\ref{fig:A2645}, Abell~2813; Fig.~\ref{fig:A2813}). In other cases, the blending of only one but extended radio source can also be challenging for a secure classification, as cluster radio shocks co-exist and blend with old AGN radio plasma, complicating the classification, such as the embedded head-tail radio source in the case of the mini-halo in Abell~4038 (Fig.~\ref{fig:A4038}). 

The third reason is the requirement for complementary data to determine the nature of the diffuse radio emission. Radio emission from halos is known to correlate with the thermal X--ray emission of the ICM, and, similarly, a central X--ray emission is necessary to determine the presence of mini-halos. The presence of X--ray emission is essential for a robust classification of radio halos, relics, and mini-halos. However, although archival X--ray data are available for almost all of the MGCLS clusters, for the systems that are classified as candidate relics and mini-halos, the majority come from low-resolution, shallow X--ray observations by \textit{ROSAT} or \textit{Chandra} telescopes. This is the case for several candidate relics (i.e., in MACS~J0257.6-2209; Fig.~\ref{fig:MACSJ0257}, MACS~J0417.5-1154; Fig.~\ref{fig:MACSJ0417}, RXC~J2351.0–1954; Fig.~\ref{fig:RXCJ2351}, J0232.2–4420; Fig.~\ref{fig:J0232.2}) whose relative location regarding thermal ICM cannot be secured for several cases due to the lack of deep ancillary X--ray data. 


Although the presence of X-ray observations allows for the confirmation of a radio halo (since the radio halo classification in our sample is mainly hindered by the first reason discussed above), on the other hand, as mentioned in \citet{2019SSRv..215...16V}, the detection of shocks in the ICM is observationally challenging. We confirmed the presence of a shock for several relic sources using their relative location to the cluster core by X--ray observations; the absence of an obvious counterpart as a second criterion aids the classification. However, despite our efforts, the classification of several radio relics will remain uncertain, and for this reason, we refer to them as candidates.

Similarly, the lack of deep X-ray emission information for the clusters that potentially host a mini-halo (e.g., cMHalo in J0145.2-6033; Fig.~\ref{fig:J0145.2}) does not allow us to classify these diffuse sources robustly. In addition, as the mini-halo emission surrounds the central source, high-resolution images are also essential to distinguish between AGN lobes and mini-halos. Due to these observational limitations, the detection and classification of mini-halos were bound to an observational selection bias dependent on the current radio telescopes' detection limit. Despite all these difficulties, we note that several fainter radio mini-halos are being detected by MGCLS because of MeerKAT's sensitivity.

Lastly, the classification may also be hampered by projection effects, i.e., a chance superposition of an AGN-origin fossil plasma may resemble a cluster radio shock if observed at the cluster periphery; however, these observation-related obstacles are minimised

\onecolumn
\begin{center}
\LTcapwidth=\textwidth
\begin{longtable}{lllclcccc} 
\caption{\label{tab:diffuse}Catalogue of the 103 diffuse cluster radio sources detected in the MGCLS. Several of the 62 MGCLS clusters in this catalogue host more than one diffuse cluster source. Column descriptions are as follows: (1) Cluster name; (2--3) The first line denotes the NED cluster position: J2000 Right ascension and Declination; The rest of the coordinates denote the position of the diffuse radio morphology detected in each cluster (4) Cluster redshift; (5) Diffuse source classification --- Halo, Relic, Mini-halo (MHalo), Phoenix, candidate (c), unknown/unclear (U); (6) local rms noise at the 15$^{\prime\prime}$ resolution image in $\mu$Jy~beam$^{-1}$; (7) Flux density at 1.28~GHz in mJy; (8) Largest physical linear size at the cluster redshift in kpc in the form of minor axis $\times$ major axis; (9) Radio power at 1.28~GHz. }\\
\hline\hline       
\multicolumn{1}{l}{(1)} & \multicolumn{1}{l}{(2)} & \multicolumn{1}{l}{(3)} & (4) & \multicolumn{1}{c}{(5)} & (6) & \multicolumn{1}{c}{(7)} & \multicolumn{1}{c}{(8)} & \multicolumn{1}{c}{(9)}\\
Cluster Name & \multicolumn{1}{l}{RA} & \multicolumn{1}{l}{Dec.} & \textit{z} & \multicolumn{1}{l}{Morphology} & rms$\_$15$''$ & \multicolumn{1}{l}{Flux Density} & \multicolumn{1}{c}{LLS} & \multicolumn{1}{c}{P$_{1.28GHz}$}\\    
 & \multicolumn{1}{l}{(${\rm J2000}$)} & \multicolumn{1}{l}{(${\rm J2000}$)} & & & \multicolumn{1}{c}{($\mu$Jy~beam$^{-1}$)} & \multicolumn{1}{c}{(mJy)} & \multicolumn{1}{c}{(kpc $\times$ kpc)} & (W~Hz$^{-1}$)\\
\hline  
\endfirsthead
\endfoot
\caption{continued.}\\
\hline\hline
\multicolumn{1}{l}{(1)} & \multicolumn{1}{l}{(2)} & \multicolumn{1}{l}{(3)} & (4) & \multicolumn{1}{c}{(5)} & (6) & \multicolumn{1}{c}{(7)} & \multicolumn{1}{c}{(8)} & \multicolumn{1}{c}{(9)}\\
Cluster Name & \multicolumn{1}{l}{RA} & \multicolumn{1}{l}{Dec.} & \textit{z} & \multicolumn{1}{l}{Morphology} & rms$\_$15$''$ & \multicolumn{1}{l}{Flux Density} & \multicolumn{1}{c}{LLS} & \multicolumn{1}{c}{P$_{1.28GHz}$}\\    
 & \multicolumn{1}{l}{(${\rm J2000}$)} & \multicolumn{1}{l}{(${\rm J2000}$)} & & & \multicolumn{1}{c}{($\mu$Jy~beam$^{-1}$)} & \multicolumn{1}{c}{ (mJy)} & \multicolumn{1}{c}{(kpc $\times$ kpc)} & (W~Hz$^{-1}$)\\
\hline     
\endhead
\hline
\endlastfoot
\footnotesize{\textit{Radio-selected sample}}  &  &   &   &   &   &   &  & \\ 
Abell 13            &  00:13:32.2   &    $-$19:30:03.6 &  0.094 &  $\hspace{21pt}$- &       -     &  - & - & - \\ 
       &  00:13:27.5   &    $-$19:30:00.2 &  & W cRelic &     12    & 31.0$\pm1.9$  & $300\times320$ & 6.4$\pm0.4$ $\times10^{23}$\\
Abell 22            &   00:20:38.6  &    $-$25:43:19.2 &  0.142 & cHalo  &    6  &  1.6$\pm0.1$ & $280\times420$ & 8.1$\pm$0.5 $\times10^{22}$\\
Abell 85   &  00:41:48.7  & $-$09:19:04.8  &  0.056 & MHalo & 8 &  5.5$\pm0.3$ & $230\times320$ & 3.5$\pm$0.2 $\times 10^{22}$\\
                    &  00:41:29.03  & $-$09:22:25.6 &   & SW Phoenix & 8 & 61.9$\pm3.7$ & $250\times300$ &   4.0$\pm$0.2 $\times10^{23}$\\
Abell 168           &  01:15:09.8  &  00:14:51.0   &  0.045 &     $\hspace{21pt}$- &       -     &  - & - & - \\ 
      &  01:14:42.0  &  00:35:19.5   &   & N Relic   &      12     & 31.1$\pm1.9$ & $150\times650$ & 1.3$\pm$0.1 $\times10^{23}$\\
Abell 209           &  01:31:57.5  &   $-$13:34:35.0  &  0.209 & Halo   &  12   &  7.0$\pm0.4$ & $600\times740$ & 8.2$\pm$0.5 $\times10^{23}$\\
Abell 370           &   02:39:50.5 &    $-$01:35:08.0  &  0.375 & cHalo  & 10 &  3.5$\pm0.2$ & $940\times960$ & 1.7$\pm$0.1 $\times10^{24}$ \\
Abell 521           &  04:54:08.6  & $-$10:14:39.0 &  0.248 & Halo  &  6  & 10.1$\pm0.6$  & $860\times1440$ & 1.9$\pm$0.1 $\times10^{24}$ \\
                    &  04:54:21.1    &  $-$10:16:46.0    &   & SE Relic   & 6  & 16.6$\pm1.0$  & $240\times1260$ & 3.2$\pm$0.2 $\times10^{24}$\\
                    &  04:53:52.3  &  $-$10:13:18.4  &       & NW Relic   & 6 & 0.5$\pm0.1$  & $125\times330$ &  9.5$\pm$0.6 $\times10^{22}$\\
Abell 545     & 05:32:24.4  &  $-$11:32:35.0  &  0.154  & Halo   &      11    &  14.1$\pm0.9$ & $590\times750$ & 8.7$\pm$0.5 $\times10^{23}$\\
Abell 2645          &  23:41:16.8 &    $-$09:01:39.0 &  0.251  & cHalo  & 14 &  0.8$\pm0.1$ & $430\times550$ & 1.5$\pm$0.1 $\times10^{23}$\\
Abell 2667          & 23:51:40.7  &  $-$26:05:01.0 &  0.232 & Halo &  5    &  6.8$\pm0.4$ & $380\times530$ & 1.0$\pm$0.1 $\times10^{24}$ \\
Abell 2744          & 00:14:16.1  &  $-$30:22:58.8 &  0.307 & Halo   &      11    & 49.5$\pm3.0$ &  $1620\times2040$  & 1.5$\pm$0.1 $\times10^{25}$\\
                    &  00:14:41.4  & $-$30:20:07.2    &     & NE Relic   &           &  17.7$\pm1.1$ & $450\times1570$ & 5.3$\pm$0.3 $\times10^{24}$\\
                    &  00:14:32.9  & $-$30:25:01.5  &      & SE Relic   &       &  3.0$\pm0.2$ & $210\times1210$ & 9.0$\pm$0.5 $\times10^{23}$\\
Abell 2751 & 00:16:13.9 & $-$31:23:18.6 &                     0.107 & $\hspace{21pt}$-  &       -     &  - & - & - \\
                    &  00:16:36.6 & $-$31:28:21.3 &   & SE cPhoenix & 10 & 0.4$\pm0.1$ & $50\times90$  & 1.1$\pm$0.1 $\times10^{22}$ \\
Abell 2811          &  00:42:08.8 &  $-$28:32:08.8 &  0.108 & Halo   &      8     & 4.1$\pm0.3$  & $300\times320$ &  1.1$\pm$0.1 $\times10^{23}$\\
Abell 2813          &  00:43:24.4 &  $-$20:37:17.0 &  0.292 & cHalo  & 12 &  3.7$\pm0.2$  &  $770\times810$ & 9.8$\pm$0.6 $\times10^{23}$\\
Abell 2895          &   01:18:11.1 &  $-$26:58:23.0 &  0.228 & $\hspace{21pt}$- &       -     &  - & - & -  \\
                    &   01:18:19.9 &  $-$26:58:29.9 &   
                    & E cPhoenix & 10 & 3.4$\pm0.2$ & $100\times170$ & 4.8$\pm$0.3 $\times10^{23}$ \\
Abell 3365          & 05:48:12.0   &$-$21:56:06.0 &  0.093 &          $\hspace{21pt}$- &       -     &  - & - & -   \\
                    & 05:49:05.7   & $-$21:46:43.3 &  & NE Relic& 15 & 40.9$\pm2.5$ & $260\times630$ & 8.3$\pm$0.5 $\times10^{23}$  \\
                    & 05:48:04.0  &  $-$21:52:56.7   &       & NW Relic   & 15 &  6.2$\pm0.4$ & $210\times230$ &  1.3$\pm$0.1 $\times10^{23}$\\ 
Abell 3376          & 06:01:42.2  & $-$39:59:06.2 &  0.047 &              $\hspace{21pt}$- &       -     &  - & - & - \\
                    & 06:03:02.4  & $-$39:55:47.6 &   & E Relic  & 12  & 156.0$\pm9.4$ & $530\times1520$ & 7.2$\pm$0.4 $\times10^{23}$\\
                    & 06:00:04.3 & $-$40:03:12.1 &    & W Relic   & 12 & 139.0$\pm8.3$ & $310\times1000$ & 6.4$\pm$0.4 $\times10^{23}$\\
Abell 3558          & 13:27:54.8 & $-$31:29:32.0 & 0.048                         &  Halo  & 7 & 11.2$\pm0.7$ & $140\times290$ & 5.9$\pm$0.4 $\times10^{22}$ \\
Abell 3562          & 13:33:35.6 & $-$31:40:27.1 & 0.050                         & Halo & 10 & 26.2$\pm1.6$ & $300\times600$ & 1.4$\pm$0.1 $\times10^{23}$ \\
                    & 13:32:07.5 & $-$31:46:06.2 &  & SW Relic  & 10 & 17.9$\pm1.1$ & $190\times450$ & 9.8$\pm$0.6 $\times10^{22}$ \\
Abell 3667 & 20:12:33.7 & $-$56:50:26.3 & 0.056                         &   $\hspace{21pt}$-  &       -     &  - & - & -  \\ 
                    & 20:14:32.0 & $-$57:04:03.5 &           &  SE Relic & 15 & 384.0$\pm23.0$ & $500\times1650$ & 2.6$\pm$0.2 $\times10^{24}$\\
                    & 20:10:32.5 & $-$56:24:38.7 &           &  NW Relic & 15 & 2302.0$\pm138.1$ & $700\times2150$ & 1.6$\pm$0.1 $\times10^{25}$\\                    
Abell 4038          &  23:47:31.1 & $-$28:12:10.0 & 0.030                        & cMHalo & 8 & 10.3$\pm0.6$ & $130\times160$ & 1.6$\pm$0.1 $\times10^{22}$\\
Abell S295          & 02:45:36.0  & $-$53:02:16.8 & 0.300                        & Halo  & 8  & 7.2$\pm0.4$ & $720\times1070$ & 2.0$\pm$0.1 $\times10^{24}$\\
Abell S1063         &  22:48:43.5 & $-$44:31:44.0 & 0.348                        &  Halo & 7 & 8.8$\pm0.5$ & $1070\times1180$ & 3.5$\pm$0.2 $\times10^{24}$\\
Abell S1121         & 23:25:08.3 & $-$41:12:42.6 &  0.190 &             $\hspace{21pt}$- &       -     &  - & - & -   \\
                    & 23:25:14.5 & $-$41:13:56.2 &      
    & SE cRelic & 12 & 3.3$\pm0.2$ & $180\times410$ &  3.2$\pm$0.2 $\times10^{23}$\\
Bullet              & 06:58:37.9 & $-$55:57:00.0 &  0.297                 & Halo & 10 & 94.8$\pm5.7$ & $1550\times2050$ & 2.6$\pm$0.2 $\times10^{25}$ \\
                    & 06:58:53.1 & $-$55:57:14.9 &      
        & E Relic & 10 & 81.2$\pm4.9$ & $330\times1010$ & 2.2$\pm$0.1 $\times10^{25}$ \\
El Gordo            & 01:02:52.5 & $-$49:14:58.1 & 0.870                & Halo & 6 & 12.1$\pm0.7$ & $1250\times1530$ & 3.3$\pm$0.2 $\times10^{25}$ \\
                    & 01:02:46.9 & $-$49:14:30.3 &          
       & NW Relic & 6 & 8.2$\pm0.5$ & $400\times720$ & 2.3$\pm$0.1 $\times10^{25}$\\
                    & 01:03:01.3 & $-$49:17:06.9 &          
       & SE Relic & 6 & 4.1$\pm0.3$ & $290\times460$ & 1.1$\pm$0.1 $\times10^{25}$\\
                    & 01:03:06.6 & $-$49:16:20.3 &          
        & E Relic & 6 & 0.7$\pm0.1$ & $250\times420$ & 1.9$\pm$0.1 $\times10^{24}$\\
\footnotesize{PLCK G200.9$-$28.2}  & 04:50:20.9 &     $-$02:56:57.6 &  0.220  &  $\hspace{21pt}$-  &       -     &  - & - & -   \\
                    & 04:50:15.8 & $-$03:01:07.5 &  &  SW Relic  & 12 & 23.7$\pm1.4$ & $300\times1150$ & 3.3$\pm$0.2 $\times10^{24}$ \\
                    & 04:50:35.0 & $-$02:57:49.1 &  & E Relic & 12 & 3.6$\pm0.2$ & $160\times1230$ & 4.9$\pm$0.3 $\times10^{23}$ \\
                    & 04:50:16.6 & $-$02:54:38.8 &  & NW cRelic & 12 & 0.6$\pm0.1$ & $110\times250$ & 8.2$\pm$0.5 $\times10^{22}$ \\
\footnotesize{MACS J0257.6$-$2209$^c$} & 02:57:40.3 & $-$22:09:46.0 & 0.322 & cHalo & 6 & 0.4$\pm0.1$ & $430\times500$ & 1.3$\pm$0.1 $\times10^{23}$  \\
                    & 02:57:37.1 & $-$22:09:59.7 &   & SW cRelic  & 6 & 1.8$\pm0.1$ & $190\times390$ & 6.0$\pm$0.4 $\times10^{23}$ \\
\footnotesize{MACS J0417.5$-$1154} & 04:17:34.6 & $-$11:54:32.0 & 0.443 & Halo & 7 & 16.2$\pm1.0$ & $740\times1420$ & 1.1$\pm$0.1 $\times10^{25}$\\
                    & 04:17:27.0 & $-$11:47:23.3 &  
                & N cRelic  & 7 & 1.6$\pm0.1$ & $220\times780$ & 1.1$\pm$0.1 $\times10^{24}$\\
                    & 04:17:13.8 & $-$11:48:21.5 &  
                    & NW cRelic  & 7 & 1.0$\pm0.1$ & $210\times490$ & 7.1$\pm$0.4 $\times10^{23}$\\
\footnotesize{RXC J0510.7$-$0801} & 05:10:44.3 &$-$08:01:12.0  & 0.220 & cHalo & 8 & 5.8$\pm0.4$ & $430\times1040$ & 7.9$\pm$0.5 $\times10^{23}$\\
\footnotesize{RXC J0520.7$-$1328$^d$} & 05:20:47.2 &  $-$13:30:08.0 & 0.336 &  $\hspace{21pt}$- &       -     &  - & - & -\\
                       & 05:21:02.2 &  $-$13:35:26.5 & & SE cRelic &10& 6.1$\pm0.4$ & $230\times1100$ & 2.3$\pm$0.1 $\times10^{24}$ \\
                       & 05:21:09.8 &  $-$13:29:07.7 & & E cRelic  & 10 & 2.5$\pm0.2$ & $280\times910$& 9.3$\pm$0.6 $\times10^{23}$\\ 
                       & 05:20:49.5 & $-$13:31:57.0 & & S cRelic  & 10 & 4.9$\pm0.3$ & $470\times1200$ & 1.8$\pm$0.1 $\times10^{24}$ \\
\footnotesize{RXC J1314.4$-$2515} & 13:14:23.7 &$-$25:15:21.0 & 0.244 & Halo & 12 & 12.5$\pm0.8$ & $840\times1080$ & 2.3$\pm$0.1 $\times10^{24}$ \\
                       & 13:14:46.0 & $-$25:15:10.0 &  
                & E Relic & 12 & 12.7$\pm0.8$ & $230\times620$ & 2.4$\pm$0.1 $\times10^{24}$ \\
                       & 13:14:17.9 & $-$25:15:50.6 & 
                & W Relic & 12 & 33.6$\pm2.0$ &$210\times1080$ & 6.2$\pm$0.4 $\times10^{24}$ \\
\footnotesize{RXC J2351.0$-$1954} & 23:51:04.9 
                &$-$19:54:48.0 &  0.248 &  
                   $\hspace{21pt}$- &       -     &  - & - & - \\
                       & 23:50:41.3 & $-$19:56:27.1 &  
                & W cRelic & 7 & 0.6$\pm0.1$ & $100\times1030$ & 1.1$\pm$0.1 $\times10^{23}$\\
                          & 23:51:29.9 & $-$20:01:01.1 &  
                & E cRelic & 7 & 2.5$\pm0.2$ & $170\times1340$ & 4.5$\pm$0.3 $\times10^{23}$\\
\footnotesize{\textit{X--ray-selected sample}}  &  &   &   &   &   &   &  & \\ 
J0027.3$-$5015      & 00:27:21.3 & $-$50:15:04.0 & 0.145 
                     & cMHalo & 6 & 0.3$\pm0.1$ & $170\times210$ & 1.6$\pm$0.1 $\times10^{22}$ \\
J0145.0$-$5300      & 01:45:02.3 & $-$53:00:50.0 & 0.117 
                    & Halo & 4 & 2.5$\pm0.2$ & $320\times620$ & 8.5$\pm$0.5 $\times10^{22}$\\ 
J0145.2$-$6033      & 01:45:16.7 & $-$60:33:54.0 & 0.181 
                    & cMHalo & 4 & 0.9$\pm0.1$ & $220\times320$ & 8.2$\pm$0.5 $\times10^{22}$\\
J0216.3$-$4816      & 02:16:19.1 & $-$48:16:23.0 & 0.163 
                    & cMHalo & 6 & 3.3$\pm0.2$ & $220\times240$ & 2.3$\pm$0.1 $\times10^{23}$\\
J0217.2$-$5244      & 02:17:12.6 & $-$52:44:49.0 & 0.343 &  $\hspace{21pt}$- &       -     &  - & - & - \\
                    & 02:17:04.0 & $-$52:41:45.0 & 
        & N cRelic & 6 & 1.1$\pm0.1$ & $170\times460$ & 4.3$\pm$0.3 $\times10^{23}$\\
J0225.9$-$4154      & 02:25:54.6 & $-$41:54:35.0 & 0.220 
                    & Halo & 7 & 14.4$\pm0.9$ & $330\times500$ & 2.0$\pm$0.1 $\times10^{24}$ \\ 
J0232.2$-$4420      & 02:32:16.8 & $-$44:20:51.0 & 0.284 
                    & Halo & 6 & 10.6$\pm0.6$ & $1120\times1240$ & 2.6$\pm$0.2 $\times10^{24}$\\
                    & 02:32:17.9 & $-$44:22:04.0 & 
                    & S cRelic & 6 & 0.6$\pm0.1$ & $200\times450$ & 1.5$\pm$0.1 $\times10^{23}$\\
                    & 02:32:42.2 & $-$44:20:51.7 & 
                    & E cRelic & 6 & 0.9$\pm0.1$ & $210\times440$ & 2.2$\pm$0.1 $\times10^{23}$\\
J0303.7$-$7752      & 03:03:46.4 & $-$77:52:09.0 & 0.274 
                    & Halo & 6 & 8.6$\pm0.5$ & $630\times950$ & 2.0$\pm$0.1 $\times10^{24}$\\
J0314.3$-$4525      & 03:14:19.8 & $-$45:25:27.0 & 0.072 
                    & cMHalo & 6 & 1.1$\pm0.1$ & $100\times150$ & 1.3$\pm$0.1 $\times10^{22}$\\
J0342.8$-$5338      & 03:42:53.9 & $-$53:38:07.0 & 0.060 
                    & MHalo & 6 & 5.6$\pm0.3$ & $280\times340$ & 4.4$\pm$0.3 $\times10^{22}$\\
J0351.1$-$8212 & 03:51:08.9 & $-$82:13:00.0 & 0.061 &  cMHalo  &  8  & 1.4$\pm0.1$ &  $110\times150$    &  1.2$\pm$0.1 $\times10^{22}$\\
                    & 03:51:52.4 & $-$82:14:31.9 &  
                    & SE Relic & 8 & 1.6$\pm0.1$ & $40\times70$  & 1.3$\pm$0.1 $\times10^{22}$\\
                    & 03:51:37.0 & $-$82:14:38.4 &  
                    & S Relic & 8 & 2.8$\pm0.2$ & $40\times90$ & 2.4$\pm$0.1 $\times10^{22}$\\
                    & 03:50:44.6 & $-$82:13:55.8 &  
                    & W Relic & 8 & 4.2$\pm0.3$ & $45\times150$ & 3.5$\pm$0.2 $\times10^{22}$\\
J0352.4$-$7401 & 03:52:29.5 & $-$74:01:51.0 & 0.127 
                    & Halo & 8 & 22.8$\pm1.4$ & $680\times1460$ & 9.2$\pm$0.6 $\times10^{23}$\\
                    & 03:54:25.2 & $-$74:05:06.5 &  
                    & SE Relic & 8 & 71.7$\pm4.3$ & $420\times2010$ & 2.9$\pm$0.2 $\times10^{24}$\\
                    & 03:50:29.0 & $-$73:57:39.0 &  
                    & NW cRelic & 8 & - & - & - \\
                    & 03:51:23.4 & $-$73:50:35.4 &  
                    & NNW Relic & 8 & 5.6$\pm0.3$ & $170\times460$ & 2.3$\pm$0.1 $\times10^{23}$\\
                    & 03:52:22.0 & $-$73:49:07.7 &  
                    & N Relic & 8 & 18.6$\pm1.1$ & $350\times660$ & 7.5$\pm$0.5 $\times10^{23}$\\
J0431.4$-$6126      & 04:31:24.1 & $-$61:26:38.0 & 0.059 &                    U      
                    & 10 & 9.7$\pm0.6$  & $170\times480$  & 7.5$\pm$0.5 $\times10^{22}$\ \\
                    & 04:32:05.7 & $-$61:40:21.5 &  
                    & SE Relic & 10 & 45.3$\pm2.7$ 
                    & $200\times670$ & 3.5$\pm$0.2 $\times10^{23}$\\
J0510.2$-$4519      & 05:10:13.8 & $-$45:19:16.0 & 0.200  
                    & cMHalo & 6 & 1.3$\pm0.1$ & $350\times430$ & 1.4$\pm$0.1 $\times10^{23}$ \\
J0516.6$-$5430      & 05:16:38.0 & $-$54:30:51.0 & 0.295 
                    & Halo & 6 & 7.9$\pm0.5$ & $970\times1540$ & 2.2$\pm$0.1 $\times10^{24}$\\
                    & 05:16:39.0 & $-$54:23:45.0 &            & N Relic & 6 & 33.6$\pm2.0$ & $650\times1500$ & 9.3$\pm$0.6 $\times10^{24}$\\
                    & 05:16:49.0 & $-$54:37:01.0 & 
                    & S cRelic & 6 & 4.8$\pm0.3$ & $270\times770$ & 1.3$\pm$0.1 $\times10^{24}$\\
J0528.9$-$3927      & 05:28:56.3 & $-$39:27:46.0 & 0.284 
                    & Halo & 6 & 3.8$\pm0.2$ & $560\times930$ & 9.5$\pm$0.6 $\times10^{23}$\\
J0627.2$-$5428      & 06:27:14.4 & $-$54:28:12.0 & 0.051 &                      $\hspace{21pt}$- &       -     &  - & - & -\\ 
                    & 06:26:11.9 & $-$54:32:07.0 & 
                    & W Relic & 12 & 78.1$\pm4.7$ & $90\times320$ & 4.5$\pm$0.3 $\times10^{23}$ \\
J0631.3$-$5610 & 06:31:20.7 & $-$56:10:20.0 & 0.054 &                   $\hspace{21pt}$- &       -     &  - & - & -\\ 
                    & 06:29:10.8 & $-$56:13:23.5 &  & W cRelic & 6 & 1.7$\pm0.1$ & $60\times400$ &  1.1$\pm$0.1 $\times10^{22}$\\
J0637.3$-$4828      & 06:37:18.9 & $-$48:28:42.0 & 0.203 
                    & Halo & 6 & 16.5$\pm1.0$ & $1150\times1480$ & 1.9$\pm$0.1 $\times10^{24}$ \\
                    & 06:37:02.0 & $-$48:25:52.0 &   
                    & NW cRelic & 6 & 0.5$\pm0.1$ & $190\times400$ & 5.7$\pm$0.3 $\times10^{22}$\\
J0638.7$-$5358      & 06:38:46.5 & $-$53:58:18.0 & 0.227  
                    & Halo & 6 & 11.6$\pm0.7$ & $800\times1090$ & 1.8$\pm$0.1 $\times10^{24}$ \\
J0645.4$-$5413      & 06:45:29.3 & $-$54:13:08.0 & 0.164 
                    & Halo & 8 & 20.4$\pm1.2$ & $990\times1240$ & 1.5$\pm$0.1 $\times10^{24}$ \\
                    & 06:45:04.0 & $-$54:18:43.0 &  
                    & SW Relic & 8 & 3.2$\pm0.2$ & $260\times450$ & 2.4$\pm$0.1 $\times10^{23}$\\
J0745.1$-$5404      & 07:45:09.6 & $-$54:04:44.0 & 0.074 
                    &  $\hspace{21pt}$- &       -     &  - & - & - \\
                    & 07:45:00.0 & $-$54:20:27.0 &  
                    & S cRelic & 8 & 3.8$\pm0.2$ & $100\times330$ & 4.8$\pm$0.3 $\times10^{22}$\\
J0820.9$-$5704      & 08:20:59.6 & $-$57:04:47.0 & 0.061 
                    &   $\hspace{21pt}$- &       -     &  - & - & -  \\
                    & 08:21:48.0 & $-$57:30:11.0 &  
                    & SE cRelic & 8 & 8.6$\pm0.5$ & $150\times480$ & 7.3$\pm$0.4 $\times10^{22}$ \\
                    & 08:19:42.0 & $-$57:28:39.0 &  
                    & SW cRelic & 8 & 20.5$\pm1.2$ & $170\times610$ & 1.7$\pm$0.1 $\times10^{23}$\\
J1130.0$-$4213      & 11:30:05.6 & $-$42:13:47.0 & 0.155 
                    &   $\hspace{21pt}$- &       -     &  - & - & -  \\
                    & 11:30:35.0 & $-$42:11:47.0 &  
                    & NE cRelic & 6 & 3.2$\pm0.2$ & $120\times340$ & 2.0$\pm$0.1 $\times10^{23}$\\
J1423.7$-$5412      & 14:23:43.3 & $-$54:12:12.0 & 0.300 
                    & cHalo  & 8 & 0.6$\pm0.1$ & - & 1.7$\pm$0.1 $\times10^{23}$\\
                    & 14:23:43.4 & $-$54:07:42.6 &  
                    & N cRelic & 8 & 2.3$\pm0.1$ & $340\times540$ & 6.5$\pm$0.4 $\times10^{23}$\\
J1539.5$-$8335      & 15:39:33.9 & $-$83:35:32.0 &  0.073 
                    & MHalo  & 6 & 10.5$\pm0.6$ & $200\times230$ & 1.3$\pm$0.1 $\times10^{23}$ \\
                    & 15:35:11.0 & $-$83:37:41.0 &         
                    & SW cRelic & 6 & 4.0$\pm0.2$ & $100\times200$ & 4.9$\pm$0.3 $\times10^{22}$\\
J1601.7$-$7544      & 16:01:46.7 & $-$75:44:46.0 & 0.153 
                    & Halo & 7 & 14.2$\pm0.9$ & $630\times980$ & 8.6$\pm$0.5 $\times10^{23}$ \\
J1840.6$-$7709      & 18:40:37.2 & $-$77:09:20.0 & 0.019 
                    & cMHalo & 20 & 28.0$\pm1.7$ & $47\times55$ & 2.2$\pm$0.1 $\times10^{22}$\\
J2023.4$-$5535      & 20:23:22.0 & $-$55:34:51.0 & 0.232 
                    & Halo & 7 & 18.2$\pm1.1$ & $700\times1200$ & 3.0$\pm$0.2 $\times10^{24}$\\
                    & 20:23:04.0 & $-$55:35:07.0 &  
                    & W Relic & 7 & 8.5$\pm0.5$ & $150\times640$ & 1.4$\pm$0.1 $\times10^{24}$\\

\hline    
\end{longtable}
\end{center}

 \twocolumn

\noindent
and can, in principle, be resolved using the available high-sensitivity and resolution MGCLS continuum radio data with the aid of ancillary data. Therefore, we underline that, despite the careful approach of our analysis, there is a level of uncertainty that cannot be overcome without a detailed multi-band analysis. Automated detections and classifications of diffuse and AGN-related radio emissions with the implementation of machine learning algorithms is an extremely helpful tool to expedite the process of initially identifying and then classifying the different cluster-extended radio sources on a morphological basis (see, e.g., \citealt{AniyanThorat17,Vavilovaetal21,Tolley24}); however, as data from newer telescopes is fundamentally different from previous surveys in terms of complexity and detail, the use of new-generation radio data as input to the algorithm is a necessity for robust classifications (see, e.g., \citealt{Stuardietal24,Alametal25,Laoetal25}). The challenge of a secure radio source classification based on their nature will still remain, but MGCLS products offer a fertile basis to test and implement machine learning algorithms on automated detections and classifications of such radio sources.

\subsection{1.4 GHz radio power $-$ M$_{500}$ scaling relation} 


In galaxy clusters, the correlation between the radio power of a giant halo and its X--ray luminosity is already well-known \citep[e.g.,][]{Liang_2000,Brunettietal09,Cassanoetal06,2013ApJ...777..141C,Kale2015EGRHS,2019SSRv..215...16V}. However, since both the X--ray emission and the Sunyaev-Zeldovich (SZ) signal detection of a cluster are valuable tools for its mass estimate, it is suggested that the observed halo radio power vs X--ray luminosity correlations arise from an already existing fundamental relation between radio power and cluster mass \citep{2021A&A...651A.115V}. This is plausible, as in a merger event, part of the gravitational energy released (that is proportional to the cluster mass) is transferred into the ICM, re-accelerating cosmic rays via turbulence \citep[e.g.,][]{CassanoBrunetti05}.




Using the new MGCLS data at 1.28~GHz we can compare with the radio power correlations at the rest frame of 1.4~GHz that have extensively been used so far \citep[e.g.,][]{2013ApJ...777..141C,2021A&A...647A..51C}. We calculated the radio power P$_{1.28\,GHz}$ from the measured radio flux densities by: 

\begin{equation}
\label{Power610}
P_{1.28\,GHz}=4 \pi D_L^2 (1+z)^{-(1 + \alpha )} S_{1.28\,GHz},
\end{equation}

\noindent
where D$_L$ is the luminosity distance to the cluster, $\alpha$ is the spectral index, \textit{z} the redshift and S$_{1.28\,GHz}$ is the flux density of the source at 1.28~GHz. 

To be consistent with the estimates reported in our sample and the literature and to perform a meaningful comparison, we determine the k--corrected 1.4~GHz radio power (P$_{1.4\,GHz}$) for every system that hosts a radio halo, mini-halo, or respective candidate diffuse emission detections. The estimated 1.28~GHz MeerKAT flux densities from this work were extrapolated to 1.4~GHz, assuming a fiducial spectral index of $-1.2$ (the typical value for cluster halo radio sources) in cases where no spectral index is known for the source, adopting a conservative uncertainty in $\alpha$ of 0.3--0.4. This assumption is reasonable since scaling relations show that the average spectral index of radio halos in the 150~MHz $-$ 1.4~GHz frequency range is $-1.2$, and is also used in \citet{2021A&A...651A.115V}. We note that the reported k--corrected 1.4~GHz radio powers are mildly underestimated only if the detected diffuse radio structures exhibit very steep spectra ($\alpha\leq-1.7$). However, even if this may be the case for some diffuse structures, this effect will not be significant as the 1.28~GHz MeerKAT reference frequency is very close to the 1.4~GHz. The derived P$_{1.4\,GHz\_ext}$ is shown in column 4 of Table~\ref{Masstable}.



The masses for the MGCLS clusters were extracted from the PSZ2 catalogue \citep{Planckcollab16}, which provides an SZ-based mass estimate (see also Table~\ref{Masstable}). We performed a cross-matching between the MGCLS clusters that present a halo, mini-halo, or respective candidate emission and the PSZ2 catalogue (SZ M$_{500}$ mass proxy) within 5~arcmins (i.e. roughly equal to the resolution of the PSZ2 survey) using TOPCAT \citep{Taylor_2015}. Our cross-match resulted in 41 clusters that present PSZ2 mass estimates.

\begin{table}
\begin{minipage}{\linewidth}
 \caption{$M_{500}$ mass details and extrapolated 1.4~GHz radio power, P$_{1.4GHz\_ext}$, for the 42 MGCLS clusters used in the analysis here. The masses for all MGCLS clusters but one were extracted by the PSZ2 catalogue (SZ M$_{500}$ mass proxy; \citealt{Planckcollab16}). The $M_{500}$ mass estimate of ACT-CL~J0216.2-4816 indicated by $^{a}$ was drawn from the ACT cluster survey catalogue \citep{Hilton_2020}. }
 \centering
 \label{Masstable}
\begin{tabular}{|c|c|c|c|}
 
 \hline \hline
  Cluster Name &  Radio  & $M_{500}$   & P$_{1.4GHz\_ext}$  \\
 
          &  Morphology &  10$^{14}$ M$_{\odot}$ &  (W~Hz$^{-1}$)  \\

 \hline
Abell 22  & cHalo&4.26$\pm$0.32& 7.3$\pm$0.4 $\times10^{22}$\\ 
Abell 85  & MHalo &4.92$\pm$0.13& 3.1$\pm$0.2 $\times10^{22}$\\
Abell 209 & Halo &8.46$\pm$0.28& 7.4$\pm$0.4 $\times10^{23}$\\
Abell 370 & cHalo &7.65$\pm$0.56& 1.5$\pm$0.1 $\times10^{24}$\\
Abell 521 & Halo&7.26$\pm$0.47& 1.7$\pm$0.1 $\times10^{24}$\\
Abell 545 & Halo &5.39$\pm$0.41& 7.8$\pm$0.5 $\times10^{23}$\\
Abell 2645 & cHalo & 4.11$\pm$0.51& 1.4$\pm$0.1 $\times10^{23}$\\
Abell 2667 & Halo &7.56$\pm$0.37& 9.0$\pm$0.5 $\times10^{23}$\\ 
Abell 2744 & Halo&9.84$\pm$0.40& 1.4$\pm$0.1 $\times10^{25}$\\
Abell 2811 & Halo &3.65$\pm$0.24& 1.2$\pm$0.1 $\times10^{24}$\\
Abell 2813 & cHalo &8.13$\pm$0.37& 8.8$\pm$0.5 $\times10^{23}$\\
Abell 3558 & Halo &4.79$\pm$0.17& 5.3$\pm$0.3 $\times10^{22}$\\
Abell 3562 & Halo& 2.44$\pm$0.21& 1.2$\pm$0.1 $\times10^{23}$\\
Abell 4038 & cMHalo & 1.48$\pm$0.12& 1.4$\pm$0.1 $\times10^{22}$\\
Abell S295 & Halo & 6.78$\pm$0.37& 1.8$\pm$0.1 $\times10^{24}$\\
Abell S1063 & Halo & 11.36$\pm$0.34& 3.1$\pm$0.2 $\times10^{24}$\\ 
Bullet & Halo &13.10$\pm$0.29& 2.0$\pm$0.1 $\times10^{25}$\\
El Gordo & Halo &10.75$\pm$0.48& 2.6$\pm$0.2 $\times10^{25}$\\
MACS J0257.6–2209 & cHalo & 6.05$\pm$0.56& 1.2$\pm$0.1 $\times10^{23}$\\
MACS J0417.5–1154 & Halo & 12.25$\pm$0.53& 9.9$\pm$0.6 $\times10^{24}$\\
RXC J0510.7–0801  & cHalo & 7.74$\pm$0.48& 7.1$\pm$0.4 $\times10^{23}$\\
RXC J1314.4–2515  & Halo & 6.72$\pm$0.53& 2.1$\pm$0.1 $\times10^{24}$\\
J0027.3–5015  & cMHalo &3.54$\pm$0.33& 1.4$\pm$0.1 $\times10^{22}$\\
J0145.0–5300  & Halo & 3.64$\pm$0.24& 7.6$\pm$0.5 $\times10^{22}$\\
J0145.2–6033  & cMHalo & 3.69$\pm$0.29& 7.4$\pm$0.4 $\times10^{22}$\\
J0216.3–4816\footnote{\citet{Hilton_2020}} & cMHalo & 3.63$^{+0.95}_{-0.8}$& 2.1$\pm$0.1 $\times10^{23}$ \\
J0225.9–4154  & Halo & 6.08$\pm$0.28& 1.8$\pm$0.1 $\times10^{24}$\\
J0232.2–4420  & Halo & 7.54$\pm$0.33& 2.3$\pm$0.1 $\times10^{24}$\\
J0303.7–7752  & Halo & 6.99$\pm$0.29& 1.8$\pm$0.1 $\times10^{24}$\\ 
J0314.3–4525  & cMHalo & 1.74$\pm$0.26& 1.2$\pm$0.1 $\times10^{22}$\\
J0342.8–5338  & MHalo & 4.22$\pm$0.15& 4.0$\pm$0.2 $\times10^{22}$\\
J0351.1–8212  & cMHalo & 2.46$\pm$0.17& 1.1$\pm$0.1 $\times10^{22}$\\
J0352.4–7401  & Halo & 6.44$\pm$0.19& 8.3$\pm$0.5 $\times10^{23}$\\
J0510.2–4519  & cMHalo & 5.73$\pm$0.29& 1.3$\pm$0.1 $\times10^{23}$\\
J0516.6–5430  & Halo & 8.76$\pm$0.24&  2.0$\pm$0.1 $\times10^{24}$\\
J0528.9–3927  & Halo & 7.41$\pm$0.37&  8.5$\pm$0.5 $\times10^{23}$\\ 
J0637.3–4828  & Halo & 5.30$\pm$0.27& 8.5$\pm$0.5 $\times10^{23}$\\
J0638.7–5358  & Halo & 6.82$\pm$0.34& 1.6$\pm$0.1 $\times10^{24}$\\
J0645.4–5413  & Halo & 7.96$\pm$0.23& 1.4$\pm$0.1 $\times10^{24}$\\
J1539.5–8335  & MHalo & 3.18$\pm$0.19& 1.2$\pm$0.1 $\times10^{23}$\\
J1601.7–7544  & Halo & 8.03$\pm$0.26& 7.7$\pm$0.5 $\times10^{23}$\\
J2023.4–5535  & Halo & 6.68$\pm$0.45& 2.7$\pm$0.2 $\times10^{24}$\\

 \hline
 \end{tabular}
\end{minipage}
 \end{table}

Using the mass estimates reported in the available literature, we have included the mass estimate of only one candidate mini-halo system (ACT-CL~J0216.2-4816) from the ACT cluster survey catalogue (M500cc; \citealt{Hilton_2020}) which also provides an SZ-based mass estimate (see Table~\ref{Masstable}) to the available MGCLS mini-halos that have cross-matched Planck masses clusters. For the ACT-CL~J0216.2-4816 system that presents a candidate mini-halo morphology, we note that there may also be unknown systematic uncertainties associated with its mass as it is estimated with a different method; however, its inclusion only serves to increase the low number of clusters with mini-halo emission examined. Hence, in total, we obtained mass estimates for 42 MGCLS clusters that present a radio halo, mini-halo, or respective candidate morphology. The masses of the 42 galaxy clusters of the sample used for the analysis are shown in Table~\ref{Masstable}. We end up with 26 systems presenting a radio halo, 6 systems presenting a candidate radio halo, 3 systems with a mini-halo and 7 systems with candidate radio mini-halo morphology. We also note that 22/42 clusters with mass estimates come from the radio-selected sample and 20/42 come from the X--ray-selected sample and are, therefore, biased towards relaxed cool-core clusters \citep[e.g.,][]{Eckertetal11,Rossettietal17}. 

\begin{figure}
\centering
\includegraphics[width=0.5\textwidth]{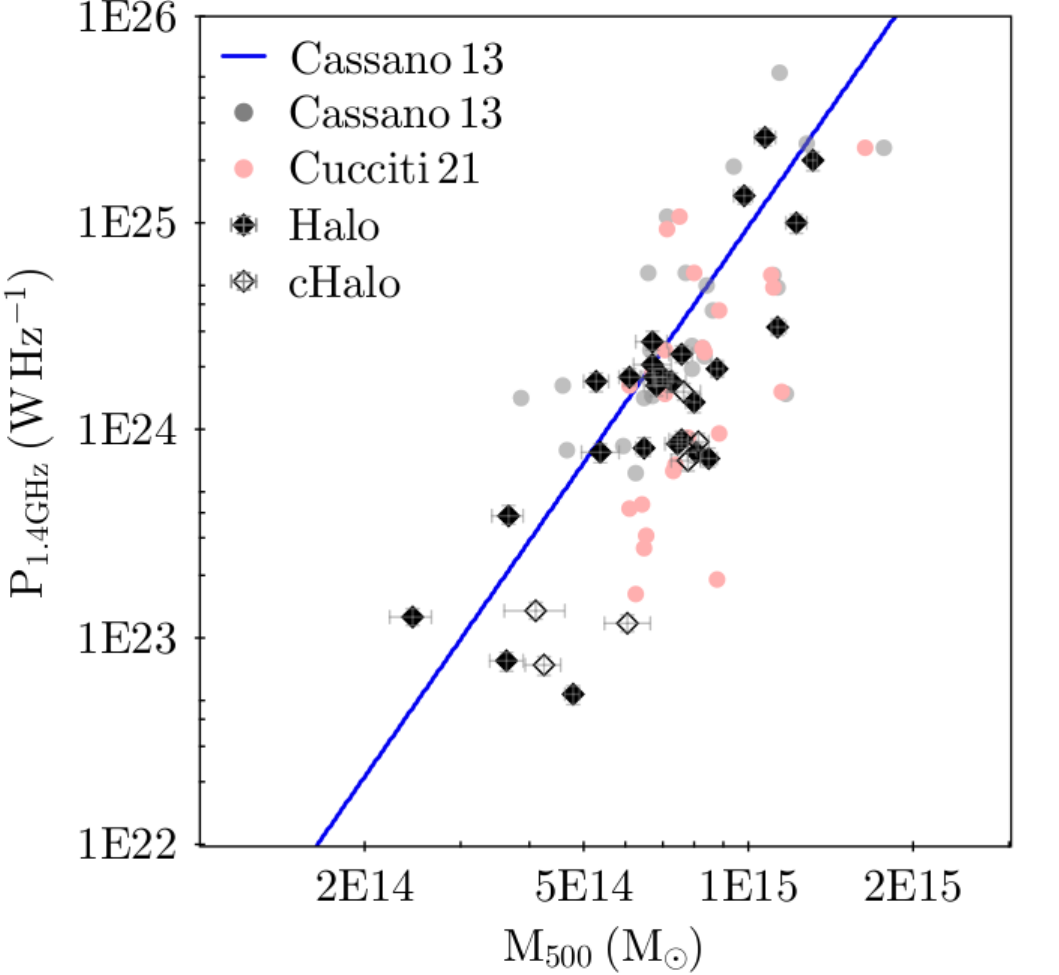}
\caption{1.4~GHz radio power versus M$_{500}$ for MGCLS radio halos (black rhombi) and candidate radio halos (open rhombi). The blue line corresponds to the scaling relation by \citet{2013ApJ...777..141C}, with the grey and pink circles corresponding to the clusters with radio halos by \citet{2013ApJ...777..141C} and \citet{2021A&A...647A..51C}, respectively. }
\label{P1400vsM500}
\end{figure}



\subsubsection{P$_{1.4\,GHz}$ $-$ M$_{500}$ for radio halos}

Figure~\ref{P1400vsM500} shows the  P$_{1.4\,GHz}$ $-$ M$_{500}$ correlation for radio halos and respective candidates detected in the MGCLS clusters along with clusters from the literature by \citet{2013ApJ...777..141C} and the more recent study by \citet{2021A&A...647A..51C}.  The blue line corresponds to the scaling relation by \citet{2013ApJ...777..141C}. The confirmed MGCLS radio halos from this work are shown in black rhombi, whereas the candidate radio halos are in open rhombi. We note here that the Cassano et al. relation did not include any Ultra Steep Spectrum Radio Halos (USSRHs), which may impact the derived spectral indices for steep diffuse radio sources. Hence, a detailed study of each individual cluster is essential for the most effective monitoring of the spectral indices in radio halos.

As expected, Figure~\ref{P1400vsM500} shows a steep correlation between the radio power of radio halos and the mass of the clusters, with the radio powers showing a scatter around the correlation that extends for over three orders of magnitude. The radio powers of the MGCLS halos are seen to lie about the Cassano et al. correlation, as is the case with literature halos. We also see that the halos from the MGCLS sample lie in a similar mass range with the halos that were used from the literature \citep{2013ApJ...777..141C,2021A&A...647A..51C}, apart from a set of 6 systems (3 with a radio halo and 3 with a candidate radio halo morphology) that present lower radio powers and masses and are mainly seen below the Cassano et al. line. These systems suggest a continuation of the steep correlation in the lower-mass and radio power regimes. 

However, extra caution should be asserted as our sample is heterogeneous without any specific selection criteria. Hence, it is biased towards more luminous radio halo systems that were previously known and were easier to find. A more carefully defined cluster sample filling the clusters' lower-mass regime is essential in examining the correlation's slope in more detail, determining its properties \citep[see, e.g.,][]{Botteonetal22,Hoangetal22}. 

In addition, the observed large scatter of the systems in the scaling relation between the radio power versus cluster masses for radio halos may also be attributed to the inclusion in the analysis of clusters that i) present different merging histories, ii) are currently in different phases of their merging events and iii) exhibit also different radio spectra \citep[e.g.,][]{Donnertetal13,2021A&A...647A..51C}.

\subsubsection{P$_{1.4\,GHz}$ $-$ M$_{500}$ for radio mini-halos} 


 
 




We also examine the relationship between the radio power P$_{1.4\,GHz}$ and the cluster mass for the MGCLS clusters with detected mini-halos and candidate mini-halos. As is already known, radio mini-halos present low-surface brightness diffuse synchrotron emission that is confined to the cool core of a relaxed massive cluster that surrounds the BCG, and usually have non-typical halo morphologies of less than 100~kpc or up to a few hundred kpcs \citep{Govonietal09,Giacintuccietal14,Giacintuccietal19,2020MNRAS.499.2934R}. About 80\% of the mini-halos are seen in the most massive systems (M$_{500}$ $\geq 6 \times 10^{14} M_{\odot}$) among the cool-core cluster population \citep{2017ApJ...841...71G,Giacintuccietal24}. 

The consensus for the origin of the mini-halo synchrotron-emitting electron population is that they are associated with the cluster thermal component and/or the BCG. On the other hand, the particle (re)-acceleration mechanism itself responsible for the production of mini-halos is considered to either be turbulent (re)-acceleration via cool-core gas sloshing \citep{Gittietal12,2013ApJ...762...78Z,2020MNRAS.499.2934R,Riseleyetal22b,Giacintuccietal24} and/or hadronic collision models \citep{Zuhoneetal15,JacobPfrommer17a,JacobPfrommer17b, Ignestietal20}. To date, only barely over 40 mini-halos (including candidates) are known \citep{Beginetal23,Giacintuccietal24} due to the difficulty in detection and imaging. However, MeerKAT has already increased their numbers and detections by $\sim$~15\%, and in the near future, the Square Kilometer Array (SKA) is predicted to detect hundreds of new mini-halos up to redshift of 1, significantly increasing the mini-halo numbers allowing detailed statistical analyses \citep{Kaleetal16,Iqbaletal17,Gittietal18}.

\begin{figure}
\centering
\includegraphics[width=0.49\textwidth]{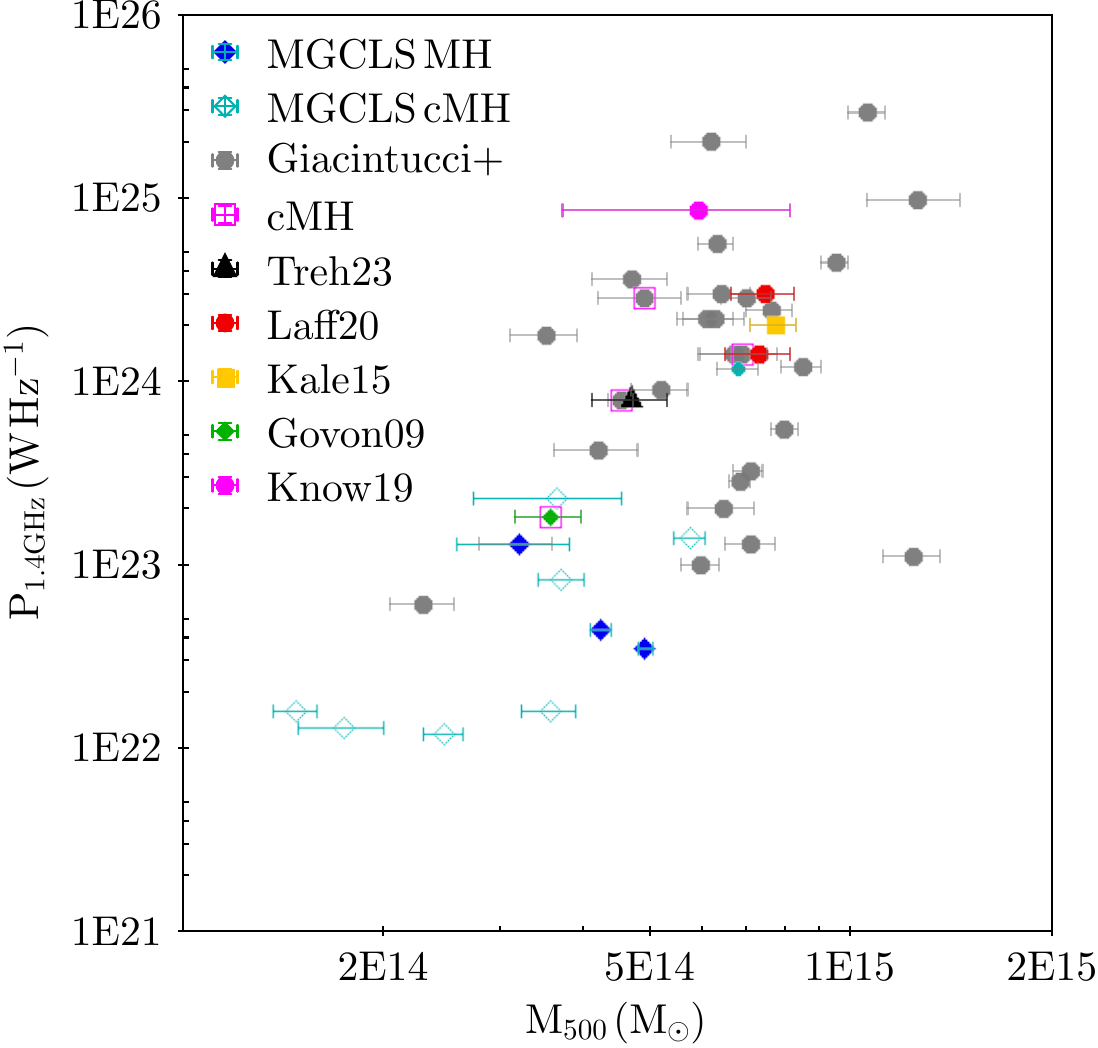}
\caption{1.4~GHz radio power versus M$_{500}$ for MGCLS radio mini-halos (blue rhombus) and candidate radio mini-halos (open rhombus). Different colours correspond to clusters from the literature. The grey circles correspond to the mini-halo sample studies by \citet{Giacintuccietal14,Giacintuccietal19}, the black triangle corresponds to a cluster by the recent study of \citet{Trehaevenetal23}, the red squares correspond to clusters by \citet{2020MNRAS.499.2934R}, the yellow square by \citet{Kale2015EGRHS}, the green rombus by \citet{Govonietal09} and the magenta square from \citet{Knowlesetal19}. The magenta open squares show the candidate radio mini-halos from the literature.} 
\label{P1400vsMhalo500}
\end{figure}

Although a close relationship between the radio properties of the BCG and the mini-halo is not expected due to the multiple AGN activity cycles that the radio galaxy may have during the lifespan of the mini-halo, \citet{Govonietal09} reported the tendency of strong mini-halos surrounding more powerful central radio galaxies. This suggests a partial relation between the mini-halo emission and the activity of the central AGN. A possible but weak trend using a larger mini-halo sample is also reported by \citet{Giacintuccietal14} between the radio power of the BCG and the radio luminosity of the surrounding mini-halos, which may be intrinsic, though with a large scatter. \citet{2020MNRAS.499.2934R} using a larger sample of all known cluster mini-halos at the time (33 mini-halos) found evidence of a strong correlation between the mini-halo radio power and the BCG's past AGN activity (steep radio power component), suggesting a link between the feedback processes of the central AGN and the mini-halo. 

The studies by \citet{2008A&A...486L..31C}, and later on by \citet{Kaleetal13},  despite the small sample and the large scatter in the relation, respectively, found signs of a possible correlation between the mini-halo radio power and the cluster total X--ray bolometric luminosity. The existence of such a correlation between radio power and cluster global and cool-core X--ray luminosity was confirmed later by several studies (e.g., \citealt{Kale2015EGRHS,Bravietal16,Gittietal18, Giacintuccietal19,2020MNRAS.499.2934R,Biavaetal24}), suggesting the existence of an underlying relationship between the thermal and non-thermal cluster properties.

In contrast to what is observed with giant radio halos in merging clusters where radio power is tightly correlated with cluster mass \citep[e.g.,][]{2013ApJ...777..141C,2019SSRv..215...16V,2021A&A...647A..51C}, in mini-halos such a correlation is not robust. Initially, \citet{Giacintuccietal14} and \citet{Giacintuccietal19} investigated this correlation using a sample of 14 and 23 mini-halos (known at the time), respectively, and found no evidence of a statistically significant correlation between the radio power of the mini-halo and the cluster global temperature or total mass. On the other hand, \citet{Yuanetal15} reported a mild correlation (using 12 mini-halos), and more recently \citet{2020MNRAS.499.2934R}, using a more up-to-date sample of 33 mini-halos, as well as more systematic estimates of M$_{500}$ from the Planck collaboration \citep{Planckcollab14}, found evidence of a general trend, between radio power at 1.4 GHz and M$_{500}$ using the BCES-orthogonal method.


Figure~\ref{P1400vsMhalo500} shows the 1.4~GHz radio power in relation to the M$_{500}$ mass for the three detected MGCLS radio mini-halos and seven candidate radio mini-halos overlaid with 33 clusters from the literature, from the mini-halo sample studies by \citet{Giacintuccietal14,Giacintuccietal19}, \citet{2020MNRAS.499.2934R} (and references therein) and the most recent study by \citet{Trehaevenetal23}. As is evident from Figure~\ref{P1400vsMhalo500}, the distribution above P$_{1.4\,GHz}$ of 10$^{23}$ W~Hz$^{-1}$ appears scattered without any trend or correlation being visible, consistent with the findings from earlier studies \citep{Giacintuccietal14,Giacintuccietal19}. We observe that mini-halos and candidate mini-halos typically lie at lower radio powers and masses, in agreement with previous studies \citep[e.g.,][]{Gittietal18}. In particular, most of the detected MGCLS radio mini-halos and candidate radio mini-halos are seen to occupy the lower mass region below M$_{500}$ of 5$\times10^{14}$ M$_{\odot}$ and the lower radio power region below P$_{1.4\,GHz}$ of 10$^{23}$ W~Hz$^{-1}$. Taking into account only the detected MGCLS radio mini-halos and candidate radio mini-halos, the Spearman test gives a rank correlation coefficient of rs = 0.59 with a p-value of 0.065, which by normal standards shows that there is no statistically significant association. However, it indicates the presence of a mild correlation between the 1.4~GHz radio power and cluster mass. 
Performing the Spearman test for all radio mini-halos and candidate radio mini-halos systems in Figure~\ref{P1400vsMhalo500}, we find a rank correlation coefficient of rs = 0.58 with a p-value $\leq0.001$, which by normal standards shows a statistically significant correlation between the 1.4~GHz radio power and the cluster mass, as previously suggested by \citet{2020MNRAS.499.2934R}. In light of this result, it is mandatory to perform a detailed multi-band study of the candidates to confirm their nature.

Adding the low-powered MGCLS candidate mini-halo systems in the diagram provides a new view of an unexplored mini-halo region thanks to the MGCLS's high sensitivity and ability to detect radio mini-halos and associate candidate systems at low-powers. The inclusion of the detected MGCLS mini-halos and respective candidate systems in the low mass region below M$_{500}$ of 5$\times10^{14}$ M$_{\odot}$ and the lower radio power region below P$_{1.4\,GHz}$ of 10$^{23}$ W~Hz$^{-1}$ provides new insights and confirms the existence of another correlation also between radio power and the mass of a cluster, in agreement with \citet{2020MNRAS.499.2934R}. This correlation appears to be intrinsic, revealing the link between the available energy deposit in cool-core clusters and the non-thermal particles that produce the emission of mini-halos.

 \section{Summary and Conclusions}
 \label{sec6}
 
In this paper, we present a follow-up study that provides detailed catalogue information on radio properties at 1.28~GHz of the MGCLS cluster-scale diffuse radio emissions that were reported in K22. The properties of the 62/115 MGCLS clusters with diffuse emission, together with relevant information on the diffuse radio sources and image quality data, are shown in Table~\ref{tab:diffuse}. We have calculated the local rms noise at the 15$^{\prime\prime}$ resolution image, the flux densities at 1.28~GHz, their largest physical size at the cluster redshift and their radio power at 1.28~GHz. Detailed literature information on 56 individual MGCLS systems is presented in Appendix~\ref{AppB} along with tentative in-band spectral index estimates for the diffuse radio sources from which it was possible to extract information. The full-resolution ($7.8''$) and tapered low-resolution ($15''$) images for the 62/115 galaxy clusters in the MGCLS that exhibit some diffuse radio emission are shown in Appendix~\ref{AppC} overlaid in optical Digital Sky Survey (DSS) images. We remind the reader that the MGCLS sample is heterogeneous and is used as a test bed for revealing the scientific potential of MeerKAT. Therefore, we note that these results are subject to this special case.




Our main conclusions from the heterogeneous MGCLS sample are summarised below. More than half ($\sim$~54\%; 62/115) of the MGCLS clusters present diffuse cluster radio emission, with the total number of detected diffuse radio sources, including candidates, being 103. This means that several of the MGCLS clusters with detected diffuse radio emission host more than one radio or candidate radio source with 58\% (60/103) of these detected diffuse radio sources reported as new in K22. Our 103 diffuse cluster emission detections can be summarised as follows: 3 mini-halos (all listed as new in K22) and eight mini-halo candidates (all new in K22), 26 halo detections (8 new in K22), and seven halo candidates (5 new in K22), 31 relics (15 new in K22) and 24 relic candidates (18 new in K22), one radio Phoenix and two new Phoenix candidates (all new in K22), and one new diffuse source with ambiguous or unknown classification (J0431.4-6126).

Including candidates, we find that $\sim$~10\% (11/115) of the 115 observed MGCLS clusters present a radio mini-halo. In contrast, radio halos in MGCLS clusters have a higher detection rate at 29\% (33/115), with clusters that exhibit only a radio halo without the presence of a relic at 16\% (18/115). The detection rate for MGCLS clusters that present at least one radio relic is 29\% (33/115), and the detection rate for MGCLS clusters that show a radio halo with double relics is $\sim$~7\% (8/115). The sources most commonly detected among the 103 diffuse radio sources in all clusters in the MGCLS sample are radio relics at 53\% (55/103), followed by radio halos at 32\% (33/103) and radio mini-halos at 11\% (11/103). Only 3\% (3/103) are found to be Phoenixes, with only 1\% of the detected radio sources listed as ambiguous/unknown.



The physical sizes of the detected MGCLS diffuse radio sources and candidates range from $\sim$~55~kpc to over 2~Mpc, with P$_{1.28GHz}$ radio power spanning from $\sim$~10$^{22}$ W~Hz$^{-1}$ to greater than 10$^{25}$ W~Hz$^{-1}$. Examining the in-band radio spectral index distribution between 908 and 1656~MHz, we find revived fossil plasma sources that exhibit very steep radio spectral indices of $\alpha_{908}^{1656}\sim-3.50$ (e.g., in Abell~13, Abell~3376, Abell~4038, Bullet, SPT-CL J2023.4-5535).


Examination of the relationship between the 1.4~GHz radio power and cluster mass shows that the radio powers of MGCLS systems with radio halos lie approximately along the  \citet{2013ApJ...777..141C} correlation in a mass range similar to that of the literature halos. 

MGCLS's high sensitivity and ability to detect radio mini-halos and associate candidate systems at low powers provide a new view of an unexplored mini-halo region. Considering only the detected MGCLS radio mini-halos and associate candidates for the same relation, we suggest the presence of a mild correlation between the 1.4~GHz radio power and the cluster mass (Spearman rank correlation coefficient rs = 0.59 with a p-value of 0.065). Considering systems also with radio mini-halos from the literature, alongside the MGCLS radio mini-halo systems, a statistically significant correlation is revealed between the 1.4~GHz radio power and cluster mass  (Spearman rank correlation coefficient rs = 0.58 with a p-value of $\leq0.001$)
. The inclusion of the detected MGCLS mini-halos and respective candidate systems at the low mass region below M$_{500}$ of 5$\times10^{14}$ M$_{\odot}$ and lower radio power region below P$_{1.4\,GHz}$ of 10$^{23}$ W~Hz$^{-1}$ provides new insights and confirms the existence of another correlation also between radio power and the mass of a cluster in agreement with \citep{2020MNRAS.499.2934R}. 

The rapid progress of radio-astronomy and the operation of new-era radio telescopes (e.g., LOFAR and MeerKAT) but also the upgrading of existing ones (such as uGMRT and JVLA) has allowed deeper observations of radio sources at very low surface brightness levels leading to increased detection of radio halos and relics in merging clusters over a wide range of cluster masses \citep[e.g.,][]{2021A&A...647A..50C,Rajpurohitetal22,Hoangetal25} and redshifts \citep[e.g.,][]{2021NatAs...5..268D}. Another example is the very faint (low luminosity) radio relic candidates detected by MeerKAT (K22), exploiting the excellent surface brightness sensitivity of the MGCLS. However, despite this rapid phase of radio evolution, current radio observational constraints still have room for improvement, as radio mini-halos often evade detection in large numbers. This indicates that the galaxy clusters observed in the MGCLS sample, a precursor project for the Square Kilometre Array (SKA), are only the tip of the iceberg of several diffuse cluster radio emission discoveries that will be revealed in the SKA era.

The classification challenges of diffuse radio sources in several MGCLS clusters, as well as the presence of diffuse radio emissions of uncertain and candidate nature, give rise to a unique opportunity in the study of galaxy clusters from MHz to GHz frequencies, opening up new areas of investigation in cluster formation and evolution introducing the need to investigate further the link between radio galaxies and the particle reservoir they deposit into the ICM, but also, in general, the mechanism of cluster merger events \citep[e.g.,][]{2017SciA....3E1634D,Schellenberger22}. The MGCLS products also provide a fertile base for testing and implementing machine learning algorithms morphologically, since the use of new-generation radio data for automated detections and classifications of diffuse and AGN-related radio emissions is a requirement for robust radio source classification (e.g. \citealt{Stuardietal24}). This will accelerate the process of identifying radio sources; however, securing a reliable radio source classification based on their nature will be challenging.

\section*{Acknowledgments}
The MeerKAT telescope is operated by the South African Radio Astronomy Observatory (SARAO), which is a facility of the National Research Foundation (NRF), an agency of the Department of Science and Innovation. K. Kolokythas and K. Knowles acknowledge funding support from the South Africa Radio Astronomy Observatory and the National Research Foundation (grant UID: 97930). T. Venturi acknowledges that this work was supported in part by the Italian Ministry of Foreign Affairs and International Cooperation, grant number ZA23GR03. M. Br\"{u}ggen acknowledges funding by the Deutsche Forschungsgemeinschaft (DFG, German Research Foundation) under Germany's Excellence Strategy -- EXC 2121 ``Quantum Universe'' -- 390833306 and the DFG Research Group "Relativistic Jets". F. de Gasperin acknowledges support from the ERC Consolidator Grant ULU 101086378. S.P. Sikhosana acknowledges the financial assistance of SARAO towards this research (www.sarao.ac.za) and also funding support from the NRF towards this research. Basic Research in radio astronomy at the U.S. Naval Research Laboratory is supported by 6.1 Base funding. The research of O.M. Smirnov  is supported by the South African Research Chairs Initiative of the DSI/NRF (grant no. 81737). S.I. Loubser is supported in part by the NRF of South Africa (NRF Grant Number: CPRR240414214079). K. Moodley acknowledges support from the NRF of South Africa. Opinions expressed, and conclusions arrived at are those of the authors and are not necessarily to be attributed to the NRF. This research has made use of the NASA/IPAC Extragalactic Database (NED) which is operated by the Jet Propulsion Laboratory, California Institute of Technology, under contract with the National Aeronautics and Space Administration. 


\section*{Data Availability}

The data underlying this article are available in the article and can be found in Table~\ref{tab:diffuse}.


\bibstyle{mnras}
\bibliography{gcls_catalog.bib} 

\begin{thebibliography}{}
\makeatletter
\relax
\def\mn@urlcharsother{\let\do\@makeother \do\$\do\&\do\#\do\^\do\_\do\%\do\~}
\def\mn@doi{\begingroup\mn@urlcharsother \@ifnextchar [ {\mn@doi@}
  {\mn@doi@[]}}
\def\mn@doi@[#1]#2{\def\@tempa{#1}\ifx\@tempa\@empty \href
  {http://dx.doi.org/#2} {doi:#2}\else \href {http://dx.doi.org/#2} {#1}\fi
  \endgroup}
\def\mn@eprint#1#2{\mn@eprint@#1:#2::\@nil}
\def\mn@eprint@arXiv#1{\href {http://arxiv.org/abs/#1} {{\tt arXiv:#1}}}
\def\mn@eprint@dblp#1{\href {http://dblp.uni-trier.de/rec/bibtex/#1.xml}
  {dblp:#1}}
\def\mn@eprint@#1:#2:#3:#4\@nil{\def\@tempa {#1}\def\@tempb {#2}\def\@tempc
  {#3}\ifx \@tempc \@empty \let \@tempc \@tempb \let \@tempb \@tempa \fi \ifx
  \@tempb \@empty \def\@tempb {arXiv}\fi \@ifundefined
  {mn@eprint@\@tempb}{\@tempb:\@tempc}{\expandafter \expandafter \csname
  mn@eprint@\@tempb\endcsname \expandafter{\@tempc}}}

\bibitem[\protect\citeauthoryear{{Abbott} et~al.,}{{Abbott}
  et~al.}{2018}]{2018ApJS..239...18A}
{Abbott} T.~M.~C.,  et~al., 2018, \mn@doi [\apjs] {10.3847/1538-4365/aae9f0},
  \href {https://ui.adsabs.harvard.edu/abs/2018ApJS..239...18A} {239, 18}

\bibitem[\protect\citeauthoryear{Abell, Corwin  \& Olowin}{Abell
  et~al.}{1989}]{Abelletal89}
Abell G.~O.,  Corwin H.~G.,   Olowin R.~P.,  1989, ApJS, 70, 1

\bibitem[\protect\citeauthoryear{{Akimoto}, {Kondou}, {Furuzawa}, {Tawara}  \&
  {Yamashita}}{{Akimoto} et~al.}{2003}]{2003ApJ...596..170A}
{Akimoto} F.,  {Kondou} K.,  {Furuzawa} A.,  {Tawara} Y.,   {Yamashita} K.,
  2003, \mn@doi [\apj] {10.1086/377629}, \href
  {https://ui.adsabs.harvard.edu/abs/2003ApJ...596..170A} {596, 170}

\bibitem[\protect\citeauthoryear{{Alam}, {Pimbblet}  \& {Gordon}}{{Alam}
  et~al.}{2025}]{Alametal25}
{Alam} A.,  {Pimbblet} K.,   {Gordon} Y.,  2025, \mn@doi [\pasa]
  {10.1017/pasa.2024.133}, \href
  {https://ui.adsabs.harvard.edu/abs/2025PASA...42...13A} {42, e013}

\bibitem[\protect\citeauthoryear{{Alvarez}, {Randall}, {Bourdin}, {Jones}  \&
  {Holley-Bockelmann}}{{Alvarez} et~al.}{2018}]{Alvarezetal18}
{Alvarez} G.~E.,  {Randall} S.~W.,  {Bourdin} H.,  {Jones} C.,
  {Holley-Bockelmann} K.,  2018, \mn@doi [\apj] {10.3847/1538-4357/aabad0},
  \href {https://ui.adsabs.harvard.edu/abs/2018ApJ...858...44A} {858, 44}

\bibitem[\protect\citeauthoryear{{Aniyan} \& {Thorat}}{{Aniyan} \&
  {Thorat}}{2017}]{AniyanThorat17}
{Aniyan} A.~K.,  {Thorat} K.,  2017, \mn@doi [\apjs]
  {10.3847/1538-4365/aa7333}, \href
  {https://ui.adsabs.harvard.edu/abs/2017ApJS..230...20A} {230, 20}

\bibitem[\protect\citeauthoryear{{B{\" o}hringer}, {Matsushita}, {Churazov},
  {Finoguenov}  \& {Ikebe}}{{B{\" o}hringer} et~al.}{2004}]{Bohringeretal04}
{B{\" o}hringer} H.,  {Matsushita} K.,  {Churazov} E.,  {Finoguenov} A.,
  {Ikebe} Y.,  2004, A\&A, \href
  {http://adsabs.harvard.edu/cgi-bin/nph-bib_query?bibcode=2004A\%26A...416L..21B&amp;db_key=AST}
  {416, L21}

\bibitem[\protect\citeauthoryear{{Bacchi}, {Feretti}, {Giovannini}  \&
  {Govoni}}{{Bacchi} et~al.}{2003}]{2003A&A...400..465B}
{Bacchi} M.,  {Feretti} L.,  {Giovannini} G.,   {Govoni} F.,  2003, \mn@doi
  [\aap] {10.1051/0004-6361:20030044}, \href
  {https://ui.adsabs.harvard.edu/abs/2003A&A...400..465B} {400, 465}

\bibitem[\protect\citeauthoryear{{Bagchi}, {Durret}, {Neto}  \&
  {Paul}}{{Bagchi} et~al.}{2006}]{2006Sci...314..791B}
{Bagchi} J.,  {Durret} F.,  {Neto} G. B.~L.,   {Paul} S.,  2006, \mn@doi
  [Science] {10.1126/science.1131189}, \href
  {https://ui.adsabs.harvard.edu/abs/2006Sci...314..791B} {314, 791}

\bibitem[\protect\citeauthoryear{{Bardelli}, {De Grandi}, {Ettori}, {Molendi},
  {Zucca}  \& {Colafrancesco}}{{Bardelli} et~al.}{2002}]{Bardellietal02}
{Bardelli} S.,  {De Grandi} S.,  {Ettori} S.,  {Molendi} S.,  {Zucca} E.,
  {Colafrancesco} S.,  2002, \mn@doi [\aap] {10.1051/0004-6361:20011665}, \href
  {https://ui.adsabs.harvard.edu/abs/2002A&A...382...17B} {382, 17}

\bibitem[\protect\citeauthoryear{{Barrena}, {Biviano}, {Ramella}, {Falco}  \&
  {Seitz}}{{Barrena} et~al.}{2002}]{Barrenaetal02}
{Barrena} R.,  {Biviano} A.,  {Ramella} M.,  {Falco} E.~E.,   {Seitz} S.,
  2002, \mn@doi [\aap] {10.1051/0004-6361:20020244}, \href
  {https://ui.adsabs.harvard.edu/abs/2002A&A...386..816B} {386, 816}

\bibitem[\protect\citeauthoryear{{Barrena}, {Girardi}, {Boschin}, {de Grandi},
  {Eckert}  \& {Rossetti}}{{Barrena} et~al.}{2011}]{Barrenaetal11}
{Barrena} R.,  {Girardi} M.,  {Boschin} W.,  {de Grandi} S.,  {Eckert} D.,
  {Rossetti} M.,  2011, \mn@doi [\aap] {10.1051/0004-6361/201016180}, \href
  {https://ui.adsabs.harvard.edu/abs/2011A&A...529A.128B} {529, A128}

\bibitem[\protect\citeauthoryear{{B{\'e}gin} et~al.,}{{B{\'e}gin}
  et~al.}{2023}]{Beginetal23}
{B{\'e}gin} T.,  et~al., 2023, \mn@doi [\mnras] {10.1093/mnras/stac3526}, \href
  {https://ui.adsabs.harvard.edu/abs/2023MNRAS.519..767B} {519, 767}

\bibitem[\protect\citeauthoryear{{Bernardi} et~al.,}{{Bernardi}
  et~al.}{2016}]{2016MNRAS.456.1259B}
{Bernardi} G.,  et~al., 2016, \mn@doi [\mnras] {10.1093/mnras/stv2589}, \href
  {https://ui.adsabs.harvard.edu/abs/2016MNRAS.456.1259B} {456, 1259}

\bibitem[\protect\citeauthoryear{{Biava} et~al.,}{{Biava}
  et~al.}{2021}]{Biavaetal21}
{Biava} N.,  et~al., 2021, \mn@doi [\mnras] {10.1093/mnras/stab2840}, \href
  {https://ui.adsabs.harvard.edu/abs/2021MNRAS.508.3995B} {508, 3995}

\bibitem[\protect\citeauthoryear{{Biava} et~al.,}{{Biava}
  et~al.}{2024}]{Biavaetal24}
{Biava} N.,  et~al., 2024, \mn@doi [\aap] {10.1051/0004-6361/202348045}, \href
  {https://ui.adsabs.harvard.edu/abs/2024A&A...686A..82B} {686, A82}

\bibitem[\protect\citeauthoryear{{B{\^\i}rzan}, {Rafferty}, {Br{\"u}ggen}  \&
  {Intema}}{{B{\^\i}rzan} et~al.}{2017}]{Birzanetal17}
{B{\^\i}rzan} L.,  {Rafferty} D.~A.,  {Br{\"u}ggen} M.,   {Intema} H.~T.,
  2017, \mn@doi [\mnras] {10.1093/mnras/stx1505}, \href
  {https://ui.adsabs.harvard.edu/abs/2017MNRAS.471.1766B} {471, 1766}

\bibitem[\protect\citeauthoryear{{Bleem} et~al.,}{{Bleem}
  et~al.}{2015}]{Bleemetal15}
{Bleem} L.~E.,  et~al., 2015, \mn@doi [\apjs] {10.1088/0067-0049/216/2/27},
  \href {https://ui.adsabs.harvard.edu/abs/2015ApJS..216...27B} {216, 27}

\bibitem[\protect\citeauthoryear{{Bock}, {Large}  \& {Sadler}}{{Bock}
  et~al.}{1999}]{SUMSS}
{Bock} D.~C.~J.,  {Large} M.~I.,   {Sadler} E.~M.,  1999, \mn@doi [\aj]
  {10.1086/300786}, \href
  {https://ui.adsabs.harvard.edu/abs/1999AJ....117.1578B} {117, 1578}

\bibitem[\protect\citeauthoryear{{B{\"o}hringer} et~al.,}{{B{\"o}hringer}
  et~al.}{2007}]{Bohringeretal07}
{B{\"o}hringer} H.,  et~al., 2007, \mn@doi [\aap] {10.1051/0004-6361:20066740},
  \href {https://ui.adsabs.harvard.edu/abs/2007A&A...469..363B} {469, 363}

\bibitem[\protect\citeauthoryear{{B{\"o}hringer}, {Chon}, {Collins}, {Guzzo},
  {Nowak}  \& {Bobrovskyi}}{{B{\"o}hringer} et~al.}{2013}]{Bohringeretal13}
{B{\"o}hringer} H.,  {Chon} G.,  {Collins} C.~A.,  {Guzzo} L.,  {Nowak} N.,
  {Bobrovskyi} S.,  2013, \mn@doi [\aap] {10.1051/0004-6361/201220722}, \href
  {https://ui.adsabs.harvard.edu/abs/2013A&A...555A..30B} {555, A30}

\bibitem[\protect\citeauthoryear{{Bonafede} et~al.,}{{Bonafede}
  et~al.}{2012}]{2012MNRAS.426...40B}
{Bonafede} A.,  et~al., 2012, \mn@doi [\mnras]
  {10.1111/j.1365-2966.2012.21570.x}, \href
  {https://ui.adsabs.harvard.edu/abs/2012MNRAS.426...40B} {426, 40}

\bibitem[\protect\citeauthoryear{{Bonafede} et~al.,}{{Bonafede}
  et~al.}{2017}]{2017MNRAS.470.3465B}
{Bonafede} A.,  et~al., 2017, \mn@doi [\mnras] {10.1093/mnras/stx1475}, \href
  {https://ui.adsabs.harvard.edu/abs/2017MNRAS.470.3465B} {470, 3465}

\bibitem[\protect\citeauthoryear{{Botteon}, {Gastaldello}, {Brunetti}  \&
  {Kale}}{{Botteon} et~al.}{2016}]{Botteonetal16}
{Botteon} A.,  {Gastaldello} F.,  {Brunetti} G.,   {Kale} R.,  2016, \mn@doi
  [\mnras] {10.1093/mnras/stw2089}, \href
  {https://ui.adsabs.harvard.edu/abs/2016MNRAS.463.1534B} {463, 1534}

\bibitem[\protect\citeauthoryear{{Botteon}, {Gastaldello}  \&
  {Brunetti}}{{Botteon} et~al.}{2018}]{Botteonetal18}
{Botteon} A.,  {Gastaldello} F.,   {Brunetti} G.,  2018, \mn@doi [\mnras]
  {10.1093/mnras/sty598}, \href
  {https://ui.adsabs.harvard.edu/abs/2018MNRAS.476.5591B} {476, 5591}

\bibitem[\protect\citeauthoryear{{Botteon} et~al.,}{{Botteon}
  et~al.}{2020}]{2020MNRAS.499L..11B}
{Botteon} A.,  et~al., 2020, \mn@doi [\mnras] {10.1093/mnrasl/slaa142}, \href
  {https://ui.adsabs.harvard.edu/abs/2020MNRAS.499L..11B} {499, L11}

\bibitem[\protect\citeauthoryear{{Botteon} et~al.,}{{Botteon}
  et~al.}{2022}]{Botteonetal22}
{Botteon} A.,  et~al., 2022, \mn@doi [\aap] {10.1051/0004-6361/202143020},
  \href {https://ui.adsabs.harvard.edu/abs/2022A&A...660A..78B} {660, A78}

\bibitem[\protect\citeauthoryear{{Botteon}, {Markevitch}, {van Weeren},
  {Brunetti}  \& {Shimwell}}{{Botteon} et~al.}{2023}]{Botteon23}
{Botteon} A.,  {Markevitch} M.,  {van Weeren} R.~J.,  {Brunetti} G.,
  {Shimwell} T.~W.,  2023, \mn@doi [\aap] {10.1051/0004-6361/202346150}, \href
  {https://ui.adsabs.harvard.edu/abs/2023A&A...674A..53B} {674, A53}

\bibitem[\protect\citeauthoryear{{Bourdin}, {Mazzotta}, {Markevitch},
  {Giacintucci}  \& {Brunetti}}{{Bourdin} et~al.}{2013}]{Bourdinetal13}
{Bourdin} H.,  {Mazzotta} P.,  {Markevitch} M.,  {Giacintucci} S.,   {Brunetti}
  G.,  2013, \mn@doi [\apj] {10.1088/0004-637X/764/1/82}, \href
  {https://ui.adsabs.harvard.edu/abs/2013ApJ...764...82B} {764, 82}

\bibitem[\protect\citeauthoryear{{Bravi}, {Gitti}  \& {Brunetti}}{{Bravi}
  et~al.}{2016}]{Bravietal16}
{Bravi} L.,  {Gitti} M.,   {Brunetti} G.,  2016, \mn@doi [\mnras]
  {10.1093/mnrasl/slv137}, \href
  {https://ui.adsabs.harvard.edu/abs/2016MNRAS.455L..41B} {455, L41}

\bibitem[\protect\citeauthoryear{{Br{\"u}ggen} et~al.,}{{Br{\"u}ggen}
  et~al.}{2021a}]{Bruggenetal21}
{Br{\"u}ggen} M.,  et~al., 2021a, \mn@doi [\aap] {10.1051/0004-6361/202039533},
  \href {https://ui.adsabs.harvard.edu/abs/2021A&A...647A...3B} {647, A3}

\bibitem[\protect\citeauthoryear{{Br{\"u}ggen} et~al.,}{{Br{\"u}ggen}
  et~al.}{2021b}]{2021A&A...647A...3B}
{Br{\"u}ggen} M.,  et~al., 2021b, \mn@doi [\aap] {10.1051/0004-6361/202039533},
  \href {https://ui.adsabs.harvard.edu/abs/2021A&A...647A...3B} {647, A3}

\bibitem[\protect\citeauthoryear{{Brunetti} \& {Jones}}{{Brunetti} \&
  {Jones}}{2014}]{2014IJMPD..2330007B}
{Brunetti} G.,  {Jones} T.~W.,  2014, \mn@doi [International Journal of Modern
  Physics D] {10.1142/S0218271814300079}, \href
  {https://ui.adsabs.harvard.edu/abs/2014IJMPD..2330007B} {23, 1430007}

\bibitem[\protect\citeauthoryear{{Brunetti} et~al.,}{{Brunetti}
  et~al.}{2008}]{2008Natur.455..944B}
{Brunetti} G.,  et~al., 2008, \mn@doi [\nat] {10.1038/nature07379}, \href
  {https://ui.adsabs.harvard.edu/abs/2008Natur.455..944B} {455, 944}

\bibitem[\protect\citeauthoryear{{Brunetti}, {Cassano}, {Dolag}  \&
  {Setti}}{{Brunetti} et~al.}{2009}]{Brunettietal09}
{Brunetti} G.,  {Cassano} R.,  {Dolag} K.,   {Setti} G.,  2009, \mn@doi [\aap]
  {10.1051/0004-6361/200912751}, \href
  {https://ui.adsabs.harvard.edu/abs/2009A&A...507..661B} {507, 661}

\bibitem[\protect\citeauthoryear{{Bulbul} et~al.,}{{Bulbul}
  et~al.}{2022}]{Bulbuletal22}
{Bulbul} E.,  et~al., 2022, \mn@doi [\aap] {10.1051/0004-6361/202142460}, \href
  {https://ui.adsabs.harvard.edu/abs/2022A&A...661A..10B} {661, A10}

\bibitem[\protect\citeauthoryear{{Bulbul} et~al.,}{{Bulbul}
  et~al.}{2024}]{Bulbuletal24}
{Bulbul} E.,  et~al., 2024, \mn@doi [\aap] {10.1051/0004-6361/202348264}, \href
  {https://ui.adsabs.harvard.edu/abs/2024A&A...685A.106B} {685, A106}

\bibitem[\protect\citeauthoryear{{Buote}}{{Buote}}{2001}]{Buote01}
{Buote} D.~A.,  2001, \mn@doi [\apjl] {10.1086/320500}, \href
  {https://ui.adsabs.harvard.edu/abs/2001ApJ...553L..15B} {553, L15}

\bibitem[\protect\citeauthoryear{Camilo, Scholz, Serylak  \& et al.}{Camilo
  et~al.}{2018}]{Camilo2018}
Camilo F.,  Scholz P.,  Serylak M.,   et al. 2018, \mn@doi [ApJ]
  {10.3847/1538-4357/aab35a}, 856, 180

\bibitem[\protect\citeauthoryear{{Cassano} \& {Brunetti}}{{Cassano} \&
  {Brunetti}}{2005}]{CassanoBrunetti05}
{Cassano} R.,  {Brunetti} G.,  2005, \mn@doi [\mnras]
  {10.1111/j.1365-2966.2005.08747.x}, \href
  {https://ui.adsabs.harvard.edu/abs/2005MNRAS.357.1313C} {357, 1313}

\bibitem[\protect\citeauthoryear{{Cassano}, {Brunetti}  \& {Setti}}{{Cassano}
  et~al.}{2006}]{Cassanoetal06}
{Cassano} R.,  {Brunetti} G.,   {Setti} G.,  2006, \mn@doi [\mnras]
  {10.1111/j.1365-2966.2006.10423.x}, \href
  {https://ui.adsabs.harvard.edu/abs/2006MNRAS.369.1577C} {369, 1577}

\bibitem[\protect\citeauthoryear{{Cassano}, {Brunetti}, {Setti}, {Govoni}  \&
  {Dolag}}{{Cassano} et~al.}{2007}]{Cassanoetal07}
{Cassano} R.,  {Brunetti} G.,  {Setti} G.,  {Govoni} F.,   {Dolag} K.,  2007,
  \mn@doi [\mnras] {10.1111/j.1365-2966.2007.11901.x}, \href
  {https://ui.adsabs.harvard.edu/abs/2007MNRAS.378.1565C} {378, 1565}

\bibitem[\protect\citeauthoryear{{Cassano}, {Gitti}  \& {Brunetti}}{{Cassano}
  et~al.}{2008}]{2008A&A...486L..31C}
{Cassano} R.,  {Gitti} M.,   {Brunetti} G.,  2008, \mn@doi [\aap]
  {10.1051/0004-6361:200810179}, \href
  {https://ui.adsabs.harvard.edu/abs/2008A&A...486L..31C} {486, L31}

\bibitem[\protect\citeauthoryear{{Cassano} et~al.,}{{Cassano}
  et~al.}{2013}]{2013ApJ...777..141C}
{Cassano} R.,  et~al., 2013, \mn@doi [\apj] {10.1088/0004-637X/777/2/141},
  \href {https://ui.adsabs.harvard.edu/abs/2013ApJ...777..141C} {777, 141}

\bibitem[\protect\citeauthoryear{{Cassano}, {Brunetti}, {Giocoli}  \&
  {Ettori}}{{Cassano} et~al.}{2016}]{Cassanoetal16}
{Cassano} R.,  {Brunetti} G.,  {Giocoli} C.,   {Ettori} S.,  2016, \mn@doi
  [\aap] {10.1051/0004-6361/201628414}, \href
  {https://ui.adsabs.harvard.edu/abs/2016A&A...593A..81C} {593, A81}

\bibitem[\protect\citeauthoryear{{Cassano} et~al.,}{{Cassano}
  et~al.}{2023}]{Cassanoetal23}
{Cassano} R.,  et~al., 2023, \mn@doi [\aap] {10.1051/0004-6361/202244876},
  \href {https://ui.adsabs.harvard.edu/abs/2023A&A...672A..43C} {672, A43}

\bibitem[\protect\citeauthoryear{{Cavagnolo}, {Donahue}, {Voit}  \&
  {Sun}}{{Cavagnolo} et~al.}{2009}]{Cavagnoloetal09}
{Cavagnolo} K.~W.,  {Donahue} M.,  {Voit} G.~M.,   {Sun} M.,  2009, \mn@doi
  [ApJS] {10.1088/0067-0049/182/1/12}, \href
  {http://adsabs.harvard.edu/abs/2009ApJS..182...12C} {182, 12}

\bibitem[\protect\citeauthoryear{{Chibueze} et~al.,}{{Chibueze}
  et~al.}{2021}]{2021Natur.593...47C}
{Chibueze} J.~O.,  et~al., 2021, \mn@doi [\nat] {10.1038/s41586-021-03434-1},
  \href {https://ui.adsabs.harvard.edu/abs/2021Natur.593...47C} {593, 47}

\bibitem[\protect\citeauthoryear{{Chon} \& {B{\"o}hringer}}{{Chon} \&
  {B{\"o}hringer}}{2012}]{2012A&A...538A..35C}
{Chon} G.,  {B{\"o}hringer} H.,  2012, \mn@doi [\aap]
  {10.1051/0004-6361/201117996}, \href
  {https://ui.adsabs.harvard.edu/abs/2012A&A...538A..35C} {538, A35}

\bibitem[\protect\citeauthoryear{{Chon}, {B{\"o}hringer}, {Dasadia}, {Kluge},
  {Sun}, {Forman}  \& {Jones}}{{Chon} et~al.}{2019}]{Chonetal19}
{Chon} G.,  {B{\"o}hringer} H.,  {Dasadia} S.,  {Kluge} M.,  {Sun} M.,
  {Forman} W.~R.,   {Jones} C.,  2019, \mn@doi [\aap]
  {10.1051/0004-6361/201833068}, \href
  {https://ui.adsabs.harvard.edu/abs/2019A&A...621A..77C} {621, A77}

\bibitem[\protect\citeauthoryear{{Chu}, {Durret}  \& {M{\'a}rquez}}{{Chu}
  et~al.}{2021}]{2021A&A...649A..42C}
{Chu} A.,  {Durret} F.,   {M{\'a}rquez} I.,  2021, \mn@doi [\aap]
  {10.1051/0004-6361/202040245}, \href
  {https://ui.adsabs.harvard.edu/abs/2021A&A...649A..42C} {649, A42}

\bibitem[\protect\citeauthoryear{{Cibirka} et~al.,}{{Cibirka}
  et~al.}{2018}]{2018ApJ...863..145C}
{Cibirka} N.,  et~al., 2018, \mn@doi [\apj] {10.3847/1538-4357/aad2d3}, \href
  {https://ui.adsabs.harvard.edu/abs/2018ApJ...863..145C} {863, 145}

\bibitem[\protect\citeauthoryear{{Cotton}}{{Cotton}}{2008}]{OBIT}
{Cotton} W.~D.,  2008, \mn@doi [PASP] {10.1086/586754}, 120, 439

\bibitem[\protect\citeauthoryear{{Covone}, {Adami}, {Durret}, {Kneib}, {Lima
  Neto}  \& {Slezak}}{{Covone} et~al.}{2006}]{2006A&A...460..381C}
{Covone} G.,  {Adami} C.,  {Durret} F.,  {Kneib} J.~P.,  {Lima Neto} G.~B.,
  {Slezak} E.,  2006, \mn@doi [\aap] {10.1051/0004-6361:20053970}, \href
  {https://ui.adsabs.harvard.edu/abs/2006A&A...460..381C} {460, 381}

\bibitem[\protect\citeauthoryear{{Coziol}, {Andernach}, {Caretta},
  {Alamo-Mart{\'\i}nez}  \& {Tago}}{{Coziol}
  et~al.}{2009}]{2009AJ....137.4795C}
{Coziol} R.,  {Andernach} H.,  {Caretta} C.~A.,  {Alamo-Mart{\'\i}nez} K.~A.,
  {Tago} E.,  2009, \mn@doi [\aj] {10.1088/0004-6256/137/6/4795}, \href
  {https://ui.adsabs.harvard.edu/abs/2009AJ....137.4795C} {137, 4795}

\bibitem[\protect\citeauthoryear{{Cruddace} et~al.,}{{Cruddace}
  et~al.}{2002}]{Cruddaceetal02}
{Cruddace} R.,  et~al., 2002, \mn@doi [\apjs] {10.1086/324519}, \href
  {https://ui.adsabs.harvard.edu/abs/2002ApJS..140..239C} {140, 239}

\bibitem[\protect\citeauthoryear{{Cuciti}, {Cassano}, {Brunetti}, {Dallacasa},
  {Kale}, {Ettori}  \& {Venturi}}{{Cuciti} et~al.}{2015}]{Cucitietal15}
{Cuciti} V.,  {Cassano} R.,  {Brunetti} G.,  {Dallacasa} D.,  {Kale} R.,
  {Ettori} S.,   {Venturi} T.,  2015, \mn@doi [\aap]
  {10.1051/0004-6361/201526420}, \href
  {https://ui.adsabs.harvard.edu/abs/2015A&A...580A..97C} {580, A97}

\bibitem[\protect\citeauthoryear{{Cuciti} et~al.,}{{Cuciti}
  et~al.}{2021a}]{2021A&A...647A..50C}
{Cuciti} V.,  et~al., 2021a, \mn@doi [\aap] {10.1051/0004-6361/202039206},
  \href {https://ui.adsabs.harvard.edu/abs/2021A&A...647A..50C} {647, A50}

\bibitem[\protect\citeauthoryear{{Cuciti} et~al.,}{{Cuciti}
  et~al.}{2021b}]{2021A&A...647A..51C}
{Cuciti} V.,  et~al., 2021b, \mn@doi [\aap] {10.1051/0004-6361/202039208},
  \href {https://ui.adsabs.harvard.edu/abs/2021A&A...647A..51C} {647, A51}

\bibitem[\protect\citeauthoryear{{Cuciti} et~al.,}{{Cuciti}
  et~al.}{2022}]{2022Natur.609..911C}
{Cuciti} V.,  et~al., 2022, \mn@doi [\nat] {10.1038/s41586-022-05149-3}, \href
  {https://ui.adsabs.harvard.edu/abs/2022Natur.609..911C} {609, 911}

\bibitem[\protect\citeauthoryear{{Cuciti} et~al.,}{{Cuciti}
  et~al.}{2023}]{Cucitietal23}
{Cuciti} V.,  et~al., 2023, \mn@doi [\aap] {10.1051/0004-6361/202346755}, \href
  {https://ui.adsabs.harvard.edu/abs/2023A&A...680A..30C} {680, A30}

\bibitem[\protect\citeauthoryear{{Cutri} \& {et al.}}{{Cutri} \& {et
  al.}}{2012}]{Cutrietal12}
{Cutri} R.~M.,  {et al.} 2012, VizieR Online Data Catalog, \href
  {https://ui.adsabs.harvard.edu/abs/2012yCat.2311....0C} {p. II/311}

\bibitem[\protect\citeauthoryear{{Dallacasa} et~al.,}{{Dallacasa}
  et~al.}{2009}]{2009ApJ...699.1288D}
{Dallacasa} D.,  et~al., 2009, \mn@doi [\apj] {10.1088/0004-637X/699/2/1288},
  \href {https://ui.adsabs.harvard.edu/abs/2009ApJ...699.1288D} {699, 1288}

\bibitem[\protect\citeauthoryear{{De Grandi} et~al.,}{{De Grandi}
  et~al.}{1999}]{DeGrandietal99}
{De Grandi} S.,  et~al., 1999, \mn@doi [\apj] {10.1086/306939}, \href
  {https://ui.adsabs.harvard.edu/abs/1999ApJ...514..148D} {514, 148}

\bibitem[\protect\citeauthoryear{{Dehghan}, {Johnston-Hollitt}, {Colless}  \&
  {Miller}}{{Dehghan} et~al.}{2017}]{Dehghanetal17}
{Dehghan} S.,  {Johnston-Hollitt} M.,  {Colless} M.,   {Miller} R.,  2017,
  \mn@doi [\mnras] {10.1093/mnras/stx582}, \href
  {https://ui.adsabs.harvard.edu/abs/2017MNRAS.468.2645D} {468, 2645}

\bibitem[\protect\citeauthoryear{{Dey} et~al.,}{{Dey} et~al.}{2019}]{Decals}
{Dey} A.,  et~al., 2019, \mn@doi [\aj] {10.3847/1538-3881/ab089d}, \href
  {https://ui.adsabs.harvard.edu/abs/2019AJ....157..168D} {157, 168}

\bibitem[\protect\citeauthoryear{{Di Gennaro} et~al.,}{{Di Gennaro}
  et~al.}{2019}]{2019ApJ...873...64D}
{Di Gennaro} G.,  et~al., 2019, \mn@doi [\apj] {10.3847/1538-4357/ab03cd},
  \href {https://ui.adsabs.harvard.edu/abs/2019ApJ...873...64D} {873, 64}

\bibitem[\protect\citeauthoryear{{Di Gennaro} et~al.,}{{Di Gennaro}
  et~al.}{2021a}]{2021NatAs...5..268D}
{Di Gennaro} G.,  et~al., 2021a, \mn@doi [Nature Astronomy]
  {10.1038/s41550-020-01244-5}, \href
  {https://ui.adsabs.harvard.edu/abs/2021NatAs...5..268D} {5, 268}

\bibitem[\protect\citeauthoryear{{Di Gennaro} et~al.,}{{Di Gennaro}
  et~al.}{2021b}]{DiGennaroetal21}
{Di Gennaro} G.,  et~al., 2021b, \mn@doi [\aap] {10.1051/0004-6361/202141510},
  \href {https://ui.adsabs.harvard.edu/abs/2021A&A...654A.166D} {654, A166}

\bibitem[\protect\citeauthoryear{{Donnert}}{{Donnert}}{2014}]{Donnert14}
{Donnert} J.~M.~F.,  2014, \mn@doi [\mnras] {10.1093/mnras/stt2291}, \href
  {https://ui.adsabs.harvard.edu/abs/2014MNRAS.438.1971D} {438, 1971}

\bibitem[\protect\citeauthoryear{{Donnert}, {Dolag}, {Brunetti}  \&
  {Cassano}}{{Donnert} et~al.}{2013}]{Donnertetal13}
{Donnert} J.,  {Dolag} K.,  {Brunetti} G.,   {Cassano} R.,  2013, \mn@doi
  [\mnras] {10.1093/mnras/sts628}, \href
  {https://ui.adsabs.harvard.edu/abs/2013MNRAS.429.3564D} {429, 3564}

\bibitem[\protect\citeauthoryear{{Dressler}, {Oemler}, {Butcher}  \&
  {Gunn}}{{Dressler} et~al.}{1994}]{1994ApJ...430..107D}
{Dressler} A.,  {Oemler} Augustus J.,  {Butcher} H.~R.,   {Gunn} J.~E.,  1994,
  \mn@doi [\apj] {10.1086/174386}, \href
  {https://ui.adsabs.harvard.edu/abs/1994ApJ...430..107D} {430, 107}

\bibitem[\protect\citeauthoryear{{Duchesne}, {Johnston-Hollitt}, {Offringa},
  {Pratt}, {Zheng}  \& {Dehghan}}{{Duchesne} et~al.}{2021a}]{Duchesneetal2021}
{Duchesne} S.~W.,  {Johnston-Hollitt} M.,  {Offringa} A.~R.,  {Pratt} G.~W.,
  {Zheng} Q.,   {Dehghan} S.,  2021a, \mn@doi [\pasa] {10.1017/pasa.2021.7},
  \href {https://ui.adsabs.harvard.edu/abs/2021PASA...38...10D} {38, e010}

\bibitem[\protect\citeauthoryear{{Duchesne}, {Johnston-Hollitt}, {Offringa},
  {Pratt}, {Zheng}  \& {Dehghan}}{{Duchesne}
  et~al.}{2021b}]{2021PASA...38...10D}
{Duchesne} S.~W.,  {Johnston-Hollitt} M.,  {Offringa} A.~R.,  {Pratt} G.~W.,
  {Zheng} Q.,   {Dehghan} S.,  2021b, \mn@doi [\pasa] {10.1017/pasa.2021.7},
  \href {https://ui.adsabs.harvard.edu/abs/2021PASA...38...10D} {38, e010}

\bibitem[\protect\citeauthoryear{{Duchesne}, {Johnston-Hollitt}  \&
  {Wilber}}{{Duchesne} et~al.}{2021c}]{2021PASA...38...31D}
{Duchesne} S.~W.,  {Johnston-Hollitt} M.,   {Wilber} A.~G.,  2021c, \mn@doi
  [\pasa] {10.1017/pasa.2021.24}, \href
  {https://ui.adsabs.harvard.edu/abs/2021PASA...38...31D} {38, e031}

\bibitem[\protect\citeauthoryear{{Duchesne}, {Johnston-Hollitt}  \&
  {Bartalucci}}{{Duchesne} et~al.}{2021d}]{Duchesneetal2021b}
{Duchesne} S.~W.,  {Johnston-Hollitt} M.,   {Bartalucci} I.,  2021d, \mn@doi
  [\pasa] {10.1017/pasa.2021.45}, \href
  {https://ui.adsabs.harvard.edu/abs/2021PASA...38...53D} {38, e053}

\bibitem[\protect\citeauthoryear{{Duchesne}, {Johnston-Hollitt}  \&
  {Bartalucci}}{{Duchesne} et~al.}{2021e}]{2021PASA...38...53D}
{Duchesne} S.~W.,  {Johnston-Hollitt} M.,   {Bartalucci} I.,  2021e, \mn@doi
  [\pasa] {10.1017/pasa.2021.45}, \href
  {https://ui.adsabs.harvard.edu/abs/2021PASA...38...53D} {38, e053}

\bibitem[\protect\citeauthoryear{{Duchesne}, {Johnston-Hollitt}, {Riseley},
  {Bartalucci}  \& {Keel}}{{Duchesne} et~al.}{2022}]{Duchesneetal22}
{Duchesne} S.~W.,  {Johnston-Hollitt} M.,  {Riseley} C.~J.,  {Bartalucci} I.,
  {Keel} S.~R.,  2022, \mn@doi [\mnras] {10.1093/mnras/stac335}, \href
  {https://ui.adsabs.harvard.edu/abs/2022MNRAS.511.3525D} {511, 3525}

\bibitem[\protect\citeauthoryear{{Duchesne} et~al.,}{{Duchesne}
  et~al.}{2024}]{Duchesneetal24}
{Duchesne} S.~W.,  et~al., 2024, \mn@doi [\pasa] {10.1017/pasa.2024.10}, \href
  {https://ui.adsabs.harvard.edu/abs/2024PASA...41...26D} {41, e026}

\bibitem[\protect\citeauthoryear{{Dwarakanath} \& {Owen}}{{Dwarakanath} \&
  {Owen}}{1999}]{Dwarakanathetal99}
{Dwarakanath} K.~S.,  {Owen} F.~N.,  1999, \mn@doi [\aj] {10.1086/300973},
  \href {https://ui.adsabs.harvard.edu/abs/1999AJ....118..625D} {118, 625}

\bibitem[\protect\citeauthoryear{{Dwarakanath}, {Malu}  \&
  {Kale}}{{Dwarakanath} et~al.}{2011}]{2011JApA...32..529D}
{Dwarakanath} K.~S.,  {Malu} S.,   {Kale} R.,  2011, \mn@doi [Journal of
  Astrophysics and Astronomy] {10.1007/s12036-011-9109-1}, \href
  {https://ui.adsabs.harvard.edu/abs/2011JApA...32..529D} {32, 529}

\bibitem[\protect\citeauthoryear{Dwarakanath, Parekh, Kale  \&
  George}{Dwarakanath et~al.}{2018}]{Dwarakanath_2018}
Dwarakanath K.~S.,  Parekh V.,  Kale R.,   George L.~T.,  2018, \mn@doi
  [\mnras] {10.1093/mnras/sty744}, 477, 957

\bibitem[\protect\citeauthoryear{{Ebeling}, {Mullis}  \& {Tully}}{{Ebeling}
  et~al.}{2002}]{Ebelingetal02}
{Ebeling} H.,  {Mullis} C.~R.,   {Tully} R.~B.,  2002, \mn@doi [\apj]
  {10.1086/343790}, \href
  {https://ui.adsabs.harvard.edu/abs/2002ApJ...580..774E} {580, 774}

\bibitem[\protect\citeauthoryear{{Ebeling}, {Edge}, {Mantz}, {Barrett},
  {Henry}, {Ma}  \& {van Speybroeck}}{{Ebeling} et~al.}{2010}]{Ebelingetal10}
{Ebeling} H.,  {Edge} A.~C.,  {Mantz} A.,  {Barrett} E.,  {Henry} J.~P.,  {Ma}
  C.~J.,   {van Speybroeck} L.,  2010, \mn@doi [\mnras]
  {10.1111/j.1365-2966.2010.16920.x}, \href
  {https://ui.adsabs.harvard.edu/abs/2010MNRAS.407...83E} {407, 83}

\bibitem[\protect\citeauthoryear{{Ebeling}, {Stephenson}  \& {Edge}}{{Ebeling}
  et~al.}{2014}]{2014ApJ...781L..40E}
{Ebeling} H.,  {Stephenson} L.~N.,   {Edge} A.~C.,  2014, \mn@doi [\apjl]
  {10.1088/2041-8205/781/2/L40}, \href
  {https://ui.adsabs.harvard.edu/abs/2014ApJ...781L..40E} {781, L40}

\bibitem[\protect\citeauthoryear{{Eckert}, {Molendi}, {Gastaldello}  \&
  {Rossetti}}{{Eckert} et~al.}{2011}]{Eckertetal11}
{Eckert} D.,  {Molendi} S.,  {Gastaldello} F.,   {Rossetti} M.,  2011, \mn@doi
  [A\&A] {10.1051/0004-6361/201015856}, \href
  {http://adsabs.harvard.edu/abs/2011A\%26A...526A..79E} {526, A79}

\bibitem[\protect\citeauthoryear{{Eckert}, {Jauzac}, {Vazza}, {Owers}, {Kneib},
  {Tchernin}, {Intema}  \& {Knowles}}{{Eckert} et~al.}{2016}]{Eckertetal16}
{Eckert} D.,  {Jauzac} M.,  {Vazza} F.,  {Owers} M.~S.,  {Kneib} J.~P.,
  {Tchernin} C.,  {Intema} H.,   {Knowles} K.,  2016, \mn@doi [\mnras]
  {10.1093/mnras/stw1435}, \href
  {https://ui.adsabs.harvard.edu/abs/2016MNRAS.461.1302E} {461, 1302}

\bibitem[\protect\citeauthoryear{{Einasto}, {Jaaniste}, {Einasto},
  {Hein{\"a}m{\"a}ki}, {M{\"u}ller}  \& {Tucker}}{{Einasto}
  et~al.}{2003}]{Einastoetal03}
{Einasto} M.,  {Jaaniste} J.,  {Einasto} J.,  {Hein{\"a}m{\"a}ki} P.,
  {M{\"u}ller} V.,   {Tucker} D.~L.,  2003, \mn@doi [\aap]
  {10.1051/0004-6361:20030632}, \href
  {https://ui.adsabs.harvard.edu/abs/2003A&A...405..821E} {405, 821}

\bibitem[\protect\citeauthoryear{{En{\ss}lin} \& {Gopal-Krishna}}{{En{\ss}lin}
  \& {Gopal-Krishna}}{2001}]{2001A&A...366...26E}
{En{\ss}lin} T.~A.,  {Gopal-Krishna} 2001, \mn@doi [\aap]
  {10.1051/0004-6361:20000198}, \href
  {https://ui.adsabs.harvard.edu/abs/2001A&A...366...26E} {366, 26}

\bibitem[\protect\citeauthoryear{{Feretti}, {Schuecker}, {B{\"o}hringer},
  {Govoni}  \& {Giovannini}}{{Feretti} et~al.}{2005}]{2005A&A...444..157F}
{Feretti} L.,  {Schuecker} P.,  {B{\"o}hringer} H.,  {Govoni} F.,
  {Giovannini} G.,  2005, \mn@doi [\aap] {10.1051/0004-6361:20052808}, \href
  {https://ui.adsabs.harvard.edu/abs/2005A&A...444..157F} {444, 157}

\bibitem[\protect\citeauthoryear{{Feretti}, {Giovannini}, {Govoni}  \&
  {Murgia}}{{Feretti} et~al.}{2012}]{Feretti2012}
{Feretti} L.,  {Giovannini} G.,  {Govoni} F.,   {Murgia} M.,  2012, \mn@doi
  [\aapr] {10.1007/s00159-012-0054-z}, \href
  {http://adsabs.harvard.edu/abs/2012A%26ARv..20...54F} {20, 54}

\bibitem[\protect\citeauthoryear{{Ferrari}, {Maurogordato}, {Cappi}  \&
  {Benoist}}{{Ferrari} et~al.}{2003}]{Ferrarietal03}
{Ferrari} C.,  {Maurogordato} S.,  {Cappi} A.,   {Benoist} C.,  2003, \mn@doi
  [\aap] {10.1051/0004-6361:20021741}, \href
  {https://ui.adsabs.harvard.edu/abs/2003A&A...399..813F} {399, 813}

\bibitem[\protect\citeauthoryear{{Ferrari}, {Arnaud}, {Ettori}, {Maurogordato}
  \& {Rho}}{{Ferrari} et~al.}{2006}]{2006A&A...446..417F}
{Ferrari} C.,  {Arnaud} M.,  {Ettori} S.,  {Maurogordato} S.,   {Rho} J.,
  2006, \mn@doi [\aap] {10.1051/0004-6361:20053946}, \href
  {https://ui.adsabs.harvard.edu/abs/2006A&A...446..417F} {446, 417}

\bibitem[\protect\citeauthoryear{{Finoguenov}, {Briel}, {Henry}, {Gavazzi},
  {Iglesias-Paramo}  \& {Boselli}}{{Finoguenov}
  et~al.}{2004}]{Finoguenovetal04}
{Finoguenov} A.,  {Briel} U.~G.,  {Henry} J.~P.,  {Gavazzi} G.,
  {Iglesias-Paramo} J.,   {Boselli} A.,  2004, preprint, \href
  {http://adsabs.harvard.edu/cgi-bin/nph-bib_query?bibcode=2004astro.ph..3216F&amp;db_key=PRE}
  {astro-ph/0403216}

\bibitem[\protect\citeauthoryear{{Finoguenov}, {B{\"o}hringer}  \&
  {Zhang}}{{Finoguenov} et~al.}{2005}]{Finoguenovetal05}
{Finoguenov} A.,  {B{\"o}hringer} H.,   {Zhang} Y.~Y.,  2005, \mn@doi [\aap]
  {10.1051/0004-6361:20053306}, \href
  {https://ui.adsabs.harvard.edu/abs/2005A&A...442..827F} {442, 827}

\bibitem[\protect\citeauthoryear{{Finoguenov}, {Davis}, {Zimer}  \&
  {Mulchaey}}{{Finoguenov} et~al.}{2006}]{Finoguenovetal06}
{Finoguenov} A.,  {Davis} D.~S.,  {Zimer} M.,   {Mulchaey} J.~S.,  2006,
  \mn@doi [ApJ] {10.1086/504697}, \href
  {http://adsabs.harvard.edu/cgi-bin/nph-bib_query?bibcode=2006ApJ...646..143F&db_key=AST}
  {646, 143}

\bibitem[\protect\citeauthoryear{{Fleenor}, {Rose}, {Christiansen}, {Hunstead},
  {Johnston-Hollitt}, {Drinkwater}  \& {Saunders}}{{Fleenor}
  et~al.}{2005}]{Fleenoretal05}
{Fleenor} M.~C.,  {Rose} J.~A.,  {Christiansen} W.~A.,  {Hunstead} R.~W.,
  {Johnston-Hollitt} M.,  {Drinkwater} M.~J.,   {Saunders} W.,  2005, \mn@doi
  [\aj] {10.1086/431972}, \href
  {https://ui.adsabs.harvard.edu/abs/2005AJ....130..957F} {130, 957}

\bibitem[\protect\citeauthoryear{{Forman}, {Bechtold}, {Blair}, {Giacconi},
  {van Speybroeck}  \& {Jones}}{{Forman} et~al.}{1981}]{Formanetal81}
{Forman} W.,  {Bechtold} J.,  {Blair} W.,  {Giacconi} R.,  {van Speybroeck} L.,
    {Jones} C.,  1981, \mn@doi [\apjl] {10.1086/183459}, \href
  {https://ui.adsabs.harvard.edu/abs/1981ApJ...243L.133F} {243, L133}

\bibitem[\protect\citeauthoryear{{Fox}, {Mahler}, {Sharon}  \& {Remolina
  Gonz{\'a}lez}}{{Fox} et~al.}{2022}]{2022ApJ...928...87F}
{Fox} C.,  {Mahler} G.,  {Sharon} K.,   {Remolina Gonz{\'a}lez} J.~D.,  2022,
  \mn@doi [\apj] {10.3847/1538-4357/ac5024}, \href
  {https://ui.adsabs.harvard.edu/abs/2022ApJ...928...87F} {928, 87}

\bibitem[\protect\citeauthoryear{{George} et~al.,}{{George}
  et~al.}{2015}]{2015MNRAS.451.4207G}
{George} L.~T.,  et~al., 2015, \mn@doi [\mnras] {10.1093/mnras/stv1152}, \href
  {https://ui.adsabs.harvard.edu/abs/2015MNRAS.451.4207G} {451, 4207}

\bibitem[\protect\citeauthoryear{{George} et~al.,}{{George}
  et~al.}{2017}]{2017MNRAS.467..936G}
{George} L.~T.,  et~al., 2017, \mn@doi [\mnras] {10.1093/mnras/stx155}, \href
  {https://ui.adsabs.harvard.edu/abs/2017MNRAS.467..936G} {467, 936}

\bibitem[\protect\citeauthoryear{{George}, {Kale}  \& {Wadadekar}}{{George}
  et~al.}{2021}]{2021MNRAS.507.4487G}
{George} L.~T.,  {Kale} R.,   {Wadadekar} Y.,  2021, \mn@doi [\mnras]
  {10.1093/mnras/stab2309}, \href
  {https://ui.adsabs.harvard.edu/abs/2021MNRAS.507.4487G} {507, 4487}

\bibitem[\protect\citeauthoryear{{Giacconi}, {Branduardi}, {Briel}, {Epstein}
  \& et al.}{{Giacconi} et~al.}{1979}]{Giacconietal79}
{Giacconi} R.,  {Branduardi} G.,  {Briel} U.,  {Epstein} A.,   et al. 1979,
  ApJ, 230, 540

\bibitem[\protect\citeauthoryear{{Giacintucci}, {Venturi}, {Bardelli},
  {Dallacasa}  \& {Zucca}}{{Giacintucci} et~al.}{2004}]{Giacintuccietal04}
{Giacintucci} S.,  {Venturi} T.,  {Bardelli} S.,  {Dallacasa} D.,   {Zucca} E.,
   2004, \mn@doi [\aap] {10.1051/0004-6361:20040071}, \href
  {https://ui.adsabs.harvard.edu/abs/2004A&A...419...71G} {419, 71}

\bibitem[\protect\citeauthoryear{{Giacintucci} et~al.,}{{Giacintucci}
  et~al.}{2005}]{2005A&A...440..867G}
{Giacintucci} S.,  et~al., 2005, \mn@doi [\aap] {10.1051/0004-6361:20053016},
  \href {https://ui.adsabs.harvard.edu/abs/2005A&A...440..867G} {440, 867}

\bibitem[\protect\citeauthoryear{{Giacintucci}, {Venturi}, {Bardelli},
  {Brunetti}, {Cassano}  \& {Dallacasa}}{{Giacintucci}
  et~al.}{2006}]{Giacintuccietal06}
{Giacintucci} S.,  {Venturi} T.,  {Bardelli} S.,  {Brunetti} G.,  {Cassano} R.,
    {Dallacasa} D.,  2006, \mn@doi [\na] {10.1016/j.newast.2005.11.001}, \href
  {https://ui.adsabs.harvard.edu/abs/2006NewA...11..437G} {11, 437}

\bibitem[\protect\citeauthoryear{{Giacintucci} et~al.,}{{Giacintucci}
  et~al.}{2008}]{2008A&A...486..347G}
{Giacintucci} S.,  et~al., 2008, \mn@doi [\aap] {10.1051/0004-6361:200809459},
  \href {https://ui.adsabs.harvard.edu/abs/2008A&A...486..347G} {486, 347}

\bibitem[\protect\citeauthoryear{{Giacintucci}, {Kale}, {Wik}, {Venturi}  \&
  {Markevitch}}{{Giacintucci} et~al.}{2013}]{Kaleetal13}
{Giacintucci} S.,  {Kale} R.,  {Wik} D.~R.,  {Venturi} T.,   {Markevitch} M.,
  2013, \mn@doi [\apj] {10.1088/0004-637X/766/1/18}, \href
  {https://ui.adsabs.harvard.edu/abs/2013ApJ...766...18G} {766, 18}

\bibitem[\protect\citeauthoryear{{Giacintucci}, {Markevitch}, {Venturi},
  {Clarke}, {Cassano}  \& {Mazzotta}}{{Giacintucci}
  et~al.}{2014}]{Giacintuccietal14}
{Giacintucci} S.,  {Markevitch} M.,  {Venturi} T.,  {Clarke} T.~E.,  {Cassano}
  R.,   {Mazzotta} P.,  2014, \mn@doi [\apj] {10.1088/0004-637X/781/1/9}, \href
  {https://ui.adsabs.harvard.edu/abs/2014ApJ...781....9G} {781, 9}

\bibitem[\protect\citeauthoryear{{Giacintucci}, {Markevitch}, {Cassano},
  {Venturi}, {Clarke}  \& {Brunetti}}{{Giacintucci}
  et~al.}{2017}]{2017ApJ...841...71G}
{Giacintucci} S.,  {Markevitch} M.,  {Cassano} R.,  {Venturi} T.,  {Clarke}
  T.~E.,   {Brunetti} G.,  2017, \mn@doi [\apj] {10.3847/1538-4357/aa7069},
  \href {https://ui.adsabs.harvard.edu/abs/2017ApJ...841...71G} {841, 71}

\bibitem[\protect\citeauthoryear{{Giacintucci}, {Markevitch}, {Cassano},
  {Venturi}, {Clarke}, {Kale}  \& {Cuciti}}{{Giacintucci}
  et~al.}{2019}]{Giacintuccietal19}
{Giacintucci} S.,  {Markevitch} M.,  {Cassano} R.,  {Venturi} T.,  {Clarke}
  T.~E.,  {Kale} R.,   {Cuciti} V.,  2019, \mn@doi [\apj]
  {10.3847/1538-4357/ab29f1}, \href
  {https://ui.adsabs.harvard.edu/abs/2019ApJ...880...70G} {880, 70}

\bibitem[\protect\citeauthoryear{{Giacintucci}, {Markevitch},
  {Johnston-Hollitt}, {Wik}, {Wang}  \& {Clarke}}{{Giacintucci}
  et~al.}{2020}]{Giacintuccietal20}
{Giacintucci} S.,  {Markevitch} M.,  {Johnston-Hollitt} M.,  {Wik} D.~R.,
  {Wang} Q.~H.~S.,   {Clarke} T.~E.,  2020, \mn@doi [\apj]
  {10.3847/1538-4357/ab6a9d}, \href
  {https://ui.adsabs.harvard.edu/abs/2020ApJ...891....1G} {891, 1}

\bibitem[\protect\citeauthoryear{{Giacintucci} et~al.,}{{Giacintucci}
  et~al.}{2022}]{Giacintuccietal22}
{Giacintucci} S.,  et~al., 2022, \mn@doi [\apj] {10.3847/1538-4357/ac7805},
  \href {https://ui.adsabs.harvard.edu/abs/2022ApJ...934...49G} {934, 49}

\bibitem[\protect\citeauthoryear{{Giacintucci}, {Venturi}, {Markevitch},
  {Brunetti}, {Clarke}  \& {Kale}}{{Giacintucci}
  et~al.}{2024}]{Giacintuccietal24}
{Giacintucci} S.,  {Venturi} T.,  {Markevitch} M.,  {Brunetti} G.,  {Clarke}
  T.~E.,   {Kale} R.,  2024, \mn@doi [\apj] {10.3847/1538-4357/ad12bc}, \href
  {https://ui.adsabs.harvard.edu/abs/2024ApJ...961..133G} {961, 133}

\bibitem[\protect\citeauthoryear{{Giovannini}, {Tordi}  \&
  {Feretti}}{{Giovannini} et~al.}{1999}]{1999NewA....4..141G}
{Giovannini} G.,  {Tordi} M.,   {Feretti} L.,  1999, \mn@doi [\na]
  {10.1016/S1384-1076(99)00018-4}, \href
  {https://ui.adsabs.harvard.edu/abs/1999NewA....4..141G} {4, 141}

\bibitem[\protect\citeauthoryear{{Giovannini}, {Feretti}, {Govoni}, {Murgia}
  \& {Pizzo}}{{Giovannini} et~al.}{2006}]{2006AN....327..563G}
{Giovannini} G.,  {Feretti} L.,  {Govoni} F.,  {Murgia} M.,   {Pizzo} R.,
  2006, \mn@doi [Astronomische Nachrichten] {10.1002/asna.200610589}, \href
  {https://ui.adsabs.harvard.edu/abs/2006AN....327..563G} {327, 563}

\bibitem[\protect\citeauthoryear{{Giovannini}, {Bonafede}, {Feretti}, {Govoni},
  {Murgia}, {Ferrari}  \& {Monti}}{{Giovannini}
  et~al.}{2009}]{2009A&A...507.1257G}
{Giovannini} G.,  {Bonafede} A.,  {Feretti} L.,  {Govoni} F.,  {Murgia} M.,
  {Ferrari} F.,   {Monti} G.,  2009, \mn@doi [\aap]
  {10.1051/0004-6361/200912667}, \href
  {https://ui.adsabs.harvard.edu/abs/2009A&A...507.1257G} {507, 1257}

\bibitem[\protect\citeauthoryear{{Giovannini} et~al.,}{{Giovannini}
  et~al.}{2020}]{2020A&A...640A.108G}
{Giovannini} G.,  et~al., 2020, \mn@doi [\aap] {10.1051/0004-6361/202038263},
  \href {https://ui.adsabs.harvard.edu/abs/2020A&A...640A.108G} {640, A108}

\bibitem[\protect\citeauthoryear{{Girardi}, {Escalera}, {Fadda}, {Giuricin},
  {Mardirossian}  \& {Mezzetti}}{{Girardi} et~al.}{1997}]{Girardietal97}
{Girardi} M.,  {Escalera} E.,  {Fadda} D.,  {Giuricin} G.,  {Mardirossian} F.,
   {Mezzetti} M.,  1997, \mn@doi [\apj] {10.1086/304113}, \href
  {https://ui.adsabs.harvard.edu/abs/1997ApJ...482...41G} {482, 41}

\bibitem[\protect\citeauthoryear{{Gitti}, {Brighenti}  \& {McNamara}}{{Gitti}
  et~al.}{2012}]{Gittietal12}
{Gitti} M.,  {Brighenti} F.,   {McNamara} B.~R.,  2012, \mn@doi [Advances in
  Astronomy] {10.1155/2012/950641}, \href
  {https://ui.adsabs.harvard.edu/abs/2012AdAst2012E...6G} {2012, 950641}

\bibitem[\protect\citeauthoryear{{Gitti}, {Brunetti}, {Cassano}  \&
  {Ettori}}{{Gitti} et~al.}{2018}]{Gittietal18}
{Gitti} M.,  {Brunetti} G.,  {Cassano} R.,   {Ettori} S.,  2018, \mn@doi [\aap]
  {10.1051/0004-6361/201832749}, \href
  {https://ui.adsabs.harvard.edu/abs/2018A&A...617A..11G} {617, A11}

\bibitem[\protect\citeauthoryear{{Golovich} et~al.,}{{Golovich}
  et~al.}{2019}]{Golovich2019}
{Golovich} N.,  et~al., 2019, \mn@doi [\apj] {10.3847/1538-4357/ab2f90}, \href
  {https://ui.adsabs.harvard.edu/abs/2019ApJ...882...69G} {882, 69}

\bibitem[\protect\citeauthoryear{{G{\'o}mez} et~al.,}{{G{\'o}mez}
  et~al.}{2012}]{Gomezetal12}
{G{\'o}mez} P.~L.,  et~al., 2012, \mn@doi [\aj] {10.1088/0004-6256/144/3/79},
  \href {https://ui.adsabs.harvard.edu/abs/2012AJ....144...79G} {144, 79}

\bibitem[\protect\citeauthoryear{{Govoni}, {Feretti}, {Giovannini},
  {B{\"o}hringer}, {Reiprich}  \& {Murgia}}{{Govoni}
  et~al.}{2001}]{2001A&A...376..803G}
{Govoni} F.,  {Feretti} L.,  {Giovannini} G.,  {B{\"o}hringer} H.,  {Reiprich}
  T.~H.,   {Murgia} M.,  2001, \mn@doi [\aap] {10.1051/0004-6361:20011016},
  \href {https://ui.adsabs.harvard.edu/abs/2001A&A...376..803G} {376, 803}

\bibitem[\protect\citeauthoryear{{Govoni}, {Murgia}, {Markevitch}, {Feretti},
  {Giovannini}, {Taylor}  \& {Carretti}}{{Govoni} et~al.}{2009}]{Govonietal09}
{Govoni} F.,  {Murgia} M.,  {Markevitch} M.,  {Feretti} L.,  {Giovannini} G.,
  {Taylor} G.~B.,   {Carretti} E.,  2009, \mn@doi [\aap]
  {10.1051/0004-6361/200811180}, \href
  {https://ui.adsabs.harvard.edu/abs/2009A&A...499..371G} {499, 371}

\bibitem[\protect\citeauthoryear{{Govoni} et~al.,}{{Govoni}
  et~al.}{2019}]{Govonietal19}
{Govoni} F.,  et~al., 2019, \mn@doi [Science] {10.1126/science.aat7500}, \href
  {https://ui.adsabs.harvard.edu/abs/2019Sci...364..981G} {364, 981}

\bibitem[\protect\citeauthoryear{{Grange}, {Costantini}, {de Plaa}, {in't
  Zand}, {Verbunt}, {Kaastra}  \& {Verrecchia}}{{Grange}
  et~al.}{2010}]{2010A&A...513A...2G}
{Grange} Y.~G.,  {Costantini} E.,  {de Plaa} J.,  {in't Zand} J.~J.~M.,
  {Verbunt} F.,  {Kaastra} J.~S.,   {Verrecchia} F.,  2010, \mn@doi [\aap]
  {10.1051/0004-6361/200913245}, \href
  {https://ui.adsabs.harvard.edu/abs/2010A&A...513A...2G} {513, A2}

\bibitem[\protect\citeauthoryear{{Gupta} et~al.,}{{Gupta}
  et~al.}{2017}]{Guptaetal17}
{Gupta} Y.,  et~al., 2017, \mn@doi [Current Science]
  {10.18520/cs/v113/i04/707-714}, \href
  {https://ui.adsabs.harvard.edu/abs/2017CSci..113..707G} {113, 707}

\bibitem[\protect\citeauthoryear{{Hicks} et~al.,}{{Hicks}
  et~al.}{2013}]{2013MNRAS.431.2542H}
{Hicks} A.~K.,  et~al., 2013, \mn@doi [\mnras] {10.1093/mnras/stt348}, \href
  {https://ui.adsabs.harvard.edu/abs/2013MNRAS.431.2542H} {431, 2542}

\bibitem[\protect\citeauthoryear{{Hilton} et~al.,}{{Hilton}
  et~al.}{2021}]{Hilton_2020}
{Hilton} M.,  et~al., 2021, \mn@doi [\apjs] {10.3847/1538-4365/abd023}, \href
  {https://ui.adsabs.harvard.edu/abs/2021ApJS..253....3H} {253, 3}

\bibitem[\protect\citeauthoryear{{Hoang} et~al.,}{{Hoang}
  et~al.}{2022}]{Hoangetal22}
{Hoang} D.~N.,  et~al., 2022, \mn@doi [\aap] {10.1051/0004-6361/202243105},
  \href {https://ui.adsabs.harvard.edu/abs/2022A&A...665A..60H} {665, A60}

\bibitem[\protect\citeauthoryear{{Hoang} et~al.,}{{Hoang}
  et~al.}{2025}]{Hoangetal25}
{Hoang} D.~N.,  et~al., 2025, \mn@doi [arXiv e-prints]
  {10.48550/arXiv.2502.09472}, \href
  {https://ui.adsabs.harvard.edu/abs/2025arXiv250209472H} {p. arXiv:2502.09472}

\bibitem[\protect\citeauthoryear{{Hogan} et~al.,}{{Hogan}
  et~al.}{2015}]{Hoganetal15}
{Hogan} M.~T.,  et~al., 2015, \mn@doi [\mnras] {10.1093/mnras/stv1517}, \href
  {https://ui.adsabs.harvard.edu/abs/2015MNRAS.453.1201H} {453, 1201}

\bibitem[\protect\citeauthoryear{{Hotan} et~al.,}{{Hotan}
  et~al.}{2021}]{Hotanetal21}
{Hotan} A.~W.,  et~al., 2021, \mn@doi [\pasa] {10.1017/pasa.2021.1}, \href
  {https://ui.adsabs.harvard.edu/abs/2021PASA...38....9H} {38, e009}

\bibitem[\protect\citeauthoryear{{Hu}, {Stuardi}, {Lazarian}, {Brunetti},
  {Bonafede}  \& {Ho}}{{Hu} et~al.}{2024}]{Huetal24}
{Hu} Y.,  {Stuardi} C.,  {Lazarian} A.,  {Brunetti} G.,  {Bonafede} A.,   {Ho}
  K.~W.,  2024, \mn@doi [Nature Communications] {10.1038/s41467-024-45164-8},
  \href {https://ui.adsabs.harvard.edu/abs/2024NatCo..15.1006H} {15, 1006}

\bibitem[\protect\citeauthoryear{Hurley-Walker et~al.,}{Hurley-Walker
  et~al.}{2016}]{HurleyWalkeretal17}
Hurley-Walker N.,  et~al., 2016, \mn@doi [MNRAS] {10.1093/mnras/stw2337}, 464,
  1146

\bibitem[\protect\citeauthoryear{{HyeongHan} et~al.,}{{HyeongHan}
  et~al.}{2020}]{2020ApJ...900..127H}
{HyeongHan} K.,  et~al., 2020, \mn@doi [\apj] {10.3847/1538-4357/aba742}, \href
  {https://ui.adsabs.harvard.edu/abs/2020ApJ...900..127H} {900, 127}

\bibitem[\protect\citeauthoryear{{Iani} et~al.,}{{Iani}
  et~al.}{2019}]{2019MNRAS.487.5593I}
{Iani} E.,  et~al., 2019, \mn@doi [\mnras] {10.1093/mnras/stz1631}, \href
  {https://ui.adsabs.harvard.edu/abs/2019MNRAS.487.5593I} {487, 5593}

\bibitem[\protect\citeauthoryear{{Ibaraki}, {Ota}, {Akamatsu}, {Zhang}  \&
  {Finoguenov}}{{Ibaraki} et~al.}{2014}]{Ibarakietal14}
{Ibaraki} Y.,  {Ota} N.,  {Akamatsu} H.,  {Zhang} Y.~Y.,   {Finoguenov} A.,
  2014, \mn@doi [\aap] {10.1051/0004-6361/201322806}, \href
  {https://ui.adsabs.harvard.edu/abs/2014A&A...562A..11I} {562, A11}

\bibitem[\protect\citeauthoryear{{Ignesti}, {Brunetti}, {Gitti}  \&
  {Giacintucci}}{{Ignesti} et~al.}{2020}]{Ignestietal20}
{Ignesti} A.,  {Brunetti} G.,  {Gitti} M.,   {Giacintucci} S.,  2020, \mn@doi
  [\aap] {10.1051/0004-6361/201937207}, \href
  {https://ui.adsabs.harvard.edu/abs/2020A&A...640A..37I} {640, A37}

\bibitem[\protect\citeauthoryear{{Iqbal}, {Kale}, {Majumdar}, {Nath}, {Pandge},
  {Sharma}, {Malik}  \& {Raychaudhury}}{{Iqbal} et~al.}{2017}]{Iqbaletal17}
{Iqbal} A.,  {Kale} R.,  {Majumdar} S.,  {Nath} B.~B.,  {Pandge} M.,  {Sharma}
  P.,  {Malik} M.~A.,   {Raychaudhury} S.,  2017, \mn@doi [Journal of
  Astrophysics and Astronomy] {10.1007/s12036-017-9491-4}, \href
  {https://ui.adsabs.harvard.edu/abs/2017JApA...38...68I} {38, 68}

\bibitem[\protect\citeauthoryear{{Jacob} \& {Pfrommer}}{{Jacob} \&
  {Pfrommer}}{2017a}]{JacobPfrommer17a}
{Jacob} S.,  {Pfrommer} C.,  2017a, \mn@doi [\mnras] {10.1093/mnras/stx131},
  \href {https://ui.adsabs.harvard.edu/abs/2017MNRAS.467.1449J} {467, 1449}

\bibitem[\protect\citeauthoryear{{Jacob} \& {Pfrommer}}{{Jacob} \&
  {Pfrommer}}{2017b}]{JacobPfrommer17b}
{Jacob} S.,  {Pfrommer} C.,  2017b, \mn@doi [\mnras] {10.1093/mnras/stx132},
  \href {https://ui.adsabs.harvard.edu/abs/2017MNRAS.467.1478J} {467, 1478}

\bibitem[\protect\citeauthoryear{{Jauzac} et~al.,}{{Jauzac}
  et~al.}{2019}]{2019MNRAS.483.3082J}
{Jauzac} M.,  et~al., 2019, \mn@doi [\mnras] {10.1093/mnras/sty3312}, \href
  {https://ui.adsabs.harvard.edu/abs/2019MNRAS.483.3082J} {483, 3082}

\bibitem[\protect\citeauthoryear{{Jee}, {Hughes}, {Menanteau}, {Sif{\'o}n},
  {Mandelbaum}, {Barrientos}, {Infante}  \& {Ng}}{{Jee}
  et~al.}{2014}]{Jeeetal14}
{Jee} M.~J.,  {Hughes} J.~P.,  {Menanteau} F.,  {Sif{\'o}n} C.,  {Mandelbaum}
  R.,  {Barrientos} L.~F.,  {Infante} L.,   {Ng} K.~Y.,  2014, \mn@doi [\apj]
  {10.1088/0004-637X/785/1/20}, \href
  {https://ui.adsabs.harvard.edu/abs/2014ApJ...785...20J} {785, 20}

\bibitem[\protect\citeauthoryear{{Johnson}, {Hibbard}, {Gallagher}, {Charlton},
  {Hornschemeier}, {Jarrett}  \& {Reines}}{{Johnson}
  et~al.}{2007}]{Johnsonetal07}
{Johnson} K.~E.,  {Hibbard} J.~E.,  {Gallagher} S.~C.,  {Charlton} J.~C.,
  {Hornschemeier} A.~E.,  {Jarrett} T.~H.,   {Reines} A.~E.,  2007, \mn@doi
  [AJ] {10.1086/520921}, \href
  {http://adsabs.harvard.edu/abs/2007AJ....134.1522J} {134, 1522}

\bibitem[\protect\citeauthoryear{{Johnston-Hollitt}, {Sato}, {Gill}, {Fleenor}
  \& {Brick}}{{Johnston-Hollitt} et~al.}{2008}]{Johnstonetal08}
{Johnston-Hollitt} M.,  {Sato} M.,  {Gill} J.~A.,  {Fleenor} M.~C.,   {Brick}
  A.~M.,  2008, \mn@doi [\mnras] {10.1111/j.1365-2966.2008.13730.x}, \href
  {https://ui.adsabs.harvard.edu/abs/2008MNRAS.390..289J} {390, 289}

\bibitem[\protect\citeauthoryear{{Jonas} \& {MeerKAT Team}}{{Jonas} \& {MeerKAT
  Team}}{2016}]{Jonas2016}
{Jonas} J.,  {MeerKAT Team} 2016, in MeerKAT Science: On the Pathway to the
  SKA. p.~1

\bibitem[\protect\citeauthoryear{{Jones} et~al.,}{{Jones}
  et~al.}{2023}]{Jonesetal23}
{Jones} A.,  et~al., 2023, \mn@doi [\aap] {10.1051/0004-6361/202245102}, \href
  {https://ui.adsabs.harvard.edu/abs/2023A&A...680A..31J} {680, A31}

\bibitem[\protect\citeauthoryear{{Juett} et~al.,}{{Juett}
  et~al.}{2008}]{Juettetal08}
{Juett} A.~M.,  et~al., 2008, \mn@doi [\apj] {10.1086/523350}, \href
  {https://ui.adsabs.harvard.edu/abs/2008ApJ...672..138J} {672, 138}

\bibitem[\protect\citeauthoryear{{Kale} \& {Dwarakanath}}{{Kale} \&
  {Dwarakanath}}{2012}]{2012ApJ...744...46K}
{Kale} R.,  {Dwarakanath} K.~S.,  2012, \mn@doi [\apj]
  {10.1088/0004-637X/744/1/46}, \href
  {https://ui.adsabs.harvard.edu/abs/2012ApJ...744...46K} {744, 46}

\bibitem[\protect\citeauthoryear{{Kale} \& {Parekh}}{{Kale} \&
  {Parekh}}{2016}]{2016MNRAS.459.2940K}
{Kale} R.,  {Parekh} V.,  2016, \mn@doi [\mnras] {10.1093/mnras/stw796}, \href
  {https://ui.adsabs.harvard.edu/abs/2016MNRAS.459.2940K} {459, 2940}

\bibitem[\protect\citeauthoryear{{Kale}, {Dwarakanath}, {Bagchi}  \&
  {Paul}}{{Kale} et~al.}{2012}]{2012MNRAS.426.1204K}
{Kale} R.,  {Dwarakanath} K.~S.,  {Bagchi} J.,   {Paul} S.,  2012, \mn@doi
  [\mnras] {10.1111/j.1365-2966.2012.21519.x}, \href
  {https://ui.adsabs.harvard.edu/abs/2012MNRAS.426.1204K} {426, 1204}

\bibitem[\protect\citeauthoryear{{Kale} et~al.,}{{Kale}
  et~al.}{2015}]{Kale2015EGRHS}
{Kale} R.,  et~al., 2015, \mn@doi [\aap] {10.1051/0004-6361/201525695}, \href
  {http://adsabs.harvard.edu/abs/2015A%26A...579A..92K} {579, A92}

\bibitem[\protect\citeauthoryear{{Kale} et~al.,}{{Kale}
  et~al.}{2016}]{Kaleetal16}
{Kale} R.,  et~al., 2016, \mn@doi [Journal of Astrophysics and Astronomy]
  {10.1007/s12036-016-9406-9}, \href
  {https://ui.adsabs.harvard.edu/abs/2016JApA...37...31K} {37, 31}

\bibitem[\protect\citeauthoryear{{Kale}, {Wik}, {Giacintucci}, {Venturi},
  {Brunetti}, {Cassano}, {Dallacasa}  \& {de Gasperin}}{{Kale}
  et~al.}{2017}]{2017MNRAS.472..940K}
{Kale} R.,  {Wik} D.~R.,  {Giacintucci} S.,  {Venturi} T.,  {Brunetti} G.,
  {Cassano} R.,  {Dallacasa} D.,   {de Gasperin} F.,  2017, \mn@doi [\mnras]
  {10.1093/mnras/stx2031}, \href
  {https://ui.adsabs.harvard.edu/abs/2017MNRAS.472..940K} {472, 940}

\bibitem[\protect\citeauthoryear{{Kale}, {Parekh}  \& {Dwarakanath}}{{Kale}
  et~al.}{2018}]{2018MNRAS.480.5352K}
{Kale} R.,  {Parekh} V.,   {Dwarakanath} K.~S.,  2018, \mn@doi [\mnras]
  {10.1093/mnras/sty2227}, \href
  {https://ui.adsabs.harvard.edu/abs/2018MNRAS.480.5352K} {480, 5352}

\bibitem[\protect\citeauthoryear{{Kale}, {Shende}  \& {Parekh}}{{Kale}
  et~al.}{2019}]{Kaleetal19}
{Kale} R.,  {Shende} K.~M.,   {Parekh} V.,  2019, \mn@doi [\mnras]
  {10.1093/mnrasl/slz061}, \href
  {https://ui.adsabs.harvard.edu/abs/2019MNRAS.486L..80K} {486, L80}

\bibitem[\protect\citeauthoryear{{Kale} et~al.,}{{Kale}
  et~al.}{2022}]{Kaleetal22}
{Kale} R.,  et~al., 2022, \mn@doi [\mnras] {10.1093/mnras/stac1649}, \href
  {https://ui.adsabs.harvard.edu/abs/2022MNRAS.514.5969K} {514, 5969}

\bibitem[\protect\citeauthoryear{{Kale}, {Botteon}, {Eckert}, {Santra},
  {Brunetti}, {Venturi}, {Cassano}  \& {Dallacasa}}{{Kale}
  et~al.}{2025}]{Kaleetal25}
{Kale} R.,  {Botteon} A.,  {Eckert} D.,  {Santra} R.,  {Brunetti} G.,
  {Venturi} T.,  {Cassano} R.,   {Dallacasa} D.,  2025, \mn@doi [arXiv
  e-prints] {10.48550/arXiv.2503.18613}, \href
  {https://ui.adsabs.harvard.edu/abs/2025arXiv250318613K} {p. arXiv:2503.18613}

\bibitem[\protect\citeauthoryear{{Kempner}, {Blanton}, {Clarke}, {En{\ss}lin},
  {Johnston-Hollitt}  \& {Rudnick}}{{Kempner} et~al.}{2004}]{Kempneretal04}
{Kempner} J.~C.,  {Blanton} E.~L.,  {Clarke} T.~E.,  {En{\ss}lin} T.~A.,
  {Johnston-Hollitt} M.,   {Rudnick} L.,  2004, in {T.~Reiprich, J.~Kempner, \&
  N.~Soker} ed., The Riddle of Cooling Flows in Galaxies and Clusters of
  galaxies. p.~335 (\mn@eprint {} {arXiv:astro-ph/0310263})

\bibitem[\protect\citeauthoryear{{Knowles} et~al.,}{{Knowles}
  et~al.}{2019}]{Knowlesetal19}
{Knowles} K.,  et~al., 2019, \mn@doi [\mnras] {10.1093/mnras/stz823}, \href
  {https://ui.adsabs.harvard.edu/abs/2019MNRAS.486.1332K} {486, 1332}

\bibitem[\protect\citeauthoryear{{Knowles}, {Manaka}, {Bietenholz}, {Cotton},
  {Hilton}, {Kolokythas}, {Loubser}  \& {Oozeer}}{{Knowles}
  et~al.}{2021a}]{2021Galax...9...89K}
{Knowles} K.,  {Manaka} S.,  {Bietenholz} M.~F.,  {Cotton} W.~D.,  {Hilton} M.,
   {Kolokythas} K.,  {Loubser} S.~I.,   {Oozeer} N.,  2021a, \mn@doi [Galaxies]
  {10.3390/galaxies9040089}, \href
  {https://ui.adsabs.harvard.edu/abs/2021Galax...9...89K} {9, 89}

\bibitem[\protect\citeauthoryear{{Knowles} et~al.,}{{Knowles}
  et~al.}{2021b}]{Knowles2020}
{Knowles} K.,  et~al., 2021b, \mn@doi [\mnras] {10.1093/mnras/stab939}, \href
  {https://ui.adsabs.harvard.edu/abs/2021MNRAS.504.1749K} {504, 1749}

\bibitem[\protect\citeauthoryear{{Knowles} et~al.,}{{Knowles}
  et~al.}{2022}]{Knowlesetal22}
{Knowles} K.,  et~al., 2022, \mn@doi [\aap] {10.1051/0004-6361/202141488},
  \href {https://ui.adsabs.harvard.edu/abs/2022A&A...657A..56K} {657, A56}

\bibitem[\protect\citeauthoryear{{Lao} et~al.,}{{Lao} et~al.}{2025}]{Laoetal25}
{Lao} B.,  et~al., 2025, \mn@doi [\apjs] {10.3847/1538-4365/ad9c6d}, \href
  {https://ui.adsabs.harvard.edu/abs/2025ApJS..276...46L} {276, 46}

\bibitem[\protect\citeauthoryear{Liang, Hunstead, Birkinshaw  \&
  Andreani}{Liang et~al.}{2000}]{Liang_2000}
Liang H.,  Hunstead R.~W.,  Birkinshaw M.,   Andreani P.,  2000, \mn@doi [\apj]
  {10.1086/317223}, 544, 686

\bibitem[\protect\citeauthoryear{{Lindner} et~al.,}{{Lindner}
  et~al.}{2014a}]{Lindner2014}
{Lindner} R.~R.,  et~al., 2014a, \mn@doi [\apj] {10.1088/0004-637X/786/1/49},
  \href {https://ui.adsabs.harvard.edu/abs/2014ApJ...786...49L} {786, 49}

\bibitem[\protect\citeauthoryear{{Lindner} et~al.,}{{Lindner}
  et~al.}{2014b}]{Lindneretal14}
{Lindner} R.~R.,  et~al., 2014b, \mn@doi [\apj] {10.1088/0004-637X/786/1/49},
  \href {https://ui.adsabs.harvard.edu/abs/2014ApJ...786...49L} {786, 49}

\bibitem[\protect\citeauthoryear{{Liu} et~al.,}{{Liu} et~al.}{2015}]{Liuetal15}
{Liu} J.,  et~al., 2015, \mn@doi [\mnras] {10.1093/mnras/stv458}, \href
  {https://ui.adsabs.harvard.edu/abs/2015MNRAS.449.3370L} {449, 3370}

\bibitem[\protect\citeauthoryear{{Loi} et~al.,}{{Loi} et~al.}{2023}]{Loietal23}
{Loi} F.,  et~al., 2023, \mn@doi [\aap] {10.1051/0004-6361/202245640}, \href
  {https://ui.adsabs.harvard.edu/abs/2023A&A...672A..28L} {672, A28}

\bibitem[\protect\citeauthoryear{{Lovisari}, {Reiprich}  \&
  {Schellenberger}}{{Lovisari} et~al.}{2015}]{Lovisarietal15}
{Lovisari} L.,  {Reiprich} T.~H.,   {Schellenberger} G.,  2015, \mn@doi [A\&A]
  {10.1051/0004-6361/201423954}, \href
  {http://adsabs.harvard.edu/abs/2015A\%26A...573A.118L} {573, A118}

\bibitem[\protect\citeauthoryear{{Lovisari} et~al.,}{{Lovisari}
  et~al.}{2017}]{2017ApJ...846...51L}
{Lovisari} L.,  et~al., 2017, \mn@doi [\apj] {10.3847/1538-4357/aa855f}, \href
  {https://ui.adsabs.harvard.edu/abs/2017ApJ...846...51L} {846, 51}

\bibitem[\protect\citeauthoryear{Lovisari et~al.,}{Lovisari
  et~al.}{2020}]{Lovisari_2020}
Lovisari L.,  et~al., 2020, \mn@doi [The Astrophysical Journal]
  {10.3847/1538-4357/ab7997}, 892, 102

\bibitem[\protect\citeauthoryear{{Lucey}, {Dickens}, {Mitchell}  \&
  {Dawe}}{{Lucey} et~al.}{1983}]{Luceyetal83}
{Lucey} J.~R.,  {Dickens} R.~J.,  {Mitchell} R.~J.,   {Dawe} J.~A.,  1983,
  \mn@doi [\mnras] {10.1093/mnras/203.2.545}, \href
  {https://ui.adsabs.harvard.edu/abs/1983MNRAS.203..545L} {203, 545}

\bibitem[\protect\citeauthoryear{{Macario} et~al.,}{{Macario}
  et~al.}{2013}]{2013A&A...551A.141M}
{Macario} G.,  et~al., 2013, \mn@doi [\aap] {10.1051/0004-6361/201220667},
  \href {https://ui.adsabs.harvard.edu/abs/2013A&A...551A.141M} {551, A141}

\bibitem[\protect\citeauthoryear{{Malu}, {Datta}  \& {Sandhu}}{{Malu}
  et~al.}{2016}]{Maluetal16}
{Malu} S.,  {Datta} A.,   {Sandhu} P.,  2016, \mn@doi [\apss]
  {10.1007/s10509-016-2844-7}, \href
  {https://ui.adsabs.harvard.edu/abs/2016Ap&SS.361..255M} {361, 255}

\bibitem[\protect\citeauthoryear{{Mann} \& {Ebeling}}{{Mann} \&
  {Ebeling}}{2012}]{2012MNRAS.420.2120M}
{Mann} A.~W.,  {Ebeling} H.,  2012, \mn@doi [\mnras]
  {10.1111/j.1365-2966.2011.20170.x}, \href
  {https://ui.adsabs.harvard.edu/abs/2012MNRAS.420.2120M} {420, 2120}

\bibitem[\protect\citeauthoryear{{Markevitch}}{{Markevitch}}{1998}]{Markevitch98}
{Markevitch} M.,  1998, ApJ, 504, 27

\bibitem[\protect\citeauthoryear{{Markevitch}}{{Markevitch}}{2002}]{Markevitch02}
{Markevitch} M.,  2002, preprint, astro-ph/0205333

\bibitem[\protect\citeauthoryear{{Markevitch}}{{Markevitch}}{2006}]{Markevitch06}
{Markevitch} M.,  2006, in {Wilson} A.,  ed.,  ESA Special Publication Vol.
  604, The X-ray Universe 2005. p.~723 (\mn@eprint {arXiv} {astro-ph/0511345}),
  \mn@doi{10.48550/arXiv.astro-ph/0511345}

\bibitem[\protect\citeauthoryear{{Markevitch} \& {Vikhlinin}}{{Markevitch} \&
  {Vikhlinin}}{1997}]{1997ApJ...474...84M}
{Markevitch} M.,  {Vikhlinin} A.,  1997, \mn@doi [\apj] {10.1086/303456}, \href
  {https://ui.adsabs.harvard.edu/abs/1997ApJ...474...84M} {474, 84}

\bibitem[\protect\citeauthoryear{{Marriage} et~al.,}{{Marriage}
  et~al.}{2011a}]{Marriageetal11}
{Marriage} T.~A.,  et~al., 2011a, \mn@doi [\apj] {10.1088/0004-637X/737/2/61},
  \href {https://ui.adsabs.harvard.edu/abs/2011ApJ...737...61M} {737, 61}

\bibitem[\protect\citeauthoryear{{Marriage} et~al.,}{{Marriage}
  et~al.}{2011b}]{2011ApJ...737...61M}
{Marriage} T.~A.,  et~al., 2011b, \mn@doi [\apj] {10.1088/0004-637X/737/2/61},
  \href {https://ui.adsabs.harvard.edu/abs/2011ApJ...737...61M} {737, 61}

\bibitem[\protect\citeauthoryear{{M}auch, {C}otton, {C}ondon  \& et
  al.}{{M}auch et~al.}{2020}]{DEEP2}
{M}auch T.,  {C}otton W.,  {C}ondon J.,   et al. 2020, \mn@doi [ApJ]
  {10.3847/1538-4357/ab5d2}, 888, 61

\bibitem[\protect\citeauthoryear{{McConnell} et~al.,}{{McConnell}
  et~al.}{2020}]{McConnelletal20}
{McConnell} D.,  et~al., 2020, \mn@doi [\pasa] {10.1017/pasa.2020.41}, \href
  {https://ui.adsabs.harvard.edu/abs/2020PASA...37...48M} {37, e048}

\bibitem[\protect\citeauthoryear{{Menanteau} et~al.,}{{Menanteau}
  et~al.}{2010a}]{2010ApJ...723.1523M}
{Menanteau} F.,  et~al., 2010a, \mn@doi [\apj] {10.1088/0004-637X/723/2/1523},
  \href {https://ui.adsabs.harvard.edu/abs/2010ApJ...723.1523M} {723, 1523}

\bibitem[\protect\citeauthoryear{{Menanteau} et~al.,}{{Menanteau}
  et~al.}{2010b}]{Menanteauetal10}
{Menanteau} F.,  et~al., 2010b, \mn@doi [\apj] {10.1088/0004-637X/723/2/1523},
  \href {https://ui.adsabs.harvard.edu/abs/2010ApJ...723.1523M} {723, 1523}

\bibitem[\protect\citeauthoryear{{Menanteau} et~al.,}{{Menanteau}
  et~al.}{2012}]{Menantauetal12}
{Menanteau} F.,  et~al., 2012, \mn@doi [\apj] {10.1088/0004-637X/748/1/7},
  \href {https://ui.adsabs.harvard.edu/abs/2012ApJ...748....7M} {748, 7}

\bibitem[\protect\citeauthoryear{{Merloni} et~al.,}{{Merloni}
  et~al.}{2012}]{Merlonietal12}
{Merloni} A.,  et~al., 2012, preprint, \href
  {http://adsabs.harvard.edu/abs/2012arXiv1209.3114M} {} (\mn@eprint {arXiv}
  {1209.3114})

\bibitem[\protect\citeauthoryear{{Merloni} et~al.,}{{Merloni}
  et~al.}{2024}]{Merlonietal24}
{Merloni} A.,  et~al., 2024, \mn@doi [\aap] {10.1051/0004-6361/202347165},
  \href {https://ui.adsabs.harvard.edu/abs/2024A&A...682A..34M} {682, A34}

\bibitem[\protect\citeauthoryear{{Merluzzi} et~al.,}{{Merluzzi}
  et~al.}{2015}]{Merluzzietal15}
{Merluzzi} P.,  et~al., 2015, \mn@doi [\mnras] {10.1093/mnras/stu2085}, \href
  {https://ui.adsabs.harvard.edu/abs/2015MNRAS.446..803M} {446, 803}

\bibitem[\protect\citeauthoryear{{Mihos}}{{Mihos}}{2004}]{2004cgpc.symp..277M}
{Mihos} J.~C.,  2004, in {Mulchaey} J.~S.,  {Dressler} A.,   {Oemler} A.,  eds,
  Clusters of Galaxies: Probes of Cosmological Structure and Galaxy Evolution.
  p.~277

\bibitem[\protect\citeauthoryear{{Miller}, {Hornschemeier}, {Mobasher},
  {Bridges}, {Hudson}, {Marzke}  \& {Smith}}{{Miller}
  et~al.}{2009}]{2009AJ....137.4450M}
{Miller} N.~A.,  {Hornschemeier} A.~E.,  {Mobasher} B.,  {Bridges} T.~J.,
  {Hudson} M.~J.,  {Marzke} R.~O.,   {Smith} R.~J.,  2009, \mn@doi [\aj]
  {10.1088/0004-6256/137/5/4450}, \href
  {https://ui.adsabs.harvard.edu/abs/2009AJ....137.4450M} {137, 4450}

\bibitem[\protect\citeauthoryear{{Molnar} \& {Broadhurst}}{{Molnar} \&
  {Broadhurst}}{2015}]{MolnarBroadhurst15}
{Molnar} S.~M.,  {Broadhurst} T.,  2015, \mn@doi [\apj]
  {10.1088/0004-637X/800/1/37}, \href
  {https://ui.adsabs.harvard.edu/abs/2015ApJ...800...37M} {800, 37}

\bibitem[\protect\citeauthoryear{{Murgia}, {Govoni}, {Markevitch}, {Feretti},
  {Giovannini}, {Taylor}  \& {Carretti}}{{Murgia} et~al.}{2009}]{Murgiaetal09}
{Murgia} M.,  {Govoni} F.,  {Markevitch} M.,  {Feretti} L.,  {Giovannini} G.,
  {Taylor} G.~B.,   {Carretti} E.,  2009, \mn@doi [\aap]
  {10.1051/0004-6361/200911659}, \href
  {https://ui.adsabs.harvard.edu/abs/2009A&A...499..679M} {499, 679}

\bibitem[\protect\citeauthoryear{{Norris} et~al.,}{{Norris}
  et~al.}{2011}]{Norrisetal11}
{Norris} R.~P.,  et~al., 2011, \mn@doi [\pasa] {10.1071/AS11021}, \href
  {https://ui.adsabs.harvard.edu/abs/2011PASA...28..215N} {28, 215}

\bibitem[\protect\citeauthoryear{{Norris} et~al.,}{{Norris}
  et~al.}{2021}]{Norrisetal21}
{Norris} R.~P.,  et~al., 2021, \mn@doi [\pasa] {10.1017/pasa.2021.42}, \href
  {https://ui.adsabs.harvard.edu/abs/2021PASA...38...46N} {38, e046}

\bibitem[\protect\citeauthoryear{{O'Neill}, {Jones}, {Nolting}  \&
  {Mendygral}}{{O'Neill} et~al.}{2019}]{2019ApJ...884...12O}
{O'Neill} B.~J.,  {Jones} T.~W.,  {Nolting} C.,   {Mendygral} P.~J.,  2019,
  \mn@doi [\apj] {10.3847/1538-4357/ab40b1}, \href
  {https://ui.adsabs.harvard.edu/abs/2019ApJ...884...12O} {884, 12}

\bibitem[\protect\citeauthoryear{{Owen}, {Ledlow}, {Keel}  \&
  {Morrison}}{{Owen} et~al.}{1999}]{Owenetal99}
{Owen} F.~N.,  {Ledlow} M.~J.,  {Keel} W.~C.,   {Morrison} G.~E.,  1999,
  \mn@doi [\aj] {10.1086/300974}, \href
  {https://ui.adsabs.harvard.edu/abs/1999AJ....118..633O} {118, 633}

\bibitem[\protect\citeauthoryear{Pal, Kale, Wang  \& Wik}{Pal
  et~al.}{2025}]{Paletal25}
Pal A.,  Kale R.,  Wang Q. H.~S.,   Wik D.~R.,  2025, \mn@doi [The
  Astrophysical Journal] {10.3847/1538-4357/ad9903}, 979, 4

\bibitem[\protect\citeauthoryear{{Panagoulia}, {Fabian}, {Sanders}  \&
  {Hlavacek-Larrondo}}{{Panagoulia} et~al.}{2014}]{Panagouliaetal14b}
{Panagoulia} E.~K.,  {Fabian} A.~C.,  {Sanders} J.~S.,   {Hlavacek-Larrondo}
  J.,  2014, \mn@doi [MNRAS] {10.1093/mnras/stu1499}, \href
  {http://adsabs.harvard.edu/abs/2014MNRAS.444.1236P} {444, 1236}

\bibitem[\protect\citeauthoryear{{Pandge}, {Monteiro-Oliveira}, {Bagchi},
  {Simionescu}, {Limousin}  \& {Raychaudhury}}{{Pandge}
  et~al.}{2019}]{2019MNRAS.482.5093P}
{Pandge} M.~B.,  {Monteiro-Oliveira} R.,  {Bagchi} J.,  {Simionescu} A.,
  {Limousin} M.,   {Raychaudhury} S.,  2019, \mn@doi [\mnras]
  {10.1093/mnras/sty2937}, \href
  {https://ui.adsabs.harvard.edu/abs/2019MNRAS.482.5093P} {482, 5093}

\bibitem[\protect\citeauthoryear{{Pandge}, {Sebastian}, {Seth}  \&
  {Raychaudhury}}{{Pandge} et~al.}{2021}]{Pandgeetal21}
{Pandge} M.~B.,  {Sebastian} B.,  {Seth} R.,   {Raychaudhury} S.,  2021,
  \mn@doi [\mnras] {10.1093/mnras/stab384}, \href
  {https://ui.adsabs.harvard.edu/abs/2021MNRAS.504.1644P} {504, 1644}

\bibitem[\protect\citeauthoryear{{Parekh}, {van der Heyden}, {Ferrari}, {Angus}
   \& {Holwerda}}{{Parekh} et~al.}{2015}]{Parekhetal15}
{Parekh} V.,  {van der Heyden} K.,  {Ferrari} C.,  {Angus} G.,   {Holwerda} B.,
   2015, \mn@doi [\aap] {10.1051/0004-6361/201424123}, \href
  {https://ui.adsabs.harvard.edu/abs/2015A&A...575A.127P} {575, A127}

\bibitem[\protect\citeauthoryear{{Parekh}, {Dwarakanath}, {Kale}  \&
  {Intema}}{{Parekh} et~al.}{2017a}]{2017MNRAS.464.2752P}
{Parekh} V.,  {Dwarakanath} K.~S.,  {Kale} R.,   {Intema} H.,  2017a, \mn@doi
  [\mnras] {10.1093/mnras/stw2521}, \href
  {https://ui.adsabs.harvard.edu/abs/2017MNRAS.464.2752P} {464, 2752}

\bibitem[\protect\citeauthoryear{{Parekh}, {Durret}, {Padmanabh}  \&
  {Pandge}}{{Parekh} et~al.}{2017b}]{Parekhetal17}
{Parekh} V.,  {Durret} F.,  {Padmanabh} P.,   {Pandge} M.~B.,  2017b, \mn@doi
  [\mnras] {10.1093/mnras/stx1457}, \href
  {https://ui.adsabs.harvard.edu/abs/2017MNRAS.470.3742P} {470, 3742}

\bibitem[\protect\citeauthoryear{{Parekh}, {Lagan{\'a}}  \& {Kale}}{{Parekh}
  et~al.}{2021}]{Parekhetal21}
{Parekh} V.,  {Lagan{\'a}} T.~F.,   {Kale} R.,  2021, \mn@doi [\mnras]
  {10.1093/mnras/stab779}, \href
  {https://ui.adsabs.harvard.edu/abs/2021MNRAS.504..610P} {504, 610}

\bibitem[\protect\citeauthoryear{{Pascut} \& {Hughes}}{{Pascut} \&
  {Hughes}}{2019}]{PascutHughes19}
{Pascut} A.,  {Hughes} J.~P.,  2019, \mn@doi [\apj] {10.3847/1538-4357/ab07b1},
  \href {https://ui.adsabs.harvard.edu/abs/2019ApJ...874...71P} {874, 71}

\bibitem[\protect\citeauthoryear{{Pasini} et~al.,}{{Pasini}
  et~al.}{2024}]{Pasini24}
{Pasini} T.,  et~al., 2024, \mn@doi [\aap] {10.1051/0004-6361/202450697}, \href
  {https://ui.adsabs.harvard.edu/abs/2024A&A...689A.218P} {689, A218}

\bibitem[\protect\citeauthoryear{Pearce et~al.,}{Pearce
  et~al.}{2017}]{Pearce2017}
Pearce C. J.~J.,  et~al., 2017, \mn@doi [\apj] {10.3847/1538-4357/aa7e2f}, 845,
  81

\bibitem[\protect\citeauthoryear{{Perley}, {Chandler}, {Butler}  \&
  {Wrobel}}{{Perley} et~al.}{2011}]{Perley11}
{Perley} R.~A.,  {Chandler} C.~J.,  {Butler} B.~J.,   {Wrobel} J.~M.,  2011,
  \mn@doi [\apjl] {10.1088/2041-8205/739/1/L1}, \href
  {https://ui.adsabs.harvard.edu/abs/2011ApJ...739L...1P} {739, L1}

\bibitem[\protect\citeauthoryear{{Piffaretti}, {Arnaud}, {Pratt},
  {Pointecouteau}  \& {Melin}}{{Piffaretti} et~al.}{2011}]{2011A&A...534A.109P}
{Piffaretti} R.,  {Arnaud} M.,  {Pratt} G.~W.,  {Pointecouteau} E.,   {Melin}
  J.-B.,  2011, \mn@doi [\aap] {10.1051/0004-6361/201015377}, \href
  {http://adsabs.harvard.edu/abs/2011A\%26A...534A.109P} {534, A109}

\bibitem[\protect\citeauthoryear{{Pimbblet}, {Edge}  \& {Couch}}{{Pimbblet}
  et~al.}{2005}]{2005MNRAS.357L..45P}
{Pimbblet} K.~A.,  {Edge} A.~C.,   {Couch} W.~J.,  2005, \mn@doi [\mnras]
  {10.1111/j.1745-3933.2005.00015.x}, \href
  {https://ui.adsabs.harvard.edu/abs/2005MNRAS.357L..45P} {357, L45}

\bibitem[\protect\citeauthoryear{{Planck Collaboration} et~al.,}{{Planck
  Collaboration} et~al.}{2011a}]{Planckcollab11a}
{Planck Collaboration} et~al., 2011a, \mn@doi [\aap]
  {10.1051/0004-6361/201116459}, \href
  {https://ui.adsabs.harvard.edu/abs/2011A&A...536A...8P} {536, A8}

\bibitem[\protect\citeauthoryear{{Planck Collaboration} et~al.,}{{Planck
  Collaboration} et~al.}{2011b}]{Planckcollab11b}
{Planck Collaboration} et~al., 2011b, \mn@doi [\aap]
  {10.1051/0004-6361/201116460}, \href
  {https://ui.adsabs.harvard.edu/abs/2011A&A...536A...9P} {536, A9}

\bibitem[\protect\citeauthoryear{{Planck Collaboration} et~al.,}{{Planck
  Collaboration} et~al.}{2013}]{2013A&A...550A.134P}
{Planck Collaboration} et~al., 2013, \mn@doi [\aap]
  {10.1051/0004-6361/201220194}, \href
  {http://adsabs.harvard.edu/abs/2013A\%26A...550A.134P} {550, A134}

\bibitem[\protect\citeauthoryear{{Planck Collaboration} et~al.,}{{Planck
  Collaboration} et~al.}{2014}]{Planckcollab14}
{Planck Collaboration} et~al., 2014, \mn@doi [A\&A]
  {10.1051/0004-6361/201321521}, 571, A20

\bibitem[\protect\citeauthoryear{{Planck Collaboration} et~al.,}{{Planck
  Collaboration} et~al.}{2015}]{Planckcollab15}
{Planck Collaboration} et~al., 2015, \mn@doi [\aap]
  {10.1051/0004-6361/201525787}, \href
  {https://ui.adsabs.harvard.edu/abs/2015A&A...581A..14P} {581, A14}

\bibitem[\protect\citeauthoryear{{Planck Collaboration} et~al.,}{{Planck
  Collaboration} et~al.}{2016}]{Planckcollab16}
{Planck Collaboration} et~al., 2016, \mn@doi [A\&A]
  {10.1051/0004-6361/201525823}, 594, A27

\bibitem[\protect\citeauthoryear{{Poole}, {Fardal}, {Babul}, {McCarthy},
  {Quinn}  \& {Wadsley}}{{Poole} et~al.}{2006}]{Pooleetal06}
{Poole} G.~B.,  {Fardal} M.~A.,  {Babul} A.,  {McCarthy} I.~G.,  {Quinn} T.,
  {Wadsley} J.,  2006, \mn@doi [MNRAS] {10.1111/j.1365-2966.2006.10916.x},
  \href {http://adsabs.harvard.edu/abs/2006MNRAS.373..881P} {373, 881}

\bibitem[\protect\citeauthoryear{{Pratt}, {Croston}, {Arnaud}  \&
  {B{\"o}hringer}}{{Pratt} et~al.}{2009}]{Prattetal09}
{Pratt} G.~W.,  {Croston} J.~H.,  {Arnaud} M.,   {B{\"o}hringer} H.,  2009,
  \mn@doi [A\&A] {10.1051/0004-6361/200810994}, \href
  {http://adsabs.harvard.edu/abs/2009A\%26A...498..361P} {498, 361}

\bibitem[\protect\citeauthoryear{{Predehl} et~al.,}{{Predehl}
  et~al.}{2021}]{2021A&A...647A...1P}
{Predehl} P.,  et~al., 2021, \mn@doi [\aap] {10.1051/0004-6361/202039313},
  \href {https://ui.adsabs.harvard.edu/abs/2021A&A...647A...1P} {647, A1}

\bibitem[\protect\citeauthoryear{{Rahaman}, {Raja}, {Datta}, {Burns}, {Alden}
  \& {Rapetti}}{{Rahaman} et~al.}{2021}]{2021MNRAS.505..480R}
{Rahaman} M.,  {Raja} R.,  {Datta} A.,  {Burns} J.~O.,  {Alden} B.,   {Rapetti}
  D.,  2021, \mn@doi [\mnras] {10.1093/mnras/stab1225}, \href
  {https://ui.adsabs.harvard.edu/abs/2021MNRAS.505..480R} {505, 480}

\bibitem[\protect\citeauthoryear{{Rajpurohit} et~al.,}{{Rajpurohit}
  et~al.}{2021}]{2021A&A...646A..56R}
{Rajpurohit} K.,  et~al., 2021, \mn@doi [\aap] {10.1051/0004-6361/202039428},
  \href {https://ui.adsabs.harvard.edu/abs/2021A&A...646A..56R} {646, A56}

\bibitem[\protect\citeauthoryear{Rajpurohit et~al.,}{Rajpurohit
  et~al.}{2022}]{Rajpurohitetal22}
Rajpurohit K.,  et~al., 2022, \mn@doi [The Astrophysical Journal]
  {10.3847/1538-4357/ac4708}, 927, 80

\bibitem[\protect\citeauthoryear{{Reiprich} \& {B{\" o}hringer}}{{Reiprich} \&
  {B{\" o}hringer}}{2002}]{ReiprichBohringer02}
{Reiprich} T.~H.,  {B{\" o}hringer} H.,  2002, ApJ, \href
  {http://adsabs.harvard.edu/cgi-bin/nph-bib_query?bibcode=2002ApJ...567..716R&db_key=AST}
  {567, 716}

\bibitem[\protect\citeauthoryear{{Reiprich} et~al.,}{{Reiprich}
  et~al.}{2021}]{Reiprichetal21}
{Reiprich} T.~H.,  et~al., 2021, \mn@doi [\aap] {10.1051/0004-6361/202039590},
  \href {https://ui.adsabs.harvard.edu/abs/2021A&A...647A...2R} {647, A2}

\bibitem[\protect\citeauthoryear{{Richard-Laferri{\`e}re}
  et~al.,}{{Richard-Laferri{\`e}re} et~al.}{2020}]{2020MNRAS.499.2934R}
{Richard-Laferri{\`e}re} A.,  et~al., 2020, \mn@doi [\mnras]
  {10.1093/mnras/staa2877}, \href
  {https://ui.adsabs.harvard.edu/abs/2020MNRAS.499.2934R} {499, 2934}

\bibitem[\protect\citeauthoryear{{Riseley}, {Scaife}, {Oozeer}, {Magnus}  \&
  {Wise}}{{Riseley} et~al.}{2015}]{Riseleyetal15}
{Riseley} C.~J.,  {Scaife} A.~M.~M.,  {Oozeer} N.,  {Magnus} L.,   {Wise}
  M.~W.,  2015, \mn@doi [\mnras] {10.1093/mnras/stu2591}, \href
  {https://ui.adsabs.harvard.edu/abs/2015MNRAS.447.1895R} {447, 1895}

\bibitem[\protect\citeauthoryear{{Riseley} et~al.,}{{Riseley}
  et~al.}{2022a}]{Riseleyetal22b}
{Riseley} C.~J.,  et~al., 2022a, \mn@doi [\mnras] {10.1093/mnras/stac672},
  \href {https://ui.adsabs.harvard.edu/abs/2022MNRAS.512.4210R} {512, 4210}

\bibitem[\protect\citeauthoryear{{Riseley} et~al.,}{{Riseley}
  et~al.}{2022b}]{Riseleyetal22}
{Riseley} C.~J.,  et~al., 2022b, \mn@doi [\mnras] {10.1093/mnras/stac1771},
  \href {https://ui.adsabs.harvard.edu/abs/2022MNRAS.515.1871R} {515, 1871}

\bibitem[\protect\citeauthoryear{{Riseley} et~al.,}{{Riseley}
  et~al.}{2023}]{Riseleyetal23}
{Riseley} C.~J.,  et~al., 2023, \mn@doi [\mnras] {10.1093/mnras/stad2218},
  \href {https://ui.adsabs.harvard.edu/abs/2023MNRAS.524.6052R} {524, 6052}

\bibitem[\protect\citeauthoryear{{Rizza}, {Burns}, {Ledlow}, {Owen}, {Voges}
  \& {Bliton}}{{Rizza} et~al.}{1998}]{1998MNRAS.301..328R}
{Rizza} E.,  {Burns} J.~O.,  {Ledlow} M.~J.,  {Owen} F.~N.,  {Voges} W.,
  {Bliton} M.,  1998, \mn@doi [\mnras] {10.1046/j.1365-8711.1998.01972.x},
  \href {https://ui.adsabs.harvard.edu/abs/1998MNRAS.301..328R} {301, 328}

\bibitem[\protect\citeauthoryear{{Rossetti}, {Ghizzardi}, {Molendi}  \&
  {Finoguenov}}{{Rossetti} et~al.}{2007}]{2007A&A...463..839R}
{Rossetti} M.,  {Ghizzardi} S.,  {Molendi} S.,   {Finoguenov} A.,  2007,
  \mn@doi [\aap] {10.1051/0004-6361:20054621}, \href
  {https://ui.adsabs.harvard.edu/abs/2007A&A...463..839R} {463, 839}

\bibitem[\protect\citeauthoryear{{Rossetti}, {Gastaldello}, {Eckert}, {Della
  Torre}, {Pantiri}, {Cazzoletti}  \& {Molendi}}{{Rossetti}
  et~al.}{2017}]{Rossettietal17}
{Rossetti} M.,  {Gastaldello} F.,  {Eckert} D.,  {Della Torre} M.,  {Pantiri}
  G.,  {Cazzoletti} P.,   {Molendi} S.,  2017, \mn@doi [\mnras]
  {10.1093/mnras/stx493}, \href
  {https://ui.adsabs.harvard.edu/abs/2017MNRAS.468.1917R} {468, 1917}

\bibitem[\protect\citeauthoryear{{Rottgering}, {Wieringa}, {Hunstead}  \&
  {Ekers}}{{Rottgering} et~al.}{1997}]{1997MNRAS.290..577R}
{Rottgering} H.~J.~A.,  {Wieringa} M.~H.,  {Hunstead} R.~W.,   {Ekers} R.~D.,
  1997, \mn@doi [\mnras] {10.1093/mnras/290.4.577}, \href
  {https://ui.adsabs.harvard.edu/abs/1997MNRAS.290..577R} {290, 577}

\bibitem[\protect\citeauthoryear{{Rudnick}}{{Rudnick}}{2002}]{2002PASP..114..427R}
{Rudnick} L.,  2002, \mn@doi [\pasp] {10.1086/342499}, \href
  {https://ui.adsabs.harvard.edu/abs/2002PASP..114..427R} {114, 427}

\bibitem[\protect\citeauthoryear{{Rudnick}, {Cotton}, {Knowles}  \&
  {Kolokythas}}{{Rudnick} et~al.}{2021}]{2021Galax...9...81R}
{Rudnick} L.,  {Cotton} W.,  {Knowles} K.,   {Kolokythas} K.,  2021, \mn@doi
  [Galaxies] {10.3390/galaxies9040081}, \href
  {https://ui.adsabs.harvard.edu/abs/2021Galax...9...81R} {9, 81}

\bibitem[\protect\citeauthoryear{{Sanders} et~al.,}{{Sanders}
  et~al.}{2022}]{Sandersetal22}
{Sanders} J.~S.,  et~al., 2022, \mn@doi [\aap] {10.1051/0004-6361/202141501},
  \href {https://ui.adsabs.harvard.edu/abs/2022A&A...661A..36S} {661, A36}

\bibitem[\protect\citeauthoryear{{Santra} et~al.,}{{Santra}
  et~al.}{2024}]{Santraetal24}
{Santra} R.,  et~al., 2024, \mn@doi [\apj] {10.3847/1538-4357/ad1190}, \href
  {https://ui.adsabs.harvard.edu/abs/2024ApJ...962...40S} {962, 40}

\bibitem[\protect\citeauthoryear{{Savini} et~al.,}{{Savini}
  et~al.}{2018}]{Savinietal18}
{Savini} F.,  et~al., 2018, \mn@doi [\mnras] {10.1093/mnras/sty1125}, \href
  {https://ui.adsabs.harvard.edu/abs/2018MNRAS.478.2234S} {478, 2234}

\bibitem[\protect\citeauthoryear{{Schellenberger} et~al.,}{{Schellenberger}
  et~al.}{2022}]{Schellenberger22}
{Schellenberger} G.,  et~al., 2022, \mn@doi [\apj] {10.3847/1538-4357/ac3b5a},
  \href {https://ui.adsabs.harvard.edu/abs/2022ApJ...925...91S} {925, 91}

\bibitem[\protect\citeauthoryear{{Shabala}, {Jurlin}, {Morganti}, {Brienza},
  {Hardcastle}, {Godfrey}, {Krause}  \& {Turner}}{{Shabala}
  et~al.}{2020}]{2020MNRAS.496.1706S}
{Shabala} S.~S.,  {Jurlin} N.,  {Morganti} R.,  {Brienza} M.,  {Hardcastle}
  M.~J.,  {Godfrey} L. E.~H.,  {Krause} M. G.~H.,   {Turner} R.~J.,  2020,
  \mn@doi [\mnras] {10.1093/mnras/staa1172}, \href
  {https://ui.adsabs.harvard.edu/abs/2020MNRAS.496.1706S} {496, 1706}

\bibitem[\protect\citeauthoryear{{Shakouri}, {Johnston-Hollitt}  \&
  {Pratt}}{{Shakouri} et~al.}{2016}]{Shakouri2016ARDES}
{Shakouri} S.,  {Johnston-Hollitt} M.,   {Pratt} G.~W.,  2016, \mn@doi [\mnras]
  {10.1093/mnras/stw812}, \href
  {http://adsabs.harvard.edu/abs/2016MNRAS.459.2525S} {459, 2525}

\bibitem[\protect\citeauthoryear{{Shimwell}, {Brown}, {Feain}, {Feretti},
  {Gaensler}  \& {Lage}}{{Shimwell} et~al.}{2014}]{Shimwelletal15}
{Shimwell} T.~W.,  {Brown} S.,  {Feain} I.~J.,  {Feretti} L.,  {Gaensler}
  B.~M.,   {Lage} C.,  2014, \mn@doi [\mnras] {10.1093/mnras/stu467}, \href
  {https://ui.adsabs.harvard.edu/abs/2014MNRAS.440.2901S} {440, 2901}

\bibitem[\protect\citeauthoryear{{Shimwell}, {Markevitch}, {Brown}, {Feretti},
  {Gaensler}, {Johnston-Hollitt}, {Lage}  \& {Srinivasan}}{{Shimwell}
  et~al.}{2015}]{Shimwell15B}
{Shimwell} T.~W.,  {Markevitch} M.,  {Brown} S.,  {Feretti} L.,  {Gaensler}
  B.~M.,  {Johnston-Hollitt} M.,  {Lage} C.,   {Srinivasan} R.,  2015, \mn@doi
  [\mnras] {10.1093/mnras/stv334}, \href
  {https://ui.adsabs.harvard.edu/abs/2015MNRAS.449.1486S} {449, 1486}

\bibitem[\protect\citeauthoryear{{Shimwell} et~al.,}{{Shimwell}
  et~al.}{2022}]{Shimwelletal22}
{Shimwell} T.~W.,  et~al., 2022, \mn@doi [\aap] {10.1051/0004-6361/202142484},
  \href {https://ui.adsabs.harvard.edu/abs/2022A&A...659A...1S} {659, A1}

\bibitem[\protect\citeauthoryear{{Sikhosana}, {Knowles}, {Hilton}, {Moodley}
  \& {Murgia}}{{Sikhosana} et~al.}{2023}]{Sikhosanaetal23}
{Sikhosana} S.~P.,  {Knowles} K.,  {Hilton} M.,  {Moodley} K.,   {Murgia} M.,
  2023, \mn@doi [\mnras] {10.1093/mnras/stac3370}, \href
  {https://ui.adsabs.harvard.edu/abs/2023MNRAS.518.4595S} {518, 4595}

\bibitem[\protect\citeauthoryear{{Sivanandam}, {Zabludoff}, {Zaritsky},
  {Gonzalez}  \& {Kelson}}{{Sivanandam} et~al.}{2009}]{2009ApJ...691.1787S}
{Sivanandam} S.,  {Zabludoff} A.~I.,  {Zaritsky} D.,  {Gonzalez} A.~H.,
  {Kelson} D.~D.,  2009, \mn@doi [\apj] {10.1088/0004-637X/691/2/1787}, \href
  {https://ui.adsabs.harvard.edu/abs/2009ApJ...691.1787S} {691, 1787}

\bibitem[\protect\citeauthoryear{{Skrutskie} et~al.,}{{Skrutskie}
  et~al.}{2006}]{Skrutskieetal06}
{Skrutskie} M.~F.,  et~al., 2006, \mn@doi [AJ] {10.1086/498708}, \href
  {http://adsabs.harvard.edu/abs/2006AJ....131.1163S} {131, 1163}

\bibitem[\protect\citeauthoryear{{Slee}, {Siegman}  \& {Wilson}}{{Slee}
  et~al.}{1983}]{1983AuJPh..36..101S}
{Slee} O.~B.,  {Siegman} C.~B.,   {Wilson} I.~R.~G.,  1983, \mn@doi [Australian
  Journal of Physics] {10.1071/PH830101}, \href
  {https://ui.adsabs.harvard.edu/abs/1983AuJPh..36..101S} {36, 101}

\bibitem[\protect\citeauthoryear{{Struble}}{{Struble}}{1988}]{Struble88}
{Struble} M.~F.,  1988, \mn@doi [\apjl] {10.1086/185197}, \href
  {https://ui.adsabs.harvard.edu/abs/1988ApJ...330L..25S} {330, L25}

\bibitem[\protect\citeauthoryear{{Stuardi} et~al.,}{{Stuardi}
  et~al.}{2019}]{2019MNRAS.489.3905S}
{Stuardi} C.,  et~al., 2019, \mn@doi [\mnras] {10.1093/mnras/stz2408}, \href
  {https://ui.adsabs.harvard.edu/abs/2019MNRAS.489.3905S} {489, 3905}

\bibitem[\protect\citeauthoryear{{Stuardi}, {Gheller}, {Vazza}  \&
  {Botteon}}{{Stuardi} et~al.}{2024}]{Stuardietal24}
{Stuardi} C.,  {Gheller} C.,  {Vazza} F.,   {Botteon} A.,  2024, \mn@doi
  [\mnras] {10.1093/mnras/stae2014}, \href
  {https://ui.adsabs.harvard.edu/abs/2024MNRAS.533.3194S} {533, 3194}

\bibitem[\protect\citeauthoryear{{Sugawara}, {Takizawa}, {Itahana}, {Akamatsu},
  {Fujita}, {Ohashi}  \& {Ishisaki}}{{Sugawara} et~al.}{2017}]{Sugawaraetal17}
{Sugawara} Y.,  {Takizawa} M.,  {Itahana} M.,  {Akamatsu} H.,  {Fujita} Y.,
  {Ohashi} T.,   {Ishisaki} Y.,  2017, \mn@doi [\pasj] {10.1093/pasj/psx104},
  \href {https://ui.adsabs.harvard.edu/abs/2017PASJ...69...93S} {69, 93}

\bibitem[\protect\citeauthoryear{{Sunyaev} \& {Zeldovich}}{{Sunyaev} \&
  {Zeldovich}}{1972}]{1972CoASP...4..173S}
{Sunyaev} R.~A.,  {Zeldovich} Y.~B.,  1972, Comments on Astrophysics and Space
  Physics, \href {https://ui.adsabs.harvard.edu/abs/1972CoASP...4..173S} {4,
  173}

\bibitem[\protect\citeauthoryear{{Tanaka}, {Inoue}  \& {Holt}}{{Tanaka}
  et~al.}{1994}]{Tanakaetal94}
{Tanaka} Y.,  {Inoue} H.,   {Holt} S.~S.,  1994, \pasj, \href
  {https://ui.adsabs.harvard.edu/abs/1994PASJ...46L..37T} {46, L37}

\bibitem[\protect\citeauthoryear{{Tarenghi} \& {Wilson}}{{Tarenghi} \&
  {Wilson}}{1989}]{TarenghiWilson89}
{Tarenghi} M.,  {Wilson} R.~N.,  1989, in {Roddier} F.~J.,  ed.,  Society of
  Photo-Optical Instrumentation Engineers (SPIE) Conference Series Vol. 1114,
  Active telescope systems. pp 302--313, \mn@doi{10.1117/12.960835}

\bibitem[\protect\citeauthoryear{{Taylor} et~al.,}{{Taylor}
  et~al.}{2015}]{Taylor_2015}
{Taylor} E.~N.,  et~al., 2015, \mn@doi [\mnras] {10.1093/mnras/stu1900}, \href
  {https://ui.adsabs.harvard.edu/abs/2015MNRAS.446.2144T} {446, 2144}

\bibitem[\protect\citeauthoryear{{Tempel, E.}, {Kipper, R.}, {Tamm, A.},
  {Gramann, M.}, {Einasto, M.}, {Sepp, T.}  \& {Tuvikene, T.}}{{Tempel, E.}
  et~al.}{2016}]{refId0}
{Tempel, E.} {Kipper, R.} {Tamm, A.} {Gramann, M.} {Einasto, M.} {Sepp, T.}
  {Tuvikene, T.} 2016, \mn@doi [A\&A] {10.1051/0004-6361/201527755}, 588, A14

\bibitem[\protect\citeauthoryear{{Tingay} et~al.,}{{Tingay}
  et~al.}{2013}]{Tingayetal23}
{Tingay} S.~J.,  et~al., 2013, \mn@doi [\pasa] {10.1017/pasa.2012.007}, \href
  {https://ui.adsabs.harvard.edu/abs/2013PASA...30....7T} {30, e007}

\bibitem[\protect\citeauthoryear{{Tittley} \& {Henriksen}}{{Tittley} \&
  {Henriksen}}{2001}]{TittleyHenriksen01}
{Tittley} E.~R.,  {Henriksen} M.,  2001, \mn@doi [\apj] {10.1086/323955}, \href
  {https://ui.adsabs.harvard.edu/abs/2001ApJ...563..673T} {563, 673}

\bibitem[\protect\citeauthoryear{{Tolley}}{{Tolley}}{2024}]{Tolley24}
{Tolley} E.,  2024, \mn@doi [URSI Radio Science Letters] {10.46620/23-0028},
  \href {https://ui.adsabs.harvard.edu/abs/2024URSL....5...28T} {5, 28}

\bibitem[\protect\citeauthoryear{{Trehaeven} et~al.,}{{Trehaeven}
  et~al.}{2023}]{Trehaevenetal23}
{Trehaeven} K.~S.,  et~al., 2023, \mn@doi [\mnras] {10.1093/mnras/stad391},
  \href {https://ui.adsabs.harvard.edu/abs/2023MNRAS.520.4410T} {520, 4410}

\bibitem[\protect\citeauthoryear{{Trehaeven} et~al.,}{{Trehaeven}
  et~al.}{2025}]{Trehaevenetal25}
{Trehaeven} K.~S.,  et~al., 2025, \mn@doi [\mnras] {10.1093/mnras/staf1136},
  \href {https://ui.adsabs.harvard.edu/abs/2025MNRAS.tmp.1099T} {}

\bibitem[\protect\citeauthoryear{{Tucker} et~al.,}{{Tucker}
  et~al.}{1998}]{Tuckeretal98}
{Tucker} W.,  et~al., 1998, \mn@doi [\apjl] {10.1086/311234}, \href
  {https://ui.adsabs.harvard.edu/abs/1998ApJ...496L...5T} {496, L5}

\bibitem[\protect\citeauthoryear{{Tully}}{{Tully}}{2015}]{Tully_2015}
{Tully} R.~B.,  2015, \mn@doi [\aj] {10.1088/0004-6256/149/5/171}, \href
  {https://ui.adsabs.harvard.edu/abs/2015AJ....149..171T} {149, 171}

\bibitem[\protect\citeauthoryear{{Urdampilleta}, {Akamatsu}, {Mernier},
  {Kaastra}, {de Plaa}, {Ohashi}, {Ishisaki}  \& {Kawahara}}{{Urdampilleta}
  et~al.}{2018}]{2018A&A...618A..74U}
{Urdampilleta} I.,  {Akamatsu} H.,  {Mernier} F.,  {Kaastra} J.~S.,  {de Plaa}
  J.,  {Ohashi} T.,  {Ishisaki} Y.,   {Kawahara} H.,  2018, \mn@doi [\aap]
  {10.1051/0004-6361/201732496}, \href
  {https://ui.adsabs.harvard.edu/abs/2018A&A...618A..74U} {618, A74}

\bibitem[\protect\citeauthoryear{{Urdampilleta}, {Simionescu}, {Kaastra},
  {Zhang}, {Di Gennaro}, {Mernier}, {de Plaa}  \& {Brunetti}}{{Urdampilleta}
  et~al.}{2021}]{2021A&A...646A..95U}
{Urdampilleta} I.,  {Simionescu} A.,  {Kaastra} J.~S.,  {Zhang} X.,  {Di
  Gennaro} G.,  {Mernier} F.,  {de Plaa} J.,   {Brunetti} G.,  2021, \mn@doi
  [\aap] {10.1051/0004-6361/201937160}, \href
  {https://ui.adsabs.harvard.edu/abs/2021A&A...646A..95U} {646, A95}

\bibitem[\protect\citeauthoryear{{Valtchanov}, {Murphy}, {Pierre}, {Hunstead}
  \& {L{\'e}monon}}{{Valtchanov} et~al.}{2002}]{Valtchanovetal02}
{Valtchanov} I.,  {Murphy} T.,  {Pierre} M.,  {Hunstead} R.,   {L{\'e}monon}
  L.,  2002, \mn@doi [\aap] {10.1051/0004-6361:20020940}, \href
  {https://ui.adsabs.harvard.edu/abs/2002A&A...392..795V} {392, 795}

\bibitem[\protect\citeauthoryear{{Vavilova}, {Dobrycheva}, {Vasylenko},
  {Elyiv}, {Melnyk}  \& {Khramtsov}}{{Vavilova} et~al.}{2021}]{Vavilovaetal21}
{Vavilova} I.~B.,  {Dobrycheva} D.~V.,  {Vasylenko} M.~Y.,  {Elyiv} A.~A.,
  {Melnyk} O.~V.,   {Khramtsov} V.,  2021, \mn@doi [\aap]
  {10.1051/0004-6361/202038981}, \href
  {https://ui.adsabs.harvard.edu/abs/2021A&A...648A.122V} {648, A122}

\bibitem[\protect\citeauthoryear{{Venturi}, {Bardelli}, {Morganti}  \&
  {Hunstead}}{{Venturi} et~al.}{2000}]{2000MNRAS.314..594V}
{Venturi} T.,  {Bardelli} S.,  {Morganti} R.,   {Hunstead} R.~W.,  2000,
  \mn@doi [\mnras] {10.1046/j.1365-8711.2000.03403.x}, \href
  {https://ui.adsabs.harvard.edu/abs/2000MNRAS.314..594V} {314, 594}

\bibitem[\protect\citeauthoryear{{Venturi}, {Bardelli}, {Dallacasa},
  {Brunetti}, {Giacintucci}, {Hunstead}  \& {Morganti}}{{Venturi}
  et~al.}{2003}]{2003A&A...402..913V}
{Venturi} T.,  {Bardelli} S.,  {Dallacasa} D.,  {Brunetti} G.,  {Giacintucci}
  S.,  {Hunstead} R.~W.,   {Morganti} R.,  2003, \mn@doi [\aap]
  {10.1051/0004-6361:20030345}, \href
  {https://ui.adsabs.harvard.edu/abs/2003A&A...402..913V} {402, 913}

\bibitem[\protect\citeauthoryear{{Venturi}, {Giacintucci}, {Brunetti},
  {Cassano}, {Bardelli}, {Dallacasa}  \& {Setti}}{{Venturi}
  et~al.}{2007}]{2007A&A...463..937V}
{Venturi} T.,  {Giacintucci} S.,  {Brunetti} G.,  {Cassano} R.,  {Bardelli} S.,
   {Dallacasa} D.,   {Setti} G.,  2007, \mn@doi [\aap]
  {10.1051/0004-6361:20065961}, \href
  {https://ui.adsabs.harvard.edu/abs/2007A&A...463..937V} {463, 937}

\bibitem[\protect\citeauthoryear{{Venturi}, {Rossetti}, {Bardelli},
  {Giacintucci}, {Dallacasa}, {Cornacchia}  \& {Kantharia}}{{Venturi}
  et~al.}{2013}]{Venturietal13}
{Venturi} T.,  {Rossetti} M.,  {Bardelli} S.,  {Giacintucci} S.,  {Dallacasa}
  D.,  {Cornacchia} M.,   {Kantharia} N.~G.,  2013, \mn@doi [\aap]
  {10.1051/0004-6361/201322023}, \href
  {https://ui.adsabs.harvard.edu/abs/2013A&A...558A.146V} {558, A146}

\bibitem[\protect\citeauthoryear{{Venturi} et~al.,}{{Venturi}
  et~al.}{2022}]{Venturietal22}
{Venturi} T.,  et~al., 2022, \mn@doi [\aap] {10.1051/0004-6361/202142048},
  \href {https://ui.adsabs.harvard.edu/abs/2022A&A...660A..81V} {660, A81}

\bibitem[\protect\citeauthoryear{{Veronica} et~al.,}{{Veronica}
  et~al.}{2022}]{Veronicaetal22}
{Veronica} A.,  et~al., 2022, \mn@doi [\aap] {10.1051/0004-6361/202141415},
  \href {https://ui.adsabs.harvard.edu/abs/2022A&A...661A..46V} {661, A46}

\bibitem[\protect\citeauthoryear{{Vikhlinin} et~al.,}{{Vikhlinin}
  et~al.}{2009}]{Vikhlininetal09}
{Vikhlinin} A.,  et~al., 2009, \mn@doi [\apj] {10.1088/0004-637X/692/2/1033},
  \href {https://ui.adsabs.harvard.edu/abs/2009ApJ...692.1033V} {692, 1033}

\bibitem[\protect\citeauthoryear{{Voges}}{{Voges}}{1992}]{Voges92}
{Voges} W.,  1992, {The ROSAT All-Sky X-ray survey.}, In ESA, Environment
  Observation and Climate Modelling Through International Space Projects. Space
  Sciences with Particular Emphasis on High-Energy Astrophysics p 9-19 (SEE
  N93-23878 08-88)

\bibitem[\protect\citeauthoryear{{Voges} et~al.,}{{Voges}
  et~al.}{1999}]{Vogesetal99}
{Voges} W.,  et~al., 1999, A\&A, 349, 389

\bibitem[\protect\citeauthoryear{{Wang}, {Xu}, {Gu}, {Gu}, {Qin}, {Wang},
  {Zhang}  \& {Wu}}{{Wang} et~al.}{2010}]{Wangetal10}
{Wang} Y.,  {Xu} H.,  {Gu} L.,  {Gu} J.,  {Qin} Z.,  {Wang} J.,  {Zhang} Z.,
  {Wu} X.-P.,  2010, \mn@doi [\mnras] {10.1111/j.1365-2966.2010.16264.x}, \href
  {https://ui.adsabs.harvard.edu/abs/2010MNRAS.403.1909W} {403, 1909}

\bibitem[\protect\citeauthoryear{{Wayth} et~al.,}{{Wayth}
  et~al.}{2018}]{Waythetal18}
{Wayth} R.~B.,  et~al., 2018, \mn@doi [\pasa] {10.1017/pasa.2018.37}, \href
  {https://ui.adsabs.harvard.edu/abs/2018PASA...35...33W} {35, e033}

\bibitem[\protect\citeauthoryear{Wilber, Johnston-Hollitt, Duchesne, Tasse,
  Akamatsu, Intema  \& Hodgson}{Wilber et~al.}{2020}]{wilber_johnston_2020}
Wilber A.~G.,  Johnston-Hollitt M.,  Duchesne S.~W.,  Tasse C.,  Akamatsu H.,
  Intema H.,   Hodgson T.,  2020, \mn@doi [Publications of the Astronomical
  Society of Australia] {10.1017/pasa.2020.34}, 37, e040

\bibitem[\protect\citeauthoryear{{Xie} et~al.,}{{Xie}
  et~al.}{2020}]{2020A&A...636A...3X}
{Xie} C.,  et~al., 2020, \mn@doi [\aap] {10.1051/0004-6361/201936953}, \href
  {https://ui.adsabs.harvard.edu/abs/2020A&A...636A...3X} {636, A3}

\bibitem[\protect\citeauthoryear{{Yoon}, {Lee}, {Jee}, {Finner}, {Smith}  \&
  {Kim}}{{Yoon} et~al.}{2020}]{Yoonetal20}
{Yoon} M.,  {Lee} W.,  {Jee} M.~J.,  {Finner} K.,  {Smith} R.,   {Kim} J.-W.,
  2020, \mn@doi [\apj] {10.3847/1538-4357/abb76d}, \href
  {https://ui.adsabs.harvard.edu/abs/2020ApJ...903..151Y} {903, 151}

\bibitem[\protect\citeauthoryear{{Yuan}, {Han}  \& {Wen}}{{Yuan}
  et~al.}{2015}]{Yuanetal15}
{Yuan} Z.~S.,  {Han} J.~L.,   {Wen} Z.~L.,  2015, \mn@doi [\apj]
  {10.1088/0004-637X/813/1/77}, \href
  {https://ui.adsabs.harvard.edu/abs/2015ApJ...813...77Y} {813, 77}

\bibitem[\protect\citeauthoryear{{Zhang}, {Yu}  \& {Lu}}{{Zhang}
  et~al.}{2015}]{Zhangetal15}
{Zhang} C.,  {Yu} Q.,   {Lu} Y.,  2015, \mn@doi [\apj]
  {10.1088/0004-637X/813/2/129}, \href
  {https://ui.adsabs.harvard.edu/abs/2015ApJ...813..129Z} {813, 129}

\bibitem[\protect\citeauthoryear{{ZuHone}, {Markevitch}  \& {Lee}}{{ZuHone}
  et~al.}{2011}]{2011ApJ...743...16Z}
{ZuHone} J.~A.,  {Markevitch} M.,   {Lee} D.,  2011, \mn@doi [\apj]
  {10.1088/0004-637X/743/1/16}, \href
  {https://ui.adsabs.harvard.edu/abs/2011ApJ...743...16Z} {743, 16}

\bibitem[\protect\citeauthoryear{{ZuHone}, {Markevitch}, {Brunetti}  \&
  {Giacintucci}}{{ZuHone} et~al.}{2013}]{2013ApJ...762...78Z}
{ZuHone} J.~A.,  {Markevitch} M.,  {Brunetti} G.,   {Giacintucci} S.,  2013,
  \mn@doi [\apj] {10.1088/0004-637X/762/2/78}, \href
  {https://ui.adsabs.harvard.edu/abs/2013ApJ...762...78Z} {762, 78}

\bibitem[\protect\citeauthoryear{{ZuHone}, {Brunetti}, {Giacintucci}  \&
  {Markevitch}}{{ZuHone} et~al.}{2015}]{Zuhoneetal15}
{ZuHone} J.~A.,  {Brunetti} G.,  {Giacintucci} S.,   {Markevitch} M.,  2015,
  \mn@doi [\apj] {10.1088/0004-637X/801/2/146}, \href
  {https://ui.adsabs.harvard.edu/abs/2015ApJ...801..146Z} {801, 146}

\bibitem[\protect\citeauthoryear{{de Gasperin} et~al.,}{{de Gasperin}
  et~al.}{2017}]{2017SciA....3E1634D}
{de Gasperin} F.,  et~al., 2017, \mn@doi [Science Advances]
  {10.1126/sciadv.1701634}, \href
  {https://ui.adsabs.harvard.edu/abs/2017SciA....3E1634D} {3, e1701634}

\bibitem[\protect\citeauthoryear{{de Gasperin} et~al.,}{{de Gasperin}
  et~al.}{2023}]{deGasperin23}
{de Gasperin} F.,  et~al., 2023, \mn@doi [\aap] {10.1051/0004-6361/202245389},
  \href {https://ui.adsabs.harvard.edu/abs/2023A&A...673A.165D} {673, A165}

\bibitem[\protect\citeauthoryear{{van Haarlem} et~al.,}{{van Haarlem}
  et~al.}{2013}]{vanHaarlemetal13}
{van Haarlem} M.~P.,  et~al., 2013, \mn@doi [\aap]
  {10.1051/0004-6361/201220873}, \href
  {https://ui.adsabs.harvard.edu/abs/2013A&A...556A...2V} {556, A2}

\bibitem[\protect\citeauthoryear{{van Weeren}, {R{\"o}ttgering}, {Br{\"u}ggen}
  \& {Hoeft}}{{van Weeren} et~al.}{2010}]{2010Sci...330..347V}
{van Weeren} R.~J.,  {R{\"o}ttgering} H. J.~A.,  {Br{\"u}ggen} M.,   {Hoeft}
  M.,  2010, \mn@doi [Science] {10.1126/science.1194293}, \href
  {https://ui.adsabs.harvard.edu/abs/2010Sci...330..347V} {330, 347}

\bibitem[\protect\citeauthoryear{{van Weeren}, {Br{\"u}ggen}, {R{\"o}ttgering},
  {Hoeft}, {Nuza}  \& {Intema}}{{van Weeren}
  et~al.}{2011}]{2011A&A...533A..35V}
{van Weeren} R.~J.,  {Br{\"u}ggen} M.,  {R{\"o}ttgering} H.~J.~A.,  {Hoeft} M.,
   {Nuza} S.~E.,   {Intema} H.~T.,  2011, \mn@doi [\aap]
  {10.1051/0004-6361/201117149}, \href
  {https://ui.adsabs.harvard.edu/abs/2011A&A...533A..35V} {533, A35}

\bibitem[\protect\citeauthoryear{{van Weeren}, {R{\"o}ttgering}, {Intema},
  {Rudnick}, {Br{\"u}ggen}, {Hoeft}  \& {Oonk}}{{van Weeren}
  et~al.}{2012}]{2012A&A...546A.124V}
{van Weeren} R.~J.,  {R{\"o}ttgering} H.~J.~A.,  {Intema} H.~T.,  {Rudnick} L.,
   {Br{\"u}ggen} M.,  {Hoeft} M.,   {Oonk} J.~B.~R.,  2012, \mn@doi [\aap]
  {10.1051/0004-6361/201219000}, \href
  {https://ui.adsabs.harvard.edu/abs/2012A&A...546A.124V} {546, A124}

\bibitem[\protect\citeauthoryear{{van Weeren}, {de Gasperin}, {Akamatsu},
  {Br{\"u}ggen}, {Feretti}, {Kang}, {Stroe}  \& {Zandanel}}{{van Weeren}
  et~al.}{2019}]{2019SSRv..215...16V}
{van Weeren} R.~J.,  {de Gasperin} F.,  {Akamatsu} H.,  {Br{\"u}ggen} M.,
  {Feretti} L.,  {Kang} H.,  {Stroe} A.,   {Zandanel} F.,  2019, \mn@doi [\ssr]
  {10.1007/s11214-019-0584-z}, \href
  {https://ui.adsabs.harvard.edu/abs/2019SSRv..215...16V} {215, 16}

\bibitem[\protect\citeauthoryear{{van Weeren} et~al.,}{{van Weeren}
  et~al.}{2021}]{2021A&A...651A.115V}
{van Weeren} R.~J.,  et~al., 2021, \mn@doi [\aap]
  {10.1051/0004-6361/202039826}, \href
  {https://ui.adsabs.harvard.edu/abs/2021A&A...651A.115V} {651, A115}

\makeatother
\end{thebibliography}

 \clearpage


\appendix


\section{Information on individual clusters}
\label{AppB}
In this appendix, we provide relevant information on the individual clusters of the MGCLS sample presented in this study that show diffuse radio or candidate radio emission, as shown in Table~\ref{tab:diffuse}. We provide detailed information on earlier radio observations of the systems and their X--ray environment, as well as optical data available where relevant. 

Information is provided here on 56/62 MGCLS systems with detected diffuse radio structures, as six systems (Abell~85, Abell~3667, RXC~J520.7-1328, J0351.1-8212, J0352.4-7401, and J0631.3-5610) were presented in detail in K22. For these six systems, to avoid repetition, we only show the radio images in \ref{AppC} and refer the reader to K22 for more details.
We also provide general estimates of the in-band $\alpha_{908}^{1656}$ spectral index values for the MGCLS structures for which it was possible to extract information by visual inspection of the MGCLS in-band spectral index maps.  Caution is advised that these values are to be treated as tentative estimates of the radio structures' observed spectral index status. For the rest of the systems where no information on the in-band $\alpha_{908}^{1656}$ spectral index value is mentioned, the diffuse emission is too faint for a spectral index estimation as the sensitivity of the sub-bands used for the in-band spectral index mapping is reduced by a factor of $\sim$ 4 and falls below detection.

\subsection{Abell 13}

Abell 13 is an X--ray luminous galaxy cluster at $\textit{z} = 0.943$. It is an unusual diffuse cluster radio source, as it displays most of the properties of a radio relic. It is located significantly closer to the cluster's centre, forming part of the subclass known as a candidate relic. An X--ray spectral study of the region was conducted using the \textit{Chandra} X--ray Observatory \citep{Juettetal08} found a broad radial filament of cooler gas connecting the cluster core to the relic radio source at approximately 300~kpc, ruling out merger shocks as its origin. The \textit{Chandra} data, along with the classification system of \citet{Kempneretal04}, resulted in \citet{Juettetal08} classifying Abell~13 as an AGN relic system. However, the authors could not conclude whether these lower temperatures were the aftereffect of a recent merger or from a cooler gas being lifted from the cluster core.

\begin{figure*}
   \centering
   \includegraphics[width=0.496\textwidth]{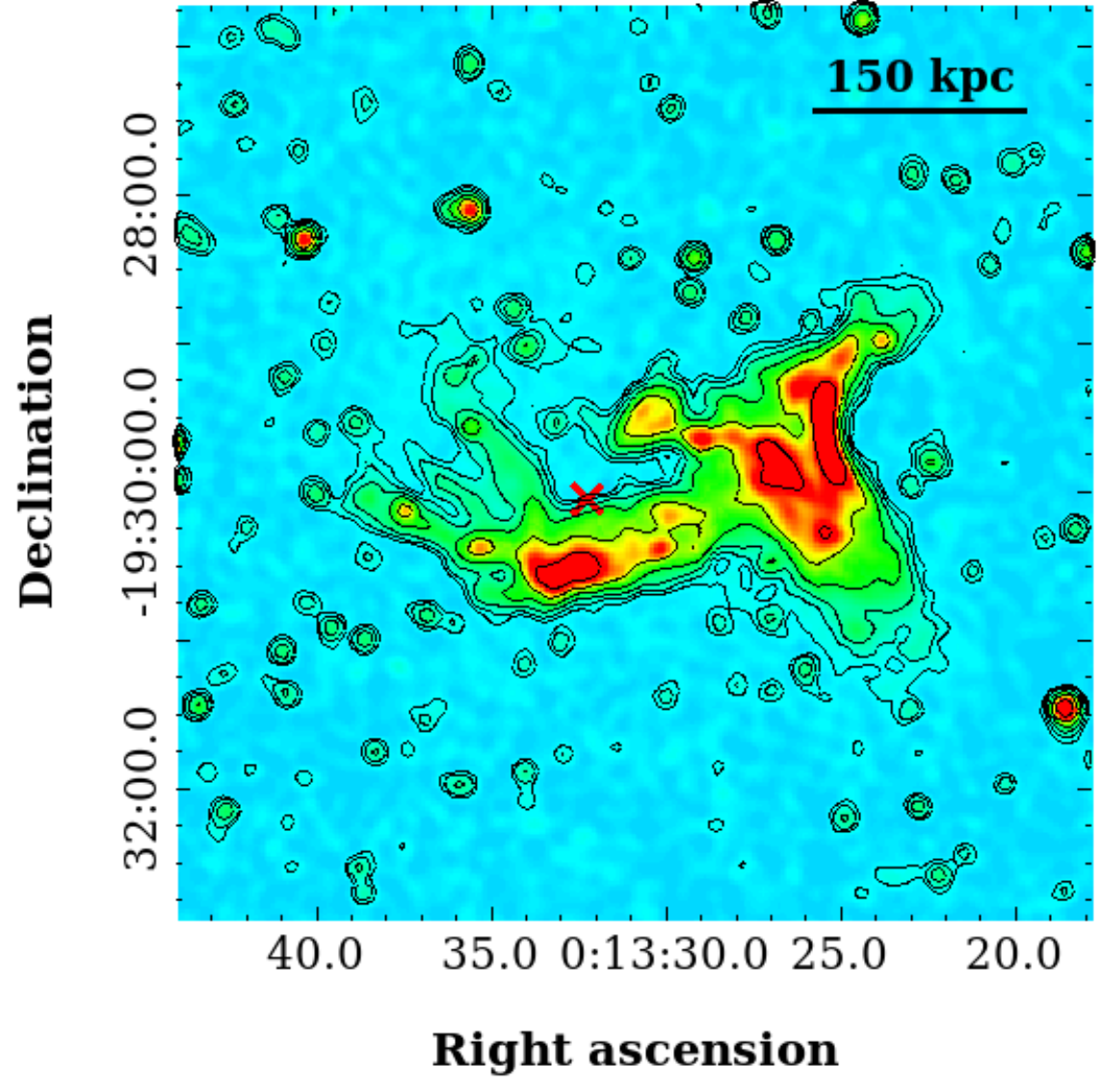}
    \includegraphics[width=0.5\textwidth]{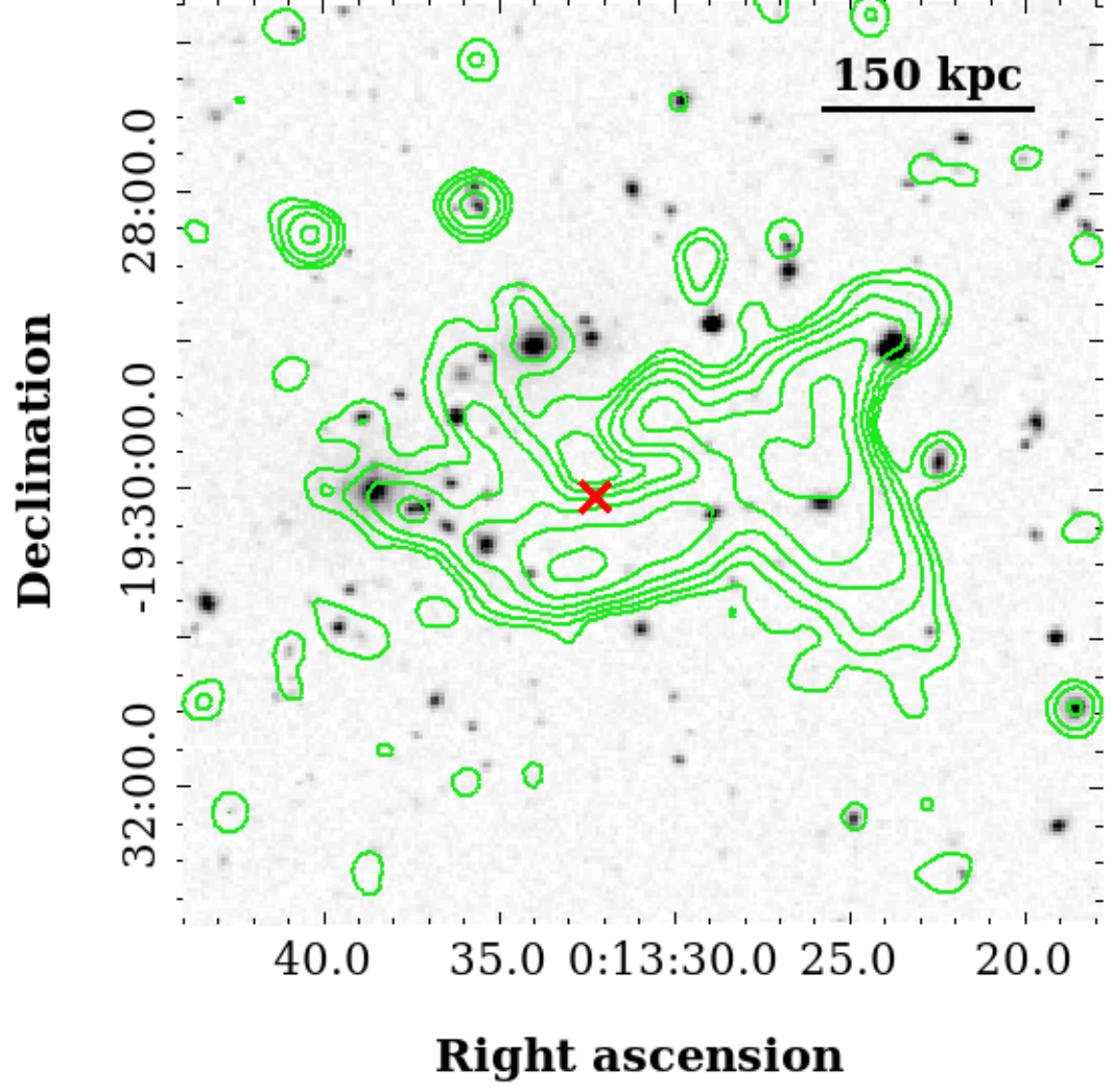}
   \caption{Abell~13 \textbf{Left:} Full-resolution (7.8\arcsec\,$\times$\,7.8\arcsec) 1.28~GHz MGCLS radio image with radio contours in black overlaid (1$\sigma$ = 3.5 $\mu$Jy beam$^{-1}$). \textbf{Right:} 1.28~GHz MGCLS low-resolution (15\arcsec\,$\times$\,15\arcsec) radio contours in green (1$\sigma$ = 12 $\mu$Jy beam$^{-1}$), overlaid on the r-band \textit{Digitized Sky Survey (DSS)} optical image. In both panels, the radio contours start at 3\,$\sigma$ and rise by a factor of 2. The physical scale at the cluster redshift is indicated on top right, and the red $\times$ indicates the NED cluster position. } 
   \label{fig:A13}%
\end{figure*}

The 1.28~GHz MeerKAT images of this cluster are presented in Figure~\ref{fig:A13}. Both full-resolution ($7.8^{\prime\prime} \times~7.8^{\prime\prime}$; rms = 3.5~$\mu$Jy~beam$^{-1}$) and low-resolution ($15^{\prime\prime} \times 15^{\prime\prime}$; rms = 12~$\mu$Jy~beam$^{-1}$) images show a similar `mushroom-like' structure that consists of two parts. The first part of the diffuse emission is elongated in the east--west direction and is embedded in several discrete radio sources that surround the cluster centre. The second part on the west side shows a structure that is perpendicular to the earlier structure and extends in the north--south direction. We report the detection of a west candidate radio relic in the system that is not associated with an optical counterpart, with a total size of $300\times320$~kpc$^2$, having a flux density of $S_{1.28~GHz}$ = 31~mJy 
 (excluding point radio sources) that corresponds to a radio power of $P_{1.28~GHz}$ = 6.4$\times$10$^{23}$ W~Hz$^{-1}$. 
 
The in-band spectral index for the candidate western relic appears to be steep, i.e., $\sim$ $-$2 to $-$2.4 along its length, which is consistent in both radio images. The eastern tail-like feature is slightly
steeper, i.e. $\sim$ $-$2.6, with the spectrum steepening eastward along the east$-$west filament, to values $\sim$ $-$3.5, with some flatter regions ($\sim$ $-$2.6 to $-$2.8) in the brighter spots. There is not enough signal-to-noise ratio in the filamentary structure east of the tail to obtain an indication of the in-band spectral index in this area.

\subsection{Abell 22}

\citet{2005MNRAS.357L..45P} used photometric
and spectroscopic data from the Las Campanas Observatory and the Anglo-Australian Telescope Rich Cluster Survey (LARCS) to study Abell~22. Their study showed that the redshift distribution of Abell 22 exhibits a foreground wall-like structure. They concluded that the wall is part of a large-scale filament that passes in front of Abell~22 and Abell~47, and connects to Abell~15, which is at a similar redshift to the filament and elongated toward it. They also suggested that this region is a supercluster candidate.

 \citet{1999NewA....4..141G} reported a detection of a diffuse radio source from the NVSS data. They were uncertain whether to label it as a relic or blended radio galaxies. \citet{2021PASA...38...10D} were also uncertain about the category of the source and labeled it as a candidate halo or a candidate relic. They were not able to subtract point sources, and hence did not report on the flux density or the size of the radio emission.

In both our 1.28~GHz MeerKAT radio images at full ($7.8^{\prime\prime} \times 7.8^{\prime\prime}$; rms = 3~$\mu$Jy~beam$^{-1}$) and low-resolution ($15^{\prime\prime} \times 15^{\prime\prime}$; $rms = 6$ $\mu$Jy~beam$^{-1}$), a faint radio halo at the centre of Abell~22 is detected in the north--south direction. The faint radio structure is embedded in several compact radio sources, making a clear classification more difficult. Hence, this source is classified as a candidate radio halo. The size of the candidate radio halo is found to be 280$\times$420 kpc$^2$ with a flux density at 1.28~GHz (after compact radio source removal) of 1.6~$\mu$Jy. We estimate the radio power at 1.28~GHz to be 8.1 $\times$ 10$^{22}$~W~Hz$^{-1}$.

\subsection{Abell 168}

Abell 168 is a low-mass merging galaxy cluster at $\textit{z} = 0.0450$. Twin relics were discovered when cross-referencing the MCXC catalogue \citep{2011A&A...534A.109P} with the GaLactic and Extragalactic All-sky MWA (GLEAM) survey at 200~MHz \citep{HurleyWalkeretal17}. The first relic (outer relic or relic B) is reported to be elongated and located around 900~kpc north of the cluster centre, while the second relic (inner relic or relic A) is ring-shaped and located near the inner edge of relic B, at a distance of around 600~kpc from the cluster centre.

\citet{Dwarakanath_2018} observed these radio relics using the Giant Meterwave Radio Telescope (GMRT) at 323~MHz and the Karl G. Jansky Very Large Array (VLA) at 600~MHz. Relic B was detected in all the aforementioned images, while relic A was detected in all except the VLA images. Relic B was found to have a radio power of 1.38 $\pm$0.14 $\times$ 10$^{23}$~W~Hz$^{-1}$ at 1.4~GHz and a spectral index, $\alpha$ = 1.1 $\pm$ 0.04. As the radio power of relic B and the galaxy cluster's mass agree well with the expected relation, Abell~168 is the lowest-mass cluster in which relics due to merger shocks have been detected. Relic A is reported to have a spectral index, $\alpha$ = 1.74 $\pm$ 0.29 and is proposed to have resulted from adiabatic compression by the same merger shocks that also produced relic B.

In both our 1.28~GHz MeerKAT radio images at full ($7.8^{\prime\prime} \times 7.8^{\prime\prime}$; rms = 5~$\mu$Jy~beam$^{-1}$) and low-resolution ($15^{\prime\prime} \times 15^{\prime\prime}$; $rms = 12$ $\mu$Jy~beam$^{-1}$) only the outer relic (so-called relic B) is detected at a distance of $\sim$ 1.1~Mpc from the NED cluster centre position with a size of 150$\times$650 kpc$^2$. We estimate the radio power at 1.28~GHz to be 1.3 $\times$ 10$^{23}$~W~Hz$^{-1}$ which is in agreement with \citet{Dwarakanath_2018}.

\subsection{Abell 209}

Abell\,209 is a radio-halo host cluster at $\textit{z} = 0.209$  \citep{2006AN....327..563G,2007A&A...463..937V} and is dominated by a central cD galaxy. \citet{2009A&A...507.1257G} reported a flux density at 1.4~GHz, of 16.9 $\pm$ 1.0 mJy, and a size of $\sim$ 1.2~Mpc. The MeerKAT image of this cluster at 1.28~GHz (see Figure~\ref{fig:A209}) shows a smaller size (600$\times$740 kpc$^2$), which is in better agreement with the size reported at 610~MHz by \citet{2007A&A...463..937V} and a lower flux density (S$_{\rm 1.28\,GHz}$ = 7~mJy), which reflects in a lower radio power. This is most likely due to the difference in the individual radio source subtraction in the estimation of the flux density.


\subsection{Abell 370}

Abell\,370 is a galaxy cluster at $\textit{z} = 0.3750$ that presents signs of strong gravitational lensing. It shows evidence of a recent merging event as each of the two primary substructures is centred on the brightest cluster
galaxies (BCG) in the north and south, respectively. The presence of X--ray surface brightness edges in the ICM also 
provides evidence of the unperturbed dynamical state of Abell\,370 and may be the result of shocks or a cold front.

\citet{2020A&A...636A...3X} observed Abell\,370 using both the VLA at 1.5~GHz (bandwidths of 1~GHz and 2~GHz), as well as the GMRT at 325~MHz (bandwidth of 33.3~MHz). X--ray observations were also made with the \textit{Chandra} X--ray Observatory using the ACIS-I and S-detectors. For the faint radio halo source discovered, the observed flux densities (for which compact radio source contributions were subtracted) at each frequency were found to be S$_{325\,MHz}$ = 20.0 $\pm$ 2.3~mJy, S$_{1.5\,GHz}$ = 3.7 $\pm$ 0.3~mJy and S$_{3\,GHz}$ $\leq$ 1.3~mJy. As the radio halo is undetectable at 3~GHz, the spectral index was obtained for frequencies between 325~MHz and 1.5~GHz, where $\alpha$ = $-$1.10 $\pm$ 0.09. The radio halo power at 1.4~GHz was estimated to be P$_{1.4\,GHz}$ = (2.00 $\pm$ 0.16) $\times$ 10$^{24}$ W~Hz$^{-1}$, which corresponds to the scaling relation.

The faint radio halo in the centre of Abell~370 is only detected in our low-resolution 1.28~GHz MeerKAT radio image ($15^{\prime\prime} \times 15^{\prime\prime}$; $rms = 10$ $\mu$Jy~beam$^{-1}$; see Figure~\ref{fig:A370}). Therefore, this source is classified as a candidate radio halo. The faint radio structure is embedded in several compact radio sources, making a clear classification more difficult. The size of the candidate radio halo is found to be 940$\times$960 kpc$^2$ with a flux density at 1.28~GHz (after compact radio source removal) of 3.5~mJy, which is very similar to the one calculated in \citet{2020A&A...636A...3X} at 1.4~GHz. We estimate the radio power at 1.28~GHz to be 1.7 $\times$ 10$^{24}$~W~Hz$^{-1}$, which is in good agreement with the one mentioned in the literature \citep{2020A&A...636A...3X}.

\subsection{Abell 521}

 Abell 521, the archetypal Ultra Steep Spectrum Radio Halo (USSRH) cluster \citep{2008Natur.455..944B}, is a rich and massive galaxy cluster (1.9$\times$10$^{15}$ M$_{\odot}$; \citealt{2008A&A...486..347G}) at $\textit{z} = 0.248$ that is characterized by a disturbed dynamic state, with multiple subclusters converging on its centre \citep{Ferrarietal03,2006A&A...446..417F}. Abell~521 belongs to a subclass of radio halos that is expected to form as a consequence of merging events between low-mass clusters (i.e., M$\le 4-6\times10^{14}$M$_{\odot}$ or between clusters with large mass ratios (i.e., M2/M1$\ge$1:4-1:5, \citealt{Cassanoetal16}). Recent HST weak lensing studies by \citet{Yoonetal20} have shown that the cluster is made up of three main distinct substructures, although their numerical simulations do not provide a conclusive picture of the merging scenario in the central region. Its giant radio halo (LLS $\sim$ 1.4~Mpc), first detected with GMRT data at 235~MHz \citep{2008Natur.455..944B}, exhibits an Ultra Steep Spectrum (USS) of $\alpha \sim -1.86\pm0.08$ between 330~MHz and 1.4~GHz (\citealt{2009ApJ...699.1288D}; VLA D-array and BnC array) with \citet{2013A&A...551A.141M} reporting a similar steepness for the halo with $\alpha = -1.81\pm0.02$. Abell~521 exhibits an arc-shaped radio relic located southeast of its centre, first detected at 610~MHz with the GMRT and at 1.4~GHz with the VLA (\citealt{Giacintuccietal06,2008A&A...486..347G,2006A&A...446..417F}, respectively) that is connected to the radio halo. The estimated integrated spectral index of the southeast relic is $\alpha$ = $-$1.45 $\pm$ 0.02 fitted using a single power law between 153~MHz and 5~GHz \citep{2013A&A...551A.141M}. The cluster is part of the MGCLS survey in which a new candidate counter-relic in the north--west direction was reported by K22.

In the X--rays, a detailed study based on {\it XMM--Newton} observations \citep{Bourdinetal13} confirms the presence of two interacting gas components in the cluster's central region, elongated in the northwest--southeast direction (in the plane of the sky) and separated by two cold fronts. In good coincidence with the relic, another shock front is detected at a large distance from the cluster's central region, in the southeast direction. The Mach number estimated for this shock (M=2.4$\pm$0.2) is consistent with the expectation from estimates made based on the relic spectrum \citep{2008A&A...486..347G}, suggesting (at least a partial) shock re-acceleration origin for the relic.

The MeerKAT image of Abell~521 (figure~\ref{fig:A521}) shows new features compared to the 1.4~GHz VLA image published in \citet{2009ApJ...699.1288D}. The observational parameters of the diffuse emission features are reported in Table~\ref{tab:diffuse}.
We detect a higher flux density in the radio halo, S$_{\rm MeerKAT}$=10.1 mJy compared to S$_{\rm VLA}=6.4\pm0.6$ mJy, over a very similar extent. We measure a size for the radio halo of $860\times1440$ kpc. Moreover, a second new candidate relic in the north--west direction of the cluster centre, which was reported for the first time in the K22 MGCLS data, is located in the opposite direction from the main relic. This second northwest candidate counter-relic, which is very faint (we measure S$_{\rm 1.28GHz}$=0.5~mJy), has a smaller extent than the main relic (125$\times$330 kpc) and is located at the northwest edge of the X--ray emission. The northwest counter-relic appears in good coincidence with the northwest--southeast plane of the sky direction of the two interacting gas components confirmed by {\it XMM--Newton} observations in \citet{Bourdinetal13}. On the other hand, the radio halo appears to be encompassed by the two relics extending from the main relic almost out to the newly detected candidate counter-relic. 

More recently, Abell~521 was examined using uGMRT band 3 (300-500~MHz), and 4 data (550-750~MHz) \citep{Santraetal24} in which the new northwest counter-relic discovered by MGCLS is confirmed, also revealing a faint extension of the southeast relic and also mentioning complex substructures in it. From the 1.28~GHz MeerKAT data, we do not confirm any further extension on the southeast relic.  Santra et al. report for the central radio halo emission an LLS of 1.3 Mpc at 400~MHz and 1 Mpc at 650~MHz, respectively. Their spatially resolved 400$-$650~MHz spectral index map revealed that the radio halo spectral index at local scales differs from the average value, also reporting a radial steepening of the average spectral index towards the edge of the halo. Examining the MeerKAT in-band radio spectral index distribution, we confirm that the halo structure has a relatively steep integrated spectral index of $\alpha_{908}^{1656}\sim-1.7$ and confirm the average spectral index steepening towards the outer parts of the radio halo.
 
The complexity of the radio emission both at the cluster centre and periphery requires further detailed investigation, complementing the existing radio data with high-frequency and sensitivity radio continuum and polarization observations to characterise the magnetic field properties at the shock area and examine the radio halo spectral characteristics over a broader range of frequencies. However, the new MeerKAT Abell~521 observations have already revealed a double relic in this less energetic merger subclass of USSRH systems that is largely unexplored due to the lack of sensitive low and high-frequency observations.

\subsection{Abell 545}

Abell 545b is an X--ray luminous, cD galaxy cluster at $\textit{z} = 0.1540$ and is dubbed the second richest cluster in the Abell catalogue. It is also host to a red, diffuse, elongated feature near the cluster centre, named the ``star pile" by \citet{Struble88}, who discovered it during the Palomar Observatory Sky Survey (POSS). The presence of a giant radio halo was first suggested by \citet{1999NewA....4..141G} in the NRAO VLA Sky Survey. The total flux density reported by \citet{1999NewA....4..141G} is 41~mJy. Abell~545 was later studied by \citet{2003A&A...400..465B} using images obtained with the Very Large Array (VLA) at 1365~MHz and 1435~MHz at C and D configurations, resulting in a resolution of $42'' \times 30''$ and a noise level of 45~$\mu$Jy~beam$^{-1}$. Applying discrete source subtraction for the first time for this radio halo source, the flux density at 1.4~GHz was found to be 23 $\pm$ 1~mJy with the equivalent radio power at 1.4~GHz P$_{1.4~GHz}= $2.52 $\times$ 10$^{24}$~W~Hz$^{-1}$ and the lower limit of the spectral index $\alpha$ between 1365~MHz and 1435~MHz being $\alpha$ $\geq$ 1.4.

The flux density derived with the MeerKAT at 1.28~GHz in this study (S$_{1.28~GHz}=$14.1~mJy; see Figure~\ref{fig:A545}) is lower than the earlier mentioned flux density estimate by \citet{2003A&A...400..465B}, most likely due to an even more accurate subtraction of discrete radio sources that are visible at the higher sensitivity and resolution of MeerKAT.

We report that the X--ray structure of the Abell~545 cluster shows an elongated structure within 0.5~Mpc, which hints at the dynamical youth of the system \citep{Buote01}. Due to the asymmetry of the star pile and its alignment with the merging direction, Abell~545b is the first case that shows a strong connection between cD galaxy formation and cluster mergers \citep{Barrenaetal11}.

\subsection{Abell 2645}
\citet{Dwarakanathetal99} and \citet{Owenetal99} used optical data from the KPNO 0.9~meter telescope and radio data from the VLA to study the radio galaxy population and star formation rates in Abell~2645 (redshift $\textit{z}=0.25$). \citet{Owenetal99} reports that Abell~2645 presents very few radio sources (only four based on complete observations to the same limiting magnitude and radio flux density) and a low star formation rate. They concluded that this is most likely due to the cluster's relaxed dynamic state. 
\par
The more recent study by \citet{2021A&A...647A..51C} used radio data from \citet{2013ApJ...777..141C} in order to derive upper limits and reported that the potential halo radio power upper limit is 0.59$\times$10$^{24}$ W~Hz$^{-1}$. GMRT observations by \citet{Kale2015EGRHS} resulted in an image with too high noise levels; therefore, they were unable to derive upper limits.

The 1.28~GHz MeerKAT image presented in Figure~\ref{fig:A2645} shows some hints of extended filamentary radio emission at the cluster centre only in the low-resolution image of 15$^{\prime\prime}$. This filamentary structure appears to connect a series of point radio sources that are embedded in the cluster centre, hence making this classification unclear and only hinting at the potential existence of a radio halo. 

The candidate radio halo presents a flux density of S$_{1.28~GHz}$ = 0.8~mJy and a radio power at 1.28~GHz of 1.5 $\times$ 10$^{23}$~W~Hz$^{-1}$ which is in good agreement and within the upper limit set by \citet{2021A&A...647A..51C}. 

\subsection{Abell 2667}

Abell~2667 is a cluster at $\textit{z}=0.23$ that hosts a radio-loud BCG and a radio mini-halo that was first detected by \citet{Giacintuccietal19} using the GMRT. The flux density of the mini-halo extracted from the GMRT 610~MHz and 1.15~GHz is 15.3$\pm$1.2~mJy and 8.3$\pm$0.7 mJy, respectively, with the noise level at 610~MHz and 1.15~GHz being 50~$\mu$Jy~beam$^{-1}$ and 35~$\mu$Jy~beam$^{-1}$, respectively.
\par
\cite{1998MNRAS.301..328R} used ROSAT High Resolution Imager
(HRI) to study the X--ray properties of Abell~2667. This cluster was the most luminous in their sample, with an X--ray luminosity of 9.36$\pm$0.21$\times$10$^{44}$ ergs~s$^{-1}$. They also report that the cluster has strong evidence of cooling flows. \cite{2006A&A...460..381C} used optical data from the Hubble Space Telescope (HST) and X--ray archival data from \textit{XMM-Newton} and \textit{Chandra} to study this cluster. From the optical data, with evidence of tidal stripping from infalling galaxies, they suggest an ongoing accretion process onto the central cluster galaxy, whereas, from their X--ray analysis, they concluded that Abell~2667 is a cool-core cluster (see also \citealt{Cavagnoloetal09}). In a more recent optical study, \citet{2019MNRAS.487.5593I} used HST and MUSE data to study the properties of the BCG, suggesting viable scenarios of chaotic cold accretion and merging with a gas-rich disc galaxy to explain the state of the BCG.
\par
The diffuse radio emission detected by MeerKAT is reported in Figure~\ref{fig:A2667}. The morphology of the radio structure at full-resolution is consistent with the 610~MHz image in \citet{Giacintuccietal19}, while the source appears to be more extended than reported at 1.15 GHz in the same paper. The diffuse radio emission detected in the MeerKAT low-resolution image (figure~\ref{fig:A2667} right panel), which we classify
as a radio halo, with a size calculated to be $380\times530~kpc^2$, has a flux density of S$_{1.28~GHz}$ = 6.8~mJy at a noise level of 5~$\mu$Jy~beam$^{-1}$ and a radio power at 1.28~GHz of 1.0 $\times$ 10$^{24}$~W~Hz$^{-1}$ which is in good agreement with the extrapolated radio power by \citet{Giacintuccietal19} using a spectral index of $-$1. 

The in-band spectral index can be estimated only in the region of the radio halo in the immediate surroundings of the radio loud cluster BCG and has an annular distribution steepening radially in all directions.
The spectral index values range from $\sim-$0.5 at the BCG location to $\sim-$1.6 at the edge of the detection. In the southwest direction, the spectral index signal is detected up to the embedded bright point radio source, with values of $\sim-$1.5 to $-$1.6.


\subsection{Abell 2744}

Abell~2744 is a galaxy cluster at $\textit{z} = 0.307$ that is undergoing two merger events: a major one near the cluster centre and a minor one in the northwest region, where a subcluster is also located. It is host to a central radio halo and a radio relic located northeast of the cluster centre, both of which were first detected in the NVSS by \citet{1999NewA....4..141G}. 

A VLA observation at 1.4~GHz conducted by \citet{2001A&A...376..803G} was the first to confirm these two radio structures and reported that the flux densities of the halo and northeast relic at this frequency were S$_{\rm 1.4~GHz}$ = 57.1~mJy and S$_{\rm 1.4~GHz}$ = 18.2~mJy, respectively making the halo one of the most powerful known (P$_{\rm 1.4~GHz}$ = 2.6 $\times$ 10$^{25}$~W~Hz$^{-1}$). They also found that their average spectral indices are $\alpha$ = $-$1.4 and $\alpha$ $\geq$ $-$2, respectively. The cluster was also observed with the GMRT at 325~MHz by \citet{Venturietal13}, which reported a flux density S$_{\rm 325~MHz} = 323\pm26$~mJy and S$_{\rm 325~MHz}=122\pm10$~mJy for the halo and relic, respectively. 

More recently, new Jansky VLA images observed in the L- (1–2~GHz) and S-bands (2$-$4~GHz) detected three new radio relic sources in Abell~2744 \citep{Pearce2017}. The flux densities of the halo in the L- and S-bands were found to be S$_{\rm 1.5~GHz}$ = 45.1 $\pm$ 2.3 mJy and S$_{\rm 3~GHz}$ = 17 $\pm$ 1.0 mJy respectively. For the northeast relic they report S$_{\rm 1.5~GHz}$ = 11.74 $\pm$ 0.62 mJy and S$_{\rm 3~GHz}$ = 4.78 $\pm$ 0.14 mJy respectively. Calculation of the largest physical linear size (LLS) in these structures from the very sensitive VLA images shows an LLS of 2.1~Mpc and 1.5~Mpc for the halo and the relic, respectively. They identified three more filamentary features, fainter and smaller than the main relic, which clearly show the complexity of the diffuse emission in this cluster.  
\par

On the X--ray side, a temperature jump found in the same region as the northeast radio relic is suggestive of an X--ray shock \citep{Ibarakietal14} whereas a more recent study by \citet{Eckertetal16} using \textit{XMM-Newton} and \textit{Suzaku} data, confirmed the detection of a surface brightness and temperature jump at the eastern edge of the northeast radio relic which corresponds to a weak shock.

The MeerKAT images of Abell~2744, reported in Figure~\ref{fig:A2744}, show a good agreement with the 1.5~GHz images presented in \citet{Pearce2017}. The largest linear sizes of the halo (2.04~Mpc) and the northeast radio relic (1.57~Mpc) are consistent, and the three new filamentary features reported are also detected. Our MeerKAT 1.28~GHz images show that the southeast relic (R2 in \citealt{Pearce2017}) is connected to the radio halo. The feature R3 (northwest filamentary extension at the edge of the radio halo) is detected with higher significance (at comparable resolution) in full and low-resolution MeerKAT images at 3$\sigma$ of 10.5 $\mu$Jy beam$^{-1}$ and 33 $\mu$Jy beam$^{-1}$ respectively. The flux density values of all the reported features are consistent within errors with the \citet{Pearce2017} values if extrapolated using the provided average spectral index values. 

For this cluster, both the main relic and a consistent part of the radio halo are detected in the in-band spectral index image at 15$''$. The radio halo shows a radial steepening, with values ranging from $\sim-$1.4 to $-$1.5 in the central part to $\sim-$2.2 in the north--west direction towards the filament. The main relic (NE) shows a spectral index structure parallel to the largest extent, with mild steepening towards the cluster centre. In particular, the flattest ridge presents $\sim-$1.2, and steepens to values $\sim-$1.8 to $-$2. Only the peak of the southeast relic is detected in the spectral index image, and the $\alpha$ value is $\sim-$1.7.

\subsection{Abell 2751} 

Abell~2751 is a cluster at redshift $\textit{z}=0.107$ neighbouring cluster APMCC~039 ($\textit{z} = 0.082$) falling in the footprint of the Dark Energy Camera Legacy Survey (DECaLS; \citealt{Decals}) and the All-sky Widefield Infrared Survey Explorer (All-WISE; \citealt{Cutrietal12}), and for this reason, its field has been recently used for the search and study of high redshift galaxies \citep{2021Galax...9...89K}. From its galaxy distribution, \citet{2021PASA...38...10D} suggests that the clusters are interacting or are part of the same system.
\par
\citet{2021PASA...38...10D} has also reported the detection of a new radio relic at 168~MHz to the east of Abell~2751 (in between Abell~2751 and APMCC~039) using the MWA's Epoch of Reionization 0-hour field observations. The EoR0 field has an overall noise rms level of $\gtrsim$ 2.3 mJy~beam$^{-1}$. The flux density of the relic was calculated to be 323$\pm$62~mJy and using the NVSS data, they derived an integrated spectral index of $\alpha_{168}^{1400}$ = 1.27$\pm$0.11. They also report a head-tail galaxy or candidate relic on the outskirts of the APMCC~039 cluster. We note that the Abell~2751 and APMCC~039 angular separation is 17.7 arcmins. 

The 1.28~GHz MeerKAT images presented in Figure~\ref{fig:A2751} show for the first time a hint of a slightly extended ($50\times90$~kpc$^2$), very faint (0.4~mJy) radio emission about 600~kpc southeast of the Abell~2751 cluster centre and its central AGN radio source. This structure is better viewed at the low-resolution image of 15$^{\prime\prime}$ and based on X--ray ROSAT imaging inspection and the absence of an optical counterpart, this faint radio emission is potentially a Phoenix. However, this classification is not certain and only hints at the potential existence of a radio Phoenix. Hence, this source is classified as a candidate radio Phoenix, as seen in Table~\ref{tab:diffuse}. We also note that the relic reported in \citet{2021PASA...38...10D} is most likely a wide-angle tail (WAT) radio source based on the inspection of the MeerKAT and optical images, hence it was not reported as a diffuse radio emission structure for this system. The diffuse radio emission is too faint for a spectral index estimation.

\subsection{Abell 2811}

Abell 2811 is a cluster of galaxies at redshift $\textit{z}= 0.108$ suggested to be undergoing merging from X--ray observations using \textit{XMM-Newton} \citep{2009ApJ...691.1787S}. \citet{2009ApJ...691.1787S} report that Abell~2811 is a significant outlier in the L$_{X}-$T relation and appears to be underluminous for its temperature, and also find that the BCG in the cluster is offset, which suggests merger activity. 
\par


Using MWA observations at 168~MHz, \citet{2021PASA...38...10D} discovered extended diffuse emission, which they classified as a candidate radio halo or a mini-halo. They measured the flux density of the halo to be 81$\pm$17~mJy. They further extrapolated the flux density at 1.4~GHz and derived an integrated spectral index $\alpha^{1400}_{168}$ $\leq$ $-$1.5$\pm$0.1, indicating that this is a USSRH.

The diffuse emission of Abell~2811 is visible only in the 15$^{\prime\prime}$ image (figure~\ref{fig:A2811}), suggesting a very low surface brightness structure ($\sim$ 0.2~$\mu$Jy~arcsec$^{-2}$ over an area with a radius of 150~kpc) that is resolved out in the full $7.8^{\prime\prime}$ resolution image. We calculate a flux density of S$_{\rm 1.28GHz}$ = 3.9~mJy for the diffuse radio component. The diffuse radio structure is too weak for an in-band spectral index analysis; thus, we cannot infer any results with certainty using the integrated flux density values that we have extracted. However, although at different resolutions and areas covered by the MWA flux density at 168~MHz, we also suggest that this is a steep radio halo with $\alpha^{1280}_{168}$ $\leq$ $-$1.49.


Comparison between our MeerKAT 1.28~GHz image and those reported in \citet{2021PASA...38...10D} clearly shows the contribution and blending of point radio sources in the extended features detected with MWA north and southeast of the cluster centre. The central BCG is a weak radio source, and this would be consistent with a mini--halo type of emission, as the extent of the diffuse emission that we detect at 1.28~GHz with the 15$^{\prime\prime}$ image at hand is smaller in size (300$\times$320 kpc) than MWA and the low-resolution image used in K22 ($\sim620~kpc$). However, the diffuse cluster source is classified as a radio halo, taking into account knowledge from existing lower-resolution and low-frequency data and the fact that we detected part of the diffuse emission at 15$^{\prime\prime}$ with MeerKAT. 



\subsection{Abell 2813}
\citet{2021A&A...649A..42C} used HST data to study the properties of the BCGs in Abell~2813. They found that this cluster hosts two BCGs with similar magnitude and size.
\par
\citet{2021A&A...647A..50C} observed Abell~2813 using JVLA at C configuration at 1.5~GHz. Extended radio emission was not detected in this field with their data reduction, resulting in an image with an rms of 0.035 mJy~beam$^{-1}$. \citet{2021A&A...647A..50C} derived an upper flux density limit of 5~mJy, corresponding to a radio power of 14$\times$10$^{23}$ W~Hz$^{-1}$ for the detection of diffuse emission in Abell~2813.
\\
MeerKAT has detected extended emission at 1.28~GHz, which is new, and we classify it as a candidate halo (see Figure~\ref{fig:A2813}). The central part of the cluster hosts a strong compact source, which strongly affects image quality. However, our 15$^{\prime\prime}$ image (rms 12~$\mu$Jy~beam$^{-1}$) clearly shows the presence of diffuse emission, whose largest linear size is $\sim$ 800~kpc, and which we classify as a candidate halo. The total flux density of the source (S$_{\rm 1.28~GHz}$ = 3.7 mJy) is consistent with the upper limit set by \citet{2021A&A...647A..50C}.

\subsection{Abell 2895}
The Abell 2895 cluster lies at $\textit{z}=0.228$, and its dynamical status has been classified as a merger system in \citet{Cucitietal15}.
\par
In the most recent radio study, \citet{2021A&A...647A..50C} observed Abell~2895 using JVLA at C configuration at 1.5~GHz. Their data reduction resulted in an image with an rms of 0.040 mJy~beam$^{-1}$. No extended emission was detected in this field. \citet{2021A&A...647A..50C} derived an upper flux density limit of 2.5~mJy, which corresponds to a radio power of 9.7$\times$10$^{23}$ W~Hz$^{-1}$.
\par
The 1.28~GHz MeerKAT images in Figure~\ref{fig:A2895} are deeper than the VLA image in \citet{2021A&A...647A..50C}. Only the low-resolution image of 15$^{\prime\prime}$ with an rms of 10~$\mu$Jy~beam$^{-1}$ shows some residual emission around the strong radio galaxy that is associated with the brightest cluster galaxy \citep{Kale2015EGRHS}. We identified an extended ($100\times170$~kpc$^2$), faint (3.4~mJy) radio emission south of the tailed radio galaxy located 1.5 arcmin east of the cluster centre. The in-band spectral index distribution of the tailed radio galaxy (at 15$''$ resolution) clearly shows the two inner jets with $\alpha$ being $\sim-$0.5 at the head and gradually steepening along the two jets to $-$0.86. Then, the two jets merge into a single tail, and the spectral index gradually increases to $-$2. 

The extended, faint radio emission that appears as a continuation of the radio tail is unlikely to be a blend of radio sources. Given the lack of an optical counterpart for this feature, we propose that it is a candidate Phoenix. The spectral index in what we refer to as the candidate Phoenix is steep, ranging from $-$2.5 to $-$2.7. The full-resolution spectral index map is consistent with the 15$''$ one and shows the flattest spectrum at the tip of the head ($\sim-$0.3), then $\alpha$ gradually steepens along the two tails up to $\sim-$0.7 and continues steepening once the two tails merge, to a value consistent with the 15$''$ image.


\subsection{Abell 3365}
Abell~3365 presents a complex galaxy and ICM distribution. The galaxy distribution has two large concentrations along the east-west axis and another smaller concentration at the far west. The ICM is also elongated along the east-west axis.

\citet{2021A&A...646A..95U} studied the cluster using X--ray data from \textit{XMM-Newton} / EPIC. They report shocks at regions coinciding with radio relics having Mach numbers of  3.5$\pm$0.8 and 3.8$\pm$0.5. \citet{2021A&A...646A..95U} also detect a cold front at a distance of $\sim$ 1.6 arcmin from the X--ray peak. They calculated the dynamical age of the main merger along the east--west direction to be $\sim$ 0.6~Gyr. 

\citet{2011A&A...533A..35V} observed Abell~3365 using the WSRT and VLA at 1.4 GHz. They reported the discovery of a double relic system in the cluster. The emission was detected in both the WSRT and VLA images. The flux density of the eastern relic was reported to be 42.4$\pm$3.0 mJy and 42.6$\pm$2.6~mJy for WSRT and VLA observations, respectively. The flux density of the western relic was reported to be 5.4$\pm$0.5 mJy and 5.3$\pm$0.5 mJy for WSRT and VLA observations, respectively.

The 1.28~GHz MeerKAT images in Figure~\ref{fig:A3365} are deeper than the \citet{2011A&A...533A..35V} images and we detect both aforementioned radio relics but also no radio halo emission (the low-resolution image of 15$^{\prime\prime}$ has an rms of 15~$\mu$Jy~beam$^{-1}$). We find that the MeerKAT 1.28~GHz radio flux density for the northeast relic is $S_{1.28~GHz}$ = 40.9~mJy and for the northwest one $S_{1.28~GHz}$ = 6.2~mJy, which are consistent within errors with the \citet{2011A&A...533A..35V} study and correspond to a radio power of 8.3$\times$10$^{23}$ W~Hz$^{-1}$ and 1.3$\times$10$^{23}$ W~Hz$^{-1}$ respectively.

The in-band spectral index of the northeast relic has a fairly homogeneous ridge with $\alpha$ values of $\sim-$1.2 to $-$1.4 at the eastern edge, presenting some patches which reflect the peaks in the total intensity and then steepening towards the cluster centre, with $\alpha$ values of $\sim-$1.8 just outside the flatter ridge that reach up to $\sim-$2.5. On the other hand, the northwest relic presents a very different spectral index distribution, which follows the peaks of the total intensity, presenting $\alpha$ values of $\sim-$1.0 to $-$1.2, with marginal steepening in the direction of the minor axis.

\subsection{Abell 3376}
Abell 3376, at redshift $\textit{z}=0.046$, is a rich cluster and has been extensively studied at multiple wavelengths. Optical studies showed that the cluster is merging, while X--ray simulations predicted that the merger occurred $\sim$ 0.5~Gyr ago. \textit{Suzaku} X--ray studies by \citet{2018A&A...618A..74U} confirmed the presence of a strong western shock (Mach number = 2.8$\pm$0.4) and a weaker eastern shock (Mach number = 1.5$\pm$0.1). They also estimated that the dynamical age of the shock front is $\sim$ 0.6~Gyr after core passage.

\citet{2006Sci...314..791B} were the first to report that the cluster hosts a double relic. They observed the cluster using the VLA at CnB and DnC configurations. Since
then, there have been multiple radio follow-ups on this cluster \citep{2012MNRAS.426.1204K,2015MNRAS.451.4207G,2021Natur.593...47C}. \citet{2012MNRAS.426.1204K} used VLA data to study the spectral properties and the polarization of the relics. They reported that the eastern and western relics have polarisation percentages of 30$\%$ and 5$-$20$\%$, respectively. They calculated the respective integrated spectral indices to be $-$1.82$\pm$0.06 and $-$1.70$\pm$0.06. In the most recent radio study, \citet{2015MNRAS.451.4207G} used the MWA to study the Abell~3376 cluster at low frequencies (88, 118, 154, 188, and 215~MHz). Even at these frequencies, they were not able to detect a radio halo in the cluster. They calculated the spectral indices to be $-$1.17$\pm$0.06 and $-$1.37$\pm$0.08 for the western and eastern relics, respectively. They further calculated the flux density in the central region as a representation of the halo upper limits. At 215~MHz, the flux density in the central region was measured to be 17$\pm$11~mJy.

The 1.28~GHz MeerKAT images presented in Figure~\ref{fig:A3376} show both radio relics but no radio halo emission. The eastern relic presents a size of $530\times1520$~kpc$^2$ and a flux density of 156~mJy which corresponds to a radio power of  7.2$\times$10$^{23}$ W~Hz$^{-1}$ whereas the western one has a size of $310\times1000$~kpc$^2$ and a flux density of 139~mJy which corresponds to a radio power of  6.4$\times$10$^{23}$ W~Hz$^{-1}$. 

Both relics are detected in the in-band spectral index map at 15$''$. The northeast relic is detected in the brightest southern part of the arc and shows the typical
flatter ridge away from the cluster centre, with $\alpha$ values in the range $\sim-$1.5 to $-$1.7. The remaining filamentary features detected here are much steeper, with
$\alpha$ values reaching $\sim-$3.4. The western relic is overall flatter, with patches in the range $\sim-$1.2 to $-$1.5 in the few regions with higher surface brightness and values steeper than $-$2 anywhere else.

\subsection{Abell 3558}

Abell~3558 is amongst the clusters at the core of the Shapley supercluster and is often referenced as the centre of the supercluster. It is the most massive system in the Shapley Supercluster. Multi-wavelength studies have shown that the cluster has complex morphology and cannot be distinctly categorised as a relaxed or merging system. 
X--ray studies carried out with \textit{ROSAT}, \textit{XMM-
Newton} and \textit{Chandra}, have confirmed that the cluster is asymmetric and that the X--ray brightness peaks at the core and radially declines \citep{1997ApJ...474...84M,2003ApJ...596..170A,2007A&A...463..839R}. 
These studies also highlighted that although the cluster has an inner `cool-core' it cannot be classified as a dynamically relaxed cluster since there are traces of in-falling subgroups.

The cluster has been observed in the radio band with VLA, ATCA and GMRT over a range of frequencies ranging from 240~MHz to 2.3~GHz \citep{2000MNRAS.314..594V,Giacintuccietal04} in search for diffuse emission at its centre. Only more recent high-sensitivity observations carried out with ASKAP at 883~MHz, and MeerKAT have led to the detection of diffuse emission at the cluster centre \citep{Venturietal22}.
The 1.28~GHz radio data used in \citet{Venturietal22} are also drawn from the MGCLS products with the flux density estimation performed in \citet{Venturietal22} using the filtering method of \citet{2002PASP..114..427R} and the subtraction of
individual galaxies from the $u - v$ plane with further convolution of the residual emission \citep{2007A&A...463..937V}. The flux density reported for the Abell~3558 diffuse emission is S$_{887MHz}$=30$\pm$3 mJy and S$_{1283MHz}$=13$\pm$1 mJy, which leads to a very steep spectrum, of $\alpha_{887}^{1283}$= –2.3$\pm$0.4. It is asymmetric in size, with the largest linear size of $\sim$ 400~kpc. Whether this is a mini-halo or a radio halo of a small size remains uncertain. The MeerKAT data in \citet{Venturietal22} were convolved with a restoring beam of 40.9$^{\prime\prime}$ $\times$ 40.4$^{\prime\prime}$ that is favorable for the detection of extended emission, resulting in an rms of $\sim$ 35 $\mu$Jy~beam$^{-1}$. More recently, the study of \citet{Trehaevenetal25} using new MeerKAT UHF-Band and uGMRT Band-3 observations, along with existing MeerKAT L-band and ASKAP 887~MHz data, perform a detailed analysis of the central diffuse emission in Abell~3558. \citet{Trehaevenetal25} find a greater extension of the diffuse radio emission by detecting a 100~kpc ultra-steep spectrum northern extension beyond the innermost X--ray cold front, filling a local X--ray cavity caused by large-scale gas sloshing, increasing the total size of the diffuse emission to 550~kpc. A strip of flatter spectrum is seen to coincide with the cold front, suggesting a direct link between the dynamics of the thermal component and the energetics of the non-thermal component of the ICM. Furthermore, \citet{Trehaevenetal25} shows from the radio-X--ray point-to-point correlation a sub-linear trend that steepens with increasing frequency and radial distance, suggesting that this is a radio mini-halo powered by turbulent (re)-acceleration induced by sloshing motions caused by the galaxy group SC~1327-312 perturbing the Abell~3558 cluster in a minor merger with mass ratio 1:5. Lastly, \citet{Trehaevenetal25} finds a flatter integrated spectrum than \citet{Venturietal22} of $\alpha_{400}^{1569}$= –1.2$\pm$0.1.

We note that using the low-resolution MGCLS image of 15$^{\prime\prime}$ at an rms of 7~$\mu$Jy~beam$^{-1}$, we have a flux density estimation of the diffuse emission in Abell~3558 of 11.2~mJy and a largest linear size of $\sim$ 300~kpc. Based on the MGCLS data reported here, the detected diffuse radio emission for this cluster is classified as a radio halo. We note that the flux density estimation is in agreement within the errors with the analysis of \citet{Venturietal22} (using the filtering method that provides slightly higher values); in which a conservative flux density uncertainty of the order of 20\% is assumed. In addition, our lower flux density and largest linear size reported here, can also be attributed to the different angular resolutions of the convolved images that provide these slight differences.


\subsection{Abell 3562}

Abell 3562 is part of the Shapley Supercluster and is located about 2 degrees east of Abell~3558. Two small groups (SC1327--312 and SC1329--313) are situated between Abell~3558 and Abell~3562, and the whole region is host to group accretion and merger \citep{Bardellietal02,Merluzzietal15}. Recently, a bridge of radio emission connecting Abell~3562 and SC1329--313 was discovered by MeerKAT, providing further evidence of the complex merger in this region of the sky \citep{Venturietal22}.
Abell~3562 itself is an intriguing cluster that has been studied at multiple wavelengths. X--ray studies confirmed that the cluster is dynamically
disturbed and shows traces of interactions with the galaxy group SC 1329–313 \citep{Finoguenovetal04}. 

In the radio band, it hosts a radio halo with a very steep spectrum ($\alpha\,\sim\,2$) in the frequency range from 240~MHz to 1.4~GHz \citep{2003A&A...402..913V}. The largest linear size of the radio halo is reported to be $\sim$ 500 kpc. Embedded in the radio halo is the head-tail radio galaxy J1333--3141, which is most likely the supplier of relativistic electrons re-accelerated by the sloshing motion at the cluster centre induced by the interaction between Abell~3562 and SC1329--313 \citep{2005A&A...440..867G}.
More recently, the radio halo was studied using pointed observations with MeerKAT \citep{Venturietal22,Giacintuccietal22}, which revealed a 
straight and thin filament of radio emission departing from the ending part of J1333--3141. The flux density detected with MeerKAT is S$_{\rm 1.28~GHz}=26\pm2$ mJy, consistent with previous spectral studies.

\citet{Venturietal22} discussed the nature of the extended source located 20$^{\prime\prime}$ south--west of the cluster centre. Regardless of its morphological classification, the origin of this source is most likely related to the complex ongoing merger in the Shapley Supercluster core \citep{Venturietal22}.  

The 1.28~GHz MeerKAT images presented in Figure~\ref{fig:A3562} here show the radio halo emission, which has a size of $300\times600$~kpc$^2$ and almost identical flux density to the \citet{Venturietal22} analysis. The south-western morphology has a size of $190\times450$~kpc$^2$ and a flux density of 17.9~mJy, which corresponds to a radio power of  9.8$\times$10$^{22}$ W~Hz$^{-1}$, which is again similar to that reported in \citet{Venturietal22}. Based on the detailed recent analysis by \citet{Venturietal22}, the distribution of the in-band spectral index supports the idea that the visible compact radio source and the extended southwestern emission are not connected, with no visible jets or other features linking the diffuse emission to the compact component. 



\subsection{Abell 4038}

Abell~4038 hosts a radio relic source, which was initially reported as a steep spectrum source in \citet{1983AuJPh..36..101S} with a polarisation percentage of 4$\%$. \citet{2012ApJ...744...46K} further studied the properties of the relic using GMRT observations at 150, 240, and 606~MHz. They detected a larger extent of $\sim$ 200~kpc and reported detecting another steep spectrum source in the field. They produced spectral index maps using the multi-frequency data, in which the relic was found to be consistent with the model of a radio galaxy lobe that is powered by adiabatic compression due to a passing shock wave.


\cite{2018MNRAS.480.5352K} observed Abell~4038 using the upgraded GMRT at band 3 (300$-$500~MHz) and 5 (1050$-$1450~MHz) to study the relic's complex morphological and spectral features. The rms noise at 1.26~GHz was 0.03 mJy~beam$^{-1}$ with the largest extent of the relic in the north$-$south direction being 102 $\times$ 58 kpc$^2$ and the calculated integrated spectral index being 2.3$\pm$0.2.

The 1.28~GHz MeerKAT images presented in Figure~\ref{fig:A4038} here reveal a faint radio halo emission in which the head-tailed radio source is embedded in the centre (mentioned earlier as a relic). The diffuse emission appears to be more extended at 15$^{\prime\prime}$ resolution, having a total size of $130\times160$~kpc$^2$. We classify this diffuse emission as a candidate mini-halo that appears to be almost circular in projection and also extends in the opposite direction of the radio source tail orientation. The average in-band spectral index of the candidate mini-halo is estimated to be $\leq -3$, with the seed electrons for this structure most likely being provided by the decaying head-tail radio source. 

The in-band spectral index distribution along the radio tail is traced properly by the spectral index map, which presents a very steep $\alpha$ value of $\sim-$2.5 at the head of the tailed source, and a steepening along the direction of the tail up to $\sim-$3.2. Images at both angular resolutions are consistent with these values and this trend.

\subsection{Abell S295}
Abell S295, also known as SPT-CL~J0245-5302 and ACT-CL J0245-5302, is a cluster at redshift $\textit{z} = 0.3$ that is going through a merging process \citep{2010ApJ...723.1523M}. \citet{2018ApJ...863..145C} used optical HST data to probe strong lensing in the Abell~S295 field. They reported that Abell~S295 shows a bimodal mass distribution, indicating that the cluster might be in a non-relaxed dynamical state and reinforcing the binary nature of the merger.

More recent X--ray studies using \textit{Chandra} observations \citep{PascutHughes19} confirm the merger scenario and show a 6~keV cool gas associated with the core of the secondary cluster, while the central gas temperature is typical of a non-cool core cluster and similar to the cluster mean value. \citet{PascutHughes19} reports the merger to be off-axis in the southeast--northwest direction with the secondary cluster heading to the apocentre.

The 1.28~GHz MeerKAT images presented in Figure~\ref{fig:AS295} reveal in both full (rms $\sim$ 3$~\mu$Jy~beam$^{-1}$) and low-resolution (rms $\sim$ 8$~\mu$Jy~beam$^{-1}$) the presence of a new radio halo that has a largest linear extent of $\sim$ 1~Mpc. We find that the MeerKAT 1.28~GHz radio flux density for the radio halo, removing the contribution from compact sources, is $S_{1.28~GHz}$ = 7.2~mJy, which corresponds to a radio power of 2.0$\times$10$^{24}$ W~Hz$^{-1}$.

\subsection{Abell S1063}

Abell~S1063 is a Hubble Frontier Field cluster known as SPT-CL~J2248-4431; RXC~J2248.7-4431; PLCKESZ~G349.46-59.94.
\citet{2020A&A...636A...3X} was the first to report the detection of a radio halo in this field from their VLA and GMRT observations. \citet{2021MNRAS.505..480R} used \textit{Chandra} and GMRT (325~MHz) observations to study Abell~S1063.

Their X--ray study indicates that the Abell~S1063 cluster is one of the hottest in the nearby Universe and presents a disturbed morphology. \citet{2021MNRAS.505..480R} reports the detection of large-scale `excess brightness' in the residual \textit{Chandra} X--ray surface brightness map, which extends $\sim$ 2.7~Mpc northeast from the cluster core (most likely due to the stripped gas from the merging counterpart) and also provides the first observational confirmation of the merging axis proposed in an earlier simulation by \citet{Gomezetal12}.

Their GMRT observations derived a radio halo flux density of 62.0$\pm$6.3~mJy and an integrated spectral index by a single power-law fit of $\alpha$=$-$1.43$\pm$0.13. From the derived halo radio power, it was found that this halo is located well below the L$_X$ - P$_{1.4}$ correlation line, concluding that the radio halo might be either of hadronic origin or, if in its `switch on' stage, it will originate from the turbulent re-acceleration mechanism (supported by the steep spectral index).


The 1.28~GHz MeerKAT images presented in Figure~\ref{fig:AS1063} confirm in both full (rms $\sim$ 3$~\mu$Jy~beam$^{-1}$) and low-resolution (rms = 7~$\mu$Jy~beam$^{-1}$) the presence of a radio halo that spans a size of $\sim$~1.2~Mpc. The BCG presents a point radio AGN emission with several more bright radio sources embedded in the radio halo along with an extended radio source (most likely a Head-Tail galaxy; \citealt{2021PASA...38...10D,2020A&A...636A...3X}) in the north--east direction. We find that the MeerKAT 1.28~GHz radio flux density for the radio halo, removing the contribution from compact sources, is $S_{1.28~GHz}$ = 8.8~mJy, which corresponds to a radio power of 3.5$\times$10$^{24}$ W~Hz$^{-1}$. This value is in good agreement with the calculated radio flux density at 325~MHz by \citet{2021MNRAS.505..480R} if extrapolated using the integrated spectral index of $\alpha=-1.43$.

\subsection{Abell S1121}

Abell~S1121 is a galaxy cluster at $\textit{z}=0.190$ \citep{2009AJ....137.4450M} and is part of the SPT-SZ 2500~deg$^2$ survey (\citealt{Bleemetal15}; SPT-CL~J2325-4111) included in several X--ray cluster studies as part of a sample without any detailed reported analysis. 

\citet{2021PASA...38...10D} reported the detection of diffuse emission in Abell~S1121 near the cluster centre, presenting also counterpart emission in SUMSS at 843~MHz. Examination of archival \textit{Chandra} data by \citet{2021PASA...38...10D} shows a core-located X--ray emitting plasma that is slightly extended, suggesting a complex dynamical state. 

From their MWA 168~MHz data \citet{2021PASA...38...10D} suggest the detection of a radio halo, or a relic with a largest linear extent of $\sim$ 1.1~Mpc using the redshift of 0.3580 \citep{Liuetal15}. They recovered a total flux density for the emission of 80$\pm$13~mJy, with an integrated spectral index of $\alpha^{843~MHz}_{168~MHz}=-1.2\pm0.2$. 

The 1.28~GHz MeerKAT images presented in Figure~\ref{fig:AS1121} do not confirm in both full and low-resolution (rms =~12~$\mu$Jy~beam$^{-1}$) the presence of a radio halo, similar to the TGSS data, where no significant emission was detected. We report here the detection of a candidate relic southeast of the cluster centre, with a total size of $180\times410$~kpc$^2$ and a flux density of $S_{1.28~GHz}$ = 3.3~mJy, corresponding to a radio power of 3.2$\times$10$^{24}$ W~Hz$^{-1}$. 

The in-band spectral index map of the candidate southeast relic shows a steepening trend along the major axis, with the southeastern edge having an $\alpha$ value of $\sim-$1.1, which steepens to $\sim-$1.6. 

\subsection{Bullet}

The Bullet cluster is a massive galaxy cluster ($\sim11.4\times10^{14}$ M$_{\odot}$) at a redshift of 0.296 \citep{Tuckeretal98} that was first detected using the ROSAT satellite \citep{Voges92}, the Advanced Satellite for Cosmology and Astrophysics \citep{Tanakaetal94}, and the Einstein IPC instrument \citep{Giacconietal79}. The Bullet cluster is known to host both a giant radio halo and a relic \citep{Liang_2000,Shimwelletal15}. Several multi-wavelength studies have been conducted on this system and have found that the cluster has recently undergone a major merger, approximately perpendicular to the line of sight \citep{Markevitch02,Markevitch06}.

In more detail, \citet{Barrenaetal02} used spectroscopic data from the ESO New Technology Telescope \citep{TarenghiWilson89} to study the dynamical state of the cluster, showing that the Bullet cluster contains a subcluster that is offset from the central cluster to the west by 0.7 Mpc, concluding that a major merger has occurred within the past 0.15~Gyr. \textit{Chandra} X--ray studies by \citet{Markevitch02} and \citet{Markevitch06} showed that the merger was the ideal case of a bow shock merger with the bow shock Mach number being M$_{\chi}$ = 3.0 $\pm$ 0.4 and the cluster velocity $\sim$ 3000$-$4000 km/s. Their observations also indicate that the subcluster traversed the central cluster 0.1 to 0.2 Gyr ago and is currently undergoing its final stage of dynamic gas instabilities. They studied the temperature profile across the shock and found that the electrons are fast-heated at the bow shock front. 

\citet{Liang_2000} were the first to report the detection of a radio halo hosted by the Bullet cluster. They observed the cluster using the Australia Telescope Compact Array (ATCA) at 1.3, 2.4, 4.9, 5.9, and 8.8 GHz. Several follow-up studies of radio emission have been conducted using deeper, higher frequency ATCA observations \citep{Shimwelletal15,Shimwell15B,Maluetal16}. The first detection of a luminous ‘toothbrush’ relic was reported by \citet{Shimwell15B}. They measured an integrated flux density of 77.8 $\pm$ 3.1 and 4.8 $\pm$ 0.6 mJy. More recently, a MeerKAT L-Band study by \citet{Sikhosanaetal23} measured the size of the recovered halo to be 1.6 $\times$ 1.3~Mpc$^2$ with the
largest linear size of the relic being $\approx$980~kpc. In the same study, a new decrement feature was detected in the southern part of the radio halo emission, appearing as a region of lower surface brightness than its surroundings. \citet{Sikhosanaetal23} derived a higher Mach number for the relic shock region of M$_{R}$ = 4.6 $\pm$ 0.9 that can be attributed to the relic orientation. \citet{Botteon23} adopting a filtering technique in identifying surface brightness edges in radio images showed for the Bullet Cluster that there is agreement between the locations of the detected four radio discontinuities and the existing X-ray edges, which indicates an interplay between thermal and non-thermal components most likely due to the existence of frozen ICM magnetic field, which enforces the advection of cosmic rays while conveyed by thermal gas motions.

The 1.28~GHz MeerKAT images presented in Figure~\ref{fig:Bullet} show in both full (rms =~4~$\mu$Jy~beam$^{-1}$) and low-resolution (rms =~10~$\mu$Jy~beam$^{-1}$) the presence of the east radio relic that is attached to the radio halo. The largest linear sizes of the radio halo are 1.55 $\times$ 2.05~Mpc$^2$, and for the east radio relic are 0.33 $\times$ 1.01~Mpc$^2$. The flux density of the radio halo is calculated to be $S_{1.28~GHz}$ $\approx$ 95~mJy, which corresponds to a radio power of 2.6$\times$10$^{25}$ W~Hz$^{-1}$ whereas for the east relic, the flux density is calculated to be $S_{1.28~GHz}$ $\approx$ 81~mJy, which corresponds to a radio power of 2.2$\times$10$^{25}$ W~Hz$^{-1}$.

The in-band spectral index map detects both the relic and the brightest part of the halo. The head-tail south of the relic has a spectrum with $\alpha$$\sim-$0.7 in the head, which steepens to $\sim-$2 along the tail westward. The in-band spectral index in the relic shows some structure along the major axis, with features having $\alpha$$\sim-$1.4 and others steepening to $\sim-$2. The halo shows a radial steepening, from $\alpha$ $\sim-$2 on the central ridge to $\alpha$$\sim-$3 at the boundary of the detection of the spectral index.

\subsection{El Gordo}

El Gordo (ACT-CL/SPTCL J0102-4915; \citealt{Menantauetal12}) is the first most massive cluster (M$_{200}$ $\sim$ 2.16$\times$10$^{15}$ M$_\odot$; \citealt{Jeeetal14}) known beyond the redshift of 0.6 that hosts a radio halo and a double radio relic system. It was discovered via the SZ effect (SZE; \citealt{1972CoASP...4..173S}) with the Atacama Cosmology Telescope (ACT; \citealt{Marriageetal11}), and its redshift is confirmed at $\textit{z}=0.87$ \citep{Menanteauetal10}. El Gordo presents an average temperature (kT) of 14.5$\pm$0.1 keV with a \textit{Chandra} X--ray surface brightness discontinuity, revealing the first signs of a merger in El Gordo  \citep{Menantauetal12} in the shape of two tails \citep{MolnarBroadhurst15}. 

\citet{Lindneretal14} using GMRT 610~MHz and 2.1~GHz ATCA observations, revealed that El Gordo hosts a bright northwest radio relic and found evidence for the presence of a faint radio halo and another relic (counter). Simulation studies suggest that El Gordo either underwent on and off-axis mergers with high initial relative velocities \citep{MolnarBroadhurst15,Zhangetal15} or was initially a highly disturbed cluster \citep{Donnert14}. Its northwest relic exhibits a highly polarised flux fraction ($\sim$ 33\% at 2.1~GHz) that reinforces the scenario of Fermi acceleration at a shock during the cluster merger. \citet{Botteonetal16} using deep \textit{Chandra} X--ray observations revealed the presence of a greater than 3 Mach shock at the location of the northwest relic, suggesting shock acceleration of electrons as a viable origin and also confirming the presence of the radio halo with 610 MHz GMRT data.

Recently, \citet{Huetal24} used the Synchrotron Intensity Gradient (SIG) and MeerKAT observations at 1.28 GHz to represent the system's magnetic field. In the latest study on this system, \citet{Kaleetal25} reported observations of El Gordo with the uGMRT in the frequency range of 300 - 1450~MHz, detecting the relics known from previous studies (northwest Relic, southeast Relic and east Relic) and adding the detection of an extension of the eastern Relic to the north. They also obtained a spectral index map between 366 and 672~MHz, with a median spectral index of the radio halo reported as $-$1.3. Using the radio power of the halo, Kale et al. measured the turbulent velocity dispersion 1112 \kms, implying a 3D turbulent Mach number of $\sim$ 0.6.

The MeerKAT image of El Gordo (figure~\ref{fig:ElGordo}) shows in detail all the previously known diffuse radio sources with the radio halo at the 15$^{''}$ image encompassing the area between the northwest, southeast and east relic. The observational parameters of the diffuse emission features are reported in Table~\ref{tab:diffuse}.
We detect a flux density in the radio halo of S$_{\rm MeerKAT}$=12.1 mJy over a size of $1.25\times1.44$ Mpc. The northwest counter-relic presents a flux density of S$_{\rm MeerKAT}$=8.2~mJy that corresponds to a radio power of 2.3$\times$10$^{25}$ W~Hz$^{-1}$ extending over 700 kpc in size. The detected southeast relic flux density is S$_{\rm MeerKAT}$=4.1~mJy extending over a size of 460~kpc, and, the east faint relic is clearly detected to have a flux density of only S$_{\rm MeerKAT}$=0.7~mJy extending over a size of 420~kpc north of the southeast one.

Examining the MeerKAT in-band radio spectral index distribution, we confirm that the halo structure has a relatively steep integrated spectral index of $\alpha_{908}^{1656}\sim-1.7$ and confirm the average spectral index steepening towards the outer parts of the radio halo. The two relics and the ridge of the radio halo are detected in the spectral index map at 15$''$. The spectral index in the southeast relic follows the trend of the compact sources embedded in it and has a value of $\alpha$ $\sim-$1 at the bottom right (western edge), steepening to $\alpha$ $\sim-$1.5 eastward. The northwestern relic has a more uniform distribution with a value of  $\sim-$1.4 to $-$1.5. However, the ridge of the radio halo is much steeper, with values around $\sim-$3. In the full-resolution spectral index map, no $\alpha$ value of the radio ridge could be measured. The northwestern relic has values consistent with the 15$''$ image, with basically no steepening toward the cluster centre.

\subsection{PLCK G200.9--28.2}
PLCK G200.9--28.2 is a cluster at a redshift of 0.22 with an X--ray luminosity at $0.1-2.4$~keV of L$_{500}\sim10^{44}$ erg s$^{-1}$ that presents a significant offset between the X--ray and the SZ position \citep{Planckcollab11a,Planckcollab11b}. The cluster is also part of the ACT-CL cluster catalogue \citep{Hilton_2020}.

\citet{Kaleetal13} identified an extended source as a candidate radio relic, first seen in the NVSS.  \citet{2017MNRAS.472..940K} using deeper GMRT observations at 235 and 610~MHz and JVLA data at 1500~MHz report the discovery of a single radio relic in the galaxy cluster that has an arc-like morphology and is located $\sim$ 0.9~Mpc from the X--ray brightness peak in the cluster. The integrated spectral index of the relic was estimated to be 1.21$\pm$0.15 with the GMRT 235-610~MHz spectral index map showing a steepening towards the inner edge of the relic, which is consistent with a cluster merger shock. Assuming diffusive shock acceleration, \citet{2017MNRAS.472..940K} calculate a radio Mach number of 3.3 $\pm$ 1.8 for the shock. 


\citet{2017MNRAS.472..940K} also re-analyzed archival \textit{XMM-Newton} data to study the nature of the radio relic and the host cluster. The analysis of the X--ray data revealed that PLCK G200.9--28.2 is a bimodal system with northern and southern subclusters that are in a state of merging, which suggests a physical origin for the offset.


The 1.28~GHz MeerKAT images presented in Figure~\ref{fig:PLCKG200} confirm in both full and low-resolution (rms =~12 $\mu$Jy~beam$^{-1}$) the presence of the southwest radio relic (the same as the one reported in \citealt{2017MNRAS.472..940K}) with a total size of $300\times1150$~kpc$^2$ and a flux density of $S_{1.28~GHz}$ = 23.7~mJy. The MeerKAT data further reveal the detection of a new radio relic on the east side with a total size of $160\times1230$~kpc$^2$ having a flux density of $S_{1.28~GHz}$ = 3.6~mJy which corresponds to a radio power of 4.9$\times$10$^{23}$ W~Hz$^{-1}$ and another new faint candidate relic northwest of the cluster centre with a total size at 1.28~GHz of $110\times250$~kpc$^2$. These latest findings confirm the dynamic state of the merging of PLCK~G200.9--28.2. We note that the 15$''$ total intensity image shows residuals between the two relics, which might suggest the presence of a very faint radio halo.

The main southwest relic and the brightest part of the newly detected eastern relic are detected in the in-band 15$''$ spectral index map. The southwest relic shows a steepening towards the cluster centre. The outer ridge has a reasonably flat spectrum with $\alpha$ values in the range of $\sim-$0.8 to $-$1.0, which steepens to $\sim-$2.5 going inwards. The central ridge of the new east relic has steeper values, ranging from approximately $-$1.9 to $-$2.3.

In a more recent study, \citet{Paletal25} present new observations of this cluster using data from the uGMRT at 400 and 650~MHz, the MGCLS at 1.28~GHz and X--rays from \textit{Chandra}. The southwest, eastern and candidate northwest relics are detected, with the respective LLSs for these relics calculated by the uGMRT data as 1.53~Mpc, 1.12~Mpc and 340~kpc. All three radio relics are polarized at 1.28~GHz with \citet{Paletal25} also reporting the detection of a radio ring in their 650~MHz uGMRT image that could be an Odd Radio Circle candidate at the core of the cluster between two point radio sources, possibly formed by the ram pressure produced in the ICM by the cluster merger. Assuming diffusive shock acceleration, \citet{Paletal25} calculate a radio shock Mach number of 3.1 $\pm$ 0.8 and 2.8 $\pm$ 0.9 for the southwest and eastern relics, respectively. The Chandra X-ray surface brightness map reveals, as in \citealt{2017MNRAS.472..940K}, two prominent subclusters. \citet{Paletal25} finds no shock presence toward both radio relic locations and that the orientation of the relics is not perpendicular to the possible merger axis, suggesting that the cluster is most likely an off-axis merger.

\subsection{MACS J0257.6--2209}

MACS J0257.6--2209 or Abell~402 is a massive cluster of galaxies at a redshift of 0.322. X--ray analysis of the cluster shows a centrally peaked X--ray emission, which presents, however, a high central entropy (k$_{0}=156\pm25 
~keV~cm^{-2}$; \citealt{Cavagnoloetal09,2017ApJ...841...71G}). The cluster was previously classified as relaxed by \citet{2012MNRAS.420.2120M} using X--ray and optical morphology; however, evidence of extreme ram pressure stripping \citep{2014ApJ...781L..40E} and the presence of a non-cool core suggest that the cluster is undergoing some kind of dynamical activity. 

On the radio side, using two different archival GMRT observations at 330~MHz, \citet{2017ApJ...841...71G} revealed evidence of diffuse emission at the cluster centre and classified it as a candidate radio halo. \citet{2020A&A...640A.108G} using short JVLA observations at the L-band of the statistically complete sample of very X--ray luminous clusters from the Massive Cluster Survey (MACS; \citealt{Ebelingetal10}) examined MACS J0257.6--2209 cluster in CnD configuration ($37^{\prime\prime} \times 37^{\prime\prime}$; rms $\sim$ 0.048~mJy~beam$^{-1}$) and C-array using the longest baselines only ($15^{\prime\prime} \times 37^{\prime\prime}$; rms $\sim$ 0.040~mJy~beam$^{-1}$). \citet{2020A&A...640A.108G} reports a complex radio morphology from the radio images, with faint diffuse radio emission being present at the cluster centre in the low-resolution image, along with many discrete radio sources that were visible in the high-resolution image. After subtracting the contribution of the discrete sources from the total flux density (43.5$\pm$2.2~mJy), \citet{2020A&A...640A.108G} reports a flux density of the radio halo at 1.5~GHz of 22$\pm$3~mJy, corresponding to a radio power of $6.7\times 10^{24}$~W~Hz$^{-1}$.

The cluster is also included in the \citet{2021A&A...647A..50C} sample where archival GMRT 330~MHz high ($13.4^{\prime\prime} \times 8.6^{\prime\prime}$; rms $\sim$~0.1~mJy~beam$^{-1}$) and low-resolution ($57^{\prime\prime} \times 42^{\prime\prime}$; rms $\sim$~1~mJy~beam$^{-1}$) data show the central diffuse emission which is co-spatial with the core of the cluster. \citet{2021A&A...647A..50C} classified the central emission as a candidate radio mini-halo based on its central surface brightness profile and its LLS in the east--west direction being $\sim$ 370~kpc. The flux density for this source is reported to be $\sim$~12~mJy by \citet{2021A&A...647A..50C}.


The 1.28~GHz MeerKAT images of this cluster are presented in Figure~\ref{fig:MACSJ0257}. The full-resolution ($7.8^{\prime\prime} \times 7.8^{\prime\prime}$; rms =~6~$\mu$Jy~beam$^{-1}$) image clearly shows only the presence of the discrete radio sources around the cluster centre without any sign of detection of diffuse radio emission. The presence of a faint radio halo is only scarcely seen in the MeerKAT low-resolution image (rms =~12~$\mu$Jy~beam$^{-1}$) with a total size of $430\times500$~kpc$^2$ confined by the presence of discrete radio sources, having a flux density of just $S_{1.28~GHz}$ = 0.4~mJy. With the current MeerKAT data, we classify the diffuse emission as a candidate halo because we are restricted by the high resolution, which does not reveal the total detection of the radio halo. In addition, the presence of residuals in the central region of the cluster also suggests the existence of a very faint and low surface brightness radio halo. However, we also report here with the MeerKAT data the detection of a new candidate radio relic on the south-west side with a total size of $190\times390$~kpc$^2$ having a flux density of $S_{1.28~GHz}$ = 1.8~mJy which corresponds to a radio power of 5.7$\times$10$^{23}$ W~Hz$^{-1}$. From the \citet{2021A&A...647A..50C} image, the candidate relic source is seen to be located at the boundary of the brightest part of the X-ray emission; therefore, even if tentative, we include the detection of a new candidate radio relic here.

The in-band spectral index map at 15$''$ shows a flat spectrum ($\alpha \sim-0.6$) at the location of the point radio source to the southeast end of the proposed southwest relic,
which steepens along the major axis to reach a value of $\sim-$1.3 at the location of the second peak in the total intensity image.

\subsection{MACS J0417.5--1154}

MACS J0417.5--1154 or ACT-CL J0417.5-1154 \citep{Hilton_2020} is a hot ($\sim$ 11~keV), X--ray luminous (at $0.3-5$~keV L$_{500}\sim29.1\pm0.5\times 10^{44}$ erg~s$^{-1}$; \citealt{2011A&A...534A.109P}), and massive cluster (M$_{500}\sim22.1\pm3.9\times 10^{14}$ M$_\odot$; \citealt{2011A&A...534A.109P}) at a redshift of $\textit{z}=0.443$ that is part of a merging system as a combination of a central
cluster plus a subcluster. In a recent study by \citet{2019MNRAS.482.5093P}, MACS~J0417.5--1154 is classified as a dissociative merger by its mass distribution, as one of the merging clusters has lost its gas content after its pericentric passage. The SZ decrement position is offset from the peak of the X--ray emission because of the merger dynamics, and this system also presents indications of strong lensing \citep{2019MNRAS.483.3082J}. MACS~J0417.5--1154 was initially part of the ROSAT All-Sky Survey Bright Source Catalogue (RASS-BSC; \citealt{Vogesetal99}) and later included in the Reionization Lensing Cluster Survey (RELICS; \citealt{2019ApJ...884...12O}).

MACS~J0417.5--1154 was first reported to host a $\sim$ 1~Mpc scale radio halo by \citet{2011JApA...32..529D} based on GMRT observations at 230 and 610~MHz. In a combined detailed \textit{Chandra} X--ray and optical study by \citet{2012MNRAS.420.2120M}, a distinct comet-like morphology of the hot gas was found, with an X--ray bright and compact core coincident with the optical core of one of the two merging clusters. The possibility of a shock front in this cluster was also reported by \citet{2012MNRAS.420.2120M}. In a later study, \citet{2017MNRAS.464.2752P} using GMRT (235/610~MHz) and JVLA (1.5~GHz) radio observations along with X--rays from \textit{Chandra}, reported that the diffuse radio emission of the cluster is co-located with the hot emitting X--ray gas. At all radio frequencies, an extended comet-like (elliptical) diffuse radio emission is visible in the northwest direction, which also suggests that it is a disturbed and merging cluster. At 610~MHz, the largest linear size of the halo is in agreement with \citet{2011JApA...32..529D}, having a flux density of 77.0$\pm$8.0~mJy, 54.0$\pm$5.5~mJy and 10.6$\pm$1.0~mJy at 235, 610 and 1500~MHz, respectively. 

\citet{2020A&A...640A.108G}, analyzed the same archival data used by \citet{2017MNRAS.464.2752P} in the L-band using C- and D-array data for the halo source and B-array data to subtract unrelated sources. \citet{2020A&A...640A.108G} reports that the radio image obtained by combining C- and D-configuration data show discrete sources at the cluster centre that are embedded in the diffuse radio emission, in very good agreement with the diffuse X--ray emission. These discrete sources are clearly shown in their B-array data, which represent the dominant galaxy of the main cluster, a head-tail galaxy near the centre of the cluster, and two other head-tail galaxies related to the second sub-cluster. Using the D-configuration to obtain an image of only diffuse emission, \citet{2020A&A...640A.108G} reports that the rms noise level of the diffuse radio halo in the centre of the cluster is rms $\sim$ 0.03 mJy (1.5~GHz; 15$^{\prime\prime}$) with a total flux density (after subtraction of discrete sources) being 33.7~mJy with maximum linear size at L-band of 1.2~Mpc. The detection of a halo source agrees with the findings of \citet{2017MNRAS.464.2752P}, but appears more extended, overlapping with the same region as the number density distribution of the red cluster galaxies and the mass distribution shown by \citet{2019MNRAS.482.5093P}.


In addition, \citet{2019MNRAS.482.5093P} confirm a surface brightness edge to the east-–south direction at a distance of $\sim$ 250~kpc from the centre, suggestive of a sloshing-induced cold front in this system, as was also reported by \citet{Botteonetal18}, which discovered a cold front, but without shocks.

The 1.28~GHz MeerKAT images presented in Figure~\ref{fig:MACSJ0417} confirm in both full and low-resolution (rms =~7~$\mu$Jy~beam$^{-1}$) the presence of the reported elongated radio halo in the southeast--northwest direction. The largest extent of the halo appears to be more extended than in earlier studies, at $\sim$ 1420~Mpc with a flux density of $S_{1.28~GHz}$ = 16.2~mJy, which corresponds to a radio power of 1.1$\times$10$^{25}$ W~Hz$^{-1}$. The MeerKAT data further reveal the detection of two new candidate radio relics on the north and northwest sides with total sizes of $220\times780$~kpc$^2$ and $210\times490$~kpc$^2$ and flux densities of 1.6~mJy and 1.0~mJy, respectively. These new candidate relics are most likely products of the system's dynamic state during a merger.

The in-band spectral index map at 15$''$ detects the brightest part of the halo with this central cometary ridge having a spectral index $\alpha$ value of $\sim-$1.8, with a radial steepening, reaching values on the order of $\sim-$2.6 to $-$2.8. The northwest part of the
cometary tail steepens to values of $\sim-$2.3. The proposed candidate relics present a patchy spectral index distribution, with values ranging from approximately $-$1 to $-$1.9.

\subsection{RXC J0510.7--0801}

RXC~J0510.7–0801 or MCXC J0510.7-0801 is a massive cluster (M$=7.4\times10^{14}$~M$_\odot$; \citealt{Planckcollab14}) at a redshift of $\textit{z} = 0.220$ that is particularly luminous in the X--rays (L$_X$ = $12.83\times10^{44}$~erg~s$^{-1}$; \citealt{Bohringeretal04}) but does not possess a cool-core \citep{2017ApJ...841...71G}. Although \citet{Kale2015EGRHS} included this cluster in their GMRT sample analysis, the poor data quality did not allow for the inspection or detection of any diffuse radio emission. 

More recently, \citet{2021MNRAS.507.4487G} analysed archival old GMRT hardware correlator 325~MHz data using the SPAM processing software and detected the presence of a radio halo emission for the first time in RXC~J0510-0801 (rms $\sim$ 0.26~mJy~beam$^{-1}$; 14.6$^{\prime\prime}$$\times$8.5$^{\prime\prime}$) without reporting any further details. \textit{XMM-Newton} data show a highly disturbed X--ray morphology which suggests the presence of a typical giant radio halo. In addition, \citet{2021MNRAS.507.4487G} find that the detected diffuse emission agrees well with the observed correlation between cluster luminosity and radio halo power.

The 1.28~GHz MeerKAT images of this cluster are presented in Figure~\ref{fig:RXCJ0510}. The low-resolution ($15^{\prime\prime} \times 15^{\prime\prime}$; rms =~8~$\mu$Jy~beam$^{-1}$) image reveals the presence of an extended and elongated diffuse emission, while only the brightest part of the emission is detected at full-resolution. Both radio images show the presence of an embedded AGN; hence, we classify this diffuse emission as a candidate radio halo. The total size of the candidate radio halo is $430\times1040$~kpc$^2$, having a flux density of $S_{1.28~GHz}$ = 5.8~mJy, which corresponds to a radio power of 7.9$\times$10$^{23}$ W~Hz$^{-1}$. We note that the diffuse radio emission in this system follows the elongation of the X--ray data, but does not fully overlap in the northwest edge (i.e., not overlap as much as in the previous case of MACS~J0417.5--1154).


\subsection{RXC J1314.4--2515}
It is a dynamically disturbed bimodal cluster at redshift $\textit{z}=0.244$. The bimodality is clear both from the redshift distribution of the galaxies and from the X--ray emission that is elongated in the east-west direction, suggesting an ongoing merger activity along this axis \citep{Valtchanovetal02}. It is one of the first clusters in which a radio halo and two relics were studied with VLA and GMRT \citep{2005A&A...444..157F,2007A&A...463..937V,Venturietal13}. More recently, the cluster has been studied in detail with the JVLA in multiple configurations in the frequency range 1-4 GHz \citep{2019MNRAS.489.3905S}. The very sensitive VLA images clearly show that the radio halo extends to the western relic, which extends in a north$-$south direction for $\sim$ 970 kpc. This relic shows a substructure with a bifurcation at the northern edge. The eastern relic is most likely an example of an AGN-relic connection, as the southern tip of the relic is associated with a cluster member, narrow-angle-tailed radio galaxy.

\textit{XMM-Newton} observations show that the western relic is coincident with an M-shaped shock front, having a Mach number of $\sim$ 2. More recently, \citet{Golovich2019} estimated a significant line-of-sight velocity difference between the merging clusters, which is not typical of double relic systems. RXC J1314.4--2515 was also observed with the Murchison Widefield Array (MWA) from 88 to 215~MHz \citep{2017MNRAS.467..936G}, leading to an estimate of the integrated spectral index of eastern and western relics of $\alpha_{118}^{1400}$ = 1.03 $\pm$ 0.12 and $\alpha_{118}^{1400}$ = 1.23 $\pm$ 0.09, respectively. These spectral index values are in good agreement with the JVLA values estimated by \citet{2019MNRAS.489.3905S} of $\alpha_{1.5GHz}^{3GHz}$ = 1.0$\pm$0.1 for the eastern relic, $\alpha_{1.5GHz}^{3GHz}$ = 1.6$\pm$0.1 for the western relic, and $\alpha_{1.5GHz}^{3GHz}$ = 1.3$\pm$0.2 for the radio halo. \citet{2019MNRAS.489.3905S} reports flux densities based on the combined 1.5~GHz (B+C+D array; rms =~35~$\mu$Jy~beam$^{-1}$; resolution: $17^{\prime\prime} \times 14^{\prime\prime}$) of 11.3$\pm$0.6, 33.0$\pm$2.0 and 5.3$\pm$0.3 for the eastern relic, western relic, and radio halo, respectively.

The 1.28~GHz MeerKAT images presented in Figure~\ref{fig:RXCJ1314} show in full and low-resolution (rms =~12~$\mu$Jy) the presence of the reported radio halo and radio relics. The greatest extents of the radio halo and the western relic are more extended than reported in \citet{2019MNRAS.489.3905S}. However, the morphology and size of the eastern relic are consistent in both images. The radio halo appears to fill the western relic area in the north--south direction having a total size of $840\times1080$~kpc$^2$ and $S_{1.28~GHz}$ = 12.5~mJy, which corresponds to a radio power of $P_{1.28~GHz}$ = 2.3$\times$10$^{24}$ W~Hz$^{-1}$. The two radio relics on the east and west sides have total sizes of $230\times620$~kpc$^2$ and $210\times1080$~kpc$^2$ and flux densities of 12.7~mJy and 33.6~mJy, respectively.


The in-band spectral index in the western relic appears flatter at the location of peak brightness in the total intensity image, with an $\alpha$ value of around $-$1.3. This value then steepens moving north, reaching values of around $-$2.7 and $-$2.8, except for a few patchy features with flatter values. No steepening is seen towards the inner part of the cluster, which is surprising considering
the edge brightening of this feature. On the other hand, the eastern relic shows a patchy distribution along the direction of the major axis, with flatter spectral index values between $-$1.2 and $-$1.8. The few spectral index patches at the western part of the halo are very steep, with values around $-$3. It is interesting to note that the deep MeerKAT image presented here does not detect the diffuse emission north of the current eastern relic, up to $Dec=-25^{\circ}12^{\prime}$, seen at 610~MHz with the GMRT \citep{2007A&A...463..937V}. This suggests a very steep spectrum for this feature. 


\subsection{RXC J2351.0--1954}

RXC~J2351.0--1954 (also ACT-CL~J2351.1-1957; \citealt{Hilton_2020} or PSZ1~G057.09-74.45) is a cluster at redshift $\textit{z}=0.248$ with an estimated M$_{500}$ Planck mass of 5.60$^{+0.59}_{-0.62}\times10^{14}$ M$_\odot$ \citep{Planckcollab15}. This system is part of the REFLEX~II cluster sample \citep{2012A&A...538A..35C} that has an X--ray luminosity of L$_X$ = $4.33\pm0.84\times10^{44}$~erg~s$^{-1}$. 

\citet{2021PASA...38...10D} report two candidate radio relics in their 168~MHz MWA EoR0 study, the first east of the cluster centre with a largest linear size of $\sim$~1.4~Mpc and flux density of S$_{168MHz}=57\pm9$~mJy. No counterpart was detected in SUMSS or NVSS, with an upper limit of S$_{1400MHz}\leq4.2$~mJy at 1400~MHz. The second is in the opposite direction (west), has an LLS of $\sim$~1.3~Mpc, being significantly brighter (flux density of S$_{168MHz}=147\pm13$~mJy) and remarkably located at an extremely far distance for a radio relic, $\sim$~2.8~Mpc away from the cluster centre. The second candidate relic is partially detected in the TGSS data, but without any detection at 1400~MHz, with the NVSS upper limit being S$_{1400MHz}\leq4.7$~mJy. \citet{2021PASA...38...10D} also estimate steep spectral indices for the candidate relics using the NVSS upper limits ($\alpha^{1400}_{168} \leq -1.2\pm0.1$ and $-1.68\pm0.04$, respectively). In addition, \citet{2021PASA...38...10D} report the possibility of faint candidate halo emission, but in a follow-up study MWA-2 (frequencies 88 to 216~MHz) and RACS ASKAP data at 887~MHz \citet{2021PASA...38...53D} show that the earlier reported candidate radio halo is a combination of blended compact radio sources. 

The 1.28~GHz MeerKAT images of this cluster are presented in Figure~\ref{fig:RXCJ2351}. The full-resolution ($7.8^{\prime\prime} \times 7.8^{\prime\prime}$; rms =~4~$\mu$Jy~beam$^{-1}$) image reveals only the presence of discrete radio sources around the cluster centre without any sign of detection of diffuse radio halo emission. The presence of the east and west candidate relics is confirmed, with the west one only marginally detected at this resolution. Similarly, in the MeerKAT low-resolution image (rms =~7~$\mu$Jy~beam$^{-1}$) there is no visible diffuse radio emission, only several compact discrete radio sources, which confirms the results by \citet{2021PASA...38...53D}. The candidate radio relic of the eastern part is detected with a total size of $170\times1340$~kpc$^2$, having a flux density of $S_{1.28~GHz}$ = 2.5~mJy, whereas the western one has a size of $100\times1030$~kpc$^2$ and a flux density of just $S_{1.28~GHz}$ = 0.6~mJy. Both candidate radio relics seem to be heavily blended by the presence of discrete radio sources. The total intensity image shows an arc-like feature at the location of the northwest candidate relic below the 3$\sigma$ level. With the current MeerKAT data, we confirm the absence of diffuse radio emission as a candidate halo in this system and also still cannot confirm whether the radio relics are real relics, remnants, or radio galaxies, thus classifying them here also as candidate radio relics. 

Only the eastern candidate relic shows features in the 15$''$ in-band spectral index map, with fairly uniform values in the range of $-$1.1 to $-$1.3 and a patchy distribution.

\subsection{J0027.3--5015}

J0027.3--5015 (also known as Abell~2777, MCXC~J0027.3-5015) is part of the REFLEX cluster sample \citep{Bohringeretal04} at redshift $\textit{z}=0.145$ having an X--ray luminosity of L$_X$ = $2.25\times10^{44}$~erg~s$^{-1}$ and an M$_{500}$ mass of 2.96$\times10^{14}$ M$_\odot$ \citep{2011A&A...534A.109P}. To date, no radio studies or upper limits have been reported for this system.

The 1.28~GHz MeerKAT images of this cluster are presented in Figure~\ref{fig:J0027}. The full-resolution ($7.8^{\prime\prime} \times 7.8^{\prime\prime}$; rms =~3~$\mu$Jy~beam$^{-1}$) image shows only the presence of compact radio sources at the cluster centre and around, without any sign of detection of diffuse radio emission. The presence of a very faint diffuse component surrounding the central radio source is only revealed at the MeerKAT low-resolution image (rms =~6~$\mu$Jy~beam$^{-1}$) with a total size of $170\times210$~kpc$^2$ having a flux density of just $S_{1.28~GHz}$ = 0.3~mJy which corresponds to a radio power of 1.6$\times$10$^{22}$ W~Hz$^{-1}$. With the current MeerKAT data, we classify this faint diffuse emission as a possible candidate radio mini-halo. The average spectral index of the candidate mini-halo is estimated to be $\leq -3$ with the seed electrons for the diffuse emission most likely provided by the central AGN radio source.

An investigation of the in-band spectral index map reveals that the compact radio emission associated with the BCG steepens from the northeast to the southwest, with spectral index values ranging from $-$0.9 to $-$1.2.

\subsection{J0145.0--5300}

J0145.0--5300 (also known as Abell~2941, MCXC~J0027.3-5015) is a galaxy cluster at redshift $\textit{z}=0.117$ that is also part of the REFLEX, REXCESS \citep{Bohringeretal04,Bohringeretal07} and ATCA REXCESS diffuse emission survey (ARDES) \citep{Shakouri2016ARDES} cluster samples. It was first detected in SZ by ACT \citep{2010ApJ...723.1523M,2011ApJ...737...61M} and then included in the Planck \citep{Planckcollab14} survey. The system has an X--ray luminosity of L$_{500}$ = $2\times10^{44}$~erg~s$^{-1}$ and an M$_{500}$ mass of 2.89$\times10^{14}$ M$_\odot$ \citep{2011A&A...534A.109P}. \citet{2013MNRAS.431.2542H} examined the X--ray properties of this system with \textit{Chandra} reporting a disturbed X--ray morphology.

\citet{Shakouri2016ARDES} using ATCA low-resolution radio observations at 1.4~GHz with a 256~MHz bandwidth ($78^{\prime\prime} \times 34^{\prime\prime}$; $rms = 200$~$\mu$Jy~beam$^{-1}$) before the telescope upgrade reported possible diffuse emission for this system. However, re-observing J0145.0--5300 (A2941) with the upgraded ATCA at 2.1~GHz at low ($102^{\prime\prime} \times 71^{\prime\prime}$; $rms = 186$~$\mu$Jy~beam$^{-1}$) and high resolution ($5^{\prime\prime} \times 4^{\prime\prime}$; $rms = 56$~$\mu$Jy~beam$^{-1}$) and also tapered 1.4~GHz ($84^{\prime\prime} \times 51^{\prime\prime}$; $rms = 176$~$\mu$Jy~beam$^{-1}$), they report only a small bright extended radio source with a core that is not associated with the cluster, not a diffuse emission.

The high sensitivity 1.28~GHz MeerKAT images of this cluster are presented in Figure~\ref{fig:J0145.0}. The full-resolution ($7.8^{\prime\prime} \times 7.8^{\prime\prime}$; rms =~3~$\mu$Jy~beam$^{-1}$) image shows only the presence of several compact radio sources in the centre and around of the cluster, without any detection of diffuse radio emission. The presence of a very faint diffuse component surrounding the central radio sources is only revealed at the MeerKAT low-resolution image (rms =~4~$\mu$Jy~beam$^{-1}$) with a total size of $320\times620$~kpc$^2$ having a flux density of just $S_{1.28~GHz}$ = 2.5~mJy which corresponds to a radio power of 8.5$\times$10$^{22}$ W~Hz$^{-1}$. Although the diffuse emission is blended with compact radio sources, including the central AGN radio source, we classify this faint diffuse emission as a radio halo. The average spectral index of the radio halo from the MeerKAT data is estimated to be $\alpha_{908}^{1656}\leq -2$.

\subsection{J0145.2--6033}


J0145.2-6033 is part of the Meta-Catalogue of X--ray-detected Clusters (MCXC; \citealt{2011A&A...534A.109P}) lying at redshift $\textit{z} = 0.181$. It was drawn from the REFLEX cluster survey \citep{Bohringeretal04} and was also initially detected by RASS \citep{Vogesetal99}. On the X--ray side, \citet{2021PASA...38...53D} suggests from the RASS data that the cluster is relaxed, since no significant offset is observed between the X--ray peak and the BCG.

\citet{2021PASA...38...53D} has also reported the detection of a candidate mini-halo using MWA-2 at several low-frequencies (88, 118, 154, 185, and 216~MHz) and a marginal detection with the Rapid ASKAP Continuum Survey \citep[RACS;][]{McConnelletal20} at
887~MHz ($\sim$~15$''$resolution, rms $\sim250-400\, \mu Jy\, beam^{-1}$). The total flux density of the candidate mini-halo at 154~MHz was estimated to be 110$\pm$10~mJy, with the largest linear extent being $\sim$~350~kpc. An integrated spectral index was also derived at $\alpha^{887}_{88} =$ $-$2.1$\pm$0.1, indicating that this is an ultra-steep spectrum radio source.


Our 1.28~GHz MeerKAT data of MCXC~J0145.2-6033 also reveal an extended diffuse emission that surrounds the central BCG radio source. The full-resolution image shows only partially this extension with the diffuse emission more clearly visible in the 15$^{\prime\prime}$ image (see Figure~\ref{fig:J0145.2}), with several point radio sources surrounding the cluster centre. Our 1.28~GHz MeerKAT image also clearly shows the detection of a candidate radio mini-halo structure with a total size of $220\times320$~kpc that appears to be elongated in the southwest$-$northeast axis as reported in \citet{2021PASA...38...53D}. This candidate radio mini-halo exhibits a very low surface brightness ($\sim$~0.1~$\mu$Jy~arcsec$^{-2}$ over an area with a radius of 150~kpc), revealing the ability of MeerKAT to detect radio emission with low signal-to-noise ratio at 15$^{\prime\prime}$ resolution.

\subsection{J0216.3--4816}

J0216.3--4816 (also known as Abell~2988, ACT-CL~J0216--4816) is a galaxy cluster at redshift $\textit{z}=0.163$ that was initially part of the ROSAT All-Sky Survey (RASS; \citealt{Cruddaceetal02}) and later included in the MCXC \citep{2011A&A...534A.109P} and ACT \citep{Hilton_2020} catalogues. The system has an X--ray luminosity of L$_{500}$ = $1.95\times10^{44}$~erg~s$^{-1}$ and an M$_{500}$ mass of 2.72$\times10^{14}$ M$_\odot$ \citep{2011A&A...534A.109P}. \citet{2009AJ....137.4795C} reports that the structure of the BCG suggests possible evidence of a past merger (if not a mere superposition). No radio observations or upper limits have been reported for this system.

Our 1.28~GHz MeerKAT images of this cluster are presented in Figure~\ref{fig:J0216.3}. Both full (rms =~3~$\mu$Jy~beam$^{-1}$) and low-resolution (rms =~6~$\mu$Jy~beam$^{-1}$) images reveal, for the first time, the detection of a faint diffuse component surrounding the central compact radio source with a total size of $220\times240$~kpc$^2$. The flux density of the diffuse emission is $S_{1.28~GHz}$ = 3.3~mJy which corresponds to a radio power of $P_{1.28~GHz}$ = 2.3$\times$10$^{23}$ W~Hz$^{-1}$. We classify this faint diffuse emission as a possible candidate radio mini-halo, with the seed electrons for the diffuse emission being provided by the central AGN radio source. 

The in-band spectral index map at 15$''$ shows the distribution of the BCG, with a homogeneous spectral index value of $\sim-$1, surrounded by a sharp transition in the east--west direction where $\alpha$ drops to $\sim-$2.2. The full-resolution spectral index map confirms the $\sim-$1 spectral index value in the core of the radio emission, which then follows the isophotes in the NE-SW direction, where $\alpha$ steepens to values around $\sim-$1.6 to $-$1.8. The average in-band spectral index of the candidate mini-halo is estimated to be $\alpha_{908}^{1656}\leq -2.5$.

\subsection{J0217.2--5244}

J0217.2--5244 (also known as ACT-CL~J0217.1--5244) is a galaxy cluster at redshift $\textit{z}=0.343$ and is part of the REFLEX cluster sample \citep{Bohringeretal04} first detected in SZ by ACT \citep{2010ApJ...723.1523M,2011ApJ...737...61M}, included in Planck \citep{Planckcollab16} survey and also in the MCXC \citep{2011A&A...534A.109P} and ACT \citep{Hilton_2020} catalogues. The system has an X--ray luminosity of L$_{500}$ = $11\times10^{44}$~erg~s$^{-1}$ and an M$_{500}$ mass of 6.84$\times10^{14}$ M$_\odot$ \citep{2011A&A...534A.109P}. 

\citet{Hoganetal15} reports that the BCG position in this cluster seems tentatively matched to a SUMSS radio source \citep{SUMSS} with the lack of higher resolution radio data making it impossible to resolve whether the source could be extended just below the SUMSS resolution limit. They report this source as an associated match, but conservatively place only upper limits ($\leq19~mJy$ at 1~GHz) for the non-core component. \citet{Birzanetal17} reports a deprojected X--ray core temperature of $\sim$~7.56 KeV for this system with a cooling time in the innermost region of $t_{cool}\sim29.1$~Gyr.

The 1.28~GHz MeerKAT images of this cluster are presented in Figure~\ref{fig:J0217.2}. The full-resolution ($7.8^{\prime\prime} \times 7.8^{\prime\prime}$; rms =~3.5~$\mu$Jy~beam$^{-1}$) image shows no signs of diffuse emission, revealing only the presence of discrete radio sources around the cluster centre. Similarly, in the MeerKAT low-resolution image (rms =~6~$\mu$Jy~beam$^{-1}$), there is no visible diffuse radio emission, only several compact discrete radio sources with the cluster centre not exactly associated with a radio source. We report the detection of a north candidate radio relic in the system that is not associated with an optical counterpart, with a total size of $170\times460$~kpc$^2$ having a flux density of $S_{1.28~GHz}$ = 1.1~mJy which corresponds to a radio power of $P_{1.28~GHz}$ = 4.3$\times$10$^{23}$ W~Hz$^{-1}$. 

Only the brightest part of the ridge of the candidate relic is visible in the 15$''$ spectral index map, showing very steep values in the range of $-$1.6 to $-$1.8, with a slightly flatter feature. The compact component at the north$-$eastern end of this structure has a very flat spectral index, with $\alpha$ $\sim-$0.4.


\subsection{J0225.9--4154 / Abell 3017}

J0225.9--4154 or ACT-CL~J0225.9--4154 system lies at redshift $\textit{z} = 0.220$ and consists of two sub-clusters, Abell~3017 and Abell~3016, that are in the process of merging. Abell~3017 was initially detected by the ROSAT All-Sky Survey (RASS; \citealt{Vogesetal99}) and is part of the ROSAT-ESO Flux Limited X--ray (REFLEX) Galaxy cluster survey sample \citep{Bohringeretal04} from where it was later identified by \citet{Bohringeretal13} as one of the most X--ray luminous clusters in the $z\leq0.22$ Southern sky with $L_X = 6.5\times10^{44} erg~s^{-1}$ (0.1 $-$ 2.4 keV). 

Abell~3017 presents an intriguing dynamical behaviour in X--rays. Using \textit{Chandra} observations, \citet{Parekhetal17} showed that the system is embedded in an Mpc-scale ICM filament (bridge) that connects Abell~3017 and Abell~3016 sub-clusters with signatures of recent cluster and filament interactions. In a more recent and detailed X--ray study of the system, \citet{Chonetal19} also detected indications of internal ICM disturbances along the filament of the main Abell~3017 cluster (a cold front and a shock front south of the core). Based on the observed X--ray morphology, \citet{Chonetal19} favoured the scenario that the system is going through its first collapse, with the filament connecting the two subclusters being of external origin, coming from a detached overdense region between the clusters. This interpretation of the evolutionary state of the cluster merger is supported by the fact that the filament was found to be too massive to have emerged from the interaction of the two sub-clusters, thus making a post-merger phase scenario with the filament being built up by entrained gas from both clusters,  least likely.

In previous radio observations of Abell~3017, \citet{Parekhetal17} using dual frequency 235/610~MHz GMRT data showed that the BCG of Abell~3017 exhibits a young AGN that presents a radio core and a symmetric extended emission (radio lobes) that is attached to the central radio source component. The AGN is also seen to have excavated two oppositely directed X--ray cavities. More recently, \citet{Pandgeetal21} using the upgraded GMRT (uGMRT) also detected at Band-3 (250$-$500 MHz) the central component and the attached radio lobes; however, at Band-5 (1200$-$1400 MHz) only the central component was detected as a point radio source without any surrounding radio emission. \citet{Parekhetal17} also provides a low-frequency radio spectral index for the central AGN at  $\alpha_{235MHz}^{610MHz}\sim-1.3$. \citet{Hoganetal15} using radio data from the Australia Telescope Compact Array (ATCA) reports Abell~3017 as a core-dominated unresolved radio source at C (5~GHz) and X (8~GHz) bands with an ultra-steep component ($\alpha\leq-1.5$) appearing in the Sydney University Molonglo Sky Survey (SUMSS, 843~MHz; \citealt{SUMSS}).


Our 1.28~GHz MeerKAT data of Abell~3017 reveal an extended diffuse radio component (reported in K22) that surrounds the central bright radio core, which has not been reported in other earlier radio observations, providing new information on the radio properties of the system. We detected a diffuse radio structure with a total size of $350\times500$~kpc that appears to be elongated in the southeast$-$northwest axis (see Figure~\ref{fig:J0225.9}). The diffuse emission in the south--west direction encompasses a background radio source that does not have an optical counterpart. We calculate the flux density of the diffuse emission only as $S_{1.28GHz}=14.4~mJy$. The central radio source presents an in-band spectral index of $\alpha_{908}^{1656}\sim-1.5$ whereas the diffuse emission appears to be much steeper at $\alpha_{908}^{1656}\sim-2.2$. In more detail, the spectral map shows a central ridge with values $\sim-$1.1 to $-$1.2. The southern edge of this ridge has a very flat value ($\sim-$0.5) and resembles a point radio source. Eastward and westward of the ridge, the spectral index steepens to values $-$2 to $-$2.4, flattening again at the location of the small-tailed radio galaxy embedded in the faint emission surrounding the BCG.

Based on the available information on the dynamical state of the Abell~3017 system that comprises a pair of AGN-related X--ray cavities at the centre, one infalling group, and several other groups interacting within the Abell~3017 system, we classify the new central diffuse radio emission detected as a radio halo. We find that the radio halo is mainly contaminated in the centre by two bright, compact radio sources and appears to be attached in the north$-$northwest direction to the radio emission of several compact radio sources that present optical identifications. From what was previously reported from the \textit{Chandra} X--ray image, the radio halo overlaps in good agreement with the central X--ray emission peak and the detection of the cold front reported at 150~kpc by \citet{Chonetal19} towards the south from the core. We find that the extension of the diffuse radio halo emission south of the core is at 150~kpc, also where the cold front is detected, confirming the signatures of cluster and filament interactions. 

Based on the available information on the dynamical state of the Abell~3017 system that comprises a pair of AGN-related X--ray cavities at the centre, one infalling group, and several other groups interacting within the Abell~3017 system, we classify the new central diffuse radio emission detected as a radio halo. We find that the radio halo is mainly contaminated in the centre by two bright, compact radio sources and appears to be attached in the north$-$northwest direction to the radio emission of several compact radio sources that present optical identifications. From what was previously reported from the \textit{Chandra} X--ray image, the radio halo overlaps in good agreement with the central X--ray emission peak and the detection of the cold front reported at 150~kpc by \citet{Chonetal19} towards the south from the core. We find that the extension of the diffuse radio halo emission south of the core is at 150~kpc, also where the cold front is detected, confirming the signatures of cluster and filament interactions.

\subsection{J0232.2--4420}

J0232.2--4420 (also known as PSZ1~G260.00-63.45, ACT-CL~J0232.2--4421) is a galaxy cluster at redshift $\textit{z}=0.284$ that was initially part of the ROSAT All-Sky Survey (RASS; \citealt{DeGrandietal99,Cruddaceetal02}) and later included in the REFLEX cluster sample \citep{Bohringeretal04} and the MCXC \citep{2011A&A...534A.109P}, SPT \citep{Bleemetal15}, Planck \citep{Planckcollab16} and ACT \citep{Hilton_2020} catalogues. The system has an X--ray luminosity of L$_{500}$ = $8.5\times10^{44}$~erg~s$^{-1}$ \citep{2011A&A...534A.109P} and a M$_{500}$ mass of 7.54$\times10^{14}$ M$_\odot$ \citep{Planckcollab16}, which also presents indications of strong lensing \citep{2022ApJ...928...87F}.

 The X--ray morphology dynamical state of the J0232.2--4420 cluster has been classified as relaxed \citep{2017ApJ...846...51L}. A more recent detailed substructure analysis by \citet{Parekhetal21}, using X--ray data from the \textit{Chandra} and \textit{XMM-Newton}, has also shown that the core of the cluster is not disturbed; however, a substructure in the south--west direction is suggested to possibly introduce energy in the system via turbulence. \citet{Kaleetal19} reports that RXC~J0232.2--4420, is host to a moderate-sized radio halo, and in combination with the relaxed dynamical state nature of the cluster in the X--rays, the system is part of a rare class of clusters with relaxed morphology hosting radio halos (e.g., \citealt{2016MNRAS.459.2940K}). As the relaxed X--ray morphology in this system is unexpected, \citet{Kaleetal19} suggests that RXC~J0232.2--4420 may be a case of a rare system undergoing a transition from mini-halo to radio halo. 

\citet{Kaleetal22} using the MGCLS data products of RXC J0232.2--4420, also confirm the detection with MeerKAT \citep{Knowlesetal22}, of the already known and detected by GMRT radio halo \citep{Kaleetal19}.  GMRT radio images (rms =~40~$\mu$Jy~beam$^{-1}$) detected a radio halo with a flux density of $S_{610~MHz}$ = 52$\pm$5~mJy and size of $800$~kpc at 610~MHz with the MeerKAT flux density reported as $S_{1.28~GHz}$ = 9.8$\pm$2.8~mJy \citep{Kaleetal22}. The spectral index of the radio halo is estimated to be $\alpha_{610}^{1683} = -1.6\pm0.4$. \citet{Kaleetal22}, following \citet{Knowlesetal22}, also reports the detection of two elongated filamentary diffuse structures, one toward the east and one to the south of the cluster, labelling them as east and south candidate relics, respectively. 

These candidate radio relics are located at distances of 1 (east) and 1.9~Mpc (south) from the cluster centre without any clear galaxy match in the Dark Energy Survey (DES; \citealt{2018ApJS..239...18A} optical images. The largest linear sizes of the east and south candidate relics are reported to be 380~kpc and 600~kpc, respectively \citep{Kaleetal22}. These faint candidate relics are also being reported as outliers from the radio power -- mass scaling relation, suggesting that they are likely not products of a cluster merger, and in combination with their large projected distances from the cluster centre, \citet{Kaleetal22} suggests that these structures may be unrelated to the cluster. Such filamentary radio sources may as well originate from remnant radio galaxies in which their core activity is relatively fast ceased unless revived (e.g., \citealt{2001A&A...366...26E,2020MNRAS.496.1706S}). 

The 1.28~GHz MeerKAT images presented in Figure~\ref{fig:J0232.2} show in both full and low-resolution (rms =~6$\mu$Jy~beam$^{-1}$) the presence of the reported radio halo and candidate radio relics. The largest extent and flux density of the halo is found to agree with \citet{Kaleetal22}, at $\sim$~1240~Mpc and $S_{1.28~GHz}$ = 10.6~mJy which corresponds to a radio power of $P_{1.28~GHz}$ = 2.6$\times$10$^{24}$ W~Hz$^{-1}$. The two new candidate radio relics on the east and south sides are also in good agreement with total sizes of $210\times440$~kpc$^2$ and $200\times450$~kpc$^2$ and flux densities of 0.9~mJy and 0.6~mJy, respectively.  

The 15$''$ spectral index map detects the radio halo up to the third contour level (80~$\mu$Jy~beam$^{-1}$). It shows a radial steepening, with values ranging from approximately $-$1.6 near the central source to approximately $-$1.9 in the southeastern filament and $-$2.3 in the outer regions. The candidate relic does not show a particular structure in the spectral index map, and the candidate eastern relic shows a ridge with an $\alpha$ value of approximately $-$1.2 along the major axis, with mild steepening inward ($-$1.5 to $-$1.6).

   
\subsection{J0303.7--7752}

J0303.7--7752 (also known as PSZ1~G294.68-37.01) is a galaxy cluster at redshift $\textit{z}=0.274$ that was initially part of the ROSAT All-Sky Survey  Bright Source Catalogue (RASS-BSC; \citealt{Vogesetal99}) and later included in the REFLEX \citep{Bohringeretal04}, MCXC \citep{2011A&A...534A.109P} and Planck \citep{Planckcollab11b} cluster samples. The system has an X--ray luminosity of L$_{500}$ = $6.34\times10^{44}$~erg~s$^{-1}$ and an M$_{500}$ mass of 5.17$\times10^{14}$ M$_\odot$ \citep{2011A&A...534A.109P}. 

In the recent X--ray study by \citet{2017ApJ...846...51L}, the dynamical state of the J0303.7--7752 cluster is classified as disturbed with no other radio observations available for this system apart from SUMSS \citep{SUMSS}. \citet{Hoganetal15} reports that the BCG appears to be associated with a SUMSS radio source with a number of other radio sources visible nearby, placing only upper limits using appropriate indices ($\leq$16.8~mJy at 1~GHz) for the non-core component.

The high sensitivity 1.28~GHz MeerKAT images of this cluster are presented in Figure~\ref{fig:J0303.7}. The full-resolution ($7.8^{\prime\prime} \times 7.8^{\prime\prime}$; rms =~3~$\mu$Jy~beam$^{-1}$) image reveals, for the first time, the detection of an extended diffuse radio emission between several compact radio sources at the cluster centre and around. The presence of a faint diffuse component surrounding the central radio sources is also confirmed and visible in the MeerKAT low-resolution image (rms =~6~$\mu$Jy~beam$^{-1}$) with a total size of $630\times950$~kpc$^2$ having a flux density of $S_{1.28~GHz}$ = 8.6~mJy which corresponds to a radio power of 2.0$\times$10$^{24}$ W~Hz$^{-1}$. Although the diffuse emission is confined by and partially blended with several compact radio sources in the central cluster area, we classify this detected faint diffuse emission as a radio halo. The average spectral index of the radio halo from the MeerKAT data is estimated to be $\alpha_{908}^{1656}\leq -2$.

\subsection{J0314.3--4525}

J0314.3--4525 (also known as Abell~3104, ACT-CL~J0314.3--4525) is a galaxy cluster at redshift $\textit{z}=0.072$ that was initially part of the ROSAT All-Sky Survey (RASS; \citealt{DeGrandietal99,Cruddaceetal02}) and later included in the REFLEX \citep{Bohringeretal04} cluster sample, and in the MCXC \citep{2011A&A...534A.109P}, Planck \citep{Planckcollab16} and ACT \citep{Hilton_2020} catalogues. The system has an X--ray luminosity of L$_{500}$ = $1.03\times10^{44}$~erg~s$^{-1}$ and an M$_{500}$ mass of 1.98$\times10^{14}$ M$_\odot$ \citep{2011A&A...534A.109P}. \citet{2009AJ....137.4795C} reports that the structure of the BCG suggests a giant dominant elliptical galaxy. To date, no radio studies or upper limits have been reported for this system.

Our 1.28~GHz MeerKAT images of this cluster are presented in Figure~\ref{fig:J0314.3}. Both full (rms =~3~$\mu$Jy~beam$^{-1}$) and low-resolution (rms =~6~$\mu$Jy~beam$^{-1}$) images reveal for the first time the detection of a faint diffuse component surrounding the central compact radio source with a total size of $100\times150$~kpc$^2$. The flux density of the diffuse emission is just $S_{1.28~GHz}$ = 1.1~mJy which corresponds to a radio power of $P_{1.28~GHz}$ = 1.3$\times$10$^{22}$ W~Hz$^{-1}$. We classify this faint diffuse emission as a possible candidate radio mini-halo, with the seed electrons for the diffuse emission most likely provided by the central BCG AGN radio source.

The spectral index map at 15$''$ detects only the compact source associated with the BCG and the inner extension eastward. The spectral index is about $-$1.2 at the location of the BCG and shows a mild flattening to $-$1 going eastward. The total intensity image suggests that the central diffuse source could be larger than the lowest contour shown in Figure~\ref{fig:J0314.3}. The average spectral index of the candidate mini-halo is estimated to be $\alpha_{908}^{1656}\leq -2.5$.

\subsection{J0342.8--5338}

J0342.8--5338 (also known as Abell~3158, ACT-CL~J0342.9--5337) is a galaxy cluster at redshift $\textit{z}=0.060$ that was initially part of the ROSAT All-Sky Survey (RASS; \citealt{DeGrandietal99}) and later included in the REFLEX cluster sample \citep{Bohringeretal04} and the MCXC \citep{2011A&A...534A.109P} and Planck \citep{Planckcollab14} catalogues. The system has an X--ray luminosity of L$_{500}$ = $2.8\times10^{44}$~erg~s$^{-1}$ and an M$_{500}$ mass of 3.65$\times10^{14}$ M$_\odot$ \citep{2011A&A...534A.109P}. It is situated in the most dynamically important position within the supercluster central region of the so-called Horologium Reticulum Supercluster (HRS; \citealt{Fleenoretal05}) which consists of two large cluster groupings \citep{Einastoetal03} with Abell~3158 moving toward the double cluster system Abell~3125/Abell~3128 \citep{Luceyetal83} that defines the supercluster centre.

\citet{Parekhetal15}, from \textit{Chandra} archival data, examined possible correlations between morphological parameters and other X--ray gas properties. Using a combination of adopted X--ray cluster morphology parameters, J0342.8--5338 was classified as a non-relaxed cluster having a relatively high cooling time value (t$_{cool}\sim$8.2~Gyr). The cluster was previously identified as relaxed by \citet{Vikhlininetal09}. In an earlier X--ray study, \citet{Wangetal10} analysing also archival \textit{Chandra} and \textit{XMM-Newton} data of Abell~3158, identified a bow-edge-shaped discontinuity in the X--ray surface brightness distribution associated with a massive, off-centre cool gas clump, suggesting a merger origin, with the cool gas clump currently what was left from the original central cool-core of the main cluster or the infalling substructure. In addition,  \citet{Johnstonetal08} also suggests that, by using the radio galaxy population as an indicator of merger history, the Abell~A3158 cluster, most likely, is currently captured in a late merger state with a significant excess of low-powered, blue galaxies aligned along the axis linking to the Abell~3125/A3128 substructure. 


On the radio side, \citet{2016MNRAS.456.1259B} using intermediate KAT-7 observations at 1.86~GHz reports the detection at the cluster centre of a barely resolved faint radio source that has a flux density of $S_{1.86~GHz}$ = 7.6$\pm$0.4~mJy (rms =~300~$\mu$Jy~beam$^{-1}$; $2.4^\prime\times2.2^\prime$)  which corresponds to a radio halo power upper limit of $P_{1.4~GHz}\leq$ 6.3$\times$10$^{23}$ W~Hz$^{-1}$. \citet{Johnstonetal08} presenting a combined optical and radio 1.4 and 2.5~GHz Australia Telescope Compact Array (ATCA) observations of the Abell~3158 cluster report the detection of an unresolved radio source at the centre that has a flux density at 1.4~GHz of 7.88$\pm$0.08~mJy (rms =~30~$\mu$Jy~beam$^{-1}$; $8.8^{\prime\prime}\times4.6^{\prime\prime}$) and at 2.5~GHz  of 5.12$\pm$0.06~mJy (rms =~30~$\mu$Jy~beam$^{-1}$; $5.0^{\prime\prime}\times2.6^{\prime\prime}$) with a spectral index for the radio source in total of $\alpha_{1.4}^{2.5} = -0.8$ \citep{Johnstonetal08} which is typical of a compact radio source. These findings are consistent with the flux density reported by the \citet{2016MNRAS.456.1259B} 1.86~GHz KAT-7 observations at the cluster centre.

Our 1.28~GHz MeerKAT images of this cluster are presented in Figure~\ref{fig:J0342.8}. Both full (rms =~3.5~$\mu$Jy~beam$^{-1}$) and low-resolution (rms =~6~$\mu$Jy~beam$^{-1}$) images reveal for the first time the detection of a faint diffuse component surrounding the central compact radio source that has been reported in previous studies. The total size of the diffuse emission is $280\times340$~kpc$^2$ with the flux density being $S_{1.28~GHz}$ = 5.6~mJy which corresponds to a radio power of $P_{1.28~GHz}$ = 4.4$\times$10$^{22}$ W~Hz$^{-1}$. We classify this faint diffuse emission as a radio mini-halo, with the seed electrons for the diffuse emission provided by the central AGN radio source. We report that the 1.28~GHz flux density of the unresolved radio source at the centre is 11.1~mJy, in good agreement with the flux density reported by \citet{2016MNRAS.456.1259B} using 1.86~GHz KAT-7 scaled with a spectral index of $-$0.8. 

The spectral index map at 15$''$ shows only the spectral index distribution associated with the brightest source, which has a flat spectrum in the centre ($\alpha\sim-$0.68), that steepens westward in the direction of the source elongation, to values
$\sim-$2.2. The average spectral index of the mini-halo is estimated to be $\alpha_{908}^{1656}\leq -2$.



\subsection{J0431.4--6126}

J0431.4--6126 (also known as Abell~3266) is a nearby massive and complex galaxy cluster at redshift $\textit{z}=0.059$ that has been extensively studied at multiple wavelengths and included in the REFLEX \citep{Bohringeretal04} cluster sample, and in the MCXC \citep{2011A&A...534A.109P} and Planck \citep{Planckcollab11a} catalogues. The system is located in the Horologium Reticulum Supercluster (HRS) \citep{Fleenoretal05} and has an X--ray luminosity of L$_{500}$ = $3.98\times10^{44}$~erg~s$^{-1}$ \citep{2011A&A...534A.109P} and an M$_{500}$ mass of 6.64$\times10^{14}$ M$_\odot$ \citep{Planckcollab16}.

In the optical study by \citet{Dehghanetal17}, the substructure of the system was explored, showing that a merger is occurring and revealing the presence of five additional individual subclusters, which are in different phases of dynamic activity. \citet{Dehghanetal17} also reports that the main core components of the merging systems most likely have already gone through a core passage. Numerous studies of Abell~3266 on the X--ray side \citep[e.g.,][]{Markevitch98,DeGrandietal99,Finoguenovetal06,Sandersetal22} reveal an elongated and asymmetric X--ray surface brightness distribution and an asymmetric temperature profile, also confirming the fact that Abell~3266 is undergoing a complex merger event. In a recent study, \citet{Sandersetal22} presented a detailed X--ray analysis of Abell~3266 using X--ray observations from the eROSITA telescope \citep{2021A&A...647A...1P}, revealing the underlying substructure within the cluster, along with two weak (Mach number M$\sim$~1.5 $-$ 1.7) peripheral shock fronts at $\sim$~1.1~Mpc from the cluster centre also confirming the ongoing complex merger state and revealing the several infalling subclusters surrounding the central system.

Using KAT-7 observations at 1.86~GHz, \citet{2016MNRAS.456.1259B} finds no evidence of diffuse radio emission at the cluster centre, reporting only the presence of several discrete radio sources.  \citet{Riseleyetal15}, using the same instrument at 1.83 and 1.32~GHz, reports the possibility of the existence of a faint central diffuse emission component after subtraction of contaminating compact radio sources. However, the very low-resolution ($\sim$~3.5$^\prime$) of the KAT-7 observations was not optimal for further investigation of the faint diffuse central component.

The first low-frequency radio observations at 88 $-$ 216~MHz of Abell~3266 were presented in \citet{Duchesneetal22} using the Phase II MWA \citep{Waythetal18} telescope, confirming the detection of the radio relic and revealing also two ultra-steep spectrum fossil plasma sources in the ICM. However, while Abell~3266 exhibits several typical characteristics of radio halo-hosting clusters, \citet{Duchesneetal22} does not report the detection of a radio halo from deep MWA observations. Most recently, also, \citet{Riseleyetal22} presented the analysis of new, deep, broad-band radio data from the Australia Telescope Compact Array (ATCA; 1.1$-$3.1~GHz) and the Australian Square Kilometre Array Pathfinder (ASKAP; 0.8$-$1.1~GHz; \citealt{Johnsonetal07,Hotanetal21}). These new deep radio observations report the detection of a `wrong-way' relic, a fossil plasma source, and an unclassified central diffuse ridge, which \citet{Riseleyetal22} suggest is the brightest part of a large-scale radio halo.

The 1.28~GHz MeerKAT images of this cluster are presented in Figure~\ref{fig:J0431.4}. Both the full-resolution ($7.8^{\prime\prime} \times 7.8^{\prime\prime}$; rms =~5~$\mu$Jy~beam$^{-1}$) and low-resolution (rms =~10~$\mu$Jy~beam$^{-1}$) images reveal the complex nature of this system with the presence of several extended radio structures around the cluster centre that have been previously reported. At the cluster core, we observe the detection of a very faint, elongated, diffuse radio emission that extends from the back of the eastern-tailed radio source (located next to the cluster core) in the northeast direction, blending with several discrete radio sources. \citet{Riseleyetal22} refers to this diffuse emission as a diffuse `ridge' detected with ASKAP but not in their 2.1~GHz ATCA maps, suggesting a steep spectral index of $\alpha_{943}^{2100}$ = $-$2.54. The LLS reported for the diffuse `ridge' is 240~kpc and the total integrated flux density at 943~MHz is $\sim$~8.0$\pm$0.8~mJy. \citet{Duchesneetal22} report an upper limit of 0.3 $-$ 0.8 $\times$ 10$^{24}$ W Hz$^{-1}$ on the luminosity of any halo that may be present, which is well below the detection limit of the Phase~II MWA data. Examining the nature of this diffuse `ridge' \citet{Riseleyetal22} suggests that it is either a product of the cluster merger turbulence, where some form of revived fossil plasma originating from a nearby tailed radio galaxy has been re-accelerated, or it is a newly detected mini-halo. However, the lack of a cool-core in the cluster and the low surface brightness of the emission make the second scenario less plausible but not unlikely.

From our MGCLS data, we detect a more extended emission for the diffuse `ridge' than the one reported by \citet{Riseleyetal22} in two regions in particular: i) south of the tailed radio galaxy, in-between the point radio source and the tailed radio galaxy, and ii) a faint elongated emission blended by the compact discrete radio sources to the north of the core. The total size of the faint diffuse emission in the centre is $170\times480$~kpc$^2$, having an estimated flux density of $S_{1.28~GHz}$ = 9.7~mJy, which corresponds to a radio power of $P_{1.28~GHz}$ = 7.5$\times$10$^{22}$ W~Hz$^{-1}$. The higher flux density reported here, compared to that reported by \citet{Riseleyetal22}, can be explained by the greater extent to which we detect it with the MGCLS data. Due to the uncertain nature of this diffuse radio source, we do not classify it and mark it as U (Unknown) in Table~\ref{tab:diffuse}. 

We also detect the southern `wrong-way' relic, an elongated and asymmetric diffuse radio source with concave morphology at $\sim$~1~Mpc south of the cluster centre. We name this structure a southeast radio relic, which is most likely the product of a shock propagating in the north--south direction following the main cluster merger \citep{Riseleyetal22}. The southeast radio relic is detected with a total size of $200\times670$~kpc$^2$, having a flux density of $S_{1.28~GHz}$ = 45.3~mJy. The sensitive MGCLS data detect a slightly greater extent for this relic than the one reported at 948~MHz by \citet{Riseleyetal22} ($\sim$~579~kpc) and at 2.1~GHz ($\sim$~394~kpc) however, the reported flux density ($S_{948~MHz}$ = 72.3$\pm$7.4~mJy) is in good agreement based on the $-$1.7 median spectral index derived for this structure between 948 and 2100~MHz. 

 We also note here that \citet{2021Galax...9...81R} using the MGCLS data for this system, presented a detailed study of the radio structure named MysTail, which is an elongated radio galaxy presenting a very complex morphology located northwest of the cluster centre, next to the faint diffuse `ridge' emission to the north. 

The in-band spectral index map reports the features of the bright-tailed radio source southwest of the cluster centre. The spectral index in the northern head tail ranges from $\alpha$  $\sim-$1 in the head to $-$2.7 at the end of the tail, while the head-tail close to the cluster centre has a flatter core ($\alpha \sim-$0.5) and a very steep spectrum at the end of the tail ($\alpha\sim-$3). The candidate relic is edge-brightened along the southern profile and has a steep parallel ridge with $\alpha$ ranging from $-$1.5 to $-$2. It steepens up to values close to $-$3 towards the cluster centre.

\subsection{J0510.2--4519}

J0510.2--4519 (also known as Abell~3322, RXC J0510.2-4519) is a galaxy cluster at redshift $\textit{z}=0.200$ that was included in the REFLEX \citep{Bohringeretal04} cluster sample, and in the MCXC \citep{2011A&A...534A.109P} and Planck \citep{Planckcollab11a} catalogues. The system has an X--ray luminosity of L$_{500}$ = $4.85\times10^{44}$~erg~s$^{-1}$ and an M$_{500}$ mass of 4.65 $\times$ 10$^{14}$ M$_\odot$ \citep{2011A&A...534A.109P}.  The X--ray morphology dynamical state of the Abell~3322 cluster has been classified as `mixed' in-between relaxed and disturbed classification by \citet{2017ApJ...846...51L}, which refers to systems that exhibit small substructures or a relatively flat X--ray distribution.

On the radio side, \citet{Kaleetal22} using GMRT 610~MHz data reports the presence of an extended component located in part in conjunction with a compact radio source, which has a total size of 248~kpc. This diffuse component has a flux density of 8.4$\pm$1.3~mJy at 610~MHz, which corresponds to an extrapolated 1.4~GHz radio power of 7.5$\times$10$^{23}$~W~Hz$^{-1}$. The peak of the \textit{Chandra} X--ray emission is found to be offset by the central radio source \citep{Kaleetal22}. The MGCLS survey paper (K22) reported the radio status of this system at 1.28~GHz with \citet{Kaleetal22} showing evidence of extended emission from the MGCLS images we also use here. However, no flux density of the diffuse emission was reported at this frequency.

Our 1.28~GHz MeerKAT images of this cluster are presented in Figure~\ref{fig:J0510.2}. Both full (rms =~4~$\mu$Jy~beam$^{-1}$) and low-resolution (rms =~6~$\mu$Jy~beam$^{-1}$) images reveal the detection of a faint diffuse component surrounding the central compact radio source with a total size of $350\times430$~kpc$^2$. The flux density of the diffuse emission was difficult to extract due to imperfect calibration (negatives) around the central radio source, however, as a lower limit, we estimated it to be $S_{1.28~GHz}$ $\geq$ 1.3~mJy which corresponds to a radio power of $P_{1.28~GHz}$ $\geq$ 1.3$\times$10$^{22}$ W~Hz$^{-1}$. We classify this faint diffuse emission as a candidate radio mini-halo, with the detected faint diffuse emission most certainly fed by the central AGN radio source. The average spectral index of the candidate mini-halo is estimated to be $\alpha_{908}^{1656}\leq -2.5$. The estimated flux density of the candidate mini-halo at 1.28~GHz is slightly lower than expected from the 610~MHz flux density extrapolation, but it is in good agreement considering the estimated steep radio spectral index.  

The compact central source shows a steep trend in the spectral index, from $\alpha\sim-$0.7 in the peak located
at the western-most edge to $\sim-$1.5. The tailed radio galaxy west of the centre exhibits a very flat core ($\alpha\sim-$0.35) and shows a gradual steepening along the jets, all the way to the spur north of the western hotspot ($\alpha\sim-$1.5 to $-$1.8).

\subsection{J0516.6--5430}  

J0516.6--5430 (also known as Abell~S0520 or RXC~J0516.6--5430) is an optically-rich \citep{Abelletal89} galaxy cluster at redshift $\textit{z}=0.295$ that was included in the REFLEX \citep{Bohringeretal04} cluster sample and in the MCXC \citep{2011A&A...534A.109P} and Planck \citep{Planckcollab11a} catalogues. The system has an X--ray luminosity of L$_{500}$ = $12.60\times10^{44}$~erg~s$^{-1}$ and an M$_{500}$ mass of 7.72$\times10^{14}$ M$_\odot$ \citep{2011A&A...534A.109P}.  The X--ray morphology dynamical state of the Abell~S0520 cluster has been classified as disturbed by \citet{2017ApJ...846...51L}, which refers to complex systems that exhibit clear evidence of merging. No radio studies have been reported for this system.

The 1.28~GHz MeerKAT images presented in Figure~\ref{fig:J0516.6} reveal for the first time in both full and low-resolution (rms =~6~$\mu$Jy~beam$^{-1}$) the presence of an elongated faint radio halo in the north--south direction that includes the central compact radio source. The diffuse emission, which appears to be blended with several discrete radio sources, is visible and more extended in the low-resolution image. The largest extent of the radio halo is $\sim$~1540~Mpc with a flux density of $S_{1.28~GHz}$ = 7.9~mJy which corresponds to a radio power of 2.2$\times$10$^{24}$ W~Hz$^{-1}$. The MeerKAT data further reveal the detection of two new extended diffuse structures located in opposite directions to the north and south of the radio halo emission, which we call the north relic and the south candidate relic. Their total sizes are $650\times1500$~kpc$^2$ and $270\times770$~kpc$^2$, and their flux densities are 33.6~mJy and 4.8~mJy, respectively. The northern relic appears to be elongated, showing the internal structure of two arc-like features. The candidate relic to the south is seen to be blended with a discrete radio source. These newly reported relic structures are most likely products of the dynamic state merging of the system. 

No signal is detected in the spectral index map for the radio halo. However, the northern relic has a double ridge and shows a steepening trend towards the cluster centre, from $\alpha\sim-$1 at the outer edge to $\alpha\sim-$2.3 towards the centre. The candidate southern relic has a peak with $\alpha\sim-$1 and steepens towards the centre to $\alpha\sim-$2.3. 

\subsection{J0528.9--3927}

J0528.9--3927 (also known as RBS~0653, RXC~J0528.9--3927) is a massive galaxy cluster at redshift $\textit{z}=0.284$ that was initially part of the ROSAT All-Sky Survey (RASS; \citealt{DeGrandietal99}) and later included in the REFLEX cluster sample \citep{Bohringeretal04} and Planck \citep{Planckcollab11b} catalogue. From X--ray studies, the ICM emission of RXC~J0528.9--3927 cluster is known to have a centrally peaked X--ray emission \citep{Finoguenovetal05} and a mixed-nature dynamical state (intermediate to relaxed and merging \citealt{2017ApJ...846...51L}). In addition, \citet{Botteonetal18} reports the presence of a cold front located to the west of its core.


On the radio side, \citet{Knowles2020} from a pilot MeerKAT observations study reported a radio halo emission at 1.28~GHz extended to the south--west direction from the BCG with a size of 540~kpc and a flux density of $\sim$~3.6$\pm$0.26~mJy. More recently, \citet{Kaleetal22} using GMRT 610~MHz data also confirms the presence of an extended radio emission surrounding the central BCG. The largest size for the halo reported from the GMRT data is 300~kpc with a flux density of 9.6$\pm$1.0~mJy with \citet{Kaleetal22} classifying this source as a mini-halo with a spectral index of $\alpha_{610}^{1280}=-1.3$. 


The 1.28~GHz MeerKAT images presented in Figure~\ref{fig:J0528.9} show the presence of the reported radio halo in both full and low-resolution (rms =~6~$\mu$Jy~beam$^{-1}$). The total size of the halo, which appears to be more extended in the low-resolution image and elongated towards the south--west direction from the BCG, is $560\times930$~kpc$^2$. The flux density of the halo is found to be  $S_{1.28~GHz}$ = 3.8~mJy which is in good agreement with \citet{Knowles2020} and corresponds to a radio power of $P_{1.28~GHz}$ = 9.5$\times$10$^{23}$ W~Hz$^{-1}$. We detect a larger extent of the radio halo mainly due to the higher sensitivity of the MGCLS data compared to the GMRT data.

There is insufficient signal to derive the in-band spectral index distribution in the radio halo. However, the trend of the spectral index is visible in the head-tail source northeast of the halo, steepening away from the cluster core, from $\alpha\sim-$0.7 in the head to $-$1.1 and up to $-$1.6 in the knot. The extended radio galaxy
southeast of the cluster core shows a nice trend in the in-band spectral index, with a flat spectrum ($-$0.7) and two hot spots with $\alpha\sim-$0.8, revealing an FRII nature. A similar FRII nature is clear in the double radio source just east of the core. The compact radio source embedded in the halo is very steep. The spectral trend confirms that it is compact with a spectral index value of $\alpha\sim-$1.2 at the peak and increases to $\alpha\sim-$2 in an annular trend.

\subsection{J0627.2--5428}

J0627.2--5428 (also known as Abell~3395) is a well-studied nearby galaxy cluster ($\textit{z}=0.051$) over a range of wavelengths observed with numerous instruments. It was first observed in the X--rays by the \textit{Einstein} satellite telescope \citep{Formanetal81} and was then part of the ROSAT All-Sky Survey (RASS; \citealt{DeGrandietal99}) and later included in the REFLEX \citep{Bohringeretal04} cluster sample and the MCXC \citep{2011A&A...534A.109P} and Planck \citep{Planckcollab11b} catalogues. The cluster is part of a triple cluster galaxy merger system, (Abell~3391/Abell~3395(north)/Abell~3395; e.g., \citealt{ReiprichBohringer02}) and presents in both its galaxy distribution \citep{Girardietal97} as well as in its X--ray emission, a double-peaked profile. The large-scale galaxy environment suggests the presence of an enormous structure that runs from north to south in projection. At $\sim$~3~Mpc to the north Abell~3395 is neighbouring to Abell~3391 ($\textit{z}=0.056$) and in between them lies a galaxy group called ESO~161-IG~006 ($\textit{z} = 0.052$; \citealt{Alvarezetal18}) that completes the triple system. Both clusters present M$_{200}$ masses of $\sim$~2$\times10^{14}$ M$_\odot$ \citep{2011A&A...534A.109P} and X--ray temperatures of kT$\sim$~5~keV \citep{Vikhlininetal09}.  

ASCA, ROSAT, and Planck observations have shown that Abell~3395 and Abell~3391 are linked via a gas bridge \citep{TittleyHenriksen01,2013A&A...550A.134P} with more recent \textit{XMM-Newton}, \textit{Chandra}, and Suzaku studies of this system confirming that the evident emission excess between the two clusters is probably a gas of cluster origin that has been tidally stripped from the clusters due to the interaction process they are involved indicating an early phase of the cluster merger \citep{Sugawaraetal17,Alvarezetal18}. Most recent X--ray observations with the extended ROentgen Survey with an Imaging Telescope Array (eROSITA) offer much greater effective area and higher energy resolution for extended sources at soft X--rays than earlier missions \citep{Merlonietal12,2021A&A...647A...1P}, reveal an impressive $\sim$~15~Mpc long continuous X--ray emission filament from the projected north toward the south of the Abell~3391/95 system. eROSITA also found that the bridge gas consists of warm gas from a filament \citep{Reiprichetal21} in addition to the hot gas, which confirms the tidally stripped gas origin scenario. The presence of this new Warm-Hot Intergalactic Medium (WHIM) filament is also confirmed by the Planck Sunyaev-Zeldovich map and the DECam galaxy density map. In addition, a few clump features have been discovered that are seen to fall into the system, two of which reside in the northern (for more details, see \citealt{Veronicaetal22}) and the Southern Filaments \citep{Reiprichetal21}, with results from the Magneticum simulation being consistent with the properties of the observed features (large-scale WHIM gas filaments with a clumpy structure; \citealt{Veronicaetal22}). 

\citet{2021A&A...647A...3B} using 1~GHz ASKAP/EMU Early Science observations studied the radio properties and population in a 30~deg$^2$ area surrounding the Abell~3391/3395 system, providing a plethora of information on extended or otherwise interesting radio sources in this field around the cluster merger. \citet{2021A&A...647A...3B} reports also the identification of seven candidates of radio relics and halos (see their Table 5), together with previously published examples \citep{wilber_johnston_2020}. However, no diffuse radio synchrotron emission has been detected in the X--ray bridge of thermal gas between the two clusters deriving an upper limit on the radio emissivity in the filament area of J$_{1GHz}$ $\leq$ 1.2 $\times$ 10$^{-44}$ W~Hz$^{-1}$~m$^{-3}$.

Our 1.28~GHz MeerKAT images of this cluster are presented in Figure~\ref{fig:J0627.2}. Both full (rms =~8~$\mu$Jy~beam$^{-1}$) and low-resolution (rms =~12~$\mu$Jy~beam$^{-1}$) images reveal the detection of the FR~I radio source in Abell~3391 and the detection of a faint diffuse L-shaped radio feature in the western region of Abell~3391 in the Abell~3395 cluster region. This L-shaped radio source is reported as source S2 and S3 in Table~2 and Figure~5 of \citet{2021A&A...647A...3B}. The west relic feature spans out in two perpendicular arms one to the north and another to the west direction with the whole structure in total being a combination of two radio galaxies whose diffuse emission has been interconnected (radio galaxy 2MASX~J06261051-5432261 for the west relic and galaxy 2MASX~J06255706-5427502 to the north--west direction for the bright radio AGN spot), presenting a filamentary structure in the north--south part, with a very steep in-band spectral index ($\alpha\sim-$3). The total size of the west relic is $90\times320$~kpc$^2$ with the flux density of the diffuse emission being $S_{1.28~GHz}$ = 78.1~mJy which corresponds to a radio power of $P_{1.28~GHz}$ = 4.5$\times$10$^{23}$ W~Hz$^{-1}$ at the cluster's mean redshift. The MeerKAT flux density of the west relic feature is in well agreement with the ASKAP/EMU flux density reported by \citet{2021A&A...647A...3B} at 1013~MHz of 72.1$\pm$7.2 at an rms of 48~$\mu$Jy~beam$^{-1}$. The integrated spectral index is $\alpha_{88}^{1013}\sim-1$. 

This west radio relic feature is a complex structure that most likely reveals the re-acceleration of plasma that originates from AGNs. The overlap of the southern part of this region with a high-temperature area, as shown from an oxygen-to-soft-band ratio in the eROSITA data \citep{Reiprichetal21}, may also imply a shock or a higher pressure area that enforces the synchrotron emission from the old AGN radio plasma.



\subsection{J0637.3--4828}

J0637.3--4828 (also known as Abell~3399) is a massive galaxy cluster at redshift $\textit{z}=0.203$ \citep{Bleemetal15} that was initially part of the ROSAT All-Sky Survey Bright Source Catalogue (RASS-BSC; \citealt{Vogesetal99}) and later included in the REFLEX \citep{Bohringeretal04}, MCXC \citep{2011A&A...534A.109P} and Planck \citep{Planckcollab11b} cluster samples. The system has an X--ray luminosity of L$_{500}$ = $3.53\times10^{44}$~erg~s$^{-1}$ \citep{2011A&A...534A.109P} and a M$_{500}$ mass of 3.19$_{-0.51}^{+0.88}$ $\times$ 10$^{14}$ M$_{\odot}$ \citep{Lovisari_2020}. 

\citet{2017ApJ...846...51L} classified the dynamical state of the J0637.3--4828 cluster as disturbed, suggesting that the cluster is undergoing a merger, with the \textit{Chandra} X--ray map (observed with \textit{Chandra-Planck} Legacy Program for massive clusters of galaxies, PI: Christine Jones) showing a southwest extension of the hot gas from the central body. On the radio side, \citet{2021PASA...38...53D} using MWA-2 (frequencies 88 to 216~MHz) and RACS ASKAP data \citep{McConnelletal20} at 887~MHz ($\sim$~15$''$resolution, rms $\sim250-400\, \mu Jy\, beam^{-1}$) reports the detection of i) a candidate radio halo at the centre of the cluster which is a combination of blended compact radio sources with a total size of $\sim$~570~kpc and ii) a candidate radio relic on the periphery of Abell~3399 with a total size of 710~kpc. \citet{2021PASA...38...53D} reports an integrated spectral index of $\alpha_{88}^{887}\sim-1.6$$\pm$0.3 for the candidate radio relic and $\alpha_{118}^{887}\sim-1.5$$\pm$0.2 for the candidate radio halo (derived only by a two-point index and not a fitted power-law model due to insufficient detection and blended sources between 154-216 MHz and the 88~MHz image).

Abell~3399 was part of the SUCCESS sample \citep{Kaleetal22} in which MeerKAT L-band observations from the MGCLS \citep{Knowlesetal22} were used and revealed a double relic and a radio halo. Using a 1$\sigma$ rms of 15~$\mu$Jy~beam$^{-1}$ in the low-resolution image at a 3$\sigma$ level of significance, a diffuse emission is visible that is noted as a southeast relic extension with a total size of $\sim$~938~kpc in \citet{Kaleetal22}. In the central region, the largest linear size of the diffuse emission is reported to be $\sim$~647~kpc, whereas a candidate northwest relic is also reported with a total size of $\sim$~404~kpc. The flux density of the southeast feature is reported to be $S_{1.28~GHz}$ = 9.5 $\pm$ 1.1~mJy, for the central radio halo $S_{1.28~GHz}$ = 2.5 $\pm$ 0.5~mJy whereas \citet{Kaleetal22} did not measure the flux density for the candidate northwest relic due to the presence of several discrete sources.


The high sensitivity 1.28~GHz MeerKAT images of this cluster are also presented here in Figure~\ref{fig:J0637.3}. The full-resolution ($7.8^{\prime\prime} \times 7.8^{\prime\prime}$; rms =~3~$\mu$Jy~beam$^{-1}$) image reveals the detection of an extended diffuse radio emission blended with several compact radio sources at the cluster centre and almost detached to the east is the diffuse emission reported as a southeast relic in \citet{Kaleetal22}. The presence of a unified and elongated diffuse component surrounding the central radio sources is visible at the MeerKAT low-resolution image (local rms =~6~$\mu$Jy~beam$^{-1}$) with a total size of $1150\times1480$~kpc$^2$ having a flux density of $S_{1.28~GHz}$ = 16.5~mJy which corresponds to a radio power of 1.9$\times$10$^{24}$ W~Hz$^{-1}$. We classify this detected and extended diffuse emission partially blended with several compact radio sources in the central cluster area, as a radio halo. It is unclear whether this structure is double-peaked or engulfs the presence of a southeast radio relic as suggested by \citet{Kaleetal22} using a much higher 1$\sigma$ rms. In the double-relic scenario then the presence of the radio halo is most likely a faint radio bridge-like morphology visible between the central diffuse halo and relic.  The flux density of the northwest candidate radio relic feature is estimated to be very faint, $S_{1.28~GHz}$ = 0.5~mJy with a size of $190\times400$~kpc$^2$. 

The sub-structure of the eastern edge of the radio halo emission is visible in the in-band spectral index map, too.  The values of $\alpha$ range from $\sim-$1 in the flattest parts to $-$1.4 and $-$1.7 in the steeper blobs. The spectral index then steepens to values of $\sim-$2.5 going
towards the cluster centre. The spur is not detected in the spectral index map. In the western part of the radio halo, only the brightest knots are detected in the spectral index map, with spectral indices ranging from $\alpha\sim-$1.2 to $-$1.5, steepening to values of $\sim-$2.5 in the fainter regions. The candidate northwest relic shows a flat spectrum ($\alpha\sim-$0.78) in the point source with only a thin and short filament visible departing from the point source toward the northeast with $\alpha\sim-$1.2 to $-$1.3.

\subsection{J0638.7--5358}

J0638.7--5358 (also known as Abell~S0592) is a massive galaxy cluster at redshift $\textit{z}=0.227$ that was initially part of the ROSAT All-Sky Survey Bright Source Catalogue (RASS-BSC; \citealt{Vogesetal99}) and later included in the REFLEX \citep{Bohringeretal04}, MCXC \citep{2011A&A...534A.109P} and Planck \citep{Planckcollab11b} cluster samples. The system has an X--ray luminosity of L$_{500}$ = $9.54\times10^{44}$~erg~s$^{-1}$ \citep{2011A&A...534A.109P} and a M$_{500}$ mass of $7.07_{-0.28}^{+0.35}$ $\times$ 10$^{14}$ M$_{\odot}$  \citep{Lovisari_2020}. 

\citet{2017ApJ...846...51L} classified the dynamical state of the J0638.7--5358 cluster between relaxed and disturbed, which suggests that it is an intermediate merging cluster with mixed morphology. A cool-core state study by \citet{Rossettietal17} showed a significant surface brightness (SB) peak but an overall disturbed X--ray morphology, confirming that it is a `disturbed cool-core' system. The \textit{Chandra} X--ray image shows a bimodal merging cluster with the ICM exhibiting an overall high temperature 
\citep{2010ApJ...723.1523M} even though it possesses a low K0 value \citep{Botteonetal18}.  \citet{Botteonetal18} also reported the presence of the two low-temperature, low-entropy cool-cores surrounded by a hot, disturbed ICM, also claiming the presence of a shock along the Southwest edge of the ICM with a Mach number between 1.61$-$1.72. The evident two X--ray peaks of the system resemble a Bullet-type merger. 


On the radio side, \cite{wilber_johnston_2020} using ASKAP’s early science observations outside the EMU Pilot Survey at 1013~MHz (resolution 20$^{\prime\prime}$; rms =~25~$\mu$Jy~beam$^{-1}$), report the presence of a diffuse intracluster emission associated with the J0638.7--5358 cluster. The diffuse radio emission that contains four bright radio galaxies at the cluster centre has a total size of $\sim$~1.04~Mpc and is classed as a giant radio halo. The source-subtracted, direction-dependent (DD) calibration ASKAP image yields an integrated flux density at 1013~MHz of 9.95 $\pm$ 2.16~mJy. ATCA observations of this cluster detect some of the diffuse emission (rms =~120~$\mu$Jy~beam$^{-1}$) which has a flux density at 2215~MHz of 3.3 $\pm$ 0.4~mJy \citep{wilber_johnston_2020}. The diffuse radio emission appears to be offset from the X--ray gas, covering only the southwest part of the ICM, as seen in projection, with the southwest border of the radio halo having a linear edge, roughly coincident with the SB edge reported by \citet{Botteonetal18}. \citet{wilber_johnston_2020} suggest that the J0638.7--5358 cluster is in the phase after core passage, with the embedded radio galaxies contributing a significant population of seed electrons that have been re-accelerated by the merger event, thus implying a turbulent re-acceleration origin for the radio halo. \cite{wilber_johnston_2020} also derives an estimate of the integrated spectral index of the radio halo, which is $\alpha^{2215}_{1013}$ = $-$1.41 $\pm$ 0.25.



\citet{Knowles2020} also reported the radio halo using early MeerKAT data with a size of 480~kpc, having a flux density of $\sim$~5~mJy, but note that this estimation is probably underestimated due to missing flux. The 1.28~GHz MeerKAT images presented in Figure~\ref{fig:J0638.7} here show the cluster's giant radio halo emission blended with several compact radio sources mainly at the cluster centre. The diffuse emission appears to be elongated in the south--west direction with a total size of $800\times1090$~kpc$^2$, having a flux density in good agreement with the ASKAP estimate of $S_{1.28~GHz}$ = 11.6~mJy, which corresponds to a radio power of 1.8$\times$10$^{24}$ W~Hz$^{-1}$. 

The average in-band spectral index of the radio halo from the MeerKAT data is estimated to be $\alpha_{908}^{1656}\leq -1.5$. In more detail, only the very central part of the halo emission is visible in the spectral index map, and it is unclear whether it is already the peak of the halo or whether the feature is connected with the compact westernmost central radio galaxy. The two central galaxies present a relatively flat spectrum, i.e., $\alpha\sim-$0.5 to $-$1.3 with
an E$-$W trend for the eastern radio galaxy and $\alpha\sim-$0.8 to $-$1 with a W$-$E trend for the western one. The spectral index west of the western central galaxy has a filamentary trend, with $\alpha\sim-$0.9 to $-$1.

\subsection{J0645.4--5413}

J0645.4--5413 (also known as Abell~3404) is a massive galaxy cluster at redshift $\textit{z}=0.164$ that was initially part of the ROSAT All-Sky Survey Bright Source Catalogue (RASS-BSC; \citealt{Vogesetal99}) and later included in the REFLEX \citep{Bohringeretal04}, MCXC \citep{2011A&A...534A.109P} and Planck \citep{Planckcollab14} cluster samples. The system has an X--ray luminosity of L$_{500}$ = $6.71\times10^{44}$~erg~s$^{-1}$ \citep{2011A&A...534A.109P} and a M$_{500}$ mass of $8.6_{-0.3}^{+0.4}$ $\times$ 10$^{14}$ M$_{\odot}$  \citep{Lovisari_2020}. 

 \textit{Chandra} X--ray observations show that the cluster is oriented from northeast to southwest. \citet{Prattetal09} using \textit{XMM-Newton} data classifies the cluster as a non-cool core, but not as morphologically disturbed based on its centroid shift \citep{Pooleetal06}. According to the most recent X--ray study of \citet{2017ApJ...846...51L}, the dynamical state of the cluster is neither fully disturbed nor relaxed, but intermediate, which means that it is a weak merging cluster with mixed X--ray morphology.



\citet{Shakouri2016ARDES} had investigated the Abell~3404 cluster in radio using ATCA as part of the ATCA REXCESS 3 Diffuse Emission Survey (ARDES). However, they report no evidence of a radio halo or any other diffuse radio source. GaLactic and Extragalactic All-sky MWA (GLEAM; \citealt{HurleyWalkeretal17}) survey data at 200~MHz revealed the presence of extended emission at the core of the Abell~3404 cluster. The extended emission remained unclassified in this study due to the combination of GLEAM's low-resolution and contamination by compact radio sources. More recently, \citet{2021PASA...38...31D} presented new observations from the MWA-2, ASKAP, and ATCA of the Abell~3404 cluster, detecting a central diffuse radio source with MWA-2 and ASKAP. They classify the extended diffuse emission as a giant radio halo with the largest linear extent of $\sim$~770 kpc. \citet{2021A&A...647A...3B} had also reported the detection of diffuse emission in this system at the periphery of the field of an ASKAP/EMU observation of Abell~3391-95. 

\citet{2021PASA...38...31D} calculated a spectral index of $\alpha_{88}^{1110}$ = $-$1.66 $\pm$ 0.07 for the radio halo which makes it the first reported Ultra Steep Spectrum Radio Halo (USSRH) detected with ASKAP that is part of a newly emerging faint radio halo population that undergoes weak mergers. \citet{2021PASA...38...31D} finds also a strong correlation in the radio–X--ray surface brightness correlation for radio halos. The detection of additional peripheral components in Abell~3404 is also reported in \citet{2021PASA...38...31D}. For the southwest structure, it is suggested that it is not associated with the central radio halo, as no signs of shocks are detected at the location in the X--rays, hence it is unlikely to be a relic. 

The 1.28~GHz MeerKAT images presented in Figure~\ref{fig:J0645.4} show in both full and low-resolution (rms =~8~$\mu$Jy~beam$^{-1}$) the presence of a complex and extended diffuse radio emission blended with several compact radio sources that we also classify as a radio halo. The total size of the halo emission, which appears to be in the form of a radio bridge and elongated from the northwest to the southeast direction in projection, is $990\times1240$~kpc$^2$. The flux density of the halo is found to be $S_{1.28~GHz}$ = 20.4~mJy corresponding to a radio power of $P_{1.28~GHz}$ = 1.5$\times$10$^{24}$ W~Hz$^{-1}$. The MeerKAT data also detect the southwest extended diffuse structure that we classify as a southwest radio relic at a distance of $\sim$~1~Mpc from the cluster centre. Its total size is $260\times450$~kpc$^2$, having a flux density of 3.2~mJy. The southwest relic appears not to have any obvious related optical counterpart, with its diffuse emission appearing to be blended by the presence of a compact-like radio source at its centre.

Only the four radio point sources at the centre of the halo are detected in the spectral index image, all with steep spectra ($\alpha$ ranging from $\sim-$1 to $-$1.4). The southwest relic has a centrally peaked spectral index distribution, with $\alpha\sim-$0.95, which steepens along the northwest to values of $\sim-$1.7.

\subsection{J0745.1--5404}

J0745.1--5404 is a system of interacting groups of galaxies identified in the 2MASS catalogue \citep{refId0,Tully_2015} at redshift $\textit{z}=0.074$ that was included in the MCXC \citep{2011A&A...534A.109P} and Planck \citep{Planckcollab16} cluster samples. The system has an X--ray luminosity of L$_{500}$ = $1.28\times10^{44}$~erg~s$^{-1}$ and an M$_{500}$ mass of 2.3 $\times$ 10$^{14}$ M$_{\odot}$  \citep{2011A&A...534A.109P}. 

No dedicated studies in radio or other wavelengths have been previously reported for this system. The 1.28~GHz MeerKAT images presented in Figure~\ref{fig:J0745.1} here show the detection of several compact radio sources at the cluster centre, with the two brightest ones coinciding with optical counterparts, therefore being of AGN origin. At a distance of $\sim$~1.3~Mpc (at the cluster redshift) south from the J0745.1--5404 cluster centre, in projection, we detect from the MGCLS images the presence of a new faint elongated diffuse radio emission with a total size of $100\times330$~kpc$^2$ having a flux density of $S_{1.28~GHz}$ = 3.8~mJy which corresponds to a radio power of 4.8$\times$10$^{22}$ W~Hz$^{-1}$. We classify this detected and extended diffuse radio emission partially coinciding with a few compact radio sources to the north as a candidate radio relic. It is unclear whether this structure is related to revived fossil emission from a radio AGN, which could also potentially classify this structure as a candidate radio Phoenix.

\subsection{J0820.9--5704}

J0820.9--5704 is a galaxy cluster at redshift $\textit{z}=0.061$ that was included in the MCXC \citep{2011A&A...534A.109P} and Planck \citep{Planckcollab14} cluster samples. The system has an X--ray luminosity of L$_{500}$ = $0.78\times10^{44}$~erg~s$^{-1}$ and an M$_{500}$ mass of 1.69 $\times$ 10$^{14}$ M$_{\odot}$  \citep{2011A&A...534A.109P}. 

As part of the Planck cluster catalogue, J0820.9--5704 was found to exhibit a filament of hot and low-density intergalactic medium between an interacting cluster pair \citep{2013A&A...550A.134P}. The 1.28~GHz MeerKAT images presented in Figure~\ref{fig:J0820.9} here show no diffuse radio emission visible related to the cluster core. We note that the MGCLS radio images for this cluster are dynamically range-limited due to the presence of a very bright AGN radio source at the cluster centre. However, a pair of radio structures without an optical counterpart is visible in the southern direction from the centre of the cluster.

The full-resolution ($7.8^{\prime\prime} \times 7.8^{\prime\prime}$; rms =~4~$\mu$Jy~beam$^{-1}$) high sensitivity 1.28~GHz MeerKAT image of the J0820.9--5704 cluster reveals at a distance of $\sim$~1.9~Mpc (at the cluster redshift) to the southeast from the cluster centre the detection of two detached diffuse radio emission blobs blended with several compact radio sources, whereas a similar extended diffuse radio structure but continuous, is detected on the opposite direction, southwest from the cluster centre. The presence of both these new elongated diffuse radio components is more clearly visible in the MeerKAT low-resolution image (local rms =~8~$\mu$Jy~beam$^{-1}$), which we classify as candidate radio relics. The detached southeast diffuse blobs from the high-resolution image appear to be connected and unified to a candidate radio relic that has a total size of $150\times480$~kpc$^2$  and a flux density of $S_{1.28~GHz}$ = 8.6~mJy which corresponds to a radio power of 7.3$\times$10$^{22}$ W~Hz$^{-1}$ at the cluster redshift. On the other side, the southwest candidate radio relic has a total size of $170\times610$~kpc$^2$ and a flux density of $S_{1.28~GHz}$ = 20.5~mJy, which corresponds to a radio power of 1.7$\times$10$^{23}$ W~Hz$^{-1}$. The orientation of these candidate relics is not consistent with their location within the cluster. They are symmetrically located about a double radio galaxy. In between the two candidate relics, there is a 2MASS galaxy group (2MASS J08205105-5728439; \citealt{Skrutskieetal06}) that most likely is the host system of these two newly detected radio structures. Further investigation is guaranteed for this system.

\subsection{J1130.0--4213}

J1130.0--4213 is a galaxy cluster at redshift $\textit{z}=0.155$ that was part of Clusters in the Zone of Avoidance (CIZA) survey \citep{Ebelingetal02} and was later included in the MCXC \citep{2011A&A...534A.109P} and Planck \citep{Planckcollab16} cluster samples. The system has an X--ray luminosity of L$_{500}$ = $2.82\times10^{44}$~erg~s$^{-1}$ and an M$_{500}$ mass of 3.45 $\times$ 10$^{14}$ M$_{\odot}$  \citep{2011A&A...534A.109P}. 

Earlier dedicated studies on this system have not been reported. The 1.28~GHz MeerKAT images presented in Figure~\ref{fig:J1130.0} here show the presence of a bright compact radio source related to the cluster centre and next to it, in projection, a tailed radio galaxy in the northeast direction. Signs of diffuse radio emission are visible surrounding the bright radio source in the cluster core with an east-west elongation; however, these are evaluated as artefacts from the high-resolution image due to imperfect calibration, also contaminated by compact radio sources. Hence, with the current MGCLS data products, we do not report the detection of any diffuse radio emission related to the cluster core, but only the presence of a new extended radio structure without an optical counterpart, which is visible in the north--east direction from the centre of the cluster. We classify this structure as a candidate radio relic due to its position with the cluster core and lack of optical identification. This northeast candidate radio relic has a total size of $120\times340$~kpc$^2$  and a flux density of $S_{1.28~GHz}$ = 3.2~mJy, which corresponds to a radio power of 2.0$\times$10$^{23}$ W~Hz$^{-1}$ at the cluster redshift.

From the investigation of the in-band spectral index map, the central compact radio source presents a spectral index with $\alpha \sim-$0.9. The spectral index of the candidate northeast relic is imaged only in the brightest central
ridge and shows a radio galaxy trend, with a central peak of $\alpha\sim-$1 and minor steepening along the major axis (up to $\alpha\sim-$1.2 to $-$1.3).

\subsection{J1423.7--5412}

J1423.7--5412 is a galaxy cluster at redshift $\textit{z}=0.300$ that was part of the Clusters in the Zone of Avoidance (CIZA) survey \citep{Ebelingetal02} and later included in the MCXC \citep{2011A&A...534A.109P} cluster sample. The system appears to be X--ray luminous with L$_{500}$ = $13.69\times10^{44}$~erg~s$^{-1}$ and massive, with M$_{500}$ mass of 8.09 $\times$ 10$^{14}$ M$_{\odot}$  \citep{2011A&A...534A.109P}. 

Earlier dedicated studies on this system have not been reported. The 1.28~GHz MeerKAT images presented in Figure~\ref{fig:J1423.7} here show the presence of a bright compact radio source in the cluster centre surrounded by several other compact radio sources. Signs of faint diffuse radio emission surrounding the bright radio source in the cluster core are visible; however, this radio emission, as is evident from the high-resolution image, is contaminated by several compact radio sources. Hence, with the current MGCLS data products, we cannot be conclusive on the presence of a faint diffuse radio emission at the cluster core. Due to the difficulty in classifying the structure in the centre of J1423.7--5412, we report the detection of this faint diffuse radio emission related to the cluster core as a candidate radio halo in Table~\ref{tab:diffuse} due to contamination by compact radio sources. 

In addition to the uncertain candidate radio structure at the core, a new extended radio structure is visible to the north of the cluster centre. We classify this structure as a candidate radio relic, as a relation with an optical counterpart is uncertain. This candidate radio relic may not be related to the cluster dynamics and could also be classified as a dying radio galaxy. This northern candidate radio relic has a total size of $340\times540$~kpc$^2$ and a flux density of $S_{1.28~GHz}$ = 2.3~mJy, which corresponds to a radio power of 6.5$\times$10$^{23}$ W~Hz$^{-1}$ at the cluster redshift. We note this northern candidate radio relic structure with caution due to its likely relation to an optical counterpart. We also report an estimate for the candidate radio halo flux density as $S_{1.28~GHz}$ = 0.6~mJy.

The spectral index map shows that the diffuse, relatively round patch of radio emission that is present north of the cluster (noted here as a candidate relic) has a spectral index value of $\alpha\sim-$1 at its peak, which steepens in annuli to $\alpha\sim-$2.3.

\subsection{J1539.5--8335}

J1539.5--8335 is a galaxy cluster at redshift $\textit{z}=0.073$ that was initially part of the REFLEX \citep{Bohringeretal04} sample and later included in the MCXC \citep{2011A&A...534A.109P} cluster catalogue. The system has an X--ray luminosity of L$_{500}$ = $2.24\times10^{44}$~erg~s$^{-1}$ and an M$_{500}$ mass of 3.18 $\times$ 10$^{14}$ M$_{\odot}$  \citep{2011A&A...534A.109P}. 

\cite{2010A&A...513A...2G} studied the \textit{XMM-Newton} data of J1539.5--8335 and found that it is a cool-core cluster. Its \textit{Chandra} image shows a relaxed cluster morphology, but also displays substructure (X--ray cavities) which suggests the existence of AGN feedback in the core. \citet{Hoganetal15} using radio data from the ATCA reports J1539.5--8335 as a core-dominated unresolved radio source at C (5~GHz) and X (8~GHz) bands, having a flat radio spectrum. The SUMSS detection at 843~MHz \citep{SUMSS} suggests a self-absorption turnover between the L and C bands instead of any additional non-core flux component at lower frequencies. \citet{Hoganetal15} place an upper limit of $\leq$52~mJy at 1~GHz for the non-core component.

Our 1.28~GHz MeerKAT images of this cluster are presented in Figure~\ref{fig:J1539.5}. Both full (rms =~3.5~$\mu$Jy~beam$^{-1}$) and low-resolution (rms =~6~$\mu$Jy~beam$^{-1}$) images reveal, for the first time, the detection of a faint diffuse component surrounding the central compact AGN radio source that has been reported in previous studies. The total size of the diffuse emission is $200\times230$~kpc$^2$ with the flux density being $S_{1.28~GHz}$ = 10.5~mJy which corresponds to a radio power of $P_{1.28~GHz}$ = 1.3$\times$10$^{23}$ W~Hz$^{-1}$. We classify this faint diffuse emission as a radio mini-halo, with the seed electrons for the diffuse emission provided by the central AGN radio source. We report that the in-band average spectral index of the mini-halo is estimated to be $\alpha_{908}^{1656}\leq -2.5$. In addition, we detect an extended diffuse radio emission in the south--west direction from the cluster centre, which we classify as a candidate radio relic. There is no obvious optical counterpart related to this structure, which on its north side appears to be contaminated by compact radio sources. The flux density of the southwest candidate radio relic feature is estimated to be $S_{1.28~GHz}$ = 4.0~mJy with a size of $100\times200$~kpc$^2$. 


The in-band spectral index map shows a strong flat spectrum ($\alpha\sim-$0.1) for the compact radio
source associated with the BCG. There appears to be a sharp increase in the spectral index between the value at the BCG location and that in the surrounding region, which steepens to values of $\sim-$1.7. The southwest candidate relic has a flat region ($\alpha\sim-$0.6) and steepens along the
major axis, reaching values of $\alpha\sim-$1. It also steepens perpendicular to the major axis, up to values of $\alpha\sim-$1.5.

\subsection{J1601.7--7544}

J1601.7--7544 (also known as PSZ1~G313.87–17.10) is a massive galaxy cluster at redshift $\textit{z}=0.153$ that was initially part of the ROSAT All-Sky Survey Bright Source Catalogue (RASS-BSC; \citealt{Vogesetal99}) and later included in the MCXC \citep{2011A&A...534A.109P} and Planck \citep{Planckcollab11b} cluster samples. The system has an X--ray luminosity of L$_{500}$ = $8.71\times10^{44}$~erg~s$^{-1}$ and an M$_{500}$ mass of 6.88$\times10^{14}$ M$_\odot$ \citep{2011A&A...534A.109P}. 

\citet{2010A&A...513A...2G} using HRI-ROSAT and \textit{XMM-Newton} data reported an elongation in the X--ray emission along with a flat temperature profile, suggesting that J1601.7--7544 is a non-cool core cluster. However, in the recent X--ray study by \citet{2017ApJ...846...51L} the dynamical state of J1601.7--7544 cluster is classified as relaxed, with the \textit{Chandra} X--ray image showing also no signs of a major merger. 


The 1.28~GHz MeerKAT images presented in Figure~\ref{fig:J1601.7} show in both full and low-resolution (rms =~7~$\mu$Jy~beam$^{-1}$) the presence of a new extended diffuse radio emission blended with several compact radio sources at the cluster core. We classify this radio structure as a radio halo. The total size of the halo emission, which appears to be more extended in the low-resolution image and elongated in the south--east direction from the BCG, is $630\times980$~kpc$^2$. The flux density of the halo is found to be  $S_{1.28~GHz}$ = 14.2~mJy corresponding to a radio power of $P_{1.28~GHz}$ = 8.6$\times$10$^{23}$ W~Hz$^{-1}$. 

The in-band spectral index map shows features only in the central part of the radio halo, in the immediate surroundings of the central galaxy, with spectral index 
values in the range of $\alpha\sim-$1.7 to $-$2. Some other patches are visible, with spectral index values of $\alpha$ in the range of $\sim-$1.8 to $-$2.4. We report that the in-band average spectral index of the radio halo is estimated to be $\alpha_{908}^{1656}\leq -2$.

\subsection{J1840.6--7709}

J1840.6--7709 (also known as MCXC J1840.6--7709) is a relatively nearby system at redshift $\textit{z}=0.019$ that is part of the REFLEX \citep{Bohringeretal04} and the MCXC \citep{2011A&A...534A.109P} cluster samples. The system has an X--ray luminosity of L$_X$ = $0.12\times10^{44}$~erg~s$^{-1}$ and an M$_{500}$ mass of 0.54$\times10^{14}$ M$_\odot$ \citep{2011A&A...534A.109P}. \citet{Panagouliaetal14b} using \textit{XMM-Newton} data report the presence of a very bright source at the cluster centre and find no evidence of X--ray cavities in this system. \citet{Lovisarietal15} using also \textit{XMM-Newton} data report a cooling time of t$_{cool}$ = 0.05~Gyr for this system.

The 1.28~GHz MeerKAT images of this cluster are presented in Figure~\ref{fig:J1840.6} with the low-resolution image exhibiting the highest rms (rms =~20~$\mu$Jy~beam$^{-1}$) in the MGCLS sample due to the presence of a very bright radio source ($\sim$ 840~mJy) at the cluster centre. Both full and low-resolution images show the presence of a bright, compact radio source, also surrounded by diffuse radio emission. The presence of a very faint, diffuse component surrounding the central radio source is detected; however, its nature is uncertain due to contamination by circular artefacts caused by improper calibration of this bright central radio source. With the current MeerKAT data, we classify this faint diffuse emission as a possible candidate radio mini-halo, with a total size of $47\times55$~kpc$^2$ having a flux density of $S_{1.28~GHz}$ = 28~mJy, which corresponds to a radio power of 2.2$\times$10$^{22}$ W~Hz$^{-1}$. The seed electrons for the diffuse emission are most likely provided by the central AGN radio source that appears to be a dying radio galaxy.

\subsection{J2023.4--5535/ SPT-CL J2023.4--5535}

SPT-CL~J2023-5535 is a merging cluster system at redshift $\textit{z} = 0.232$ detected by RASS \citep{Vogesetal99} and included in the Meta-Catalogue of X--ray detected Clusters (MCXC; \citealt{2011A&A...534A.109P}) and the REFLEX cluster survey \citep{Bohringeretal04}. Based on 843~MHz SUMSS images but restricted by survey limits, \citet{Birzanetal17} reported the existence of possible relics in SPT-CL~J2023-5535, which also showed evidence of a subcluster on its periphery. 

SPT-CL~J2023-5535 belongs to the category of clusters that present an elongated radio halo. \citet{2020ApJ...900..127H} using the deep-wide-field radio continuum pilot survey, ASKAP-EMU (800-1088 MHz) reports the presence of an elongated radio also the discovery of a radio relic. The deep, high-resolution ASKAP-EMU radio data reveal a $\sim$~2~Mpc scale radio halo elongated in the east-west direction, coinciding with the X--ray ICM gas. The location of the radio relic is in the western part of the detected radio halo with an LLS of $\sim$ 500~kpc and an in-band integrated spectral index in the band of $\alpha_{800MHz}^{1088MHz}=-0.76\pm0.06$ \citep{2020ApJ...900..127H}. Weak-lensing analysis shows that the system consists of at least three subclusters with \citet{2020ApJ...900..127H}, suggesting that the observed radio structures are the product of the collision between the eastern and middle subclusters. 

Our 1.28~GHz MeerKAT data of the SPT-CL~J2023-5535 reveal the two earlier observed radio structures in greater detail: i) the east-west elongated diffuse radio halo emission and ii) the western radio relic. The 15$^{\prime\prime}$ resolution image (see Figure~\ref{fig:J2023.4}) shows the radio halo with a total size of $700\times1200$~kpc that has a flux density of S$_{1.28GHz}$ = 18.2~mJy with several embedded point radio sources that are probably associated with galaxies. The average surface brightness of the radio halo structure is low ($\sim$~2.2~$\mu$Jy~arcsec$^{-2}$) over an area with a radius of 600~kpc. On the other hand, the western relic appears in the inner part of the western edge of the radio halo, having a concave shape bending inwards. It presents a flux density of S$_{1.28GHz}$ = 8.5~mJy with a total size of $150\times640$~kpc. The western radio relic is connected to the radio halo and linked southwest to a head-tail radio source.  

From the MeerKAT in-band spectral index maps, only the most compact features are bright enough to show structure in the spectral index image. The few patches of radio halo emission have a very steep spectrum, i.e., up to approximately $-$3.5. The western relic presents a spectrum with an $\alpha$ value in the range of $-$1.2 to $-$1.4 in the brightest part, which then steepens to values of up to $-$2.5 in the northern direction and even to $-$3 in the southern direction. We also report that the head-tail radio source just south of the west relic has a spectral index of $\sim-$0.6 to $-$0.7 in the head, which steepens gradually to $\sim-$2.3 along the tail that connects with the west relic. On the eastern side, only the brightest ridge is detected in the spectral map, with an index of  $\sim-$2.5. The presence of a candidate relic southeast of the radio halo is possible. However, the elongated radio structure consists of several blending point radio sources, as seen in the 7.8$^{\prime\prime}$ resolution image, with some of them located in the central part of the structure and having optical counterparts. In addition, the in-band spectral index map detects only the bright point radio source at the centre of the structure, with a spectral index of approximately $-$2 to $-$2.5. This also argues against a candidate relic structure and points towards an AGN-related source with a superposition of other point radio sources.


\clearpage

\onecolumn
\FloatBarrier
\section{MGCLS RADIO IMAGES}
\label{AppC}
In this appendix, we provide MeerKAT 1.28~GHz radio and Digitized Sky Survey optical images for the 56 clusters of the MGCLS that present some kind of diffuse radio emission. In all Figures, the full-resolution (7.8\arcsec\,$\times$\,7.8\arcsec) radio images are shown in the left panels, and the optical images in the right panel are overlaid by the 1.28~GHz low-resolution contours (15\arcsec\,$\times$\,15\arcsec) in green. The red (or white) $\times$, when shown, indicates the NED optical position of the cluster.

\mbox{} \mbox{} \mbox{} \mbox{} \mbox{} \mbox{} \mbox{} \mbox{} \mbox{} \mbox{} \mbox{} \mbox{} \mbox{} \mbox{} \mbox{} \mbox{} \mbox{} \mbox{} \mbox{} \mbox{} \mbox{} \mbox{} \mbox{} \mbox{} \mbox{} \mbox{} \mbox{} \mbox{} \mbox{} \mbox{} \mbox{} \mbox{} \mbox{} \mbox{} \mbox{} \mbox{} \mbox{} \mbox{} \mbox{} \mbox{} \mbox{} \mbox{} \mbox{} \mbox{} \mbox{} \mbox{} \mbox{} \mbox{} \mbox{} \mbox{} \mbox{} \mbox{} \mbox{} \mbox{} \mbox{} \mbox{} \mbox{} \mbox{} \mbox{} \mbox{} \mbox{} \mbox{} \mbox{} \mbox{} \mbox{} \mbox{} \mbox{} \mbox{} \mbox{} \mbox{} \mbox{} \mbox{} \mbox{} \mbox{} \mbox{} \mbox{} \mbox{} \mbox{} \mbox{} \mbox{} \mbox{} \mbox{} \mbox{} \mbox{} \mbox{} \mbox{} \mbox{} \mbox{} \mbox{} \mbox{} \mbox{} \mbox{} \mbox{} \mbox{} \mbox{} \mbox{} \mbox{} \mbox{} \mbox{} \mbox{} \mbox{} \mbox{} \mbox{} \mbox{} \mbox{} \mbox{} \mbox{} \mbox{} \mbox{} \mbox{} \mbox{} \mbox{} \mbox{} \mbox{} \mbox{} \mbox{} \mbox{} \mbox{} \mbox{} \mbox{} \mbox{} \mbox{} \mbox{} \mbox{} \mbox{} \mbox{} \mbox{} \mbox{} \mbox{} \mbox{} \mbox{} \mbox{} \mbox{} \mbox{} \mbox{} \mbox{} \mbox{} \mbox{} \mbox{} \mbox{} \mbox{} \mbox{} \mbox{} \mbox{} \mbox{} \mbox{} \mbox{} \mbox{} \mbox{} \mbox{} \mbox{} \mbox{} \mbox{} \mbox{} \mbox{} \mbox{} \mbox{} \mbox{} \mbox{} \mbox{} \mbox{} \mbox{} \mbox{} \mbox{} \mbox{} \mbox{} \mbox{} \mbox{} \mbox{} \mbox{} \mbox{} \mbox{} \mbox{} \mbox{} \mbox{} \mbox{} \mbox{} \mbox{} \mbox{} \mbox{} \mbox{} \mbox{} \mbox{} \mbox{} \mbox{} \mbox{} \mbox{} \mbox{} \mbox{} \mbox{} \mbox{} \mbox{} \mbox{} \mbox{} \mbox{} \mbox{} \mbox{} \mbox{} \mbox{} \mbox{} \mbox{} \mbox{} \mbox{} \mbox{} \mbox{} \mbox{} \mbox{} \mbox{} \mbox{} \mbox{} \mbox{} \mbox{} \mbox{} \mbox{} \mbox{} \mbox{} \mbox{} \mbox{} \mbox{} \mbox{} \mbox{} \mbox{} \mbox{} \mbox{} \mbox{} \mbox{} \mbox{} \mbox{} \mbox{} \mbox{} \mbox{} \mbox{} \mbox{} \mbox{} \mbox{} \mbox{} \mbox{} \mbox{} \mbox{} \mbox{} \mbox{} \mbox{} \mbox{} \mbox{} \mbox{} \mbox{} \mbox{} \mbox{} \mbox{} \mbox{} \mbox{} \mbox{} \mbox{} \mbox{} \mbox{} \mbox{} \mbox{} \mbox{} \mbox{} \mbox{} \mbox{} \mbox{} \mbox{} \mbox{} \mbox{} \mbox{} \mbox{} \mbox{} \mbox{} \mbox{} \mbox{} \mbox{} \mbox{} \mbox{} \mbox{} \mbox{} \mbox{} \mbox{} \mbox{} \mbox{} \mbox{} \mbox{} \mbox{} \mbox{} \mbox{} \mbox{} \mbox{} \mbox{} \mbox{} \mbox{} \mbox{} \mbox{} \mbox{} \mbox{} \mbox{} \mbox{} \mbox{} \mbox{} \mbox{} \mbox{} \mbox{} \mbox{} \mbox{} \mbox{} \mbox{} \mbox{} \mbox{} \mbox{} \mbox{} \mbox{} \mbox{} \mbox{} \mbox{} \mbox{} \mbox{} \mbox{} \mbox{} \mbox{} \mbox{} \mbox{} \mbox{} \mbox{} \mbox{} \mbox{} \mbox{} \mbox{} \mbox{} \mbox{} \mbox{} \mbox{} \mbox{} \mbox{} \mbox{} \mbox{} \mbox{} \mbox{}

\begin{figure*}
   \centering
   \includegraphics[width=0.496\textwidth]{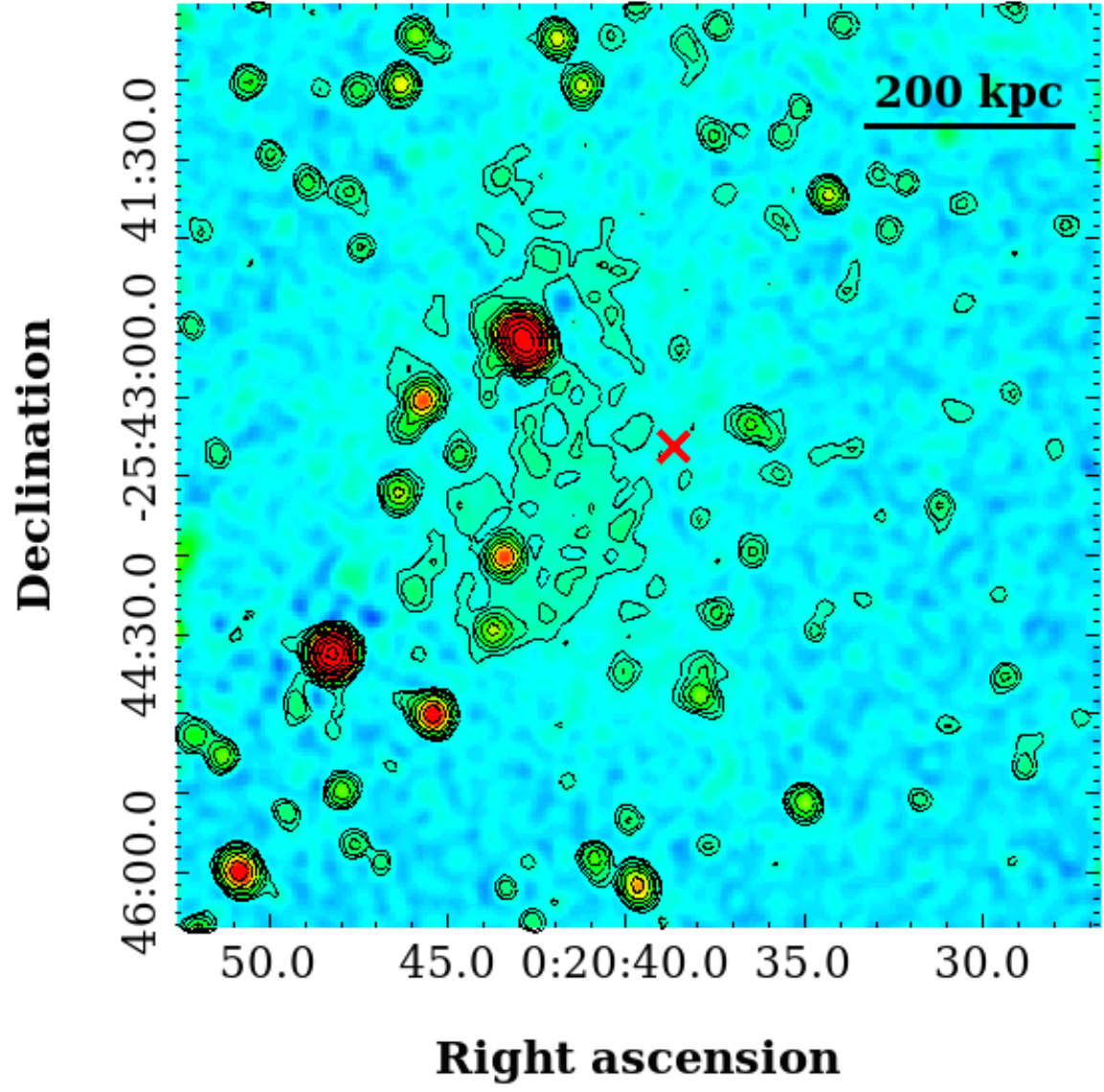}
    \includegraphics[width=0.5\textwidth]{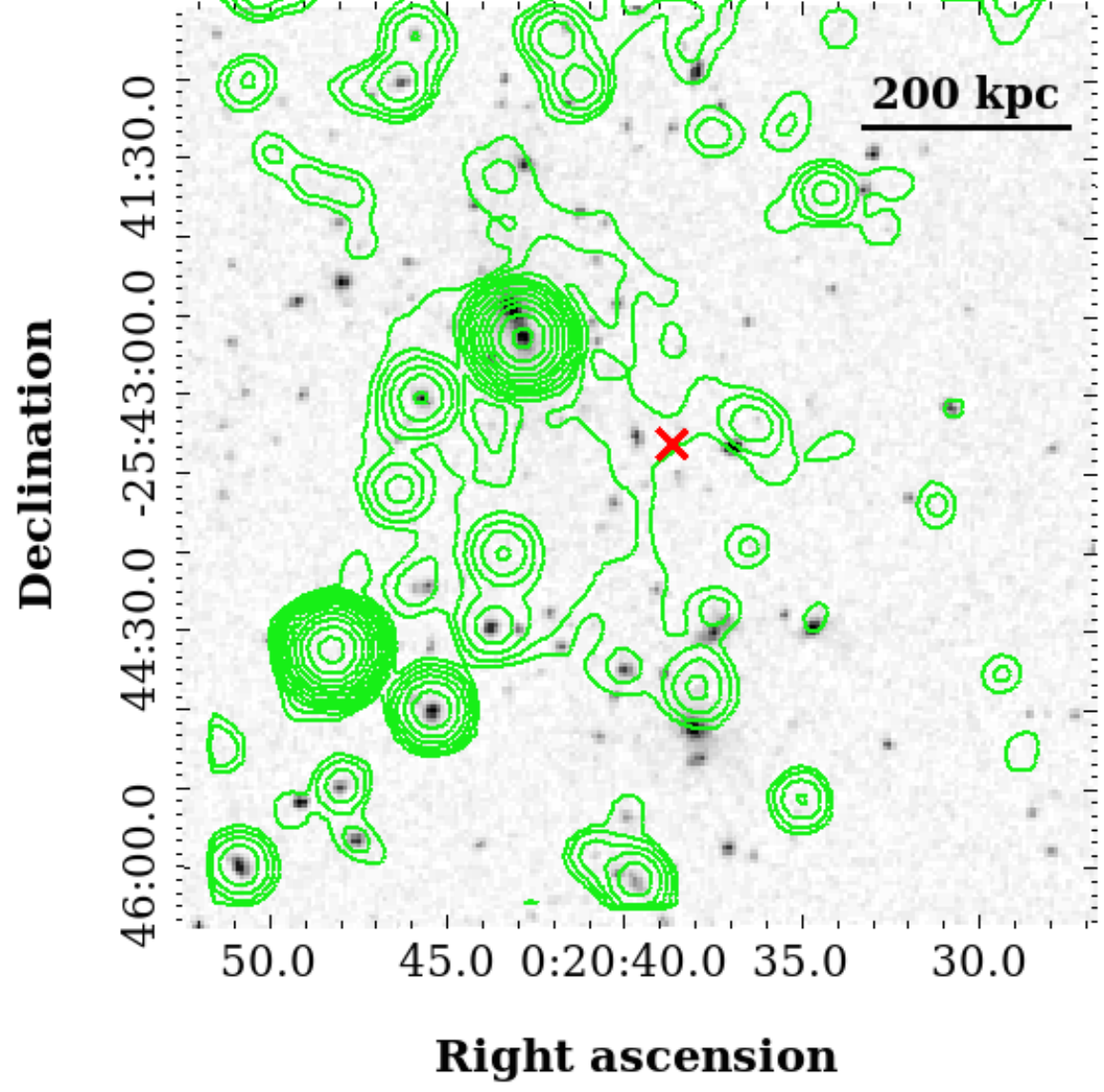}
   \caption{Abell~22 \textbf{Left:} Full-resolution (7.8\arcsec\,$\times$\,7.8\arcsec) 1.28~GHz MGCLS radio image with radio contours in black overlaid (1$\sigma$ = 3 $\mu$Jy beam$^{-1}$). \textbf{Right:} 1.28~GHz MGCLS low-resolution  (15\arcsec\,$\times$\,15\arcsec) radio contours in green (1$\sigma$ = 6 $\mu$Jy beam$^{-1}$), overlaid on the r-band \textit{Digitized Sky Survey (DSS)} optical image. In both panels, the radio contours start at 3\,$\sigma$ and rise by a factor of 2. The physical scale at the cluster redshift is indicated on the top right, and the red $\times$ indicates the NED cluster position. } 
   \label{fig:A22}%
\end{figure*}

\begin{figure*}
   \centering
   \includegraphics[width=0.496\textwidth]{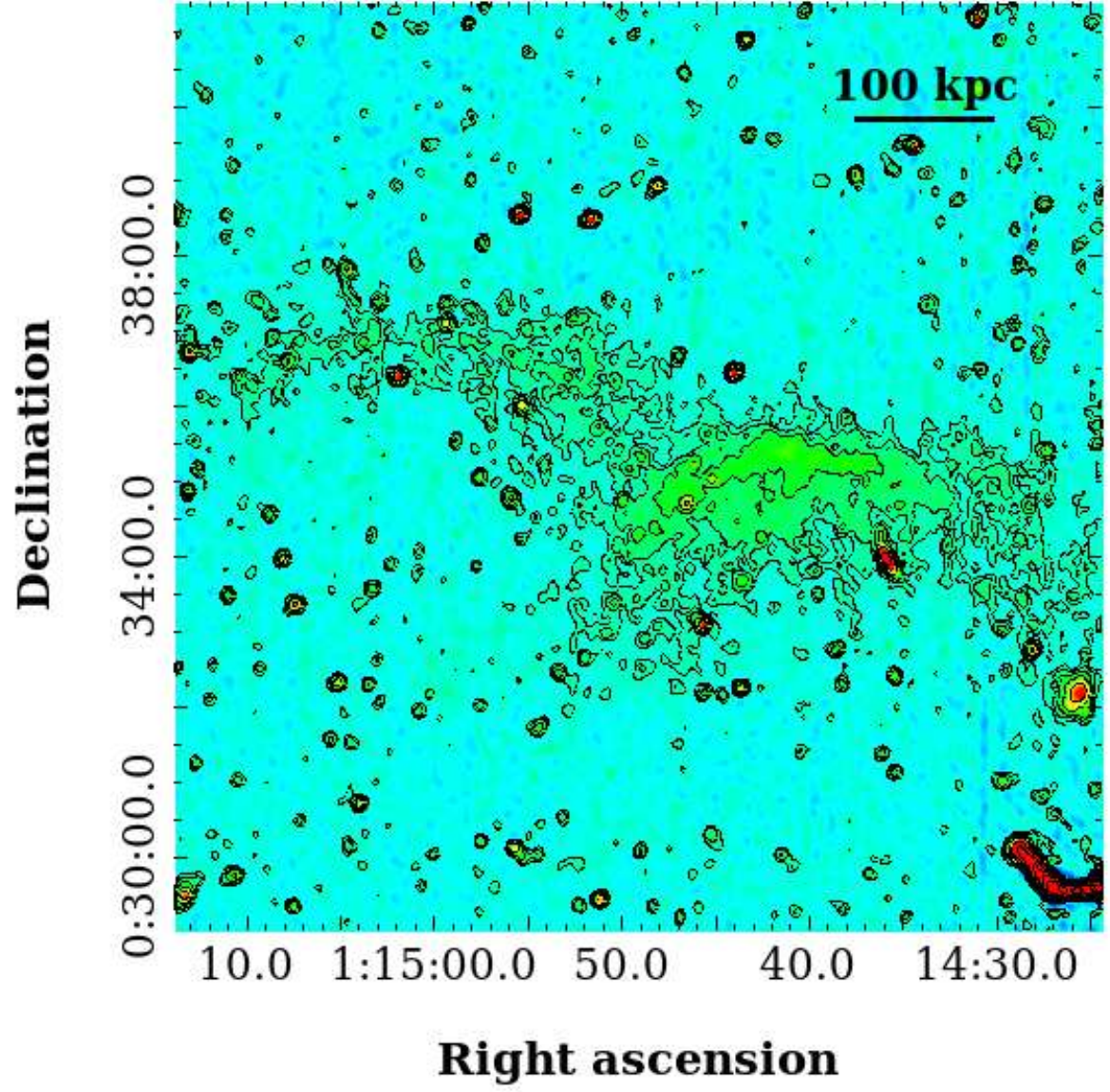}
    \includegraphics[width=0.5\textwidth]{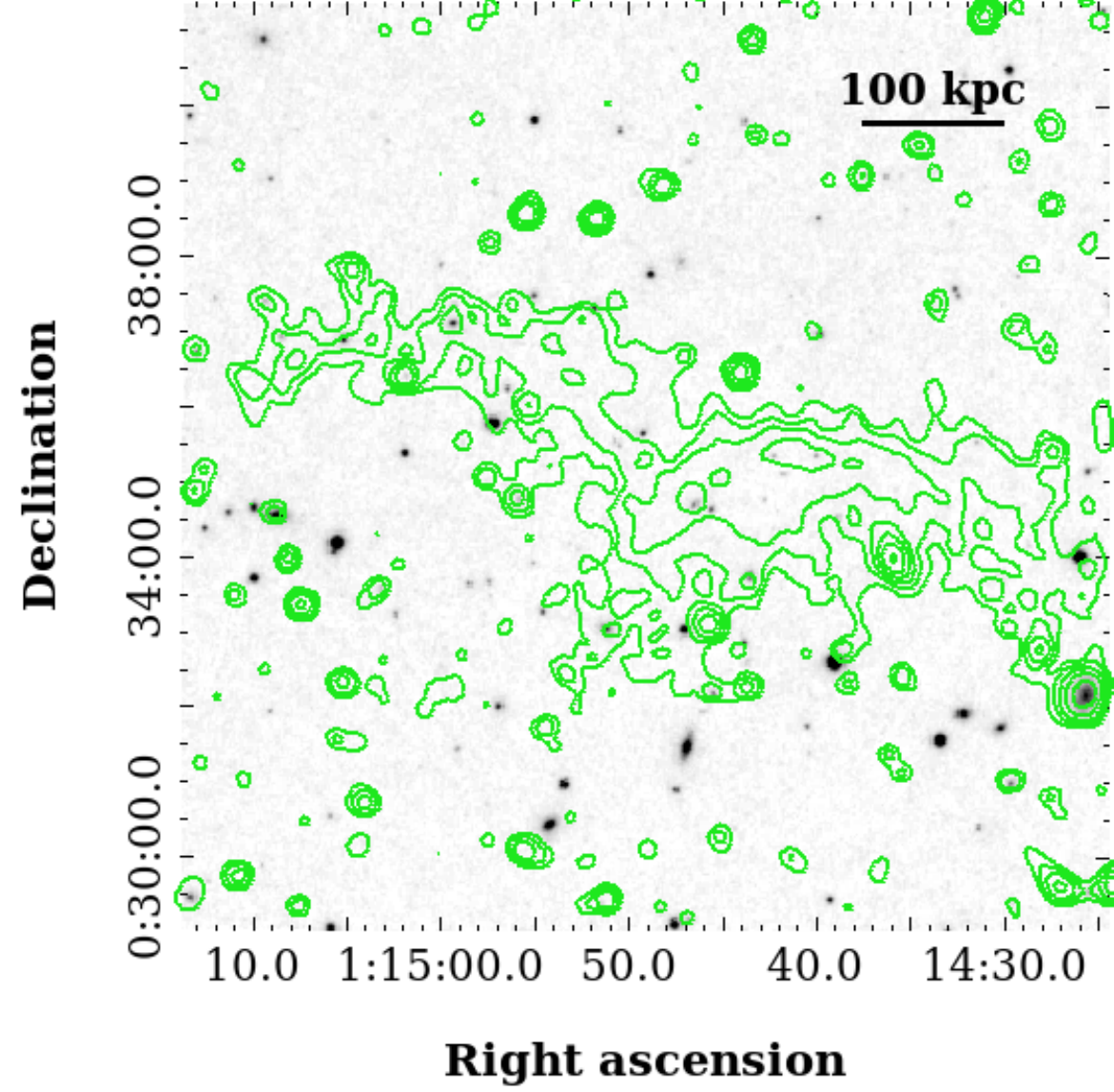}
   \caption{Abell~168 \textbf{Left:} Full-resolution (7.8\arcsec\,$\times$\,7.8\arcsec) 1.28~GHz MGCLS radio image with radio contours in black overlaid (1$\sigma$ = 5 $\mu$Jy beam$^{-1}$). \textbf{Right:} 1.28~GHz MGCLS low-resolution  (15\arcsec\,$\times$\,15\arcsec) radio contours in green (1$\sigma$ = 12 $\mu$Jy beam$^{-1}$), overlaid on the r-band \textit{Digitized Sky Survey (DSS)} optical image. In both panels, the radio contours start at 3\,$\sigma$ and rise by a factor of 2. The physical scale at the cluster redshift is indicated on top right. We note that the relic shown here is located $\sim$~1.1~Mpc north from the NED cluster position. } 
   \label{fig:A168}%
\end{figure*}

\begin{figure*}
   \centering
   \includegraphics[width=0.496\textwidth]{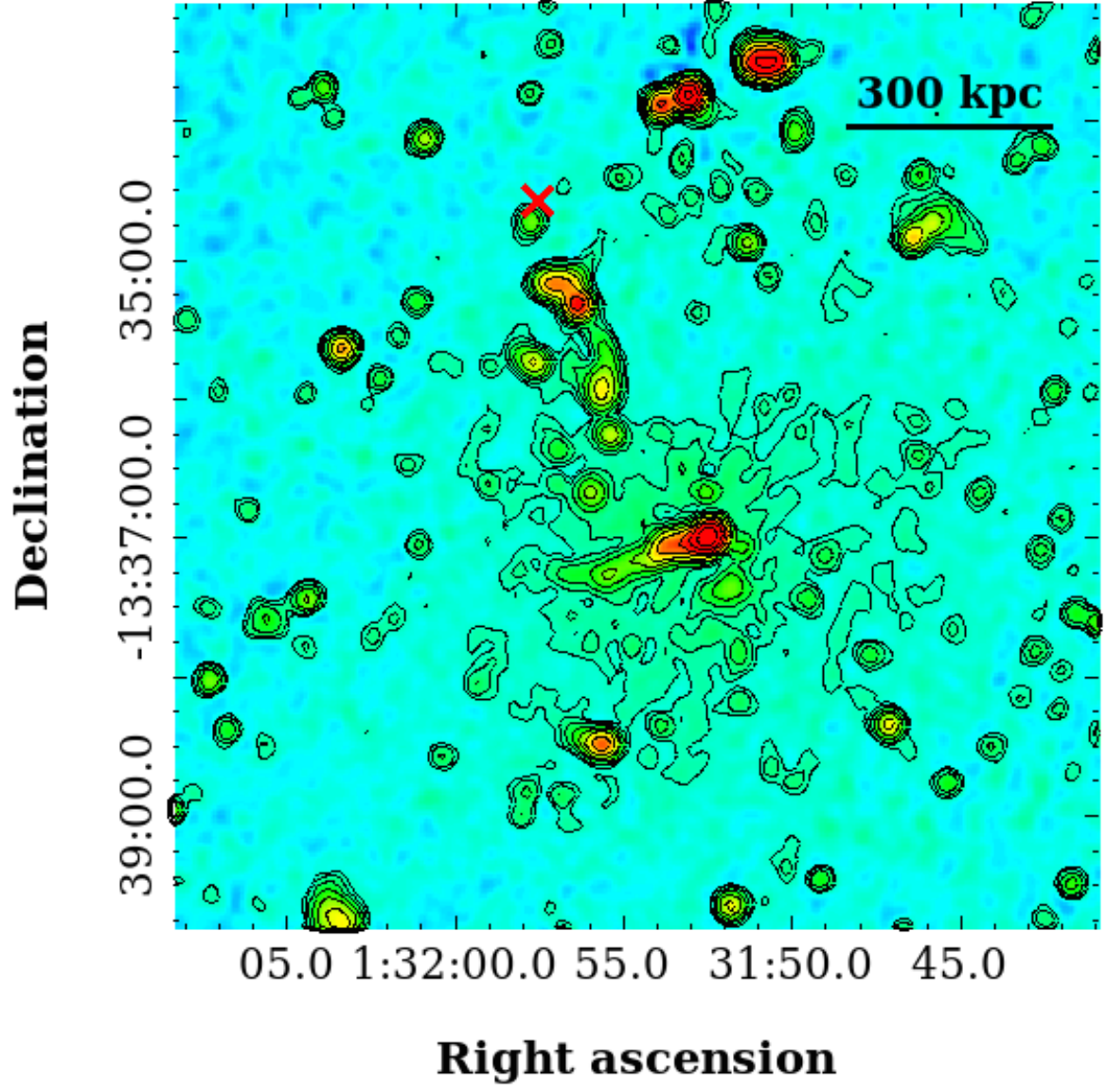}
    \includegraphics[width=0.5\textwidth]{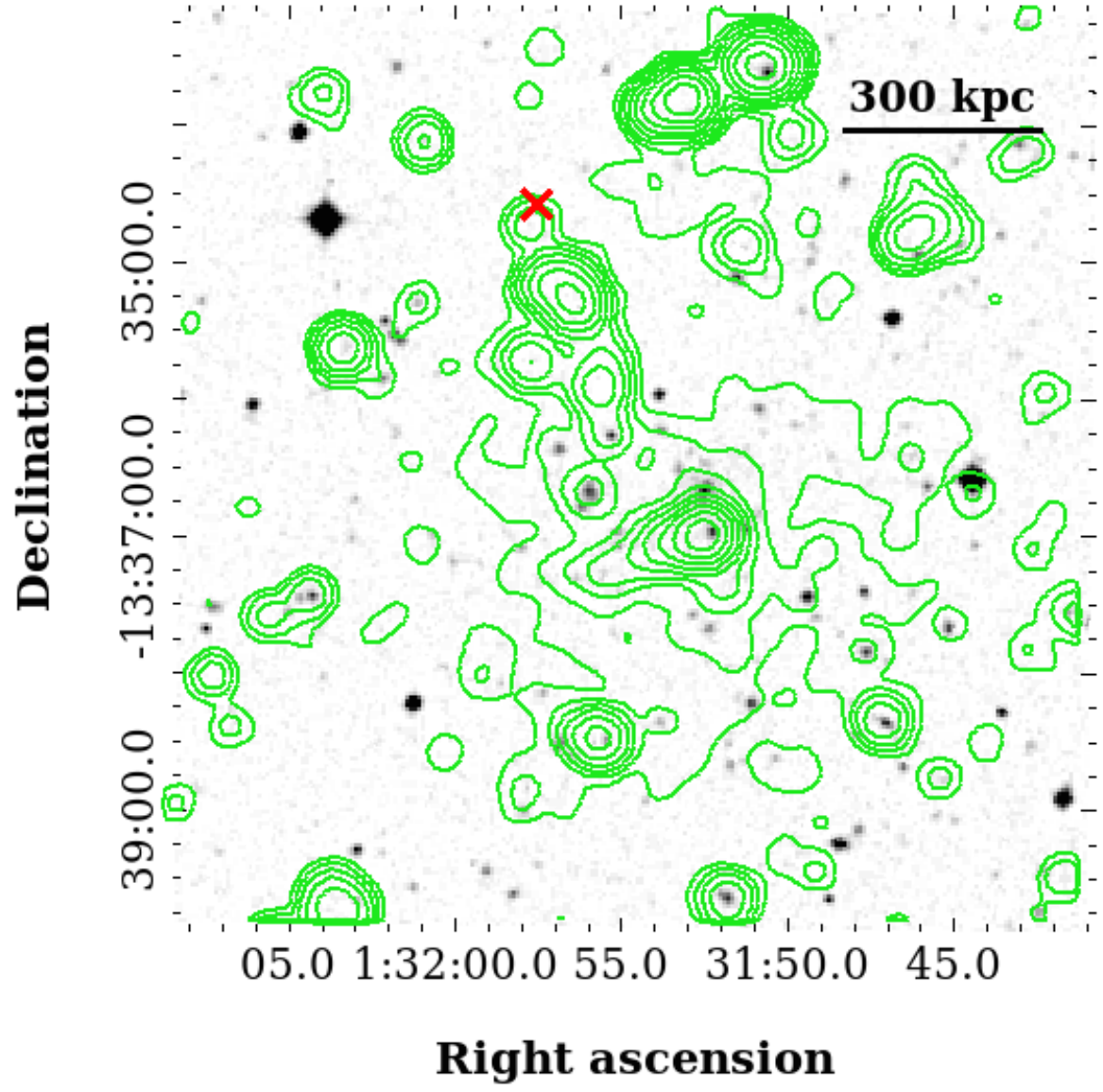}
   \caption{Abell~209 \textbf{Left:}  Full-resolution (7.8\arcsec\,$\times$\,7.8\arcsec) 1.28~GHz MGCLS radio image with radio contours in black overlaid (1$\sigma$ = 4 $\mu$Jy beam$^{-1}$). \textbf{Right:} 1.28~GHz MGCLS low-resolution  (15\arcsec\,$\times$\,15\arcsec) radio contours in green (1$\sigma$ = 12 $\mu$Jy beam$^{-1}$), overlaid on the r-band \textit{Digitized Sky Survey (DSS)} optical image. In both panels, the radio contours start at 3\,$\sigma$ and rise by a factor of 2. The physical scale at the cluster redshift is indicated on top right, and the red $\times$ indicates the NED cluster position. } 
   \label{fig:A209}%
\end{figure*}

\begin{figure*}
   \centering
   \includegraphics[width=0.496\textwidth]{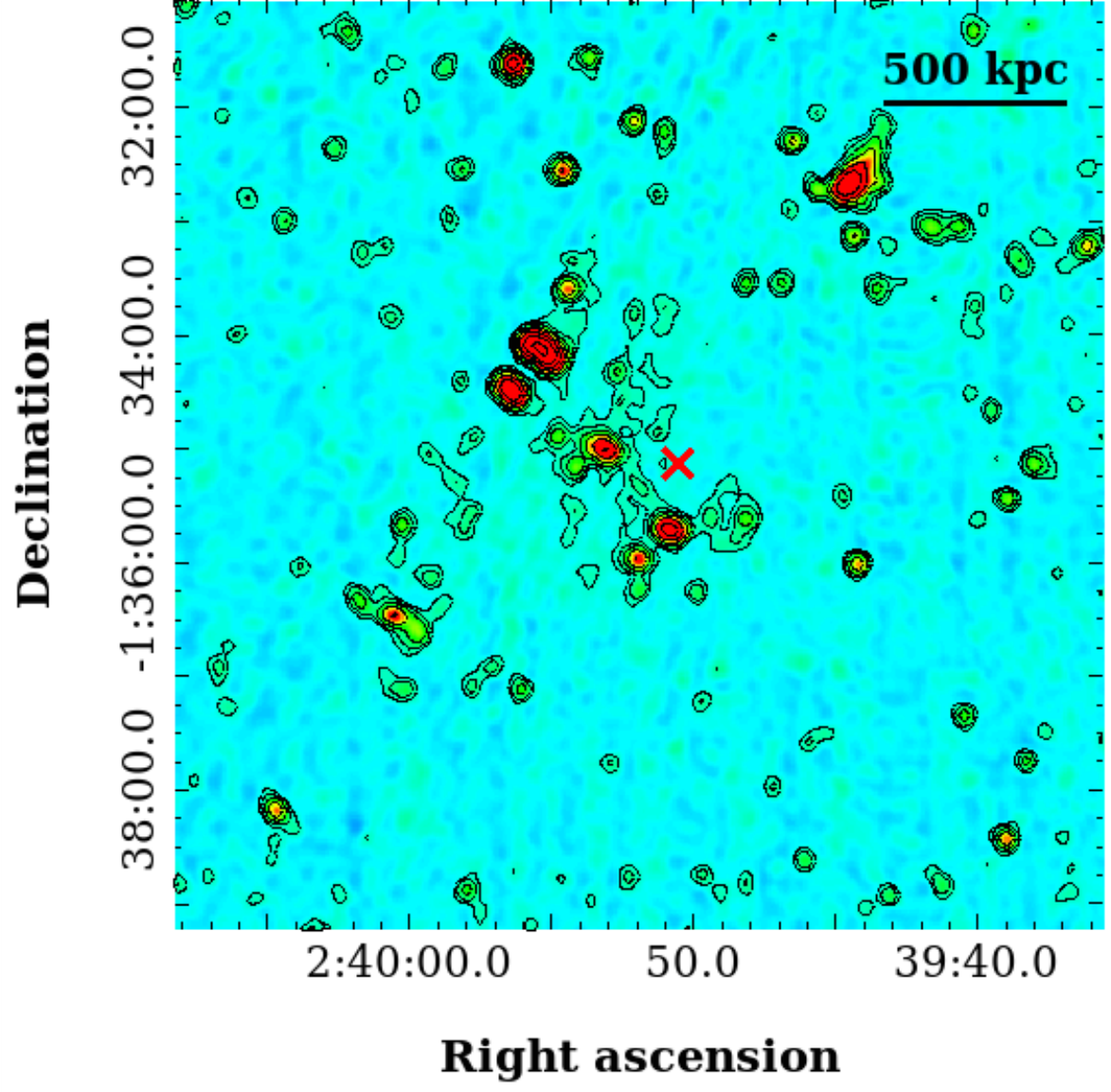}
    \includegraphics[width=0.5\textwidth]{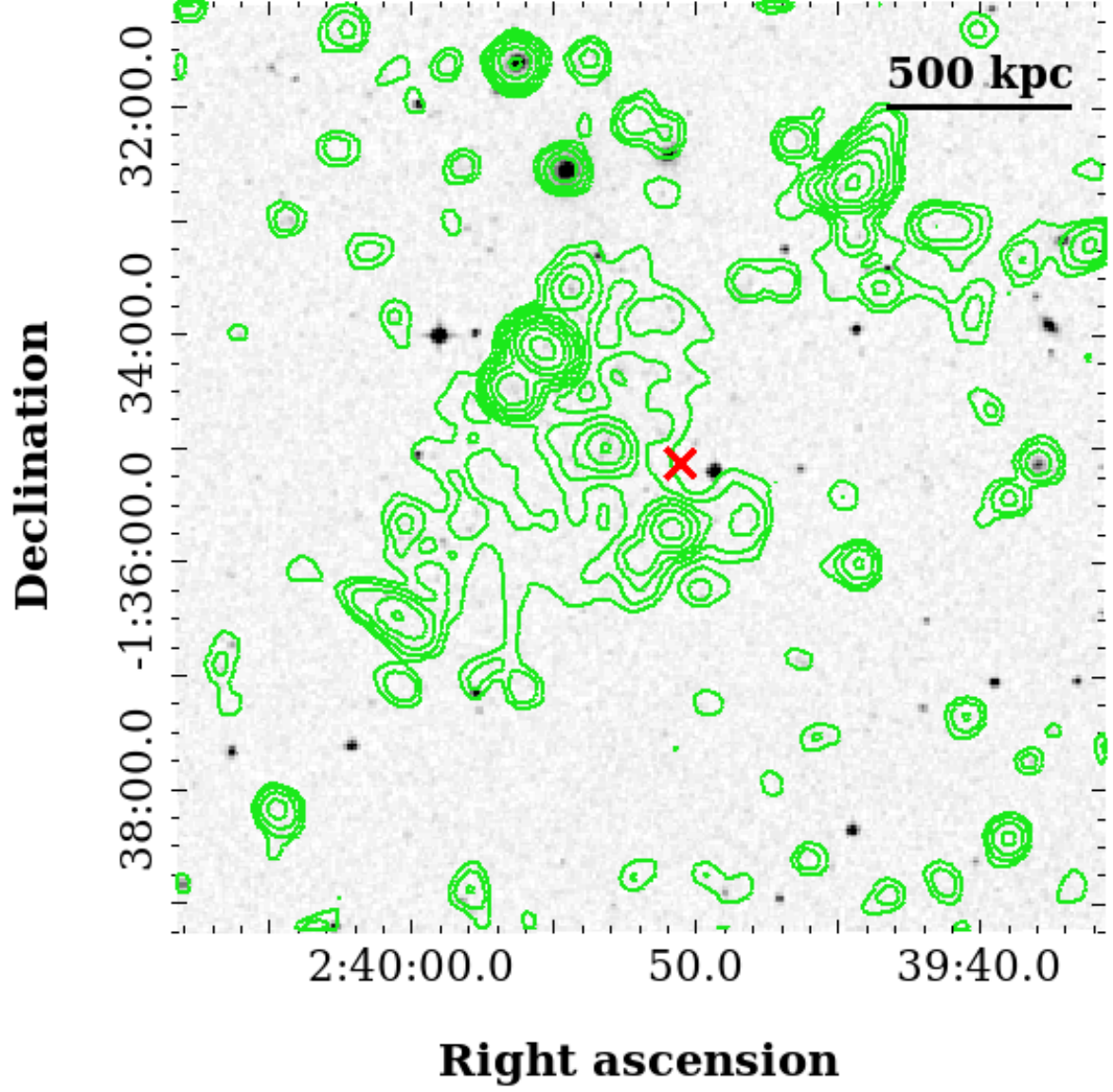}
   \caption{Abell~370 \textbf{Left:}  Full-resolution (7.8\arcsec\,$\times$\,7.8\arcsec) 1.28~GHz MGCLS radio image with radio contours in black overlaid (1$\sigma$ = 7 $\mu$Jy beam$^{-1}$). \textbf{Right:} 1.28~GHz MGCLS low-resolution  (15\arcsec\,$\times$\,15\arcsec) radio contours in green (1$\sigma$ = 10 $\mu$Jy beam$^{-1}$), overlaid on the r-band \textit{Digitized Sky Survey (DSS)} optical image. In both panels, the radio contours start at 3\,$\sigma$ and rise by a factor of 2. The physical scale at the cluster redshift is indicated on top right, and the red $\times$ indicates the NED cluster position. } 
   \label{fig:A370}%
\end{figure*}

\begin{figure*}
   \centering
   \includegraphics[width=0.496\textwidth]{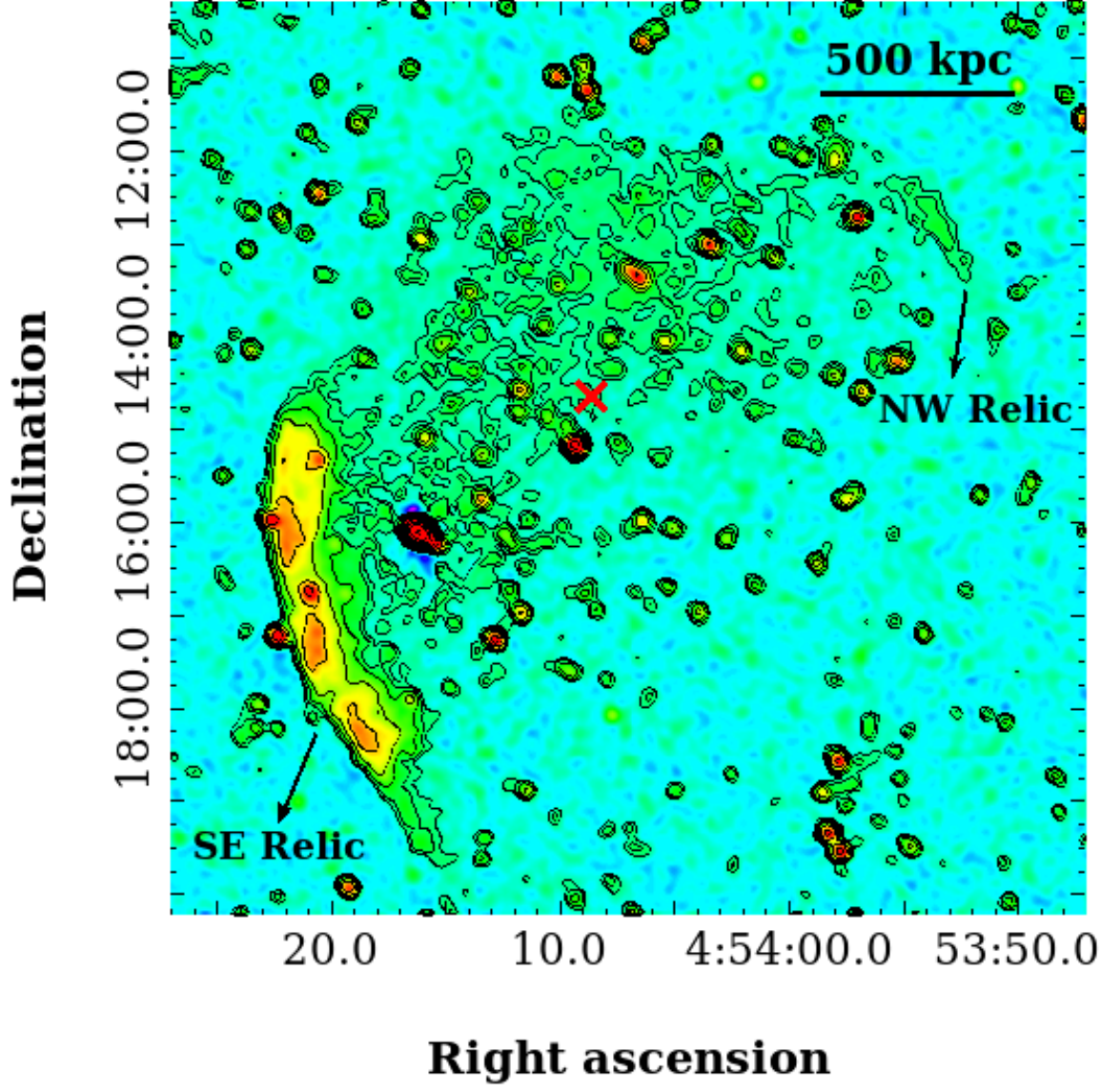}
    \includegraphics[width=0.5\textwidth]{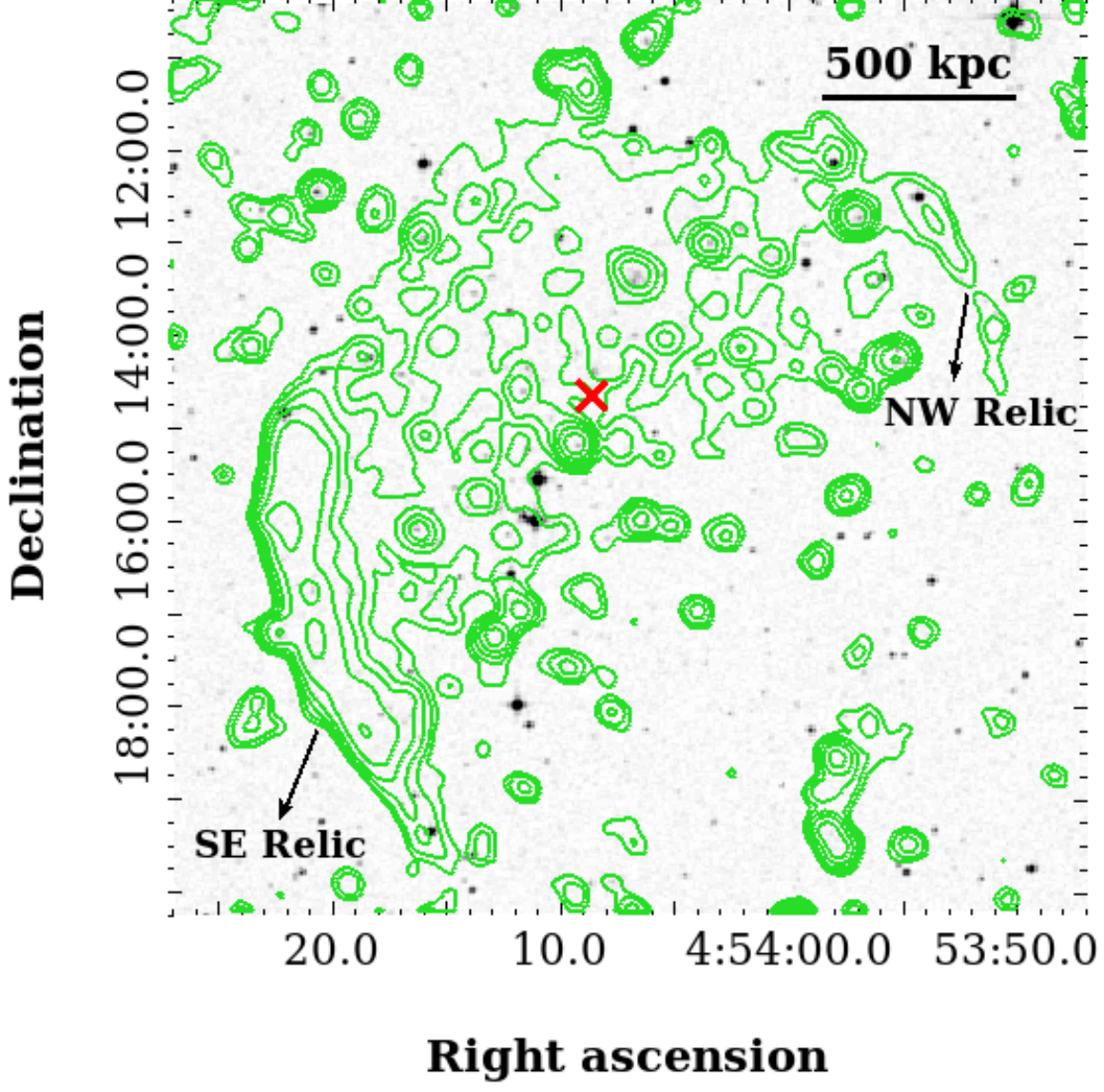}
   \caption{Abell~521 \textbf{Left:}  Full-resolution (7.8\arcsec\,$\times$\,7.8\arcsec) 1.28~GHz MGCLS radio image with radio contours in black overlaid (1$\sigma$ = 3.5 $\mu$Jy beam$^{-1}$). \textbf{Right:} 1.28~GHz MGCLS low-resolution  (15\arcsec\,$\times$\,15\arcsec) radio contours in green (1$\sigma$ = 6 $\mu$Jy beam$^{-1}$), overlaid on the r-band \textit{Digitized Sky Survey (DSS)} optical image. In both panels, the radio contours start at 3\,$\sigma$ and rise by a factor of 2. The physical scale at the cluster redshift is indicated on the top right, and the red $\times$ indicates the NED cluster position. } 
   \label{fig:A521}%
\end{figure*}

\begin{figure*}
   \centering
   \includegraphics[width=0.496\textwidth]{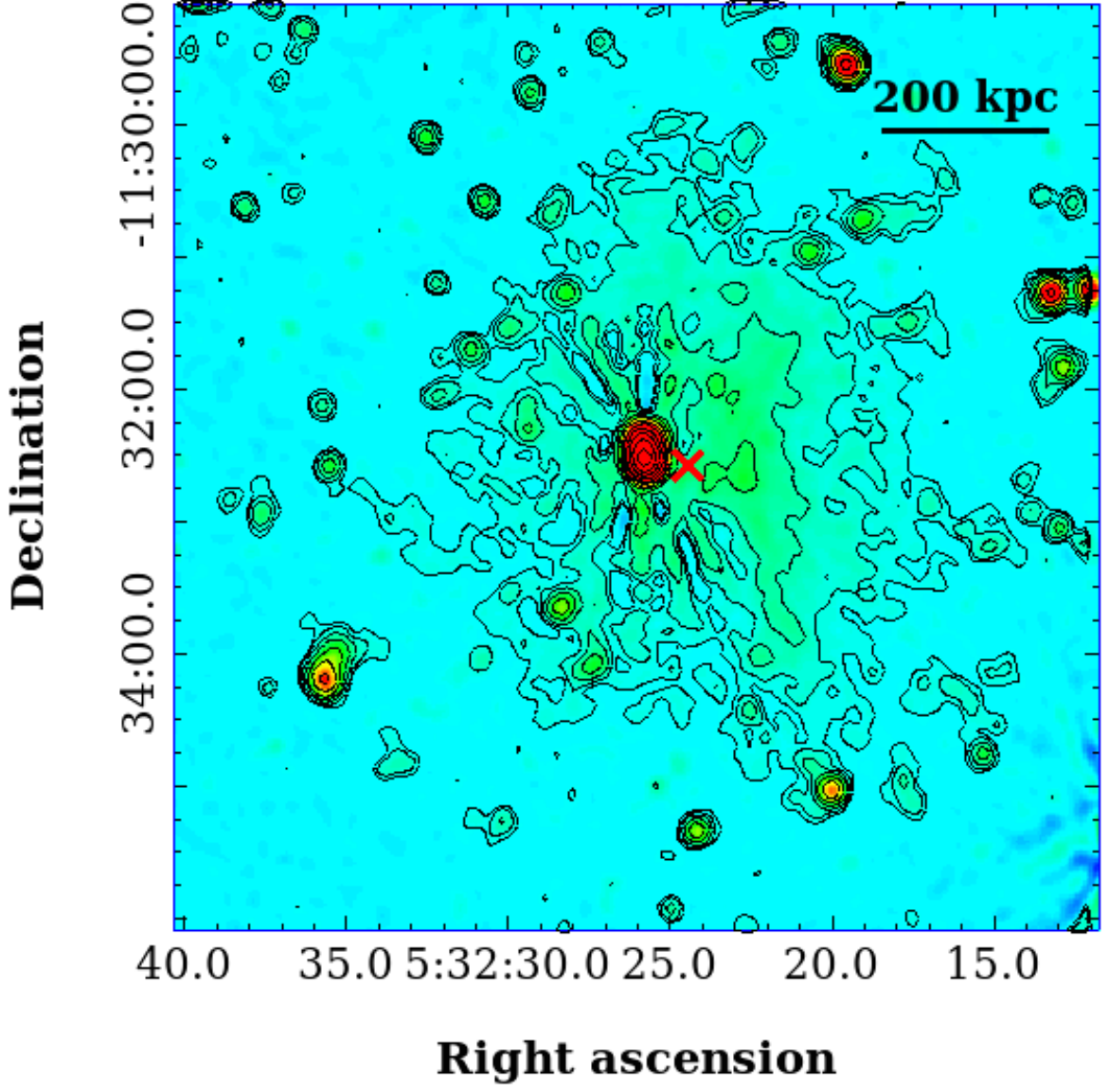}
    \includegraphics[width=0.5\textwidth]{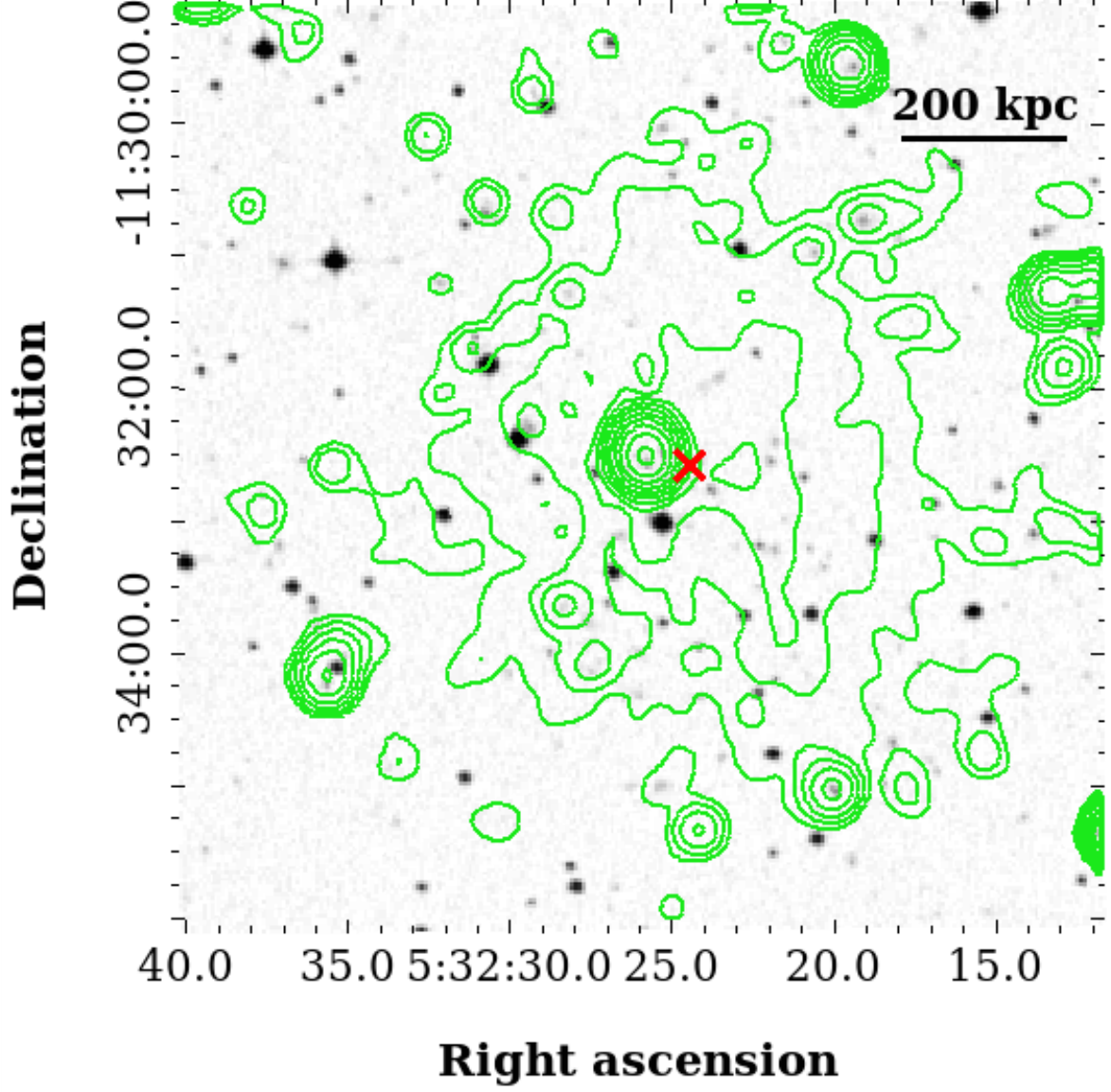}
   \caption{Abell~545 \textbf{Left:}  Full-resolution (7.8\arcsec\,$\times$\,7.8\arcsec) 1.28~GHz MGCLS radio image with radio contours in black overlaid (1$\sigma$ = 3.5 $\mu$Jy beam$^{-1}$). \textbf{Right:} 1.28~GHz MGCLS low-resolution  (15\arcsec\,$\times$\,15\arcsec) radio contours in green (1$\sigma$ = 11 $\mu$Jy beam$^{-1}$), overlaid on the r-band \textit{Digitized Sky Survey (DSS)} optical image. In both panels, the radio contours start at 3\,$\sigma$ and rise by a factor of 2. The physical scale at the cluster redshift is indicated on top right, and the red $\times$ indicates the NED cluster position. } 
   \label{fig:A545}%
\end{figure*}

\begin{figure*}
   \centering
   \includegraphics[width=0.496\textwidth]{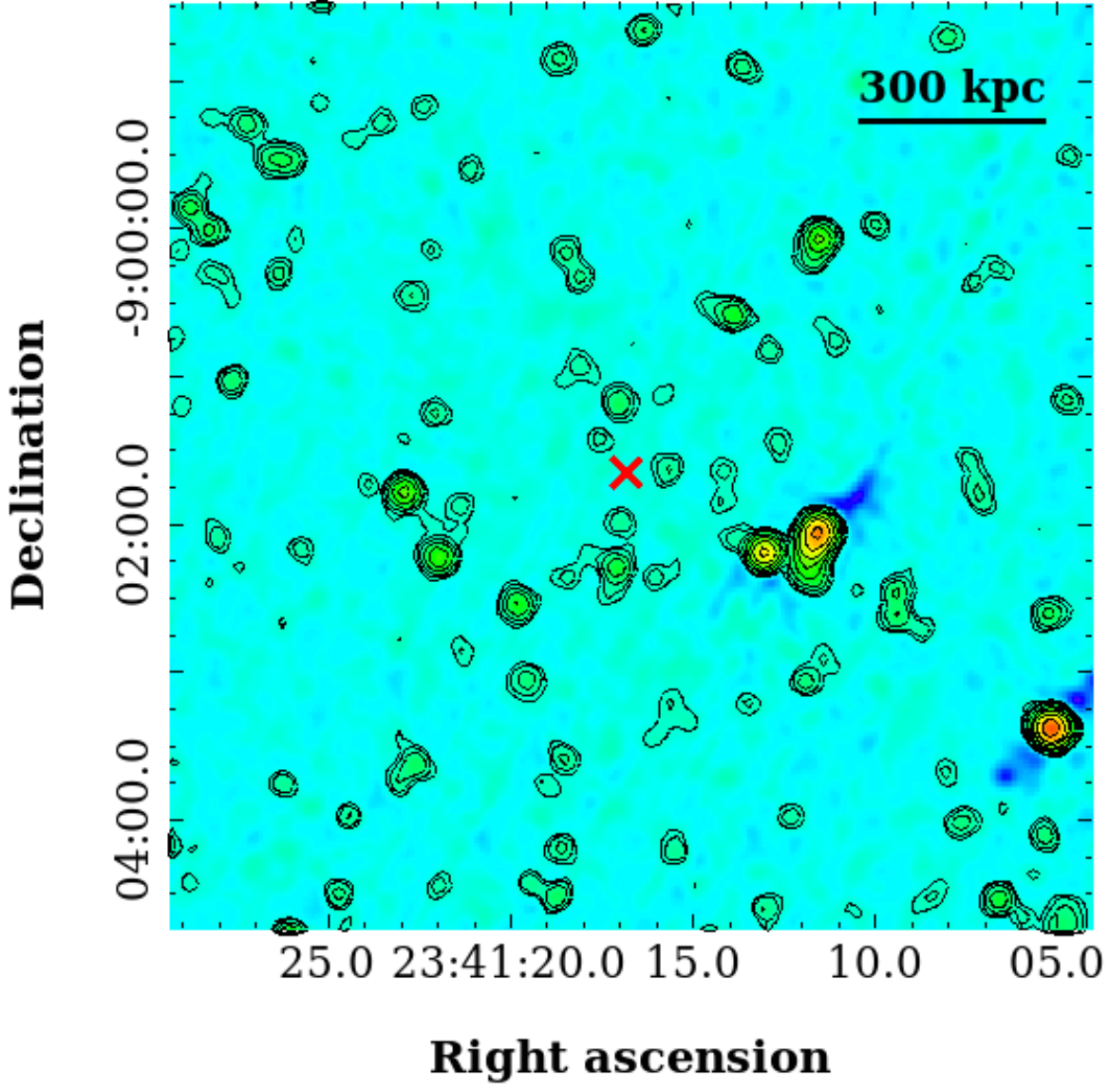}
    \includegraphics[width=0.5\textwidth]{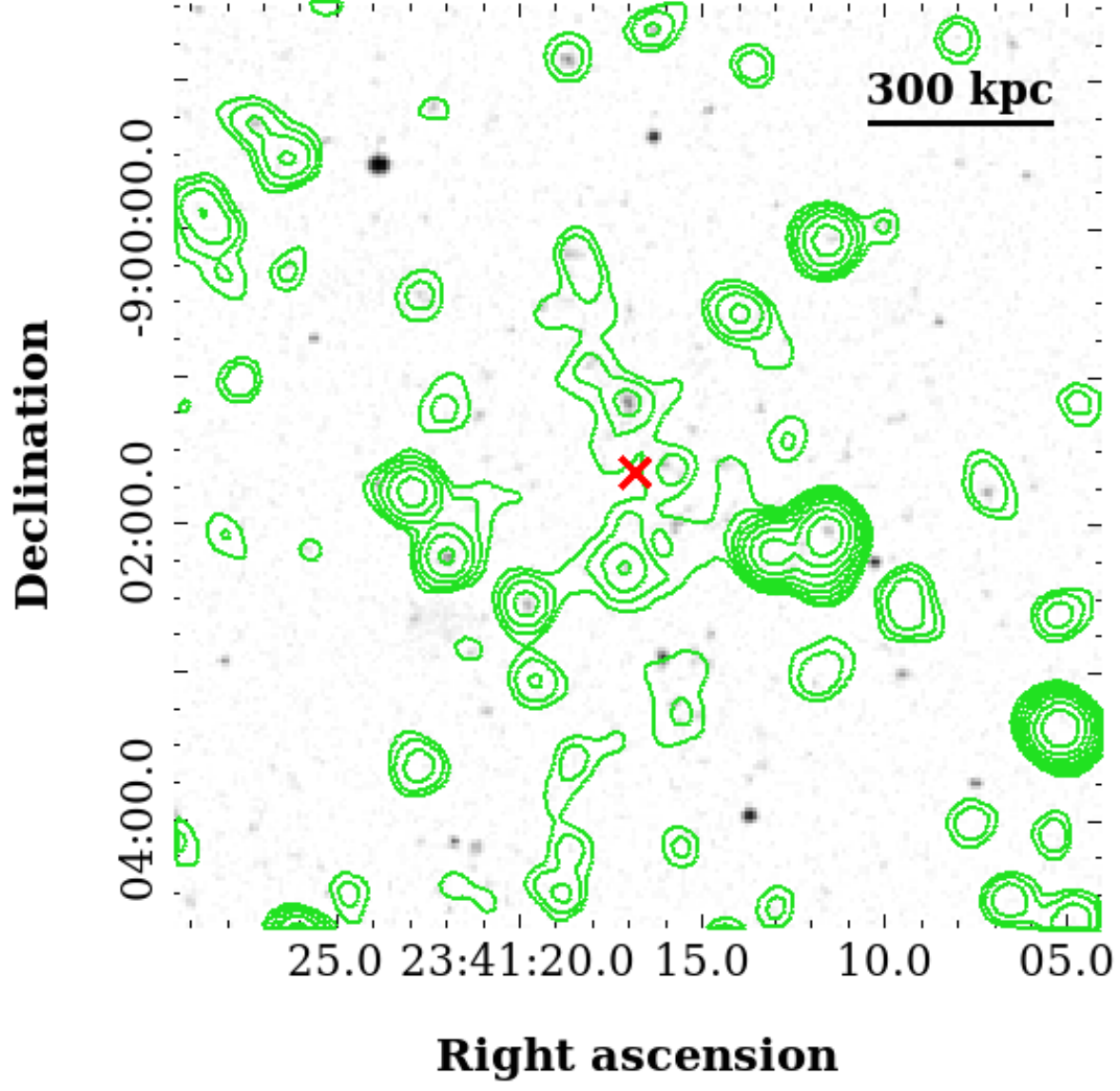}
   \caption{Abell~2645 \textbf{Left:}  Full-resolution (7.8\arcsec\,$\times$\,7.8\arcsec) 1.28~GHz MGCLS radio image with radio contours in black overlaid (1$\sigma$ = 4 $\mu$Jy beam$^{-1}$). \textbf{Right:} 1.28~GHz MGCLS low-resolution  (15\arcsec\,$\times$\,15\arcsec) radio contours in green (1$\sigma$ = 7 $\mu$Jy beam$^{-1}$), overlaid on the r-band \textit{Digitized Sky Survey (DSS)} optical image. In both panels, the radio contours start at 3\,$\sigma$ and rise by a factor of 2. The physical scale at the cluster redshift is indicated on top right, and the red $\times$ indicates the NED cluster position. } 
   \label{fig:A2645}%
\end{figure*}

\begin{figure*}
   \centering
   \includegraphics[width=0.496\textwidth]{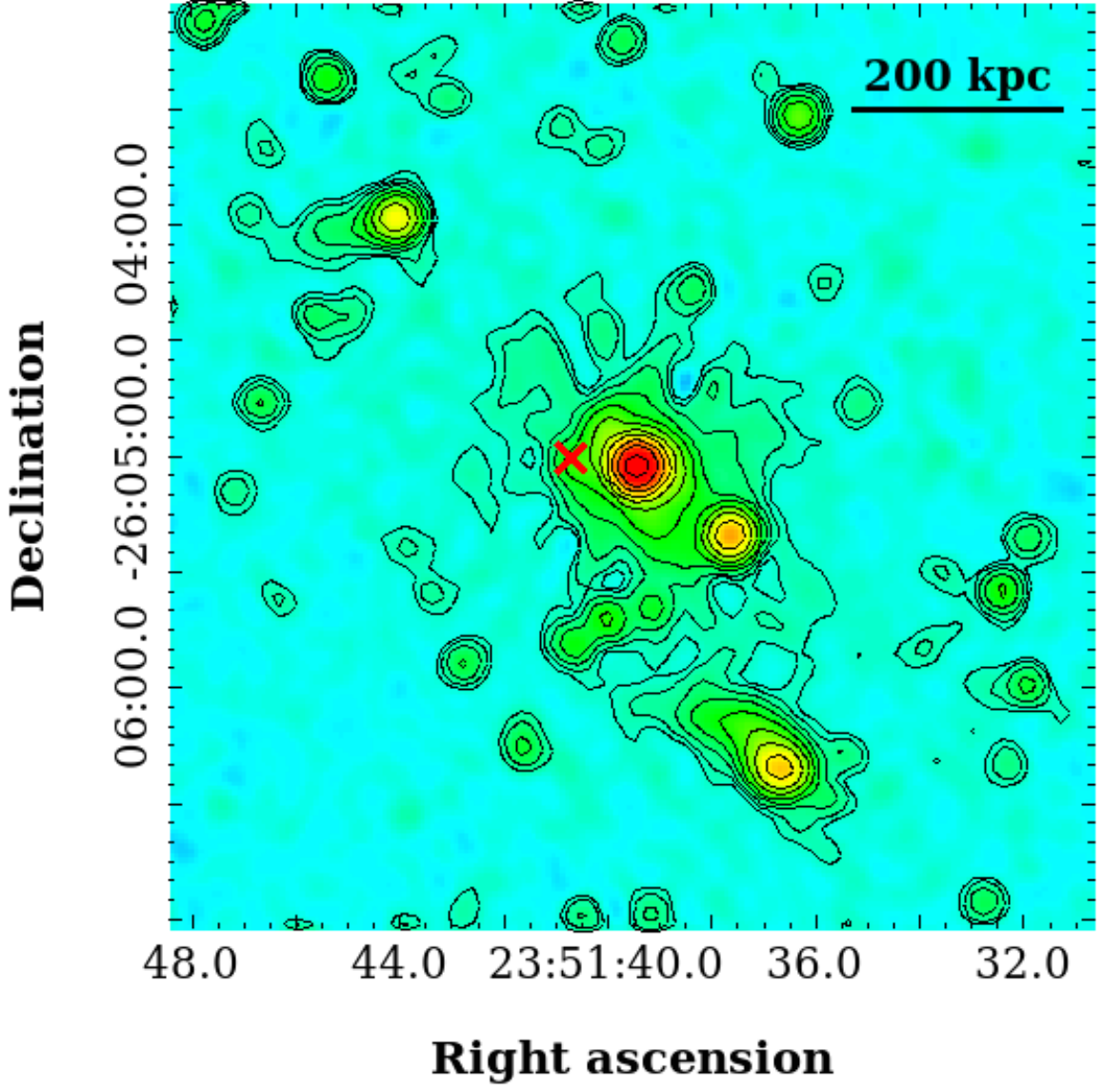}
    \includegraphics[width=0.5\textwidth]{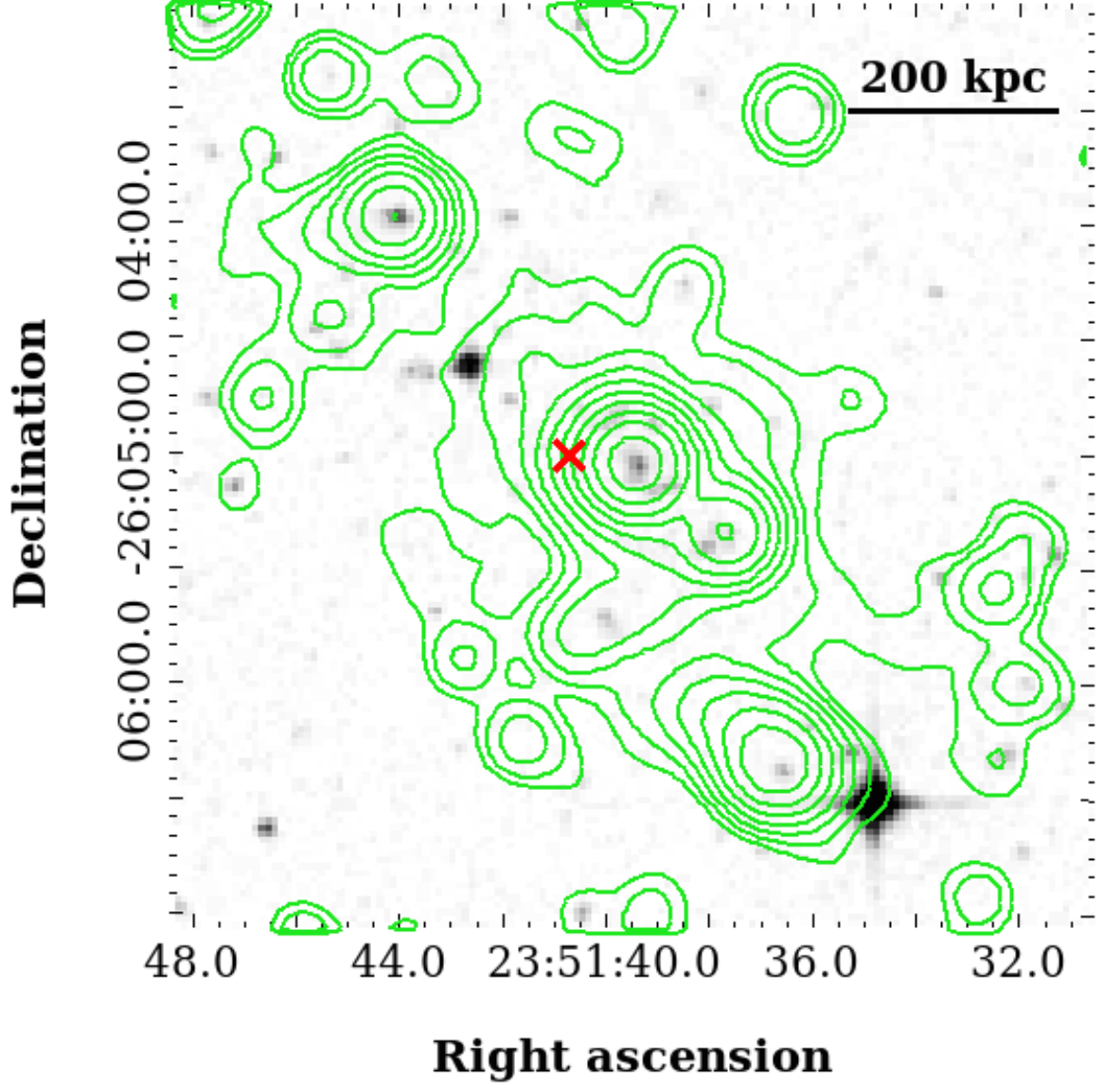}
   \caption{Abell~2667 \textbf{Left:}  Full-resolution (7.8\arcsec\,$\times$\,7.8\arcsec) 1.28~GHz MGCLS radio image with radio contours in black overlaid (1$\sigma$ = 3 $\mu$Jy beam$^{-1}$). \textbf{Right:} 1.28~GHz MGCLS low-resolution  (15\arcsec\,$\times$\,15\arcsec) radio contours in green (1$\sigma$ = 6 $\mu$Jy beam$^{-1}$), overlaid on the r-band \textit{Digitized Sky Survey (DSS)} optical image. In both panels, the radio contours start at 3\,$\sigma$ and rise by a factor of 2. The physical scale at the cluster redshift is indicated on top right, and the red $\times$ indicates the NED cluster position. } 
   \label{fig:A2667}%
\end{figure*}

\begin{figure*}
   \centering
   \includegraphics[width=0.496\textwidth]{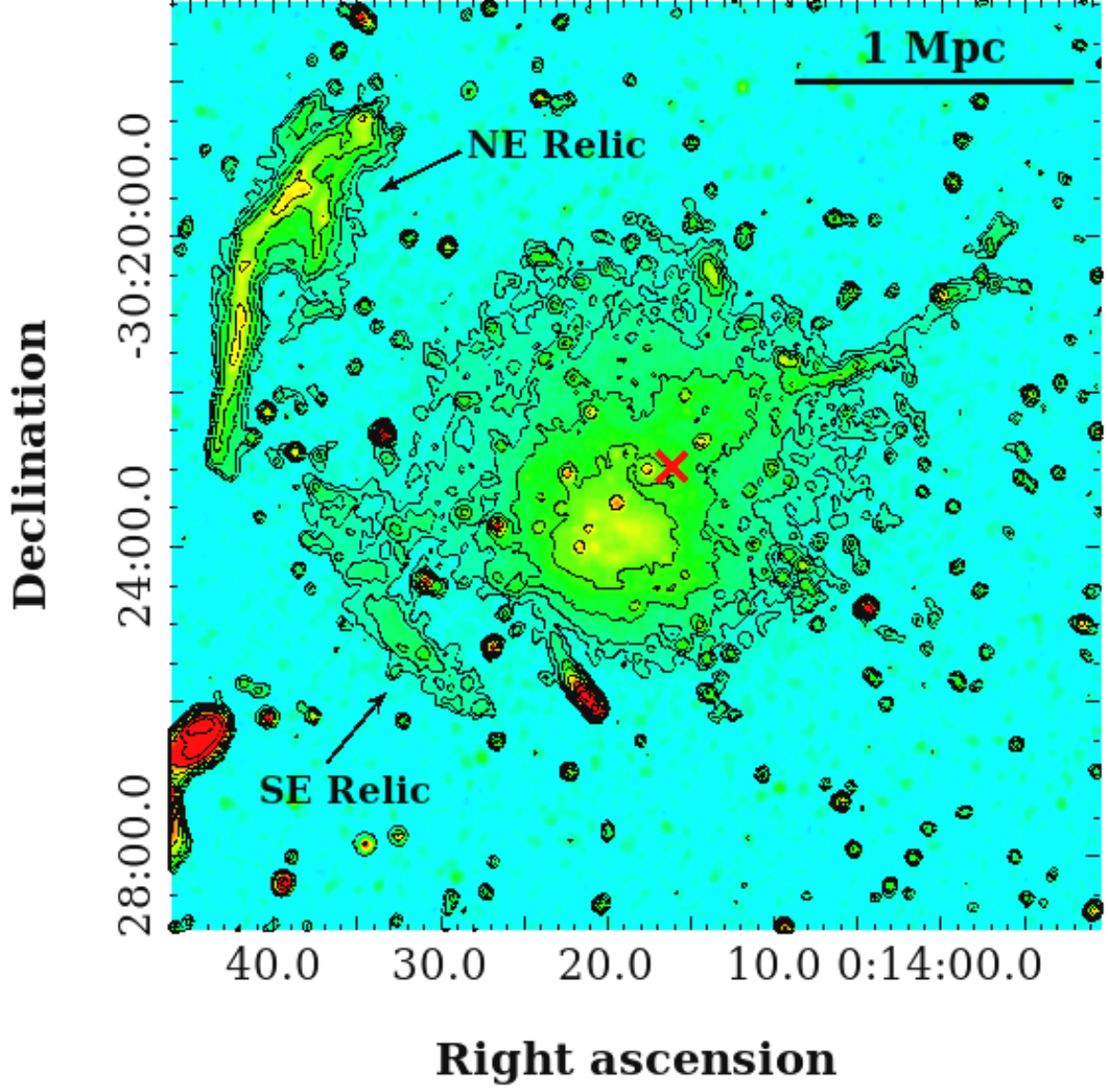}
    \includegraphics[width=0.5\textwidth]{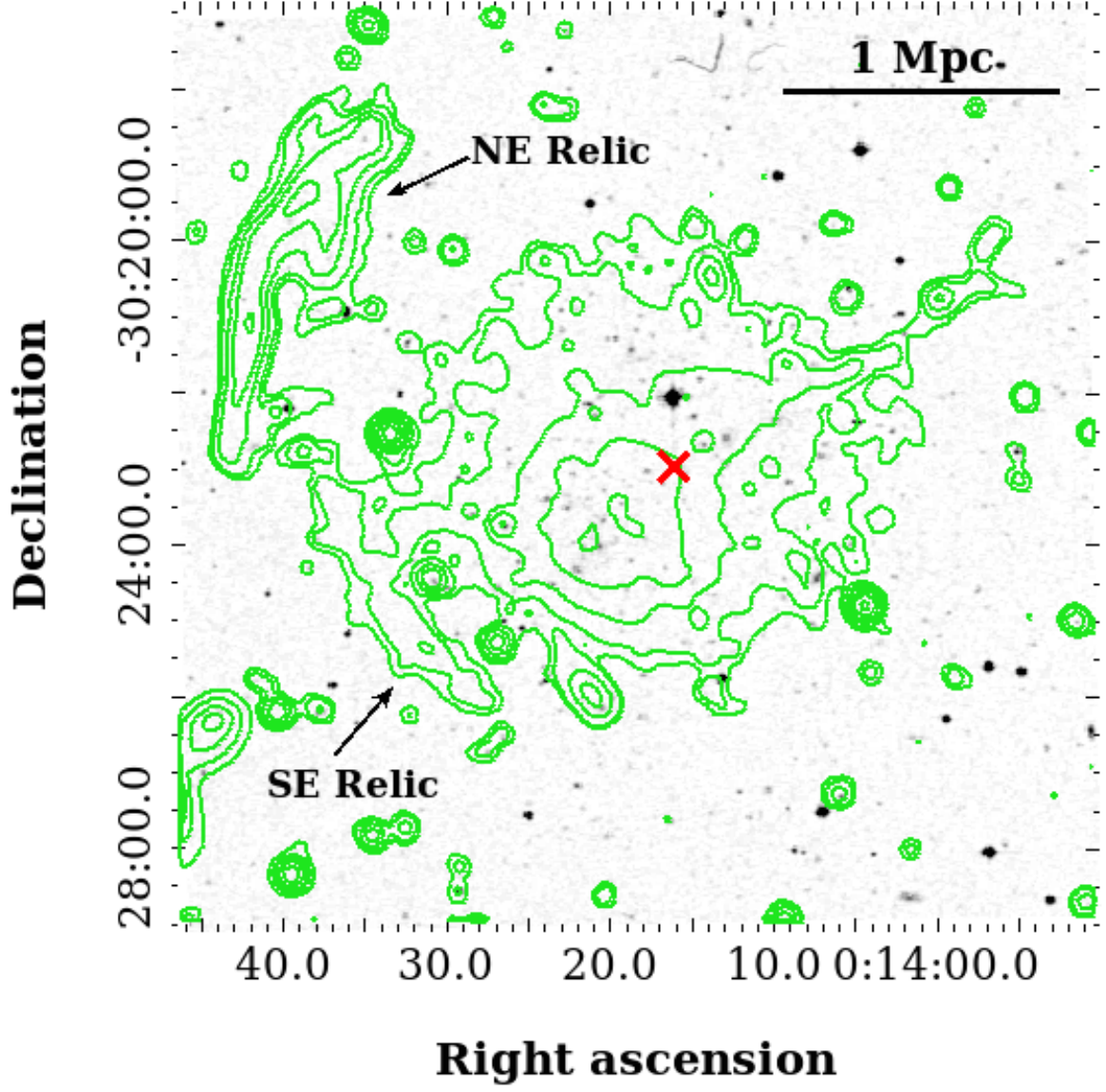}
   \caption{Abell~2744 \textbf{Left:}  Full-resolution (7.8\arcsec\,$\times$\,7.8\arcsec) 1.28~GHz MGCLS radio image with radio contours in black overlaid (1$\sigma$ = 3.5 $\mu$Jy beam$^{-1}$). \textbf{Right:} 1.28~GHz MGCLS low-resolution  (15\arcsec\,$\times$\,15\arcsec) radio contours in green (1$\sigma$ = 11 $\mu$Jy beam$^{-1}$), overlaid on the r-band \textit{Digitized Sky Survey (DSS)} optical image. In both panels, the radio contours start at 3\,$\sigma$ and rise by a factor of 2. The physical scale at the cluster redshift is indicated on top right, and the red $\times$ indicates the NED cluster position. } 
   \label{fig:A2744}%
\end{figure*}


\begin{figure*}
   \centering
   \includegraphics[width=0.496\textwidth]{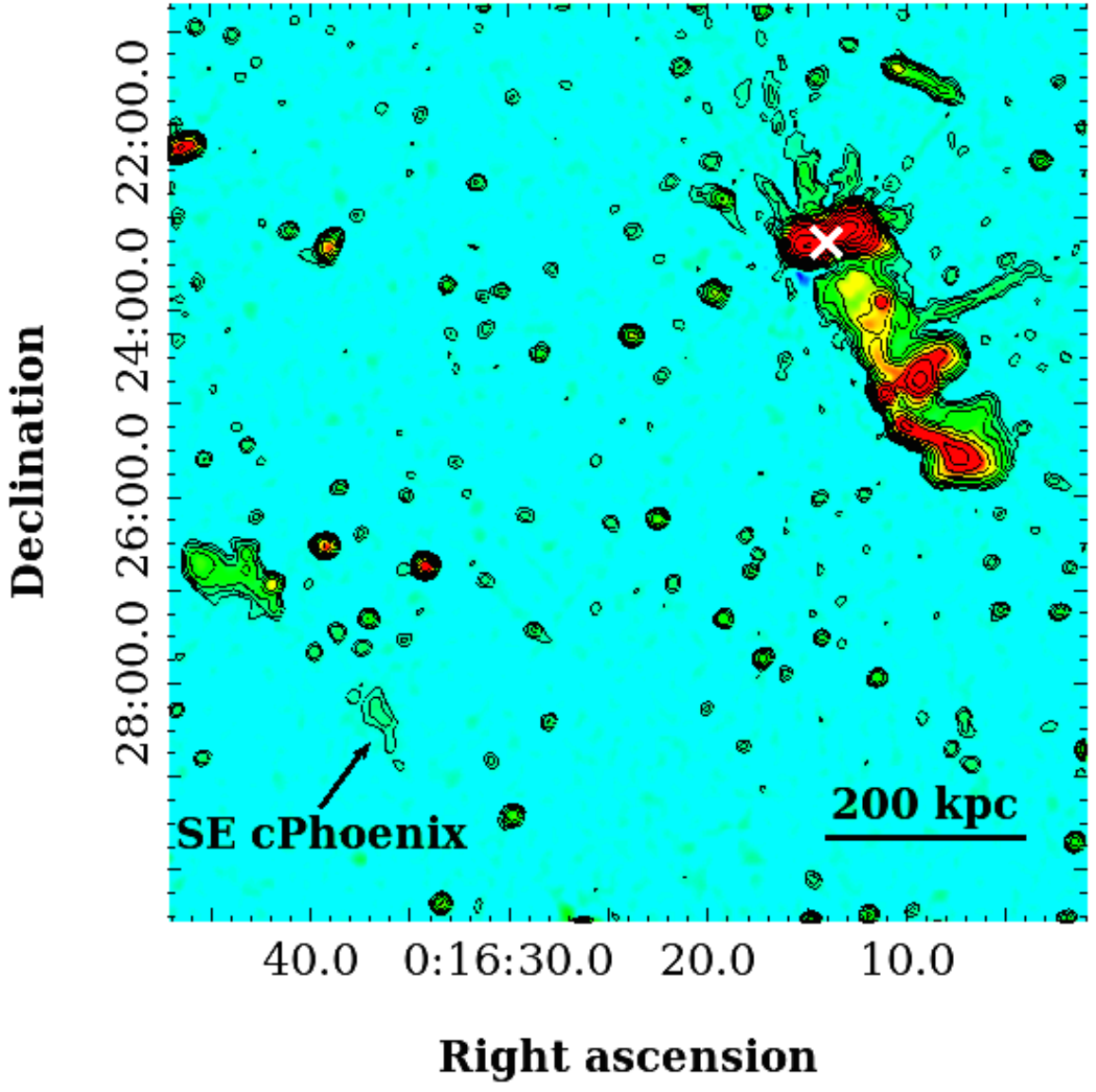}
    \includegraphics[width=0.5\textwidth]{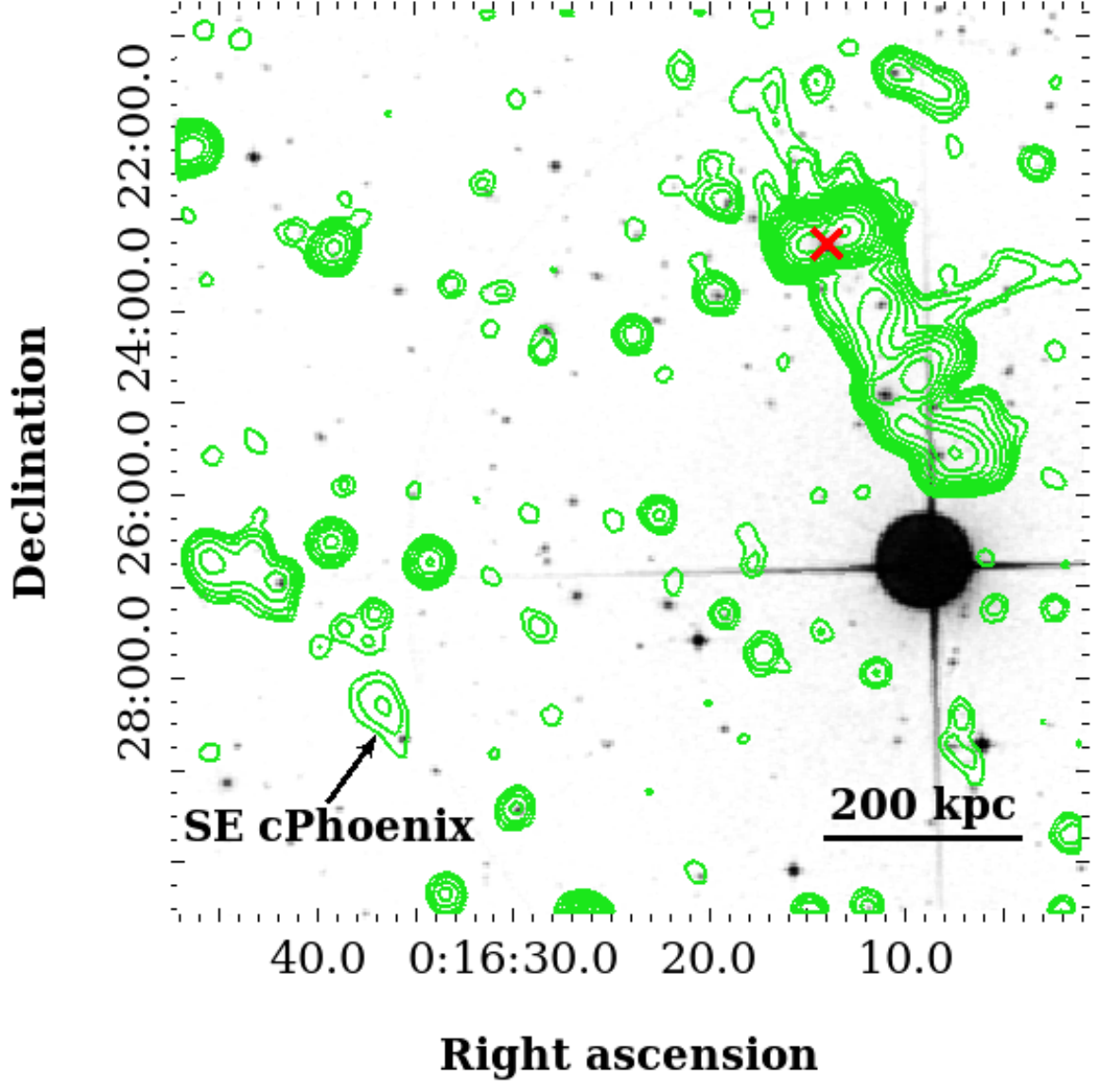}
   \caption{Abell~2751 \textbf{Left:}  Full-resolution (7.8\arcsec\,$\times$\,7.8\arcsec) 1.28~GHz MGCLS radio image with radio contours in black overlaid (1$\sigma$ = 6 $\mu$Jy beam$^{-1}$). \textbf{Right:} 1.28~GHz MGCLS low-resolution  (15\arcsec\,$\times$\,15\arcsec) radio contours in green (1$\sigma$ = 10 $\mu$Jy beam$^{-1}$), overlaid on the r-band \textit{Digitized Sky Survey (DSS)} optical image. In both panels, the radio contours start at 3\,$\sigma$ and rise by a factor of 2. The physical scale at the cluster redshift is indicated on bottom right, and the white/red $\times$ indicate the NED cluster position. } 
   \label{fig:A2751}%
\end{figure*}

\begin{figure*}
   \centering
   \includegraphics[width=0.496\textwidth]{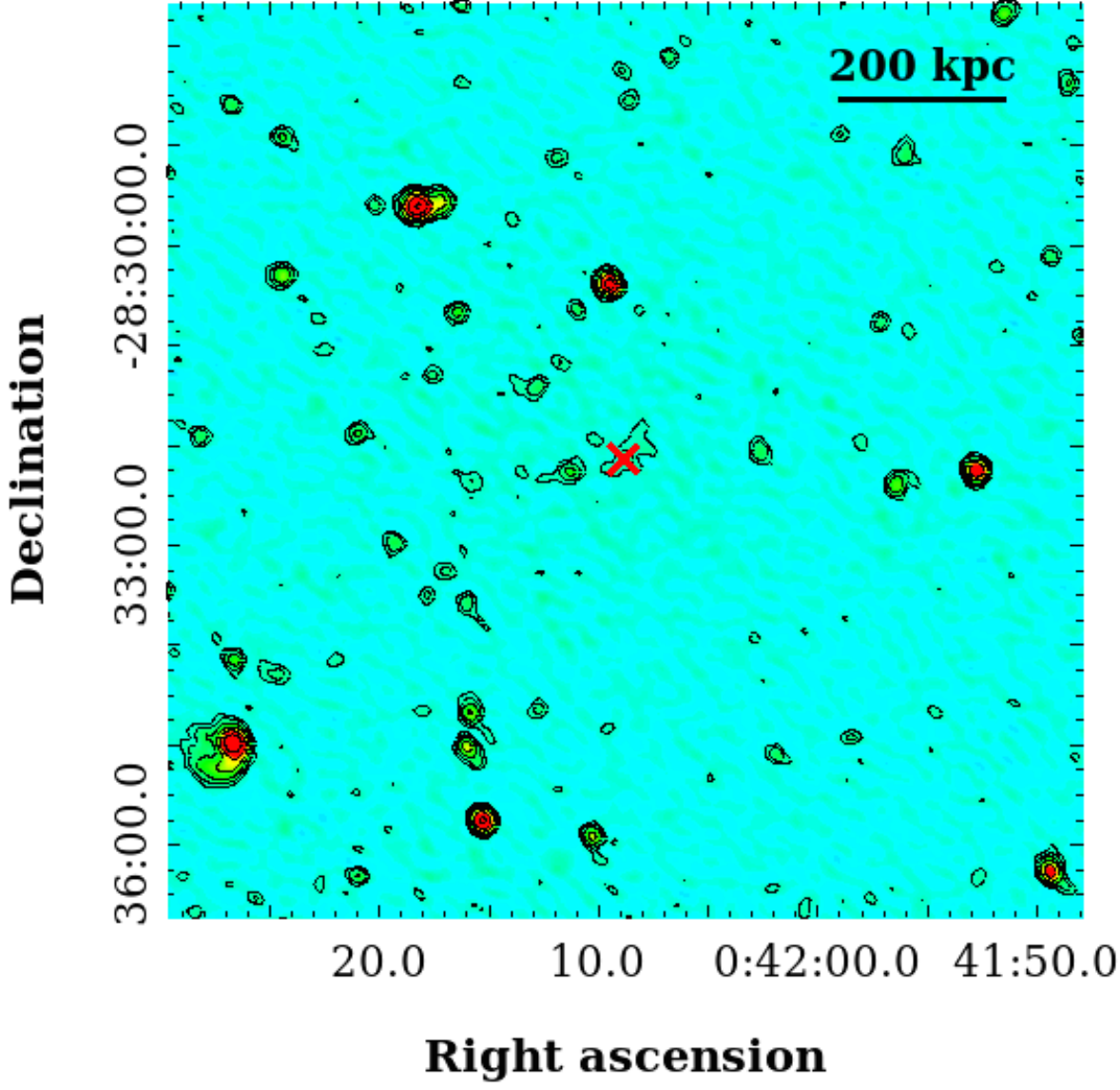}
    \includegraphics[width=0.5\textwidth]{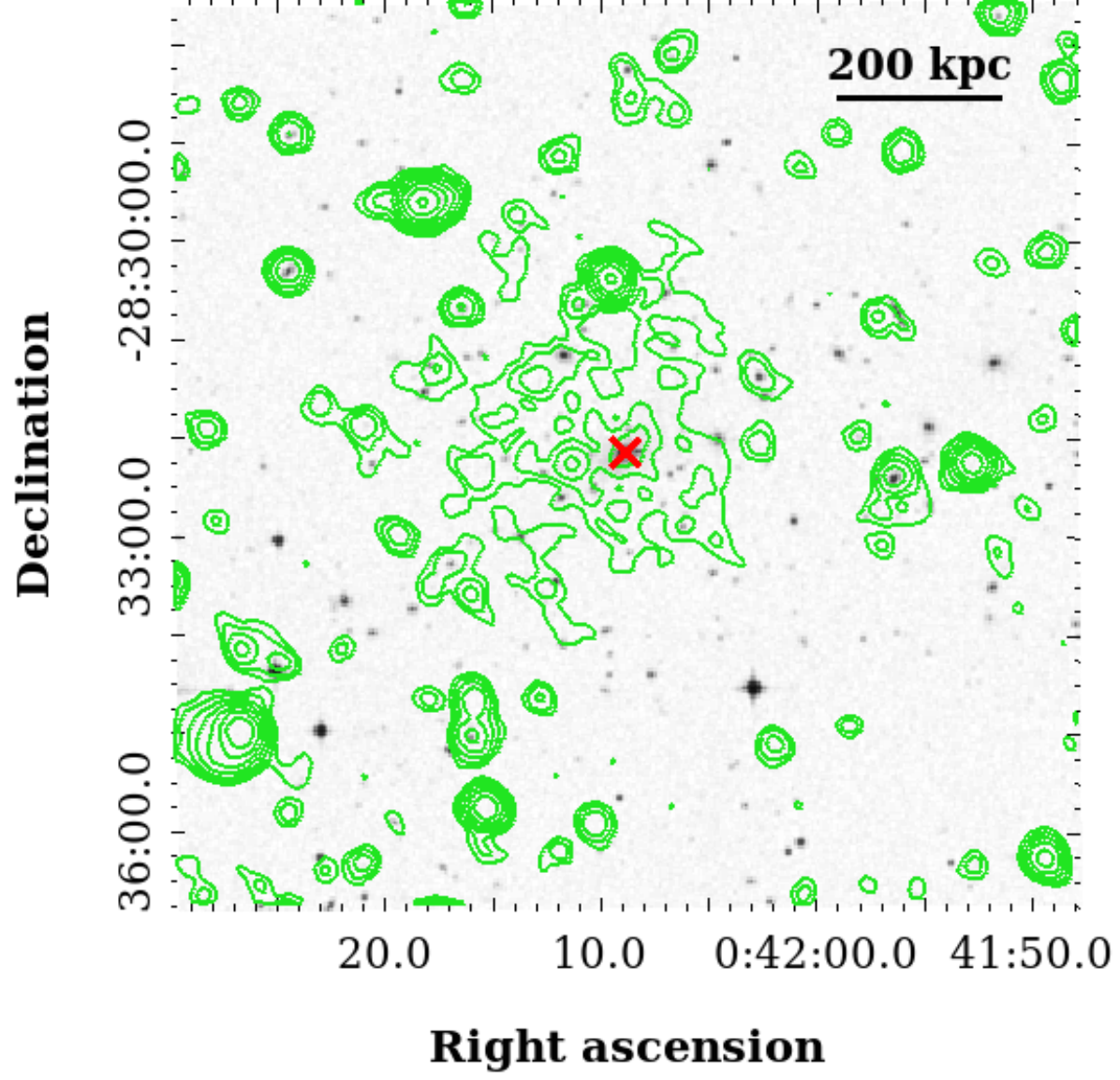}
   \caption{Abell~2811 \textbf{Left:}  Full-resolution (7.8\arcsec\,$\times$\,7.8\arcsec) 1.28~GHz MGCLS radio image with radio contours in black overlaid (1$\sigma$ = 5 $\mu$Jy beam$^{-1}$). \textbf{Right:} 1.28~GHz MGCLS low-resolution  (15\arcsec\,$\times$\,15\arcsec) radio contours in green (1$\sigma$ = 8 $\mu$Jy beam$^{-1}$), overlaid on the r-band \textit{Digitized Sky Survey (DSS)} optical image. In both panels, the radio contours start at 3\,$\sigma$ and rise by a factor of 2. The physical scale at the cluster redshift is indicated on top right, and the red $\times$ indicates the NED cluster position. } 
   \label{fig:A2811}%
\end{figure*}

\begin{figure*}
   \centering
   \includegraphics[width=0.496\textwidth]{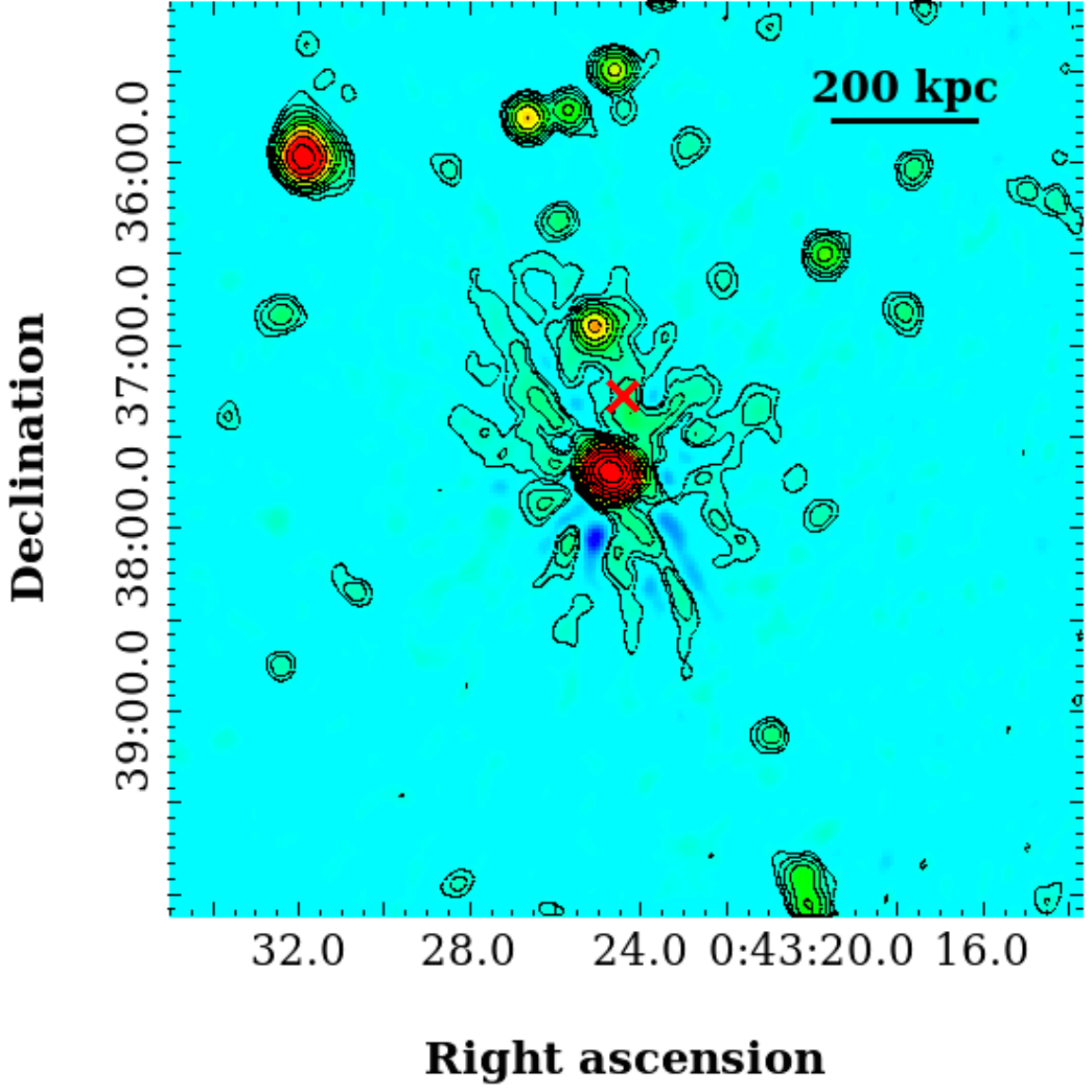}
    \includegraphics[width=0.5\textwidth]{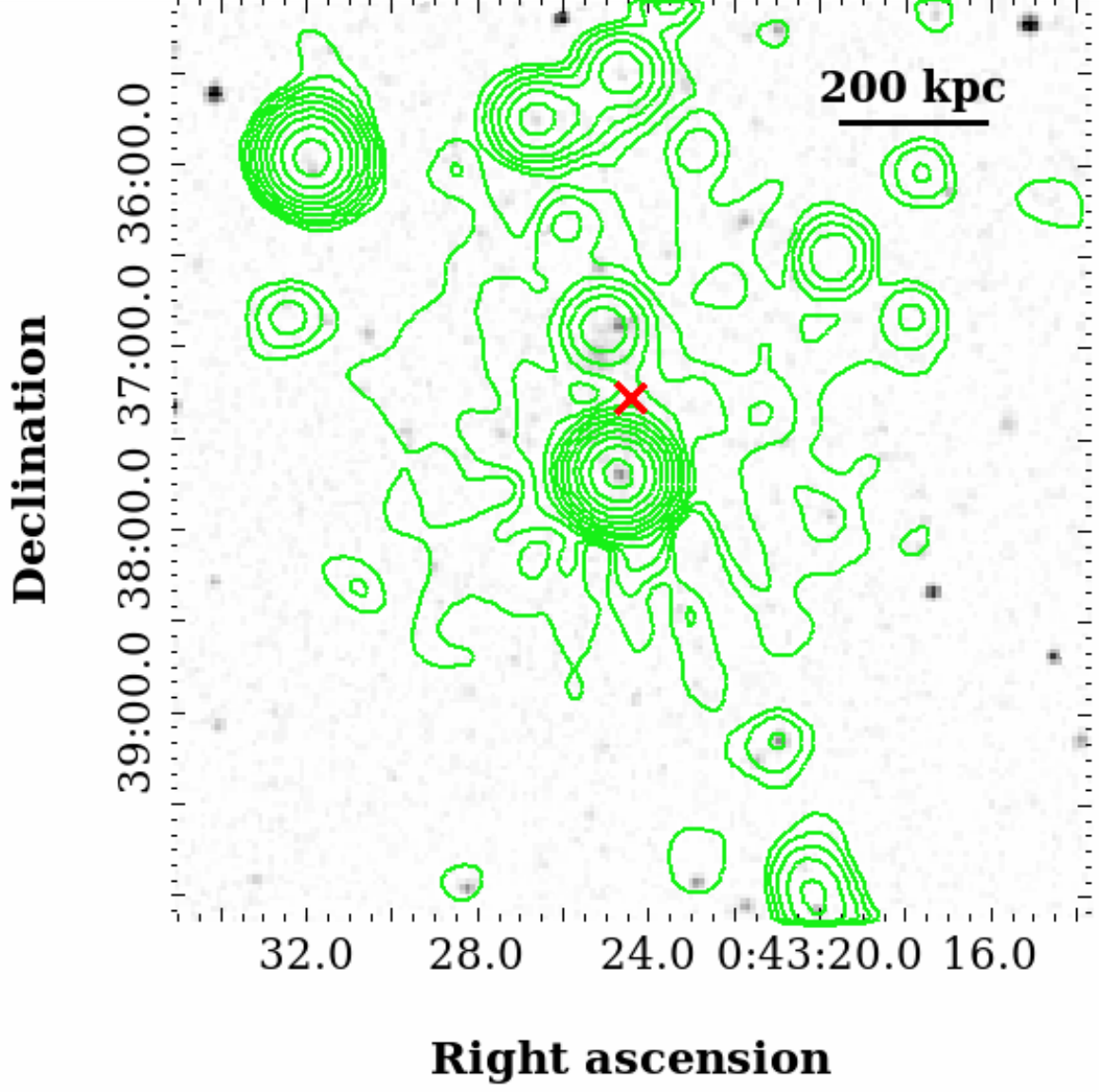}
   \caption{Abell~2813 \textbf{Left:}  Full-resolution (7.8\arcsec\,$\times$\,7.8\arcsec) 1.28~GHz MGCLS radio image with radio contours in black overlaid (1$\sigma$ = 8 $\mu$Jy beam$^{-1}$). \textbf{Right:} 1.28~GHz MGCLS low-resolution  (15\arcsec\,$\times$\,15\arcsec) radio contours in green (1$\sigma$ = 12 $\mu$Jy beam$^{-1}$), overlaid on the r-band \textit{Digitized Sky Survey (DSS)} optical image. In both panels, the radio contours start at 3\,$\sigma$ and rise by a factor of 2. The physical scale at the cluster redshift is indicated on top right, and the red $\times$ indicates the NED cluster position. } 
   \label{fig:A2813}%
\end{figure*}

\begin{figure*}
   \centering
   \includegraphics[width=0.496\textwidth]{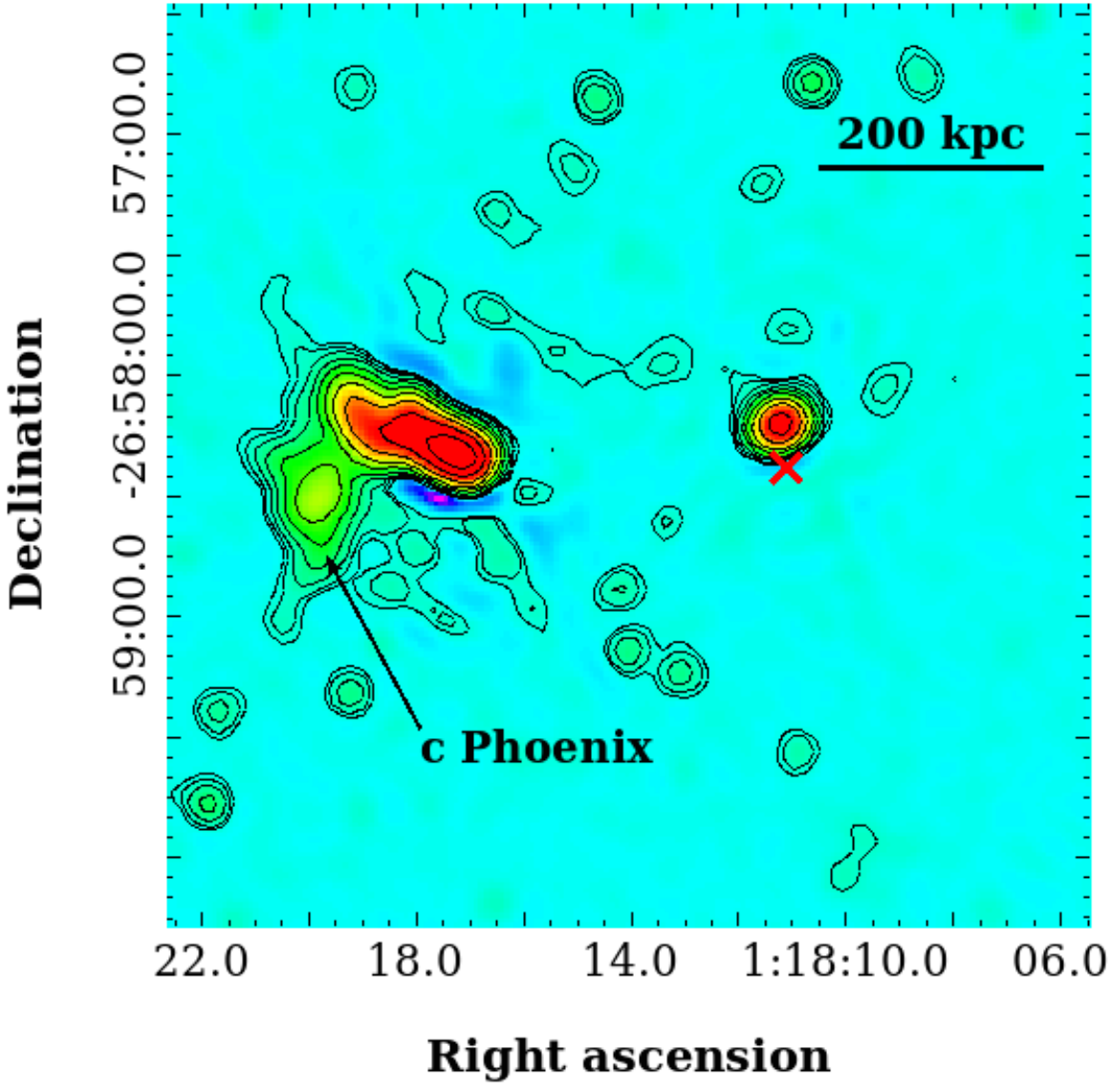}
    \includegraphics[width=0.5\textwidth]{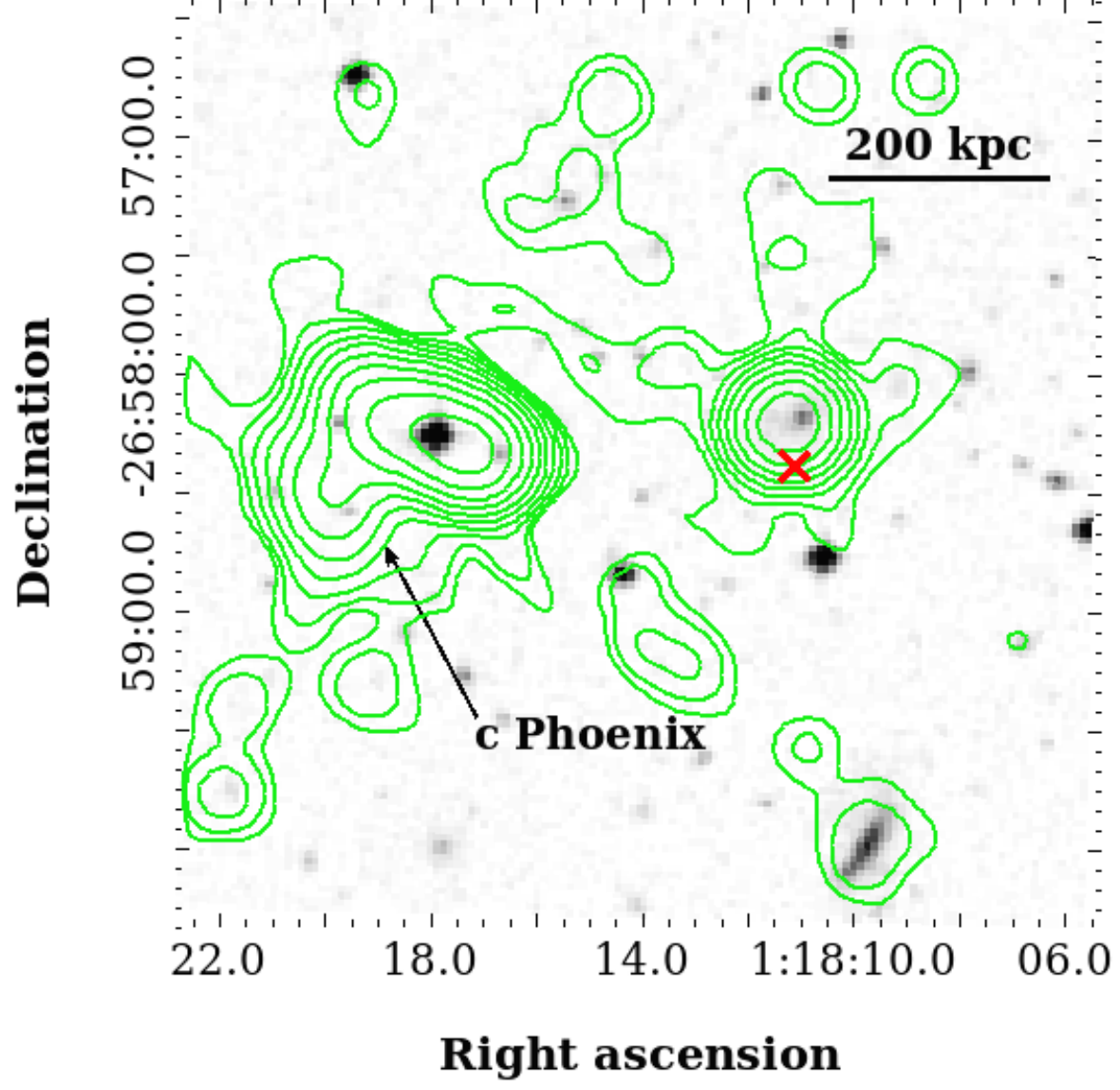}
   \caption{Abell~2895 \textbf{Left:}  Full-resolution (7.8\arcsec\,$\times$\,7.8\arcsec) 1.28~GHz MGCLS radio image with radio contours in black overlaid (1$\sigma$ = 6 $\mu$Jy beam$^{-1}$). \textbf{Right:} 1.28~GHz MGCLS low-resolution  (15\arcsec\,$\times$\,15\arcsec) radio contours in green (1$\sigma$ = 10 $\mu$Jy beam$^{-1}$), overlaid on the r-band \textit{Digitized Sky Survey (DSS)} optical image. In both panels, the radio contours start at 3\,$\sigma$ and rise by a factor of 2. The physical scale at the cluster redshift is indicated on top right, and the red $\times$ indicates the NED cluster position. } 
   \label{fig:A2895}%
\end{figure*}

\begin{figure*}
   \centering
   \includegraphics[width=0.496\textwidth]{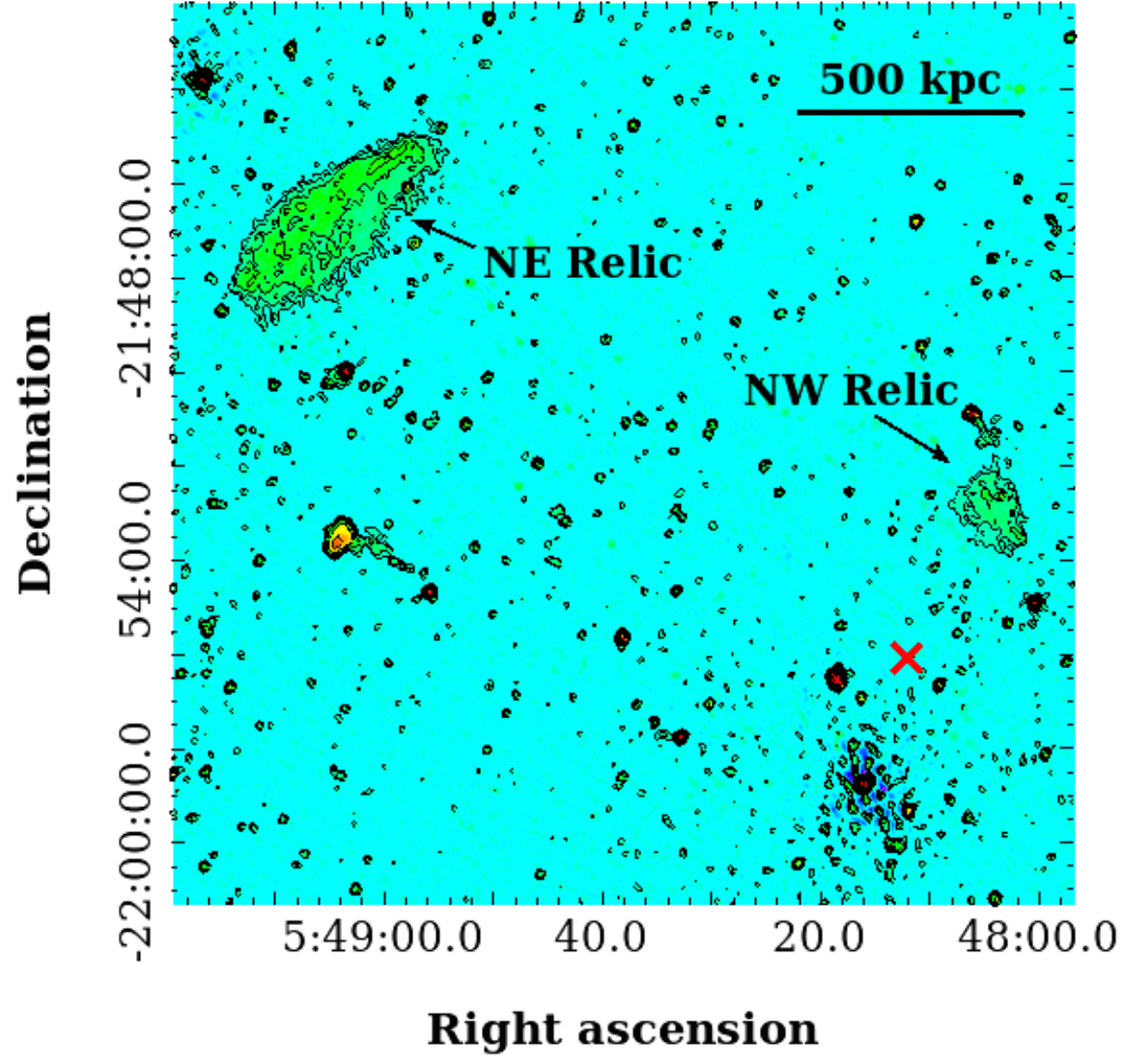}
    \includegraphics[width=0.5\textwidth]{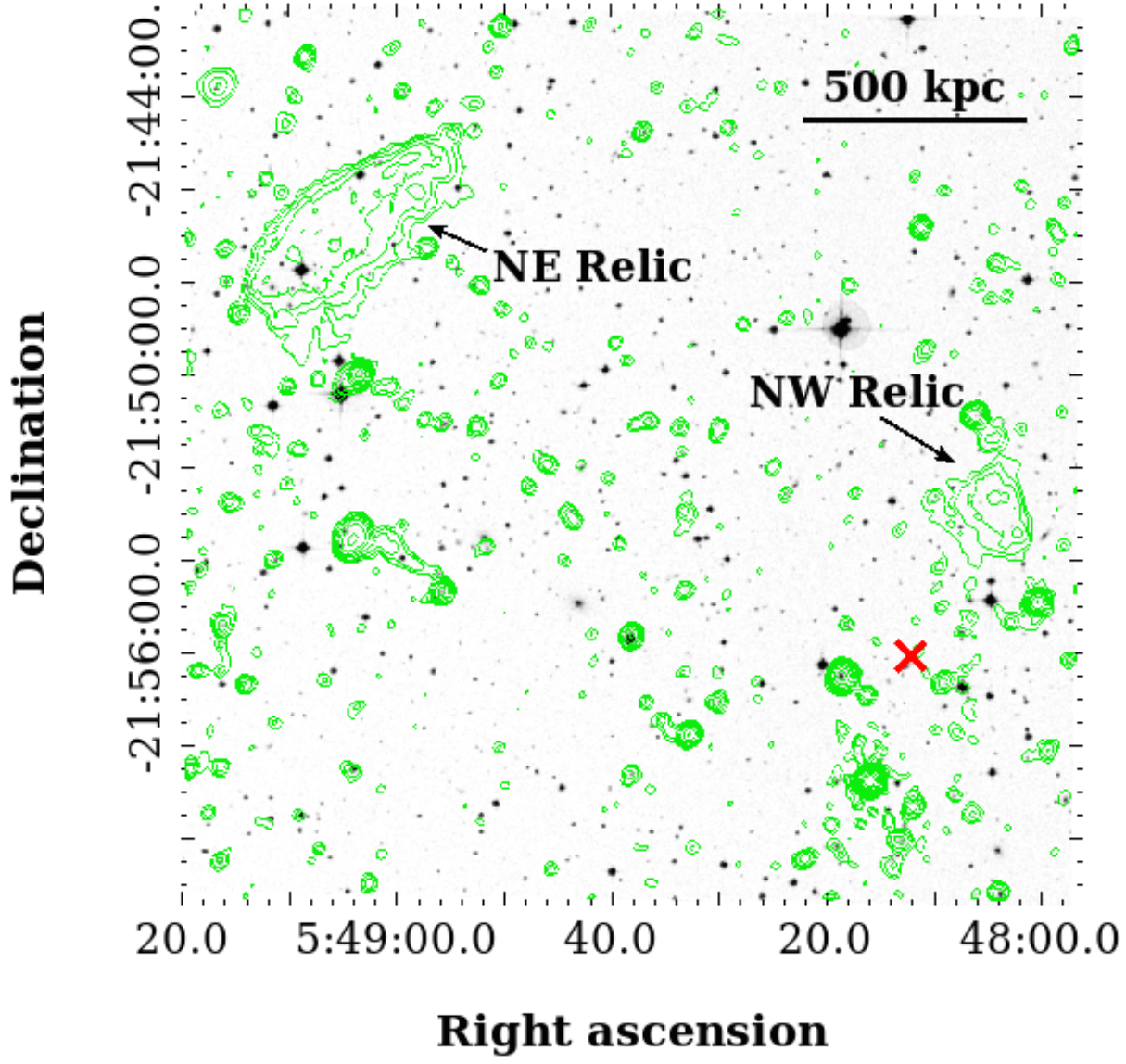}
   \caption{Abell~3365 \textbf{Left:}  Full-resolution (7.8\arcsec\,$\times$\,7.8\arcsec) 1.28~GHz MGCLS radio image with radio contours in black overlaid (1$\sigma$ = 8 $\mu$Jy beam$^{-1}$). \textbf{Right:} 1.28~GHz MGCLS low-resolution  (15\arcsec\,$\times$\,15\arcsec) radio contours in green (1$\sigma$ = 12 $\mu$Jy beam$^{-1}$), overlaid on the r-band \textit{Digitized Sky Survey (DSS)} optical image. In both panels, the radio contours start at 3\,$\sigma$ and rise by a factor of 2. The physical scale at the cluster redshift is indicated on top right, and the red $\times$ indicates the NED cluster position. } 
   \label{fig:A3365}%
\end{figure*}

\begin{figure*}
   \centering
   \includegraphics[width=0.497\textwidth]{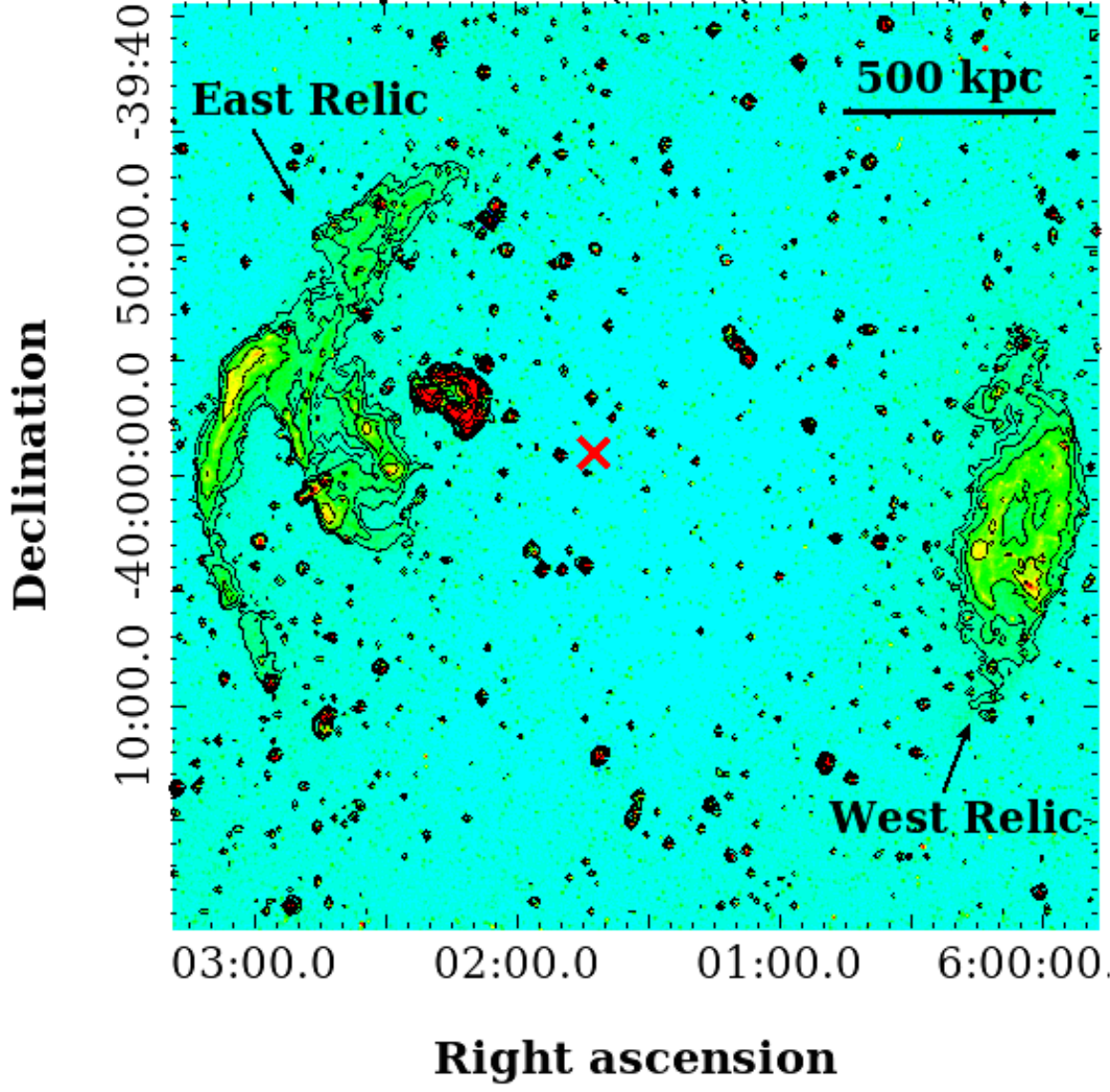}
    \includegraphics[width=0.498\textwidth]{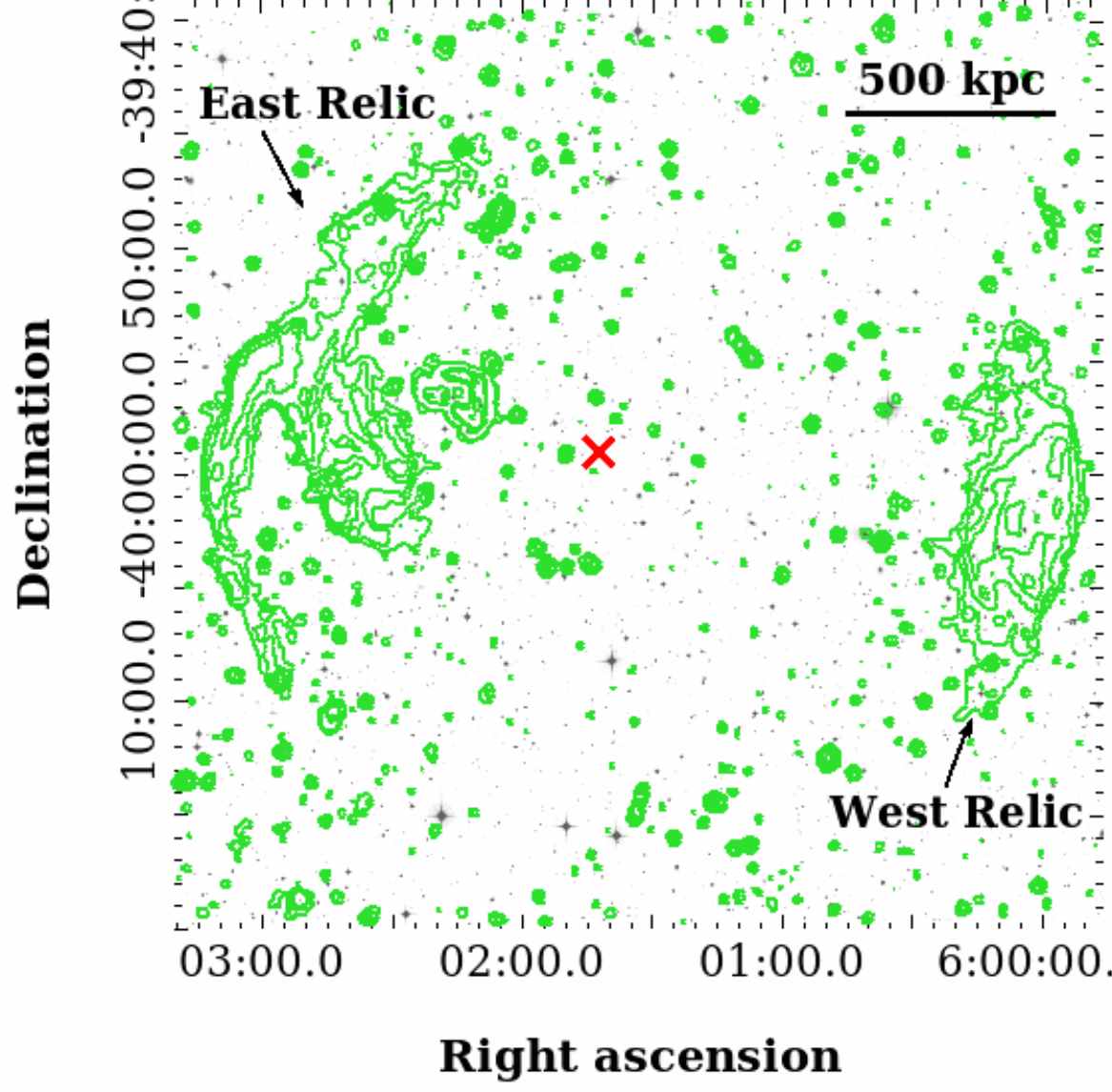}
   \caption{Abell~3376 \textbf{Left:}  Full-resolution (7.8\arcsec\,$\times$\,7.8\arcsec) 1.28~GHz MGCLS radio image with radio contours in black overlaid (1$\sigma$ = 4 $\mu$Jy beam$^{-1}$). \textbf{Right:} 1.28~GHz MGCLS low-resolution  (15\arcsec\,$\times$\,15\arcsec) radio contours in green (1$\sigma$ = 12 $\mu$Jy beam$^{-1}$), overlaid on the r-band \textit{Digitized Sky Survey (DSS)} optical image. In both panels, the radio contours start at 3\,$\sigma$ and rise by a factor of 2. The physical scale at the cluster redshift is indicated on top right, and the red $\times$ indicates the NED cluster position. } 
   \label{fig:A3376}%
\end{figure*}

\begin{figure*}
   \centering
   \includegraphics[width=0.496\textwidth]{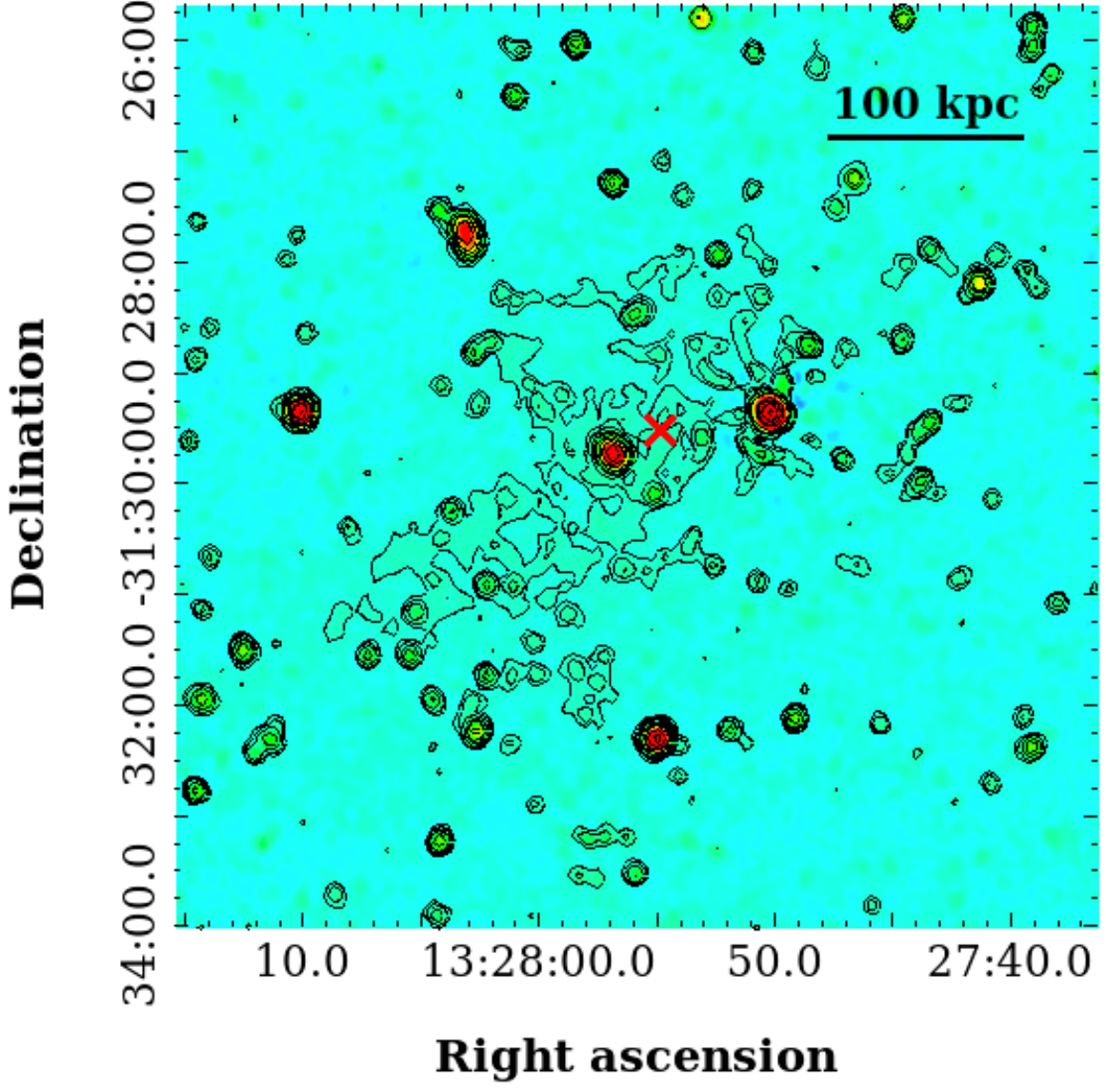}
    \includegraphics[width=0.5\textwidth]{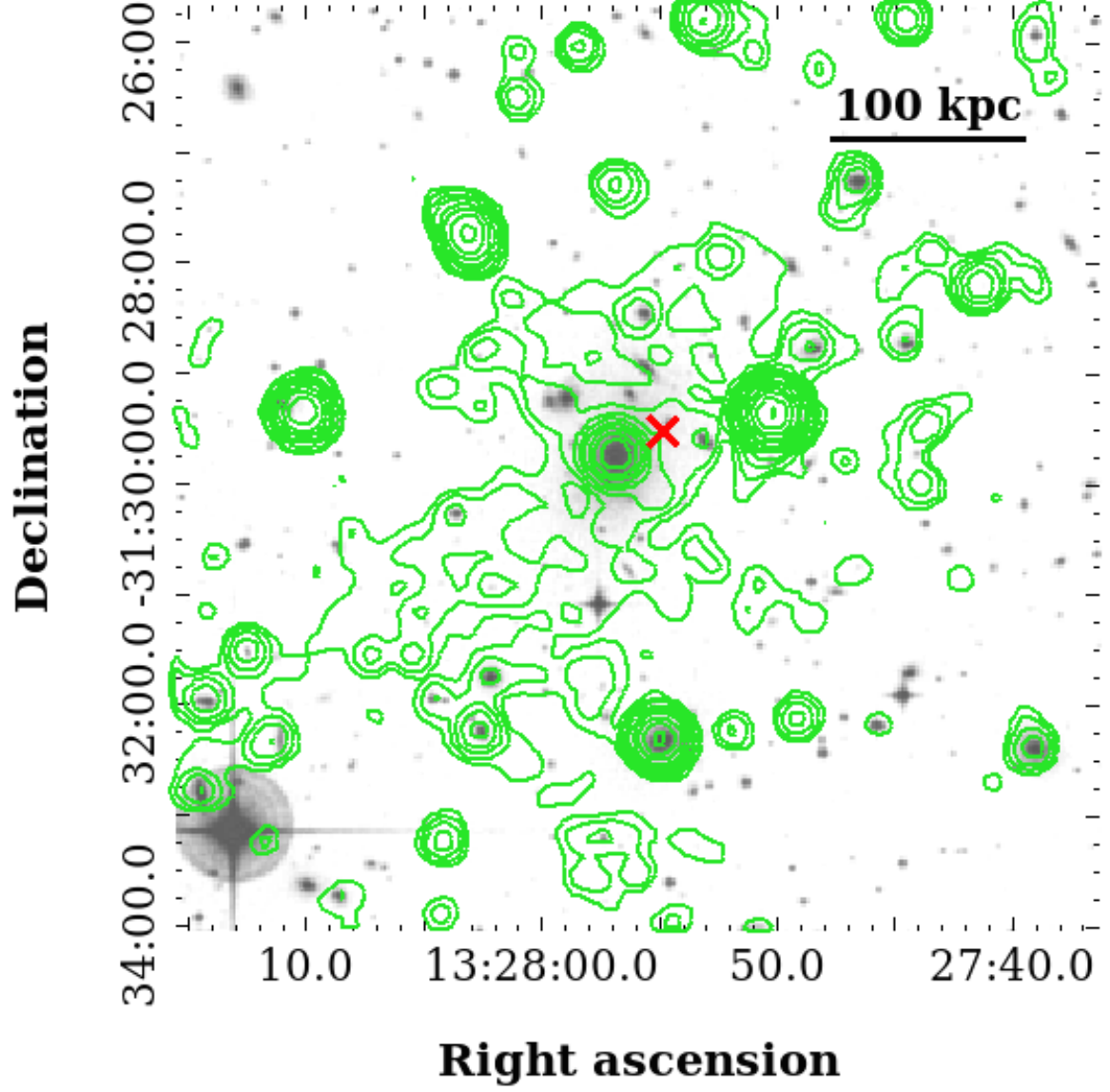}
   \caption{Abell~3558 \textbf{Left:}  Full-resolution (7.8\arcsec\,$\times$\,7.8\arcsec) 1.28~GHz MGCLS radio image with radio contours in black overlaid (1$\sigma$ = 3 $\mu$Jy beam$^{-1}$). \textbf{Right:} 1.28~GHz MGCLS low-resolution  (15\arcsec\,$\times$\,15\arcsec) radio contours in green (1$\sigma$ = 7 $\mu$Jy beam$^{-1}$), overlaid on the r-band \textit{Digitized Sky Survey (DSS)} optical image. In both panels, the radio contours start at 3\,$\sigma$ and rise by a factor of 2. The physical scale at the cluster redshift is indicated on top right, and the red $\times$ indicates the NED cluster position. } 
   \label{fig:A3558}%
\end{figure*}

\begin{figure*}
   \centering
   \includegraphics[width=0.496\textwidth]{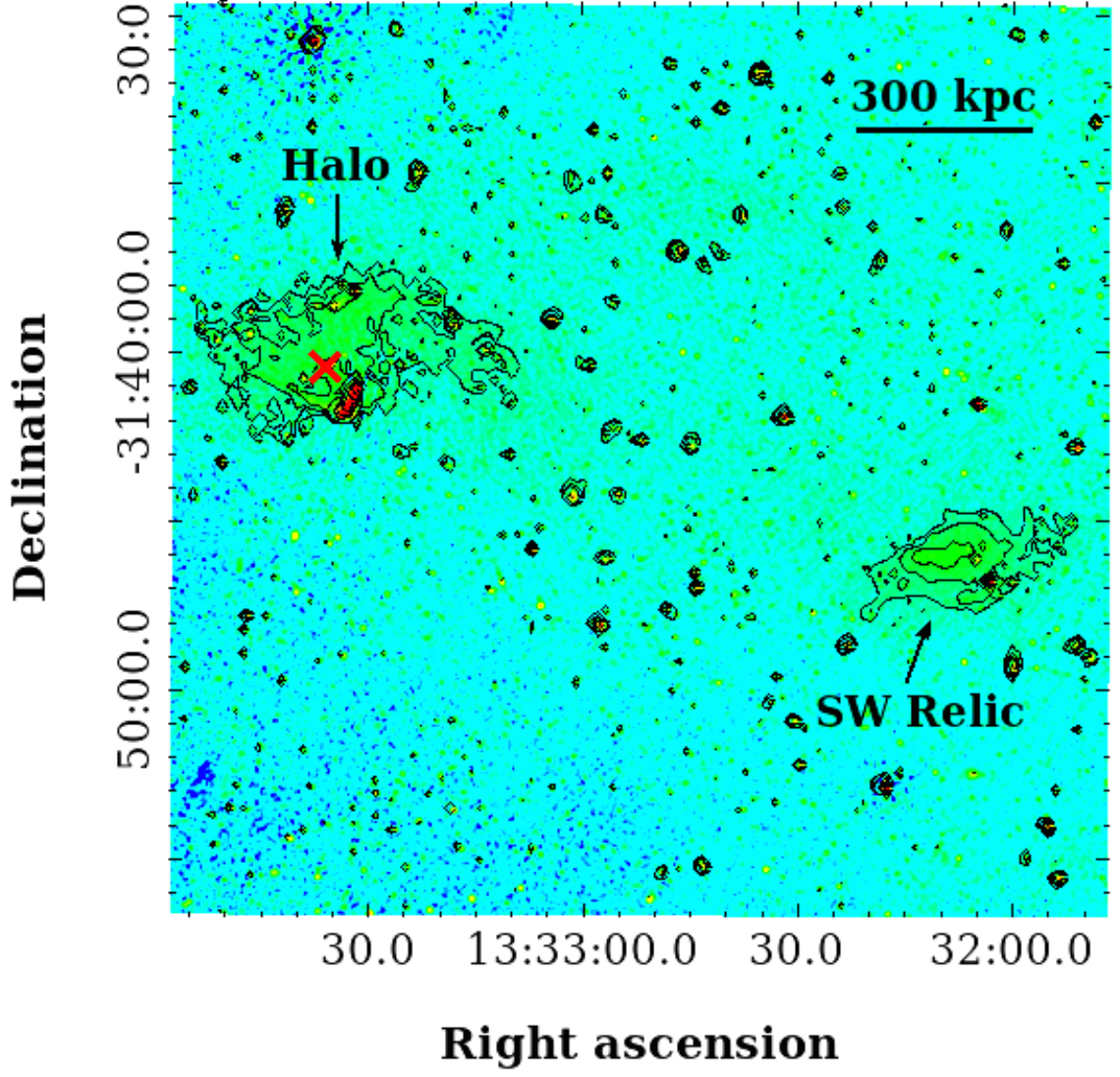}
    \includegraphics[width=0.5\textwidth]{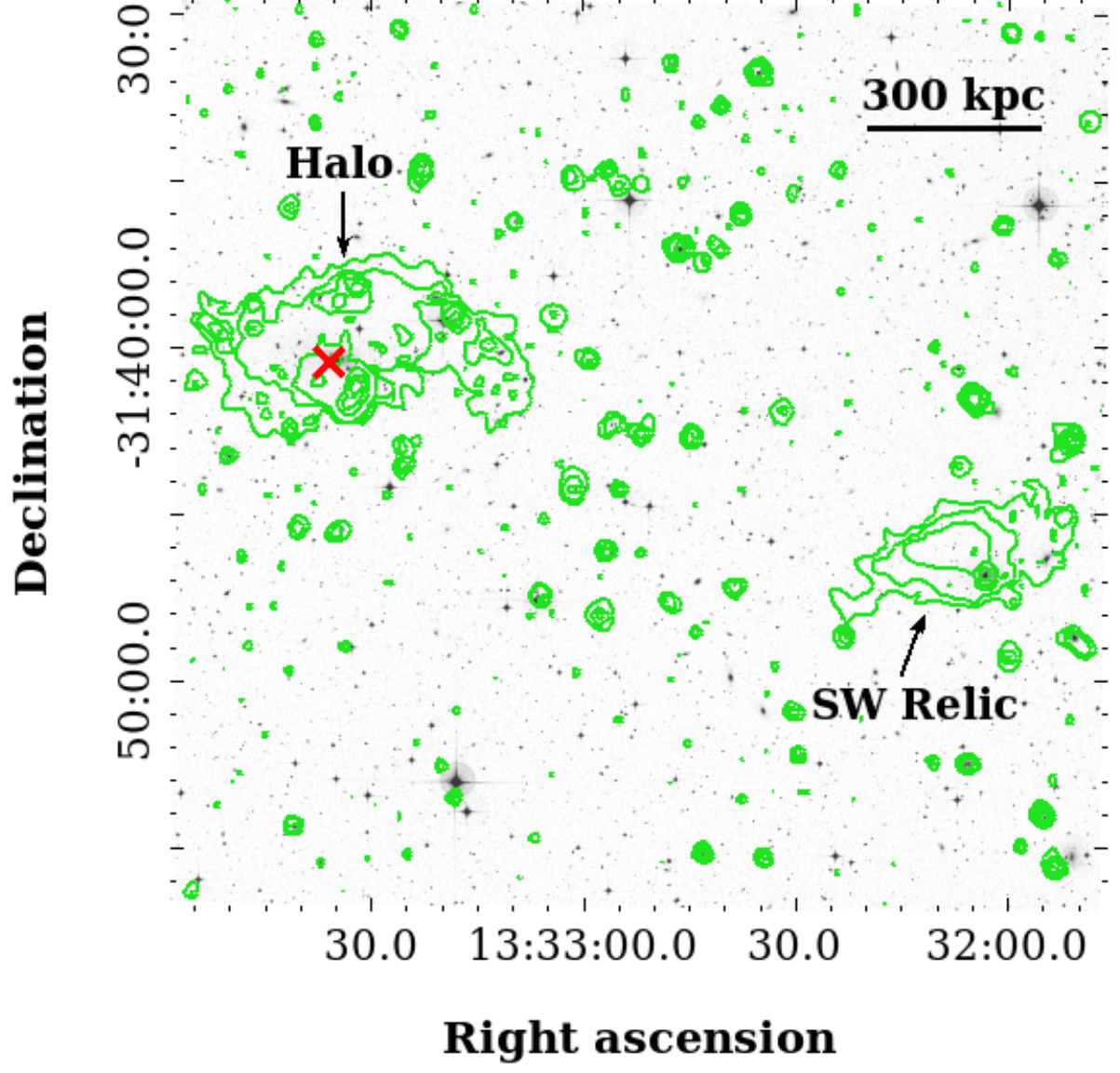}
   \caption{Abell~3562 \textbf{Left:}  Full-resolution (7.8\arcsec\,$\times$\,7.8\arcsec) 1.28~GHz MGCLS radio image with radio contours in black overlaid (1$\sigma$ = 4 $\mu$Jy beam$^{-1}$). \textbf{Right:} 1.28~GHz MGCLS low-resolution  (15\arcsec\,$\times$\,15\arcsec) radio contours in green (1$\sigma$ = 10 $\mu$Jy beam$^{-1}$), overlaid on the r-band \textit{Digitized Sky Survey (DSS)} optical image. In both panels, the radio contours start at 3\,$\sigma$ and rise by a factor of 2. The physical scale at the cluster redshift is indicated on top right, and the red $\times$ indicates the NED cluster position. } 
   \label{fig:A3562}%
\end{figure*}

\begin{figure*}
   \centering
   \includegraphics[width=0.496\textwidth]{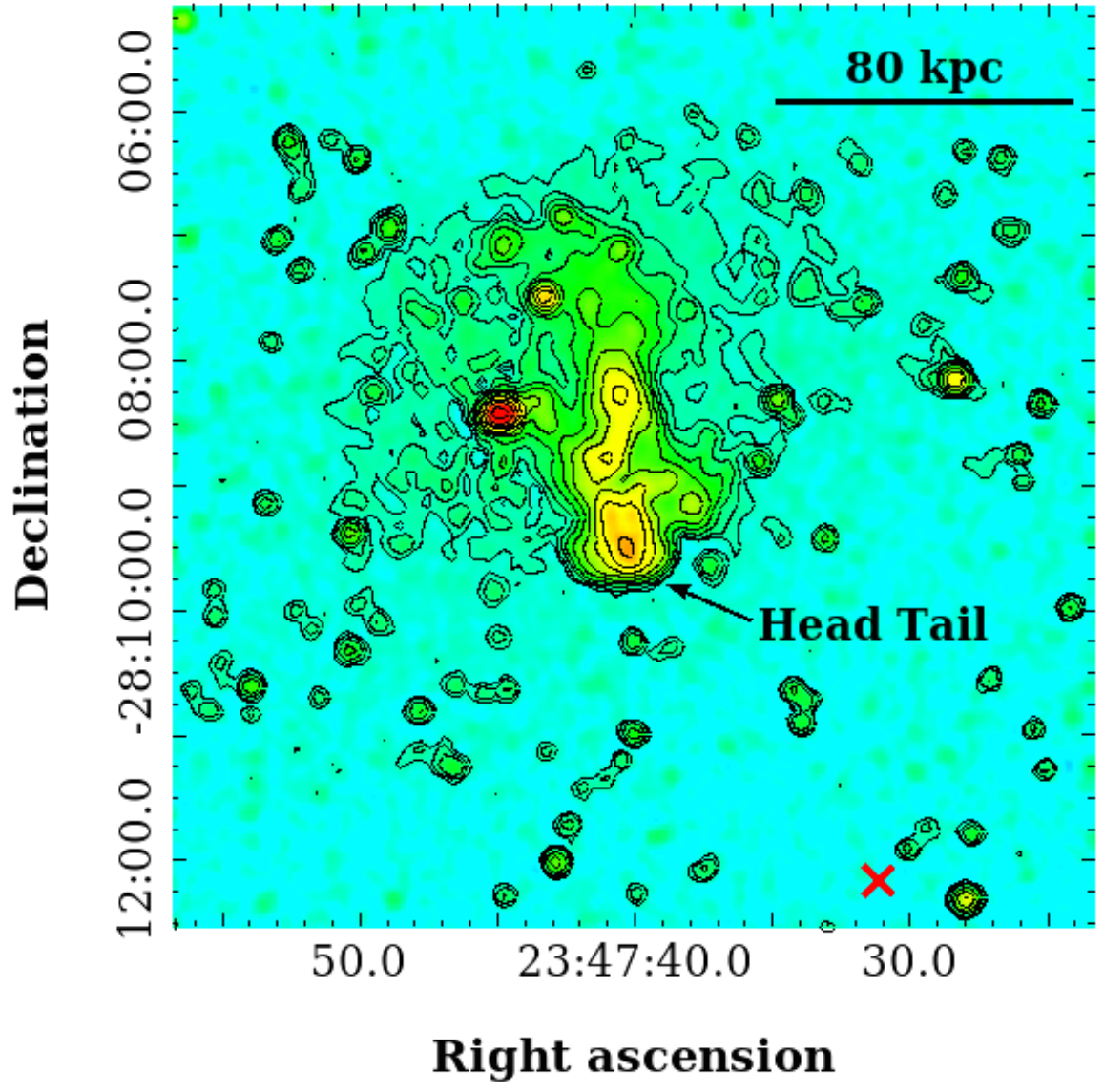}
    \includegraphics[width=0.5\textwidth]{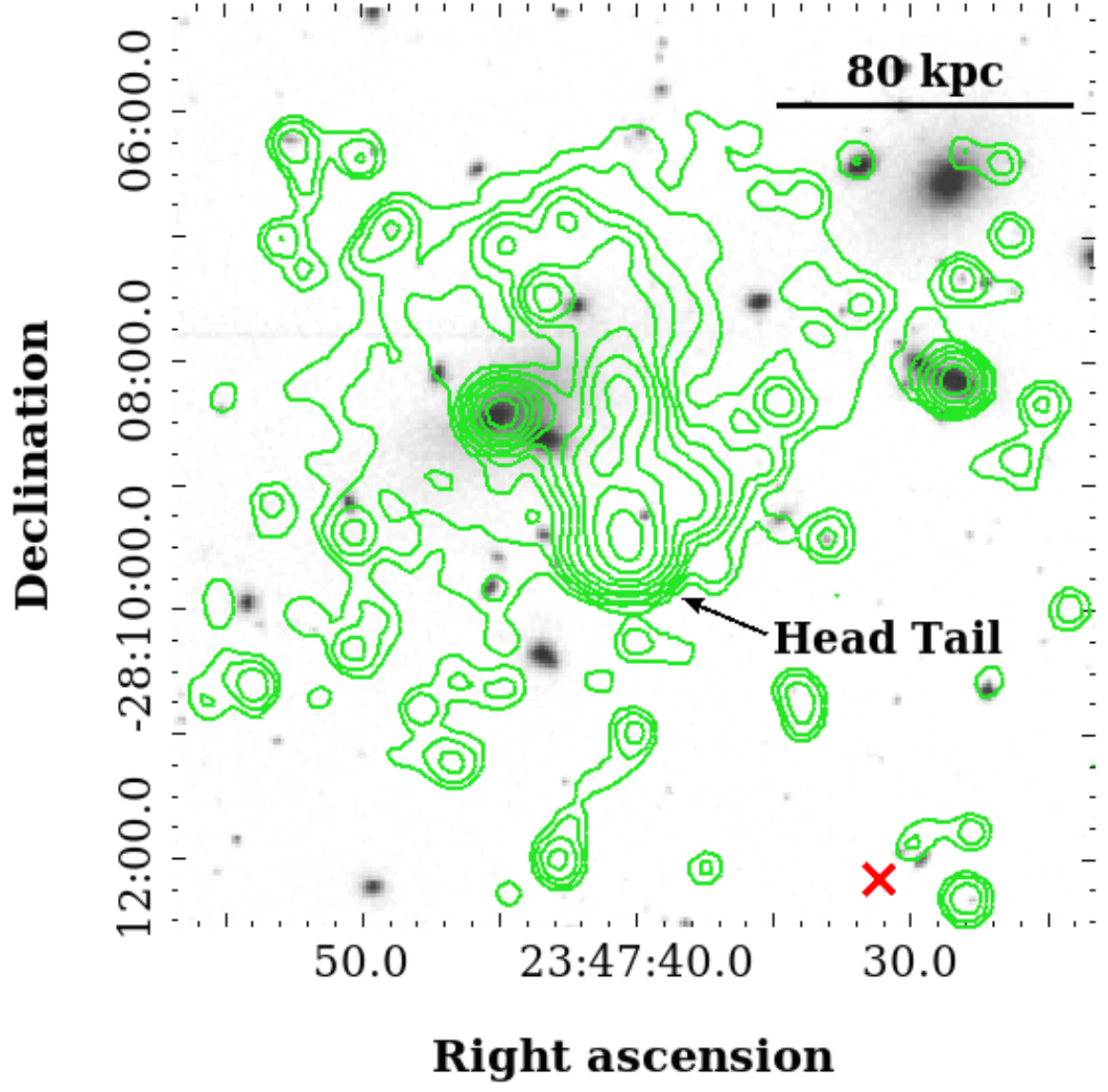}
   \caption{Abell~4038 \textbf{Left:}  Full-resolution (7.8\arcsec\,$\times$\,7.8\arcsec) 1.28~GHz MGCLS radio image with radio contours in black overlaid (1$\sigma$ = 3.5 $\mu$Jy beam$^{-1}$). \textbf{Right:} 1.28~GHz MGCLS low-resolution  (15\arcsec\,$\times$\,15\arcsec) radio contours in green (1$\sigma$ = 8 $\mu$Jy beam$^{-1}$), overlaid on the r-band \textit{Digitized Sky Survey (DSS)} optical image. In both panels, the radio contours start at 3\,$\sigma$ and rise by a factor of 2. The physical scale at the cluster redshift is indicated on top right, and the red $\times$ indicates the NED cluster position. } 
   \label{fig:A4038}%
\end{figure*}

\begin{figure*}
   \centering
   \includegraphics[width=0.496\textwidth]{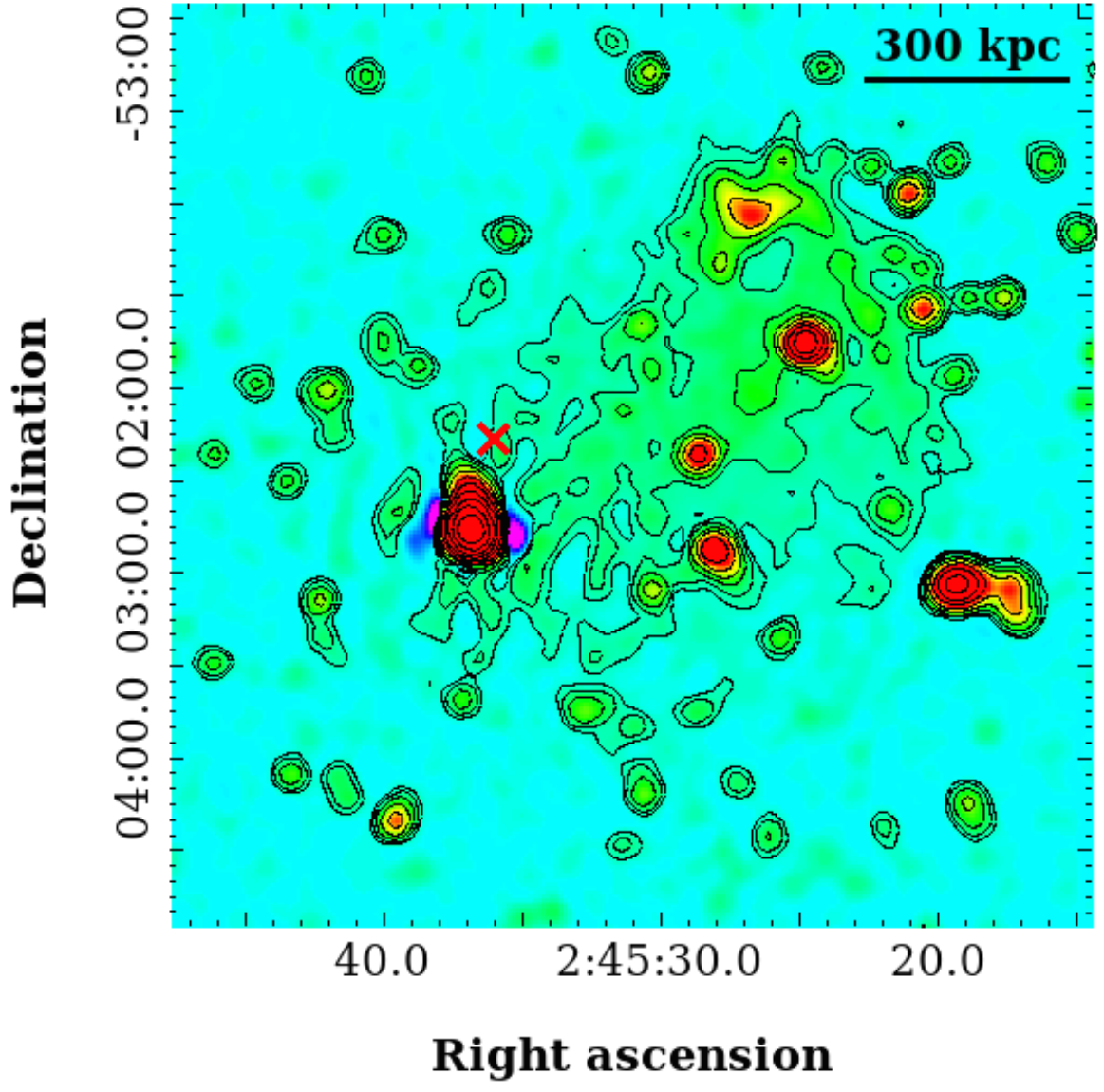}
    \includegraphics[width=0.5\textwidth]{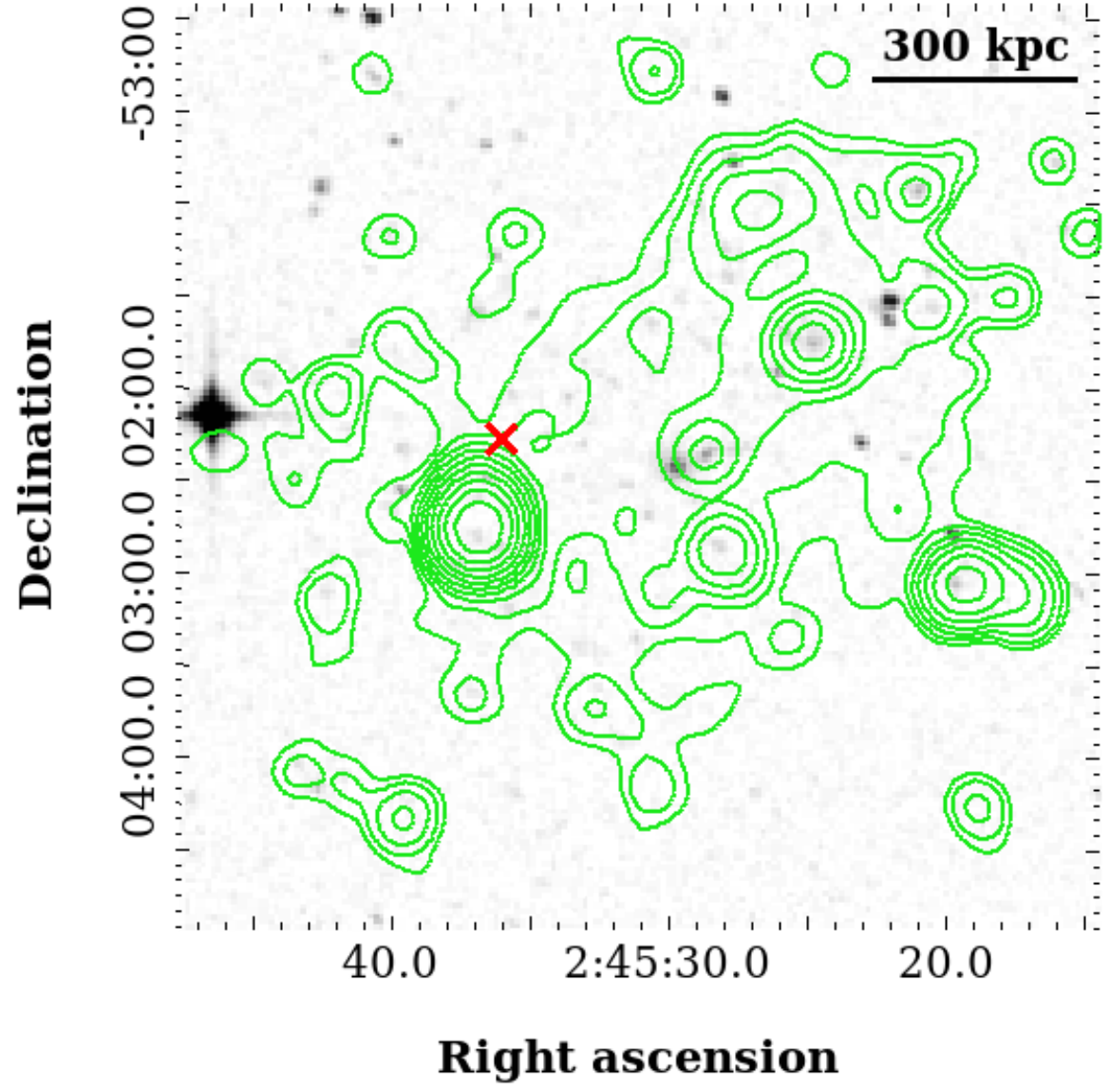}
   \caption{AS295 \textbf{Left:}  Full-resolution (7.8\arcsec\,$\times$\,7.8\arcsec) 1.28~GHz MGCLS radio image with radio contours in black overlaid (1$\sigma$ = 3 $\mu$Jy beam$^{-1}$). \textbf{Right:} 1.28~GHz MGCLS low-resolution  (15\arcsec\,$\times$\,15\arcsec) radio contours in green (1$\sigma$ = 8 $\mu$Jy beam$^{-1}$), overlaid on the r-band \textit{Digitized Sky Survey (DSS)} optical image. In both panels, the radio contours start at 3\,$\sigma$ and rise by a factor of 2. The physical scale at the cluster redshift is indicated on top right, and the red $\times$ indicates the NED cluster position. } 
   \label{fig:AS295}%
\end{figure*}

\begin{figure*}
   \centering
   \includegraphics[width=0.496\textwidth]{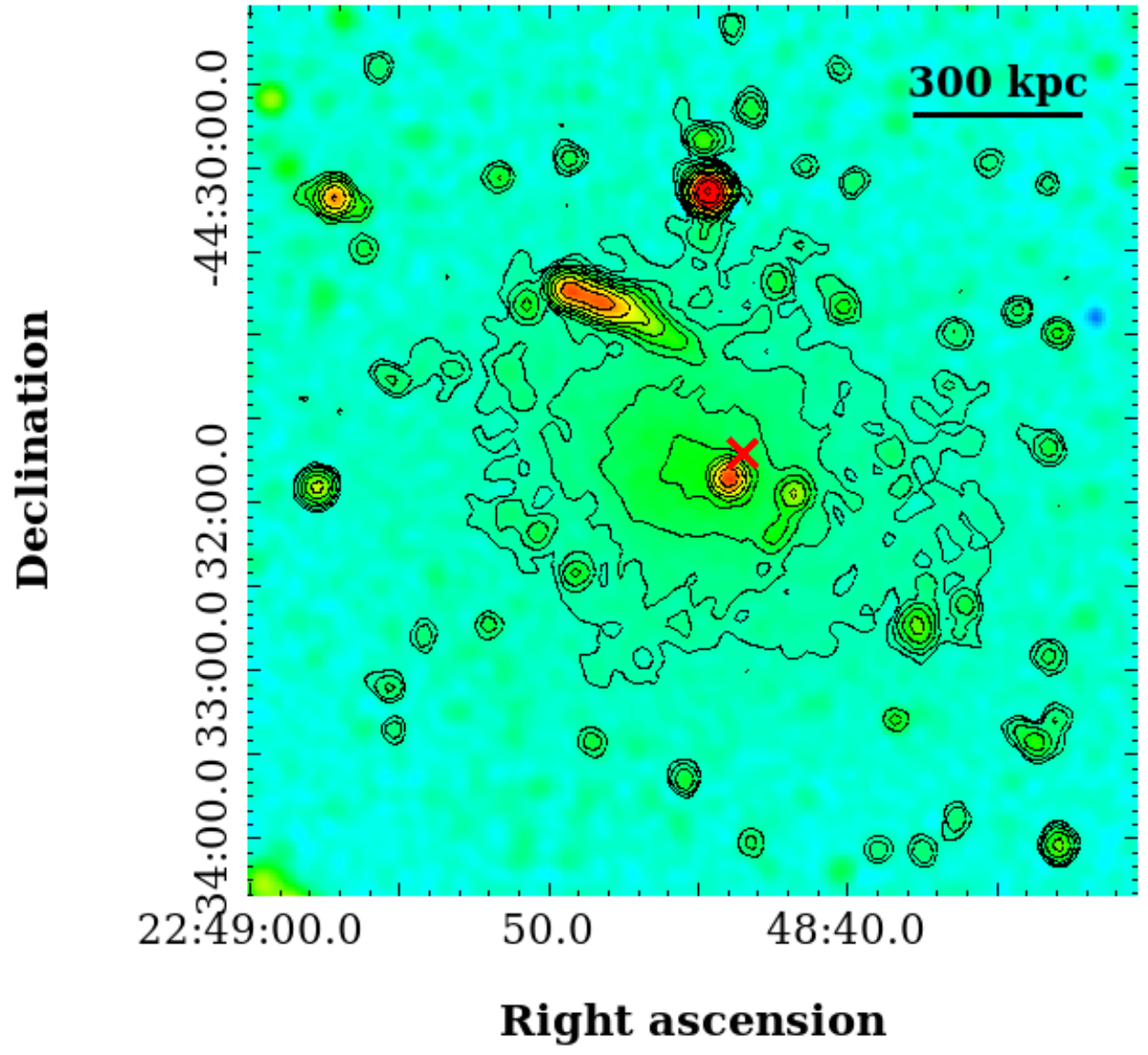}
    \includegraphics[width=0.5\textwidth]{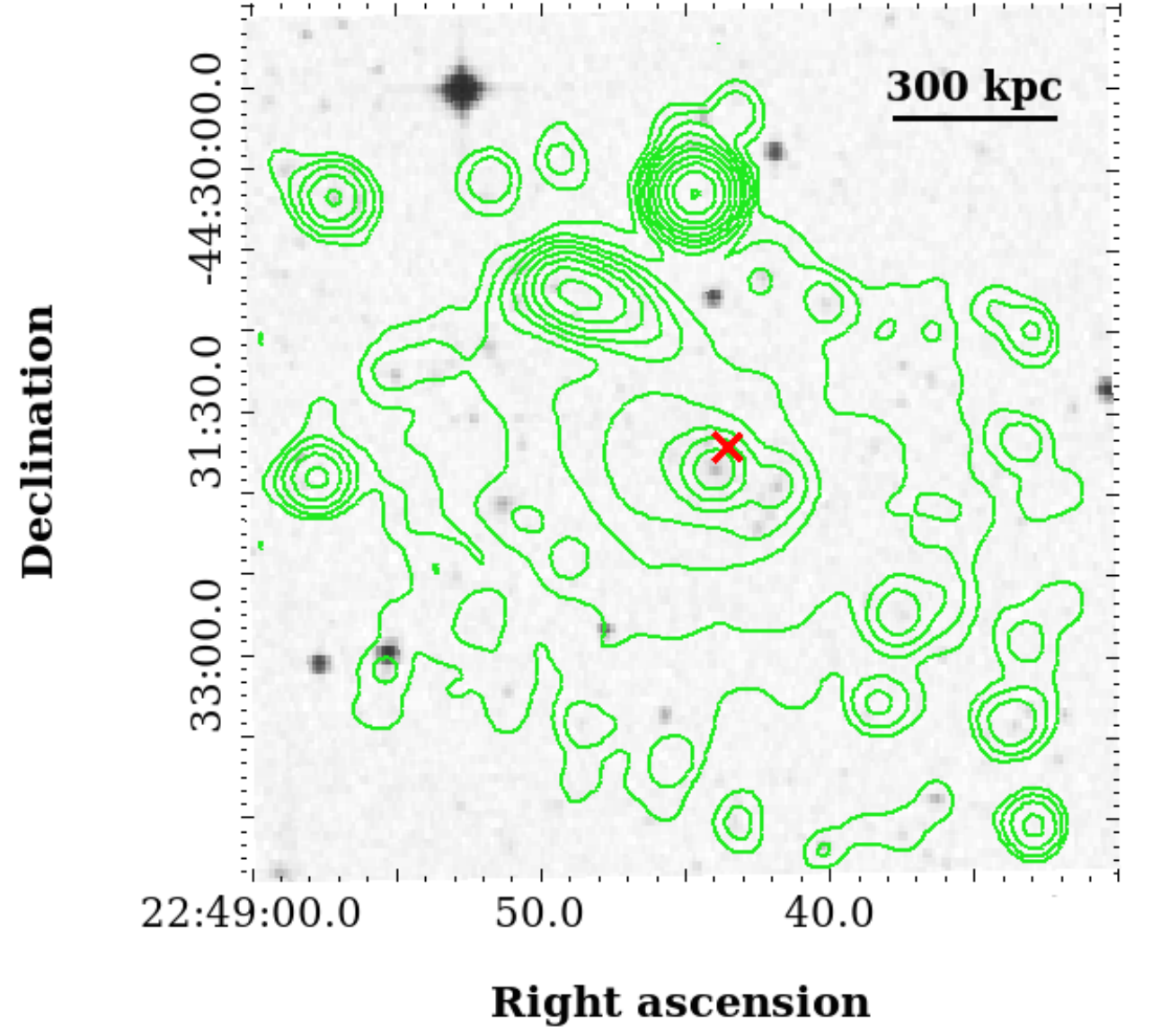}
   \caption{AS1063 \textbf{Left:}  Full-resolution (7.8\arcsec\,$\times$\,7.8\arcsec) 1.28~GHz MGCLS radio image with radio contours in black overlaid (1$\sigma$ = 3 $\mu$Jy beam$^{-1}$). \textbf{Right:} 1.28~GHz MGCLS low-resolution  (15\arcsec\,$\times$\,15\arcsec) radio contours in green (1$\sigma$ = 7 $\mu$Jy beam$^{-1}$), overlaid on the r-band \textit{Digitized Sky Survey (DSS)} optical image. In both panels, the radio contours start at 3\,$\sigma$ and rise by a factor of 2. The physical scale at the cluster redshift is indicated on top right, and the red $\times$ indicates the NED cluster position. } 
   \label{fig:AS1063}%
\end{figure*}

\begin{figure*}
   \centering
   \includegraphics[width=0.496\textwidth]{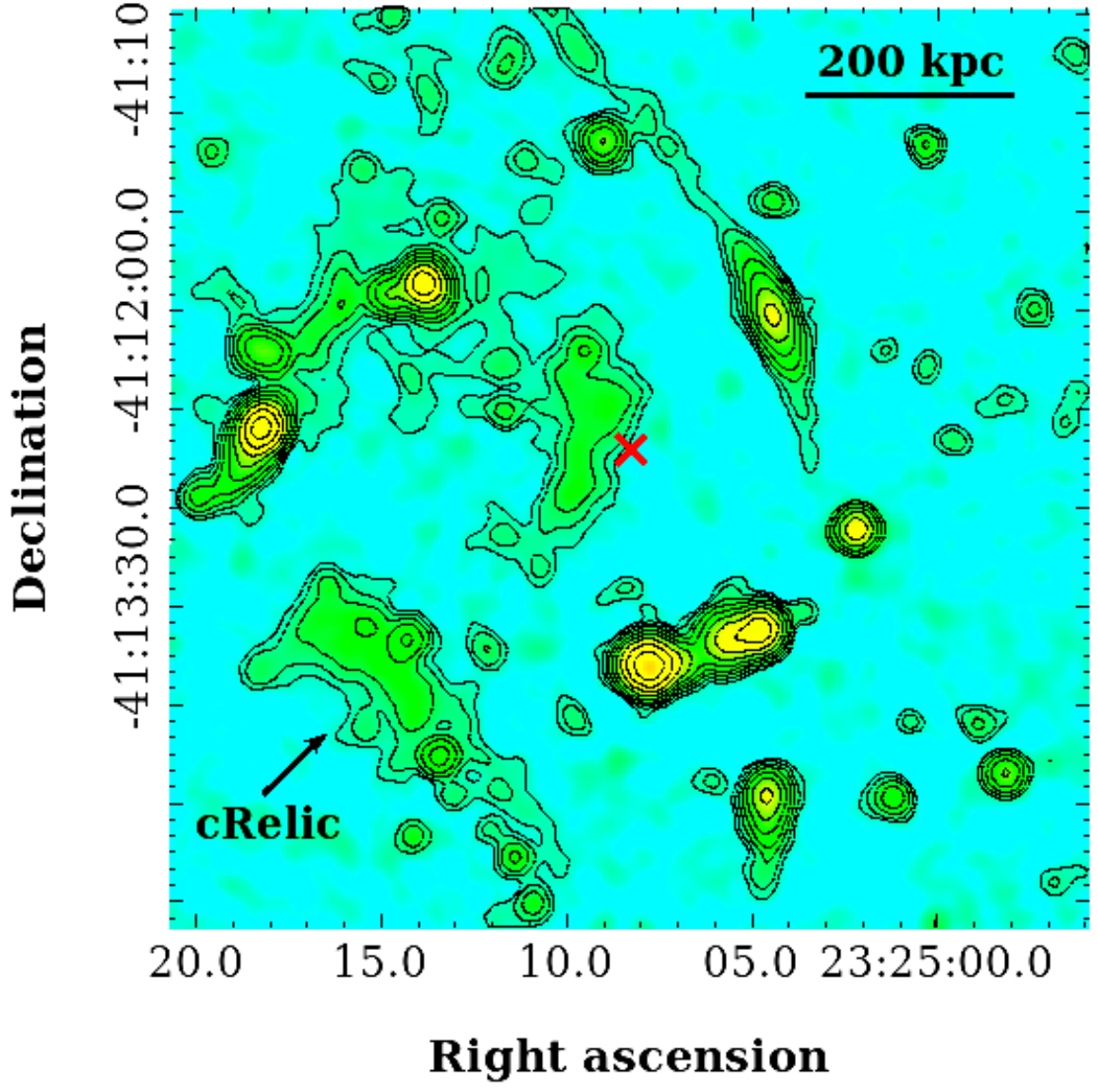}
    \includegraphics[width=0.5\textwidth]{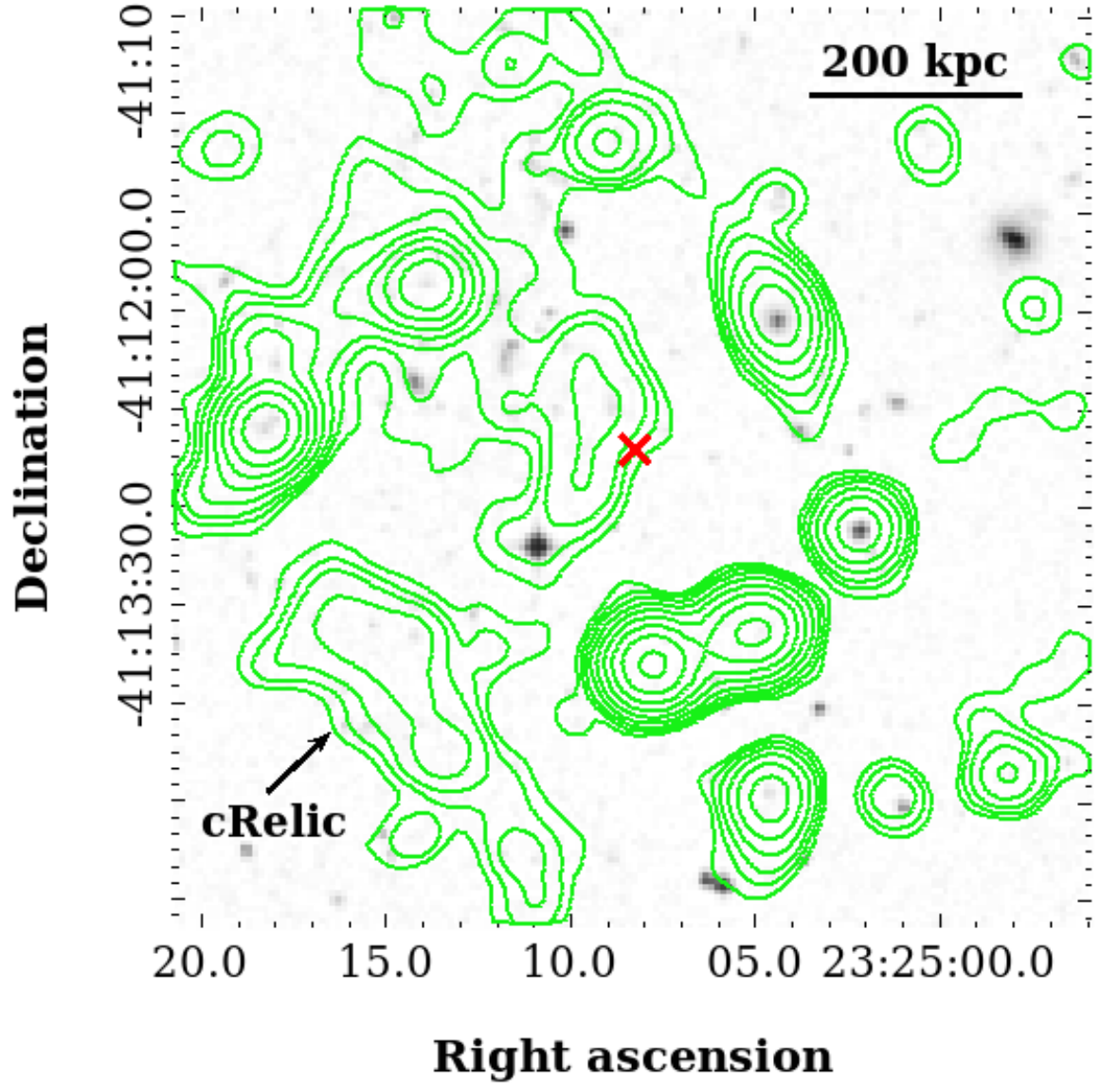}
   \caption{AS1121 \textbf{Left:}  Full-resolution (7.8\arcsec\,$\times$\,7.8\arcsec) 1.28~GHz MGCLS radio image with radio contours in black overlaid (1$\sigma$ = 6 $\mu$Jy beam$^{-1}$). \textbf{Right:} 1.28~GHz MGCLS low-resolution  (15\arcsec\,$\times$\,15\arcsec) radio contours in green (1$\sigma$ = 12 $\mu$Jy beam$^{-1}$), overlaid on the r-band \textit{Digitized Sky Survey (DSS)} optical image. In both panels, the radio contours start at 3\,$\sigma$ and rise by a factor of 2. The physical scale at the cluster redshift is indicated on top right, and the red $\times$ indicates the NED cluster position. } 
   \label{fig:AS1121}%
\end{figure*}

\begin{figure*}
   \centering
   \includegraphics[width=0.496\textwidth]{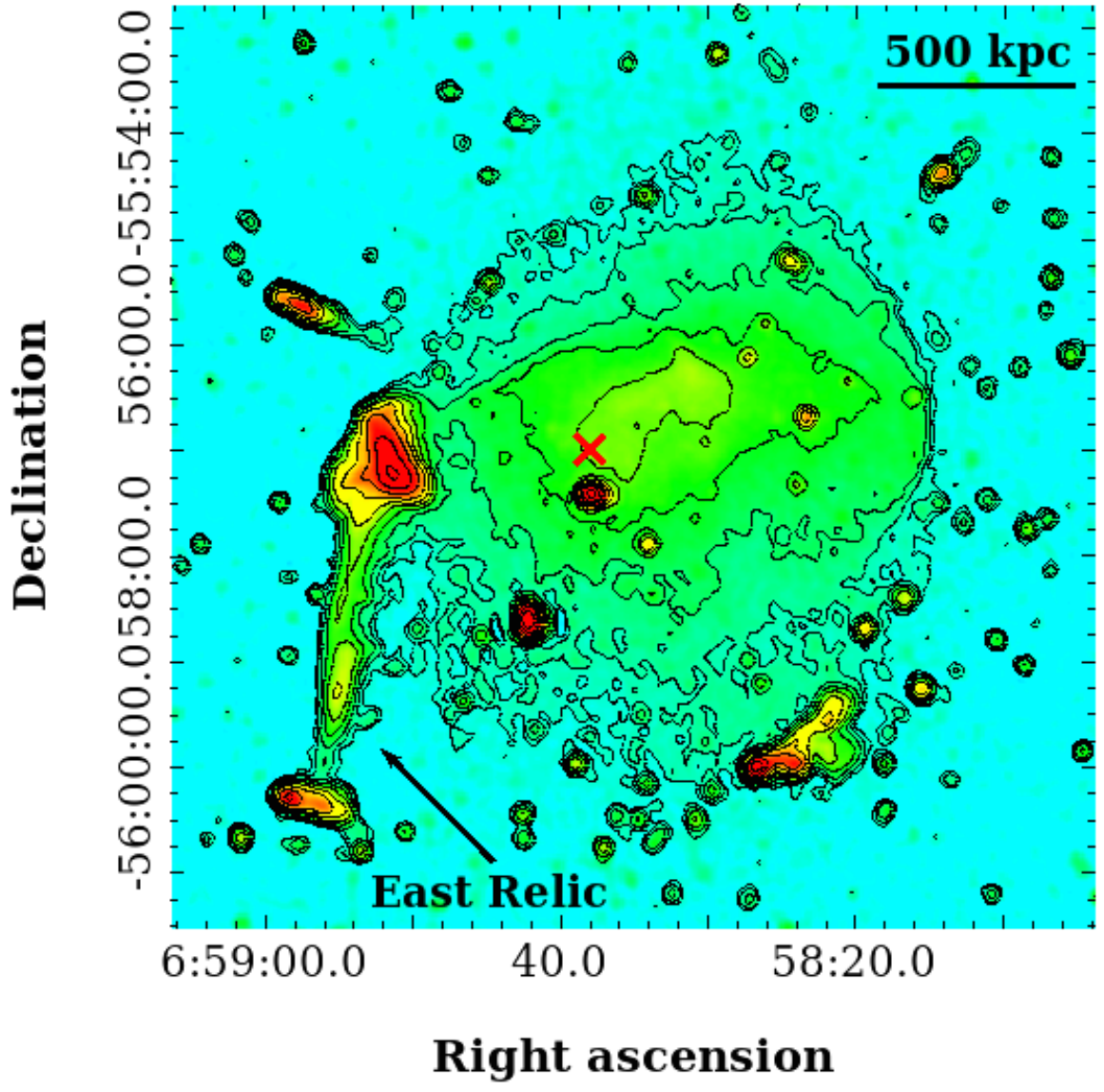}
    \includegraphics[width=0.5\textwidth]{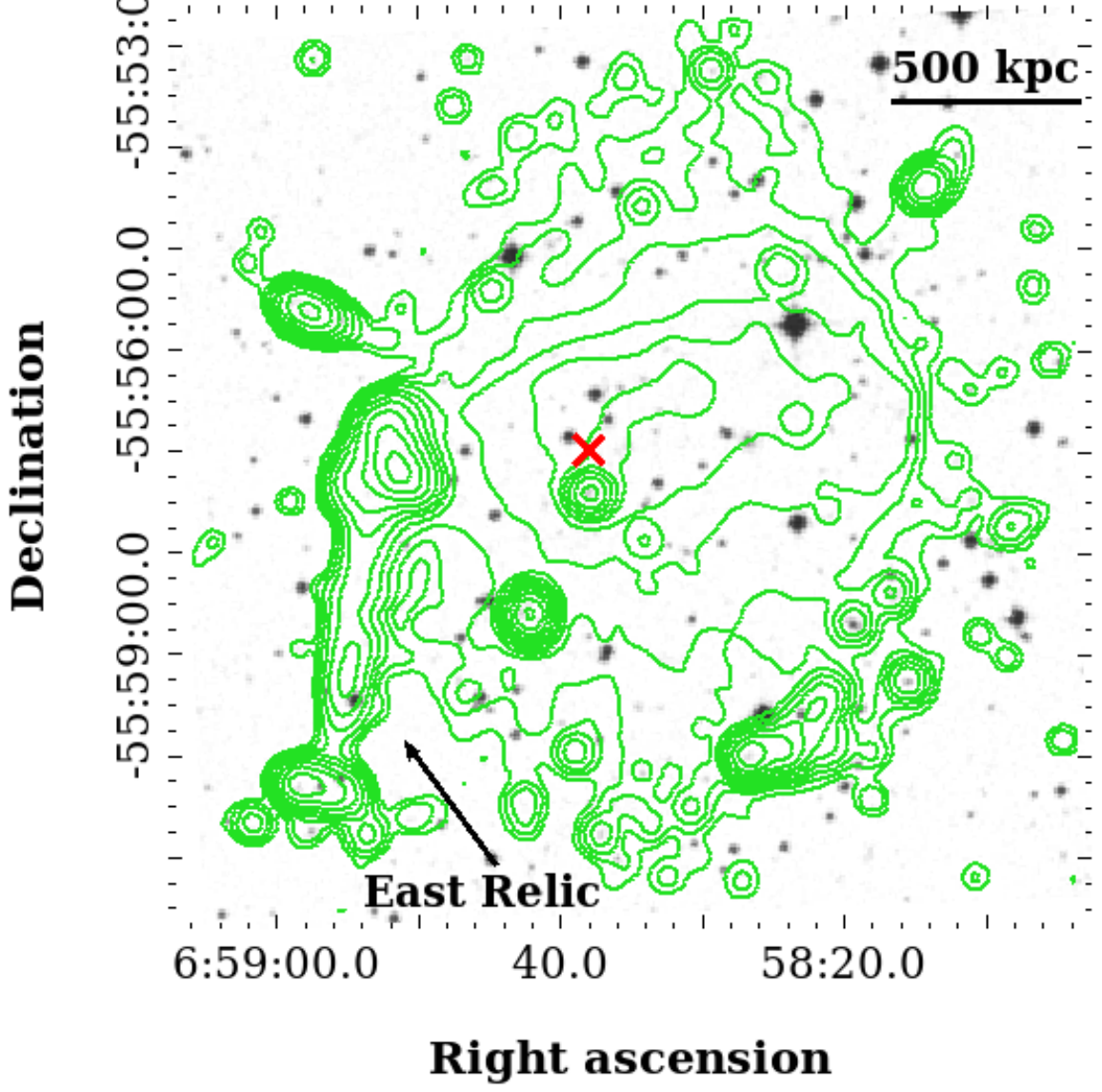}
   \caption{Bullet \textbf{Left:}  Full-resolution (7.8\arcsec\,$\times$\,7.8\arcsec) 1.28~GHz MGCLS radio image with radio contours in black overlaid (1$\sigma$ = 4 $\mu$Jy beam$^{-1}$). \textbf{Right:} 1.28~GHz MGCLS low-resolution  (15\arcsec\,$\times$\,15\arcsec) radio contours in green (1$\sigma$ = 10 $\mu$Jy beam$^{-1}$), overlaid on the r-band \textit{Digitized Sky Survey (DSS)} optical image. In both panels, the radio contours start at 3\,$\sigma$ and rise by a factor of 2. The physical scale at the cluster redshift is indicated on the top right, and the red $\times$ indicates the NED cluster position. } 
   \label{fig:Bullet}%
\end{figure*}

\begin{figure*}
   \centering
   \includegraphics[width=0.496\textwidth]{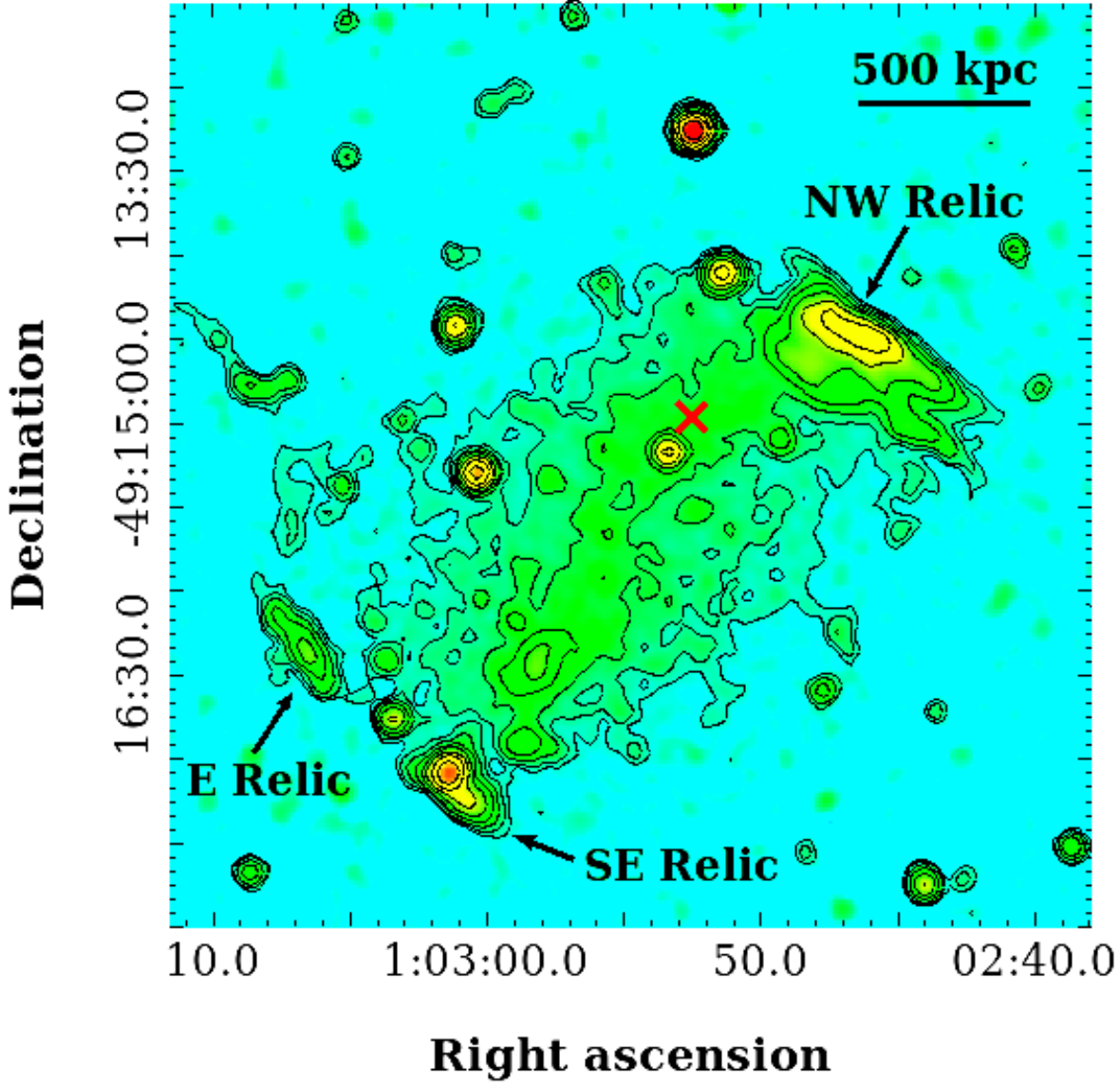}
    \includegraphics[width=0.5\textwidth]{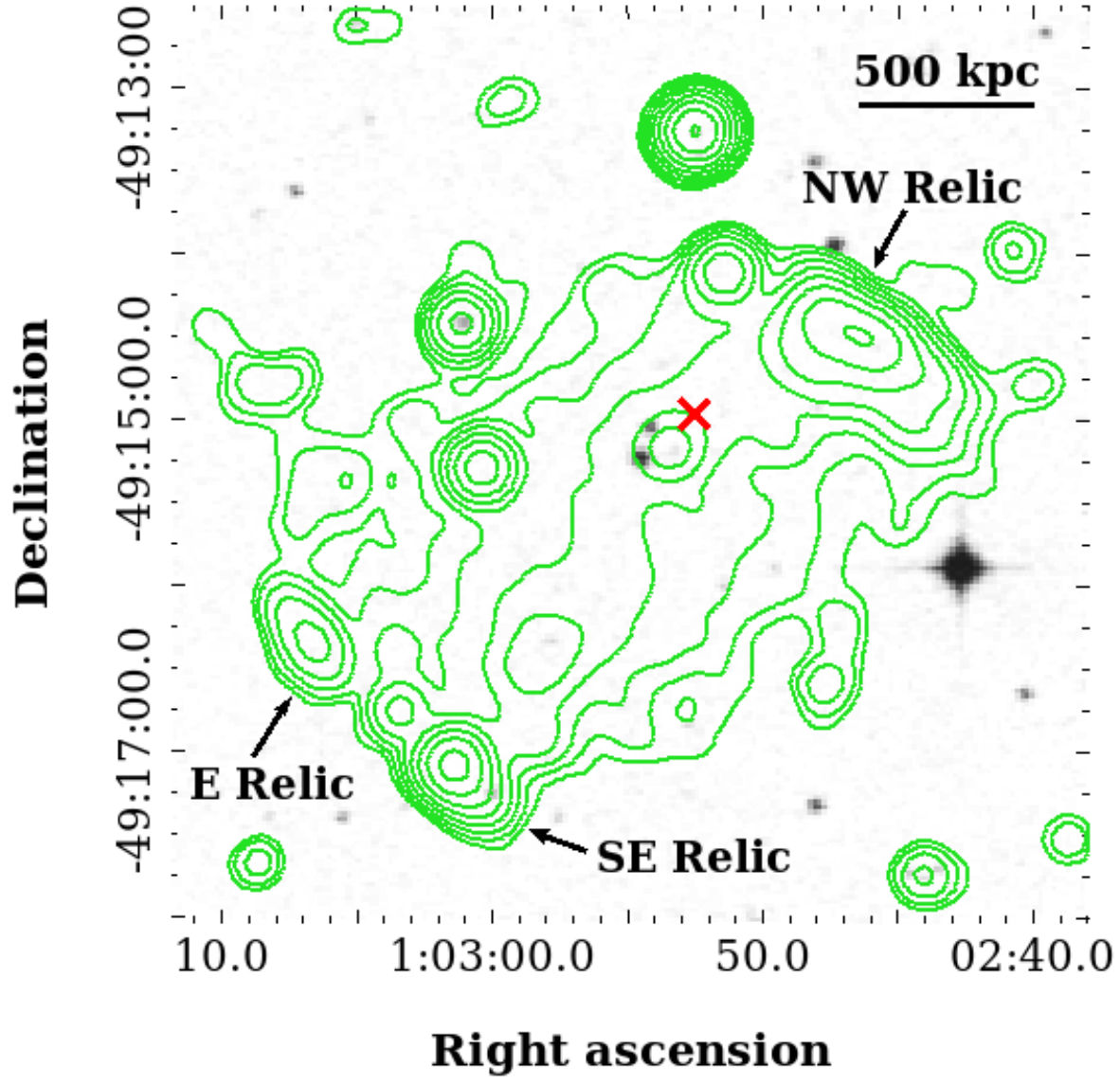}
   \caption{El Gordo \textbf{Left:}  Full-resolution (7.8\arcsec\,$\times$\,7.8\arcsec) 1.28~GHz MGCLS radio image with radio contours in black overlaid (1$\sigma$ = 3 $\mu$Jy beam$^{-1}$). \textbf{Right:} 1.28~GHz MGCLS low-resolution  (15\arcsec\,$\times$\,15\arcsec) radio contours in green (1$\sigma$ = 6 $\mu$Jy beam$^{-1}$), overlaid on the r-band \textit{Digitized Sky Survey (DSS)} optical image. In both panels, the radio contours start at 3\,$\sigma$ and rise by a factor of 2. The physical scale at the cluster redshift is indicated on top right, and the red $\times$ indicates the NED cluster position. } 
   \label{fig:ElGordo}%
\end{figure*}

\begin{figure*}
   \centering
   \includegraphics[width=0.496\textwidth]{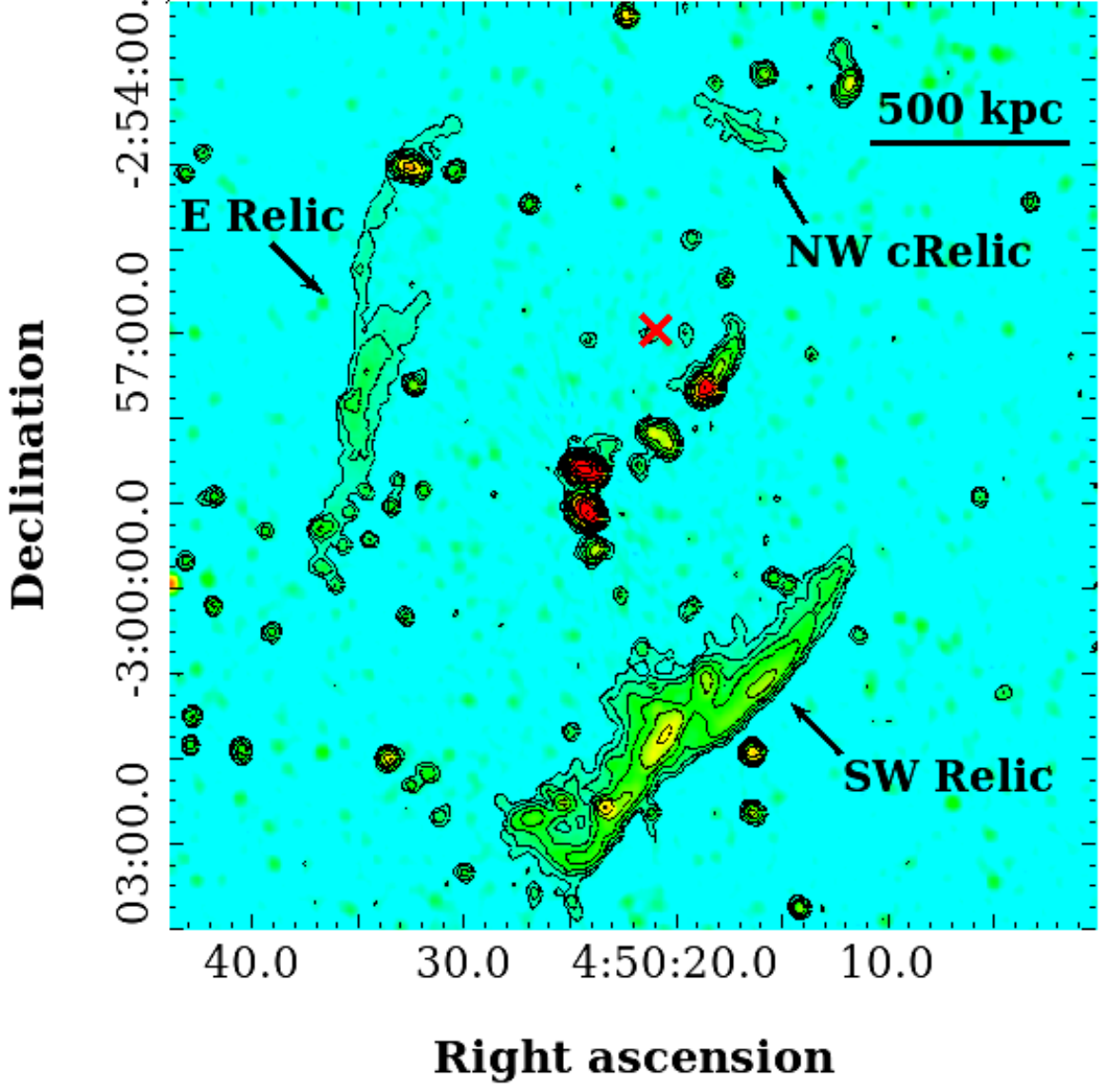}
    \includegraphics[width=0.5\textwidth]{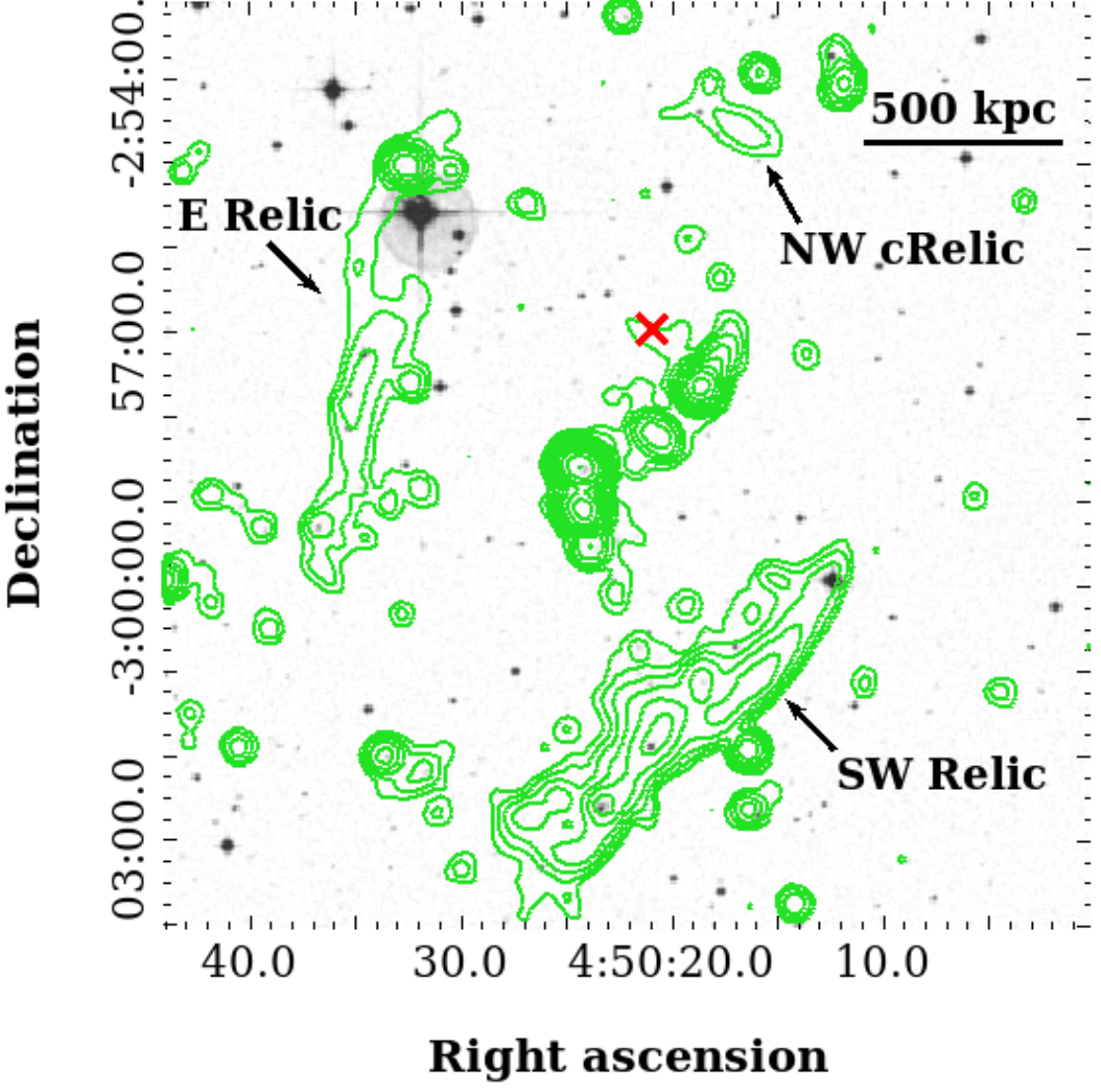}
   \caption{PLCK G200.9$-$28.2 \textbf{Left:}  Full-resolution (7.8\arcsec\,$\times$\,7.8\arcsec) 1.28~GHz MGCLS radio image with radio contours in black overlaid (1$\sigma$ = 5 $\mu$Jy beam$^{-1}$). \textbf{Right:} 1.28~GHz MGCLS low-resolution  (15\arcsec\,$\times$\,15\arcsec) radio contours in green (1$\sigma$ = 12 $\mu$Jy beam$^{-1}$), overlaid on the r-band \textit{Digitized Sky Survey (DSS)} optical image. In both panels, the radio contours start at 3\,$\sigma$ and rise by a factor of 2. The physical scale at the cluster redshift is indicated on top right, and the red $\times$ indicates the NED cluster position. } 
   \label{fig:PLCKG200}%
\end{figure*}

\begin{figure*}
   \centering
   \includegraphics[width=0.496\textwidth]{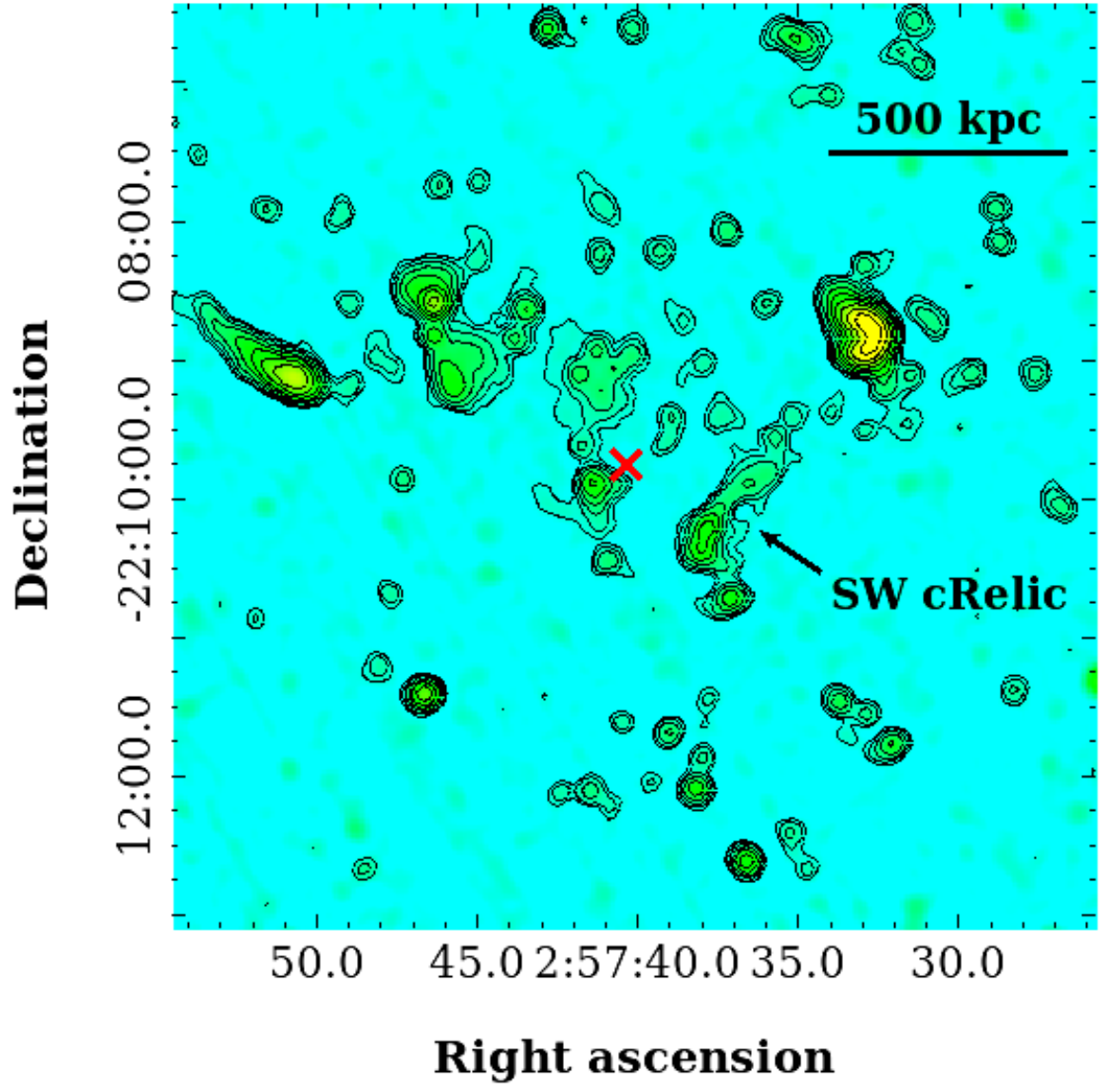}
    \includegraphics[width=0.5\textwidth]{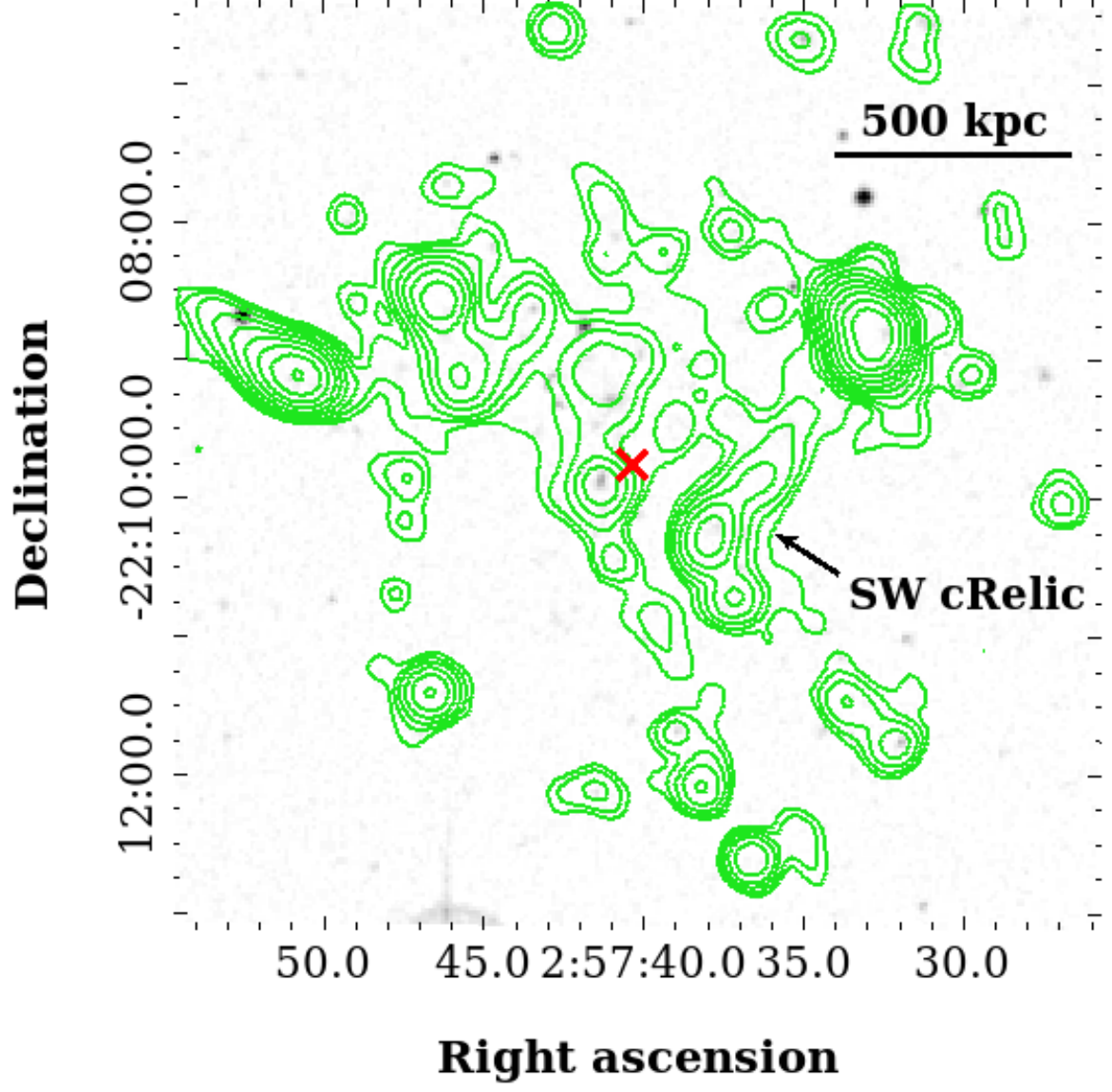}
   \caption{MACS J0257.6$-$2209 \textbf{Left:}  Full-resolution (7.8\arcsec\,$\times$\,7.8\arcsec) 1.28~GHz MGCLS radio image with radio contours in black overlaid (1$\sigma$ = 3.5 $\mu$Jy beam$^{-1}$). \textbf{Right:} 1.28~GHz MGCLS low-resolution  (15\arcsec\,$\times$\,15\arcsec) radio contours in green (1$\sigma$ = 6 $\mu$Jy beam$^{-1}$), overlaid on the r-band \textit{Digitized Sky Survey (DSS)} optical image. In both panels, the radio contours start at 3\,$\sigma$ and rise by a factor of 2. The physical scale at the cluster redshift is indicated on top right, and the red $\times$ indicates the NED cluster position. } 
   \label{fig:MACSJ0257}%
\end{figure*}

\begin{figure*}
   \centering
   \includegraphics[width=0.496\textwidth]{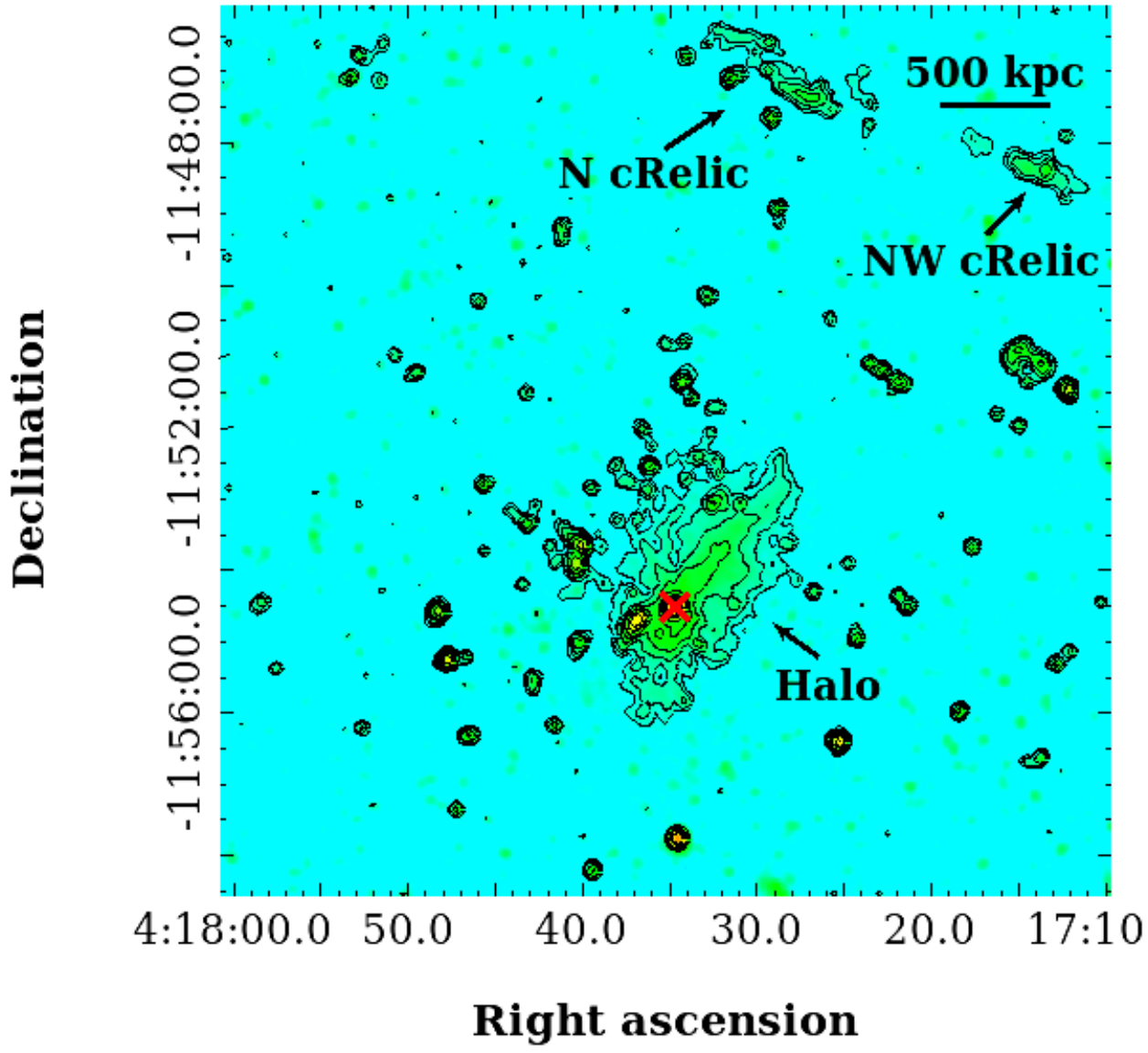}
    \includegraphics[width=0.5\textwidth]{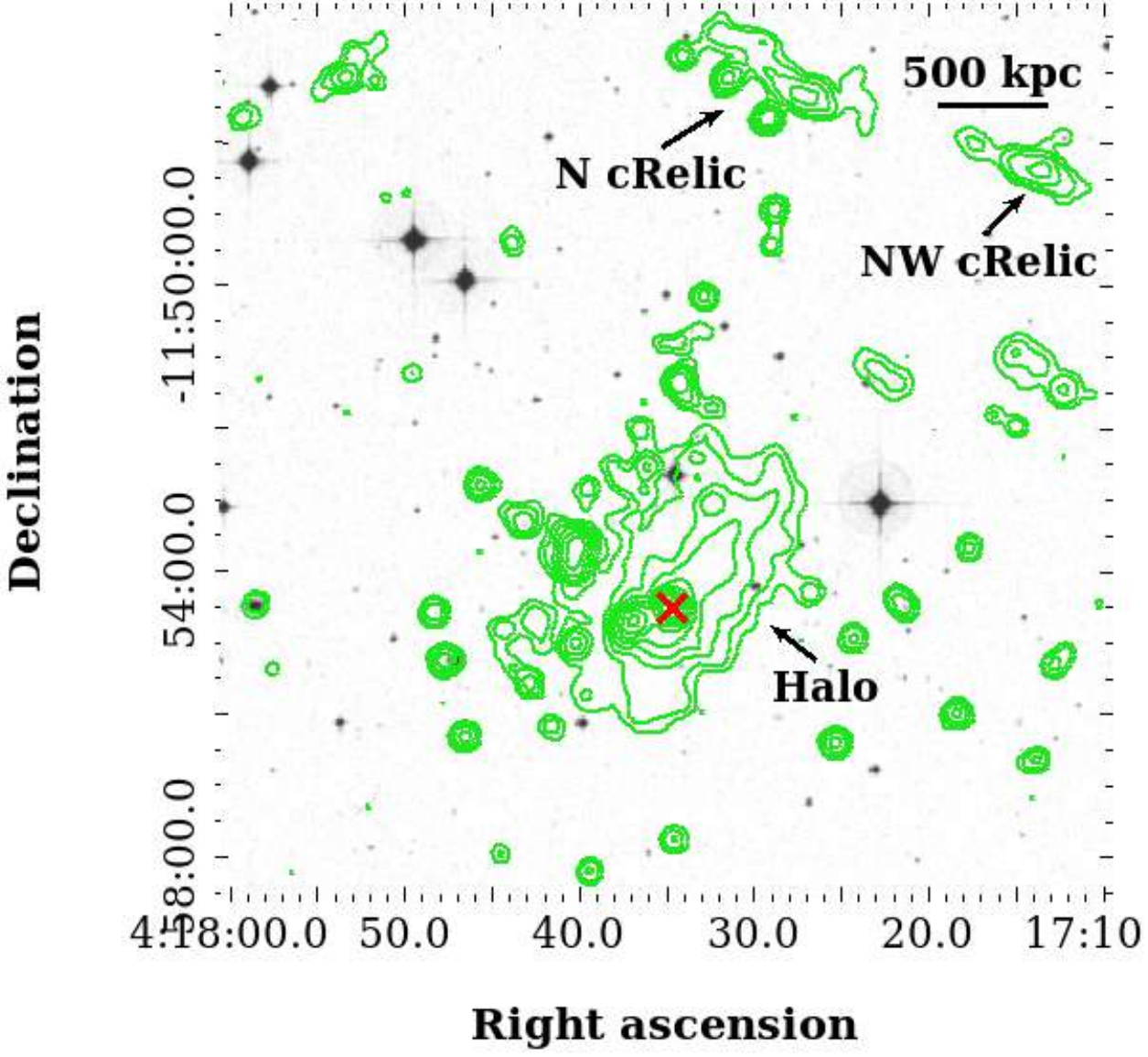}
   \caption{MACS J0417.5$-$1154 \textbf{Left:}  Full-resolution (7.8\arcsec\,$\times$\,7.8\arcsec) 1.28~GHz MGCLS radio image with radio contours in black overlaid (1$\sigma$ = 4 $\mu$Jy beam$^{-1}$). \textbf{Right:} 1.28~GHz MGCLS low-resolution  (15\arcsec\,$\times$\,15\arcsec) radio contours in green (1$\sigma$ = 9 $\mu$Jy beam$^{-1}$), overlaid on the r-band \textit{Digitized Sky Survey (DSS)} optical image. In both panels, the radio contours start at 3\,$\sigma$ and rise by a factor of 2. The physical scale at the cluster redshift is indicated on top right, and the red $\times$ indicates the NED cluster position. } 
   \label{fig:MACSJ0417}%
\end{figure*}

\begin{figure*}
   \centering
   \includegraphics[width=0.496\textwidth]{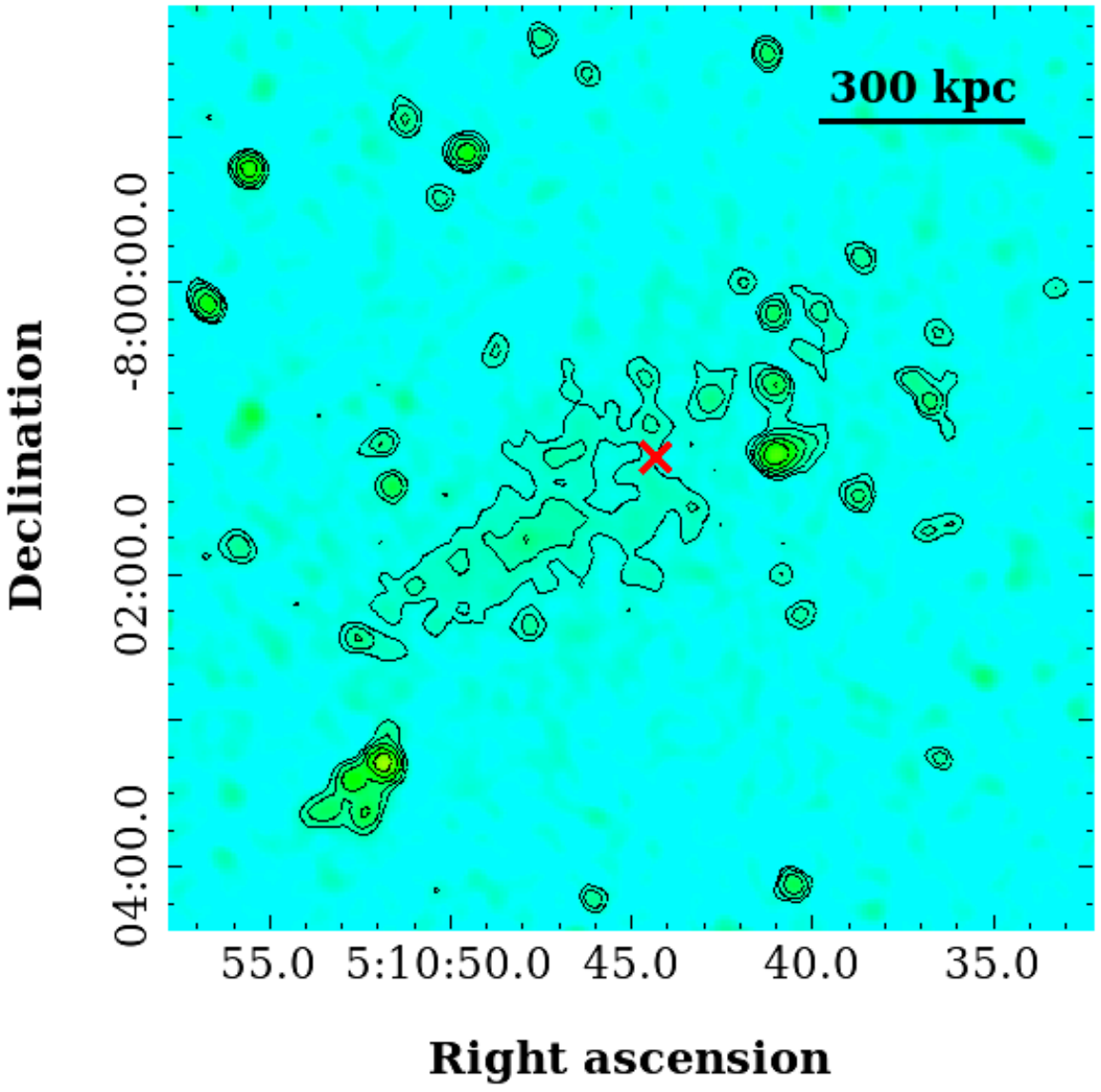}
    \includegraphics[width=0.5\textwidth]{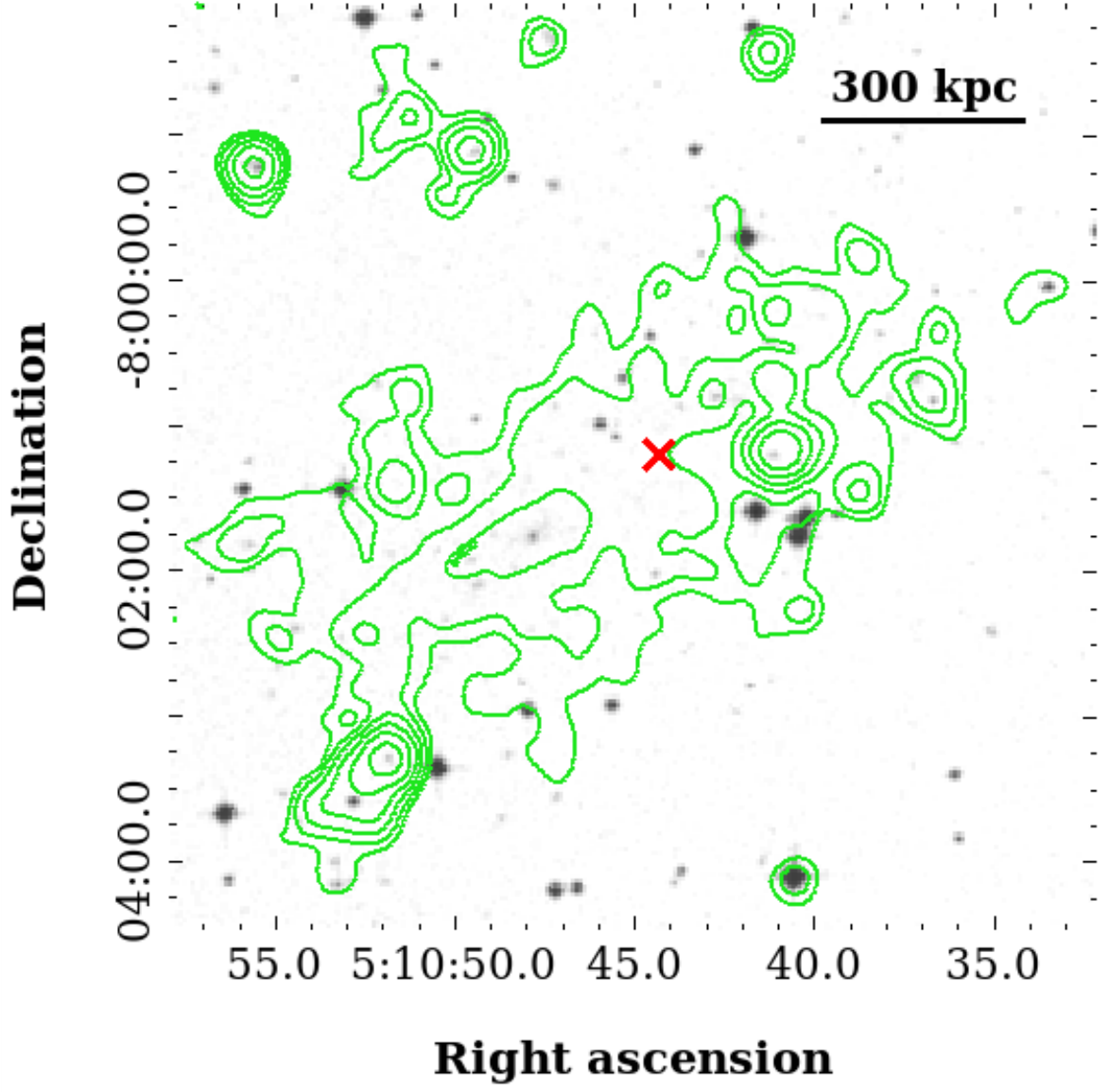}
   \caption{RXC J0510.7$-$0801 \textbf{Left:}  Full-resolution (7.8\arcsec\,$\times$\,7.8\arcsec) 1.28~GHz MGCLS radio image with radio contours in black overlaid (1$\sigma$ = 6 $\mu$Jy beam$^{-1}$). \textbf{Right:} 1.28~GHz MGCLS low-resolution  (15\arcsec\,$\times$\,15\arcsec) radio contours in green (1$\sigma$ = 8 $\mu$Jy beam$^{-1}$), overlaid on the r-band \textit{Digitized Sky Survey (DSS)} optical image. In both panels, the radio contours start at 3\,$\sigma$ and rise by a factor of 2. The physical scale at the cluster redshift is indicated on top right, and the red $\times$ indicates the NED cluster position. } 
   \label{fig:RXCJ0510}%
\end{figure*}

\begin{figure*}
   \centering
   \includegraphics[width=0.496\textwidth]{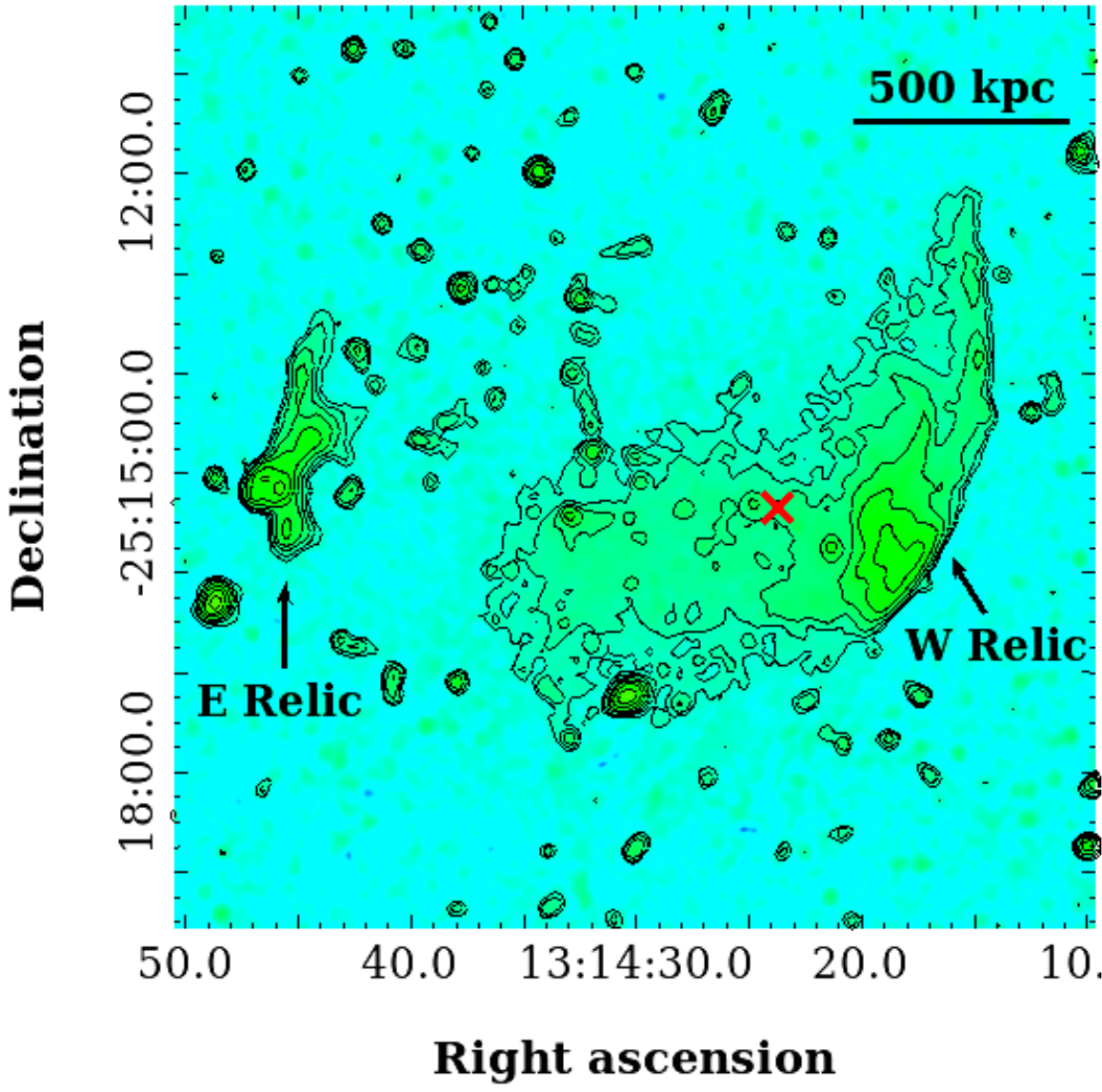}
    \includegraphics[width=0.5\textwidth]{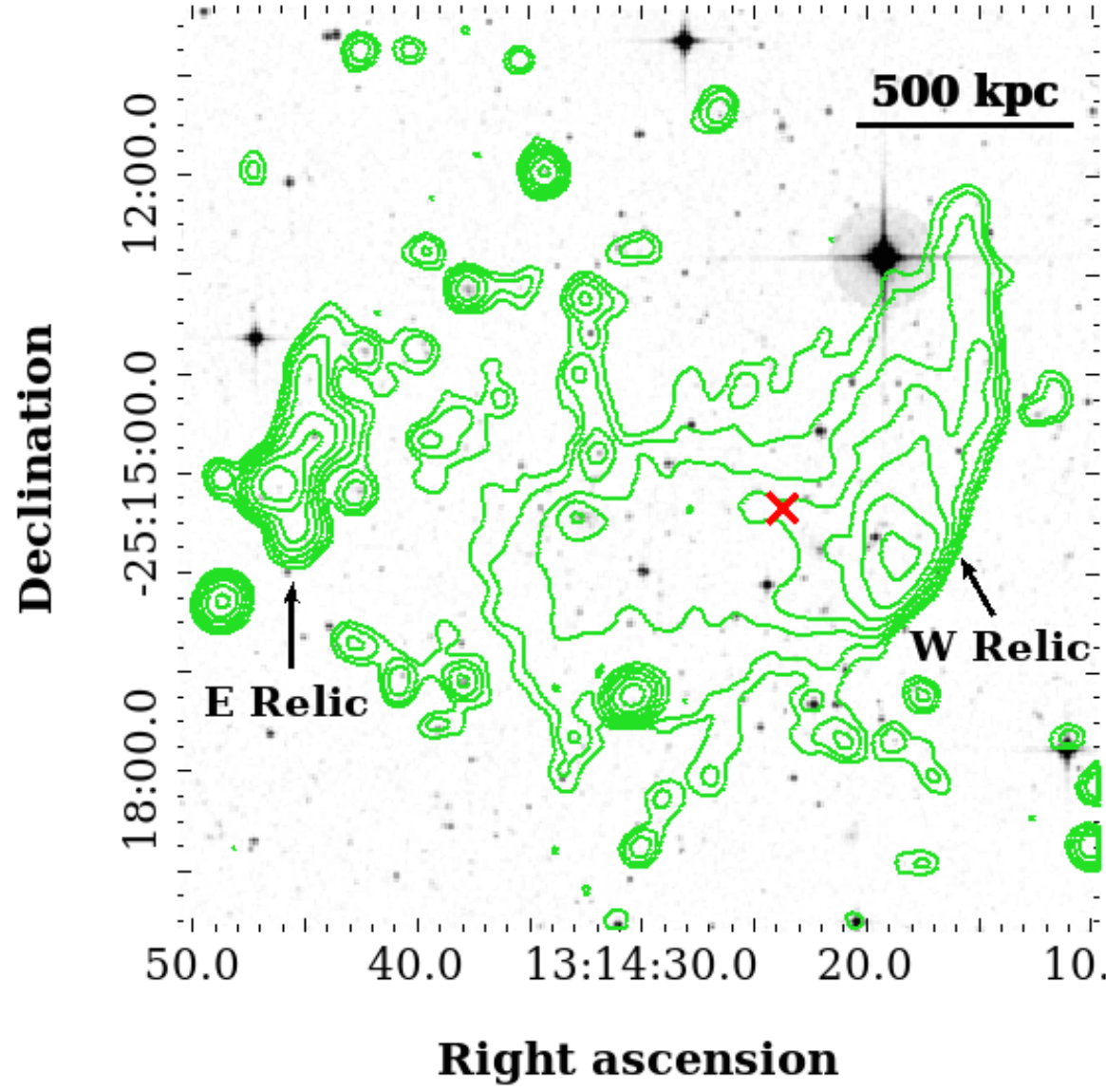}
   \caption{RXC J1314.4$-$2515 \textbf{Left:}  Full-resolution (7.8\arcsec\,$\times$\,7.8\arcsec) 1.28~GHz MGCLS radio image with radio contours in black overlaid (1$\sigma$ = 5 $\mu$Jy beam$^{-1}$). \textbf{Right:} 1.28~GHz MGCLS low-resolution  (15\arcsec\,$\times$\,15\arcsec) radio contours in green (1$\sigma$ = 10 $\mu$Jy beam$^{-1}$), overlaid on the r-band \textit{Digitized Sky Survey (DSS)} optical image. In both panels, the radio contours start at 3\,$\sigma$ and rise by a factor of 2. The physical scale at the cluster redshift is indicated on top right, and the red $\times$ indicates the NED cluster position. } 
   \label{fig:RXCJ1314}%
\end{figure*}

\begin{figure*}
   \centering
   \includegraphics[width=0.496\textwidth]{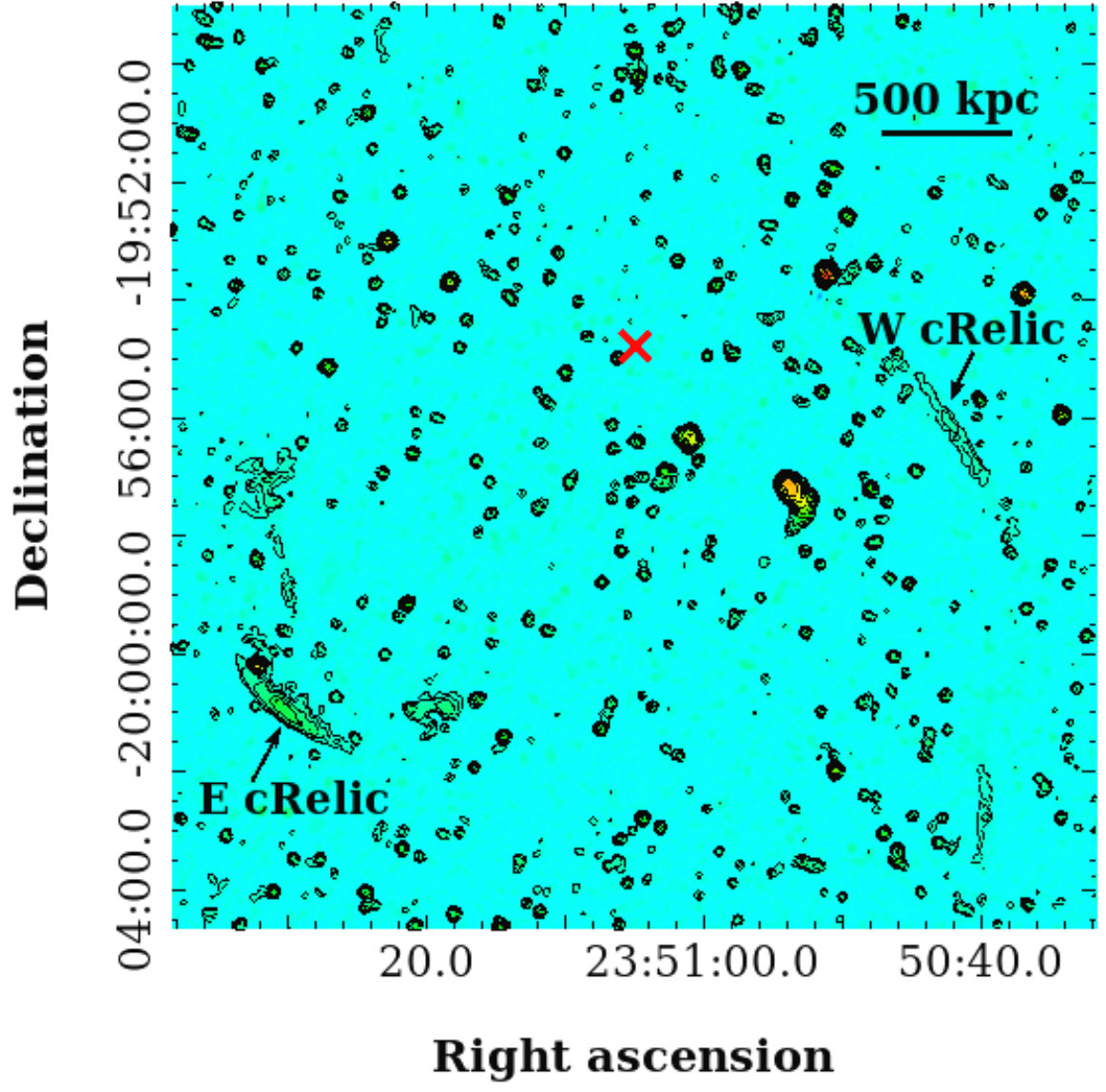}
    \includegraphics[width=0.5\textwidth]{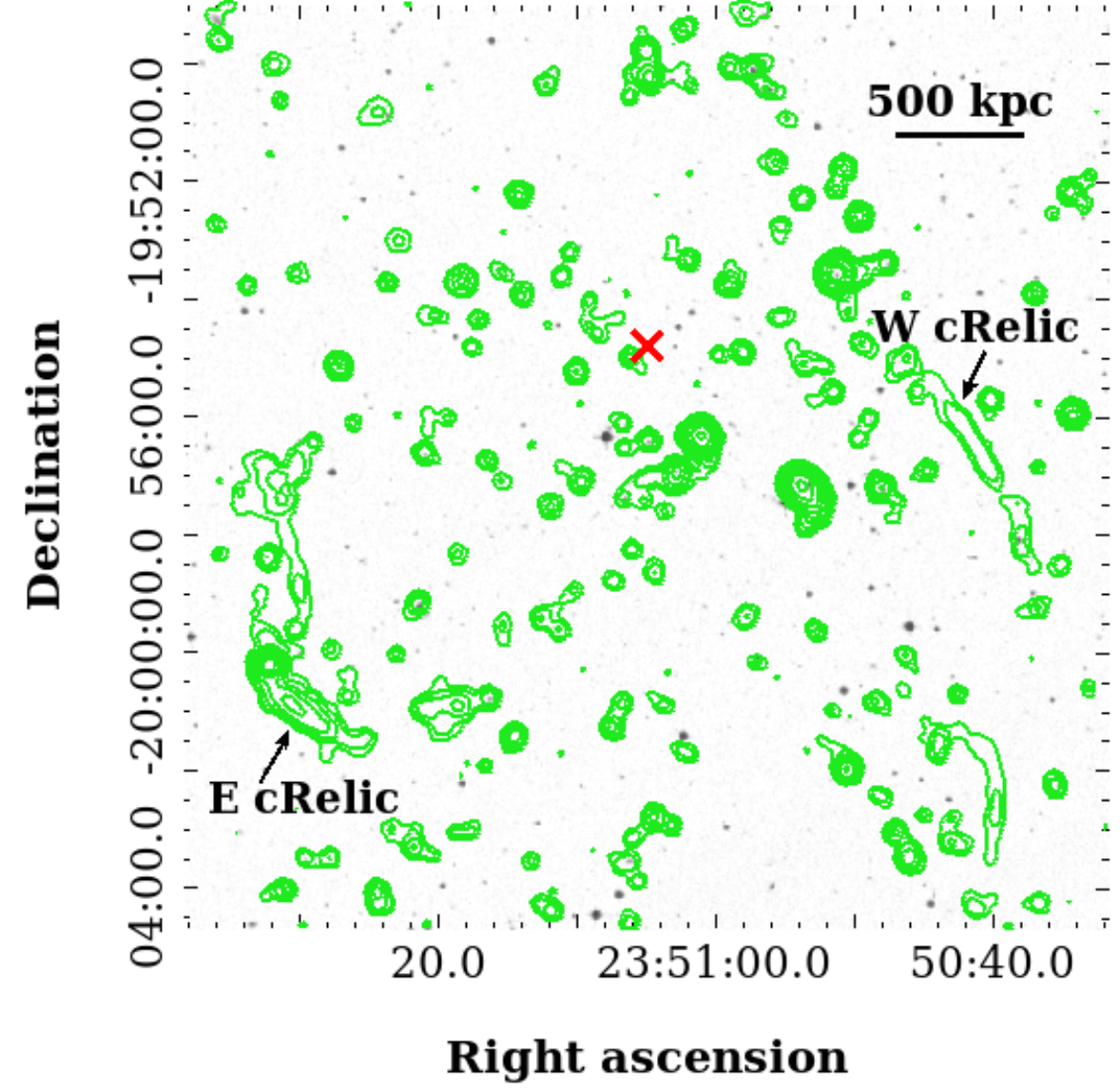}
   \caption{RXC J2351.0$-$1954 \textbf{Left:}  Full-resolution (7.8\arcsec\,$\times$\,7.8\arcsec) 1.28~GHz MGCLS radio image with radio contours in black overlaid (1$\sigma$ = 4 $\mu$Jy beam$^{-1}$). \textbf{Right:} 1.28~GHz MGCLS low-resolution  (15\arcsec\,$\times$\,15\arcsec) radio contours in green (1$\sigma$ = 7 $\mu$Jy beam$^{-1}$), overlaid on the r-band \textit{Digitized Sky Survey (DSS)} optical image. In both panels, the radio contours start at 3\,$\sigma$ and rise by a factor of 2. The physical scale at the cluster redshift is indicated on top right, and the red $\times$ indicates the NED cluster position. } 
   \label{fig:RXCJ2351}%
\end{figure*}

\begin{figure*}
   \centering
   \includegraphics[width=0.496\textwidth]{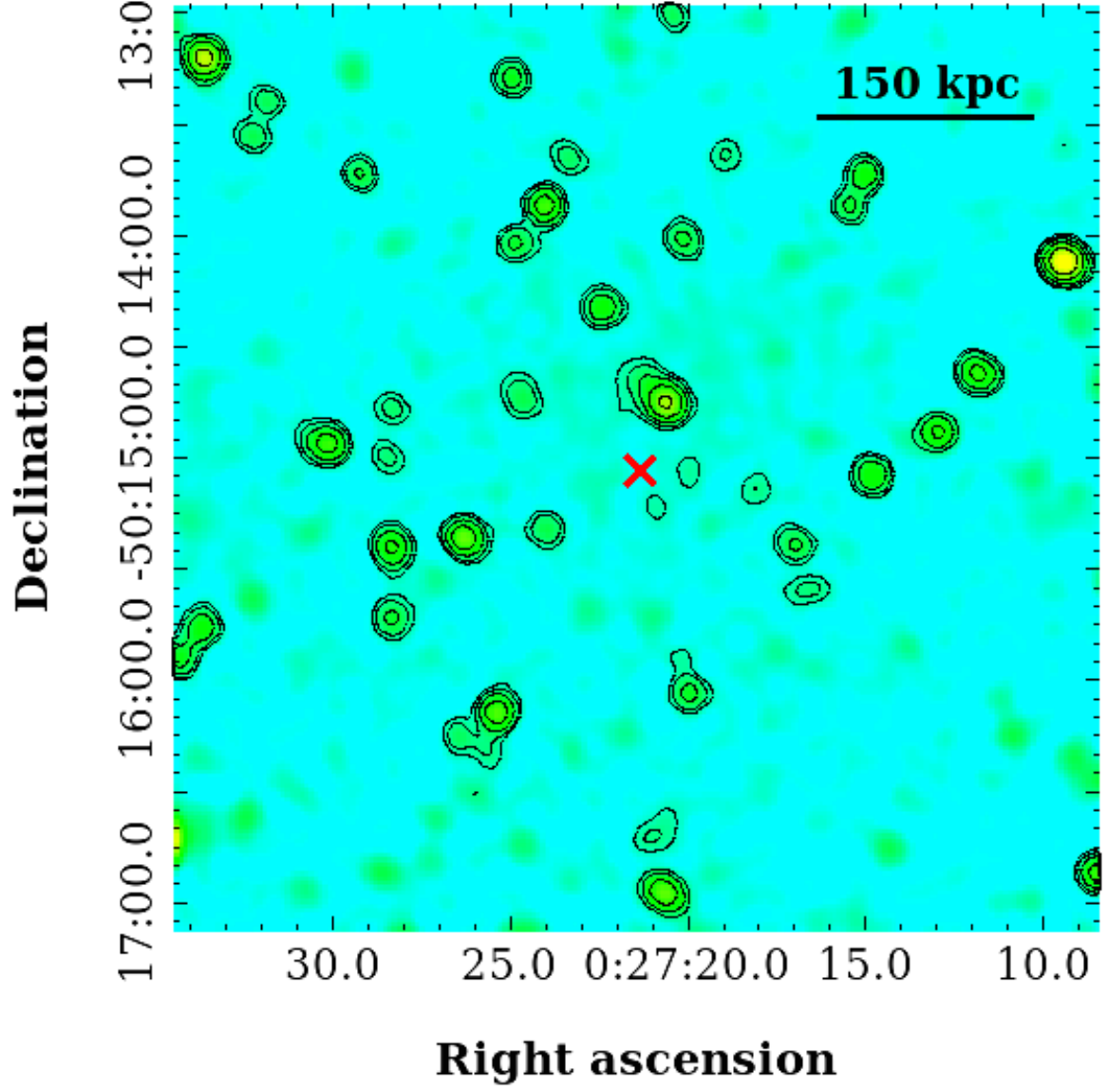}
    \includegraphics[width=0.5\textwidth]{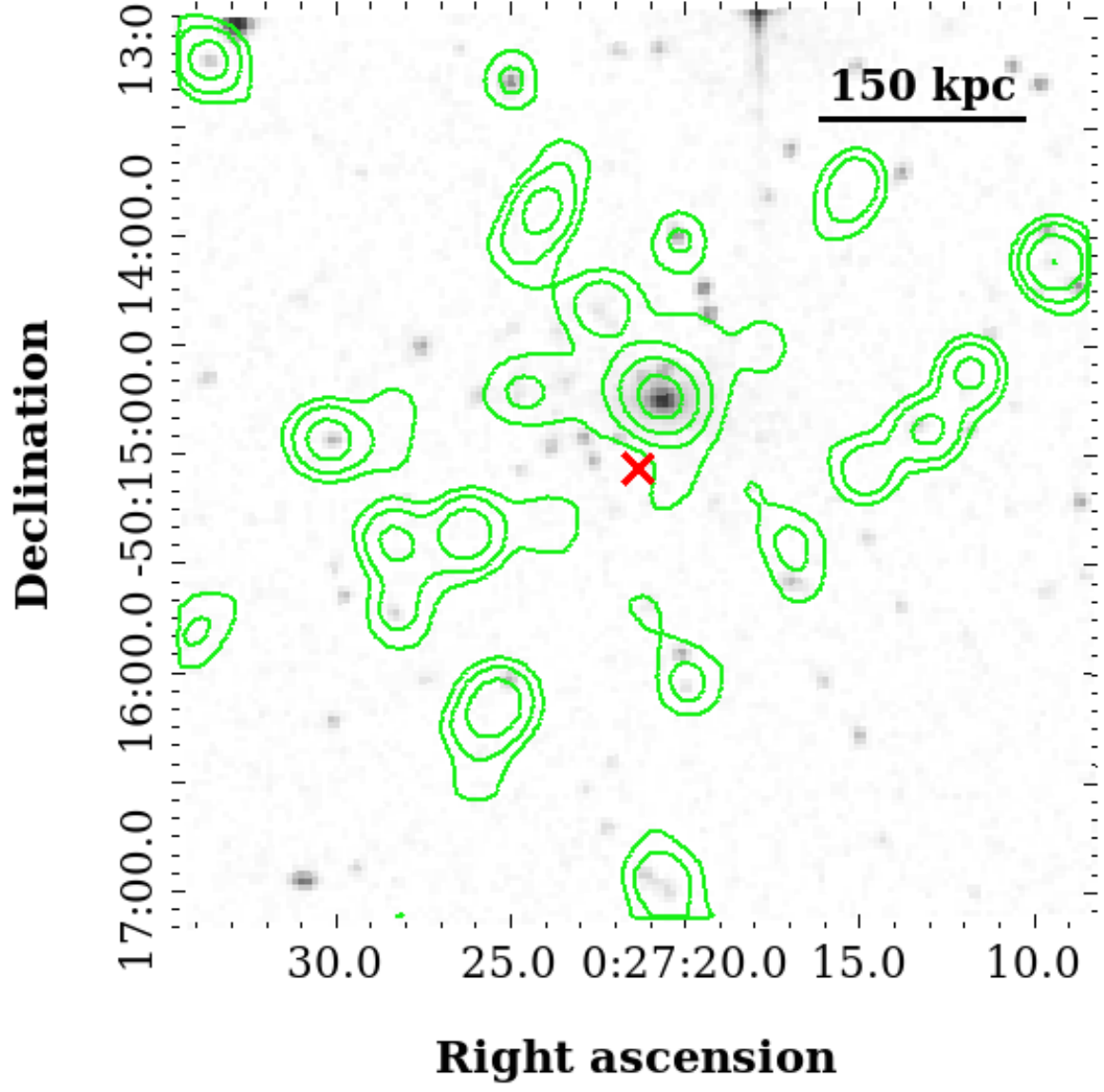}
   \caption{J0027.3$-$5015 \textbf{Left:}  Full-resolution (7.8\arcsec\,$\times$\,7.8\arcsec) 1.28~GHz MGCLS radio image with radio contours in black overlaid (1$\sigma$ = 3 $\mu$Jy beam$^{-1}$). \textbf{Right:} 1.28~GHz MGCLS low-resolution  (15\arcsec\,$\times$\,15\arcsec) radio contours in green (1$\sigma$ = 6 $\mu$Jy beam$^{-1}$), overlaid on the r-band \textit{Digitized Sky Survey (DSS)} optical image. In both panels, the radio contours start at 3\,$\sigma$ and rise by a factor of 2. The physical scale at the cluster redshift is indicated on top right, and the red $\times$ indicates the NED cluster position. } 
   \label{fig:J0027}%
\end{figure*}

\begin{figure*}
   \centering
   \includegraphics[width=0.496\textwidth]{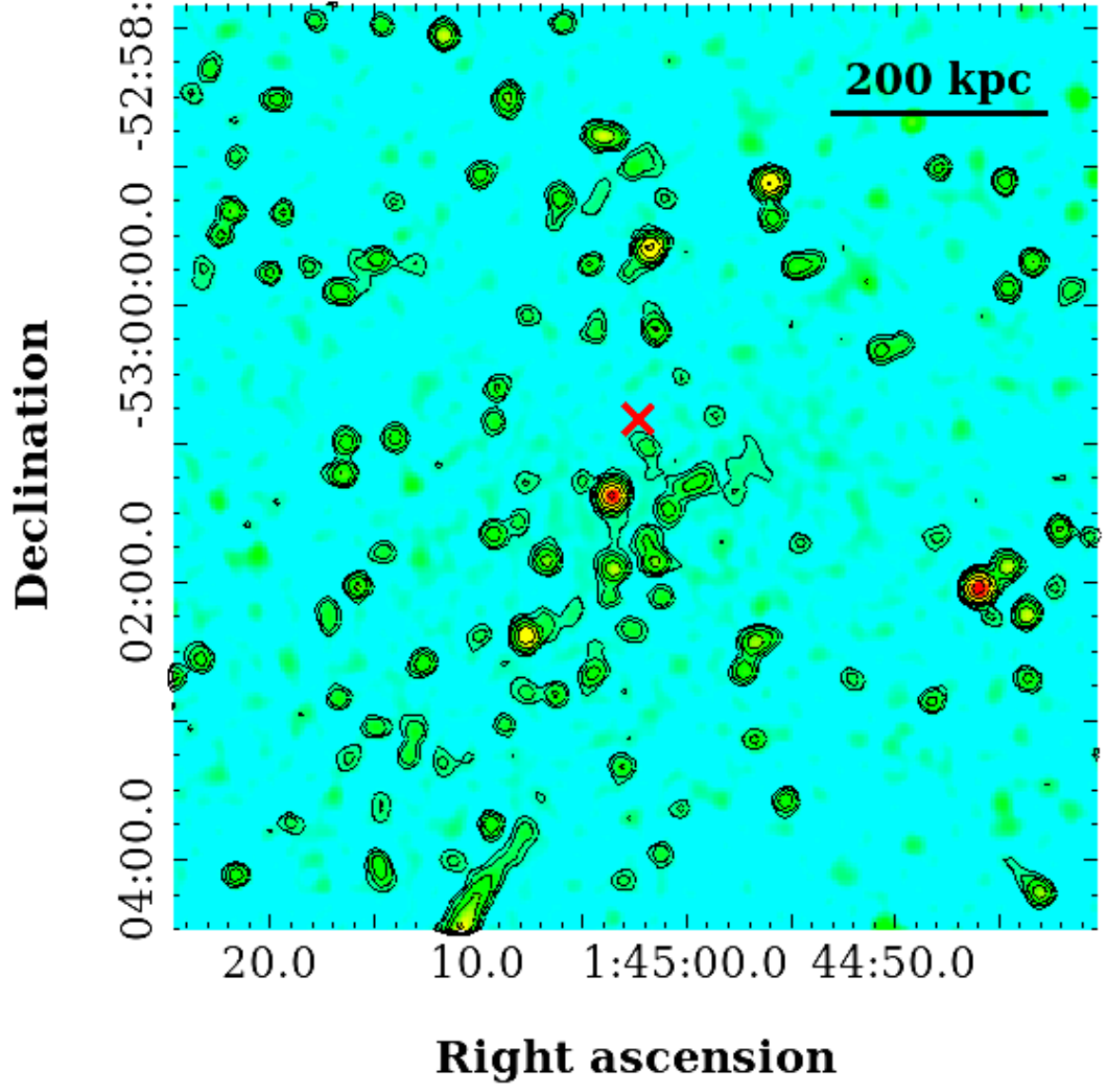}
    \includegraphics[width=0.5\textwidth]{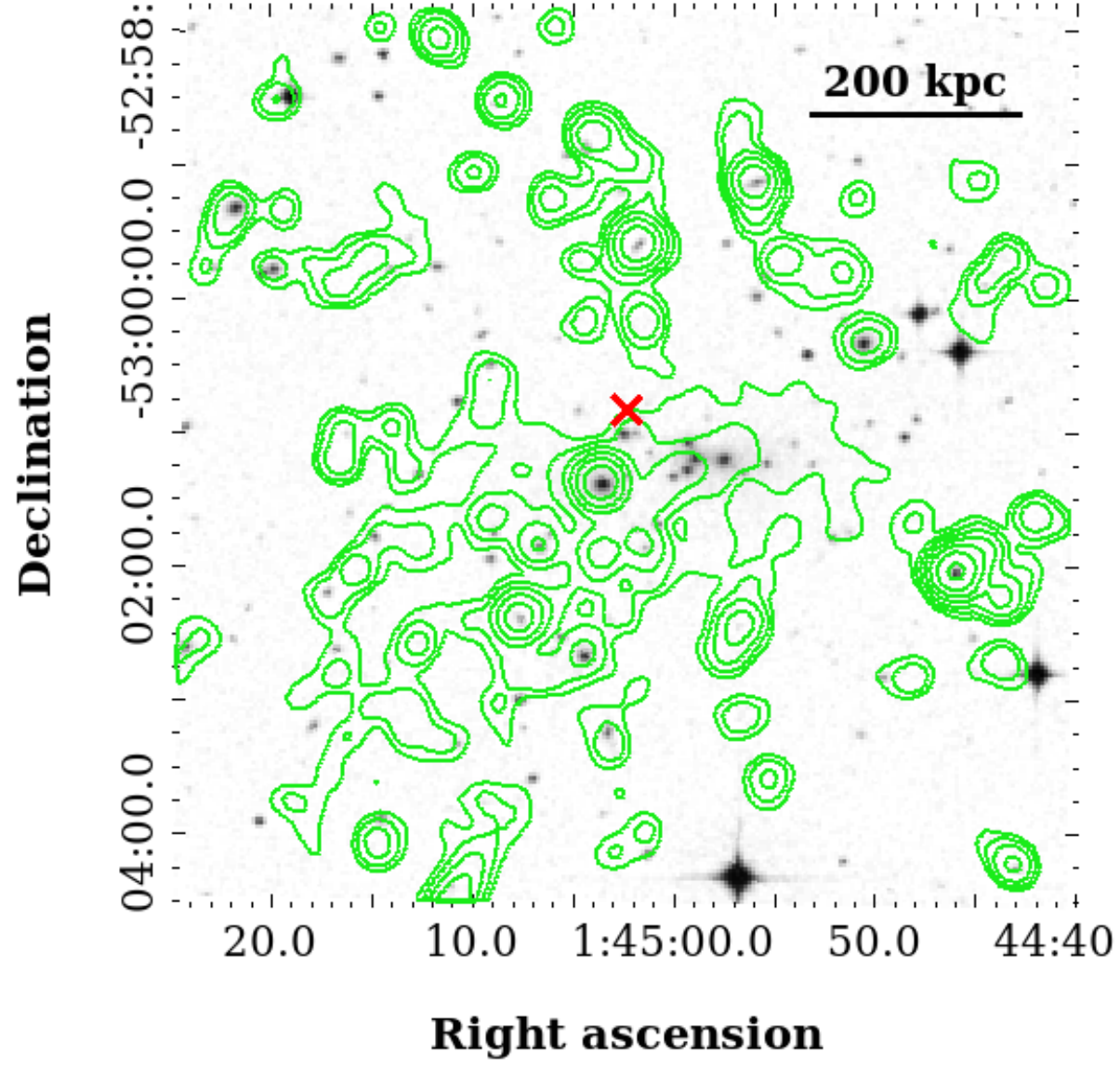}
   \caption{J0145.0$-$5300 \textbf{Left:}  Full-resolution (7.8\arcsec\,$\times$\,7.8\arcsec) 1.28~GHz MGCLS radio image with radio contours in black overlaid (1$\sigma$ = 3 $\mu$Jy beam$^{-1}$). \textbf{Right:} 1.28~GHz MGCLS low-resolution  (15\arcsec\,$\times$\,15\arcsec) radio contours in green (1$\sigma$ = 4 $\mu$Jy beam$^{-1}$), overlaid on the r-band \textit{Digitized Sky Survey (DSS)} optical image. In both panels, the radio contours start at 3\,$\sigma$ and rise by a factor of 2. The physical scale at the cluster redshift is indicated on top right, and the red $\times$ indicates the NED cluster position. } 
   \label{fig:J0145.0}%
\end{figure*}

\begin{figure*}
   \centering
   \includegraphics[width=0.496\textwidth]{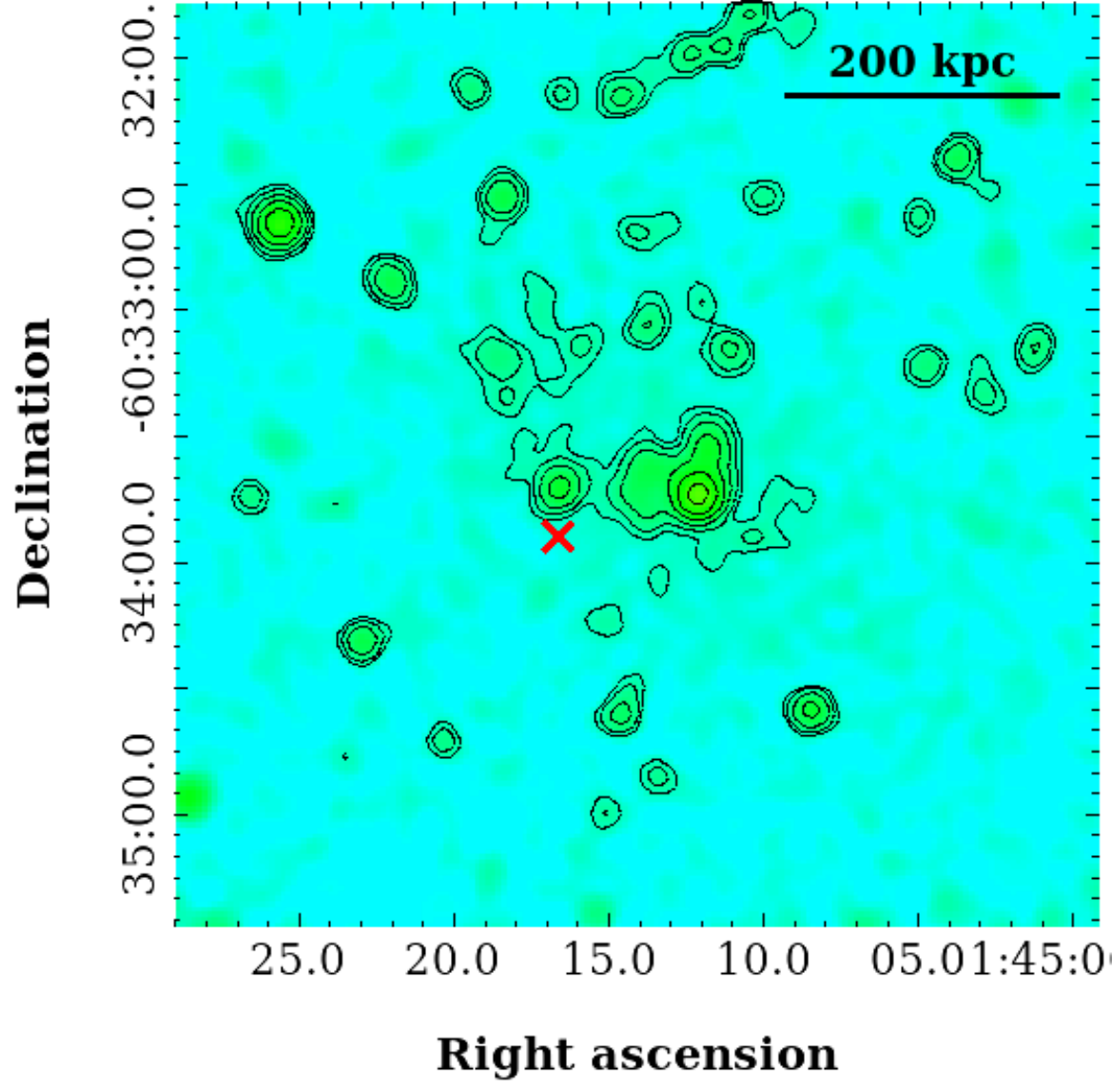}
    \includegraphics[width=0.5\textwidth]{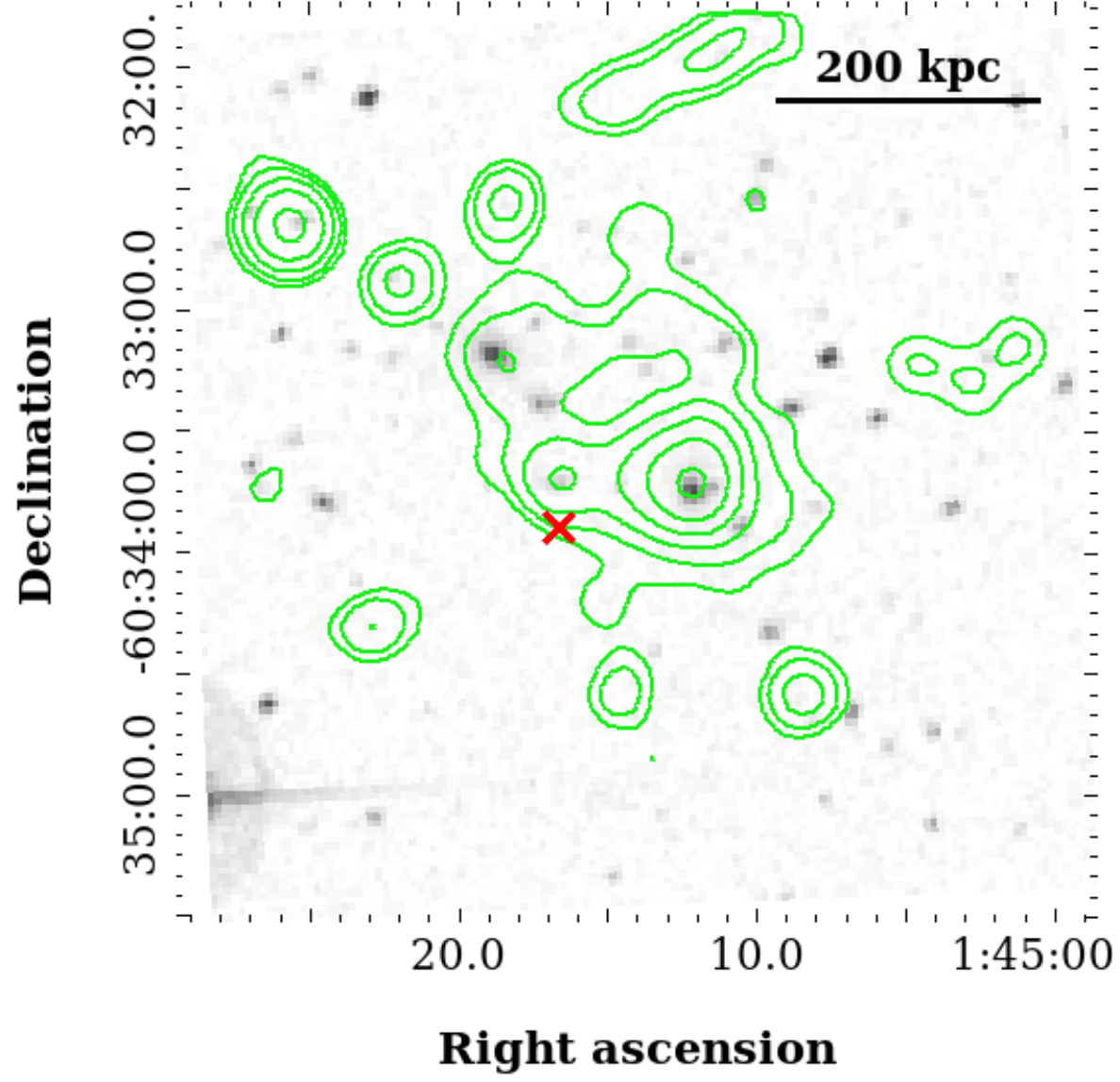}
   \caption{J0145.2$-$6033 \textbf{Left:}  Full-resolution (7.8\arcsec\,$\times$\,7.8\arcsec) 1.28~GHz MGCLS radio image with radio contours in black overlaid (1$\sigma$ = 2.5 $\mu$Jy beam$^{-1}$). \textbf{Right:} 1.28~GHz MGCLS low-resolution  (15\arcsec\,$\times$\,15\arcsec) radio contours in green (1$\sigma$ = 4 $\mu$Jy beam$^{-1}$), overlaid on the r-band \textit{Digitized Sky Survey (DSS)} optical image. In both panels, the radio contours start at 3\,$\sigma$ and rise by a factor of 2. The physical scale at the cluster redshift is indicated on top right, and the red $\times$ indicates the NED cluster position. } 
   \label{fig:J0145.2}%
\end{figure*}

\begin{figure*}
   \centering
   \includegraphics[width=0.496\textwidth]{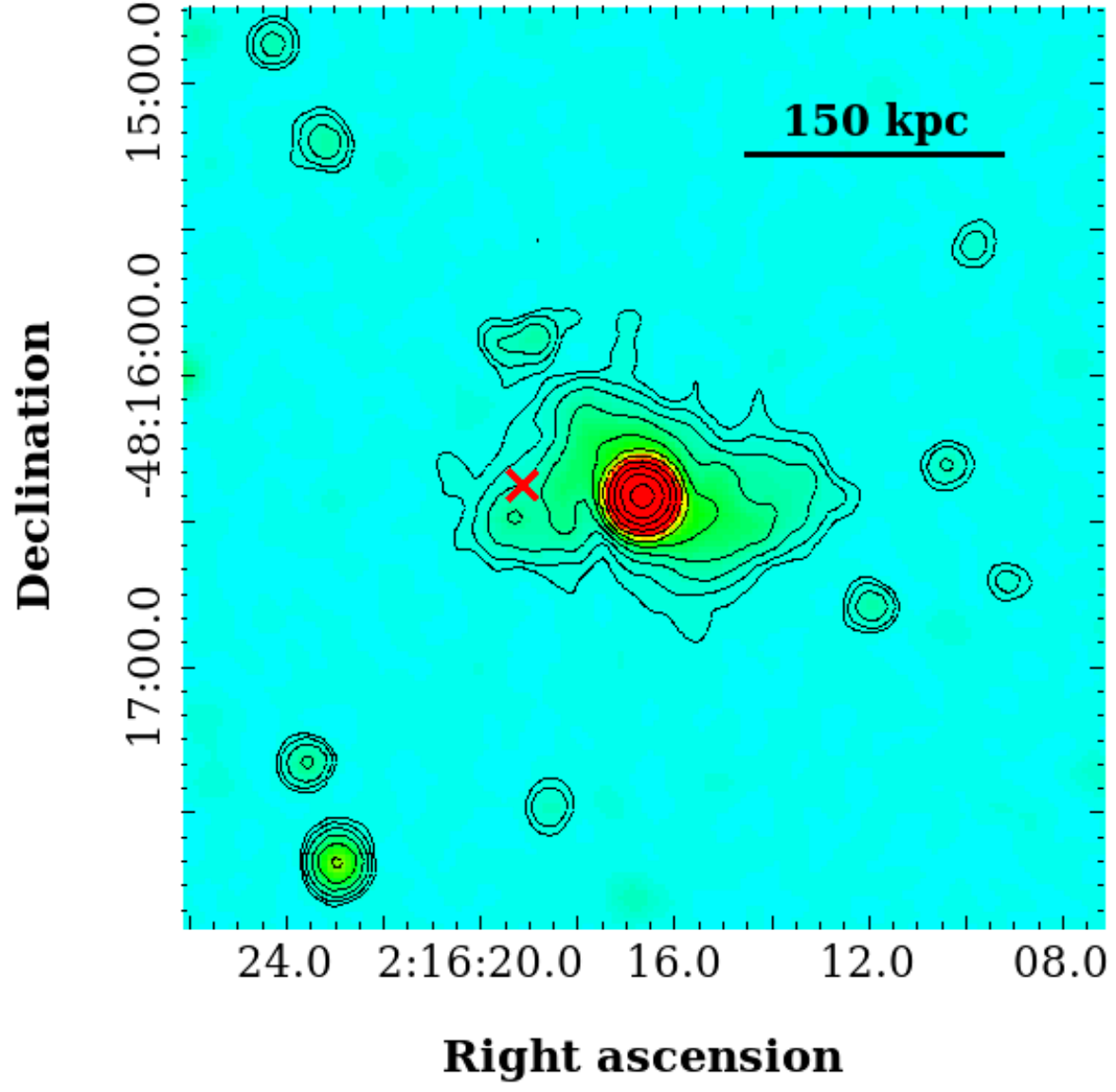}
    \includegraphics[width=0.5\textwidth]{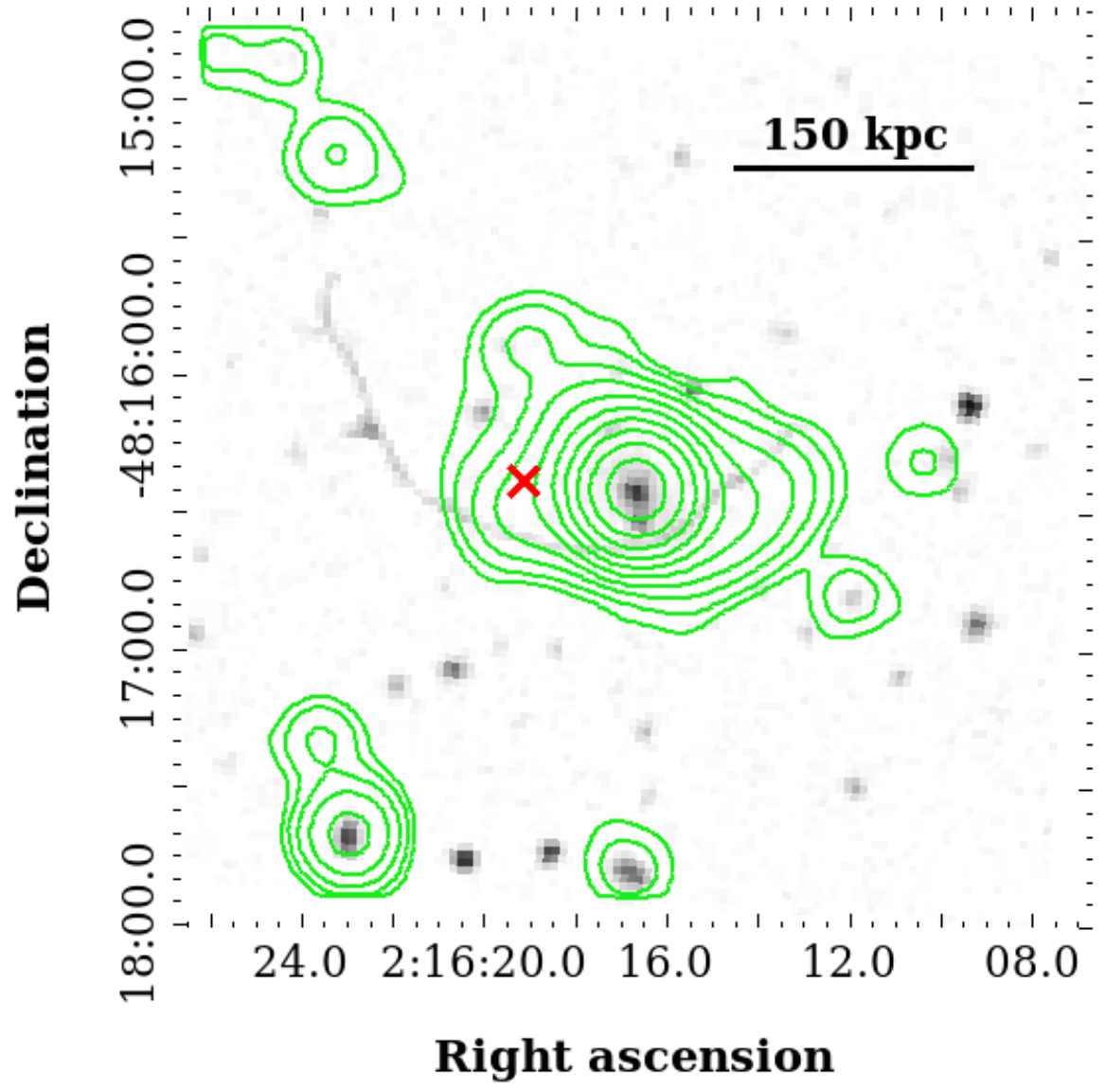}
   \caption{J0216.3$-$4816 \textbf{Left:}  Full-resolution (7.8\arcsec\,$\times$\,7.8\arcsec) 1.28~GHz MGCLS radio image with radio contours in black overlaid (1$\sigma$ = 3 $\mu$Jy beam$^{-1}$). \textbf{Right:} 1.28~GHz MGCLS low-resolution  (15\arcsec\,$\times$\,15\arcsec) radio contours in green (1$\sigma$ = 6 $\mu$Jy beam$^{-1}$), overlaid on the r-band \textit{Digitized Sky Survey (DSS)} optical image. In both panels, the radio contours start at 3\,$\sigma$ and rise by a factor of 2. The physical scale at the cluster redshift is indicated on top right, and the red $\times$ indicates the NED cluster position. } 
   \label{fig:J0216.3}%
\end{figure*}

\begin{figure*}
   \centering
   \includegraphics[width=0.496\textwidth]{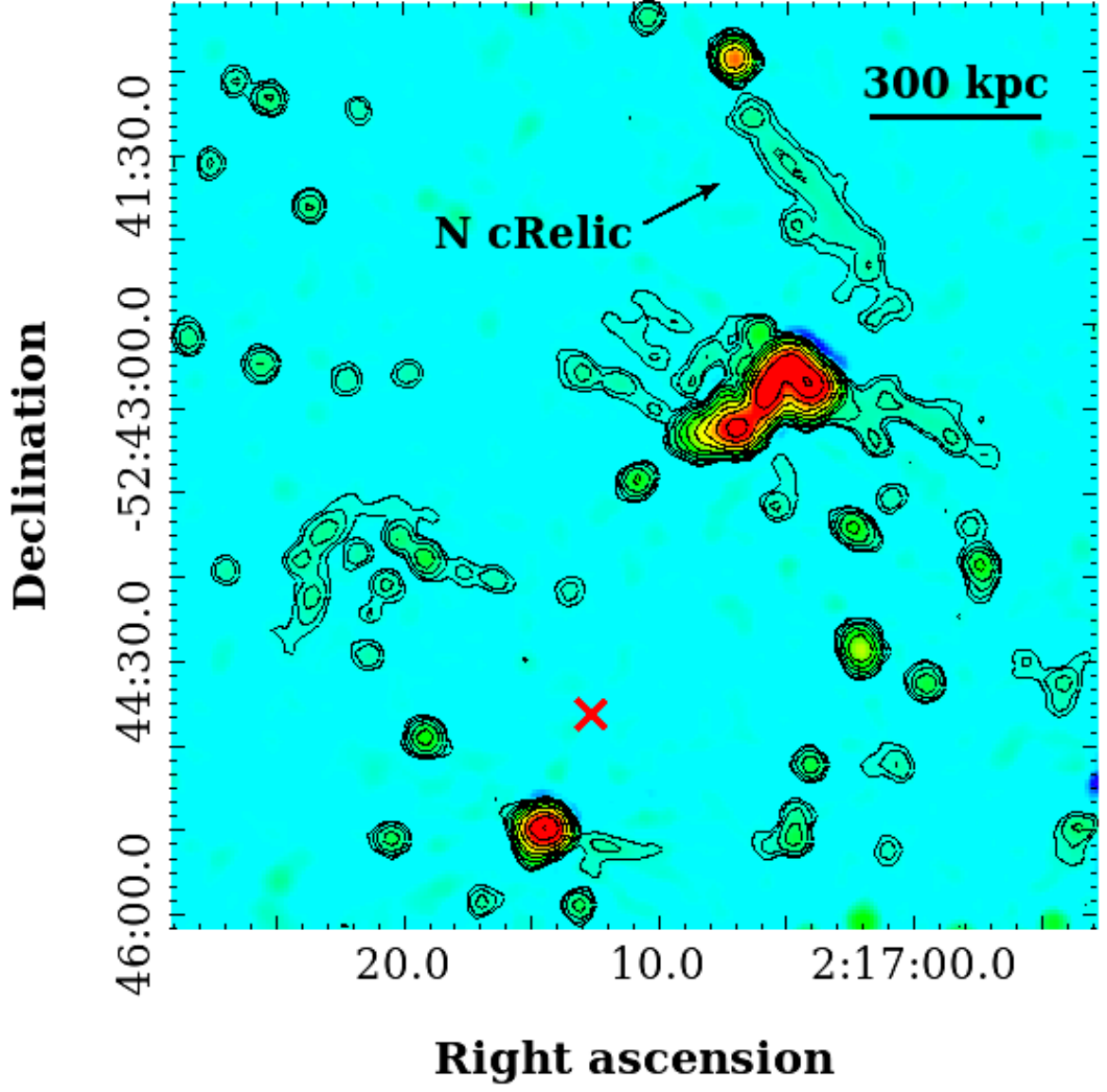}
    \includegraphics[width=0.5\textwidth]{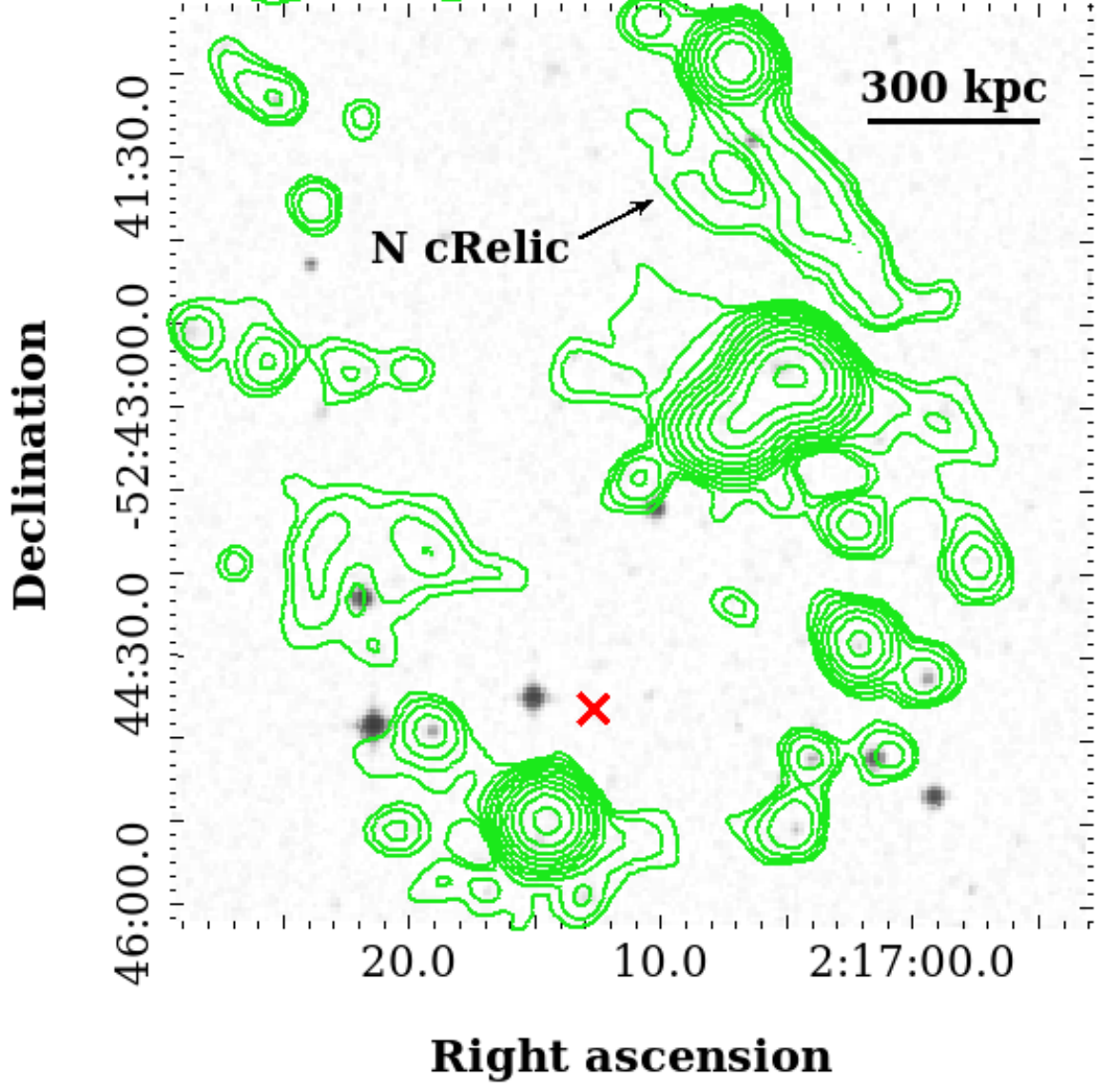}
   \caption{J0217.2$-$5244 \textbf{Left:}  Full-resolution (7.8\arcsec\,$\times$\,7.8\arcsec) 1.28~GHz MGCLS radio image with radio contours in black overlaid (1$\sigma$ = 3.5 $\mu$Jy beam$^{-1}$). \textbf{Right:} 1.28~GHz MGCLS low-resolution  (15\arcsec\,$\times$\,15\arcsec) radio contours in green (1$\sigma$ = 6 $\mu$Jy beam$^{-1}$), overlaid on the r-band \textit{Digitized Sky Survey (DSS)} optical image. In both panels, the radio contours start at 3\,$\sigma$ and rise by a factor of 2. The physical scale at the cluster redshift is indicated on top right, and the red $\times$ indicates the NED cluster position. } 
   \label{fig:J0217.2}%
\end{figure*}

\begin{figure*}
   \centering
   \includegraphics[width=0.496\textwidth]{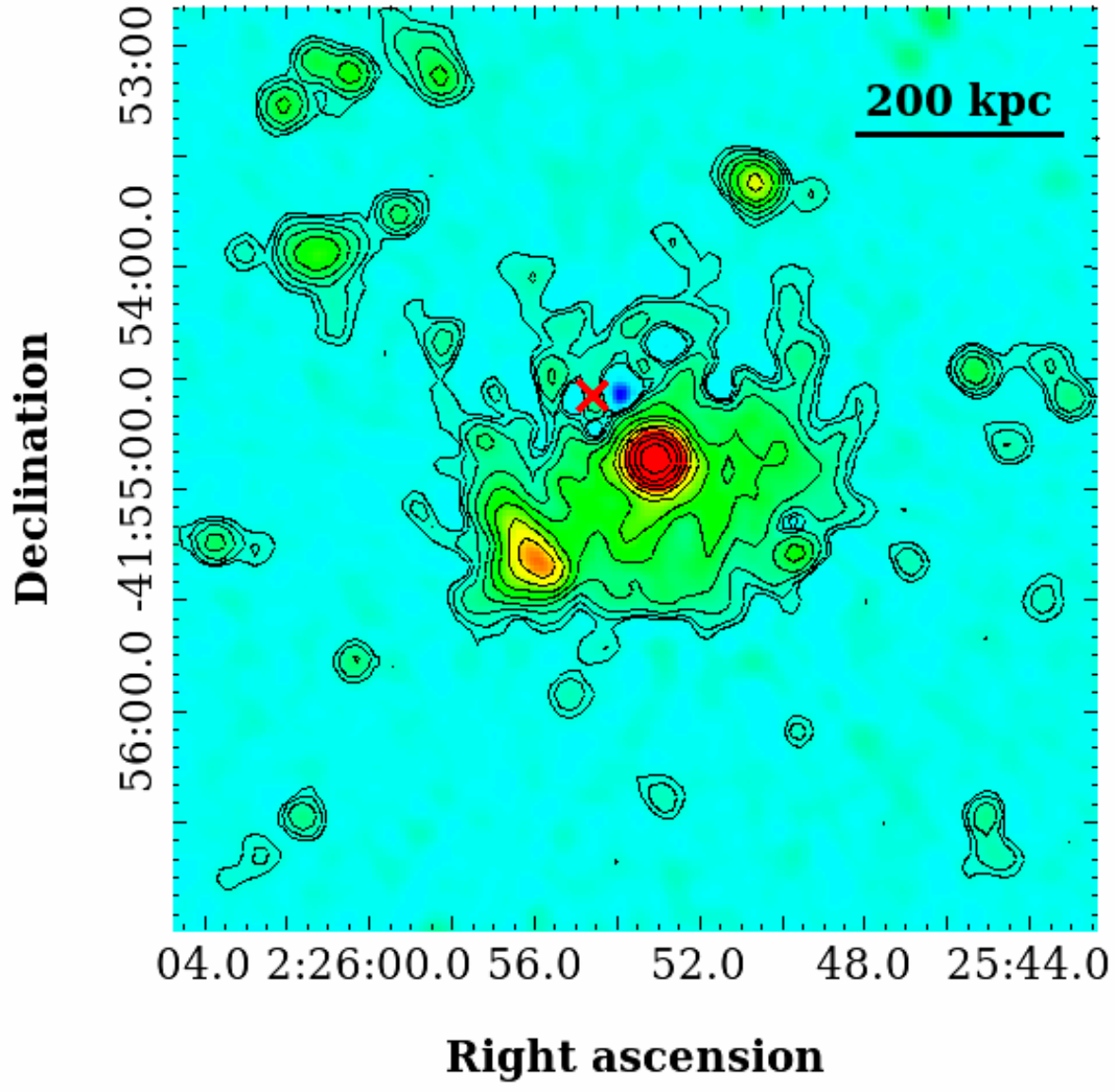}
    \includegraphics[width=0.5\textwidth]{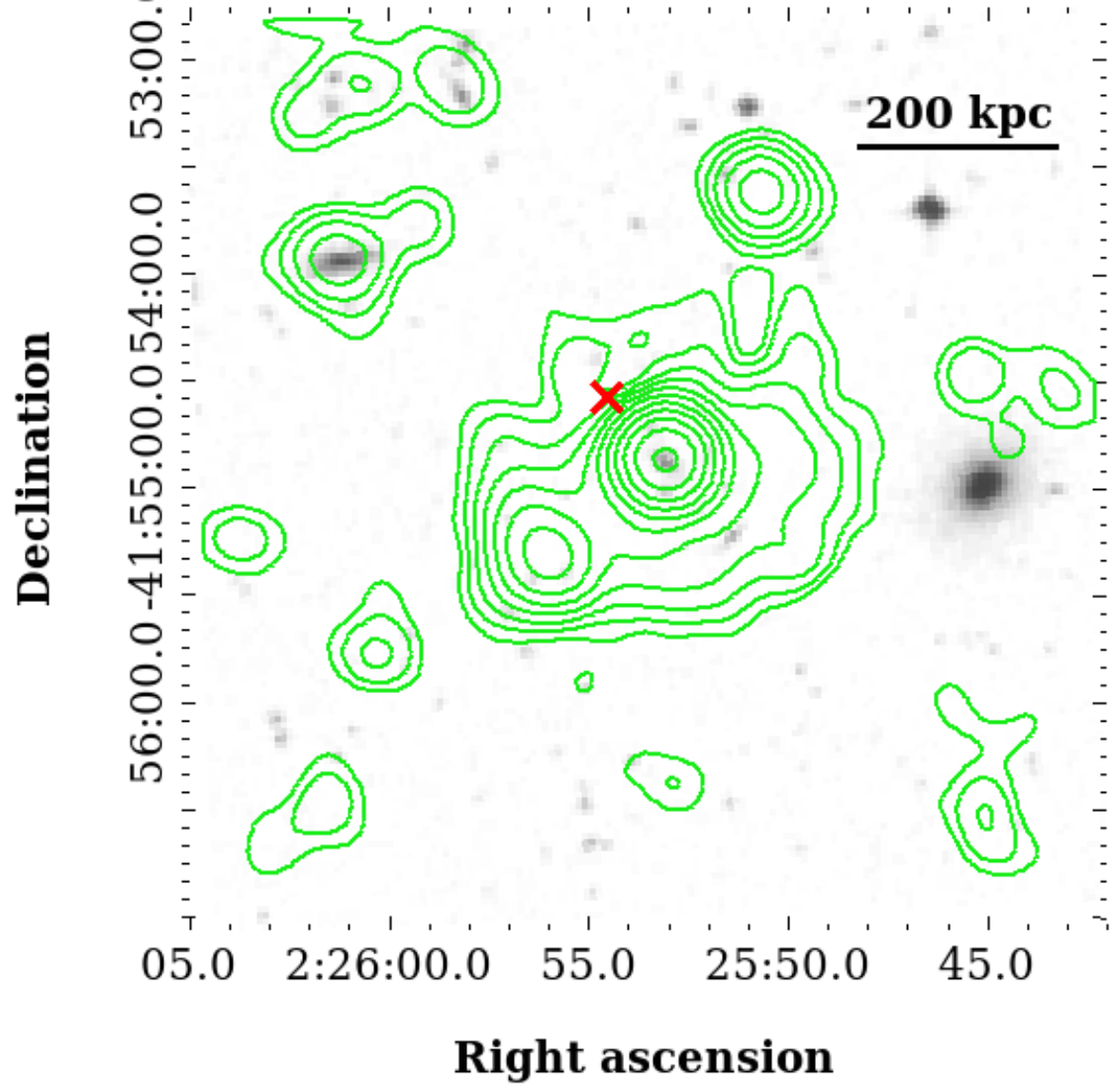}
   \caption{J0225.9$-$4154 \textbf{Left:}  Full-resolution (7.8\arcsec\,$\times$\,7.8\arcsec) 1.28~GHz MGCLS radio image with radio contours in black overlaid (1$\sigma$ = 3.5 $\mu$Jy beam$^{-1}$). \textbf{Right:} 1.28~GHz MGCLS low-resolution  (15\arcsec\,$\times$\,15\arcsec) radio contours in green (1$\sigma$ = 7 $\mu$Jy beam$^{-1}$), overlaid on the r-band \textit{Digitized Sky Survey (DSS)} optical image. In both panels, the radio contours start at 3\,$\sigma$ and rise by a factor of 2. The physical scale at the cluster redshift is indicated on top right, and the red $\times$ indicates the NED cluster position. } 
   \label{fig:J0225.9}%
\end{figure*}

\begin{figure*}
   \centering
   \includegraphics[width=0.496\textwidth]{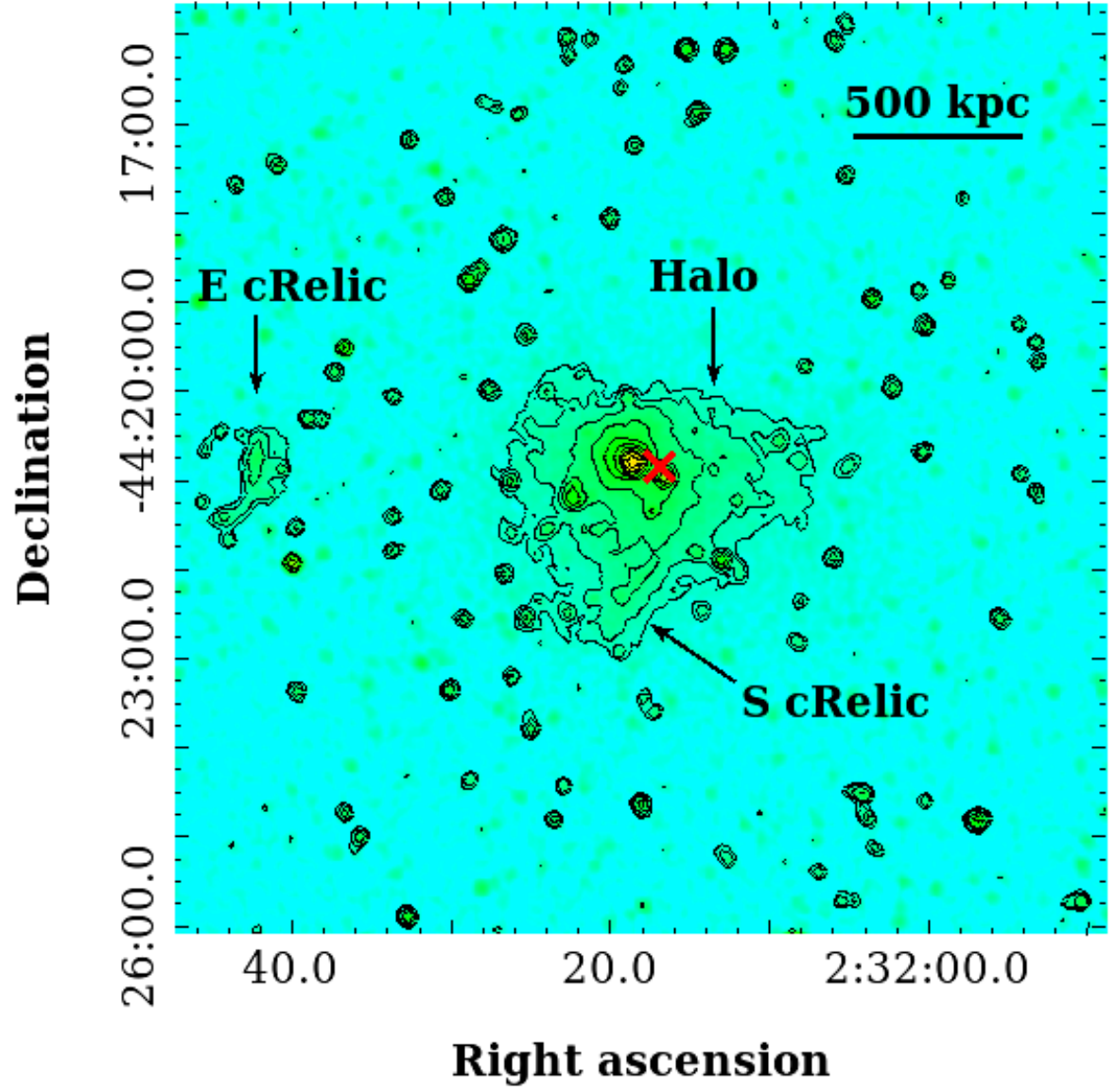}
    \includegraphics[width=0.5\textwidth]{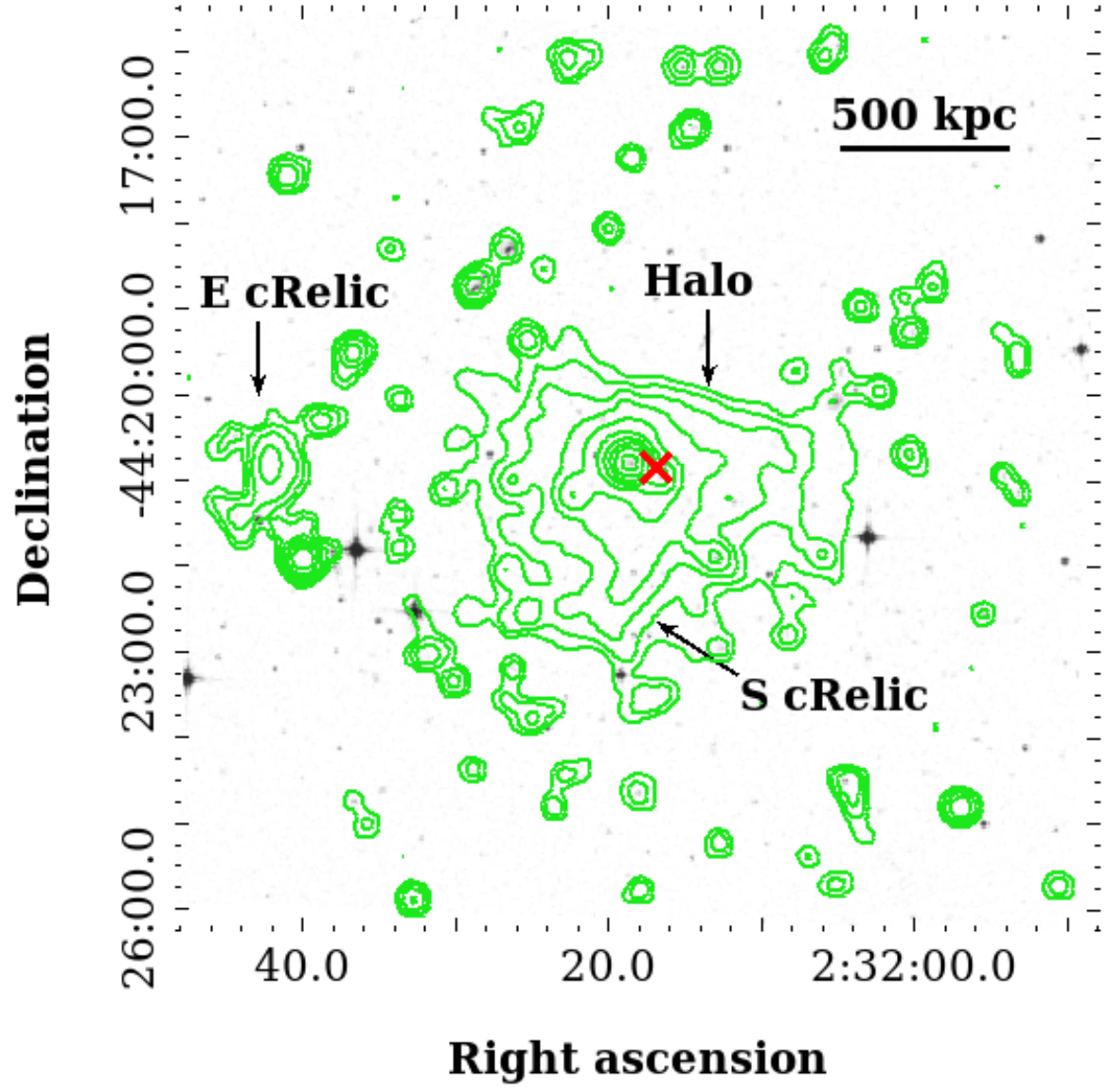}
   \caption{J0232.2$-$4420 \textbf{Left:}  Full-resolution (7.8\arcsec\,$\times$\,7.8\arcsec) 1.28~GHz MGCLS radio image with radio contours in black overlaid (1$\sigma$ = 3.5 $\mu$Jy beam$^{-1}$). \textbf{Right:} 1.28~GHz MGCLS low-resolution  (15\arcsec\,$\times$\,15\arcsec) radio contours in green (1$\sigma$ = 6 $\mu$Jy beam$^{-1}$), overlaid on the r-band \textit{Digitized Sky Survey (DSS)} optical image. In both panels, the radio contours start at 3\,$\sigma$ and rise by a factor of 2. The physical scale at the cluster redshift is indicated on top right, and the red $\times$ indicates the NED cluster position. } 
   \label{fig:J0232.2}%
\end{figure*}

\begin{figure*}
   \centering
   \includegraphics[width=0.496\textwidth]{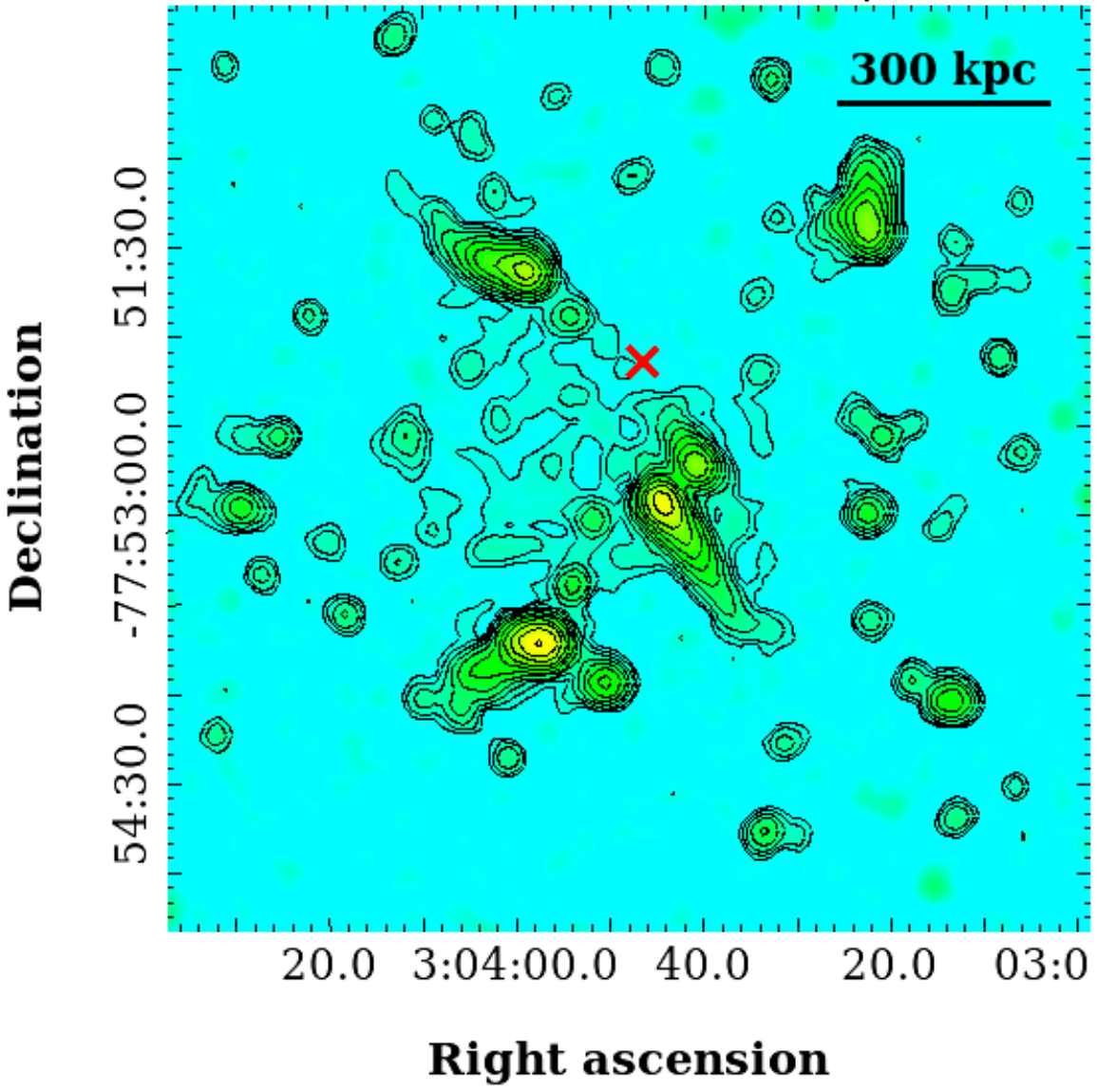}
    \includegraphics[width=0.5\textwidth]{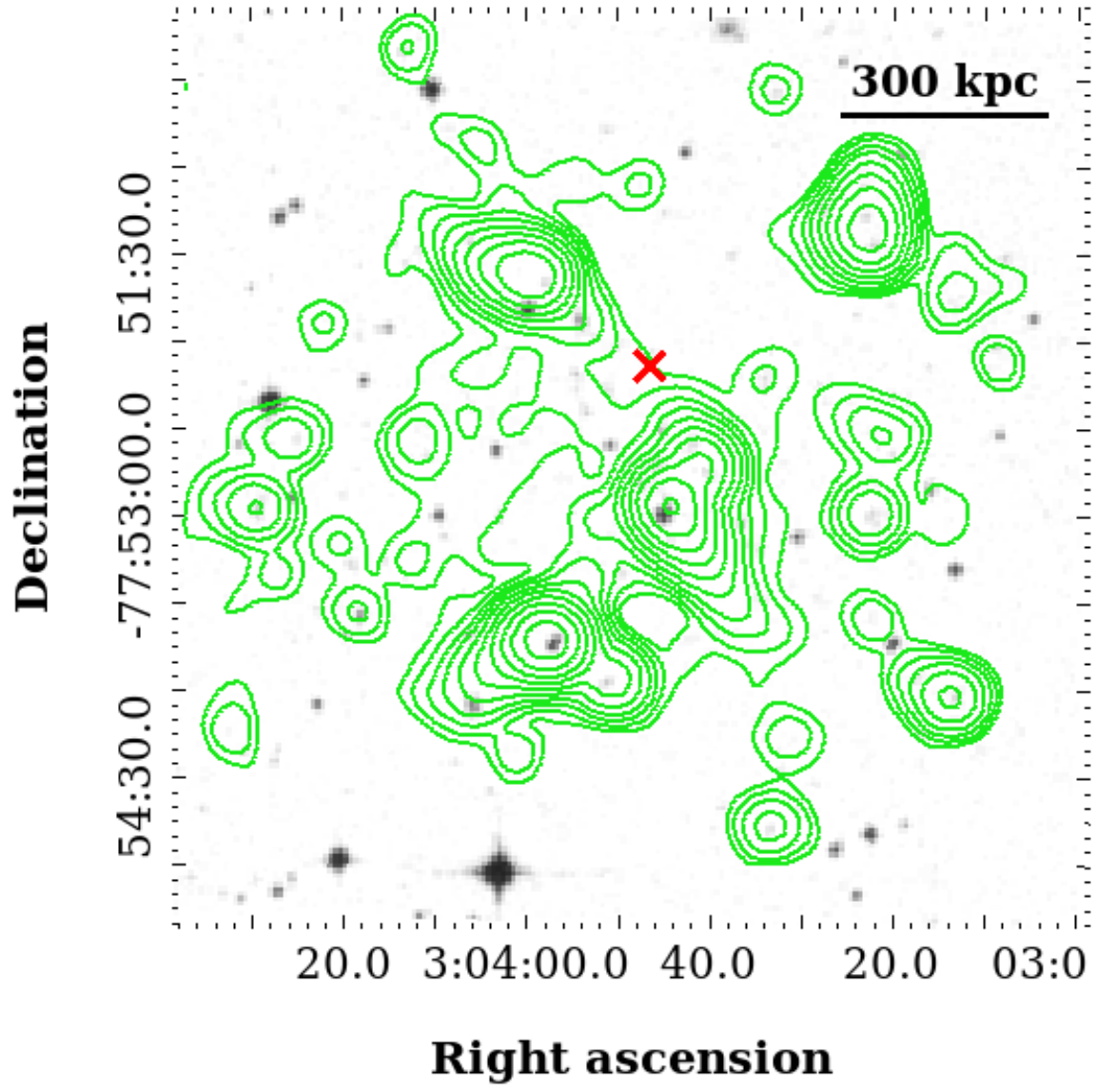}
   \caption{J0303.7$-$7752 \textbf{Left:}  Full-resolution (7.8\arcsec\,$\times$\,7.8\arcsec) 1.28~GHz MGCLS radio image with radio contours in black overlaid (1$\sigma$ = 3 $\mu$Jy beam$^{-1}$). \textbf{Right:} 1.28~GHz MGCLS low-resolution  (15\arcsec\,$\times$\,15\arcsec) radio contours in green (1$\sigma$ = 6 $\mu$Jy beam$^{-1}$), overlaid on the r-band \textit{Digitized Sky Survey (DSS)} optical image. In both panels, the radio contours start at 3\,$\sigma$ and rise by a factor of 2. The physical scale at the cluster redshift is indicated on top right, and the red $\times$ indicates the NED cluster position. } 
   \label{fig:J0303.7}%
\end{figure*}

\begin{figure*}
   \centering
   \includegraphics[width=0.496\textwidth]{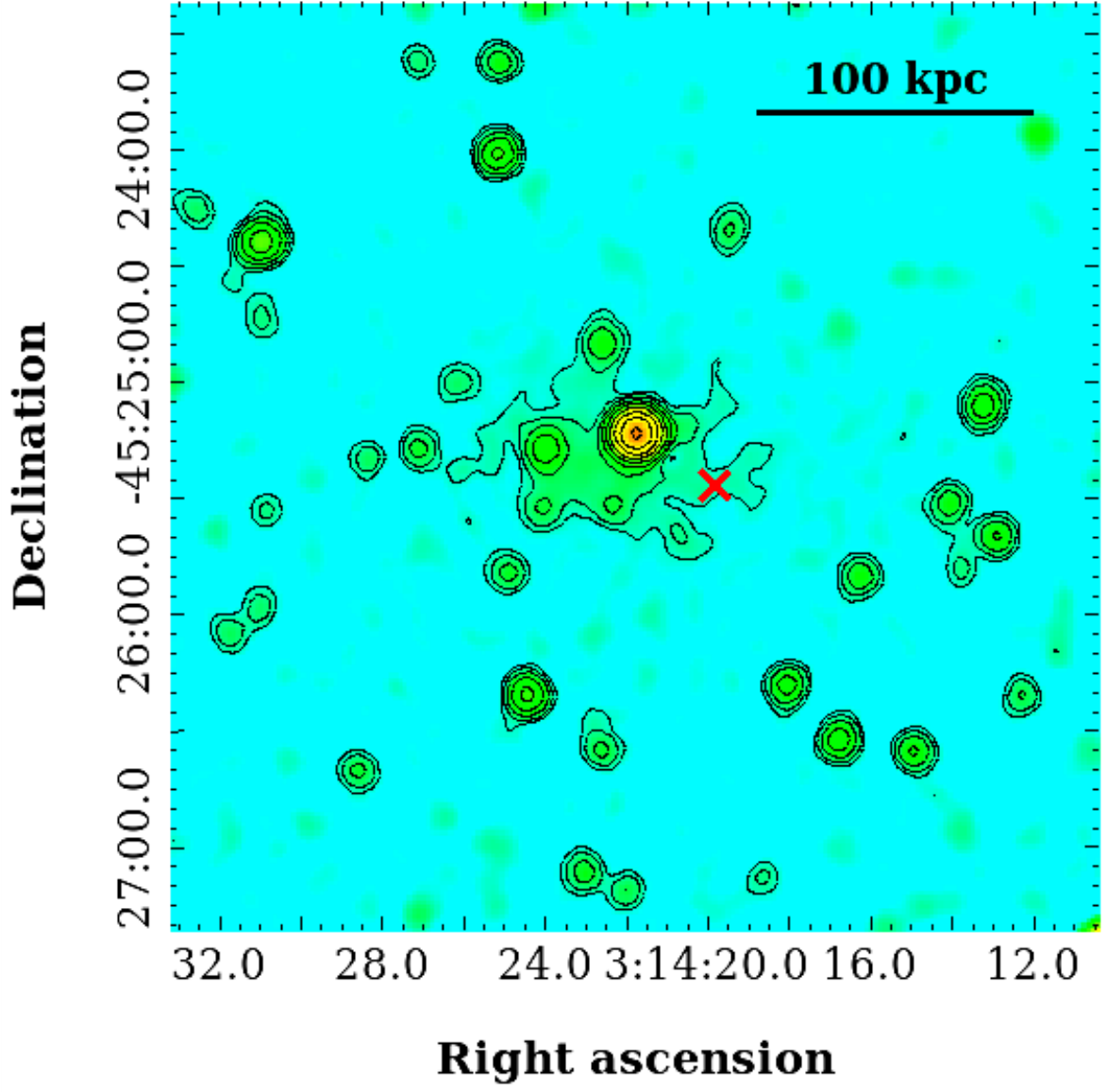}
    \includegraphics[width=0.5\textwidth]{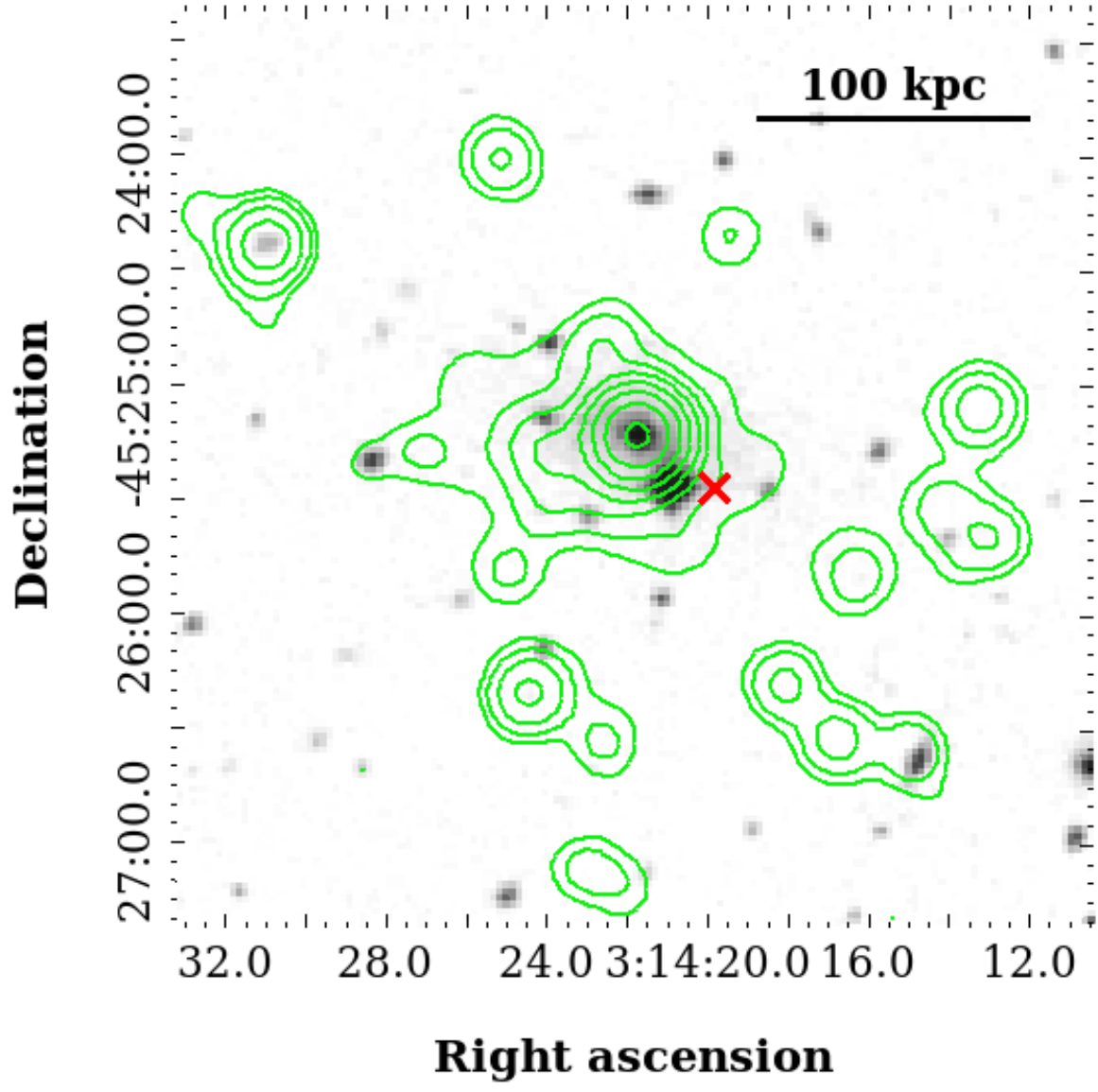}
   \caption{J0314.3$-$4525 \textbf{Left:}  Full-resolution (7.8\arcsec\,$\times$\,7.8\arcsec) 1.28~GHz MGCLS radio image with radio contours in black overlaid (1$\sigma$ = 3 $\mu$Jy beam$^{-1}$). \textbf{Right:} 1.28~GHz MGCLS low-resolution  (15\arcsec\,$\times$\,15\arcsec) radio contours in green (1$\sigma$ = 6 $\mu$Jy beam$^{-1}$), overlaid on the r-band \textit{Digitized Sky Survey (DSS)} optical image. In both panels, the radio contours start at 3\,$\sigma$ and rise by a factor of 2. The physical scale at the cluster redshift is indicated on top right, and the red $\times$ indicates the NED cluster position. } 
   \label{fig:J0314.3}%
\end{figure*}

\begin{figure*}
   \centering
   \includegraphics[width=0.496\textwidth]{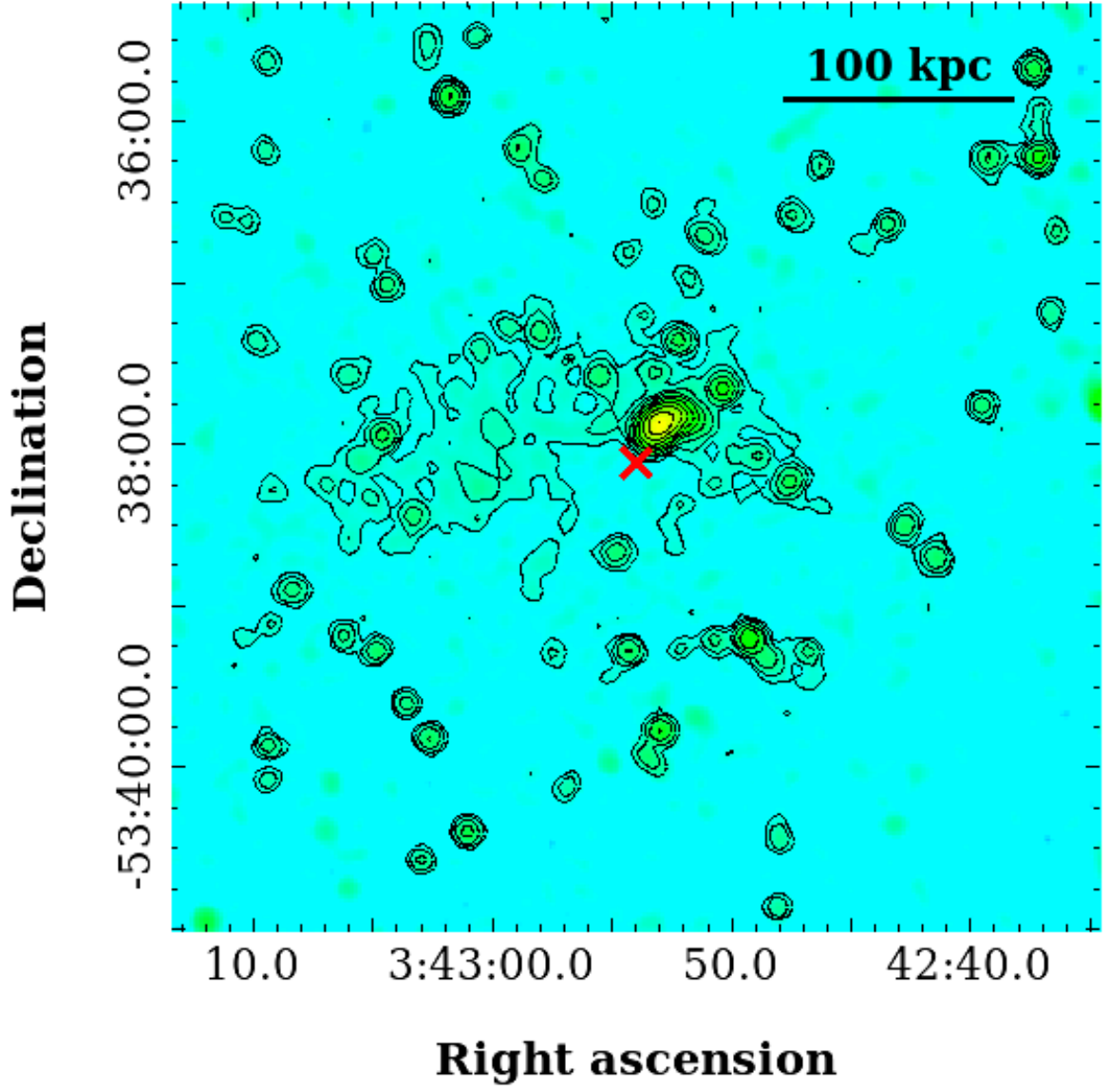}
    \includegraphics[width=0.5\textwidth]{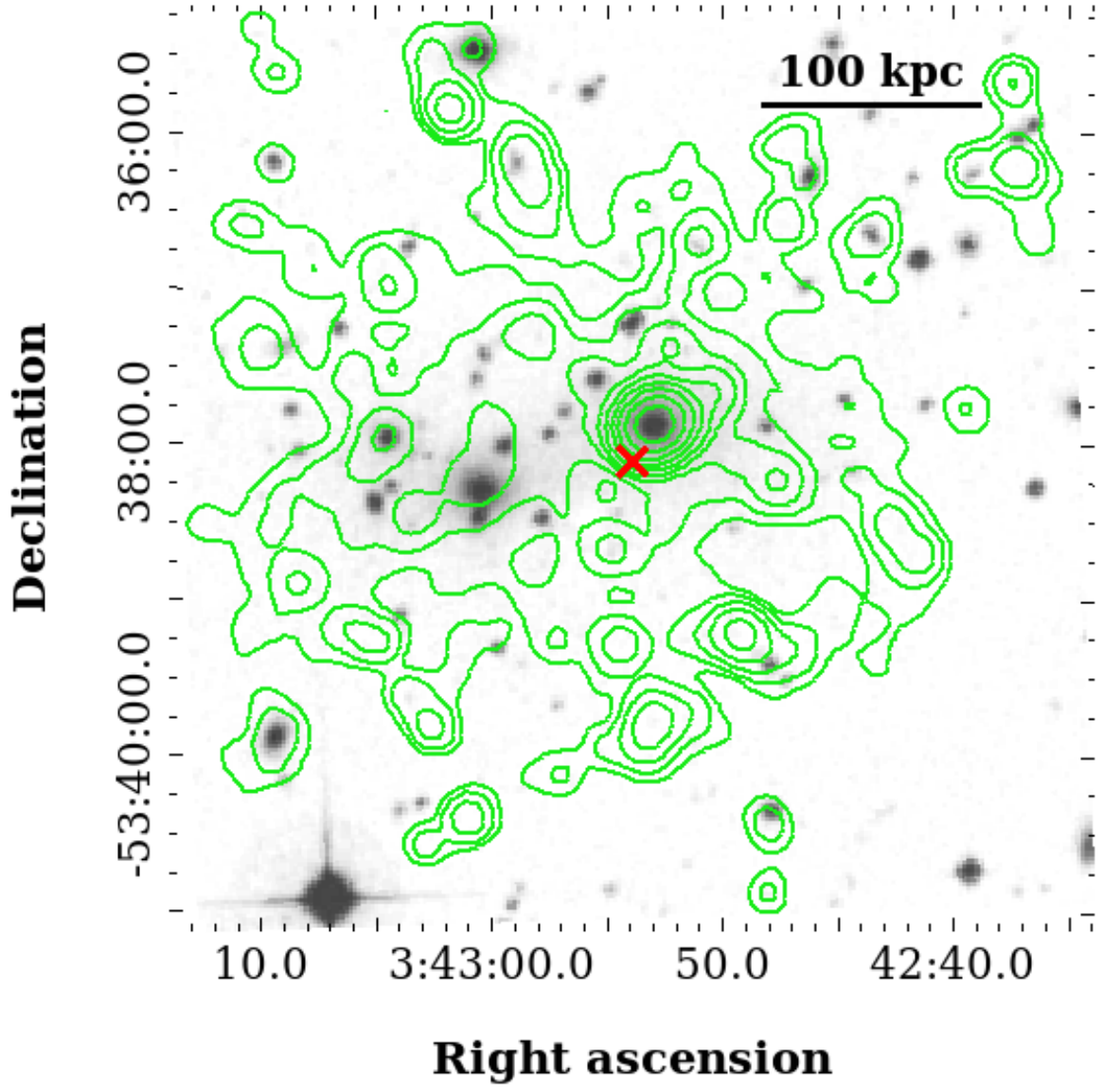}
   \caption{J0342.8$-$5338 \textbf{Left:}  Full-resolution (7.8\arcsec\,$\times$\,7.8\arcsec) 1.28~GHz MGCLS radio image with radio contours in black overlaid (1$\sigma$ = 3.5 $\mu$Jy beam$^{-1}$). \textbf{Right:} 1.28~GHz MGCLS low-resolution  (15\arcsec\,$\times$\,15\arcsec) radio contours in green (1$\sigma$ = 6 $\mu$Jy beam$^{-1}$), overlaid on the r-band \textit{Digitized Sky Survey (DSS)} optical image. In both panels, the radio contours start at 3\,$\sigma$ and rise by a factor of 2. The physical scale at the cluster redshift is indicated on the top right, and the red $\times$ indicates the NED cluster position. } 
   \label{fig:J0342.8}%
\end{figure*}

\begin{figure*}
   \centering
   \includegraphics[width=0.496\textwidth]{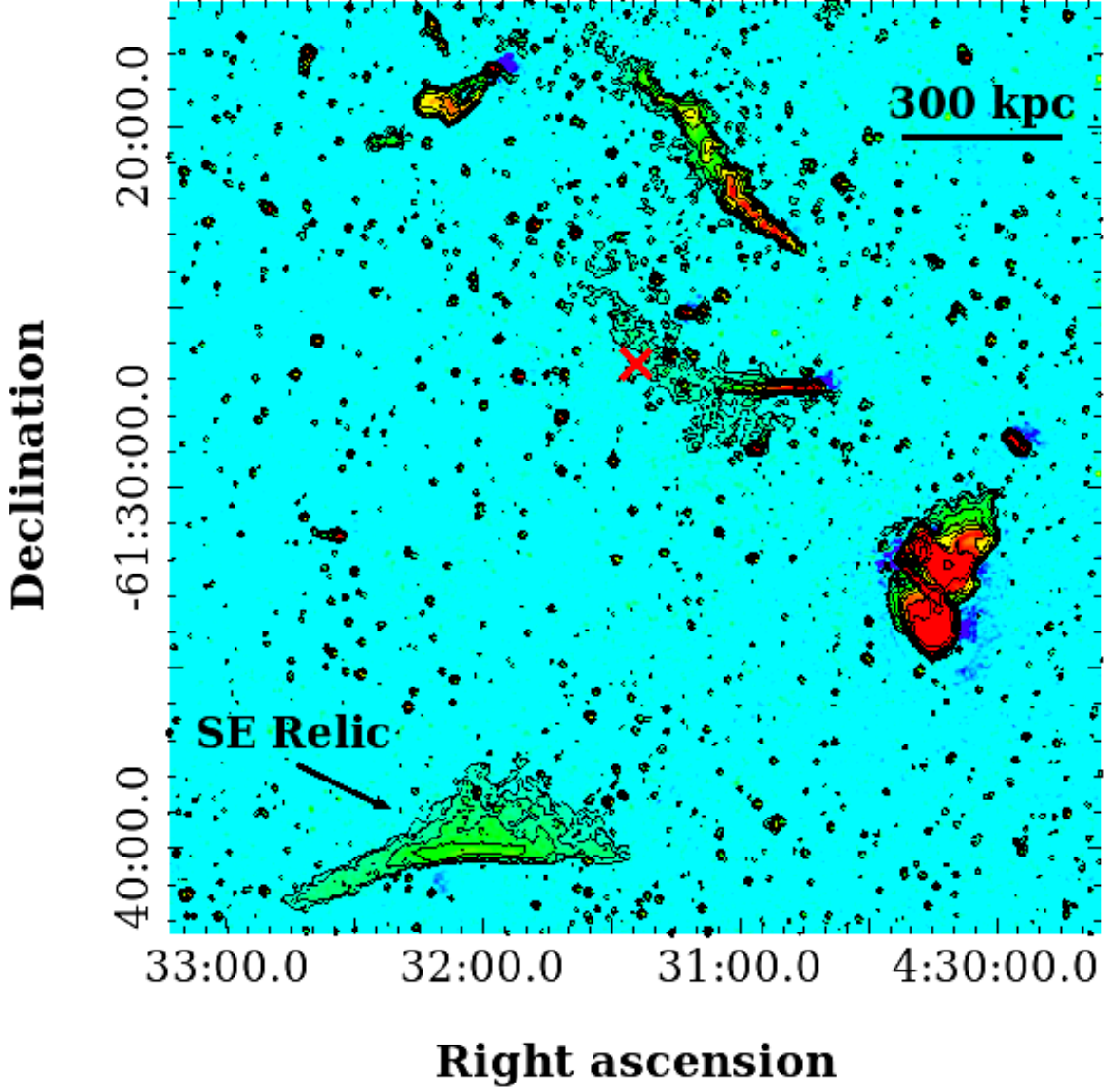}
    \includegraphics[width=0.5\textwidth]{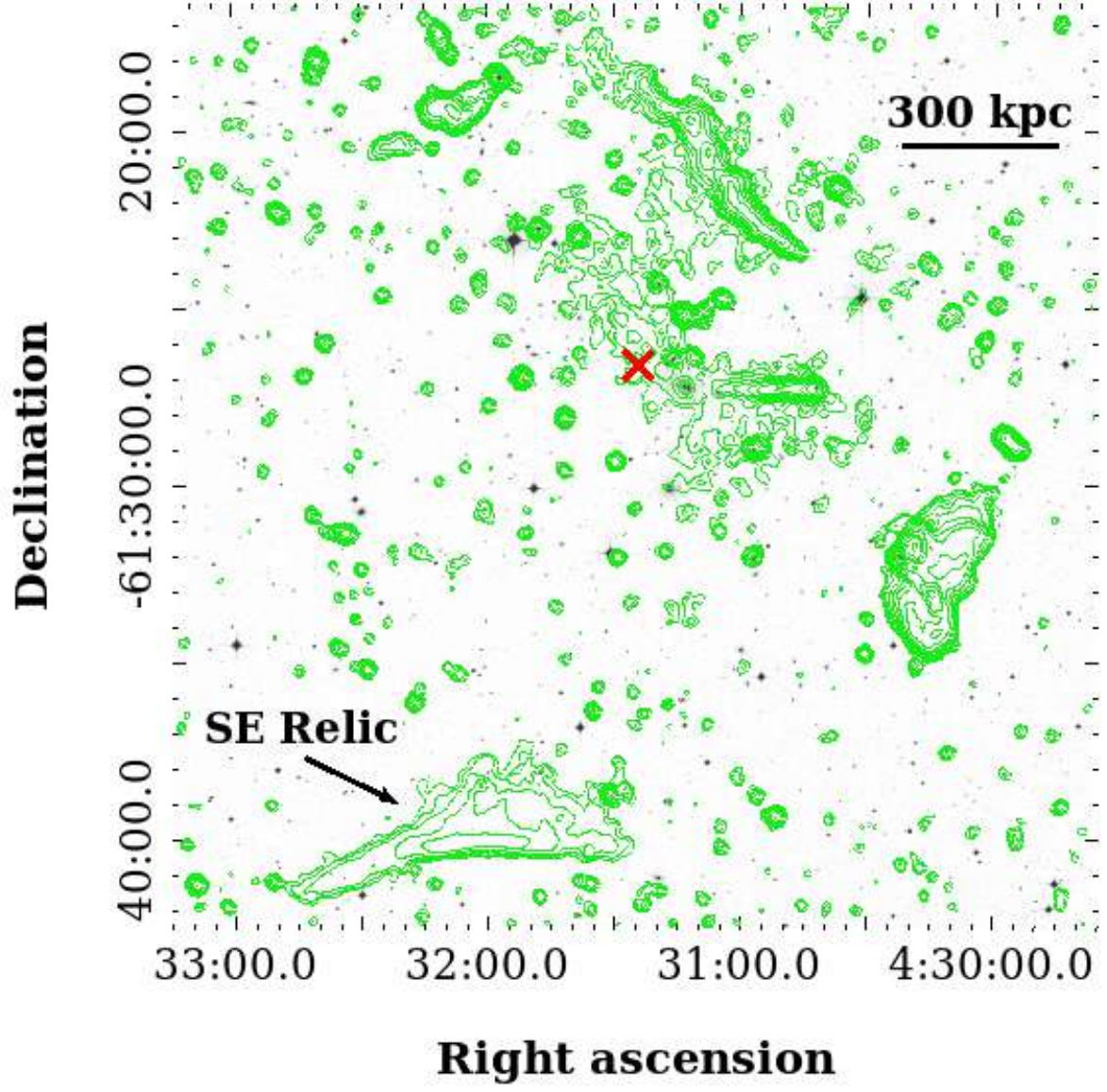}
   \caption{J0431.4$-$6126 \textbf{Left:}  Full-resolution (7.8\arcsec\,$\times$\,7.8\arcsec) 1.28~GHz MGCLS radio image with radio contours in black overlaid (1$\sigma$ = 5 $\mu$Jy beam$^{-1}$). \textbf{Right:} 1.28~GHz MGCLS low-resolution  (15\arcsec\,$\times$\,15\arcsec) radio contours in green (1$\sigma$ = 10 $\mu$Jy beam$^{-1}$), overlaid on the r-band \textit{Digitized Sky Survey (DSS)} optical image. In both panels, the radio contours start at 3\,$\sigma$ and rise by a factor of 2. The physical scale at the cluster redshift is indicated on top right, and the red $\times$ indicates the NED cluster position. } 
   \label{fig:J0431.4}%
\end{figure*}

\begin{figure*}
   \centering
   \includegraphics[width=0.496\textwidth]{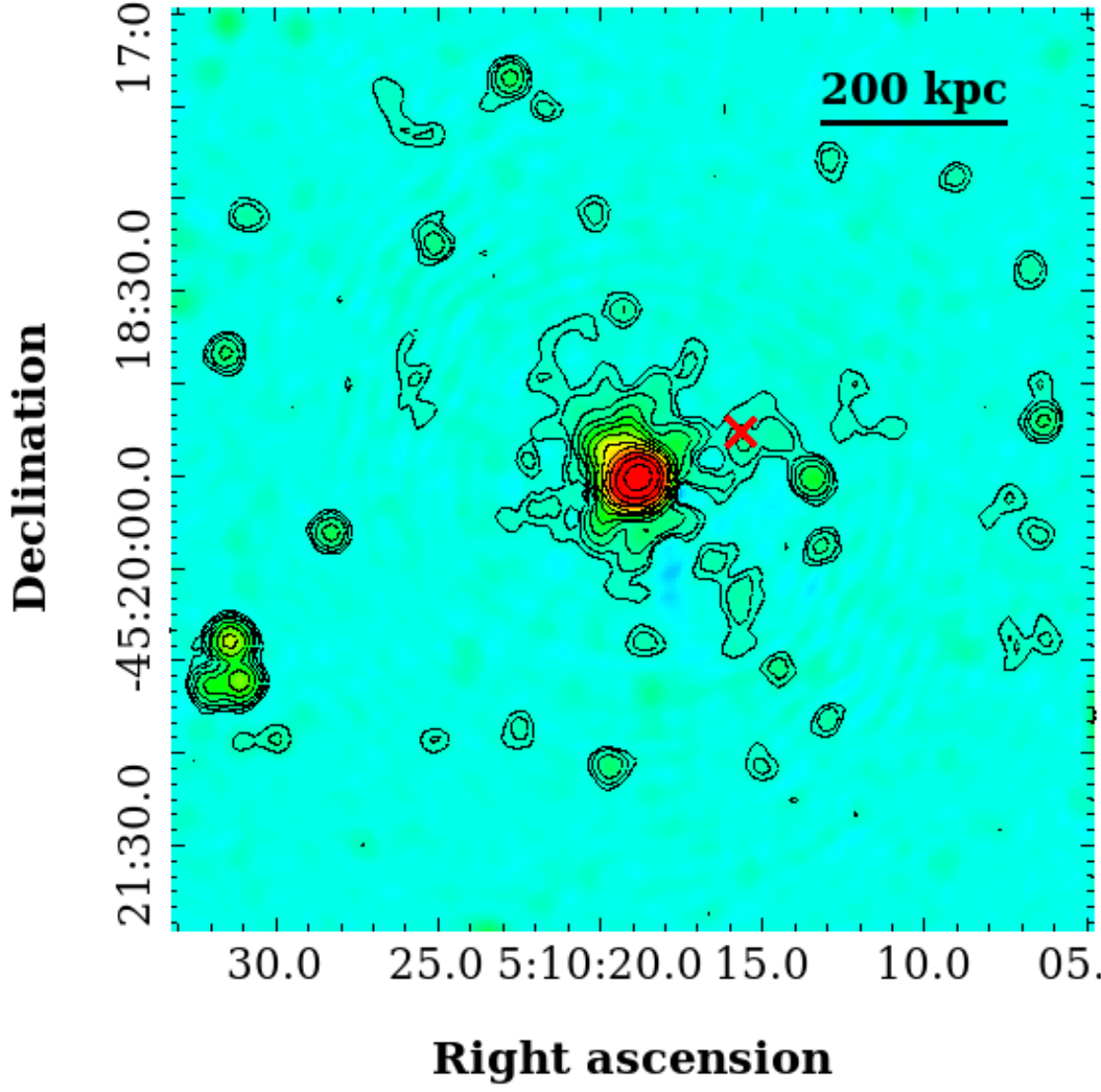}
    \includegraphics[width=0.5\textwidth]{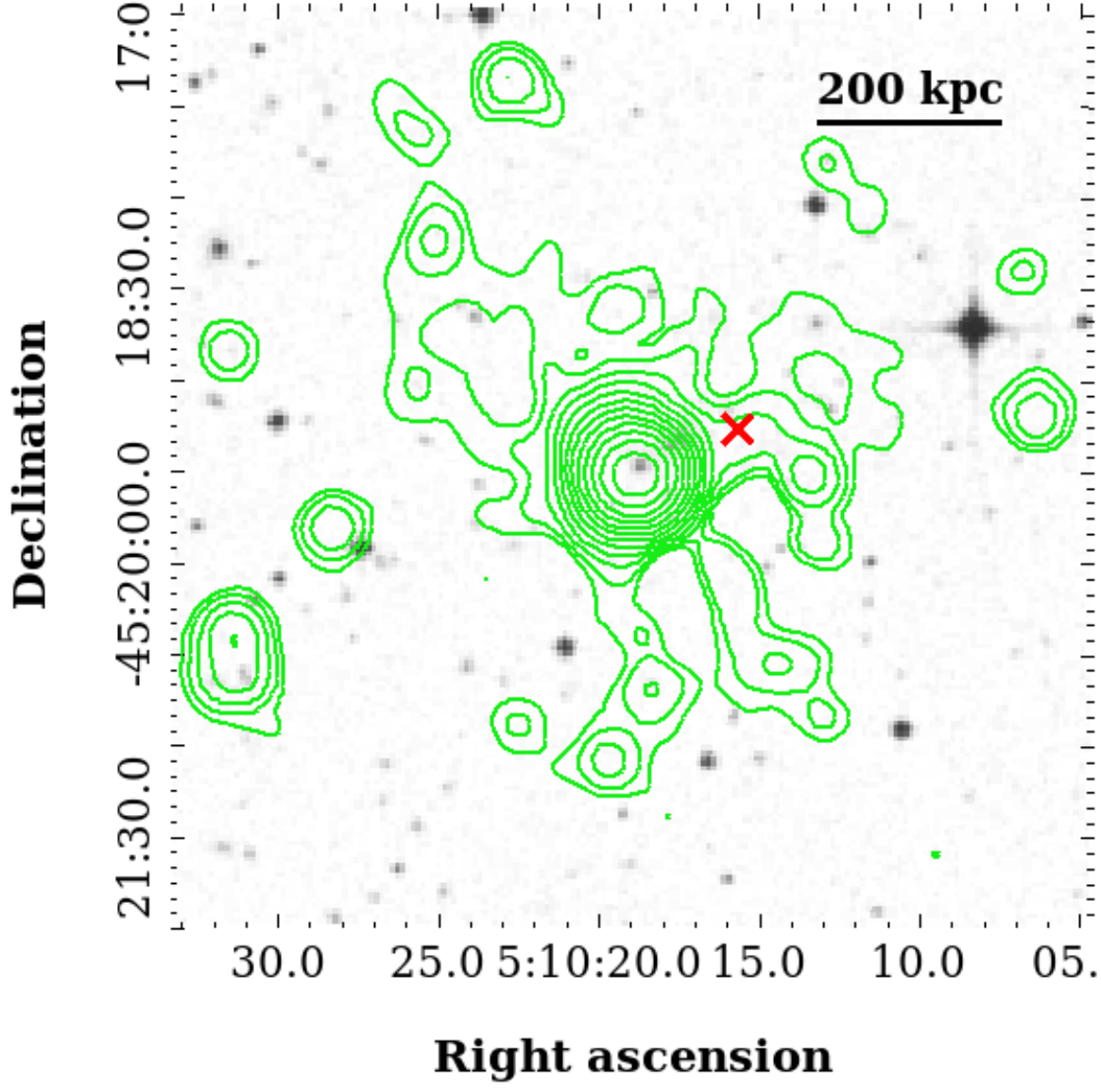}
   \caption{J0510.2$-$4519 \textbf{Left:}  Full-resolution (7.8\arcsec\,$\times$\,7.8\arcsec) 1.28~GHz MGCLS radio image with radio contours in black overlaid (1$\sigma$ = 4 $\mu$Jy beam$^{-1}$). \textbf{Right:} 1.28~GHz MGCLS low-resolution  (15\arcsec\,$\times$\,15\arcsec) radio contours in green (1$\sigma$ = 6 $\mu$Jy beam$^{-1}$), overlaid on the r-band \textit{Digitized Sky Survey (DSS)} optical image. In both panels, the radio contours start at 3\,$\sigma$ and rise by a factor of 2. The physical scale at the cluster redshift is indicated on top right, and the red $\times$ indicates the NED cluster position. } 
   \label{fig:J0510.2}%
\end{figure*}

\begin{figure*}
   \centering
   \includegraphics[width=0.496\textwidth]{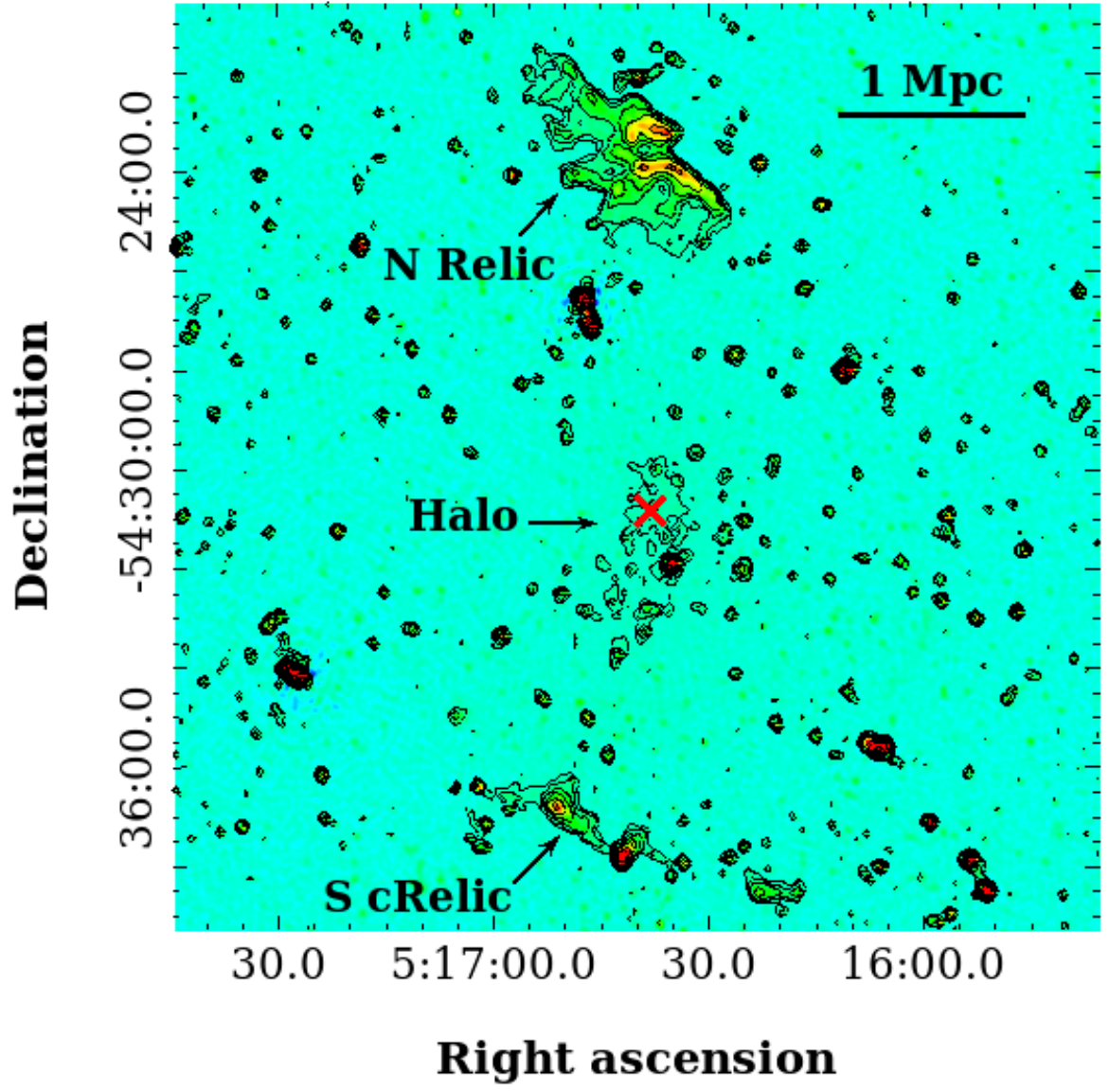}
    \includegraphics[width=0.5\textwidth]{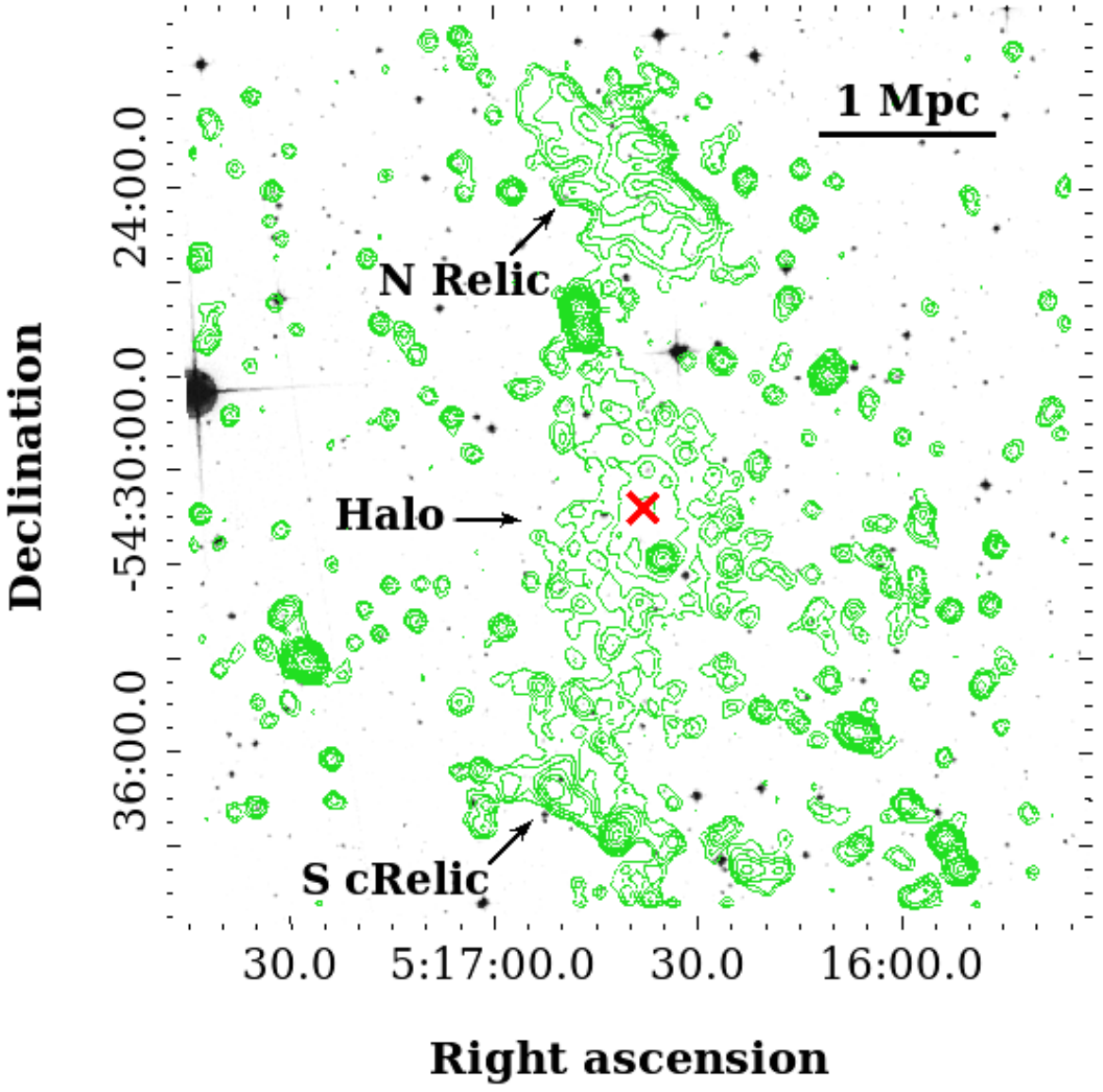}
   \caption{J0516.6$-$5430 \textbf{Left:}  Full-resolution (7.8\arcsec\,$\times$\,7.8\arcsec) 1.28~GHz MGCLS radio image with radio contours in black overlaid (1$\sigma$ = 4 $\mu$Jy beam$^{-1}$). \textbf{Right:} 1.28~GHz MGCLS low-resolution  (15\arcsec\,$\times$\,15\arcsec) radio contours in green (1$\sigma$ = 6 $\mu$Jy beam$^{-1}$), overlaid on the r-band \textit{Digitized Sky Survey (DSS)} optical image. In both panels, the radio contours start at 3\,$\sigma$ and rise by a factor of 2. The physical scale at the cluster redshift is indicated on the top right, and the red $\times$ indicates the NED cluster position. } 
   \label{fig:J0516.6}%
\end{figure*}

\begin{figure*}
   \centering
   \includegraphics[width=0.496\textwidth]{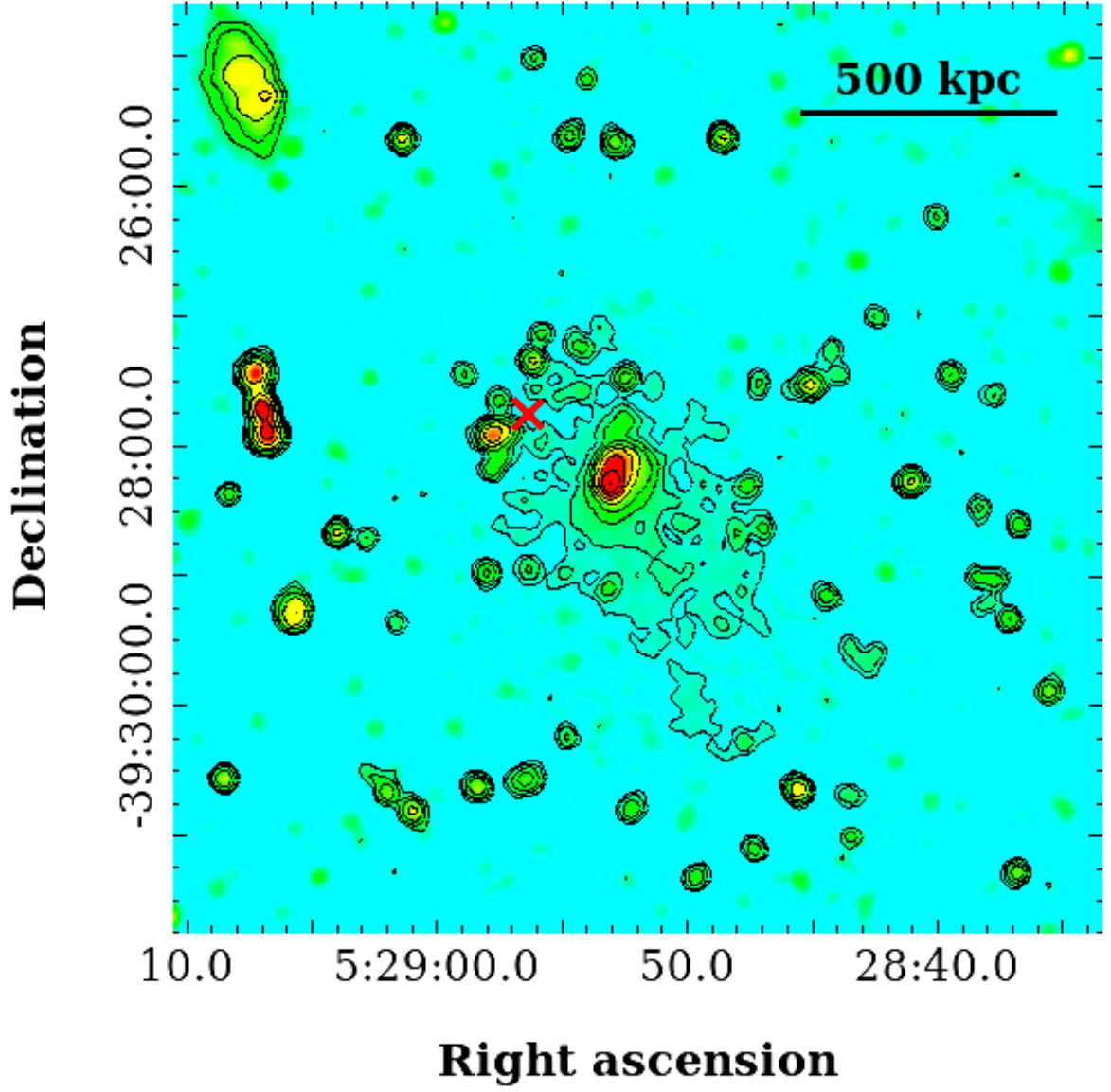}
    \includegraphics[width=0.5\textwidth]{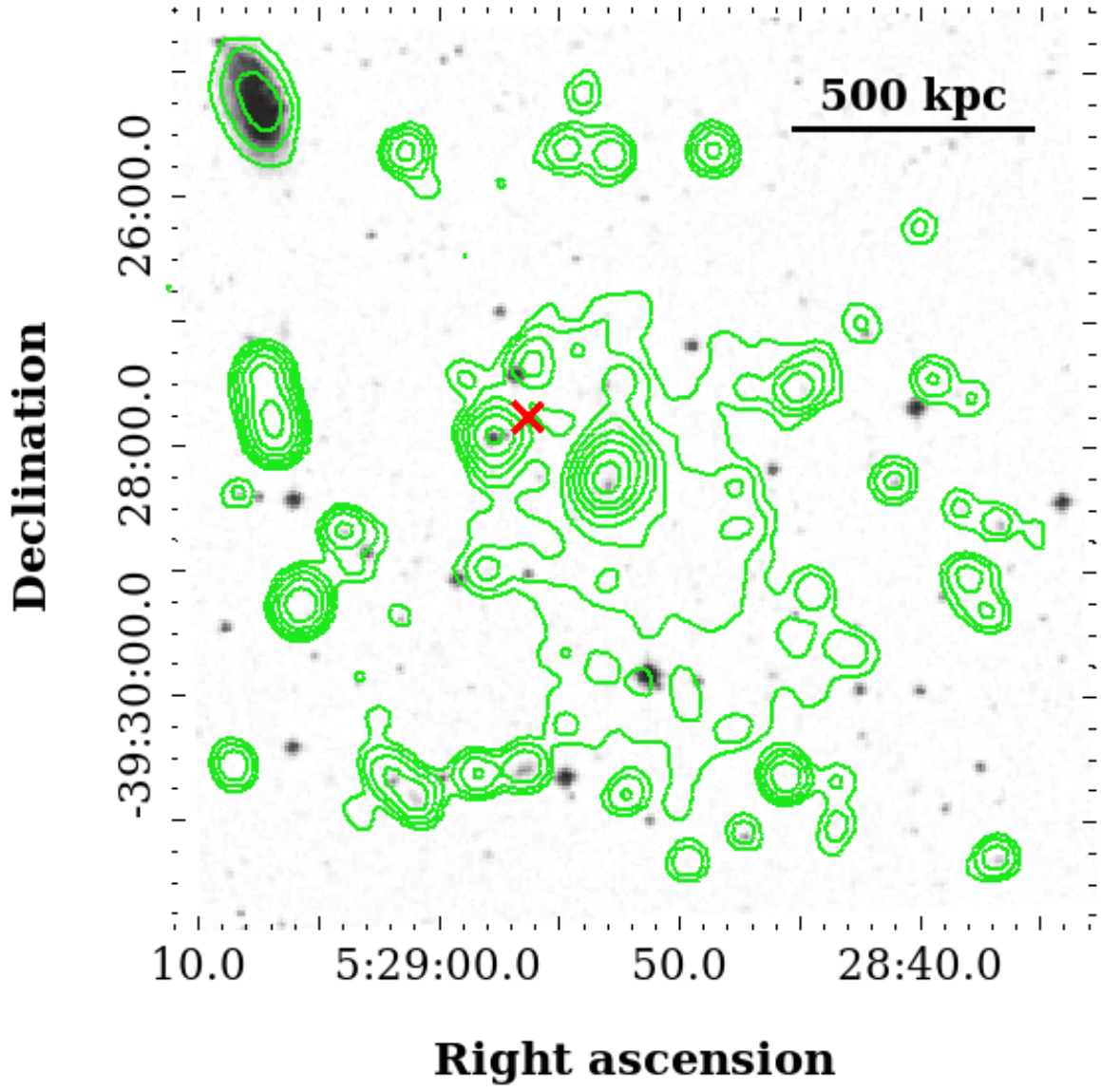}
   \caption{J0528.9$-$3927 \textbf{Left:}  Full-resolution (7.8\arcsec\,$\times$\,7.8\arcsec) 1.28~GHz MGCLS radio image with radio contours in black overlaid (1$\sigma$ = 3 $\mu$Jy beam$^{-1}$). \textbf{Right:} 1.28~GHz MGCLS low-resolution  (15\arcsec\,$\times$\,15\arcsec) radio contours in green (1$\sigma$ = 6 $\mu$Jy beam$^{-1}$), overlaid on the r-band \textit{Digitized Sky Survey (DSS)} optical image. In both panels, the radio contours start at 3\,$\sigma$ and rise by a factor of 2. The physical scale at the cluster redshift is indicated on the top right, and the red $\times$ indicates the NED cluster position. } 
   \label{fig:J0528.9}%
\end{figure*}

\begin{figure*}
   \centering
   \includegraphics[width=0.496\textwidth]{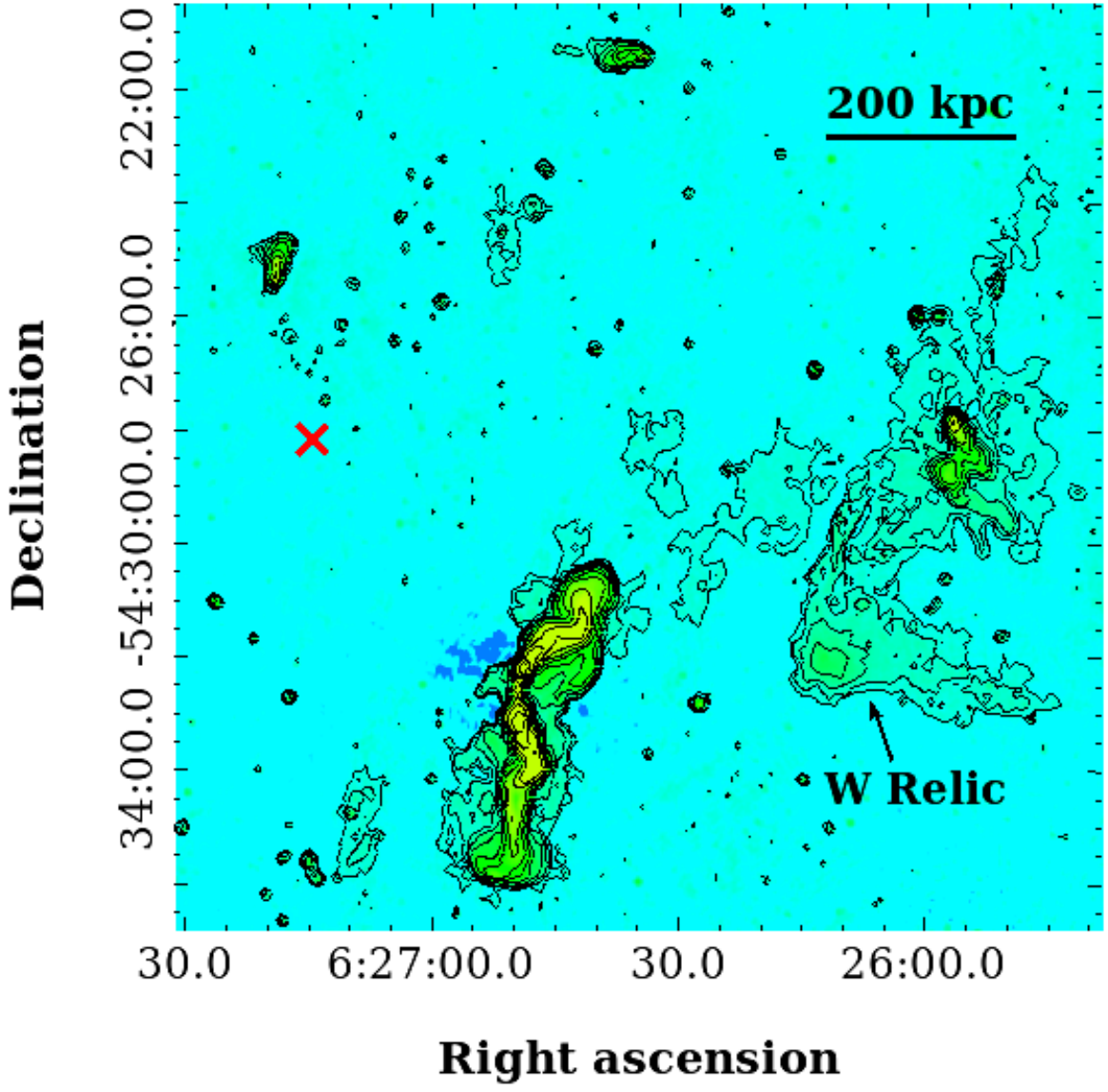}
    \includegraphics[width=0.5\textwidth]{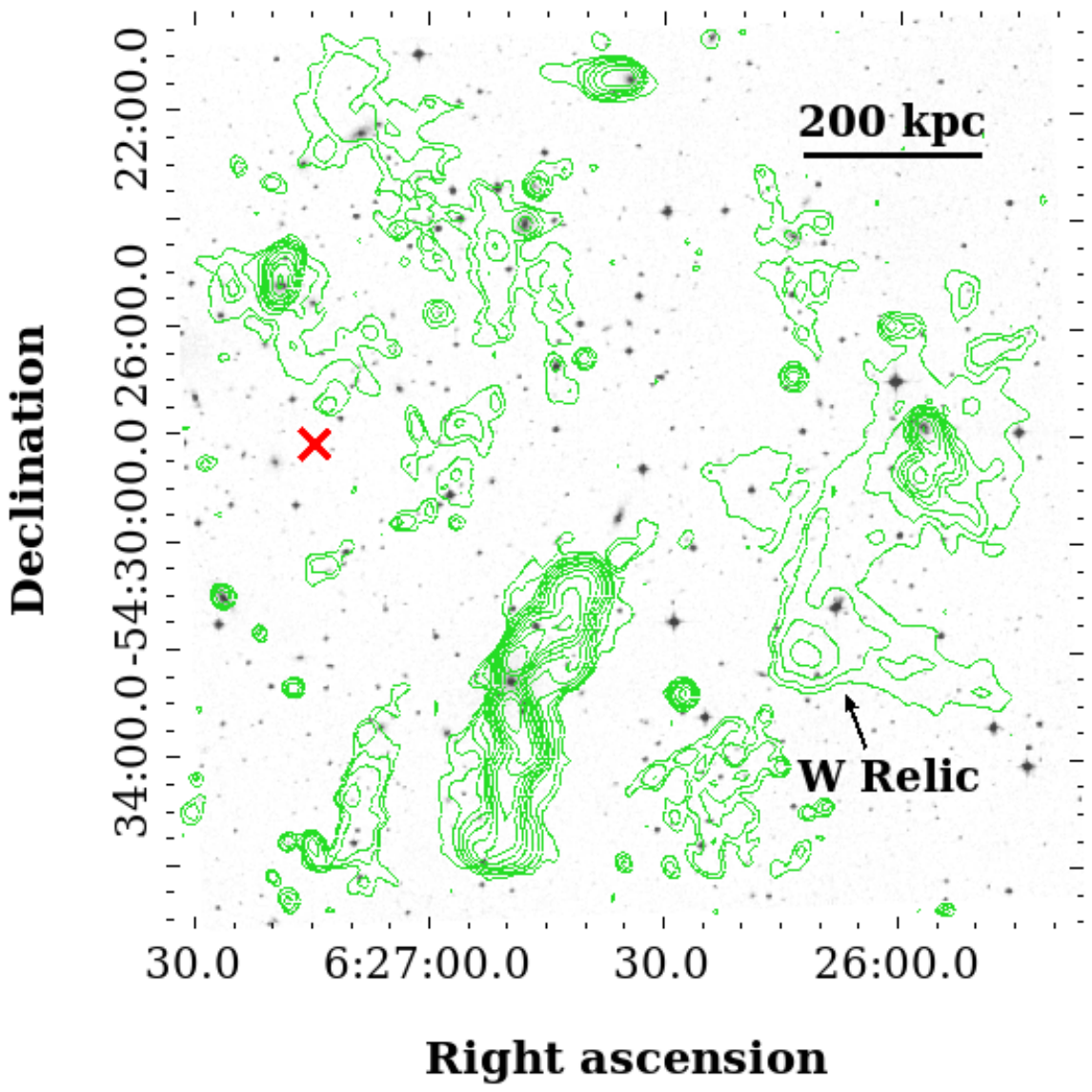}
   \caption{J0627.2$-$5428 \textbf{Left:}  Full-resolution (7.8\arcsec\,$\times$\,7.8\arcsec) 1.28~GHz MGCLS radio image with radio contours in black overlaid (1$\sigma$ = 8 $\mu$Jy beam$^{-1}$). \textbf{Right:} 1.28~GHz MGCLS low-resolution  (15\arcsec\,$\times$\,15\arcsec) radio contours in green (1$\sigma$ = 12 $\mu$Jy beam$^{-1}$), overlaid on the r-band \textit{Digitized Sky Survey (DSS)} optical image. In both panels, the radio contours start at 3\,$\sigma$ and rise by a factor of 2. The physical scale at the cluster redshift is indicated on top right, and the red $\times$ indicates the NED cluster position. } 
   \label{fig:J0627.2}%
\end{figure*}

\begin{figure*}
   \centering
   \includegraphics[width=0.496\textwidth]{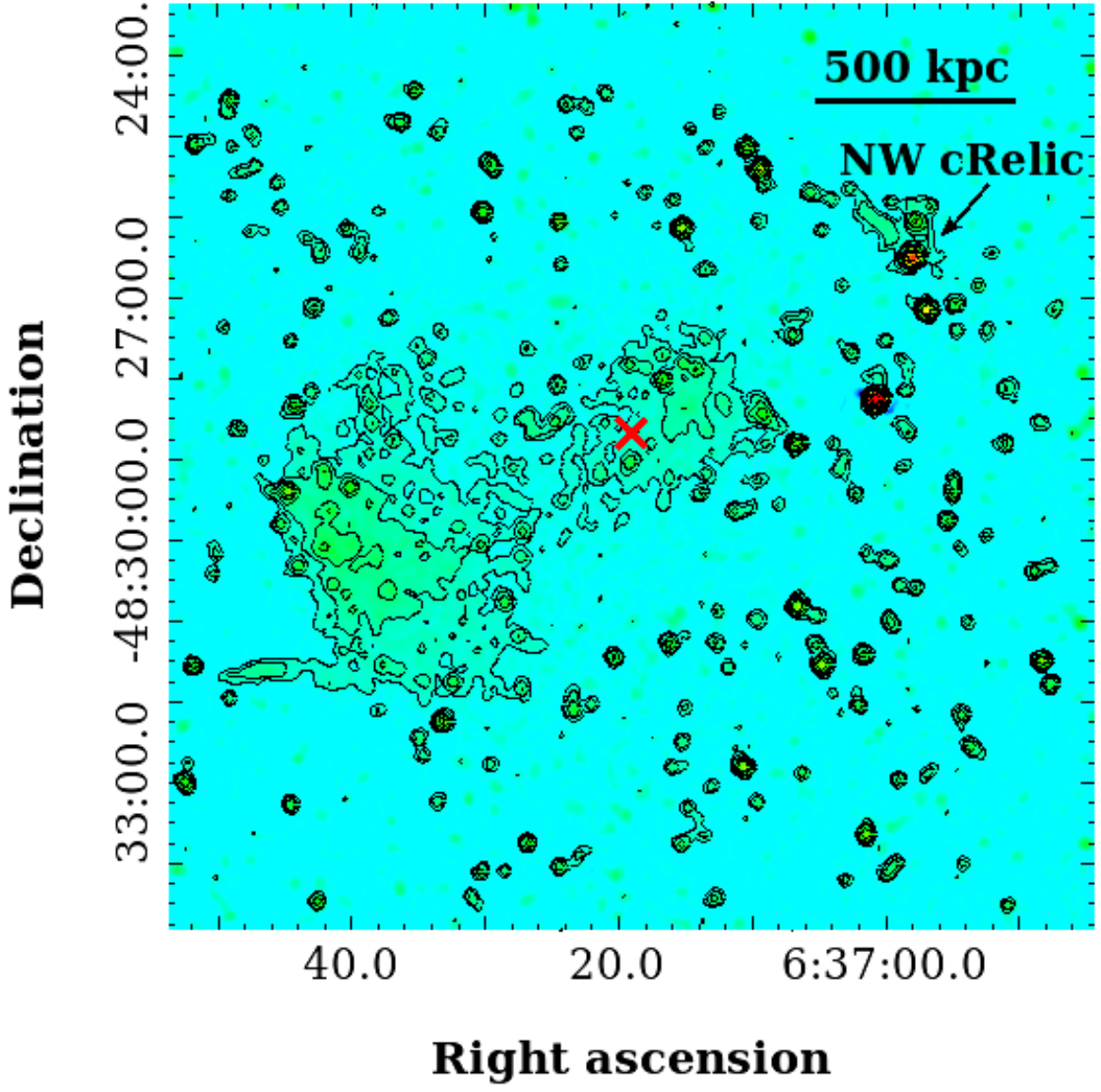}
    \includegraphics[width=0.5\textwidth]{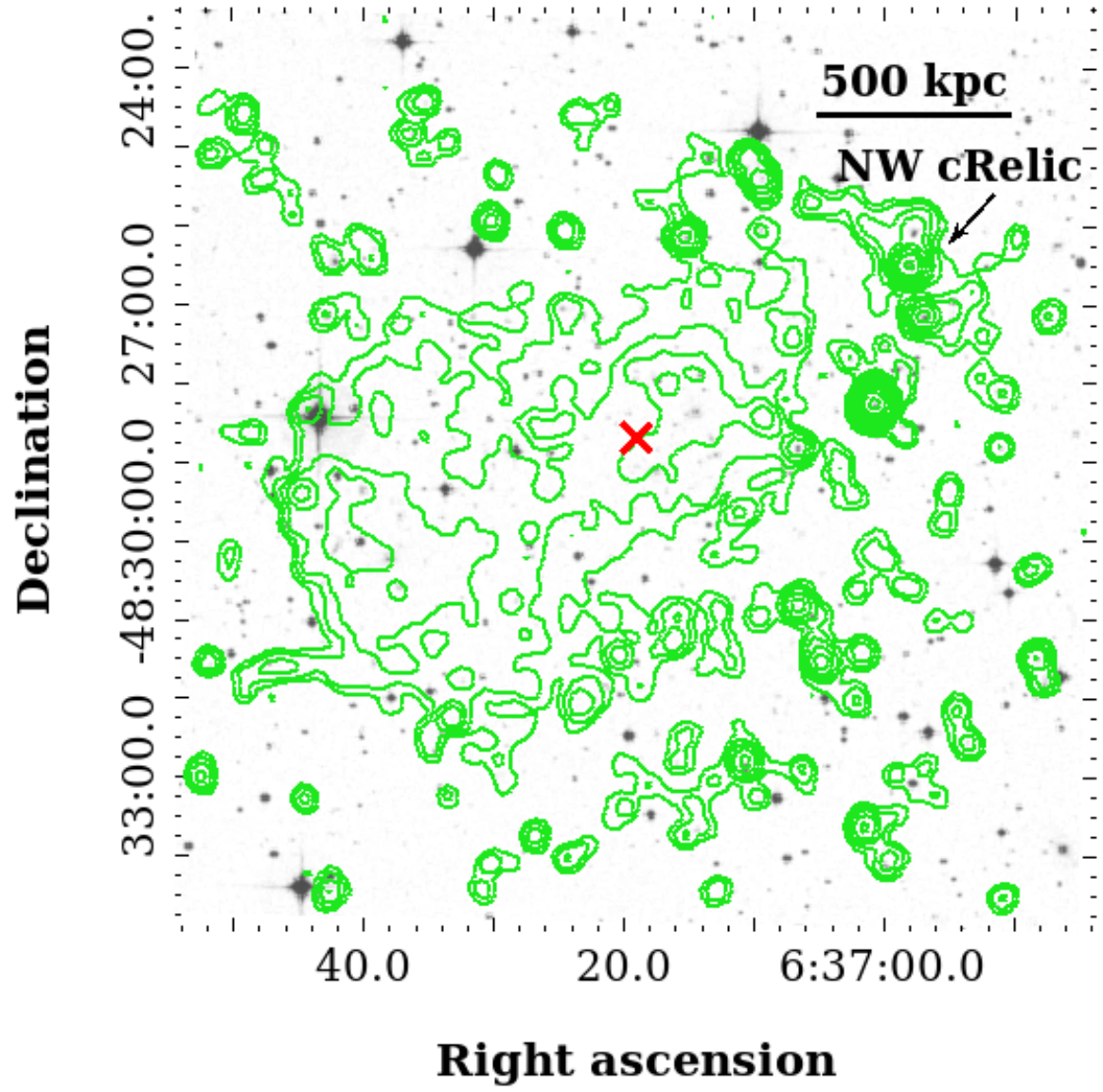}
   \caption{J0637.3$-$4828 \textbf{Left:}  Full-resolution (7.8\arcsec\,$\times$\,7.8\arcsec) 1.28~GHz MGCLS radio image with radio contours in black overlaid (1$\sigma$ = 3 $\mu$Jy beam$^{-1}$). \textbf{Right:} 1.28~GHz MGCLS low-resolution  (15\arcsec\,$\times$\,15\arcsec) radio contours in green (1$\sigma$ = 6 $\mu$Jy beam$^{-1}$), overlaid on the r-band \textit{Digitized Sky Survey (DSS)} optical image. In both panels, the radio contours start at 3\,$\sigma$ and rise by a factor of 2. The physical scale at the cluster redshift is indicated on top right, and the red $\times$ indicates the NED cluster position. } 
   \label{fig:J0637.3}%
\end{figure*}

\begin{figure*}
   \centering
   \includegraphics[width=0.496\textwidth]{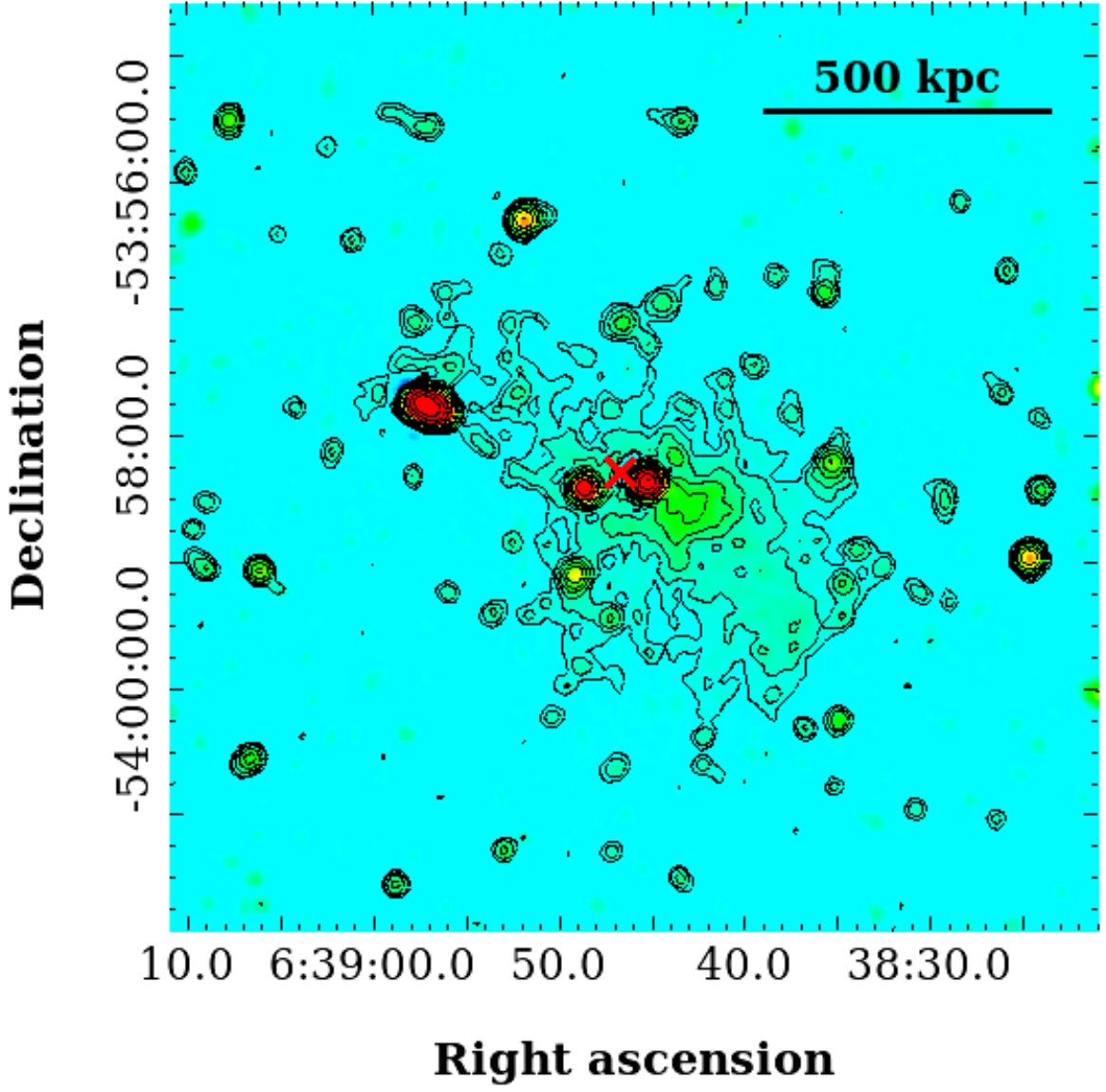}
    \includegraphics[width=0.5\textwidth]{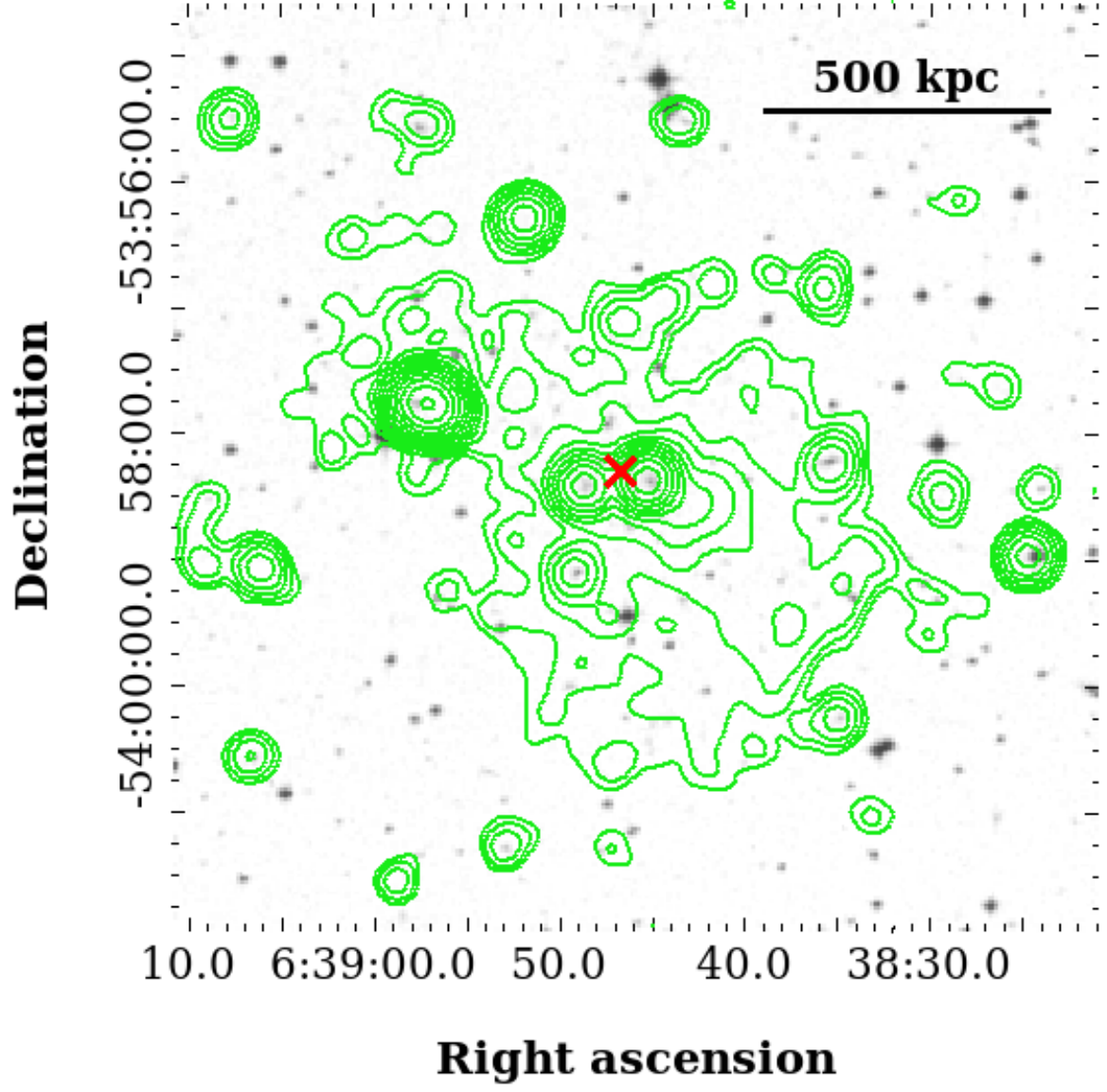}
   \caption{J0638.7$-$5358 \textbf{Left:}  Full-resolution (7.8\arcsec\,$\times$\,7.8\arcsec) 1.28~GHz MGCLS radio image with radio contours in black overlaid (1$\sigma$ = 4 $\mu$Jy beam$^{-1}$). \textbf{Right:} 1.28~GHz MGCLS low-resolution  (15\arcsec\,$\times$\,15\arcsec) radio contours in green (1$\sigma$ = 6 $\mu$Jy beam$^{-1}$), overlaid on the r-band \textit{Digitized Sky Survey (DSS)} optical image. In both panels, the radio contours start at 3\,$\sigma$ and rise by a factor of 2. The physical scale at the cluster redshift is indicated on top right, and the red $\times$ indicates the NED cluster position. } 
   \label{fig:J0638.7}%
\end{figure*}

\begin{figure*}
   \centering
   \includegraphics[width=0.496\textwidth]{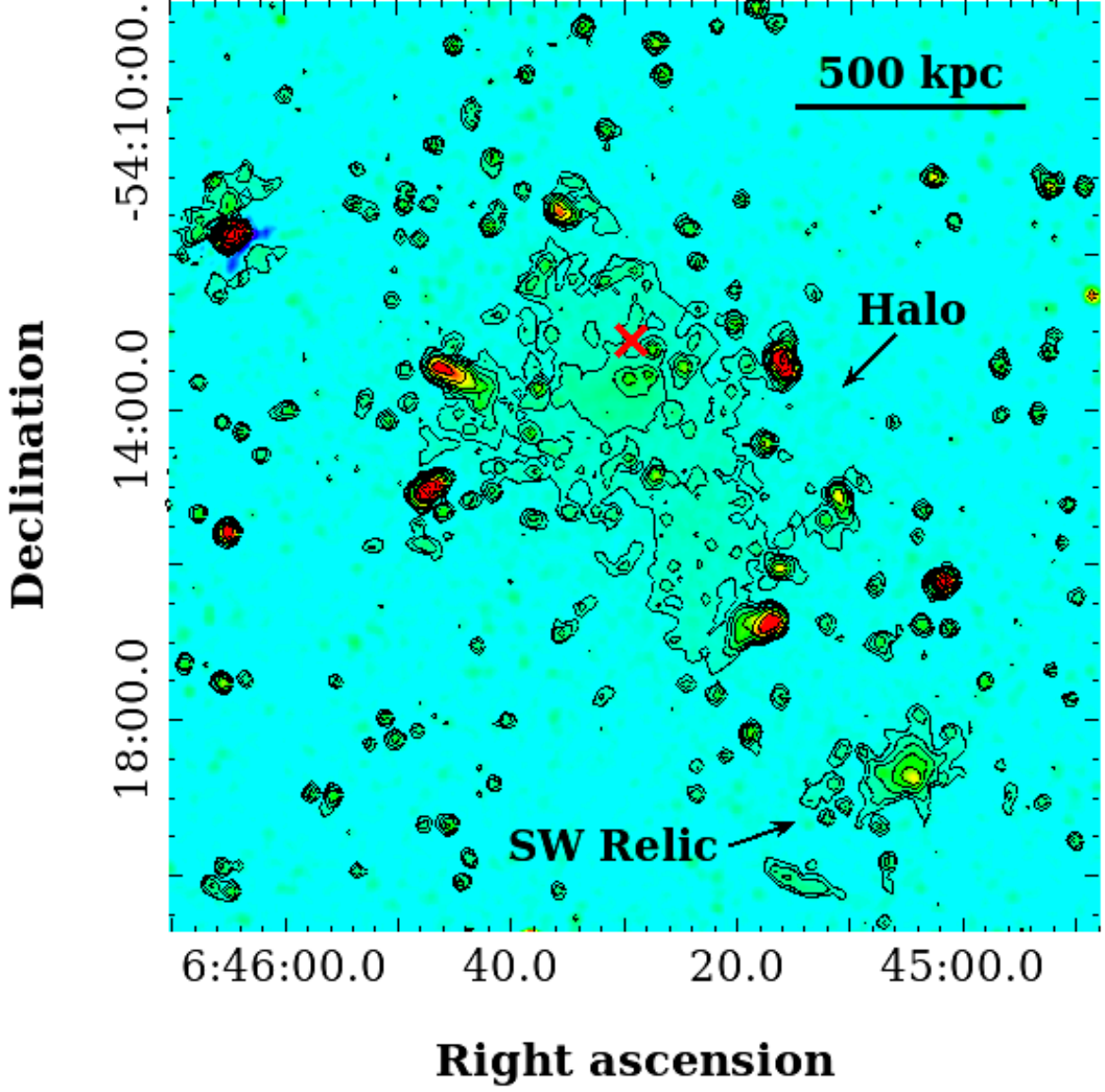}
    \includegraphics[width=0.5\textwidth]{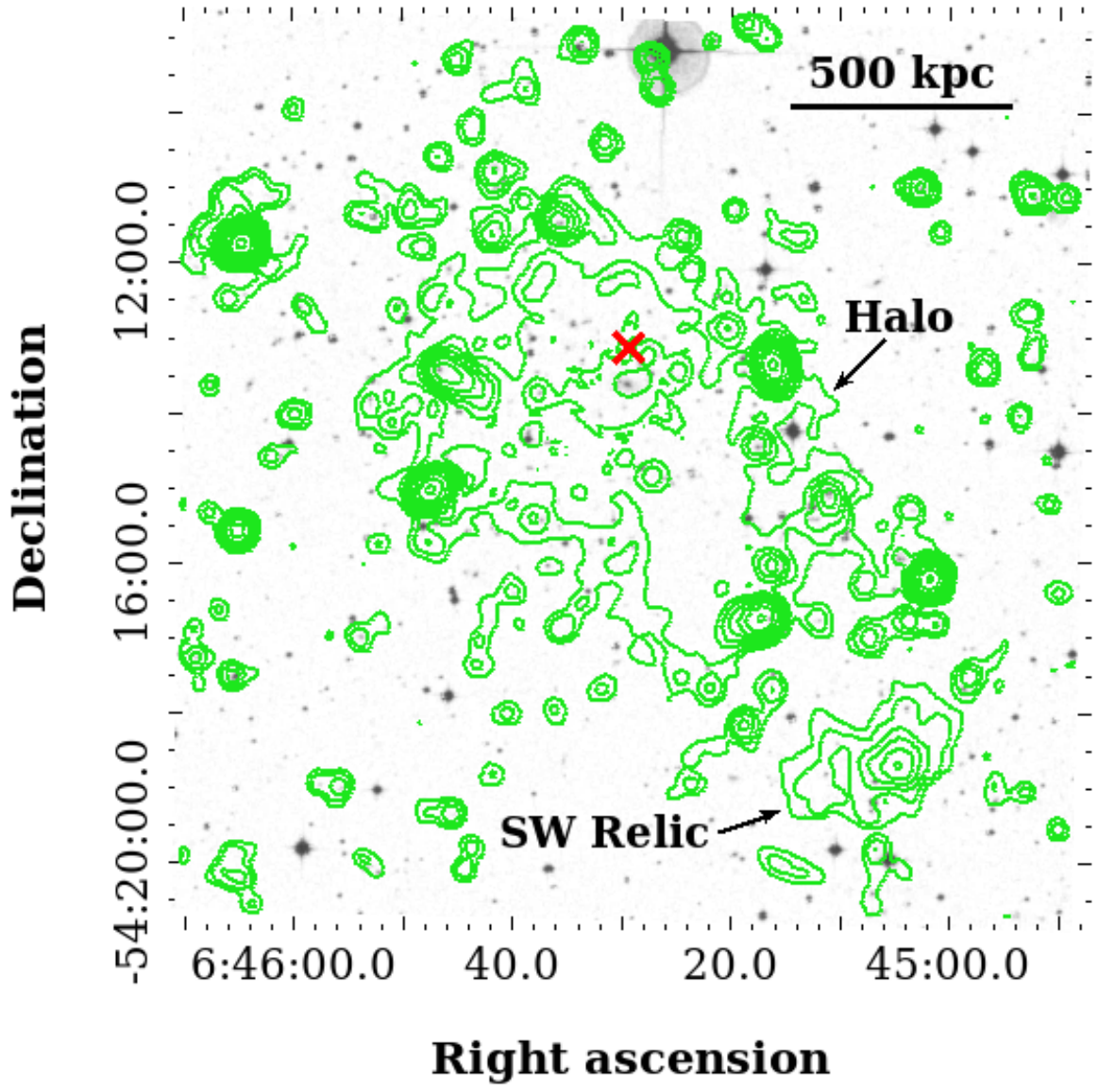}
   \caption{J0645.4$-$5413 \textbf{Left:}  Full-resolution (7.8\arcsec\,$\times$\,7.8\arcsec) 1.28~GHz MGCLS radio image with radio contours in black overlaid (1$\sigma$ = 4 $\mu$Jy beam$^{-1}$). \textbf{Right:} 1.28~GHz MGCLS low-resolution  (15\arcsec\,$\times$\,15\arcsec) radio contours in green (1$\sigma$ = 8 $\mu$Jy beam$^{-1}$), overlaid on the r-band \textit{Digitized Sky Survey (DSS)} optical image. In both panels, the radio contours start at 3\,$\sigma$ and rise by a factor of 2. The physical scale at the cluster redshift is indicated on top right, and the red $\times$ indicates the NED cluster position. } 
   \label{fig:J0645.4}%
\end{figure*}

\begin{figure*}
   \centering
   \includegraphics[width=0.496\textwidth]{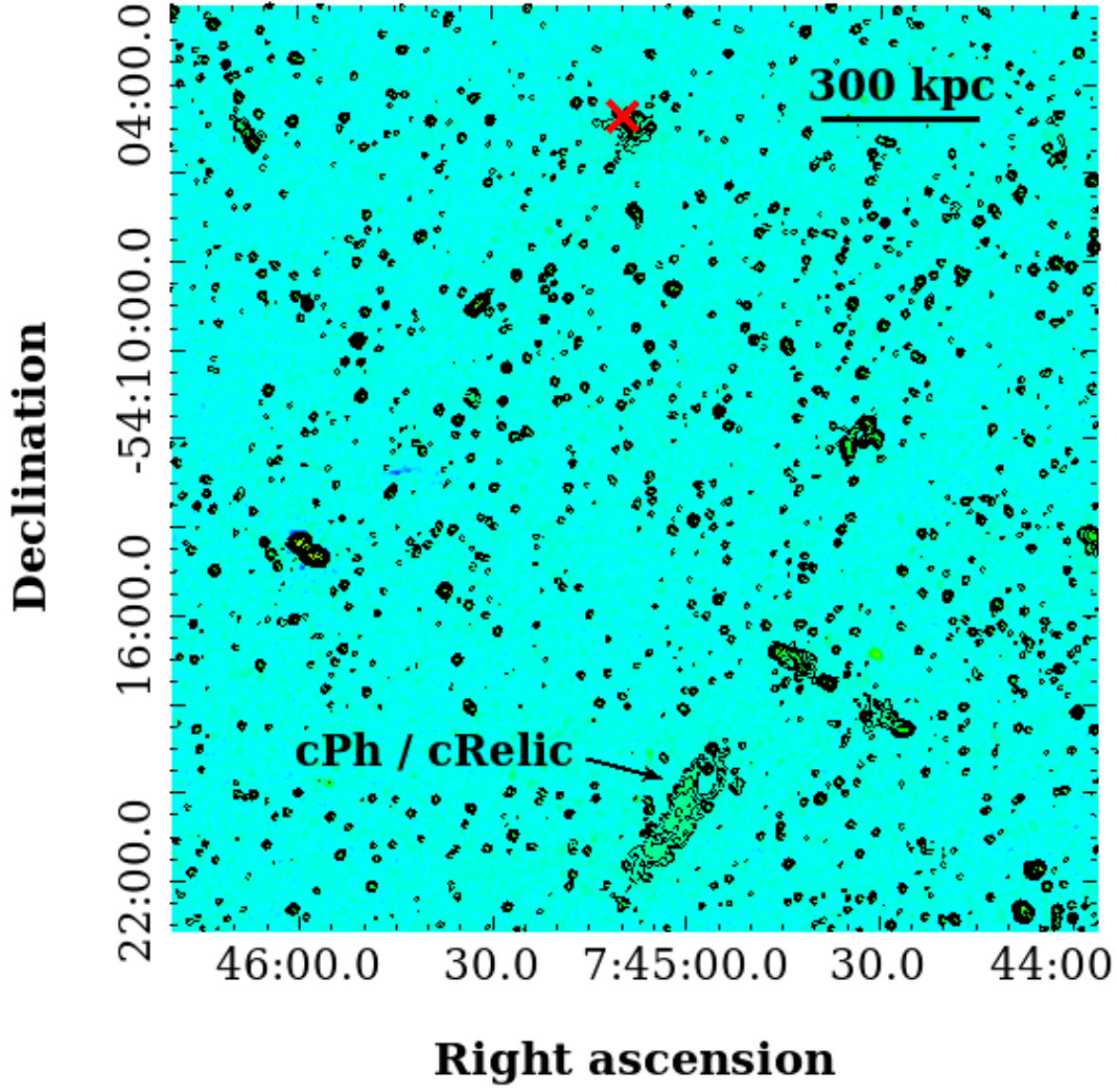}
    \includegraphics[width=0.5\textwidth]{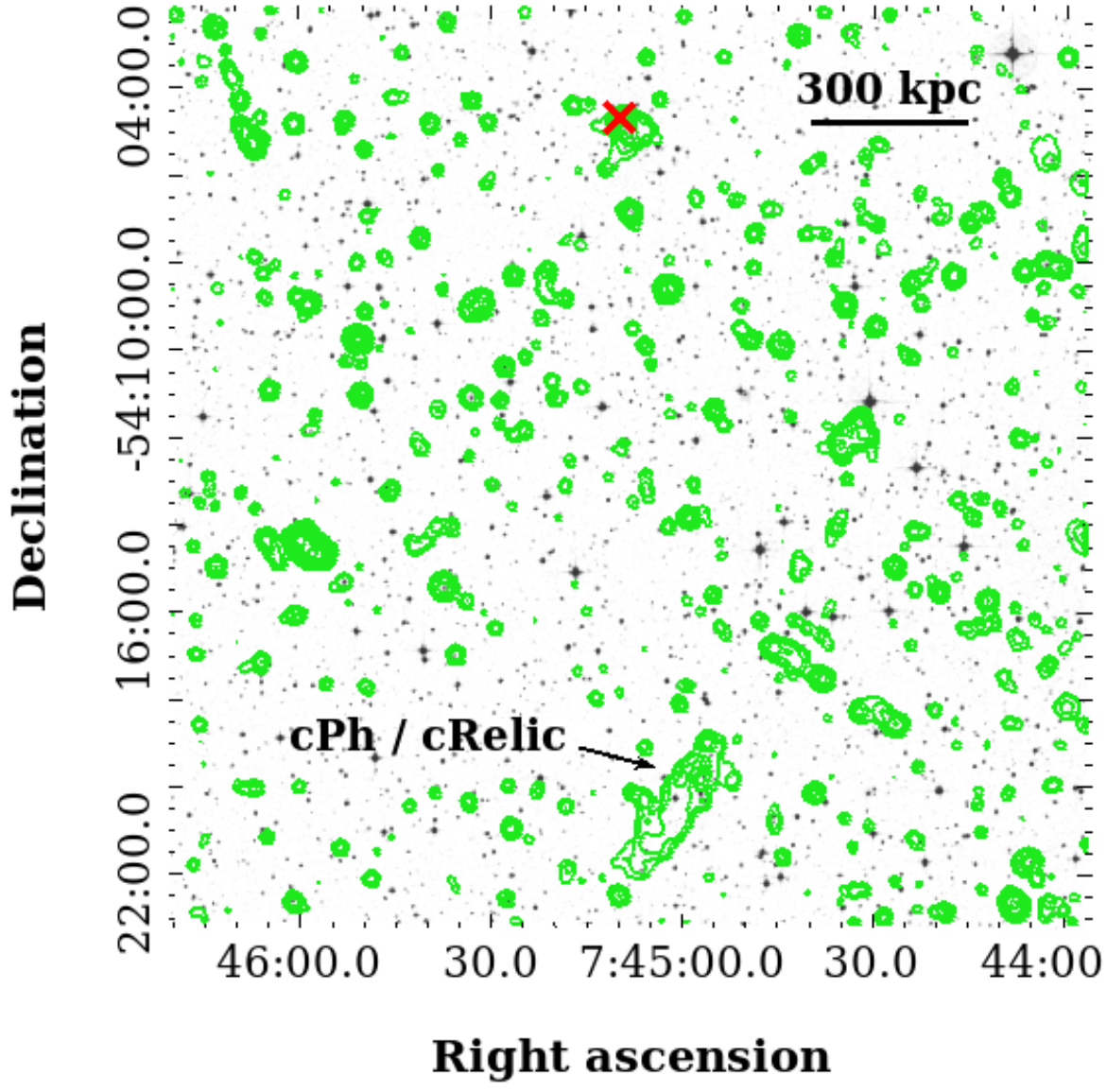}
   \caption{J0745.1$-$5404 \textbf{Left:}  Full-resolution (7.8\arcsec\,$\times$\,7.8\arcsec) 1.28~GHz MGCLS radio image with radio contours in black overlaid (1$\sigma$ = 4 $\mu$Jy beam$^{-1}$). \textbf{Right:} 1.28~GHz MGCLS low-resolution  (15\arcsec\,$\times$\,15\arcsec) radio contours in green (1$\sigma$ = 8 $\mu$Jy beam$^{-1}$), overlaid on the r-band \textit{Digitized Sky Survey (DSS)} optical image. In both panels, the radio contours start at 3\,$\sigma$ and rise by a factor of 2. The physical scale at the cluster redshift is indicated on top right, and the red $\times$ indicates the NED cluster position. } 
   \label{fig:J0745.1}%
\end{figure*}

\begin{figure*}
   \centering
   \includegraphics[width=0.496\textwidth]{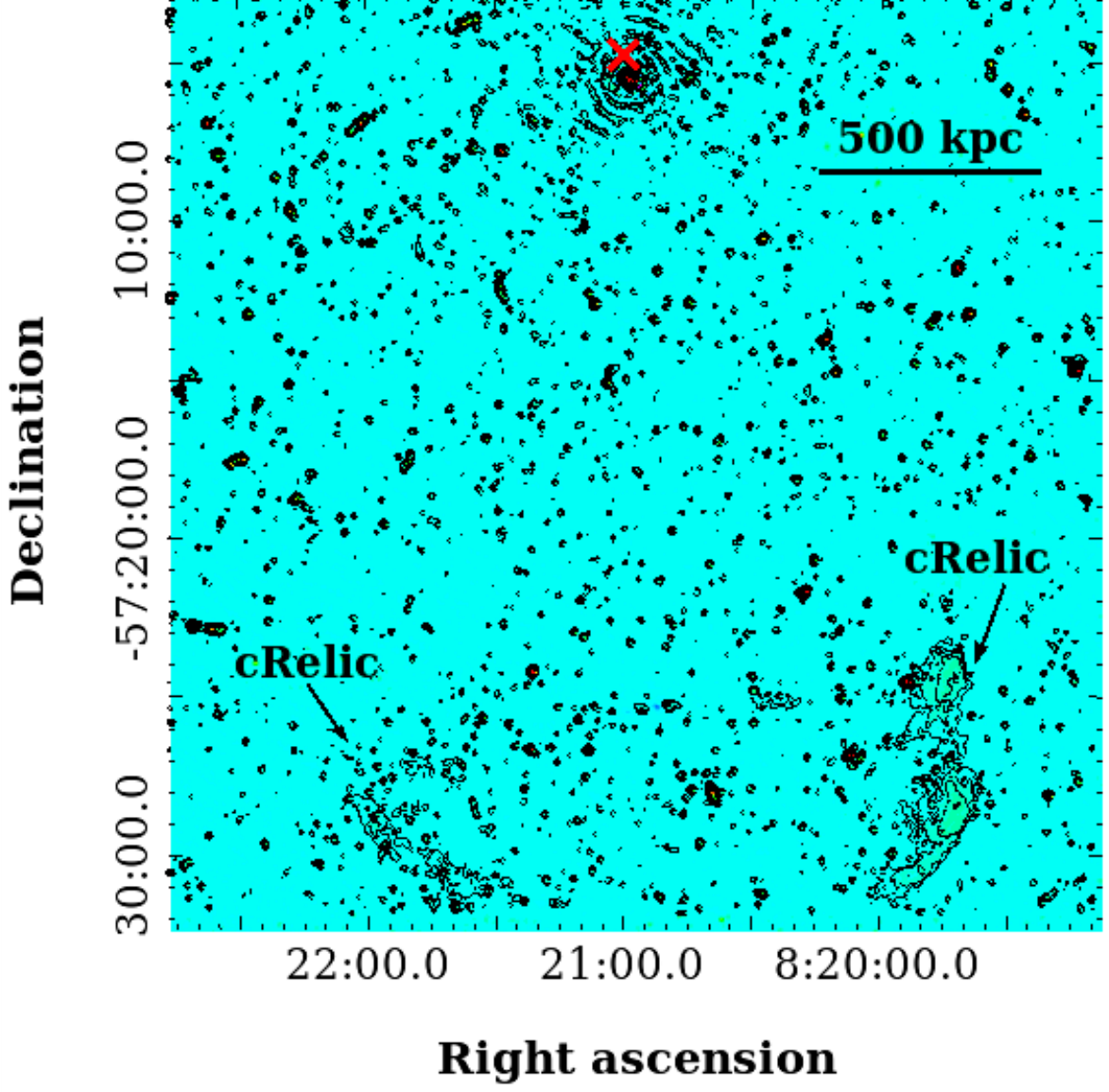}
    \includegraphics[width=0.5\textwidth]{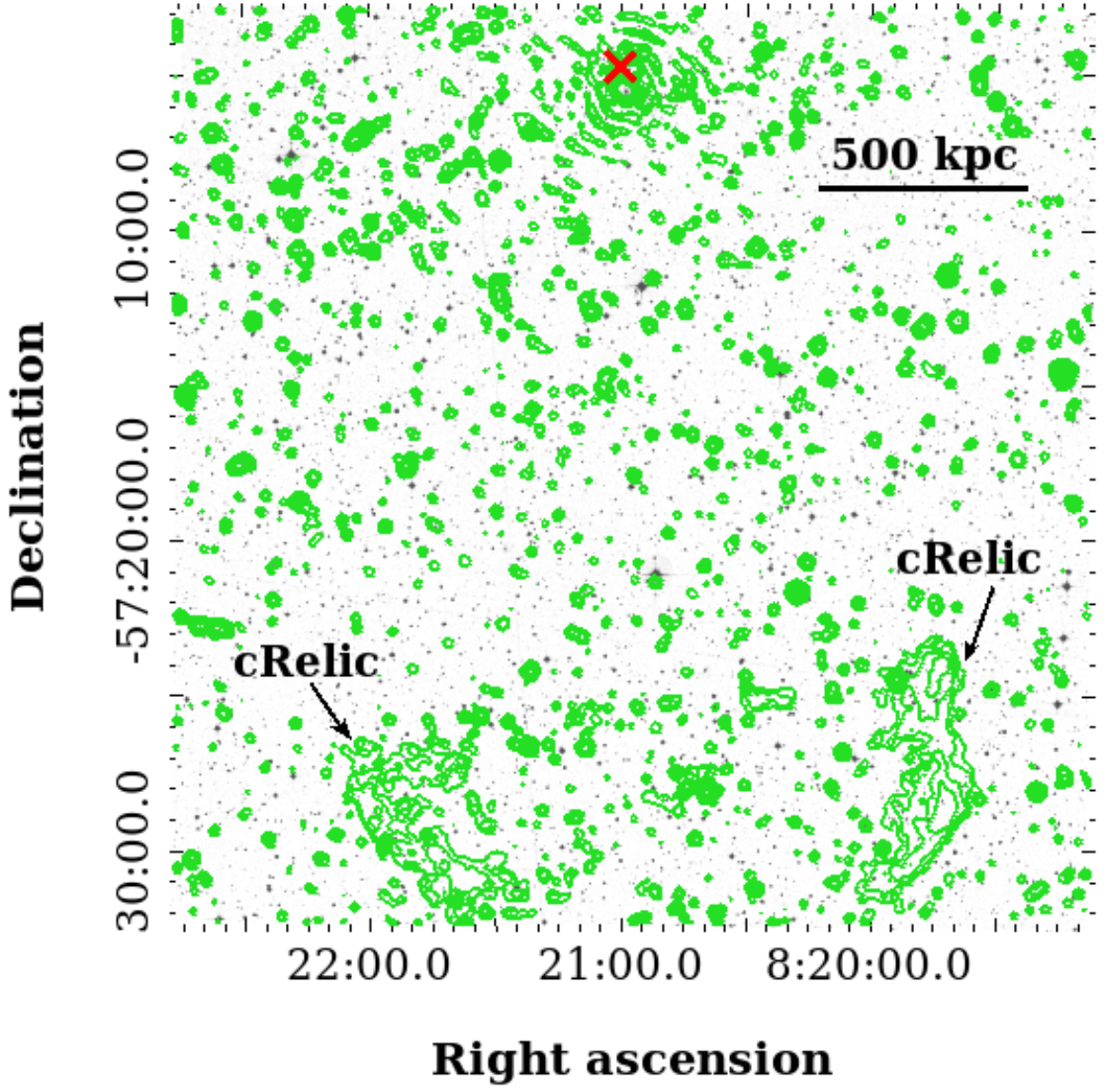}
   \caption{J0820.9$-$5704 \textbf{Left:}  Full-resolution (7.8\arcsec\,$\times$\,7.8\arcsec) 1.28~GHz MGCLS radio image with radio contours in black overlaid (1$\sigma$ = 4 $\mu$Jy beam$^{-1}$). \textbf{Right:} 1.28~GHz MGCLS low-resolution  (15\arcsec\,$\times$\,15\arcsec) radio contours in green (1$\sigma$ = 8 $\mu$Jy beam$^{-1}$), overlaid on the r-band \textit{Digitized Sky Survey (DSS)} optical image. In both panels, the radio contours start at 3\,$\sigma$ and rise by a factor of 2. The physical scale at the cluster redshift is indicated on top right, and the red $\times$ indicates the NED cluster position. } 
   \label{fig:J0820.9}%
\end{figure*}

\begin{figure*}
   \centering
   \includegraphics[width=0.496\textwidth]{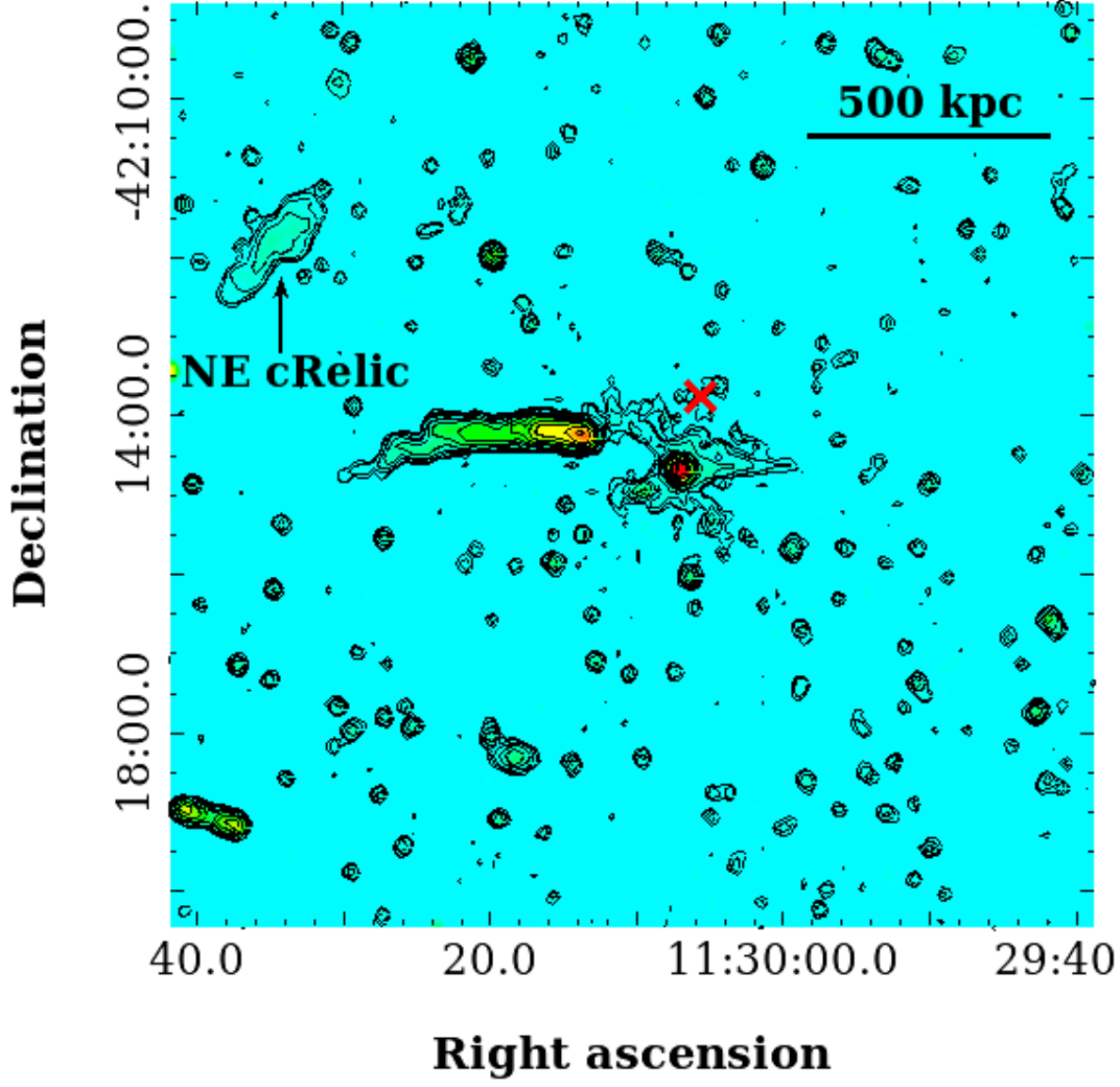}
    \includegraphics[width=0.5\textwidth]{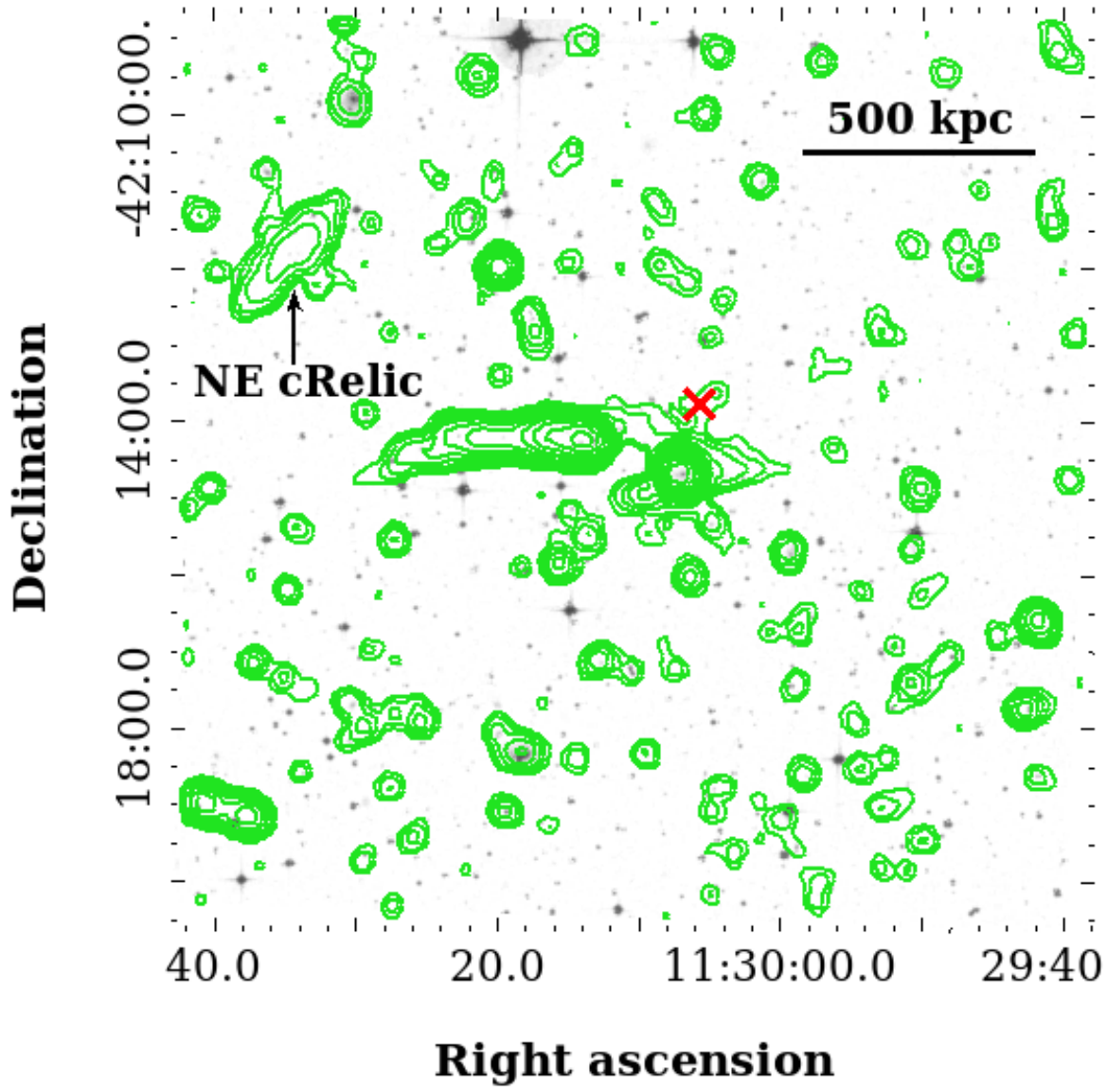}
   \caption{J1130.0$-$4213 \textbf{Left:}  Full-resolution (7.8\arcsec\,$\times$\,7.8\arcsec) 1.28~GHz MGCLS radio image with radio contours in black overlaid (1$\sigma$ = 3.5 $\mu$Jy beam$^{-1}$). \textbf{Right:} 1.28~GHz MGCLS low-resolution  (15\arcsec\,$\times$\,15\arcsec) radio contours in green (1$\sigma$ = 6 $\mu$Jy beam$^{-1}$), overlaid on the r-band \textit{Digitized Sky Survey (DSS)} optical image. In both panels, the radio contours start at 3\,$\sigma$ and rise by a factor of 2. The physical scale at the cluster redshift is indicated on top right, and the red $\times$ indicates the NED cluster position. } 
   \label{fig:J1130.0}%
\end{figure*}

\begin{figure*}
   \centering
   \includegraphics[width=0.496\textwidth]{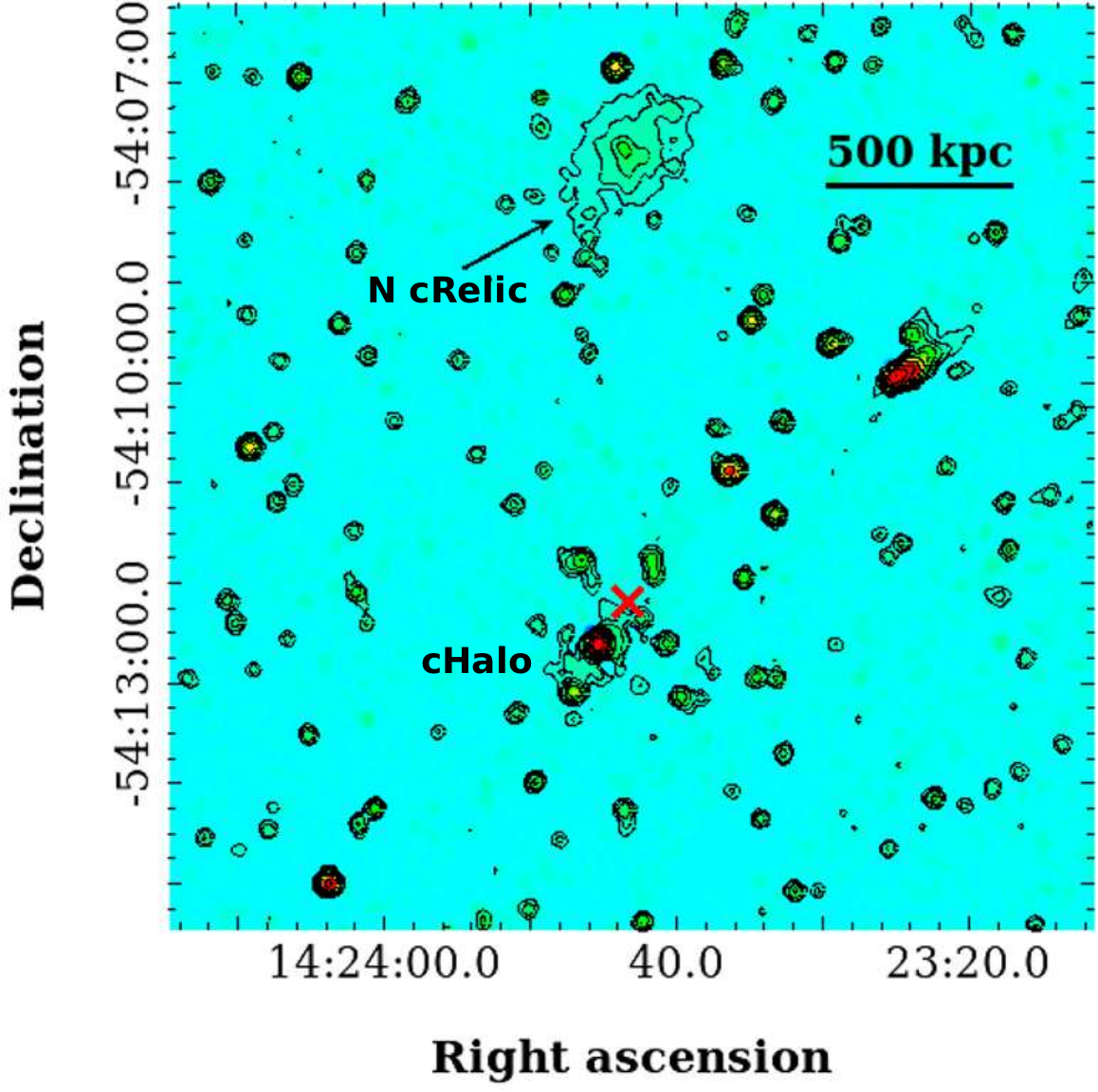}
    \includegraphics[width=0.5\textwidth]{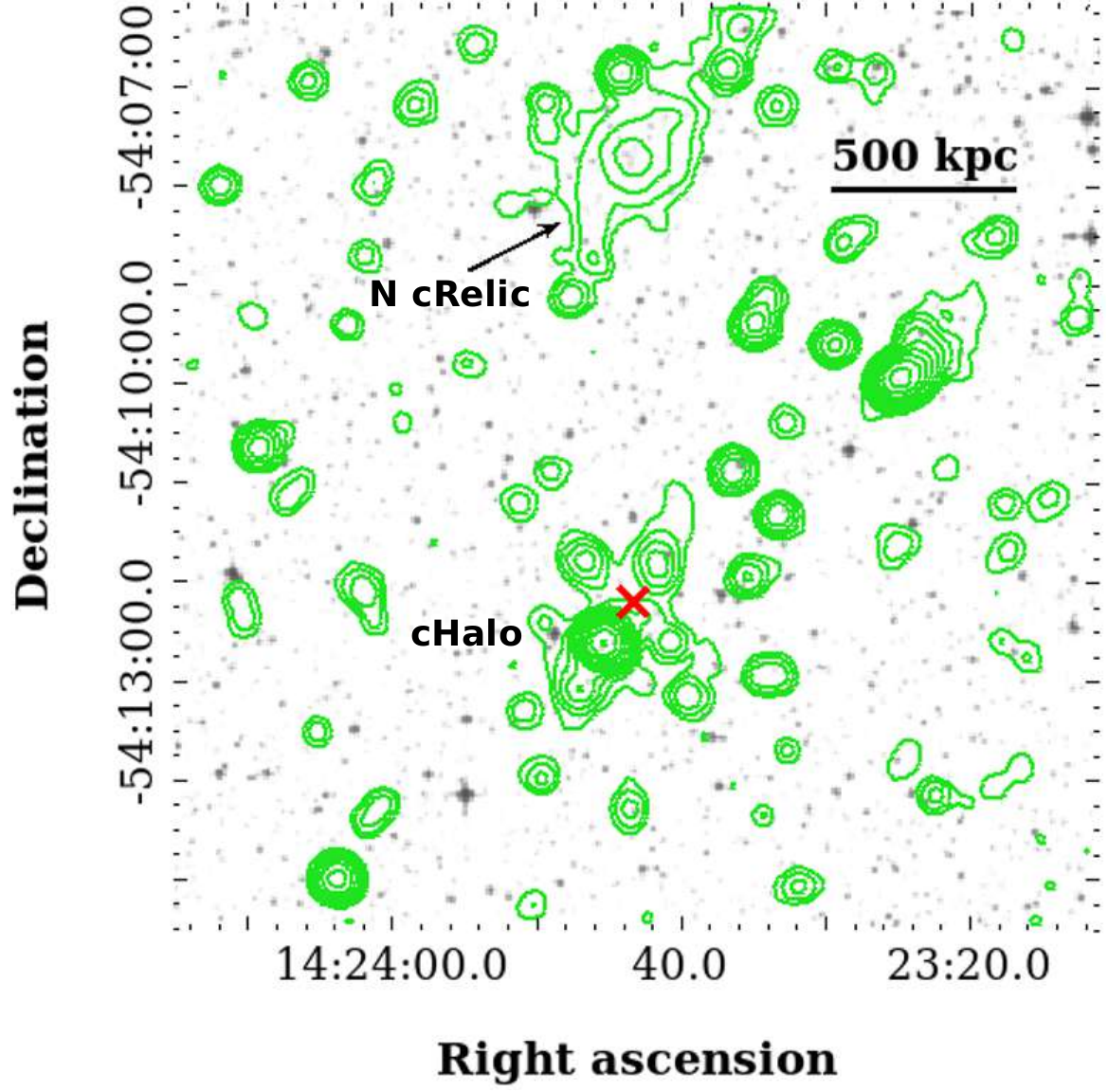}
   \caption{J1423.7$-$5412 \textbf{Left:}  Full-resolution (7.8\arcsec\,$\times$\,7.8\arcsec) 1.28~GHz MGCLS radio image with radio contours in black overlaid (1$\sigma$ = 4 $\mu$Jy beam$^{-1}$). \textbf{Right:} 1.28~GHz MGCLS low-resolution  (15\arcsec\,$\times$\,15\arcsec) radio contours in green (1$\sigma$ = 8 $\mu$Jy beam$^{-1}$), overlaid on the r-band \textit{Digitized Sky Survey (DSS)} optical image. In both panels, the radio contours start at 3\,$\sigma$ and rise by a factor of 2. The physical scale at the cluster redshift is indicated on top right, and the red $\times$ indicates the NED cluster position. } 
   \label{fig:J1423.7}%
\end{figure*}

\begin{figure*}
   \centering
   \includegraphics[width=0.496\textwidth]{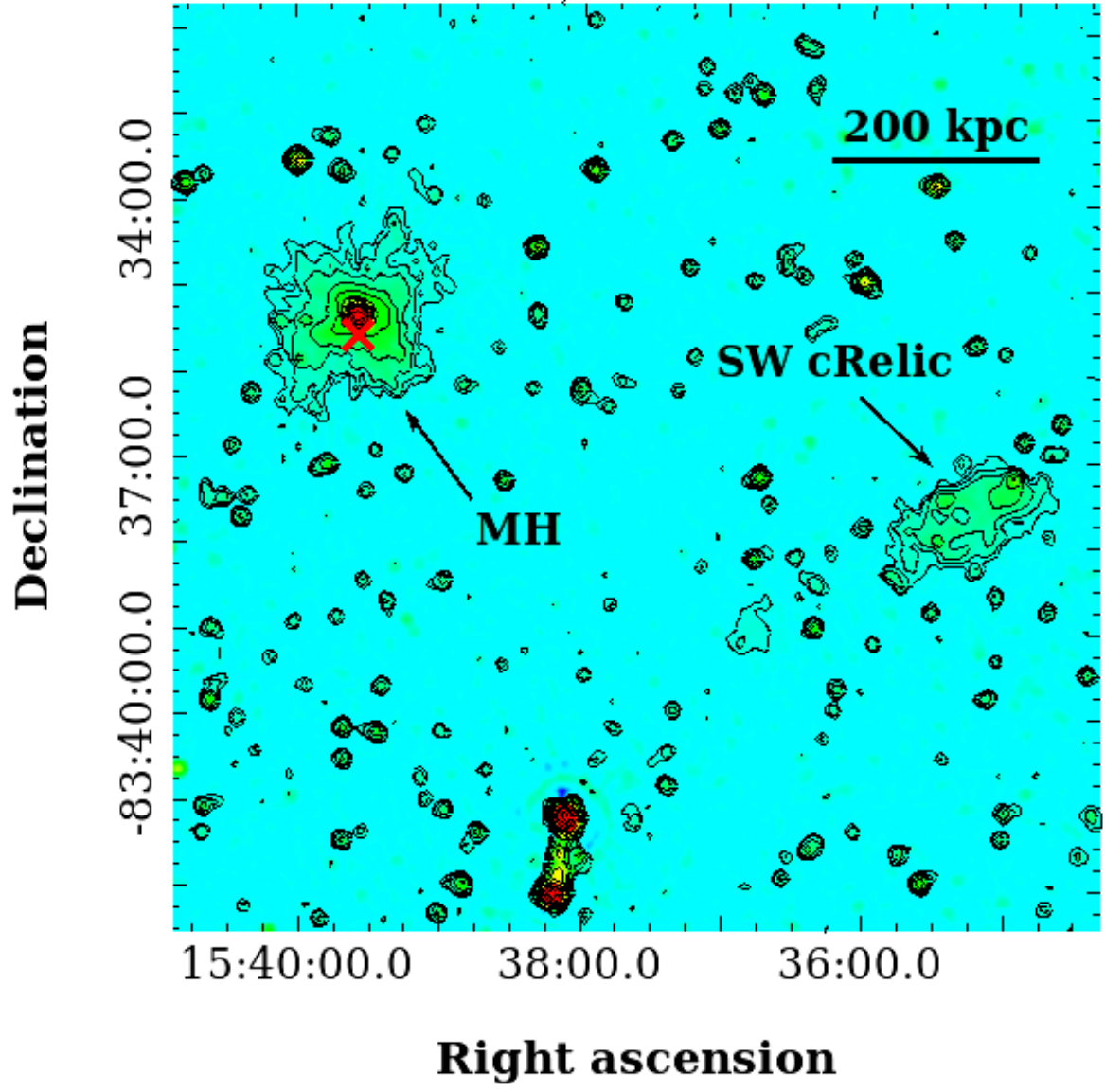}
    \includegraphics[width=0.5\textwidth]{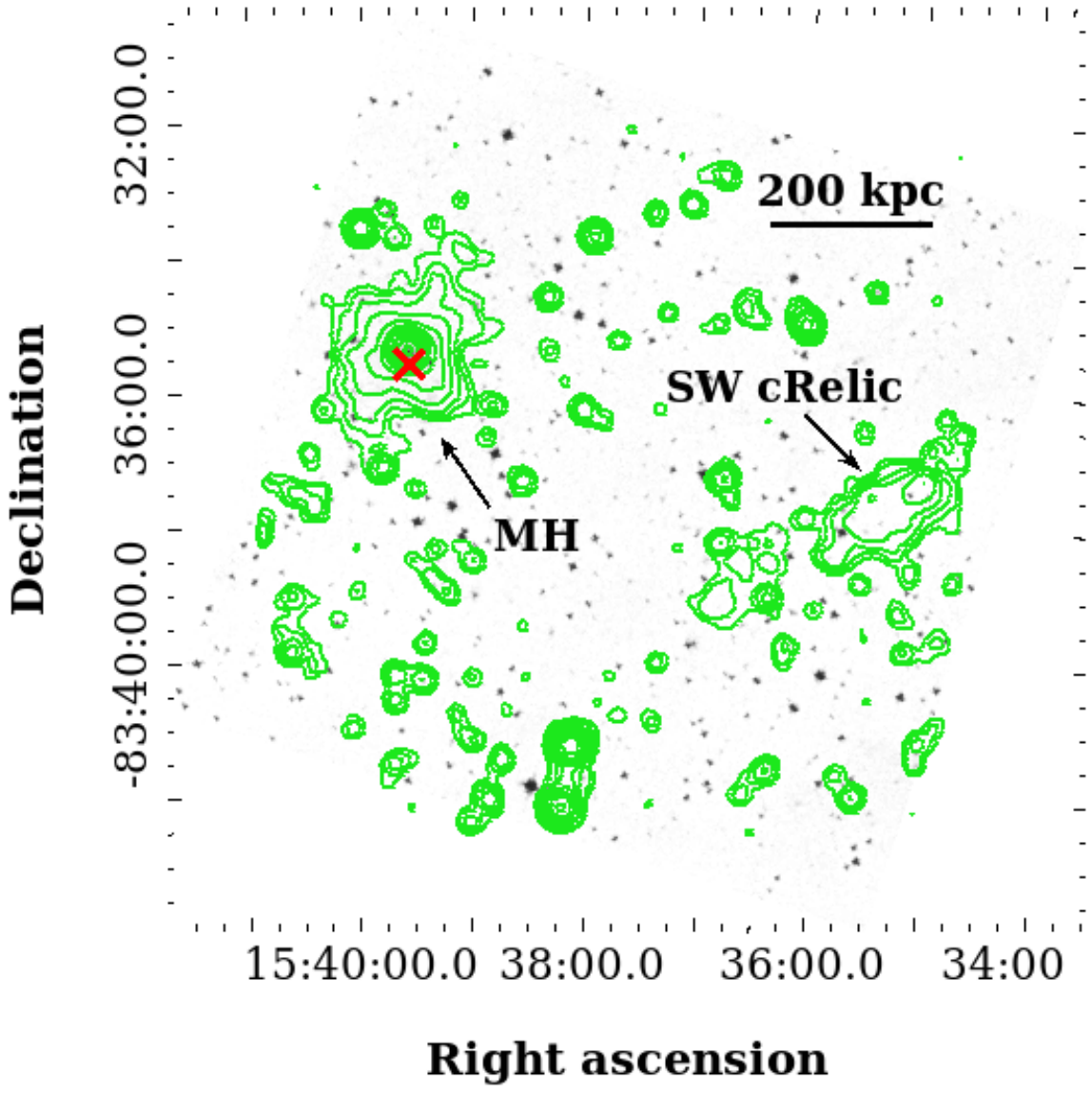}
   \caption{J1539.5$-$8335 \textbf{Left:}  Full-resolution (7.8\arcsec\,$\times$\,7.8\arcsec) 1.28~GHz MGCLS radio image with radio contours in black overlaid (1$\sigma$ = 3.5 $\mu$Jy beam$^{-1}$). \textbf{Right:} 1.28~GHz MGCLS low-resolution  (15\arcsec\,$\times$\,15\arcsec) radio contours in green (1$\sigma$ = 6 $\mu$Jy beam$^{-1}$), overlaid on the r-band \textit{Digitized Sky Survey (DSS)} optical image. In both panels, the radio contours start at 3\,$\sigma$ and rise by a factor of 2. The physical scale at the cluster redshift is indicated on top right, and the red $\times$ indicates the NED cluster position. } 
   \label{fig:J1539.5}%
\end{figure*}

\begin{figure*}
   \centering
   \includegraphics[width=0.496\textwidth]{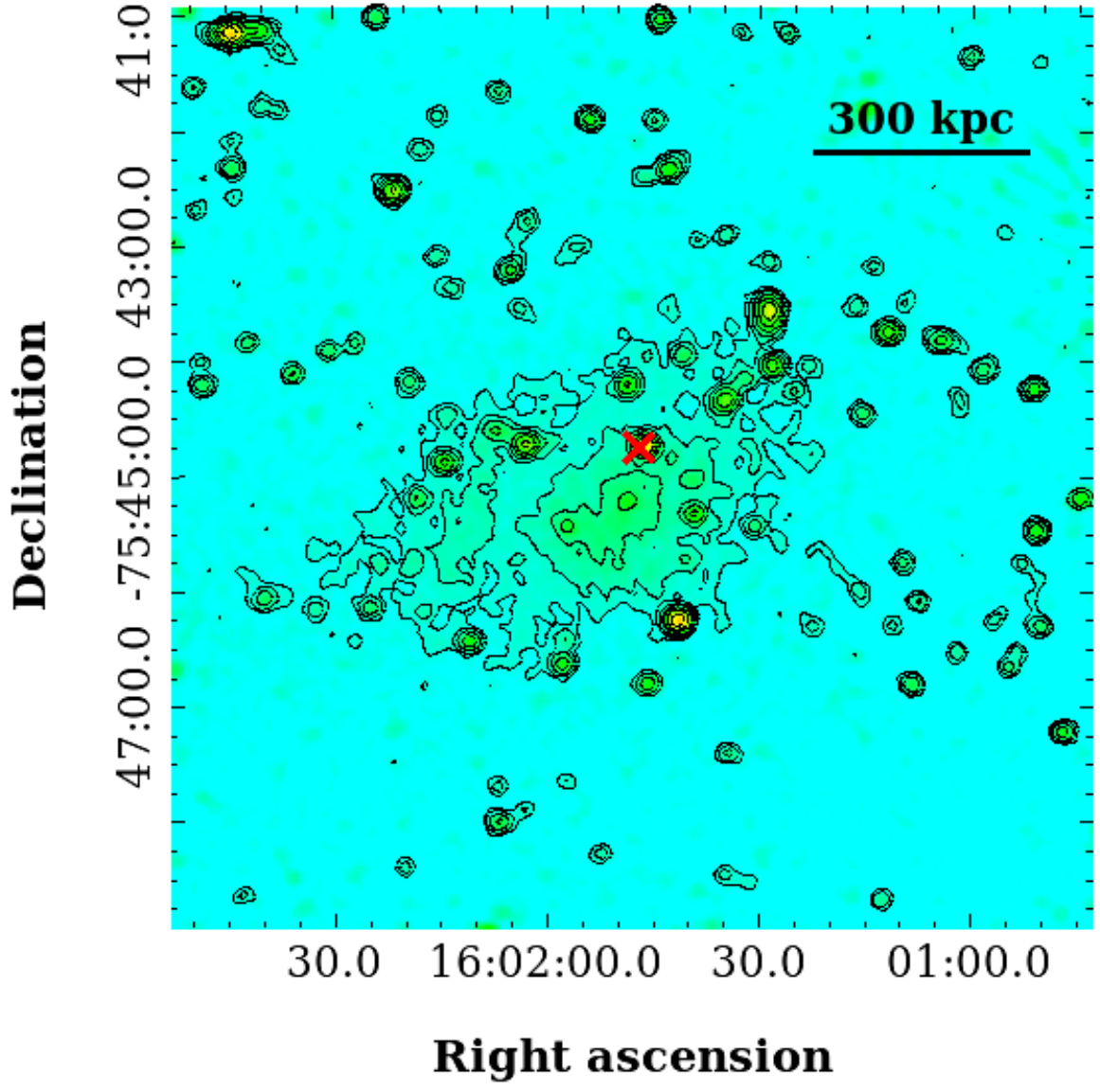}
    \includegraphics[width=0.5\textwidth]{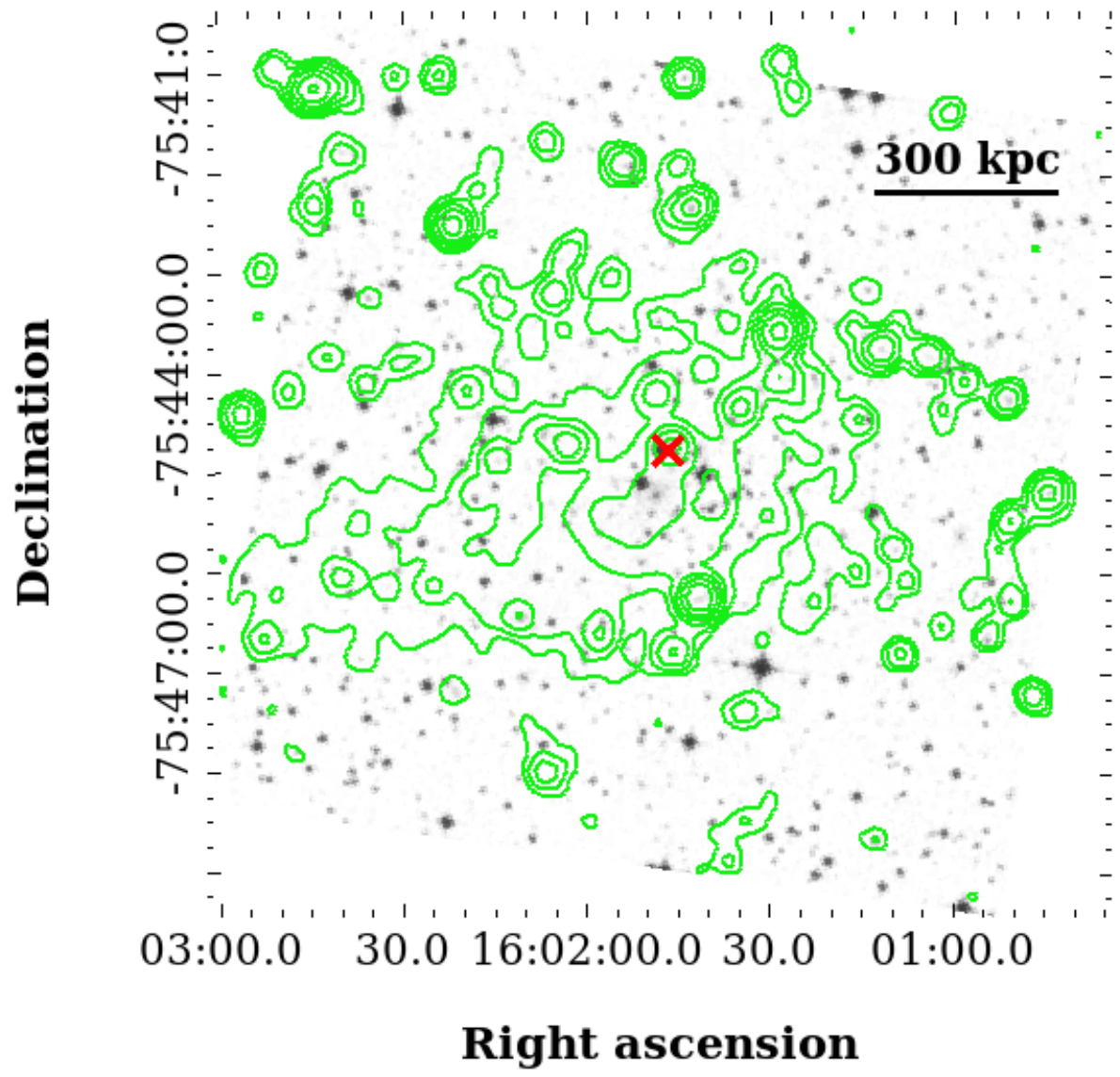}
   \caption{J1601.7$-$7544 \textbf{Left:}  Full-resolution (7.8\arcsec\,$\times$\,7.8\arcsec) 1.28~GHz MGCLS radio image with radio contours in black overlaid (1$\sigma$ = 4 $\mu$Jy beam$^{-1}$). \textbf{Right:} 1.28~GHz MGCLS low-resolution  (15\arcsec\,$\times$\,15\arcsec) radio contours in green (1$\sigma$ = 7 $\mu$Jy beam$^{-1}$), overlaid on the r-band \textit{Digitized Sky Survey (DSS)} optical image. In both panels, the radio contours start at 3\,$\sigma$ and rise by a factor of 2. The physical scale at the cluster redshift is indicated on top right, and the red $\times$ indicates the NED cluster position. } 
   \label{fig:J1601.7}%
\end{figure*}

\begin{figure*}
   \centering
   \includegraphics[width=0.496\textwidth]{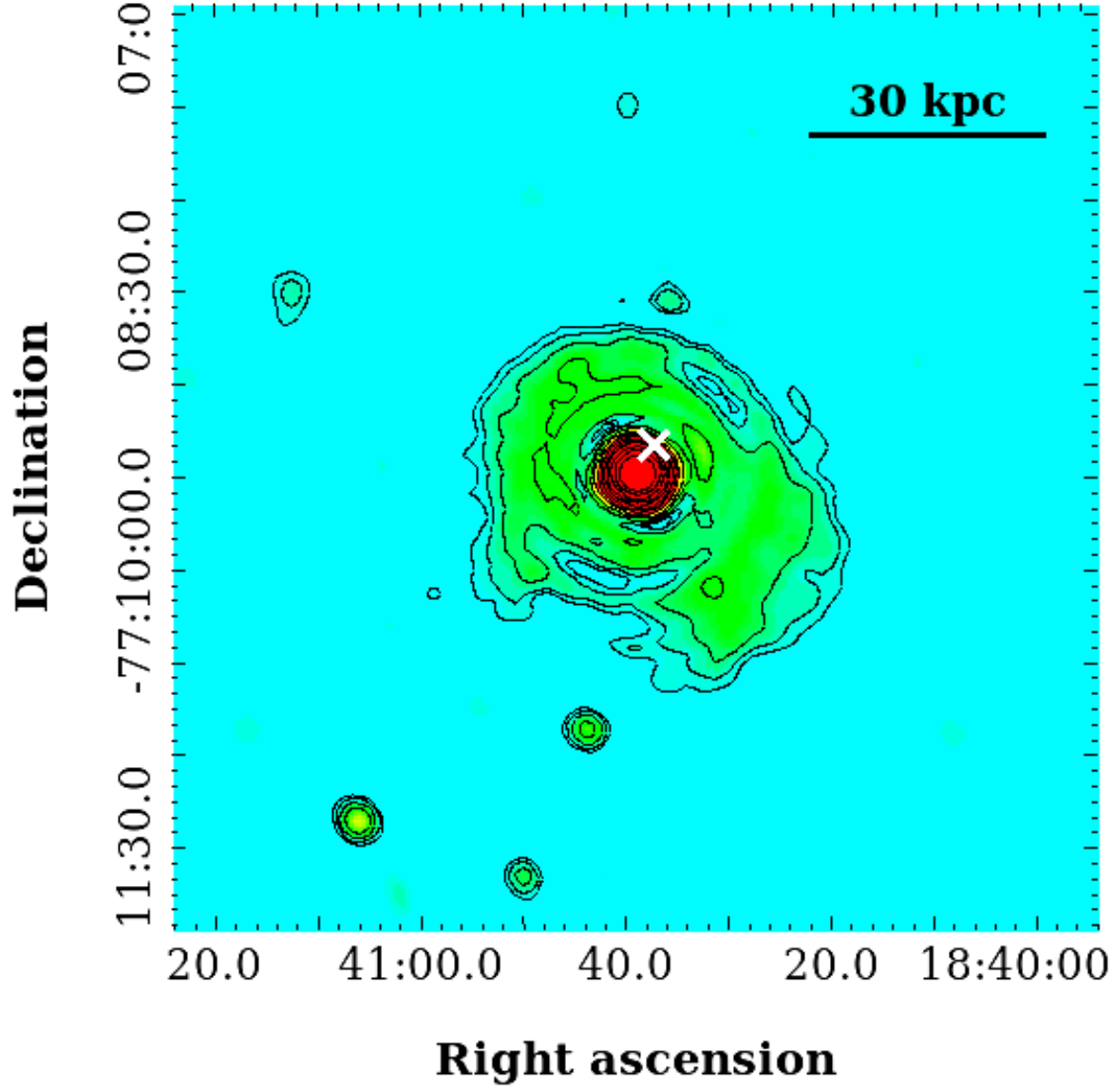}
    \includegraphics[width=0.5\textwidth]{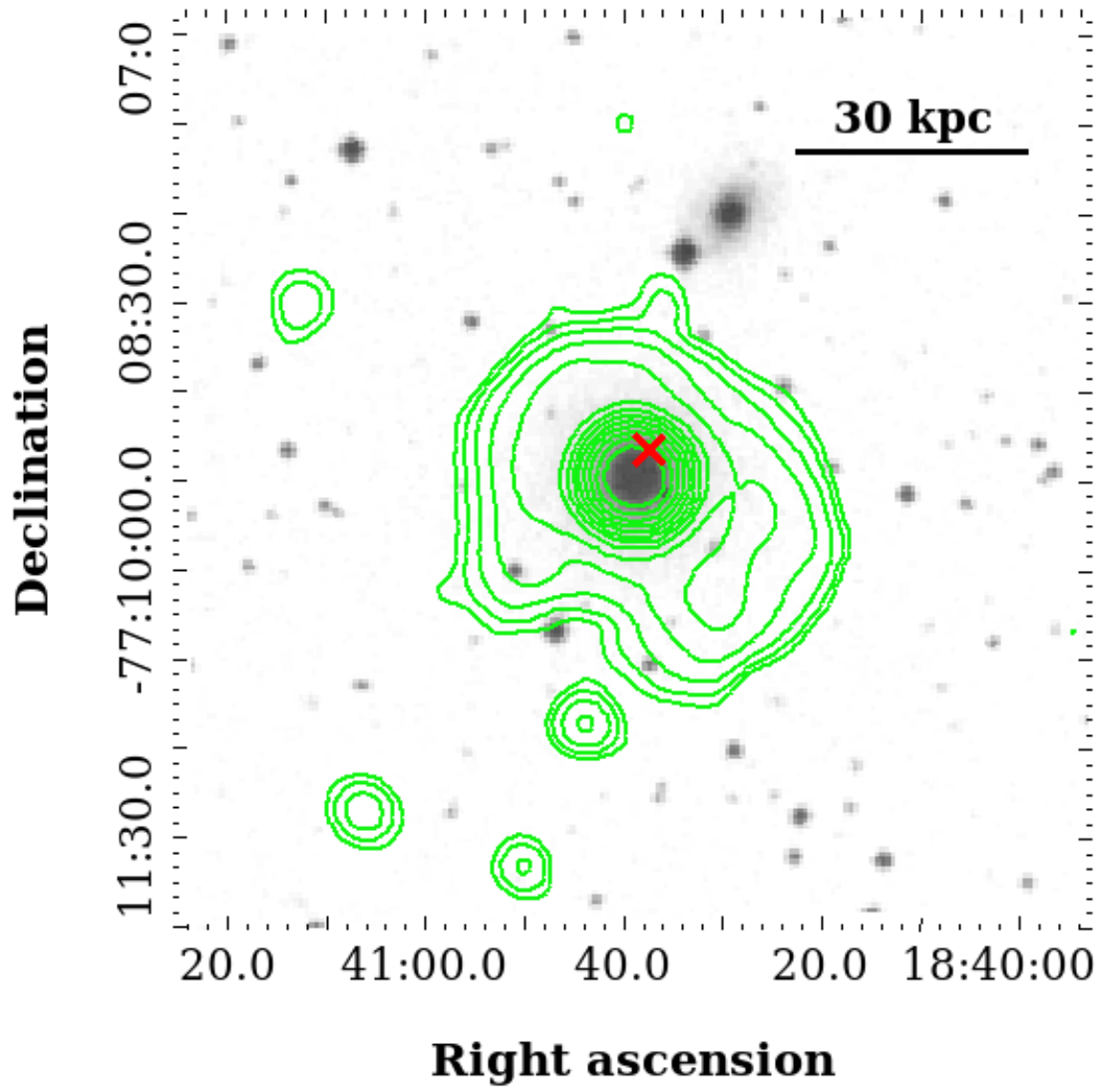}
   \caption{J1840.6$-$7709 \textbf{Left:}  Full-resolution (7.8\arcsec\,$\times$\,7.8\arcsec) 1.28~GHz MGCLS radio image with radio contours in black overlaid (1$\sigma$ = 16 $\mu$Jy~beam$^{-1}$). \textbf{Right:} 1.28~GHz MGCLS low-resolution  (15\arcsec\,$\times$\,15\arcsec) radio contours in green (1$\sigma$ = 20 $\mu$Jy~beam$^{-1}$), overlaid on the r-band \textit{Digitized Sky Survey (DSS)} optical image. In both panels, the radio contours start at 3\,$\sigma$ and rise by a factor of 2. The physical scale at the cluster redshift is indicated on top right, and the white/red $\times$ indicates the NED cluster position. } 
   \label{fig:J1840.6}%
\end{figure*}

\begin{figure*}
   \centering
   \includegraphics[width=0.498\textwidth]{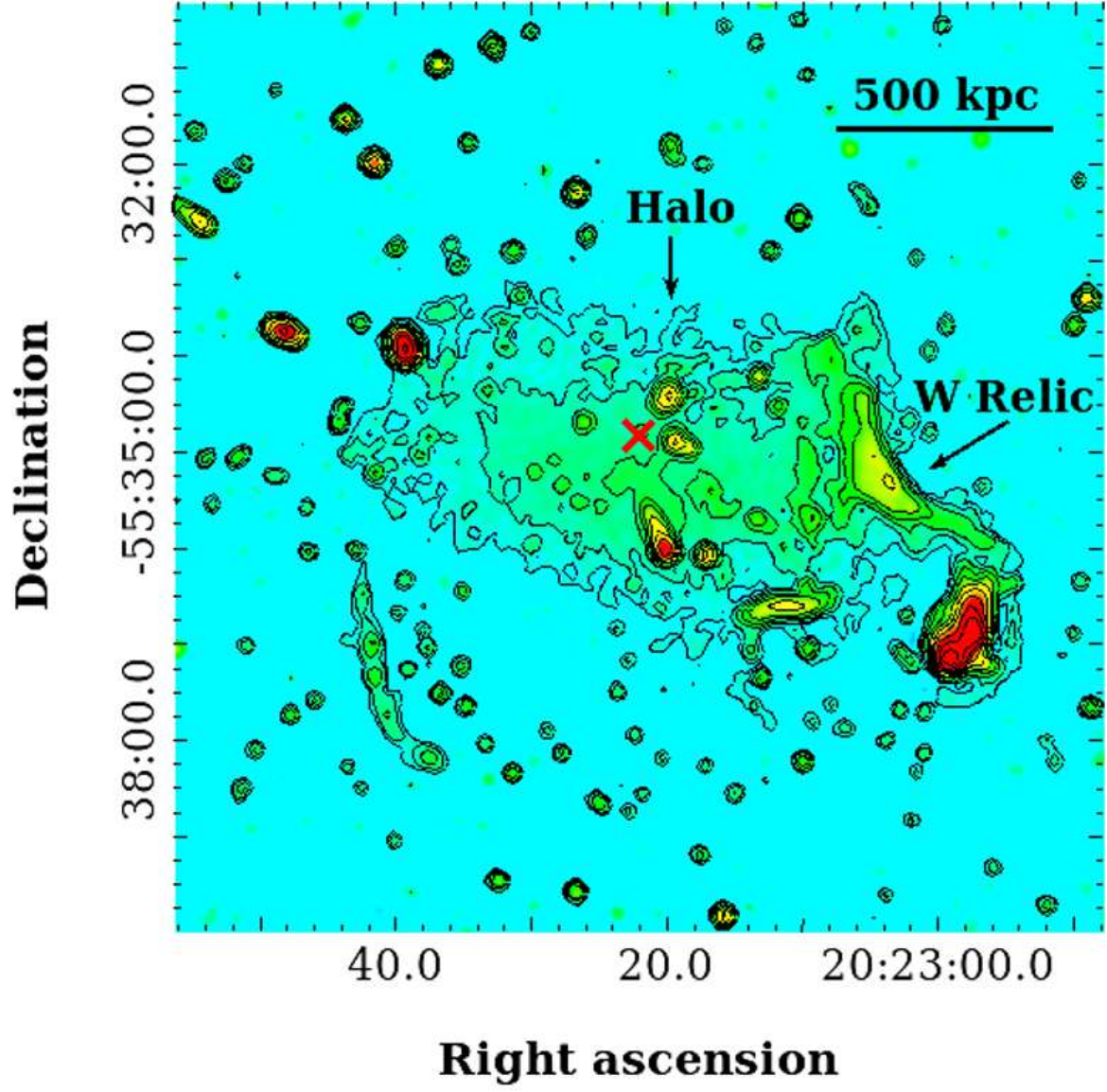}
    \includegraphics[width=0.488\textwidth]{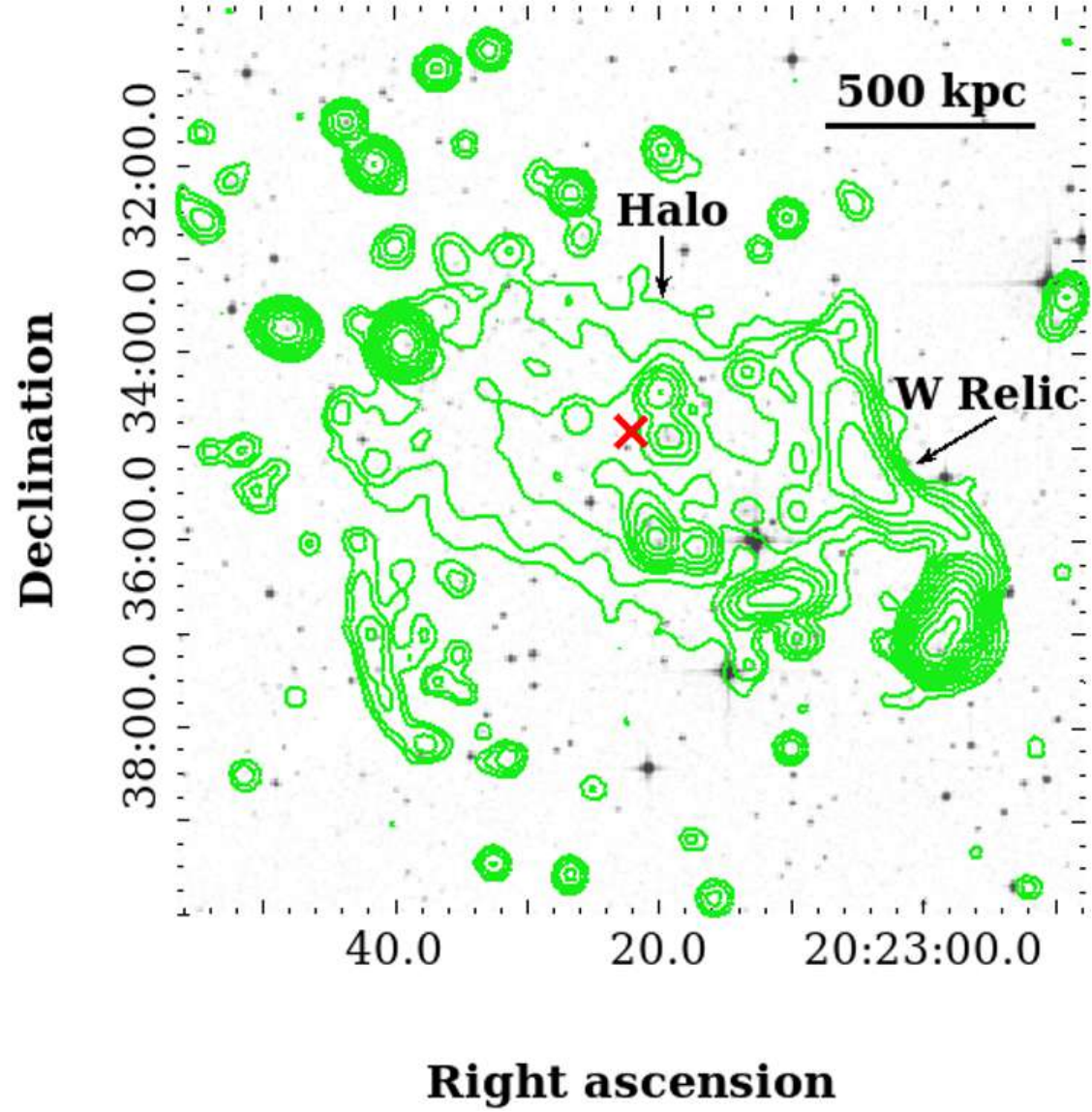}
   \caption{J2023.4$-$5535 \textbf{Left:}  Full-resolution (7.8\arcsec\,$\times$\,7.8\arcsec) 1.28~GHz MGCLS radio image with radio contours in black overlaid (1$\sigma$ = 3.5 $\mu$Jy~beam$^{-1}$). \textbf{Right:} 1.28~GHz MGCLS low-resolution  (15\arcsec\,$\times$\,15\arcsec) radio contours in green (1$\sigma$ = 7 $\mu$Jy~beam$^{-1}$), overlaid on the r-band \textit{Digitized Sky Survey (DSS)} optical image. In both panels, the radio contours start at 3\,$\sigma$ and rise by a factor of 2. The physical scale at the cluster redshift is indicated on top right, and the red $\times$ indicates the NED cluster position. } 
   \label{fig:J2023.4}%
\end{figure*}

\label{lastpage}

\end{document}